\documentstyle[12pt]{report}
\newtheorem{definition}{Definition}[chapter]
\newtheorem{theorem}{Theorem}[chapter]
\newtheorem{proposition}{Proposition}[chapter]
\newtheorem{lemma}{Lemma}[chapter]
\newtheorem{corolarry}{Corolarry}[chapter]
\input tcilatex

\begin{document}

\date{}

\begin{titlepage}
\vspace*{1mm}
\vbox{\begin{center} { HADRONIC PRESS MONOGRAPHS IN MATHEMATICS}
\end{center}}
\vspace*{2mm}
\vskip80pt
\vbox{\begin{center}
\Large{\bf Interactions, Strings and
Isotopies in Higher Order Anisotropic Superspaces}
\end{center}}
\vskip1.0cm
{\begin{center}
\bf Sergiu Ion Vacaru
\end{center}}
\vskip 0.5cm
{\begin{center}
Institute of Applied Physics \\
Academy of Sciences of Moldova \\
 5, Academy Street \\
Chi\c sin\v au MD2028 \\
Republic of Moldova
\end{center}}
\vskip 0.3cm
{\begin{center}
and \end{center}}
\vskip 0.3cm
{\begin{center}
The Institute for Basic Research \\
Box 1577 \\
Palm Harbor, FL 34682--1577
\end{center}}
\vskip 1cm
\vbox{\begin{center}{\large HADRONIC PRESS} \end{center}}
\newpage
{\bf Copyright\ \copyright\ 1998 by Sergiu Ion Vacaru}
\vskip 1cm
{\begin{center} \it All rights reserved. \\
No part of this book may be reproduced, stored in a retrieval system\\
or transmitted in any form or by any means without the written permision \\
of the copyright owner.\end{center}}
\vskip 2cm
{\bf U. \ S. Library of Congress

Cataloging in Publication Data:}
\vskip 1cm

Vacaru, Sergiu Ion
\vskip 0.5cm

{\it Interactions, Strings and Isotopies in Higher Order

Anisotropic Superspaces}
\vskip 0.5cm

Bibliography

\vskip 0.5cm
Additional data supplied on request

\vskip 0.5cm
I S B N \qquad 1 -- 5 7 4 8 5 -- 0 3 2 -- 6

\vskip 1cm

PRINTED IN THE U.S.A.

\vskip 1cm

The cover illustration is from the fronticpiece of Leibniz's

{\it Dissertation de arte combinatoria}

\newpage
{\begin{center} \bf ABOUT THE BOOK \end{center}}
This is the first monograph on the geometry of locally anisotropic phy\-sics.
The main subjects are in the theory of field interactions, strings and
diffusion processes on spaces, superspaces and isospaces with higher order
anisotropy and inhomogeneity. The approach proceeds by developing the
concept of higher order anisotropic (super)space which unify the
logical and manthematical aspects of modern Kaluza--Klein theories and
 generalized Lagrange and Finsler geometry and leads to modelling of
 physical processes on higher order fiber (super)bundles provided with
 nonlinear and distinguished connections and metric structures. The
view adopted is that a general field theory should incorporate all
 possible anisotropic and stochastic manifestations of classical and
 quantum interactions and, in consequence, a  corresponding modification
of basic principles and mathematical methods in formulation of
 physical theories. This book can be also considered as a pedagogical
 survey on the mentioned subjects.
\vskip 0.1cm
{\begin{center} \bf ABOUT THE AUTHOR \end{center}}
{\bf Sergiu Ion Vacaru} was born in 1958 in the Republic of Moldova.
He was educated at the Universities and Research Institutes of the
 former URSS (in Tomsk, Moscow, Dubna and Kiev). After receiving
 his Ph.D. in theoretical physics in 1994 at ''Al. I. Cuza'' University,
 Ia\c si, Romania, he was employed as scientific worker and associate
 professor at various academic institutions. At the present time he is
 senior researcher at the Institute of Applied Physics, Academy of
 Sciences of Moldova, and full professor of theoretical physics at
 the Institute for Basic Research (Palm Harbour, FL, U. S. A. and
 Molise, Italy). He is a member of the American Mathematical and
 Physical Societies, and a referee and editor of various International
 Scientific Journals. He has published more than hundred scientific
 works and communications on generalized Finsler and Kaluza--Klein
 (super)gravity and strings, differential (super)geometry, spinors
 and twistors, locally anisotropic stochastic processes and conservation
 laws in curved spaces, quantum field and geometric methods in
 condensed matter physics, isogeometry and isogravity, geometric
 thermodynamics and kinetics. Dr. Vacaru is also deeply involved in the
 problems of Human Rights in the post--soviet New Independent States.
\newpage
-
\vskip5cm
{\begin{center} \Large {\it In memory of Viorica P. Vacaru (Fagurel)}
  \end{center}}

\end{titlepage}
\tableofcontents
\newpage
\section{Preface}

This monograph gives a general geometric background of
the theory of field interactions, strings and diffusion processes on
spaces, superspaces and isospaces with higher order anisotro\-py and
 inhomogenity.
 The last fifteen years have been attempted
 a number of extensions of Finsler geometry with applications
 in theoretical and mathematical physics and biology which go far beyond the
 original scope. Our approach proceeds by  developing the concept of higher
 order anisotropic superspace which unify the logical and mathematical aspects
 of modern Kaluza--Klein theories and generalized Lagrange and Finsler
 geometry and leads to
 modelling of physical processes on higher order fiber bundles provided
 with nonlinear and distingushed connections and metric structures. The view
adopted here is that a general field theory should incorporate
 all possible anisotropic and stochastic manifestations of classical and
 quantum interactions and, in consequence, a corresponding modification
 of basic principles and mathematical methods in formulation of physical
 theories. This monograph can be also considered as a pedagogical survey on
 the mentioned subjects.

There are established
three approaches for modeling  field interactions and spaces
anisotropies. The first one is to deal with a usual locally isotropic
physical theory and to consider anisotropies as a consequence of the
anisotropic structure of sources in field equations (for instance, a number
of cosmological models are proposed in the framework of the Einstein theory
with the energy--momentum generated by anisotropic matter, as a general
reference see 
 [165]). The second approach to
anisotropies originates from the Finsler geometry
 [78,55,213,159] and its generalizations
 [17,18,19,160,161,13,29, 256, 255, 264, 272]
 with a general
imbedding into Kaluza--Klein (super) gravity and string theories
 [269,270,260,265,266,267],
and speculates a generic
anisotropy of the space--time structure and of fundamental field of
interactions. The Santilli's approach
 [217,219,220,218,221,\\
222,223,224] is more radical by proposing a
generalization of Lie theory and introducing isofields, isodualities and
related mathematical structures. Roughly speaking, by using corresponding
partitions of the unit we can model possible metric anisotropies as in
Finsler or generalized Lagrange geometry but the problem is also to take
into account classes of anisotropies generated by nonlinear and
distinguished connections.

In its present version this book addresses itself both to mathematicians and
physicists, to researches and graduate students which are interested in
geometrical aspects of fields theories, extended (super)gravity and
(super)strings and supersymmetric diffusion. It presupposes a general
background in the mentioned divisions of modern theoretical physics and
 assumes some familiarity with differential geometry, group theory, complex
 analysis and stochastic calculus.

The monograph is divided into three parts:

 The first five Chapters cover
the  higher order anisotropic supersymmetric theories:  Chapter 1
 is devoted to the geometry of higher order anisotropic supersaces with
an extension to supergravity models in Chapter 2. The supersymmetric nearly
autoparallel maps of superbundles and higher order anisotropic conservation
 laws are considered in Chapter 3. Higher order anisotropic superstrings
and anomalies are studied in Chapter 4. Chapter 5 contains an introduction
 into the theory of supersymmetric locally anisotropic stochastic processes.

 The next five Chapters are devoted to the (non supersymmetric) theory
of higher order anisotropic interactions and  stochastic processes. Chapter 6
 concerns the spinor formulation of field theories with locally
 anisotropic interactions and Chapter 7 considers anisotropic gauge field and
 gauge gravity models. Defining nearly autoparallel maps as generalizations
 of conformal transforms we analyze the problem of formulation of conservation
 laws in higher order anisotropic spaces in Chapter 8. Nonlinear sigma models
 and strings in locally anisotropic backgrounds are studied in Chapter 9.
 Chapter 10 is devoted to the theory of stochastic differential equations for
 locally anisotropic diffusion processes.

The rest four Chapters presents a study on Santilli's locally anisotropic
 and inhomogeneous isogeometries, namely, an introduction into the
 theory of isobuncles and generalized isofinsler gravity.
Chapter 11 is devoted to basic notations and definitions on Santilli and
coauthors isotheory. We introduce the bundle isospaces in Chapter 12 where
some necessary properties of Lie--Santilli isoalgebras and isogroups and
corresponding isotopic extensions of manifolds are applied in order to
define fiber isospaces and consider their such (being very important for
modeling of isofield interactions) classes of principal isobundles and
vector isobundles. In that Chapter there are also studied the
 isogeometry of nonlinear isoconnections in vector
 isobundles, the isotopic distinguishing of geometric
objects, the isocurvatures and isotorsions of nonlinear and distinguished
 isoconnections and the
structure equations and invariant conditions.  The
next Chapter 13 is devoted to the isotopic extensions of generalized Lagrange
and Finsler geometries. In Chapter 14 the isofield equations of locally
 anisotropic and inhomogeneous interactions will be analyzed and  an outlook
 and conclusions will be presented.

We have not attempted to give many details on previous knowledge of the
subjects or complete list of references. Each Chapter contains a brief
introduction, the first section  reviews the basic results, original papers
and monographs. If it is considered necessary, outlook and discussion are
presented at the end of the Chapter.

We hope that the reader will not suffer too much from our insufficient
 mastery of the English language.
\vskip15pt

{\bf Acknowledgments.}

 It is a pleasure for the author to give many thanks
especially to Professors R. Miron, M.Anastasiei, R. M. Santilli and
 A. Bejancu  for valuable discussions, collaboration
 and necessary offprints.

 The warmest thanks are extended to Drs
  E.  Seleznev  and L. Konopko for their help and support.

The author wish to express generic thanks to the referees for a detailed
control and numerous constructive suggestions as well he
should like to express his deep gratitude to the publishers for their
unfailing support.
\vskip 10pt

{ {\it Sergiu I. Vacaru,}}\ June 1997;\
 e-mail: vacaru@lises.asm.md

\vskip5pt

{\sl Academy of Sciences, Institute of Applied Physics,}

{\sf Chi\c sin\v au MD2028,} {\bf Republic of Moldova}

and

{\sl Institute for Basic Research,}

{\sf Palm Harbor,} {\bf USA}\ \& \ {\sf Molise,} {\bf Italy}

\newpage

\section{Notation}

(1) {\it Indices.} Unfortunately it is impossible to elaborate a general
system of labels of geometrical objects and coordinates when we a dealing
with geometrical constructions in Finsler spaces and theirs higher order
anisotropic, inhomogeneous and isotopic extensions.
 A number of various superspace, spinor, gauge field
and another indices add complexity to this problem. The reader will have to
consult the first sections in every Chapter in order to understand the
meaning of various types of boldface and/or underlined Greek or Latin
letters for operators, distinguished spinors and tensors. Here we present
only the most important denotations for local parametizations of spaces,
superspaces  and isospaces which are used in our book.

The first Part of this monograph is devoted to higher order anisotropic
superspaces (s--space). The base $G^\infty $--supermanifold $\widetilde{M}$
of s--spaces is locally parametrized by coordinates $x=(\widehat{x},\theta
)=\{x^I=(\widehat{x}^i,\theta ^{\widehat{i}})\},$ with a corresponding
splitting of indices of type $I=(i,\widehat{i}),J\left( j,\widehat{j}\right)
,...,$ where $i,j,...=1,2,...,n,\theta ^{\widehat{i}}$ are anticommuting
coordinates and $\widehat{i}=1,2,...,k$ ($\left( n,k\right) $ is the
(even,odd) dimension of s--space.

For typical vector superspaces (vs--spaces) ${\cal F}$ of dimension $\left(
m,l\right) $ we shall use local coordinates $y=(\widehat{y},\zeta )=\{y^A=(%
\widehat{y}^a,\zeta ^{\widehat{a}})\},$ with splitting of indices of type $%
A=(a,\widehat{a}),B=(b,\widehat{b}),$ where $a,b,...=1,2,...,m$ and $%
\widehat{a},\widehat{b},...=1,2,...,l,\zeta ^{\widehat{a}}$ are
anticommuting coordinates. There are also used sets of vs--spaces $$%
{\cal F}_1,{\cal F}_2,...,{\cal F}_p,...,{\cal F}_z\  (p=1,2,...,z)$$
prametrized correspondingly by coordinates $$\{...,y_{(p)},...\}=\{...,(%
\widehat{y}_{(p)},\zeta _{(p)}),...\}=\{...,y_{(p)}^A=(\widehat{y}%
_{(p)}^a,\zeta _{(p)}^{\widehat{a}}),...\}$$ with splitting of indices of
type $$\{...,A_p,...\}=\{...,(a_p,\widehat{a}_p),...\},\{...,B_p,...\}=%
\{...,(b_p,\widehat{b}_p),...\},$$ where $a_p,b_p,...=1,2,...,m_p$ and $%
\widehat{a}_p,\widehat{b}_p,...=1,2,...,l_p$ for every value of $p.$

Distinguished vector superspaces (dvs--spaces) $\widetilde{{\cal E}}^{<z>}$
are parametrized by local coordinates
$$
u=\left( x=y_{(0)},y_{(1)},y_{(2)},...,y_{(p)},...,y_{(z)}\right) =\left(
u_{(p)},\widehat{y}_{(p+1)},\zeta _{(p+1),...,}\widehat{y}_{(z)},\zeta
_{(z)}\right) =
$$
$$
\left( \widehat{x}^i,x^{\widehat{i}},\widehat{y}^{<a>},y^{<\widehat{a}%
>}\right) =\left( x^I=y^{A_0},y^{A_1},...,y^{A_p},...,y^{A_z}\right) =
\left(x^I,y^{<A>}\right) =u^{<\alpha >} $$ $$=
\left( u^{<\alpha _p>},y^{A_{p+1}},...,y^{A_z}\right) =
\left( \widehat{x}^i=%
\widehat{y}^{a_0},\theta ^{\widehat{i}} =
\zeta ^{\widehat{a}_0},\widehat{y}%
^{a_1},\zeta ^{\widehat{a}_1},...,
\widehat{y}^{a_z},\zeta ^{\widehat{a}_z}\right)
$$

The terms distinguished superbundles, distinguished
 geometric objects, higher order anisotropic (super)bundle, locally
anisotropic (super)space and so on (geometrical constructions distinguished
 by a N--connection structure and characterized by corresponding
 local anisotropies) will be used in theirs respective brief forms
 (d--superbundles,  d--objects, ha--(super)bundle, la--superspace and so on).
(see subsections 1.1.1 and 2.1.1 for details).

The second Part of the monograph contains geometric constructions and
physical theories on higher order spaces (not superspaces) models on
distinguished vector bundles ${\cal E}^{<z>},$ dv--bundles (we omit
''tilde'' on geometric objects), with local coordinates%
$$
u=\left( x=y_{(0)},y_{(1)},y_{(2)},...,y_{(p)},...,y_{(z)}\right) =\left(
u_{(p)},\widehat{y}_{(p+1)},...,\widehat{y}_{(z)}\right) =
$$
$$
\left( \widehat{x}^i=x^i,\widehat{y}^{<a>}\right) =\left(
x^i=y^{a_0},y^{a_1},...,y^{a_p},...,y^{a_z}\right) =$$ $$\left(
x^i,y^{<a>}\right) =u^{<\alpha >}=\widehat{u}^{<\alpha >}=
\left( u^{<\alpha _p>}= \widehat{u}^{<\alpha
_p>},y^{a_{p+1}},...,y^{a_z}\right) =$$
$$\left( \widehat{x}^i=\widehat{y}^{a_0},%
\widehat{y}^{a_1},...,\widehat{y}^{a_p},...,\widehat{y}^{a_z}\right)
$$
(in the first Part of the book we shall use ''hats'' on geometrical objects
in order to emphasize that we consider just the components of the even part
of s--space; as a rule we shall omit such ''hats'', but maintain cumulative
Greek indices of type $<\alpha >=(i=a_0,a_1,...,a_p,...a_z),$ in the second
Part of our work were we are dealing with the geometric background base on
vector bundles).

The underlined indices will be used for marks enumerating components with
respect to distinguished locally adapted to N--connection frame
decompositions; for instance, with respect to frames (for superspaces one
uses the term supervielbeins, in brief, s--vielbeins) $l_{<\alpha >}^{<%
\underline{\alpha }>}(u)$ we shall consider decompositions of metric of type%
$$
G_{<\alpha ><\beta >}(u)=l_{<\alpha >}^{<\underline{\alpha }>}(u)l_{<\beta
>}^{<\underline{\beta }>}(u)G_{<\underline{\alpha }><\underline{\beta }>}
$$
with a corresponding splitting of indices $<\underline{\alpha }>=(\underline{%
I}=\underline{A}_0,\underline{A}_1,...,\underline{A}_p,...,\underline{A}_z).$
Different types of underlined indices for Maiorana, or Dirac spinors and
objects of higher order anisotropic Clifford and (super)vector fibrations
(defined for corresponding locally adapted spinor or s--vector bases) will
be also considered.

We shall use such type of spinor parametrizations of anticommuting variables
in higher order anisotropic supergravity:%
$$
\theta ^{\underline{i}}=\zeta ^{\underline{a}_0}=(\theta ^{\underleftarrow{i}%
}=\zeta ^{\underleftarrow{a_0}},\theta ^{\underrightarrow{i}}=\zeta ^{%
\underrightarrow{a_0}}),\zeta ^{\underline{a}_1}=(\zeta ^{\underleftarrow{a_1%
}},\zeta ^{\underrightarrow{a_1}}),...,%
$$
$$
\zeta ^{\underline{a}_p}=(\zeta ^{\underleftarrow{a_p}},\zeta ^{%
\underrightarrow{a_p}}),...,\zeta ^{\underline{a}_z}=(\zeta ^{%
\underleftarrow{a_z}},\zeta ^{\underrightarrow{a_z}}).
$$

The two dimensional world sheets used in the theory of higher order
an\-isot\-rop\-ic (super)strings are provided with local coordinates $%
z=z^{\ddot u},$ two--metric $\gamma _{\ddot a\ddot e}$ and zweibein $%
e_{\ddot e}^{\underline{\ddot e}},$ where $\ddot u,\ddot e,\ddot a,...=1,2.$
The Dirac matrices on the worlds sheets are denoted $\gamma _{\ddot e};$ we
shall also use matrix $\gamma _5$ with the property $\left( \gamma _5\right)
^2=-1$ (to follow usual denotations we use $\gamma $-index 5 as well for two
dimensions). Maiorana spinors are denoted as $\theta _{\tilde a},\chi
_{\tilde n}.$

On (1,0)--superspaces we shall use two Bose coordinates $\left( z^{\pm
},z^{=}\right) $ and one Fermi coordinate $\theta ^{+},~\ddot u=(z^{\pm
},z^{=},\theta ^{+}),$ (see details in subsection 4.1,2); standard
derivatives on (1,0) locally anisotropic superspaces are denoted $D_{\ddot
A}=(D_{+},\partial _{\ddagger },\partial _{=}),$ where index $\ddot
A=(+,\ddagger ,=);$ s--vielbeins are written $E_{\ddot A}^{\underline{\ddot U%
}}.$ As $\Psi ^{|\underline{I}|},$ with indices of type $|\underline{I}|,|%
\underline{J}|,...,$ we shall denote heterotic Maiorana--Weyl fermi\-ons (see
section 4).

The Santilli's isotheory considered in the Part III of that monograph
 is based on the concept of fundamental  isotopy
which is the lifting $I\rightarrow \widehat{I}$ of the $n$--dimensional unit
$I=diag\left( 1,1,...,1\right) $ of the Lie's theory into an $n\times n$%
--dimensional matrix
$$
\widehat{I}=\left( I_j^i\right) =\widehat{I}\left( t,x,\dot x,\ddot x,\psi
,\psi ^{+},\partial \psi ,\partial \psi ^{+},\partial \partial \psi
,\partial \partial \psi ^{+},...\right)
$$
called the isounit. "Hats" on symbols in the third part will be used for
 distinguishing the isotopic objects from usual ones (non isotopic).

(2){\it Equations.} For instance, equation (4.6) is the 6th equation in
Chapter 4, so the first number of an equation points to the Chapter where
 that equation is introduced and the second numbers enumerates the equation
 into consideration with respect to that Chapter. The same holds true
 for theorems, definitions and so on (for example, Theorem 3.1; Definition
 4.5).

(3) {\it Differentiation.} Ordinary partial differentiation with respect to
a coordinate $x^i$ will either be denoted by the operator ${\partial}_i$ or
by subscript $i$ following a comma, for instance, ${\frac{\partial A^i }{%
\partial x^j}} \equiv \partial _j A^i \equiv {A^i}_{,j} .$ We shall use the
denotation ${\frac{\delta A^i }{\delta x^j}} \equiv \delta _j A^i $ for
partial derivations locally adapted to a nonlinear connection structure.

(4) {\it Summation convention.} We shall follow the Einstein summation rule
for spinor and tensor indices.

(5) {\it References. } In the bibliography we cite the scientific journals
in a generally accepted abbreviated form, give the volume, the year and the
first page of the authors' articles; the monographs and collections of works
are cited completely. For the author's works and communications, a part of
them been published in journals and collection of papers from Countries of
Easten Europe, the extended form (with the titles of articles and
communications) is presented.

(6) {\it Introductions and Conclusions.} As a rule every Chapter starts with
an introduction into the subject and ends, if it is necessary,
 with concluding remarks.

(7) {\it Terminology ambiguities.} \
 We emphasize that the term ''higher order'' is used as a
 general one for higher order tangent bundles 
 [295], or
 higher order extensions of vector superbundles
 [270,260,265,266,267], in a number of lines
 alternative to jet bundles 
 [226,225], and only
 under corresponding constraints one obtains the geometry of higher order
 Lagrangians 
 [162]).

(8) {\it The end of Proofs} are denoted by $\Box .$


\part{ Higher Or\-der Anisotropic Supersymmetry}

\chapter{HA--Superspaces}

The differential supergeometry have been formulated with the aim
of getting a geometric framework for the supersymmetric field
theories (see the theory
of graded manifolds 
 [37,146,147,144], the theory of supermanifolds
 [290,203,27,127] and, for detailed considerations of geometric and
topological aspects of supermanifolds and formulation of
superanalysis,
 [63,49,157,\\ 114,281,283]). In this Chapter we apply the supergeometric
formalism for a study of a new class of (higher order anisotropic)
superspaces.

The concept of local anisotropy is largely used in some divisions
of
theoretical and mathematical physics 
 [282,119,122,163] (see also
 possible applications in physics and biology in
 [14,13]). The first
models of locally anisotropic (la) spaces (la--spaces) have been
proposed by
P.Finsler
 [78] and E.Cartan
 [55] (early approaches and modern
treatments of Finsler geometry and its extensions can be found,
for
instance, in
 [213,17,18,159]). In our works
 [256,255,258,259,260,264,272,279,276] we try to formulate the
geometry of la--spaces in a manner as to include both variants of
Finsler and Lagrange, in general supersymmetric, extensions and
higher dimensional Kaluza--Klein (super)spaces as well to propose
general principles and
methods of construction of models of classical and quantum field
interactions and stochastic processes on spaces with generic
anisotropy.

We cite here the works
 [31,33] by A. Bejancu where a new
viewpoint on differential geometry of supermanifolds is
considered. The author introduced the nonlinear connection
(N--connection) structure and
developed a corresponding distinguished by N--connection
supertensor
covariant differential calculus in the frame of De Witt 
 [290] approach
to supermanifolds in the framework of the geometry of
superbundles with typical fibres parametrized by noncommutative
coordinates. This was the first example of superspace with local
anisotropy. In our turn we have given a general definition of
locally anisotropic superspaces (la--superspaces)
 [260]. It should be noted here that in our supersymmetric
generalizations  we were inspired by the R. Miron, M. Anastasiei
and Gh. Atanasiu works on the geometry of nonlinear connections
in vector bundles
and higher order La\-gran\-ge spaces 
 [160,161,162].
In this Chapter we shall formulate the theory of higher order
vector and tangent superbundles provided with nonlinear and
distinguished connections
and metric structures (a generalized model of la--superspaces).
Such superbundles contain as particular cases the supersymmetric
extensions and various higher order prolongations of Riemann,
Finsler and Lagrange spaces. We shall use instead the terms
distinguished superbundles, distinguished
 geometric objects and so on (geometrical constructions distinguished
 by a N--connection structure) theirs corresponding brief denotations
 (d--superbundles,  d--objects and so on).

The Chapter is organized as follows: Section 1.1 contains a brief
review on supermanifolds and superbundles and an introduction
into the geometry of higher order distinguished vector
superbundles. Section 1.2 deals with the geometry of nonlinear
and linear distinguished connections in vector superbundles and
distinguished vector superbundles. The geometry of the total
space of distinguished vector superbundles is studied in section
1.3; distinguished connection and metric structures,  their
torsions, curvatures and structure equations are considered.
Generalized Lagrange and Finsler superspaces there higher order
prolongations are  defined in section 1.4 . Concluding remarks on
Chapter 1 are contained in section 1.5.

\section{Supermanifolds and D--Superbundles}

In this section we establish the necessary terminology on
supermanifolds (s--manifolds)
 [290,203,204,127,281,114,157,27,49,63] and
present an introduction into the geometry of distinguished vector
superbundles (dvs--bundles) 
 [265]. Here we note that a number of
different approaches to supermanifolds are broadly equivalent for
local considerations. For simplicity, we shall restrict our study
only with geometric constructions on locally trivial superspaces.

\subsection{Supermanifolds and superbundles}

To build up s--manifolds (see 
 [203,127,281]) one uses as basic
structures \ Grassmann algebra and Banach space. A Grassmann
algebra is
introduced as a real associative algebra $\Lambda $ (with unity)
possessing
a finite (canonical) set of anticommutative generators $\beta _{\hat A}$, ${{%
[{\beta _{\hat A}},{\beta _{\hat B}}]}_{+}}={{\beta _{\hat
A}}}{{\beta
_{\hat C}}}+{{\beta _{\hat C}}}{{\beta _{\hat A}}}=0$, where ${{\hat A},{%
\hat B},...}=1,2,...,{\hat L}$. In this case it is also defined a ${Z_2}$%
-graded commutative algebra ${{\Lambda }_0}+{{\Lambda }_1}$,
whose even part
${{\Lambda }_0}$ (odd part ${{\Lambda }_1}$) is a ${2^{{\hat L}-1}}$%
--dimensional real vector space of even (odd) products of
generators ${\beta
}_{\hat A}$.After setting ${{\Lambda }_0}={\cal R}+{{\Lambda }_0}^{\prime }$%
, where ${\cal R}$ is the real number field and ${{\Lambda
}_0}^{\prime }$ is the subspace of ${\Lambda }$ consisting of
nilpotent elements, the
projections ${\sigma }:{\Lambda }\to {\cal R}$ and $s:{\Lambda }\to {{%
\Lambda }_0}^{\prime }$ are called, respectively, the body and
soul maps.

A Grassmann algebra can be provided with both structures of a
Banach algebra
and Euclidean topological space by the norm 
 [203]
$$
{\Vert }{\xi }{\Vert }={{\Sigma }_{{\hat A}_i}}{|}a^{{{\hat A}_1}...{{\hat A}%
_k}}{|},{\xi }={{\Sigma }_{r=0}^{\hat L}}a^{{{\hat A}_1}...{{\hat A}_r}}{{%
\beta }_{{\hat A}_1}}...{{\beta }_{{\hat A}_r}}.
$$
A superspace is introduced as a product
$$
{\Lambda }^{n,k}={\underbrace{{{\Lambda }_0}{\times }...{\times }{{\Lambda }%
_0}}_n{\times }{\underbrace{{{\Lambda }_1}{\times }...{\times }{{\Lambda }_1}%
}_k}}
$$
which is the $\Lambda $-envelope of a $Z_2$-graded vector space ${V^{n,k}}={%
V_0}{\otimes }{V_1}={{\cal R}^n}\oplus {{\cal R}^k}$ is obtained
by
multiplication of even (odd) vectors of $V$ on even (odd) elements of ${%
\Lambda }$. The superspace (as the ${\Lambda }$-envelope) posses
$(n+k)$
basis vectors $\{{\hat {{\beta }_i}},{\quad }i=0,1,...,n-1,$ and ${\quad }{{%
\beta }_{\hat i}},{\quad }{\hat i}=1,2,...k\}$. Coordinates of
even (odd) elements of $V^{n,k}$ are even (odd) elements of
$\Lambda $. We can consider
equivalently a superspace $V^{n,k}$ as a $({2^{{\hat L}-1}})(n+k)$%
-dimensional real vector spaces with a basis $\{{{\hat \beta }_{i({\Lambda }%
)}},{{\beta }_{{\hat i}({\Lambda })}}\}$.

Functions of superspaces, differentiation with respect to
Grassmann coordinates, supersmooth (superanalytic) functions and
mappings are introduced by analogy with the ordinary case, but
with a glance to certain specificity caused by changing of real
(or complex) number field into
Grassmann algebra $\Lambda $. Here we remark that functions on a superspace $%
{\Lambda }^{n,k}$ which takes values in Grassmann algebra can be
considered as mappings of the space ${\cal R}^{{({2^{({\hat
L}-1)}})}{(n+k)}}$ into the space ${\cal R}^{2{\hat L}}$.
Functions differentiable on Grassmann coordinates can be
rewritten via derivatives on real coordinates, which obey a
generalized form of Cauchy-Riemann conditions.

A $(n,k)$-dimensional s--manifold $\tilde M$ can be defined as a
Banach
manifold (see, for example, 
 [148]) modeled on ${\Lambda }^{n,k}$
endowed with an atlas ${\psi }={\{}{U_{(i)}},{{\psi }_{(i)}}:{U_{(i)}}\to {{%
\Lambda }^{n,k}},(i)\in J{\}}$ whose transition functions ${\psi
}_{(i)}$
are supersmooth 
 [203,127]. Instead of supersmooth functions we can
use $G^\infty $-functions 
 [203] and introduce $G^\infty $%
-supermanifolds ($G^\infty $ denotes the class of
superdifferentiable functions). The local structure of a
$G^\infty $--supermanifold is built very
much as on a $C^\infty $--manifold. Just as a vector field on a $n$%
-dimensional $C^\infty $-manifold written locally as%
$$
{\Sigma }_{i=0}^{n-1}{\quad }{f_i}{(x^j)}{\frac{{\partial }}{{{\partial }x^i}%
}},
$$
where $f_i$ are $C^\infty $-functions, a vector field on an $(n,k)$%
--dimensional $G^\infty $--su\-per\-ma\-ni\-fold $\tilde M$ can
be expressed locally on an open region $U{\subset }\tilde M$ as
$$
{\Sigma }_{I=0}^{n-1+k}{\quad }{f_I}{(x^J)}{\frac{{\partial }}{{{\partial }%
x^I}}}= {\Sigma }_{i=0}^{n-1}{\quad }{f_i}{(x^j,{{\theta }^{\hat j}})}{\frac{%
{\partial }}{{{\partial }x^i}}}+{\Sigma }_{{\hat i}=1}^k{\quad }{f_{\hat i}}{%
(x^j,{{\theta }^{\hat j}})}{\frac \partial {\partial {{\theta
}^{\hat i}}}},
$$
where $x=({\hat x},{\theta })=\{{x^I}=({{\hat x}^i},{\theta
}^{\hat i})\}$
are local (even, odd) coordinates. We shall use indices $I=(i,{\hat i}),J=(j,%
{\hat j}),K=(k,{\hat k}),...$ for geometric objects on $\tilde
M$. A vector field on $U$ is an element $X{\subset }End[{G^\infty
}(U)]$ (we can also consider supersmooth functions instead of
$G^\infty $-functions) such that
$$
X(fg)=(Xf)g+{(-)}^{{\mid }f{\mid }{\mid }X{\mid }}fXg,
$$
for all $f,g$ in $G^\infty (U)$, and
$$
X(af)={(-)}^{{\mid }X{\mid }{\mid }a{\mid }}aXf,
$$
where ${\mid }X{\mid }$ and ${\mid }a{\mid }$ denote
correspondingly the parity $(=0,1)$ of values $X$ and $a$ and for
simplicity in this work we
shall write ${(-)}^{{\mid }f{\mid }{\mid }X{\mid }}$ instead of ${(-1)}^{{%
\mid }f{\mid }{\mid }X{\mid }}.$

A super Lie group (sl--group) 
 [204] is both an abstract group and a
s--manifold, provided that the group composition law fulfills a
suitable smoothness condition (i.e. to be superanalytic, for
short, $sa$
 [127]).

In our further considerations we shall use the group of automorphisms of ${%
\Lambda}^{(n,k)}$, denoted as $GL(n,k,{\Lambda})$, which can be
parametrized as the super Lie group of invertible matrices
$$
Q={\left(
\begin{array}{cc}
A & B \\
C & D
\end{array}
\right) } ,%
$$
where A and D are respectively $(n{\times}n)$ and $(k{\times}k)$
matrices consisting of even Grassmann elements and B and C are
rectangular matrices consisting of odd Grassmann elements. A
matrix Q is invertible as soon as maps ${\sigma}A$ and
${\sigma}D$ are invertible matrices. A sl-group
represents an ordinary Lie group included in the group of linear transforms $%
GL(2^{{\hat L}-1}(n+k),{\cal R})$. For matrices of type Q one
defines
 [37,146,147] the superdeterminant, $sdetQ$, supertrace, $strQ$, and
superrank, $srankQ$.

A Lie superalgebra (sl--algebra) is a
$Z_2$-graded algebra $A={A_0}\oplus A_1$ endowed with product
$[,\}$ satisfying the following properties:
$$
[I,I^{\prime }\}=-{(-)}^{{\mid }I{\mid }{\mid }I^{\prime }{\mid
}}[I^{\prime },I\},
$$
$$
[I,[I^{\prime },I^{\prime \prime }\}\}=[[I,I^{\prime
}\},I^{\prime \prime }\}+{(-)}^{{\mid }I{\mid }{\mid }I^{\prime
}{\mid }}[I^{\prime }[I,I^{\prime \prime }\}\},
$$
$I{\in }A_{{\mid }I{\mid }},{\quad }I^{\prime }{\in }A_{{\mid }I^{\prime }{%
\mid }}$, where ${\mid }I{\mid },{\mid }I^{\prime }{\mid }=0,1$
enumerates, respectively, the possible parity of elements
$I,I^{\prime }$. The even part $A_0$ of a sl-algebra is a usual
Lie algebra and the odd part $A_1$ is a
representation of this Lie algebra. This enables us to classify
sl--algebras
following the Lie algebra classification 
 [128]. We also point out that
irreducible linear representations of Lie superalgebra A are realized in $%
Z_2 $-graded vector spaces by matrices $\left(
\begin{array}{cc}
A & 0 \\
0 & D
\end{array}
\right) $ for even elements and $\left(
\begin{array}{cc}
0 & B \\
C & 0
\end{array}
\right) $ for odd elements and that, roughly speaking, A is a
superalgebra of generators of a sl--group.

A sl--module $W$ (graded Lie module) 
 [203] is introduced as a
$Z_2$--graded left $\Lambda $-module endowed with a product
 $[,\}$ which satisfies
the graded Jacobi identity and makes $W$ into a
graded-anticommuta\-ti\-ve Banach algebra over $\Lambda $. One
calls the Lie module {\cal G} the set of
the left-invariant derivatives of a sl--group $G$.

The tangent superbundle (ts-bundle) $T\tilde M$ over a s-manifold $\tilde M$%
, ${\pi }:T\tilde M\to {\tilde M}$ is constructed in a usual
manner (see,
for instance, 
 [148]) by taking as the typical fibre the superspace ${%
\Lambda }^{n,k}$ and as the structure group the group of
automorphisms, i.e. the sl-group $GL(n,k,{\Lambda }).$

Let us denote by ${\cal F}$ a vector superspace ( vs--space ) of dimension $%
(m,l)$ (with respect to a chosen base we parametrize an element $y\in {\cal E%
}$ as $y=({\hat y},\zeta )=\{{y^A}=({\hat {y^a}},{\zeta }^{\hat
a})\}$,
where $a=1,2,...,m$ and ${\hat a}=1,2,...,l$). We shall use indices $A=(a,{%
\hat a}),B=(b,{\hat b}),...$ for objects on vs--spaces. A vector
superbundle
(vs--bundle) $\tilde {{\cal E}}$ over base $\tilde M$ with total superspace $%
\tilde E$, standard fibre ${\hat {{\cal F}}}$ and surjective projection ${{%
\pi }_E}:\tilde E{\to }\tilde M$ is defined (see details and
variants in
 [49,283]) as in the case of ordinary manifolds (see, for instance,
 [148,160,161]). A section of $\tilde {{\cal E}}$ is a supersmooth map
$s:U{\to }\tilde E$ such that ${{\pi }_E}{\cdot }s=id_U.$

A subbundle of ${\tilde {{\cal E}}}$ is a triple $(\tilde {{\cal B}%
},f,f^{\prime})$, where $\tilde {{\cal B}}$ is a vs--bundle on
$\tilde M$,
maps $f: \tilde {{\cal B}} {\to} \tilde {{\cal E}}$ and $f^{\prime}: \tilde M%
{\to} \tilde M$ are supersmooth, and $(i) {\quad}{{\pi}_E}{\circ}f=f^{\prime}%
{\circ}{{\pi}_B};$ $(ii) {\quad} f:{\pi}^{-1}_B {(x)} {\to} {\pi}^{-1}_E {%
\circ} f^{\prime}(x)$ is a vs--space homomorphism.
We denote by
$$
u=(x,y)=({\hat x},{\theta },{\hat y},{\zeta })=\{u^\alpha
=(x^I,y^A)=({{\hat
x}^i},{\theta }^{\hat i},{{\hat y}^a},{\zeta }^{\hat a})=({{\hat x}^i}%
,x^{\hat i},{{\hat y}^a},y^{\hat a})\}
$$
the local coordinates in ${\tilde {{\cal E}}}$ and write their
transformations as
$$
x^{I^{\prime }}=x^{I^{\prime }}({x^I}),{\quad }srank({\frac{{\partial }%
x^{I^{\prime }}}{{\partial }x^I}})=(n,k),\eqno(1.1)
$$
$y^{A^{\prime }}=Y_A^{A^{\prime }}(x){y}^A,$ where
$Y_A^{A^{\prime }}(x){\in }G(m,l,\Lambda ).$

For local coordinates and geometric objects on ts-bundle $T\tilde
M$ we shall not distinguish indices of coordinates on the base
and in the fibre and write, for instance,
$$
u=(x,y)=({\hat x},{\theta },{\hat y},{\zeta })=\{u^\alpha =({x^I},{y^I})=({{%
\hat x}^i},{\theta }^{\hat i},{{\hat y}^i},{\zeta }^{\hat i})=({{\hat x}^i}%
,x^{\hat i},{{\hat y}^i},y^{\hat i})\}.
$$
We shall use Greek indices for marking local coordinates on both
s--vector and usual vector bundles (see the second part of the
monograph).

\subsection{Distinguished vector superbundles}
Some recent considerations in mathematical physics are based on
the so--called k--jet spaces (see, for instance,
 [226,225,19]). In order
to formulate a systematic theory of connections and of geometric
structures on k--jet bundles,
 in a manner following the approaches 
 [295] and 
 [160,161] R. Miron and Gh. Atanasiu 
 [162] introduced the concept
of k--osculator bundle for which a fiber of k-jets is changed
into a
k--osculator fiber representing an element of k--order curve. Such
considerations are connected with geometric constructions on
tangent bundles of higher order. On the other hand for
developments in modern supersymmetric
Kaluza--Klein theories (see, for instance, 
 [215]) a substantial
interest would present a variant of ''osculator'' space for which
the higher order tangent s--space distributions are of different
dimensions. The second part of this section is devoted to the
definition of such type distinguished vector superbundle spaces.

A vector superspace ${\cal F}^{<z>}$ of dimension $(m,l)$ is a
distinguished
vector superspace ( dvs--space ) if it is decomposed into an
invariant oriented
direct sum ${\cal F}^{<z>}={\cal F}_{(1)}\oplus {\cal
F}_{(2)}\oplus
...\oplus {\cal F}_{(z)}$ of vs--spaces ${\cal F}_{(p)},\dim {\cal F}%
_{(p)}=(m_{(p)},l_{(p)}),$ where $(p)=(1),(2),...,(z),%
\sum_{p=1}^{p=z}m_{(p)}=m,\sum_{p=1}^{p=z}l_{(p)}=l.$

Coordinates on ${\cal F}^{<p>}$ will be parametrized as
$$
{y}^{<p>}=(y_{(1)},y_{(2)},...,y_{(p)})=({\hat y}_{(1)},{\zeta
}_{(1)},{\hat y}_{(2)},{\zeta }_{(2)},...,{\hat y}_{(p)},{\zeta
}_{(p)})=
$$
$$
\{y^{<A>}=({{\hat y}^{<a>}},{\zeta }^{<\hat a>})=({{\hat
y}^{<a>}},y^{<\hat a>})\},
$$
where bracketed indices are correspondingly split on ${\cal F}_{(p)}$%
--components:
$$
<A>=\left( A_{(1)},A_{(2)},...,A_{(p)}\right)
,<a>=(a_{(1)},a_{(2)},...,a_{(p)})
$$
$$
\mbox{ and }<\widehat{a}>=(\widehat{a}_{(1)}\widehat{a}_{(2)},...,\widehat{a}%
_{(p)}),\eqno(1.2)
$$
For simplicity, we shall also write (1.2) as $<A>=\left(
A_1,A_2,...,A_p\right) ,<a>=(a_1,a_2,...,a_p)$ and $<\widehat{a}>=(\widehat{a%
}_1\widehat{a}_2,...,\widehat{a}_p)$ if this will give not rise to
ambiguities.

A distinguished vector superbundle ( dvs--bundle )\\ $\widetilde{{\cal E}}%
^{<z>}=(\tilde E^{<z>},\pi ^{<d>},{\cal F}^{<d>},\tilde M),$ with
surjective projection $\pi ^{<z>}:\tilde E^{<z>}\rightarrow
\tilde M,$ where $\tilde M$ and $\tilde E^{<z>}$ are respectively
base and total s--spaces and the dvs--space ${\cal F}^{<z>}$ is
the standard fibre.

A dvs--bundle $\widetilde{{\cal E}}^{<z>}$ is constructed as an
oriented set of vs--bundles $\pi ^{<p>}:\tilde E^{<p>}\rightarrow
\tilde E^{<p-1>}$ (with typical fiber ${\cal
F}^{<p>},p=1,2,...,z);$ $\tilde E^{<0>}=\tilde M.$ We shall use
index $z~(p)$ as to denote the total (intermediate) numbers of
consequent vs--bundle coverings of $\tilde M.$

Local coordinates on $\widetilde{{\cal E}}^{<p>}$ are denoted as%
$$
u_{(p)}=(x,y_{<p>})=(x,y_{(1)},y_{(2)},...,y_{(p)})=
$$
$$
({\hat x},{\theta },{\hat y}_{<p>},{\zeta }_{<p>})=({\hat
x},{\theta },{\hat
y}_{(1)},{\zeta }_{(1)},{\hat y}_{(2)},{\zeta }_{(2)},...,{\hat y}_{(p)},{%
\zeta }_{(p)})=
$$
$$
\{u^{<\alpha >}=(x^I,y^{<A>})=({{\hat x}^i},{\theta }^{\hat i},{{\hat y}%
^{<a>}},{\zeta }^{<\hat a>})=({{\hat x}^i},x^{\hat i},{{\hat y}^{<a>}}%
,y^{<\hat a>}) =$$
$$(x^I = y^{A_{0}}, y^{A_{1}},..., y^{A_{p}},...,y^{A_{z}}) \}$$
(in our further considerations we shall consider different
variants of splitting of indices of geometric objects).

Instead of (1.1) the coordinate transforms for dvs--bundles\\
$\{u^{<\alpha
>}=(x^I,y^{<A>})\}\rightarrow \{u^{<\alpha ^{\prime }>}=(x^{I^{\prime
}},y^{<A^{\prime }>})\}$ are given by recurrent maps:%
$$
x^{I^{\prime }}=x^{I^{\prime }}({x^I}),{\quad }srank({\frac{{\partial }%
x^{I^{\prime }}}{{\partial }x^I}})=(n,k),\eqno(1.3)
$$
$$
y_{(1)}^{A_1^{\prime }}=K_{A_1}^{A_1^{\prime }}(x){y}%
_{(1)}^{A_1},K_{A_1}^{A_1^{\prime }}(x){\in
}G(m_{(1)},l_{(1)},\Lambda ),
$$
$$
..................................................
$$
$$
y_{(p)}^{A_p^{\prime }}=K_{A_p}^{A_p^{\prime }}(u_{(p-1)}){y}%
_{(p)}^{A_p},K_{A_p}^{A_p^{\prime }}(u_{(p-1)}){\in }G(m_{(p)},l_{(p)},%
\Lambda ),
$$
$$
.................................................
$$
$$
y_{(z)}^{A_z^{\prime }}=K_{A_z}^{A_z^{\prime }}(u_{(z-1)}){y}%
_{(z)}^{A_z},K_{A_z}^{A_z^{\prime }}(u_{(z-1)}){\in }G(m_{(z)},l_{(z)},%
\Lambda ).
$$
In brief we write transforms (1.3) as
$$
x^{I^{\prime }}=x^{I^{\prime }}(x^I),~y^{<A^{\prime
}>}=K_{<A>}^{<A^{\prime }>}y^{<A>}.
$$
More generally, we shall consider matrices $K_{<\alpha
>}^{<\alpha ^{\prime }>}=(K_I^{I^{\prime }},K_{<A>}^{<A^{\prime
}>}),$ where $K_I^{I^{\prime }}\doteq \frac{\partial x^{I^{\prime
}}}{\partial x^I}.$

In consequence the local coordinate bases of the module of ds--vector fields $%
\Xi (\widetilde{{\cal E}}^{<z>}),$
$$
\partial _{<\alpha >}=(\partial _I,\partial _{<A>})=(\partial _I,\partial
_{(A_1)},\partial _{(A_2)},...,\partial _{(A_z)})=
$$
$$
\frac \partial {\partial u^{<\alpha >}}=(\frac \partial {\partial
x^I},\frac
\partial {\partial y_{(1)}^{A_1}},\frac \partial {\partial
y_{(2)}^{A_2}},...,\frac \partial {\partial
y_{(z)}^{A_z}})\eqno(1.4)
$$
(the dual coordinate bases are denoted as%
$$
d^{<\alpha
>}=(d^I,d^{<A>})=(d^I,d^{(A_1)},d^{(A_2)},...,d^{(A_z)})=
$$
$$
du^{<\alpha >}=(dx^I,dy^{(A_1)},dy^{(A_2)},...,dy^{(A_z)})\quad
)\eqno(1.5)
$$
are transformed as%
$$
\partial _{<\alpha >}=(\partial _I,\partial _{<A>})=(\partial _I,\partial
_{(A_1)},\partial _{(A_2)},...,\partial _{(A_z)})\rightarrow
\partial _{<\alpha >}=
$$
$$
(\partial _I,\partial _{<A>})=(\partial _I,\partial
_{(A_1)},\partial _{(A_2)},...,\partial _{(A_z)})
$$
$$
\frac \partial {\partial x^I}=K_I^{I^{\prime }}\frac \partial
{\partial x^{I^{\prime }}}+Y_{(1,0)I}^{A_1^{\prime }}\frac
\partial {\partial y_{(1)}^{A_1^{\prime
}}}+Y_{(2,0)I}^{A_2^{\prime }}\frac \partial {\partial
y_{(2)}^{A_2^{\prime }}}+...+Y_{(z,0)I}^{A_z^{\prime }}\frac
\partial {\partial y_{(z)}^{A_z^{\prime }}},\eqno(1.6)
$$
$$
\frac \partial {\partial y_{(1)}^{A_1}}=K_{A_1}^{A_1^{\prime
}}\frac
\partial {\partial y_{(1)}^{A_1^{\prime }}}+Y_{(2,1)A_1}^{A_2^{\prime
}}\frac \partial {\partial y_{(2)}^{A_2^{\prime
}}}+...+Y_{(z,1)A_1}^{A_z^{\prime }}\frac \partial {\partial
y_{(z)}^{A_z^{\prime }}},
$$
$$
\frac \partial {\partial y_{(2)}^{A_2}}=K_{A_2}^{A_2^{\prime
}}\frac
\partial {\partial y_{(2)}^{A_2^{\prime }}}+Y_{(3,2)A_2}^{A_3^{\prime
}}\frac \partial {\partial y_{(3)}^{A_3^{\prime
}}}+...+Y_{(z,2)A_2}^{A_z^{\prime }}\frac \partial {\partial
y_{(z)}^{A_z^{\prime }}},
$$
$$
........................................................
$$
$$
\frac \partial {\partial
y_{(z-1)}^{A_{z-1}}}=K_{A_{z-1}}^{A_{z-1}^{\prime }}\frac
\partial {\partial y_{(z-1)}^{A_{z-1}^{\prime
}}}+Y_{(z,z-1)A_{s-1}}^{A_z^{\prime }}\frac \partial {\partial
y_{(z)}^{A_z^{\prime }}},
$$
$$
\frac \partial {\partial y_{(z)}^{A_z}}=K_{A_z}^{A_z^{\prime
}}\frac
\partial {\partial y_{(z)}^{A_z^{\prime }}}.
$$

$Y$--matrices from (1.6) are partial derivations of corresponding
combinations of $K$--coefficients from coordinate transforms
(1.3),
$$
Y_{A_f}^{A_p^{\prime }}=\frac{\partial (K_{A_p}^{A_p^{\prime }}~y^{A_p})}{%
\partial y^{A_f}},~f<p.
$$

In brief we denote respectively ds--coordinate transforms of
coordinate bases
(1.4)\ and (1.5)\ as%
$$
\partial _{<\alpha >}=(K_{<\alpha >}^{<\alpha ^{\prime }>}+Y_{<\alpha
>}^{<\alpha ^{\prime }>})~\partial _{<\alpha ^{\prime }>}\mbox{ and }%
~d^{<\alpha >}=(K_{<\alpha ^{\prime }>}^{<\alpha >}+Y_{<\alpha
^{\prime }>}^{<\alpha >})d^{<\alpha ^{\prime }>},
$$
where matrix $K_{<\alpha >}^{<\alpha ^{\prime }>}$, its s-inverse $%
K_{<\alpha ^{\prime }>}^{<\alpha >}$, as well $Y_{<\alpha
>}^{<\alpha ^{\prime }>}$ and $Y_{<\alpha ^{\prime }>}^{<\alpha
>}$ are paramet\-riz\-ed according to (1.6). In order to
illustrate geometric properties of some of our transforms it is
useful to introduce matrix operators and to consider in explicit
form the parametrizations of matrices under consideration. For
instance, in operator form the transforms (1.6)
$$
{\bf \partial =}\widehat{{\bf Y}}{\bf \partial }^{\prime },
$$
are characterized by matrices of type
$$
{\bf \partial =}\partial _{<\alpha >}=\left(
\begin{array}{c}
\partial _I \\
\partial _{A_1} \\
\partial _{A_2} \\
... \\
\partial _{A_z}
\end{array}
\right) =\left(
\begin{array}{c}
\frac \partial {\partial x^I} \\
\frac \partial {\partial y_{(1)}^{A_1}} \\
\frac \partial {\partial y_{(2)}^{A_2}} \\
... \\
\frac \partial {\partial y_{(z)}^{A_z}}
\end{array}
\right) ,{\bf \partial }^{\prime }{\bf =}\partial _{<\alpha
^{\prime }>}=\left(
\begin{array}{c}
\partial _{I^{\prime }} \\
\partial _{A_1^{\prime }} \\
\partial _{A_2^{\prime }} \\
... \\
\partial _{A_z^{\prime }}
\end{array}
\right) =\left(
\begin{array}{c}
\frac \partial {\partial x^{I^{\prime }}} \\
\frac \partial {\partial y_{(1)}^{A_1^{\prime }}} \\
\frac \partial {\partial y_{(2)}^{A_2^{\prime }}} \\
... \\
\frac \partial {\partial y_{(z)}^{A_z^{\prime }}}
\end{array}
\right)
$$
and
$$
\widehat{{\bf Y}}{\bf =}\widehat{Y}_{<\alpha >}^{<\alpha ^{\prime
}>}=\left(
\begin{array}{ccccc}
K_I^{I^{\prime }} & Y_{(1,0)I}^{A_1^{\prime }} &
Y_{(2,0)I}^{A_2^{\prime }}
& ... & Y_{(z,0)I}^{A_z^{\prime }} \\
0 & K_{A_1}^{A_1^{\prime }} & Y_{(2,1)A_1}^{A_2^{\prime }} & ... &
Y_{(z,1)A_1}^{A_z^{\prime }} \\
0 & 0 & K_{A_2}^{A_2^{\prime }} & ... & Y_{(z,2)A_2}^{A_z^{\prime }} \\
... & ... & ... & ... & ... \\
0 & 0 & 0 & ... & K_{A_z}^{A_z^{\prime }}
\end{array}
\right) .
$$

We note that we obtain a supersimmetric generalization of the
Miron--Atanasiu 
 [162] osculator bundle
$\left( Osc^z\tilde M,\pi ,\tilde M\right) $ if the fiber space
in Definition 1.2 is taken to be a direct sum of $z$ vector
s-spaces of the same dimension $\dim {\cal F}=\dim
\widetilde{M},$ i.e. ${\cal F}^{<d>}={\cal F}\oplus {\cal
F}\oplus ...\oplus {\cal F}.$ In this case the $K$ and $Y$
matrices from (1.3) and (1.6)
satisfy identities:%
$$
K_{A_1}^{A_1^{\prime }}=K_{A_2}^{A_2^{\prime
}}=...=K_{A_z}^{A_z^{\prime }},
$$
$$
Y_{(1,0)A}^{A^{\prime }}=Y_{(2,1)A}^{A^{\prime
}}=...=Y_{(z,z-1)A}^{A^{\prime }},
$$
$$
.............................
$$
$$
Y_{(p,0)A}^{A^{\prime }}=Y_{(p+1,1)A}^{A^{\prime
}}=...=Y_{(z,z-1)A}^{A^{\prime }},\quad (p=2,...,z-1).
$$
For $s=1$ the $Osc^1\widetilde{M}$ is the ts--bundle
$T\widetilde{M}.$

Introducing projection $\pi _0^z\doteq \pi ^{<z>}:\widetilde{{\cal E}}%
^{<z>}\rightarrow \widetilde{M}$ we can also consider projections
$\pi
_{p_2}^{p_1}:\widetilde{{\cal E}}^{<p_1>}\rightarrow \widetilde{{\cal E}}%
^{<p_2>}~\quad (p_2<p_1)$ defined as
$$
\pi
_{s_2}^{s_1}(x,y^{(1)},...,y^{(p_1)})=(x,y^{(1)},...,y^{(p_2)}).
$$
The s-differentials $d\pi _{p_2}^{p_1}:T(\widetilde{{\cal E}}%
^{<p_1>})\rightarrow T(\widetilde{{\cal E}}^{<p_2>})$ of maps $\pi
_{p_2}^{p_1}$ in turn define vertical dvs-subbundles
$V_{h+1}=Kerd\pi
_h^{p_1}~(h=0,1,..,p_1-1)$ of the tangent dvs-bundle $T(\widetilde{{\cal E}}%
^{<z>})~$( the dvs-space $V_1=V$ is the vertical dvs-subbundle on $%
\widetilde{{\cal E}}^{<z>}.$ The local fibres of dvs-subbundles
$V_h$
determines this regular s-distribution $V_{h+1}:u\in \widetilde{{\cal E}}%
^{<z>}\rightarrow V_{h+1}(u)\subset T(\widetilde{{\cal
E}}^{<z>})$ for which one holds inclusions $V_z\subset
V_{z-1}\subset ...\subset V_1.\,$ The enumerated properties of
vertical dvs--subbundles are explicitly illustrated by
transformation laws (1.6) for distinguished local bases.

\section{Nonlinear Connections in DVS--Bund\-les}

The purpose of this section is to present an introduction into
geometry of the nonlinear connection structures in dvs--bundles.
The concept of nonlinear
connection ( N--connection ) was introduced in the fra\-me\-work
of Finsler geometry
 [56,55,134] (the global definition of N--connec\-ti\-on is
given in
 [26]). It should be noted here that the N--connection
( splitting ) field could play an important role in modeling
various variants of dynamical reduction from higher dimensional
to lower dimensional s--spaces with (or not) different types of
local anisotropy. In monographs
[160,161] there are contained detailed investigations of
geometrical properties of N-connection structures in v--bundles
and different generalizations of Finsler geometry and some
proposals (see Chapter XII in
 [160], written by S. Ikeda) on physical interpretation of
N--connection in the framework of ''unified'' field theory with
interactions nonlocalized by y--dependencies are discussed. We
emphasize that N--connection is a different geometrical object
from that introduced by using nonlinear realizations of gauge
groups and supergroups (see, for instance, the
collection of works on supergravity 
 [215] and approaches to gauge
gravity 
 [240,194]).To make the presentation to aid rapid assimilation we
shall have realized our geometric constructions firstly for
vs--bundles then we shall extend them for higher order
extensions, i.e. for general dvs--bundles.

\subsection{N-connections in vs--bundles}

Let consider the definitions of N--connection structure 
 [260] in a
vs-bundle $\tilde {{\cal E}}=(\tilde E,{\pi }_E,\tilde M)$ whose
type fibre is ${\hat {{\cal F}}}$ and ${{\pi }^T}:T\tilde {{\cal
E}}{\to }T\tilde M$ is the superdifferential of the map ${{\pi
}_E}$ (${{\pi }^T}$ is a
fibre-preserving morphism of the ts-bundle $(T\tilde {{\cal E}},{{\tau }_E}%
,\tilde M)$ to $\tilde E$ and of ts-bundle $(T\tilde M,{\tau
},\tilde M)$ to
$\tilde M$). The kernel of this vs-bundle morphism being a subbundle of $%
(T\tilde E,{{\tau }_E},\tilde E)$ is called the vertical subbundle over $%
\tilde {{\cal E}}$ and denoted by $V\tilde {{\cal E}}=(V\tilde E,{{\tau }_V}%
,\tilde E)$. Its total space is $V\tilde {{\cal E}}={{\bigcup
}_{u\in \tilde
{{\cal E}}}}{\quad }{V_u},{\quad }$ where ${V_u}={ker}{{\pi }^T},{\quad }{u{%
\in }\tilde {{\cal E}}}.$ A vector
$$
Y={Y^\alpha }{\frac{{\partial }}{{{\partial }{u^\alpha }}}}={Y^I}{\frac{{%
\partial }}{{{\partial }{x^I}}}}+{Y^A}{\frac{{\partial }}{{{\partial }{y^A}}}%
}={Y^i}{\frac{{\partial }}{{{\partial }{x^i}}}}+{Y^{\hat
i}}{\frac{{\partial
}}{{{\partial }{{\theta }^{\hat i}}}}}+{Y^a}{\frac{{\partial }}{{{\partial }{%
y^a}}}}+{Y^{\hat a}}{\frac \partial {\partial {{\zeta }^{\hat
a}}}}
$$
tangent to $\tilde {{\cal E}}$ in the point $u\in \tilde {{\cal
E}}$ is locally represented as
$$
(u,Y)=({u^\alpha },{Y^\alpha })=({x^I},{y^A},{Y^I},{Y^A})=({{\hat x}^i},{{%
\theta }^{\hat i}},{{\hat y}^a},{{\zeta }^{\hat a}},{{\hat Y}^i},{Y^{\hat i}}%
,{{\hat Y}^a},{Y^{\hat a}}).
$$

A nonlinear connection, N--connection, in vs--bundle $\tilde
{{\cal E}}$ is a splitting on the left of the exact sequence
$$
0{\longmapsto }{V\tilde {{\cal E}}}\stackrel{i}{\longmapsto }{T\tilde {{\cal %
E}}}{\longmapsto }{{T\tilde {{\cal E}}{/}V\tilde {{\cal E}}}}{\longmapsto }0,%
\eqno(1.7)
$$
i.e. a morphism of vs-bundles $N:T\tilde {{\cal E}}\in {V\tilde
{{\cal E}}}$ such that $N{\circ }i$ is the identity on $V\tilde
{{\cal E}}$.

The ker\-nel of the mor\-phism $N$ is called the hor\-i\-zon\-tal
sub\-bun\-dle and denoted by
$$
(H\tilde E,{{\tau }_E},\tilde E).
$$
From the exact sequence (1.7) one follows that N--connection
structure can be
equivalently defined as a distribution $\{{{{\tilde E}_u}\to {{H_u}\tilde E}}%
,{{T_u}\tilde E}={{H_u}\tilde E}{\oplus }{{V_u}\tilde E}\}$ on
$\tilde E$ defining a global decomposition, as a Whitney sum,
$$
T{\tilde {{\cal E}}}=H{\tilde {{\cal E}}}+V{\tilde {{\cal
E}}}.\eqno(1.8)
$$

To a given N-connection we can associate a covariant s-derivation on ${%
\tilde M}:$

$$
{\bigtriangledown }_X{Y}=X^I{\{{\frac{{\partial Y^A}}{{\partial x^I}}}+{N_I^A%
}(x,Y)\}}s_A,\eqno(1.9)
$$
where $s_A$ are local independent sections of $\tilde {{\cal E}},{\quad }Y={%
Y^A}s_A$ and $X={X^I}s_I$.

S--differentiable func\-tions $N^{A}_{I}$ from (1.3) writ\-ten as
func\-tions on $x^I$ and $y^{A},\\ N^{A}_{I}(x,y),$ are called
the coefficients of the N--connection and satisfy these
transformation laws under coordinate transforms (1.1) in ${\cal
E}$:
$$
N^{A^{\prime}}_{I^{\prime}}{\frac{{\partial x^{I^{\prime}}}}{{\partial x^{I}}%
}}=M^{A^{\prime}}_{A} N^{A}_{I}- {\frac{\partial {M^{A^{\prime}}_{A}{(x)}}}{%
\partial x^I}} {y^A}.%
$$

If coefficients of a given N-connection are s--differentiable
with respect to coordinates $y^A$ we can introduce (additionally
to covariant nonlinear s-derivation (1.9)) a linear covariant
s-derivation $\hat D$ (which is a generalization for vs--bundles
of the Berwald connection
 [39]) given
as follows:
$$
{{\hat D}_{({\frac{{\partial }}{{\partial x^I}}})}}({\frac{{\partial }}{{%
\partial y^A}}})={{{\hat N}^B}_{AI}}({\frac{{\partial }}{{\partial y^B}}}),{%
\quad }{{\hat D}_{({\frac{{\partial }}{{\partial y^A}}})}}({\frac{{\partial }%
}{{\partial y^B}}})=0,
$$
where
 $$
{{\hat N}^A}_{BI}(x,y)={\frac{{{\partial }{{N^A}_I}{(x,y)}}}{{\partial y^B}}}%
\eqno(1.10)
$$
and
$$
{{{\hat N}^A}_{BC}}{(x,y)}=0.
$$
For a vector field on ${\tilde {{\cal E}}}{\quad }Z={Z^I}{\frac \partial {{%
\partial x^I}}}+{Y^A}{\frac \partial {{\partial y^A}}}$ and $B={B^A}{(y)}{%
\frac \partial {{\partial y^A}}}$ being a section in the vertical s--bundle $%
(V\tilde E,{{\tau }_V},\tilde E)$ the linear connection (1.10)
defines s--derivation (compare with (1.9)):
$$
{{\hat D}_Z}B=[{Z^I}({\frac{\partial B^A}{\partial x^I}}+{\hat N}%
_{BI}^AB^B)+Y^B{\frac{\partial B^A}{\partial y^B}}]{\frac
\partial {\partial y^A}}.
$$

Another important characteristic of a N--connection is its curvature \\
 ( N--connection curvature ):
$$
\Omega ={\frac 12}{\Omega }_{IJ}^A{dx^I}\land {dx^J}\otimes
{\frac \partial {\partial y^A}}
$$
with local coefficients
$$
{\Omega }_{IJ}^A={\frac{\partial N_I^A}{\partial x^J}}-{(-)}^{|IJ|}{\frac{%
\partial N_J^A}{\partial x^I}}+N_I^B{\hat N}_{BJ}^A-{(-)}^{|IJ|}N_J^B{\hat N}%
_{BI}^A,\eqno(1.11)
$$
where for simplicity we have written ${(-)}^{{\mid K\mid }{\mid J\mid }}={(-)%
}^{\mid {KJ}\mid }.$

We note that lin\-ear con\-nec\-tions are par\-tic\-u\-lar cases
of
N--connections locally pa\-ra\-met\-rized as $N_I^A{(x,y)}=N_{BI}^A{(x)}%
x^Iy^B,$ where functions $N_{BI}^A{(x)}$, defined on $\tilde M,$
are called the Christoffel coefficients.

\subsection{N--connections in dvs--bundles}


In order to define a N--connection into dvs--bundle $\widetilde{{\cal E}}%
^{<z>}$ we consider a s--sub\-bund\-le $N\left( \widetilde{{\cal E}}%
^{<z>}\right) $ of the ts-bundle $T\left( \widetilde{{\cal E}}^{<z>}\right) $%
for which one holds (see 
 [226] and 
 [162] respectively for jet
and osculator bundles) the Whitney sum (compare with (1.8))%
$$
T\left( \widetilde{{\cal E}}^{<z>}\right) =N\left( \widetilde{{\cal E}}%
^{<z>}\right) \oplus V\left( \widetilde{{\cal E}}^{<z>}\right) .
$$
$N\left( \widetilde{{\cal E}}^{<z>}\right) $ can be also
interpreted as a regular s--distribution (horizontal distribution
being supplementary to the vertical s-distribution $V\left(
\widetilde{{\cal E}}^{<z>}\right) )$ determined by maps $N:u\in
\widetilde{{\cal E}}^{<z>}\rightarrow N(u)\subset T_u\left(
\widetilde{{\cal E}}^{<z>}\right) .$

The condition of existence of a N--connection in a dvs--bundle $\widetilde{%
{\cal E}}^{<z>}$ can be proved as in 
 [160,161,162]: It is required
that $\widetilde{{\cal E}}^{<z>}$ is a paracompact
s--differentiable (in our case) manifold.

Locally a N--connection in $\widetilde{{\cal E}}^{<z>}$ is given
by its coefficients
$$
N_{(01)I}^{A_1}(u),(N_{(02)I}^{A_2}(u),
N_{(12)A_1}^{A_2}(u)),...(N_{(0p)I}^{A_p}(u),N_{(1p)A_1}^{A_p}(u),...
N_{(p-1p)A_{p-1}}^{A_p}(u)),
$$
$$
...,(N_{(0z)I}^{A_z}(u),N_{(1z)A_1}^{A_z}(u),...,
N_{(pz)A_p}^{A_z}(u),...,N_{(z-1z)A_{z-1}}^{A_z}(u)),
$$
where, for instance, $$%
(N_{(0p)I}^{A_p}(u),N_{(1p)A_1}^{A_p}(u),...,N_{(p-1p)A_{p-1}}^{A_p}(u))$$
are components of N--connection in vs-bundle $\pi ^{<p>}:\tilde
E^{<p>}\rightarrow \tilde E^{<p-1>}.$ Here we note that if a
N-con\-nect\-i\-on structure is defined we must correlate to it
the local partial derivatives on $\widetilde{{\cal E}}^{<z>}$ by
considering instead of local coordinate bases (1.4) and (1.5) the
so--called locally adapted bases
(la--bases) 
$$
\delta _{<\alpha >}=(\delta _I,\delta _{<A>})=(\delta _I,\delta
_{(A_1)},\delta _{(A_2)},...,\delta _{(A_s)})=
$$
$$
\frac \delta {\partial u^{<\alpha >}}=(\frac \delta {\partial
x^I},\frac \delta {\partial y_{(1)}^{A_1}},\frac \delta {\partial
y_{(2)}^{A_2}},...,\frac \delta {\partial
y_{(z)}^{A_z}})\eqno(1.12)
$$
(the dual la--bases are denoted as 
$$
\delta ^{<\alpha >}=(\delta ^I,\delta ^{<A>})=(\delta ^I,\delta
^{(A_1)},\delta ^{(A_2)},...,\delta ^{(A_z)})=
$$
$$
\delta u^{<\alpha >}=(\delta x^I,\delta y^{(A_1)},\delta
y^{(A_2)},...,\delta y^{(A_z)})\quad )\eqno(1.13)
$$
with components parametrized as
$$
\delta _I=\partial _I-N_I^{A_1}\partial _{A_1}-N_I^{A_2}\partial
_{A_2}-...-N_I^{A_{z-1}}\partial _{A_{z-1}}-N_I^{A_z}\partial _{A_z},%
\eqno(1.14)
$$
$$
\delta _{A_1}=\partial _{A_1}-N_{A_1}^{A_2}\partial
_{A_2}-N_{A_1}^{A_3}\partial _{A_3}-...-N_{A_1}^{A_{z-1}}\partial
_{A_{z-1}}-N_{A_1}^{A_z}\partial _{A_z},
$$
$$
\delta _{A_2}=\partial _{A_2}-N_{A_2}^{A_3}\partial
_{A_3}-N_{A_2}^{A_4}\partial _{A_4}-...-N_{A_2}^{A_{z-1}}\partial
_{A_{z-1}}-N_{A_2}^{A_z}\partial _{A_z},
$$
$$
..............................................................
$$
$$
\delta _{A_{z-1}}=\partial _{A_{z-1}}-N_{A_{z-1}}^{A_z}\partial
_{A_z},
$$
$$
\delta _{A_z}=\partial _{A_z},
$$
or, in matrix form, as%
$$
{\bf \delta }_{\bullet }{\bf =}\widehat{{\bf N}}(u)\times {\bf \partial }%
_{\bullet },$$
where%
$$
{\bf \delta }_{\bullet }{\bf =}\delta _{<\alpha >}=\left(
\begin{array}{c}
\delta _I \\
\delta _{A_1} \\
\delta _{A_2} \\
... \\
\delta _{A_z}
\end{array}
\right) =\left(
\begin{array}{c}
\frac \delta {\partial x^I} \\
\frac \delta {\partial y_{(1)}^{A_1}} \\
\frac \delta {\partial y_{(2)}^{A_2}} \\
... \\
\frac \delta {\partial y_{(z)}^{A_z}}
\end{array}
\right) ,{\bf \partial }_{\bullet }{\bf =}\partial _{<\alpha
>}=\left(
\begin{array}{c}
\partial _I \\
\partial _{A_1} \\
\partial _{A_2} \\
... \\
\partial _{A_z}
\end{array}
\right) =\left(
\begin{array}{c}
\frac \partial {\partial x^I} \\
\frac \partial {\partial y_{(1)}^{A_1}} \\
\frac \partial {\partial y_{(2)}^{A_2}} \\
... \\
\frac \partial {\partial y_{(z)}^{A_z}}
\end{array}
\right) .
$$
and%
$$
\widehat{{\bf N}}{\bf =}\left(
\begin{array}{ccccc}
1 & -N_I^{A_1} & -N_I^{A_2} & ... & -N_I^{A_z} \\
0 & 1 & -N_{A_1}^{A_2} & ... & -N_{A_1}^{A_z} \\
0 & 0 & 1 & ... & -N_{A_2}^{A_z} \\
... & ... & ... & ... & ... \\
0 & 0 & 0 & ... & 1
\end{array}
\right) $$ In generalized index form we write the matrix (1.6) as
$\widehat{N}_{<\beta
>}^{<\alpha >},$ where, for instance, $\widehat{N}_J^I=\delta _J^I,\widehat{N%
}_{B_1}^{A_1}=\delta _{B_1}^{A_1},...,\widehat{N}_I^{A_1}=-N_I^{A_1},...,%
\widehat{N}_{A_1}^{A_z}=-N_{A_1}^{A_z},\widehat{N}%
_{A_2}^{A_z}=-N_{A_2}^{A_z},...~.$

So in every point $u\in \widetilde{{\cal E}}^{<z>}$ we have this
invariant
decomposition:%
$$
T_u\left( \widetilde{{\cal E}}^{<d>}\right) =N_0(u)\oplus
N_1(u)\oplus ...\oplus N_{z-1}(u)\oplus V_z(u),\eqno(1.15)
$$
where $\delta _I\in N_0,\delta _{A_1}\in N_1,...,\delta
_{A_{z-1}}\in N_{z-1},\partial _{A_z}\in V_z.$

We note that for the osculator s--bundle $\left( Osc^z\tilde
M,\pi ,\tilde M\right) $ there is an additional (we consider the
N--adapted variant)
s--tangent structure
$$
J:\chi \left( Osc^z\tilde M\right) \rightarrow \chi \left(
Osc^z\tilde M\right)
$$
defined as
$$
\frac \delta {\partial y_{(1)}^I}=J\left( \frac \delta {\partial
x^I}\right) ,...,\frac \delta {\partial y_{(z-1)}^I}=J\left(
\frac \delta {\partial y_{(z-2)}^I}\right) ,\frac \partial
{\partial y_{(z)}^I}=J\left( \frac \delta {\partial
y_{(z-1)}^I}\right) \eqno(1.16)
$$
(in this case $I$- and $A$--indices take the same values and we
can not distinguish them), by considering vertical
$J$--distributions
$$
N_0=N,N_1=J\left( N_0\right) ,...,N_{z-1}=J\left( N_{z-2}\right) .
$$
In consequence, for the la-adapted bases on $\left( Osc^z\tilde
M,\pi
,\tilde M\right) $ there is written this N--connection matrix:%
$$
{\bf N=}N_{<I>}^{<J>}=\left(
\begin{array}{ccccc}
1 & -N_{(1)I}^J & -N_{(2)I}^J & ... & -N_{(z)I}^J \\
0 & 1 & -N_{(1)I}^J & ... & -N_{(z-1)I}^J \\
0 & 0 & 1 & ... & -N_{(z-2)I}^J \\
... & ... & ... & ... & ... \\
0 & 0 & 0 & ... & 1
\end{array}
\right) .\eqno(1.17)
$$

There is a unique distinguished local decomposition of every
s--vector $X\in
\chi \left( \widetilde{{\cal E}}^{<z>}\right) $ on la--base (1.12):%
$$
X=X^{(H)}+X^{(V_1)}+...+X^{(V_z)},\eqno(1.18)
$$
by using the horizontal, $h,$ and verticals, $v_1,v_2,...,v_z$ projections:%
$$
X^{(H)}=hX=X^I\delta _I,~X^{(V_1)}=v_1X=X^{(A_1)}\delta
_{A_1},...,X^{(V_z)}=v_zX=X^{(A_z)}\delta _{A_z}.
$$

With respect to coordinate transforms (1.4 ) the la--bases (1.12)
and ds--vector components (1.18) are correspondingly transformed
as
$$
\frac \delta {\partial x^I}=\frac{{\partial }x^{I^{\prime }}}{{\partial }x^I}%
\frac \delta {\partial x^{I^{\prime }}},~\frac \delta {\partial
y_{(p)}^{A_p}}=K_{A_p}^{A_p^{\prime }}\frac \delta {\partial
y_{(p)}^{A_p^{\prime }}},\eqno(1.19)
$$
and%
$$
X^{I^{\prime }}=\frac{{\partial }x^{I^{\prime }}}{{\partial }x^I}%
X^I,~X^{(A_p^{\prime })}=K_{A_p}^{A_p^{\prime }}X^{(A_p^{\prime
})},\forall p=1,2,...z.
$$

Under changing of coordinates (1.3) the local coefficients of a
nonlinear
connection transform as follows:%
$$
Y_{<\alpha >}^{<\alpha ^{\prime }>}\widehat{N}_{<\alpha ^{\prime
}>}^{<\beta ^{\prime }>}=\widehat{N}_{<\alpha >}^{<\beta
>}(K_{<\beta >}^{<\beta ^{\prime }>}+Y_{<\beta >}^{<\beta
^{\prime }>})
$$
(we can obtain these relations by putting (1.19) and (1.6) into
(1.14) where $\widehat{N}_{<\alpha ^{\prime }>}^{<\beta ^{\prime
}>}$ satisfy $\delta _{<\alpha ^{\prime }>}=\widehat{N}_{<\alpha
^{\prime }>}^{<\beta ^{\prime }>}\partial _{<\beta ^{\prime }>}).$

For dual la-bases (1.13) we have these N--connection
''prolongations of
differentials'':%
$$
\delta x^I=dx^I,
$$
$$
\delta y^{A_1}=dy_{(1)}^{A_1}+M_{(1)I}^{A_1}dx^I,\eqno(1.20)
$$
$$
\delta
y^{A_2}=dy_{(2)}^{A_2}+M_{(2)A_1}^{A_2}dy_{(1)}^{A_1}+M_{(2)I}^{A_2}dx^I,
$$
$$
................................
$$
$$
\delta
y^{A_s}=dy_{(s)}^{A_s}+M_{(s)A_1}^{As}dy_{(1)}^{A_1}+M_{(s)A_2}^{As}dy_{(2)}^{A_2}+...+M_{(z)I}^{A_z}dx^I,
$$
where $M_{(\bullet )\bullet }^{\bullet }$ are the dual
coefficients of the N-connection which can be expressed
explicitly by recurrent formulas through the components of
N--connection $N_{<A>}^{<I>}.$ To do this we shall rewrite
formulas (1.20) in matrix form:%
$$
{\bf \delta }^{\bullet }={\bf d}^{\bullet }\times {\bf M}(u),
$$
where
$$
{\bf \delta }^{\bullet }=\left(
\begin{array}{ccccc}
\delta x^I & \delta y^{A_1} & \delta y^{A_2} & ... & \delta
y^{A_s}
\end{array}
\right) ,~{\bf d}^{\bullet }=\left(
\begin{array}{ccccc}
dx^I & dy_{(1)}^{A_1} & \delta y_{(2)}^{A_2} & ... & \delta
y_{(s)}^{A_s}
\end{array}
\right)
$$
and%
$$
{\bf M=}\left(
\begin{array}{ccccc}
1 & M_{(1)I}^{A_1} & M_{(2)I}^{A_2} & ... & M_{(z)I}^{A_z} \\
0 & 1 & M_{(2)A_1}^{A_2} & ... & M_{(z)A_1}^{A_z} \\
0 & 0 & 1 & ... & M_{(z)A_2}^{A_z} \\
... & ... & ... & ... & ... \\
0 & 0 & 0 & ... & 1
\end{array}
\right) ,
$$
and then, taking into consideration that bases ${\bf \partial }_{\bullet }(%
{\bf \delta }_{\bullet })$ and ${\bf d}^{\bullet }({\bf \delta
}^{\bullet })$ are mutually dual, to compute the components of
matrix ${\bf M}$ being s--inverse to matrix ${\bf N}$ (see (1.17
)). We omit these simple but tedious calculus for general
dvs-bundles and, for simplicity, we present the basic formulas
for osculator s--bundle $\left( Osc^z\tilde M,\pi ,\tilde
M\right) $ when $J$--distribution properties (1.16) and (1.17)
alleviates the problem. For common type of indices on $\tilde M$
and higher order extensions on $Osc^z\tilde M$ the dual la--base
is expressed as
$$
\delta x^I=dx^I,
$$
$$
\delta y_{(1)}^I=dy_{(1)}^I+M_{(1)J}^Idx^J,
$$
$$
\delta y_{(2)}^I=dy_{(2)}^I+M_{(1)J}^Idy_{(1)}^J+M_{(2)J}^Idx^J,
$$
$$
................................
$$
$$
\delta
y_{(z)}^I=dy_{(z)}^I+M_{(1)J}^Idy_{(s-1)}^J+M_{(2)J}^Idy_{(z-2)}^J+...+M_{(z)J}^Idx^J,
$$
with $M$--coefficients computed by recurrent formulas:%
$$
M_{(1)J}^I=N_{(1)J}^I,\eqno(1.21)
$$
$$
M_{(2)J}^I=N_{(2)J}^I+N_{(1)K}^IM_{(1)J}^K,
$$
$$
..............
$$
$$
M_{(s)J}^I=N_{(s)J}^I+N_{(s-1)K}^IM_{(1)J}^K+...+N_{(2)K}^IM_{(z-2)J}^K+N_{(1)K}^IM_{(z-1)J}^K.
$$
One holds these transformation law for dual coefficients (1.21)
with respect to coordinate transforms (1.3) :
$$
M_{(1)J}^KY_{(0,0)K}^{I^{\prime }}=M_{(1)K^{\prime }}^{I^{\prime
}}Y_{(0,0)J}^{K^{\prime }}+Y_{(1,0)J}^{I^{\prime }},
$$
$$
M_{(2)J}^KY_{(0,0)K}^{I^{\prime }}=M_{(2)K^{\prime }}^{I^{\prime
}}Y_{(0,0)J}^{K^{\prime }}+M_{(1)K^{\prime }}^{I^{\prime
}}Y_{(1,0)J}^{K^{\prime }}+Y_{(2,0)J}^{I^{\prime }},
$$
$$
................................
$$
$$
M_{(z)J}^KY_{(0,0)K}^{I^{\prime }}=M_{(z)K^{\prime }}^{I^{\prime
}}Y_{(0,0)J}^{K^{\prime }}+M_{(z-1)K^{\prime }}^{I^{\prime
}}Y_{(1,0)J}^{K^{\prime }}+...+M_{(1)K^{\prime }}^{I^{\prime
}}Y_{(z-1,0)J}^{K^{\prime }}+Y_{(z,0)J}^{I^{\prime }}.
$$
(the proof is a straightforward regroupation of terms after we
have put (1.3) into (1.21)).

Finally, we note that curvatures of a N-connection in a dvs-bundle $%
\widetilde{{\cal E}}^{<z>}$ can be introduced in a manner similar
to that for usual vs-bundles (see (1.11) by a consequent step by
step inclusion of higher dimension anisotropies :
$$
\Omega _{(p)}={\frac 12}{\Omega }_{(p)\alpha _{p-1}\beta _{p-1}}^{A_p}{%
\delta u}^{\alpha _{p-1}}\land {\delta u}^{\beta _{p-1}}\otimes
\frac \delta {\partial y_{(p)}^{A_p}},~p=1,2,...,z,
$$
with local coefficients
$$
{\Omega }_{(p)\beta _{p-1}\gamma _{p-1}}^{A_p}=\frac{\delta
N_{\beta _{p-1}}^{A_p}}{\partial u_{(p-1)}^{\gamma
_{p-1}}}-{(-)}^{|\beta _{p-1}\gamma _{p-1}|}\frac{\delta
N_{\gamma _{p-1}}^{A_p}}{\partial u_{(p-1)}^{\beta _{p-1}}}+
$$
$$
N_{\beta _{p-1}}^{D_p}{\hat N}_{D_p\gamma
_{p-1}}^{A_p}-{(-)}^{|\beta
_{p-1}\gamma _{p-1}|}N_{\gamma _{p-1}}^{D_p}{\hat N}_{D_p\beta _{p-1}}^{A_p},%
$$
where ${\hat N}_{D_p\gamma _{p-1}}^{A_p}=\frac{\delta N_{\gamma
_{p-1}}^{A_p}}{\partial y_{(p)}^{D_p}}$ (we consider
$y^{A_0}\simeq x^I).$

\section{Geometric Objects in DVS--Bundles}

The geometry of the dvs--bundles is very rich and could have
various applications in theoretical and mathematical physics. In
this section we shall present the main results from the geometry
of total spaces of dvs-bundles.

\subsection{D--tensors and d--connections in dvs--bundles}

By using adapted bases (1.12) and (1.13) one introduces algebra
$DT({\tilde {%
{\cal E}}}^{<z>})$ of distinguished tensor s--fields (ds--fields,
ds--tensors,
ds--objects) on \\ $\tilde {{\cal E}}^{<z>},{\quad }{\cal T}={\cal T}%
_{qq_1q_2...q_z}^{pp_1p_{2....}p_z},$ which is equivalent to the
tensor algebra of vs-bundle ${\pi }_{hv_1v_2...v_z}:H\tilde
{{\cal E}}^{<z>}{\oplus
}V_1\tilde {{\cal E}}^{<z>}{\oplus }V_2\tilde {{\cal E}}^{<z>}\oplus ...{%
\oplus }V_s\tilde {{\cal E}}^{<z>}{\to }\tilde {{\cal E}}^{<z>},$
hereafter
briefly denoted as ${{\tilde {{\cal E}}}_{dz}}.$ An element $Q{\in {\cal T}}%
_{qq_1q_2...q_z}^{pp_1p_{2....}p_z},$ , ds-field of type\\
$\left(
\begin{array}{ccccc}
p & p_1 & p_2 & ... & p_z \\
q & q_1 & q_2 & ... & q_z
\end{array}
\right) ,$ can be written in local form as
$$
Q={Q}_{{J_1}{\dots }{J_q}{B_1}{\dots }{B}_{q_1}{C}%
_1...C_{q_2}...F_1...F_{q_s}}^{{I_1}{\dots }{I_p}{A_1}{\dots }{A}%
_{p_1}E_1...E_{p_2}...D_1....D_{p_s}}{(u)}{{\delta }_{I_1}}\otimes {\dots }%
\otimes {{\delta }_{I_p}}\otimes {d^{J_1}}\otimes {\dots }\otimes {d^{J_q}}%
\otimes
$$
$$
{{\partial }_{A_1}}\otimes {\dots }\otimes {\partial }_{A_{p_1}}\otimes {{%
\delta }^{B_1}{\delta }^{B_1}}\otimes {\dots }\otimes {\delta }%
^{B_{q_1}}\otimes {{\partial }_{E_1}}\otimes {\dots }\otimes {\partial }%
_{E_{p_2}}\otimes {{\delta }^{C_1}}\otimes {\dots }\otimes {\delta }%
^{C_{q_2}}\otimes ...
$$
$$
\otimes {{\partial }_{D_1}}\otimes {\dots }\otimes {\partial }%
_{D_{p_z}}\otimes {{\delta }^{F_1}}\otimes {\dots }\otimes {\delta }%
^{F_{qz}}.\eqno(1.22)
$$

In addition to ds--tensors we can introduce ds--objects with
various s--group
and coordinate transforms adapted to global splitting (1.15).

A linear distinguished connection, d--connection, in dvs--bundle $\tilde {%
{\cal E}}^{<z>}$ is a linear connection $D$ on $\tilde {{\cal
E}}^{<z>}$
which preserves by parallelism the horizontal and vertical distributions in $%
\tilde {{\cal E}}^{<z>}$.

By a linear connection of a s--manifold we understand a linear
connection in its tangent bundle.

Let denote by $\Xi (\tilde M)$ and $\Xi (\tilde {{\cal
E}}^{<p>}),$ respectively, the modules of vector fields on
s-manifold $\tilde M$ and
vs-bundle $\tilde {{\cal E}}^{<p>}$ and by ${\cal F}{(\tilde M)}$ and ${\cal %
F}{(\tilde {{\cal E}}}^{<p>}{)},$ respectively, the s-modules of
functions on $\tilde M$ and on $\tilde {{\cal E}}^{<p>}.$

It is clear that for a given global splitting into horizontal and
verticals s--subbund\-les (1.15) we can associate operators of
horizontal and vertical
covariant derivations (h- and v--derivations, denoted respectively as $%
D^{(h)} $ and $D^{(v_1v_2...v_z)}$) with properties:
$$
{D_X}Y=(XD)Y={D_{hX}}Y+{D}_{v_1X}Y+{D}_{v_2X}Y+...+{D}_{v_zX}Y,
$$
where
$$
D_X^{(h)}{Y}=D_{hX}{Y},{\quad }D_X^{(h)}f=(hX)f
$$
and
$$
D_X^{(v_p)}{Y}=D_{v_pX}{Y},{\quad
}D_X^{(v_p)}f=(v_pX)f,~(p=1,...,z)
$$
for every $f\in {\cal F}(\tilde M)$ with decomposition of vectors $X,Y\in {%
\Xi }(\tilde {{\cal E}}^{<z>})$ into horizontal and vertical parts, $$%
X=hX+v_1X+....+v_zX{\quad } \mbox{and}{\quad
}Y=hY+v_1Y+...+v_zY.$$

The local coefficients of a d--connection $D$ in $\tilde {{\cal
E}}^{<z>}$ with respect to the local adapted frame (1.5) separate
into corresponding
distinguished groups. We introduce horizontal local coefficients $$%
(L_{JK}^I,L_{<B>K}^{<A>})=({L^I}_{JK}(u),{L}_{B_1K}^{A_1}{(}u),{L}%
_{B_2K}^{A_2}{(}u),...,{L}_{B_zK}^{A_z}{(}u))$$ of $D^{(h)}$ such
that
$$
D_{({\frac \delta {\delta x^K}})}^{(h)}{\frac \delta {\delta x^J}}={L^I}%
_{JK}(u){\frac \delta {\delta x^I}},D_{({\frac \delta {\delta x^K}}%
)}^{(h)}\frac \delta {\delta
y_{(p)}^{B_p}}={L}_{B_pK}^{A_p}{(u)}\frac \delta {\delta
y_{(p)}^{A_p}},(p=1,...,z),
$$
$$
D_{({\frac \delta {\delta x^k}})}^{(h)}q={\frac{{\delta
q}}{\delta x^K}},
$$
and $p$--vertical local coefficients $$(C_{J<C>}^I,C_{<B><C>}^{<A>})=({C}%
_{JC_p}^I(u),{C}_{B_1C_p}^{A_1}(u),{C}_{B_2C_p}^{A_2}(u),...,{C}%
_{B_zC_p}^{A_z}(u))$$ $(p=1,...,z)$ such that
$$
D_{(\frac \delta {\delta y^{C_p}})}^{(v_p)}{\frac \delta {\delta x^J}}={C}%
_{JC_p}^I(u){\frac \delta {\delta x^I}},D_{(\frac \delta {\delta
y^{C_p}})}^{(v_p)}\frac \delta {\delta y_{(f)}^{B_f}}={C}_{B_fC_p}^{A_f}%
\frac \delta {\delta y_{(f)}^{A_f}},D_{(\frac \delta {\delta
y^{C_p}})}^{(v_p)}q=\frac{\delta q}{\partial y^{C_p}},
$$
where $q\in {\cal F}(\tilde {{\cal E}}^{<z>}),$ $f=1,...,z.$

The covariant ds--de\-ri\-va\-ti\-on along vector
$$X=X^I{\frac \delta {\delta x^I}}%
+Y^{A_1}\frac \delta {\delta y^{A_1}}+...+Y^{A_z}\frac \delta
{\delta y^{A_z}}$$ of a ds--tensor field $Q,$ for instance, of
type $\left(
\begin{array}{cc}
p & p_r \\
q & q_r
\end{array}
\right) ,1\leq r\leq z,{\quad }$ see (1.22), can be written as
$$
{D_X}Q=D_X^{(h)}Q+D_X^{(v_1)}Q+...+D_X^{(v_z)}Q,
$$
where h-covariant derivative is defined as
$$
D_X^{(h)}Q=X^KQ_{JB_r{\mid }K}^{IA_r}{\delta }_I{\otimes \delta }_{A_r}{%
\otimes }d^I{\otimes }{\delta }^{B_r},
$$
with components
$$
Q_{JB_r{\mid }K}^{IA_r}={\frac{\delta Q_{JB_r}^{IA_r}}{\partial x^K}}+{L^I}%
_{HK}Q_{JB_R}^{HA_r}+{L}_{C_iK}^{A_r}Q_{JB_i}^{IC_r}-{L^H}%
_{JK}Q_{HB_r}^{IA_r}-{L}_{B_rK}^{C_r}Q_{JC_r}^{IA_r},
$$
and v$_p$-covariant derivatives defined as
$$
{D}_X^{(v_p)}Q={X}^{C_p}{Q}_{JB_r\perp C_p}^{IA_r}{\delta }_I{\otimes }{%
\partial }_{A_r}{\otimes }d^I{\otimes }{\delta }^{B_r},
$$
with components
$$
{Q}_{JB_r\perp C_p}^{IA_r}=\frac{\delta Q_{JB_R}^{IA_r}}{\partial y^{C_p}}+{%
C^I}_{HC_p}Q_{JB_R}^{HA_r}+{C}_{F_rC_p}^{A_r}Q_{JB_R}^{IF_r}-{C}%
_{JC_p}^HQ_{HF_R}^{IA_r}-{C}_{B_rC_p}^{F_r}Q_{JF_R}^{IA_r}..
$$

The above presented formulas show that
$$D{\Gamma }=(L,...,{L}_{(p)},...,{C},...,C_{(p)},...)$$
 are the local coefficients of the d-connection $D$ with
respect to the local frame $({\frac \delta {\delta x^I}},{\frac
\partial
{\partial y^a}}).$ If a change (1.3) of local coordinates on $\tilde {{\cal E%
}}^{<z>}$ is performed, by using the law of transformation of
local frames (1.19),we obtain the following transformation laws
of the local coefficients of a d--connect\-i\-on:
$$
{L^{I^{\prime }}}_{J^{\prime }M^{\prime }}={\frac{\partial x^{I^{\prime }}}{%
\partial x^I}}{\frac{\partial x^J}{\partial x^{J^{\prime }}}}{\frac{\partial
x^M}{\partial x^{M^{\prime }}}}{L^I}_{JM}+{\frac{\partial x^{I^{\prime }}}{%
\partial x^M}}{\frac{{\partial }^2x^M}{{{\partial x^{J^{\prime }}}{\partial
x^{M^{\prime }}}}}},\eqno(1.23)
$$
$$
{L}_{(f)B_f^{\prime }M^{\prime }}^{A_f^{\prime
}}=K_{A_f}^{A_f^{\prime
}}K_{B_f^{\prime }}^{B_f}{\frac{\partial x^M}{\partial x^{M^{\prime }}}L}%
_{(f)B_fM}^{A_f}+K_{C_f}^{A_f^{\prime }}\frac{\partial
K_{B_f^{\prime }}^{C_f}}{\partial x^{M^{\prime }}},
$$
$$
................
$$
$$
{C_{(p)J^{\prime }C_p^{\prime }}^{I^{\prime }}}={\frac{\partial
x^{I^{\prime
}}}{\partial x^I}}{\frac{\partial x^J}{\partial x^{J^{\prime }}}}%
K_{C_p}^{C_p^{\prime }}{C_{(p)JC_p}^I},...,{C_{B_f^{\prime
}C_p^{\prime }}^{A_f^{\prime }}}=K_{A_f}^{A_f^{\prime
}}{K}_{B_f^{\prime }}^{B_f}K_{C_p^{\prime
}}^{C_p}{C_{B_fC_p}^{A_f},...}.
$$
As in the usual case of tensor calculus on locally isotropic
spaces the transformation laws (1.23) for d--connections differ
from those for ds-tensors, which are written (for instance, we
consider transformation laws for ds--tensor (1.22)) as
$$
{Q}_{{J}^{\prime }{_1}{\dots }{J}^{\prime }{_q}{B}^{\prime }{_1}{\dots }{B}%
_{q_1}^{\prime }{C}_1^{\prime }...C_{q_2}^{\prime }...F_1^{\prime
}...F_{q_s}^{\prime }}^{{I}^{\prime }{_1}{\dots }{I}^{\prime }{_p}{A}%
^{\prime }{_1}{\dots }{A}_{p_1}^{\prime }E_1^{\prime
}...E_{p_2}^{\prime }...D_1^{\prime }....D_{p_s}^{\prime }}=
$$
$$
{\frac{\partial x^{I_1^{\prime }}}{\partial x^{I_1}}{\frac{\partial x^{J_1}}{%
\partial x^{J_1^{\prime }}}}\dots }K_{A_1}^{A_1^{\prime }}K_{B_1^{\prime
}}^{B_1}{\dots K}_{D_{p_s}}^{D_{p_s}^{\prime
}}{K}_{F_{p_s}^{\prime
}}^{F_{p_s}}{Q}_{{J_1}{\dots }{J_q}{B_1}{\dots }{B}_{q_1}{C}%
_1...C_{q_2}...F_1...F_{q_s}}^{{I_1}{\dots }{I_p}{A_1}{\dots }{A}%
_{p_1}E_1...E_{p_2}...D_1....D_{p_s}}.
$$

To obtain local formulas on usual higher order anisotropic spaces
we have to restrict us with even components of geometric objects
by changing, formally, capital indices $(I,J,K,...)$ into
$(i,j,k,a,..)$ and s--derivation and s--commutation rules into
those for real number fields on usual manifolds. We shall
consider various applications in the theoretical and mathematical
physics of the differential geometry of distinguished vector
bundles in the second Part of this monograph.

\subsection{ Torsions and curvatures of d--connections}

Let $\tilde {{\cal E}}^{<z>}$ be a dvs--bundle endowed with
N-connection and d--connec\-ti\-on structures. The torsion of a
d--connection is introduced as
$$
T(X,Y)=[X,DY\}-[X,Y\},{\quad }X,Y{\subset }{\Xi }{(\tilde M)}.
$$
One holds the following invariant decomposition (by using h-- and
v--pro\-jec\-ti\-ons associated to N):
$$
T(X,Y)=T(hX,hY)+T(hX,v_1Y)+T(v_1X,hX)+T(v_1X,v_1Y)+...
$$
$$
+T(v_{p-1}X,v_{p-1}Y)+T(v_{p-1}X,v_pY)+T(v_pX,v_{p-1}X)+T(v_pX,v_pY)+...
$$
$$
+T(v_{z-1}X,v_{z-1}Y)+T(v_{z-1}X,v_zY)+T(v_zX,v_{z-1}X)+T(v_zX,v_zY).
$$
Taking into ac\-count the skew\-su\-per\-sym\-me\-try of $T$ and
the equa\-tions
$$
h[v_pX,v_pY\}=0,...,v_f[v_pX,v_pY\}=0,f\neq p,
$$
we can verify that the torsion of a d--connection is completely
determined by the following ds-tensor fields:
$$
hT(hX,hY)=[X(D^{(h)}{h})Y\}-h[hX,hY\},...,
$$
$$
v_pT(hX,hY)=-v_p[hX,hY\},...,
$$
$$
hT(hX,v_pY)=-D_Y^{(v_p)}{hX}-h[hX,v_pY\},...,
$$
$$
v_pT(hX,v_pY)=D_X^{(h)}{v}_p{Y}-v_p[hX,v_pY\},...,
$$
$$
v_fT(v_fX,v_fY)=[X(D^{(v_f)}{v}_f)Y\}-v_f[v_fX,v_fY\},...,
$$
$$
v_pT(v_fX,v_fY)=-v_p[v_fX,v_fY\},...,
$$
$$
v_fT(v_fX,v_pY)=-D_Y^{(v_p)}{v}_f{X}-v_f[v_fX,v_pY\},...,
$$
$$
v_pT(v_fX,v_pY)=D_X^{(v_f)}{v}_p{Y}-v_p[v_fX,v_pY\},...,f<p,
$$
$$
v_{z-1}T(v_{z-1}X,v_{x-1}Y)=[X(D^{(v_{z-1})}{v}_{z-1})Y%
\}-v_{z-1}[v_{z-1}X,v_{z-1}Y\},...,
$$
$$
v_zT(v_{z-1}X,v_{z-1}Y)=-v_z[v_{z-1}X,v_{z-1}Y\},
$$
$$
v_{z-1}T(v_{z-1}X,v_zY)=-D_Y^{(v_z)}{v}_{z-1}{X}-v_{z-1}[v_{z-1}X,v_zY%
\},...,
$$
$$
v_zT(v_{z-1}X,v_zY)=D_X^{(v_{z-1})}{v}_z{Y}-v_z[v_{z-1}X,v_zY\}.
$$
where $X,Y\in {{\Xi }(\tilde {{\cal E}}^{<z>})}.$ In order to get
the local
form of the ds--tensor fields which determine the torsion of d--connection $D{%
\Gamma }$ (the torsions of $D{\Gamma }$) we use equations
$$
[{\frac \delta {\partial x^J}},{\frac \delta {\partial x^K}}\}={R}%
_{JK}^{<A>}\frac \delta {\partial y^{<A>}},~[{\frac \delta {\partial x^J}},{%
\frac \delta {\partial y^{<B>}}}\}={\frac{\delta
{N_J^{<A>}}}{\partial y^{<B>}}}{\frac \delta {\partial y^{<A>}}},
$$
where
$$
{R}_{JK}^{<A>}=\frac{\delta N_J^{<A>}}{\partial x^K}-{(-)}^{\mid {KJ}\mid }%
\frac{\delta N_K^{<A>}}{\partial x^J},
$$
and introduce notations
$$
hT({\frac \delta {\delta x^K}},{\frac \delta {\delta
x^J}})={T^I}_{JK}{\frac \delta {\delta x^I}},\eqno(1.24)
$$
$$
v_1T{({\frac \delta {\delta x^K}},{\frac \delta {\delta x^J}})}={\tilde T}%
_{KJ}^{A_1}\frac \delta {\partial y^{A_1}},
$$
$$
hT({\frac \delta {\partial y^{<A>}}},{\frac \delta {\partial x^J}})={\tilde P%
}_{J<A>}^I{\frac \delta {\delta x^I}},{...,}
$$
$$
v_pT(\frac \delta {\partial y^{B_p}},{\frac \delta {\delta x^J}})={P}%
_{JB_p}^{<A>}{\frac \delta {\partial y^{<A>}}},...,
$$
$$
v_pT(\frac \delta {\partial y^{C_p}},\frac \delta {\partial y^{B_f}})={S}%
_{B_fC_p}^{<A>}{\frac \delta {\partial y^{<A>}}}.
$$

Now we can compute the local components of the torsions,
introduced in
(1.24), with respect to a la--frame of a d--connection\\ $D{\Gamma }=(L,...,{L}%
_{(p)},...,{C},...,C_{(p)},...)$ $:$
$$
{T^I}_{JK}={L^I}_{JK}-{(-)}^{\mid {JK}\mid }{L^I}_{KJ},\eqno(1.25)
$$
$$
{{\tilde T}^{<A>}}_{JK}={R^{<A>}}_{JK},{{\tilde
P}^I}_{J<B>}={C^I}_{J<B>},
$$
$$
{P^{<A>}}_{J<B>}={\frac{\delta {N_J^{<A>}}}{\partial y^{<B>}}}-{L^{<A>}}%
_{<B>J},
$$
$$
{S^{<A>}}_{<B><C>}={C^{<A>}}_{<B><C>}-{(-)}^{\mid <B><C>\mid }{C^{<A>}}%
_{<C><B>}.
$$
The even and odd components of torsions (1.25) can be specified
in explicit form by using decompositions of indices into even and
odd parts $(I=(i,{\hat i}),J=(j,{\hat j}),..)$, for instance,
$$
{T^i}_{jk}={L^i}_{jk}-{L^i}_{kj},{\quad }{T^i}_{j{\hat k}}={L^i}_{j{\hat k}}+%
{L^i}_{{\hat k}j},
$$
$$
{T^{\hat i}}_{jk}={L^{\hat i}}_{jk}-{L^{\hat i}}_{kj},{\dots },
$$
and so on (see 
 [160,161] and the second Part of this monograph for
explicit formulas (6.24),(6.25) on torsions on a ha--space ${\cal
E}^{<z>}$, we shall omit ''tilde'' for usual manifolds and
bundles).

Another important characteristic of a d--connection $D\Gamma $ is
its curvature:
$$
R(X,Y)Z=D_{[X}D_{Y\}}-D_{[X,Y\}}Z,
$$
where $X,Y,Z\in {\Xi }(\tilde E^{<z>}).$ From the properties of
h- and v-projections it follows that
$$
v_pR(X,Y)hZ=0,...,hR(X,Y)v_pZ=0,v_fR(X,Y)v_pZ=0,f\neq
p,\eqno(1.26)
$$
and
$$
R(X,Y)Z=hR(X,Y)hZ+v_1R(X,Y)v_1Z+...+v_zR(X,Y)v_zZ,
$$
where $X,Y,Z\in {\Xi }(\tilde E^{<z>}).$ Tak\-ing into ac\-count
prop\-er\-ties (1.26) and the equa\-tions
$$
R(X,Y)=-{(-)}^{\mid XY\mid }R(Y,X)
$$
we prove that the curvature of a d-connection $D$ in the total
space of a vs-bundle $\tilde E^{<z>}$ is completely determined by
the following ds-tensor fields:
$$
R(hX,hY)hZ=({D_{[X}^{(h)}}D_{Y\}}^{(h)}-D_{[hX,hY\}}^{(h)}\eqno(1.27)
$$
$$
-D_{[hX,hY\}}^{(v_1)}-...D_{[hX,hY\}}^{(v_{z-1})}-D_{[hX,hY\}}^{(v_z)})hZ,
$$

$$
R(hX,hY)v_pZ=(D_{[X}^{(h)}D_{Y\}}^{(h)}-D_{[hX,hY\}}^{(h)}-
$$
$$
D_{[hX,hY\}}^{(v_1)})v_pZ-...-D_{[hX,hY\}}^{(v_{p-1})})v_pZ-D_{[hX,hY%
\}}^{(v_p)})v_pZ,
$$
$$
R(v_pX,hY)hZ=(D_{[X}^{(v_p)}D_{Y\}}^{(h)}-D_{[v_pX,hY\}}^{(h)}-
$$
$$
D_{[v_pX,hY\}}^{(v_{_1})})hZ-...-D_{[v_pX,hY\}}^{(v_{p-_1})}-D_{[v_pX,hY%
\}}^{(v_p)})hZ,
$$
$$
R(v_fX,hY)v_pZ=(D_{[X}^{(v_f)}D_{Y\}}^{(h)}-D_{[v_fX,hY\}}^{(h)}-
$$
$$
D_{[v_fX,hY\}}^{(v_1)})v_1Z-...-D_{[v_fX,hY%
\}}^{(v_{p-1})})v_{p-1}Z-D_{[v_fX,hY\}}^{(v_p)})v_pZ,
$$
$$
R(v_fX,v_pY)hZ=(D_{[X}^{(v_f)}D_{Y\}}^{(v_p)}-D_{[v_fX,v_pY%
\}}^{(v_1)})hZ-...
$$
$$
-D_{[v_fX,v_pY\}}^{(v_{z-1})}-D_{[v_fX,v_pY\}}^{(v_z)})hZ,
$$
$$
R(v_fX,v_qY)v_pZ=(D_{[X}^{(v_f)}D_{Y\}}^{(v_q)}-D_{[v_fX,v_qY%
\}}^{(v_1)})v_1Z-...
$$
$$
-D_{[v_fX,v_qY\}}^{(v_{p-1})})v_{p-1}Z-D_{[v_fX,v_qY\}}^{(v_p)})v_pZ,
$$
where
$$
{D_{[X}^{(h)}}{D_{Y\}}^{(h)}}={D_X^{(h)}}{D_Y^{(h)}}-{{(-)}^{\mid XY\mid }}{%
D_Y^{(h)}}{D_X^{(h)}},~
$$
$$
{D_{[X}^{(h)}}{D}_{Y\}}^{(v_p)}={D_X^{(h)}}{D}_Y^{(v_p)}-{(-)}^{|Xv_pY|}{D}%
_Y^{(v_p)}{D_X^{(h)},}
$$
$$
{D}_{[X}^{(v_p)}{D_{Y\}}^{(h)}}={D}_X^{(v_p)}{D_Y^{(h)}}-{(-)}^{\mid
v_pXY\mid }{D_Y^{(h)}}{D}_X^{(v_p)},~
$$
$$
~{D}_{[X}^{(v_f)}{D}_{Y\}}^{(v_p)}={D}_X^{(v_f)}{D}_Y^{(v_p)}-{(-)}%
^{|v_fXv_pY|}{D}_Y^{(v_p)}{D}_X^{(v_f)}.
$$
The local components of the ds--tensor fields (1.27) are
introduced as follows:
$$
R({{\delta }_K},{{\delta }_J}){{\delta }_H}={{{R_H}^I}_{JK}}{{\delta }_I},{~}%
R({{\delta }_K},{{\delta }_J})\delta
{_{<B>}}={R}_{<B>.JK}^{.<A>}\delta _{<A>},\eqno(1.28)
$$
$$
R({\delta _{<C>}},{{\delta }_K}){{\delta }_J}=P_{JK<C>}^I{{{\delta }_I},~}%
R\left( \delta _{<C>},\delta _K\right) \delta
_{<B>}=P_{<B>.K<C>}^{.<A>}\delta _{<A>},
$$
$$
R({\delta _{<C>}},{{\delta }_{<B>}}){{\delta }_J}=
S_{J.<B><C>}^{.I}{{{\delta}_I}},$$ $$
R({\delta _{<D>}},{{\delta }_{<C>}}){{\delta }_{<B>}}%
=S_{<B>.<C><D>}^{.<A>}{{{\delta }_{<A>}}}.
$$
Putting the components of a d--connection $$D{\Gamma }=(L,...,{L}_{(p)},...,{C}%
,...,C_{(p)},...)$$ in (1.28), by a direct computation, we obtain
these locally adapted components of the curvature (curvatures):
$$
{{R_H}^I}_{JK}={{\delta }_K}{L^I}_{HJ}-{(-)}^{\mid KJ\mid }{{\delta }_J}{L^I}%
_{HK}+
$$
$$
{L^M}_{HJ}{L^I}_{MK}-{(-)}^{\mid KJ\mid }{L^M}_{HK}{L^I}_{MJ}+{C^I}_{H<A>}{%
R^{<A>}}_{JK},
$$
$$
{R_{<B>\cdot JK}^{\cdot <A>}}={{\delta
}_K}{L^{<A>}}_{<B>J}-{(-)}^{\mid KJ\mid }{{\delta
}_J}{L^{<A>}}_{<B>K}+
$$
$$
{L^{<C>}}_{<B>J}{L^{<A>}}_{<C>K}-{(-)}^{\mid KJ\mid }{L^{<C>}}_{<B>K}+{%
C^{<A>}}_{<B><C>}{R^{<C>}}_{JK},
$$
$$
{P_{J\cdot K<A>}^{\cdot I}}={\delta _{<A>}}{L^I}_{JK}-{C^I}_{J<A>{\mid }K}+{%
C^I}_{J<B>}{P^{<B>}}_{K<A>},\eqno(1.29)
$$
$$
{{P_{<B>}}^{<A>}}_{K<C>}={\delta _{<C>}}{L^{<A>}}_{<B>K}-$$
$${C^{<A>}}_{<B><C>{\mid }K}+{C^{<A>}}_{<B><D>}{P^{<D>}}_{K<C>},
$$
$$
{S_{J\cdot <B><C>}^{\cdot I}}={\delta
_{<C>}}{C^I}_{J<B>}-{(-)}^{\mid <B><C>\mid }{\delta
_{<B>}}{C^I}_{J<C>}+
$$
$$
{C^{<H>}}_{J<B>}{C^I}_{<H><C>}-{(-)}^{\mid <B><C>\mid }{C^{<H>}}_{J<C>}{C^I}%
_{<H><B>},
$$
$$
{{S_{<B>}}^{<A>}}_{<C><D>}={\delta
_{<D>}}{C^{<A>}}_{<B><C>}-{(-)}^{\mid <C><D>\mid }{\delta
_{<C>}}{C^{<A>}}_{<B><D>}+
$$
$$
{C^{<E>}}_{<B><C>}{C^{<A>}}_{<E><D>}-{(-)}^{\mid <C><D>\mid }{C^{<E>}}%
_{<B><D>}{C^{<A>}}_{<E><C>}.
$$
Even and odd components of curvatures (1.29) can be written out
by splitting indices into even and odd parts, for instance,
$$
{{R_h}^i}_{jk}={{\delta }_k}{L^i}_{hj}-{{\delta }_j}{{L^i}_{hk}+{{L^m}_{hj}}{%
{L^i}_{mk}}-{{L^m}_{hk}}{{L^i}_{mj}}}+{{C^i}_{h<a>}}{{R^{<a>}}_{jk}},
$$
$$
{{R_h}^i}_{j{\hat k}}={{\delta }_{\hat k}}{L^i}_{hj}+{{\delta }_j}{L^i}_{h{%
\hat k}}+{L^m}_{hj}{L^i}_{m{\hat k}}+{L^m}_{h{\hat k}}{L^i}_{mj}+{C^i}_{h<a>}%
{R^{<a>}}_{j{\hat k}}{\quad },{\dots }.
$$
(formulas for even components of curvatures are given in the Part
II of this monograph, see (6.26)--(6.28)); we omit formulas for
the even--odd components of curvatures because we shall not use
them in this work).

\subsection{Bianchi and Ricci identities}
The torsions and curvatures of every linear connection $D$ on a vs--bundle $%
\tilde {{\cal E}}^{<z>}$ satisfy the following generalized Bianchi
identities:
$$
\sum_{SC}{[(D_X{T})(Y,Z)-R(X,Y)Z+T(T(X,Y),Z)]}=0,
$$

$$
\sum_{SC}{[(D_X{R})(U,Y,Z)+R(T(X,Y)Z)U]}=0,\eqno(1.30)
$$
where $\sum_{SC}$ means the respective supersymmretric cyclic sum over $%
X,Y,Z $ and $U.$ If $D$ is a d--connection, then by using (1.26)
and
$$
v_p(D_X{R})(U,Y,hZ)=0,{~}h({D_X}R(U,Y,v_pZ)=0,v_f(D_X{R})(U,Y,v_pZ)=0,
$$
the identities (1.30) become
$$
\sum_{SC}[h({D_X}T)(Y,Z)-hR(X,Y)Z+hT(hT(X,Y),Z)+
$$
$$
hT(v_1T(X,Y),Z)+...+hT(v_zT(X,Y),Z)]=0,
$$
$$
\sum_{SC}{[v}_f{({D_X}T)(Y,Z)}-{v}_f{R(X,Y)Z+}
$$
$$
{v}_f{T(hT(X,Y),Z)+}\sum\limits_{p\geq
f}{v}_f{T(v}_p{T(X,Y),Z)]}=0,
$$
$$
\sum_{SC}{[h({D_X}R)(U,Y,Z)+hR(hT(X,Y),Z)U+}
$$
$$
{hR(v}_1{T(X,Y),Z)U+...+{hR(v}_z{T(X,Y),Z)U}]}=0,
$$
$$
\sum_{SC}{[v}_f{({D_X}R)(U,Y,Z)+v}_f{R(hT(X,Y),Z)U+}
$$
$$
\sum\limits_{p\geq f}{v}_f{R(v}_p{T(X,Y),Z)U]}=0.\eqno(1.31)
$$

In order to get the component form of these identities we insert
correspondingly in (1.31) these values of triples $(X,Y,Z)$,\ ($=({{\delta }%
_J},{{\delta }_K},{{\delta }_L}),$ or\\ $({\delta _{<D>}},{\delta _{<C>}},{%
\delta _{<B>}})$), and put successively $U={\delta }_H$ and $U={\delta }%
_{<A>}.$ Taking into account (1.24),(1.25) and (1.27),(1.28) we
obtain:
$$
\sum_{SC[L,K,J\}}[{T^I}_{JK{\mid }H}+{T^M}_{JK}{T^J}_{HM}+{R^{<A>}}_{JK}{C^I}%
_{H<A>}-{{R_J}^I}_{KH}]=0,
$$
$$
\sum_{SC[L,K,J\}}[{{R^{<A>}}_{JK{\mid }H}}+{T^M}_{JK}{R^{<A>}}_{HM}+{R^{<B>}}%
_{JK}{P^{<A>}}_{H<B>}]=0,
$$
{\cal
$$
{C^I}_{J<B>{\mid }K}-{(-)}^{\mid JK\mid }{C^I}_{K<B>{\mid
}J}-{T^I}_{JK{\mid <}B>}+{C^M}_{J<B>}{T^I}_{KM}-
$$
}%
$$
{(-)}^{\mid JK\mid }{C^M}_{K<B>}{T^I}_{JM}+{T^M}_{JK}{C^I}_{M<B>}+{P^{<D>}}%
_{J<B>}{C^I}_{K<D>}
$$
{\cal
$$
-{(-)}^{\mid KJ\mid }{P^{<D>}}_{K<B>}{C^I}_{J<D>}+{{P_J}^I}_{K<B>}-{(-)}%
^{\mid KJ\mid }{{P_K}^I}_{J<B>}=0,
$$
$$
{P^{<A>}}_{J<B>{\mid }K}-{(-)}^{\mid KJ\mid }{P^{<A>}}_{K<B>{\mid }J}-{%
R^{<A>}}_{JK\perp <B>}+{C^M}_{J<B>}{R^{<A>}}_{KM}-
$$
}%
$$
{(-)}^{\mid KJ\mid }{C^M}_{K<B>}{R^{<A>}}_{JM}+{T^M}_{JK}{P^{<A>}}_{M<B>}+{%
P^{<D>}}_{J<B>}{P^{<A>}}_{K<D>}-
$$
{\cal
$$
{(-)}^{\mid KJ\mid }{P^{<D>}}_{K<B>}{P^{<A>}}_{J<D>}-{{R^{<D>}}_{JK}}{{%
S^{<A>}}_{B<D>}}+{R_{<B>\cdot JK}^{\cdot <A>}}=0,
$$
$$
{C^I}_{J<B>\perp <C>}-{(-)}^{\mid <B><C>\mid }{C^I}_{J<C>\perp <B>}+{C^M}%
_{J<C>}{C^I}_{M<B>}-
$$
}%
$$
{(-)}^{\mid <B><C>\mid }{C^M}_{J<B>}{C^I}_{M<C>}+{S^{<D>}}_{<B><C>}{C^I}%
_{J<D>}-{S_{J\cdot <B><C>}^{\cdot I}}=0,
$$
$$
{P^{<A>}}_{J<B>\perp <C>}-{(-)}^{\mid <B><C>\mid }{P^{<A>}}_{J<C>\perp <B>}+%
$$
$$
{S^{<A>}}_{<B><C>\mid J}+ {C^M}_{J<C>}{P^{<A>}}_{M<B>}-%
$$
$$
{(-)}^{\mid <B><C>\mid }{C^M}_{J<B>}{P^{<A>}}_{M<C>}+{P^{<D>}}_{J<B>}{S^{<A>}%
}_{<C><D>}-
$$
{\cal
$$
{(-)}^{\mid <C><B>\mid }{P^{<D>}}_{J<C>}{S^{<A>}}_{<B><D>}+{S^{<D>}}_{<B><C>}%
{P^{<A>}}_{J<D>}+
$$
$$
{{P_{<B>}}^{<A>}}_{J<C>}-{(-)}^{\mid <C><B>\mid
}{{P_{<C>}}^{<A>}}_{J<B>}=0,
$$
$$
\sum_{SC[<B>,<C>,<D>\}}[{S^{<A>}}_{<B><C>\perp <D>}+
$$
}%
$$
{S^{<F>}}_{<B><C>}{S^{<A>}}_{<D><F>}-{{S_{<B>}}^{<A>}}_{<C><D>}]=0,
$$
{\cal
$$
\sum_{SC[H,J,L\}}[{{R_K}^I}_{HJ\mid L}-{T^M}_{HJ}{{R_K}^I}_{LM}-{{R^{<A>}}%
_{HJ}P}{_{K\cdot L<A>}^{\cdot I}}]=0,
$$
$$
\sum_{SC[H,J,L\}}[{R_{<D>\cdot HJ\mid L}^{\cdot <A>}}-{{T^M}_{HJ}R}{%
_{<D>\cdot LM}^{\cdot
<A>}}-{{R^{<C>}}_{HJ}}{{{P_{<D>}}^{<A>}}_{L<C>}}]=0,
$$
$$
{P_{K\cdot J<D>\mid L}^{\cdot I}}-{(-)}^{\mid LJ\mid }{P_{K\cdot
L<D>\mid J}^{\cdot I}}+{{R_K}^I}_{LJ\perp
<D>}+{C^M}_{L<D>}{{R_K}^I}_{JM}-
$$
}%
$$
{(-)}^{\mid LJ\mid
}{C^M}_{J<D>}{{R_K}^I}_{LM}-{T^M}_{JL}{P_{K\cdot M<D>}^{\cdot I}}+
$$
$$
{P^{<A>}}_{L<D>}{P_{K\cdot J<A>}^{\cdot I}}-{(-)}^{\mid LJ\mid }{{P^{<A>}}%
_{J<D>}P}{_{K\cdot L<A>}^{\cdot I}}-{{R^{<A>}}_{JL}S}{_{K\cdot
<A><D>}^{\cdot I}}=0,
$$
$$
{{P_{<C>}}^{<A>}}_{J<D>\mid L}-{(-)}^{\mid LJ\mid }{{P_{<C>}}^{<A>}}%
_{L<D>\mid J}+{R_{<C>\cdot LJ\mid <D>}^{\cdot <A>}}+
$$
$$
{C^M}_{L<D>}{{R_{<C>}}^{<A>}}_{JM}-{(-)}^{\mid LJ\mid }{C^M}_{J<D>}{{R_{<C>}}%
^{<A>}}_{LM}-
$$
$$
{T^M}_{JL}{{P_{<C>}}^{<A>}}_{M<D>}+{P^{<F>}}_{L<D>}{{P_{<C>}}^{<A>}}_{J<F>}-
$$
{\cal
$$
{(-)}^{\mid LJ\mid }{P^{<F>}}_{J<D>}{{P_{<C>}}^{<A>}}_{L<F>}-{R^{<F>}}_{JL}{{%
S_{<C>}}^{<A>}}_{F<D>}=0,
$$
}%
$$
{P_{K\cdot J<D>\perp <C>}^{\cdot I}-(-)}^{\mid <C><D>\mid
}{P_{K\cdot J<C>\perp <D>}^{\cdot I}+{S_K}_{\ <D><C>|J}^I+}
$$
$$
{C^M}_{J<D>}{P_{K\cdot M<C>}^{\cdot I}}-{(-)}^{\mid <C><D>\mid }{C^M}_{J<C>}{%
P_{K\cdot M<D>}^{\cdot I}}+
$$
$$
{P^{<A>}}_{J<C>}{S_{K\cdot <D><A>}^{\cdot I}}-{(-)}^{\mid <C><D>\mid }{%
P^{<A>}}_{J<D>}{S_{K\cdot <C><A>}^{\cdot I}}+
$$
$$
{S^{<A>}}_{<C><D>}{P_{K\cdot J<A>}^{\cdot I}}=0,
$$
$$
{{P_{<B>}}^{<A>}}_{J<D>\perp <C>}-{(-)}^{\mid <C><D>\mid }{{P_{<B>}}^{<A>}}%
_{J<C>\perp <D>}+$$ $${{S_{<B>}}^{<A>}}_{<C><D>\mid J}+
{C^M}_{J<D>}{{P_{<B>}}^{<A>}}_{M<C>}-$$
$${(-)}^{\mid <C><D>\mid }{C^M}_{J<C>}{{P_{<B>}}^{<A>}}_{M<D>}+
$$
$$
{P^{<F>}}_{J<C>}{{S_{<B>}}^{<A>}}_{<D><F>}-{(-)}^{\mid <C><D>\mid }{P^{<F>}}%
_{J<D>}{{S_{<B>}}^{<A>}}_{<C><F>}+
$$
$$
{S^{<F>}}_{<C><D>}{{P_{<B>}}^{<A>}}_{J<F>}=0,
$$
$$
\sum\limits_{SC[<B>,<C>,<D>\}}[S_{K.<B><C>\perp
<D>}^{.I}-S_{.<B><C>}^{<A>}S_{K.<D><A>}^{.I}]=0,
$$
$$
\sum\limits_{SC[<B>,<C>,<D>\}}[S_{<F>.<B><C>\perp
<D>}^{.<A>}-S_{.<B><C>}^{<E>}S_{<F>.<E><A>}^{.<A>}]=0,
$$
where $\sum\limits_{SC[<B>,<C>,<D>\}}i$s the supersymmetric
cyclic sum over indices $<B>,$ $<C>,<D>.$

As a consequence of a corresponding arrangement of (1.27) we
obtain the Ricci identities (for simplicity we establish them
only for ds--vector fi\-elds, although they may be written for
every ds-tensor field):
$$
{D_{[X}^{(h)}}{D_{Y\}}^{(h)}}hZ=R(hX,hY)hZ+{D_{[hX,hY\}}^{(h)}}%
hZ+\sum\limits_{f=1}^z{D}_{[hX,hY\}}^{(v_f)}hZ,\eqno(1.32)
$$
$$
{D}_{[X}^{(v_p)}{D_{Y\}}^{(h)}}hZ=R(v_pX,hY)hZ+{D}_{[v_pX,hY\}}^{(h)}hZ+\sum%
\limits_{f=1}^z{D}_{[v_pX,hY\}}^{(v_f)}hZ,
$$
$$
{D}_{[X}^{(v_p)}D_{Y\}}^{(v_P)}=R(v_pX,v_pY)hZ+\sum\limits_{f=1}^z{D}%
_{[v_pX,v_pY\}}^{(v_f)}hZ
$$
and
$$
{D_{[X}^{(h)}}{D_{Y\}}^{(h)}}v_pZ=R(hX,hY)v_pZ+{D_{[hX,hY\}}^{(h)}}%
v_pZ+\sum\limits_{f=1}^z{D}_{[hX,hY\}}^{(v_f)}v_pZ,\eqno(1.33)
$$
$$
D_{[X}^{(v_f)}D_{Y\}}^{(h)}v_pZ=R(v_fX,hY)v_pZ+\sum%
\limits_{q=1}^zD_{[v_fX,hY\}}^{(v_q)}v_pZ+\sum\limits_{q=1}^zD_{[v_fX,hY%
\}}^{(v_q)}v_pZ,
$$
$$
D_{[X}^{(v_q)}D_{Y\}}^{(v_f)}v_pZ=R(v_qX,v_fY)v_pZ+\sum%
\limits_{s=1}^zD_{[v_fX,v_fY\}}^{(v_s)}v_pZ.
$$
Putting $X={X^I}(u){\frac \delta {\delta x^I}}+{X^{<A>}}(u){\frac
\delta {\partial y^{<A>}}}$ and taking into account the local
form of the h- and v-covariant s-derivatives and
(1.24),(1.25),(1.27),(1.28) we can express respectively
identities (1.32) and (1.33) in this form:
$$
{X^{<A>}}_{\mid K\mid L}-{(-)}^{\mid KL\mid }{X^{<A>}}_{\mid
L\mid K}=
$$
$$
{{{R_{<B>}}^{<A>}}_{KL}}{X^{<B>}}-{T^H}_{KL}{X^{<A>}}_{\mid H}-{R^{<B>}}_{KL}%
{X^{<A>}}_{\perp <B>},
$$
$$
{X^I}_{\mid K\perp <D>}-{(-)}^{\mid K<D>\mid }{X^I}_{\perp
<D>\mid K}=
$$
$$
{P_{H\cdot K<D>}^{\cdot I}}{X^H}-{C^H}_{K<D>}{X^I}_{\mid H}-{P^{<A>}}_{K<D>}{%
X^I}_{\perp <A>},
$$
$$
{X^I}_{\perp <B>\perp <C>}-{(-)}^{\mid <B><C>\mid }{X^I}_{\perp
<C>\perp <B>}=
$$
$$
{S_{H\cdot <B><C>}^{\cdot I}}{X^H}-{S^{<A>}}_{<B><C>}{X^I}_{\perp
<A>}
$$
and
$$
{X^{<A>}}_{\mid K\mid L}-{(-)}^{\mid KL\mid }{X^{<A>}}_{\mid
L\mid K}=
$$
$$
{{R_{<B>}}^{<A>}}_{KL}{X^{<B>}}-{T^H}_{KL}{X^{<A>}}_{\mid H}-{R^{<B>}}_{KL}{%
X^{<A>}}_{\perp <B>},
$$
$$
{X^{<A>}}_{\mid K\perp <B>}-{(-)}^{\mid <B>K\mid
}{X^{<A>}}_{\perp <B>\mid K}=
$$
$$
{{P_{<B>}}^{<A>}}_{KC}{X^C}-{C^H}_{K<B>}{X^{<A>}}_{\mid H}-{P^{<D>}}_{K<B>}{%
X^{<A>}}_{\perp <D>},
$$
$$
{X^{<A>}}_{\perp <B>\perp <C>}-{(-)}^{\mid <C><B>\mid
}{X^{<A>}}_{\perp <C>\perp <B>}=
$$
$$
{{S_{<D>}}^{<A>}}_{<B><C>}{X^{<D>}}-{S^{<D>}}_{<B><C>}{X^{<A>}}_{\perp
<D>}.
$$
We note that the above presented formulas 
 [265] generalize for
higher order anisotropy the similar ones for locally anisotropic
superspaces
 [260].

\subsection{Cartan structure equations in dvs--bundles}
Let consider a ds--tensor field on $\tilde {{\cal E}}^{<z>}$:
$$
t={t_{<A>}^I}{\delta }_I{\otimes }{\delta ^{<A>}}.
$$
The d-connection 1-forms ${\omega }_J^I$ and ${{\tilde \omega
}_{<B>}^{<A>}}$
are introduced as%
$$
Dt=(D{t_{<A>}^I}){\delta }_I{\otimes }{\delta }^{<A>}
$$
with
$$
Dt_{<A>}^I=dt_{<A>}^I+{\omega }_J^I{t_{<A>}^J}-{{\tilde \omega }_{<A>}^{<B>}}%
{t_{<B>}^I}=t_{<A>\mid J}^I{dx^J}+t_{<A>\perp <B>}^I{\delta
}y^{<B>}.
$$
For the d-connection 1-forms of a d-connection $D$ on $\tilde {{\cal E}%
}^{<z>}$ defined by ${{\omega }_J^I}$ and ${{\tilde \omega
}_{<B>}^{<A>}}$ one holds the following structure equations:
$$
d({d^I})-{d^H}\wedge {\omega }_H^I=-{\Omega },~d{({{\delta }^{<A>}})}-{{%
\delta }^{<B>}}\wedge {{\tilde \omega }_{<B>}^{<A>}}=-{{\tilde \Omega }^{<A>}%
},
$$
$$
d{{\omega }_J^I}-{{\omega }_J^H}\wedge {{\omega }_H^I}=-{{\Omega }_J^I},~d{{%
\tilde \omega }_{<B>}^{<A>}}-{{\tilde \omega }_{<B>}^{<C>}}\wedge
{{\tilde \omega }_{<C>}^{<A>}}=-{{\tilde \Omega }_{<B>}^{<A>}},
$$
in which the torsion 2-forms ${\Omega }^I$ and ${{\tilde \Omega
}^{<A>}}$ are given respectively by formulas:
$$
{{\Omega }^I}={\frac 12}{T^I}_{JK}{d^J}\wedge {d^K}+{\frac 12}{C^I}_{J<C>}{%
d^J}\wedge {{\delta }^{<C>}},
$$
$$
{{\tilde \Omega }^{<A>}}={\frac 12}{R^{<A>}}_{JK}{d^J}\wedge
{d^K}+$$
$${\frac 12}{P^{<A>}}_{J<C>}{d^J}\wedge {{\delta }^{<C>}}+
{\frac 12}{S^{<A>}}_{<B><C>}{{\delta }^{<B>}}\wedge {{\delta
}^{<C>}},
$$
and
$$
{{\Omega }_J^I}={\frac 12}{{R_J}^I}_{KH}{d^K}\wedge {d^H}+{\frac 12P}{%
_{J\cdot K<C>}^{\cdot I}}{d^K}\wedge {{\delta }^{<C>}}+ {\frac
12S}{_{J\cdot K<C>}^{\cdot I}}{{\delta }^{<B>}}\wedge {{\delta
}^{<C>}},
$$
$$
{{\tilde \Omega }_{<B>}^{<A>}}={\frac 12R}{_{<B>\cdot KH}^{\cdot <A>}}{d^K}%
\wedge {d^H}+%
$$
$$
{\frac 12}{{P_{<B>}}^{<A>}}_{K<C>}{d^K}\wedge {{\delta }^{<C>}}+{\frac 12}{{%
S_{<B>}}^{<A>}}_{<C><D>}{{\delta }^{<C>}}\wedge {{\delta }^{<D>}}.
$$
We have defined the exterior product on s--space to satisfy the
property
$$
{{\delta }^{<\alpha >}}\wedge {{\delta }^{<\beta >}}=-{(-)}^{\mid
<\alpha
><\beta >\mid }{{\delta }^{<\beta >}}\wedge {{\delta }^{<\alpha >}}.%
$$
\subsection{Metrics in dvs--bundles}

The base $\tilde M$ of dvs--bundle $\tilde {{\cal E}}^{<z>}$is
considered to be a connected and paracompact s--manifold.

A metric structure on the total space $\tilde E^{<z>}$ of a dvs-bundle $%
\tilde {{\cal E}}^{<z>}$is a supersymmetric, second order,
covariant s--tensor field
$$
G=G_{<\alpha ><\beta >}\partial ^{<\alpha >}\otimes \partial
^{<\beta >}
$$
which in every point $u\in \tilde {{\cal E}}^{<z>}$ is given by
nondegenerate supersymmetric matrix $G_{<\alpha ><\beta >}=G({{\partial }%
_{<\alpha >}},{{\partial }_{<\beta >}}){\quad }$ (with
nonvanishing superdeterminant, $sdetG\not =0).$

The metric and N--connection structures on $\tilde {{\cal
E}}^{<z>}$ are compatible if there are satisfied conditions:
$$
G({{\delta }_I},{{\partial }_{<A>}})=0,G(\delta _{A_f},{\partial }%
_{A_p})=0,~z\geq p>f\geq 1,
$$
or, in consequence,
$$
{G_{I<A>}}-{N_I^{<B>}}{h_{<A><B>}}=0,{G}_{A_fA_p}-{N}_{A_f}^{B_p}{h}%
_{A_pB_p}=0,\eqno(1.34)
$$
where
$$
{G_{I<A>}}=G({{\partial }_I},{{\partial }_{<A>}}),{G}_{A_fA_p}=G({\partial }%
_{A_f},{\partial }_{A_p}).
$$
From (1.34) one follows
$$
{N_I^{<B>}}={h^{<B><A>}}{G_{I<A>}},~{N}_{A_f}^{A_p}={h}^{A_pB_p}{G}%
_{A_fB_p},...,
$$
where matrices $h^{<A><B>},{h}^{A_pB_p},...$ are respectively
s--inverse to matrices
$$
h_{<A><B>}=G({{\partial }_{<A>}},{{\partial
}_{<B>}}),h_{A_pB_p}=G({\partial }_{A_p},{\partial }_{B_p}).
$$
So, in this case, the coefficients of N-connection are uniquely
determined by the components of the metric on $\tilde {{\cal
E}}^{<z>}.$

A compatible with N--connection metric on $\tilde {{\cal
E}}^{<z>}$ is written in irreducible form as
$$
G(X,Y)=G(hX,hY)+G(v_1X,v_1Y)+...+G(v_zX,v_zY),{\quad }X,Y\in {\Xi (\tilde {%
{\cal E}}^{<z>})},
$$
and looks locally as
$$
G=g_{{\alpha }{\beta }}{(u)}{{\delta }^\alpha }\otimes {{\delta }^\beta }%
=g_{IJ}{d^I}\otimes {d^J}+h_{<A><B>}{{\delta }^{<A>}}\otimes {{\delta }^{<B>}%
}=
$$
$$
g_{IJ}{d^I}\otimes {d^J}+h_{A_1B_1}{\delta }^{A_1}\otimes {\delta }%
^{B_1}+h_{A_2B_2}{\delta }^{A_2}\otimes {\delta }^{B_2}+...+h_{A_zB_z}{%
\delta }^{A_z}\otimes {\delta }^{B_z}.\eqno(1.35)
$$

A d--connection $D$ on $\tilde {{\cal E}}^{<z>}$ is metric, or
compatible
with metric $G$, if conditions
$$
{D_{<\alpha >}}{G_{<{\beta ><}{\gamma >}}}=0
$$
are satisfied.

A d--connection $D$ on $\tilde {{\cal E}}^{<z>}$ provided with a
metric $G$ is a metric d--con\-nec\-ti\-on if and only if
$$
{D_X^{(h)}}{(hG)}=0,{D_X^{(h)}}{(v}_p{G)}=0,{D}_X^{(v_p)}{(hG)}=0,{D}%
_X^{(v_f)}{(v}_p{G)}=0\eqno(1.36)
$$
for every $,f,p=1,2,...,z,$ and $X\in {\Xi (\tilde {{\cal
E}}^{<z>})}.$ Conditions (1.36) are written in locally adapted
form as
$$
g_{IJ\mid K}=0,g_{IJ\perp <A>}=0,h_{<A><B>\mid
K}=0,h_{<A><B>\perp <C>}=0.
$$

In every dvs--bundle provided with compatible N--connection and
metric structures one exists a metric d-connection (called the
canonical d--connection associated to $G)$ depending only on
components of G-metric and
N--connection $.$ Its local coefficients $C{\Gamma }=({{\grave L}^I}_{JK},{{%
\grave L}^{<A>}}_{<B>K},{{\grave C}^I}_{J<C>},{{\grave
C}^{<A>}}_{<B><C>})$ are as follows:
$$
{{\grave L}^I}_{JK}={\frac 12}{g^{IH}}({{{\delta }_K}g_{HJ}+{{\delta }_J}%
g_{HK}-{{\delta }_H}g_{JK}}),\eqno(1.37)
$$
$$
{{\grave L}^{<A>}}_{<B>K}={\delta _{<B>}}{N_K^{<A>}}+ {\frac
12}{h^{<A><C>}}\times $$ $$[{{{\delta }_{<K>}}{h_{<B><C>}}-
(\delta }_{<B>}N_K^{<D>})h_{<\dot D><C>}{ -{(\delta
}_{<C>}N_K^{<D>})h_{<\dot D><B>}}],
$$
 $$
{{\grave C}^I}_{J<C>}={\frac 12}{g^{IK}\delta }{_{<C>}}{g_{JK}},
$$
$$
{{\grave C}^{<A>}}_{<B><C>}={\frac 12}{h^{<A><D>}(\delta
}_{<C>}h_{<D><B>}+$$
$${\delta }_{<B>}h_{<D><C>}-{\delta }_{<D>}h_{<B><C>}.
$$
We emphasize that, in general, the torsion of $C\Gamma
$--connection (1.37) does not vanish.

It should be noted here that on dvs-bundles provided with
N-connection and d-connection and metric really it is defined a
multiconnec\-ti\-on ds--struc\-tu\-re,
i.e. we can use in an equivalent geometric manner different types
of d- connections with various properties. For example, for
modeling of some physical processes we can use an extension of
the Berwald d--connection
$$
B{\Gamma }=({{L^I}_{JK}},\delta _{<B>}N_K^{<A>},0,{C}_{<B><C>}^{<A>}),%
\eqno(1.38)
$$
where ${L^I}_{JK}={{\grave L}^I}_{JK}$ and ${C^{<A>}}_{<B><C>}={{\grave C}%
^{<A>}}_{<B><C>},$ which is hv-metric, i.e. satisfies conditions:
$$
{D_X^{(h)}}hG=0,...,{D}_X^{(v_p)}v_pG=0,...,{D}_X^{(v_z)}v_zG=0
$$
for every $X\in \Xi {(\tilde {{\cal E}}^{<z>})},$ or in locally
adapted coordinates,
$$
g_{IJ\mid K}=0,h_{<A><B>\perp <C>}=0.
$$

As well we can introduce the Levi--Civita connection%
$$
\{{\frac{<{\alpha >}}{<{{\beta ><}{\gamma >}}}}\}=$$
$${\frac 12}{G^{<{\alpha ><}{\beta >}}({{\partial }_{<\beta >}}
{G_{<\tau ><\gamma >}}+{{\partial }_{<\gamma >}}{G_{<\tau ><\beta
>}}- {{\partial }_{<\tau >}}{G_{<\beta><\gamma >}})},
$$
constructed as in the Riemann geometry from components of metric\\ $G_{<{%
\alpha ><}{\beta >}}$ by using partial derivations
$${{\partial }_{<\alpha >}}
={\frac \partial {\partial u^{<\alpha >}}}= ({\frac \partial
{\partial x^I}},{\frac \partial {\partial y^{<A>}}}),$$
 which is metric but not a d-connection.

In our further considerations we shall largely use the Christoffel
d--symbols defined similarly as components of Levi--Civita
connection but by
using la--partial de\-ri\-va\-ti\-ons,
$$
{{{\tilde \Gamma }^{<\alpha >}}_{<\beta ><\gamma >}}={\frac
12}{G^{<\alpha
><\tau >}}({{\delta }_{<\beta >}}{G_{<\tau ><\gamma >}}+{{\delta }_{<\gamma
>}}{G_{<\tau ><\beta >}}-{{\delta }_{<\tau >}}{G_{<\beta ><\gamma >}}),%
\eqno(1.39)
$$
having components
$$
C{\tilde \Gamma }=({L^I}_{JK},0,0,{C^{<A>}}_{<B><C>}),
$$
where coefficients ${L^I}_{JK}$ and ${C^{<A>}}_{<B><C>}$ must be
com\-puted as in formu\-las (1.38).

We can express arbitrary d--connection as a deformation of the
background d--connection (1.38):
$$
{{{\Gamma }^{<\alpha >}}_{<\beta ><\gamma >}}={{\tilde \Gamma
}_{\cdot
<\beta ><\gamma >}^{<\alpha >}}+{{P^{<\alpha >}}_{<\beta ><\gamma >}},%
\eqno(1.40)
$$
where ${{P^{<\alpha >}}_{<\beta ><\gamma >}}$ is called the
deformation ds-tensor. Putting splitting (1.40) into (1.25) and
(1.29) we can express
torsion ${T^{<\alpha >}}_{<\beta ><\gamma >}$ and curvature\\ ${{R_{<\beta >}%
}^{<\alpha >}}_{<\gamma ><\delta >}$ of a d-connection ${{\Gamma
}^{<\alpha
>}}_{<\beta ><\gamma >}$ as respective deformations of torsion ${{\tilde T}%
^{<\alpha >}}_{<\beta ><\gamma >}$ and torsion ${\tilde
R}_{<\beta >\cdot
<\gamma ><\delta >}^{\cdot <\alpha >}$ for connection ${{\tilde \Gamma }%
^{<\alpha >}}_{<\beta ><\gamma >}{\quad }:$

$$
{{T^{<\alpha >}}_{<\beta ><\gamma >}}={{\tilde T}_{\cdot <\beta
><\gamma
>}^{<\alpha >}}+{{\ddot T}_{\cdot <\beta ><\gamma >}^{<\alpha >}}
$$
and
$$
{{{R_{<\beta >}}^{<\alpha >}}_{<\gamma ><\delta >}}={{\tilde
R}_{<\beta
>\cdot <\gamma ><\delta >}^{\cdot <\alpha >}}+{{\ddot R}_{<\beta >\cdot
<\gamma ><\delta >}^{\cdot <\alpha >}},
$$
where
$$
{{\tilde T}^{<\alpha >}}_{<\beta ><\gamma >}={{\tilde \Gamma }^{<\alpha >}}%
_{<\beta ><\gamma >}-{(-)}^{\mid <\beta ><\gamma >\mid }{{\tilde \Gamma }%
^{<\alpha >}}_{<\gamma ><\beta >}+{w^{<\alpha >}}_{<\gamma
><\delta >},
$$
$$
~{{\ddot T}^{<\alpha >}}_{<\beta ><\gamma >}={{\ddot \Gamma }^{<\alpha >}}%
_{<\beta ><\gamma >}-{(-)}^{\mid <\beta ><\gamma >\mid }{{\ddot \Gamma }%
^{<\alpha >}}_{<\gamma ><\beta >},
$$
and%
$$
{{\tilde R}_{<\beta >\cdot <\gamma ><\delta >}^{\cdot <\alpha >}}={{\delta }%
_{<\delta >}}{{\tilde \Gamma }^{<\alpha >}}_{<\beta ><\gamma
>}-{(-)}^{\mid
<\gamma ><\delta >\mid }{{\delta }_{<\gamma >}}{{\tilde \Gamma }^{<\alpha >}}%
_{<\beta ><\delta >}+
$$
$$
{{{\tilde \Gamma }^{<\varphi >}}_{<\beta ><\gamma >}}{{{\tilde \Gamma }%
^{<\alpha >}}_{<\varphi ><\delta >}}-{(-)}^{\mid <\gamma ><\delta >\mid }{{{%
\tilde \Gamma }^{<\varphi >}}_{<\beta ><\delta >}}{{{\tilde \Gamma }%
^{<\alpha >}}_{<\varphi ><\gamma >}}+{{\tilde \Gamma }^{<\alpha
>}}_{<\beta
><\varphi >}{w^{<\varphi >}}_{<\gamma ><\delta >},
$$
$$
{{\ddot R}_{<\beta >\cdot <\gamma ><\delta >}^{\cdot <\alpha >}}={{\tilde D}%
_{<\delta >}}{{P^{<\alpha >}}_{<\beta ><\gamma >}}-{(-)}^{\mid
<\gamma
><\delta >\mid }{{\tilde D}_{<\gamma >}}{{P^{<\alpha >}}_{<\beta ><\delta >}}%
+
$$
$$
{{P^{<\varphi >}}_{<\beta ><\gamma >}}{{P^{<\alpha >}}_{<\varphi ><\delta >}}%
-{(-)}^{\mid <\gamma ><\delta >\mid }{{P^{<\varphi >}}_{<\beta ><\delta >}}{{%
P^{<\alpha >}}_{<\varphi ><\gamma >}}
$$
$$
+{{P^{<\alpha >}}_{<\beta ><\varphi >}}{{w^{<\varphi >}}_{<\gamma ><\delta >}%
},
$$
the nonholonomy coefficients ${w^{<\alpha >}}_{<\beta ><\gamma
>}$ are defined as
$$
[{\delta }_{<\alpha >},{\delta }_{<\beta >}\}={{\delta }_{<\alpha >}}{{%
\delta }_{<\beta >}}-{(-)}^{|<\alpha ><\beta >|}{{\delta }_{<\beta >}}{{%
\delta }_{<\alpha >}}={w^{<\tau >}}_{<\alpha ><\beta >}{{\delta }_{<\tau >}}.%
$$

We emphasize that if from geometric point of view all considered
d--con\-nect\-i\-ons are ''equ\-al in rights'', the
con\-struct\-ion of physical models on la--spaces requires an
explicit fixing of the type of d--con\-nect\-ion and metric
structures.

\section{Higher Order Tangent S--bun\-dles}
The aim of this section is to present a study of supersymmetric
extensions from $\widetilde{M}$ to $T\tilde M$ and
$Osc^{(z)}\widetilde{M}$ and to consider corresponding
prolongations of Riemann and ge\-ne\-ra\-li\-zed Fins\-ler
structures (on classical and new approaches to Finsler geometry,
its generalizations and applications in physics se, for example,
 [78,56,213,160,161,17,18,159,163,13,14,41,29]).

The presented in the previous section basic results on dvs-bundles ${\tilde {%
{\cal E}}^{<z>}}$ pro\-vid\-ed with N-connection, d-connection
and metric structures can be correspondingly adapted to the
osculator s--bundle $\left( Osc^z\tilde M,\pi ,\tilde M\right) .$
In this case the dimension of the base space and typical higher
orders fibre coincides and we shall not distinguish indices of
geometrical objects.

Coefficients of a d--connection $D\Gamma
(N)=(L_{JM}^I,C_{(1)JM}^I,...,C_{(z)JM}^I)$ in $Osc^z\tilde M,$
with respect
to a la--base are introduced as to satisfy equations%
$$
D_{\frac \delta {\delta x^I}}\frac \delta {\delta
y_{(f)}^I}=L_{IJ}^M\frac \delta {\delta y_{(f)}^M},~D_{\frac
\delta {\delta y_{(p)}^J}}\frac \delta
{\delta y_{(f)}^I}=C_{(p)IJ}^M\frac \delta {\delta y_{(f)}^M}, \eqno(1.41)%
$$
$$
(f=0,1,...,z;p=1,...,z,\mbox{ and }y_{(0)}^I=x^I).
$$

A metric structure on $Osc^z\tilde M$ is ds--tensor s--symmetric field $%
g_{IJ}(u_{(z)})=g_{IJ}(x,y_{(1)},y_{(2)},...,y_{(z)})$ of type $%
(0,2),srank|g_{ij}|=(n,m).$ The N--lift of Sasaki type of
$g_{IJ}$ is given
by (see (1.35)) defines a global Riemannian s--structure (if
$\widetilde{M}$
is a s--differentiable, paracompact s-manifold):%
$$
G=g_{IJ}(u_{(z)})dx^I\otimes dx^J+g_{IJ}(u_{(z)})dy_{(1)}^I\otimes
dy_{(1)}^J+...+g_{IJ}(u_{(z)})dy_{(z)}^I\otimes
dy_{(z)}^J.\eqno(1.42)
$$
The condition of compatibility of a d--connection (1.41) with
metric (1.42) is expressed as
$$
D_XG=0,\forall X\in \Xi (Osc^z\tilde M),
$$
or, by using d--covariant partial derivations $|_{(p)}$ defined by
coefficients\\ $(L_{JM}^I,C_{(1)JM}^I,...,C_{(z)JM}^I),$
$$
g_{IJ|M=0,~}g_{IJ|_{(p)M}}=0,(p=1,...,z).
$$
An example of compatible with metric d--connection is given by
Christof\-fel
d--sym\-bols (see (1.39)):%
$$
L_{IJ}^M=\frac 12g^{MK}\left( \frac{\delta g_{KJ}}{\partial x^I}+\frac{%
\delta g_{IK}}{\partial x^J}-\frac{\delta g_{IJ}}{\partial x^K}\right) ,%
$$
$$
C_{(p)IJ}^M=\frac 12g^{MK}\left( \frac{\delta g_{KJ}}{\partial y_{(p)}^I}+%
\frac{\delta g_{IK}}{\partial y_{(p)}^J}-\frac{\delta
g_{IJ}}{\partial y_{(p)}^K}\right) ;p=1,2,...,z.
$$

\subsection{Supersymmetric ex\-ten\-si\-ons of Fins\-ler spaces}
We start our considerations with the ts-bundle $T\tilde M.$ An s-vector $%
X\in \Xi (T\tilde M)$ is decomposed with respect to la--bundles as%
$$
X=X(u)^I\delta _I+Y(u)^I\partial _I,
$$
where $u=u^\alpha =(x^I,y^J)$ local coordinates. The s--tangent
structures (1.16) are transformed into a global map
$$
J:\Xi (T\tilde M)\to \Xi (T\tilde M)
$$
which does not depend on N-connection structure:
$$
J({\frac \delta {\delta x^I}})={\frac \partial {\partial y^I}}
\mbox{ and } J({\frac \partial {\partial y^I}})=0.
$$
This endomorphism is called the natural ( or canonical ) almost
tangent
structure on $T\tilde M;$ it has the properties:
$$
1)J^2=0,{\quad }2)ImJ=KerJ=VT\tilde M
$$
and 3) the Nigenhuis s--ten\-sor,     
$$
{N_J}(X,Y)=[JX,JY\}-J[JX,Y\}-J[X,JY]
$$
$$
(X,Y\in \Xi (TN))
$$
identically vanishes, i.e. the natural almost tangent structure $J$ on $%
T\tilde M$ is integrable.

A generalized Lagrange superspace, GLS--space, is a pair\\ ${GL}%
^{n,m}=(\tilde M,g_{IJ}(x,y))$, \quad where \quad $g_{IJ}(x,y)$
\quad is a
ds--tensor field on \\ \quad ${{\tilde {T{\tilde M}}}=T\tilde M-\{0\}},$%
\quad s--symmetric of superrank $(n,m).$

We call $g_{IJ}$ as the fundamental ds--tensor, or metric
ds--tensor, of GLS--space.

There exists an unique d-connection $C\Gamma (N)$ which is compatible with $%
g_{IJ}{(u)}$ and has vanishing torsions ${T^I}_{JK}$ and
${S^I}_{JK}$ (see formulas (1.25) rewritten for ts-bundles). This
connection, depending only on $g_{IJ}{(u)}$ and ${N_J^I}{(u)}$ is
called the canonical metric d-connection of GLS-space. It has
coefficients
$$
{L^I}_{JK}={\frac 12}{g^{IH}}({\delta }_J{g_{HK}}+{\delta }_H{g_{JK}}-{%
\delta }_H{g_{JK}}),
$$
$$
{C^I}_{JK}={\frac 12}{g^{IH}}({\partial }_J{g_{HK}}+{\partial }_H{g_{JK}}-{%
\partial }_H{g_{JK}}).
$$
There is a unique normal d-connection $D\Gamma (N)=({\bar L}_{\cdot JK}^I,{%
\bar C}_{\cdot JK}^I)$ which is metric and has a priori given torsions ${T^I}%
_{JK}$ and ${S^I}_{JK}.$ The coefficients of $D\Gamma (N)$ are
the following ones:
$$
{\bar L}_{\cdot JK}^I={L^I}_{JK}-\frac 12g^{IH}(g_{JR}{T^R}_{HK}+g_{KR}{T^R}%
_{HJ}-g_{HR}{T^R}_{KJ}),
$$
$$
{\bar C}_{\cdot JK}^I={C^I}_{JK}-\frac 12g^{IH}(g_{JR}{S^R}_{HK}+g_{KR}{S^R}%
_{HJ}-g_{HR}{S^R}_{KJ}),
$$
where ${L^I}_{JK}$ and ${C^I}_{JK}$ are the same as for the $C\Gamma (N)$%
--connection (1.41).

The Lagrange spaces were introduced 
 [136] in order to geometrize the
concept of Lagrangian in mechanics (the Lagrange geometry is
studied in
details in 
 [160,161]). For s-spaces we present this generalization:

A Lagrange s--space, LS--space, $L^{n,m}=(\tilde M,g_{IJ}),$ is
defined as a particular case of GLS-space when the ds--metric on
$\tilde M$ can be expressed as
$$
g_{IJ}{(u)}={\frac 12}{\frac{{\partial }^2L}{{{\partial y^I}{\partial y^J}}}}%
,\eqno(1.43)
$$
where $L:T\tilde M\to \Lambda ,$ is a s-differentiable function
called a s-Lagrangian on $\tilde M.$

Now we consider the supersymmetric extension of Fins\-ler space:
A Finsler s--metric on $\tilde M$ is a function $F_S:T\tilde M\to
\Lambda $ having the properties:

1. The restriction of $F_S$ to ${\tilde {T\tilde M}}=T\tilde
M\setminus \{0\} $ is of the class $G^\infty $ and F is only
supersmooth on the image of the null cross--section in the
ts-bundle to $\tilde M.$

2. The restriction of F to ${\tilde {T\tilde M}}$ is positively
homogeneous
of degree 1 with respect to ${(y^I)}$, i.e. $F(x,{\lambda }y)={\lambda }%
F(x,y),$ where ${\lambda }$ is a real positive number.

3. The restriction of F to the even subspace of $\tilde {T\tilde
M}$ is a positive function.

4. The quadratic form on ${\Lambda }^{n,m}$ with the coefficients
$$
g_{IJ}{(u)}={\frac 12}{\frac{{\partial }^2F^2}{{{\partial y^I}{\partial y^J}}%
}}
$$
defined on $\tilde {T\tilde M}$ is nondegenerate.

A pair $F^{n,m}=(\tilde M,F)$ which consists from a supersmooth
s-ma\-ni\-fold $\tilde M$ and a Finsler s--metric is called a
Finsler
superspace, FS--space. 

It's obvious that FS--spaces form a particular class of
LS--spaces with s-Lagran\-gi\-an $L={F^2}$ and a particular class
of GLS--spaces with metrics of type (1.58).

For a FS--space we can introduce the supersymmetric variant of
nonlinear
Car\-tan con\-nec\-ti\-on 
 [56,213]:
$$
N_J^I{(x,y)}={\frac \partial {\partial y^J}}G^{*I},
$$
where
$$
G^{*I}={\frac 14}g^{*IJ}({\frac{{\partial }^2{\varepsilon }}{{\partial y^I}{%
\partial x^K}}}{y^K}-{\frac{\partial {\varepsilon }}{\partial x^J}}),{\quad }%
{\varepsilon }{(u)}=g_{IJ}{(u)}y^Iy^J,
$$
and $g^{*IJ}$ is inverse to $g_{IJ}^{*}{(u)}={\frac 12}{\frac{{\partial }%
^2\varepsilon }{{{\partial y^I}{\partial y^J}}}}.$ In this case
the coefficients of canonical metric d-connection (1.25) gives the
supersymmetric variants of coefficients of the Cartan connection
of Finsler
spaces. A similar remark applies to the Lagrange superspaces.

\subsection{Prolongations of  La\-gran\-ge s--spaces}

The geometric constructions on $T\widetilde{M}$ from the previus
subsection have corresponding generalizations to the
$Osc^{(z)}\widetilde{M}$ s--bundle. The basic idea is similar to
that used for prolongations of
geometric structures (see 
 [168] for prolongations on tangent bundle).
Having defined a metric structure $g_{IJ}(x)$ on a s-manifold
$\widetilde{M}$
we can extend it to the $Osc^z\tilde M$ s--bundle by considering $%
g_{IJ}(u_{(z)})=g_{IJ}(x)$ in (1.42). R. Miron and Gh. Atanasiu 
 [162]
solved the problem of prolongations of Finsler and Lagrange
structures on osculator bundle. In this subsection we shall
analyze supersymmetric extensions of Finsler and Lagrange
structures as well present a brief introduction into geometry of
higher order Lagrange s-spaces.

Let $F^{n,m}=(\tilde M,F)$ be a FS--space with the fundamental function $%
F_S:T\tilde M\to \Lambda $ on $\widetilde{M}.$ A prolongation of $F$ on $%
Osc^z\tilde M$ is given by a map%
$$
(F\circ \pi _1^z)(u_{(z)})=F(u_{(1)})
$$
and corresponding fundamental tensor
$$
g_{IJ}(u_{(1)})=\frac 12\frac{\partial ^2F^2}{\partial
y_{(1)}^I\partial y_{(1)}^J},
$$
for which
$$
(g_{IJ}\circ \pi _1^z)(u_{(z)})=g_{IJ}(u_{(1)}).
$$
So, $g_{IJ}(u_{(1)})$ is a ds--tensor on
$$\widetilde{Osk^z\widetilde{M}}%
=Osc^z\tilde M/\{0\}=\{(u_{(z)})\in Osc^z\tilde
M,srank|y_{(1)}^I|=1\}.$$

The Christoffel d--symbols
$$
\gamma _{IJ}^M(u^{(1)})=\frac 12g^{MK}(u_{(1)})(\frac{\partial
g_{KI}(u_{(1)})}{\partial x^J}+\frac{\partial g_{JK}(u_{(1)})}{\partial x^I}-%
\frac{\partial g_{IJ}(u_{(1)})}{\partial x^K})
$$
define the Cartan nonlinear connection 
 [55]:
$$
G_{(N)J}^I=\frac 12\frac \partial {\partial y_{(1)}^J}(\gamma
_{KM}^Iy_{(1)}^Ky_{(1)}^M).\eqno(1.44)
$$
The dual coefficients for the N-connection (1.21) are recurrently
computed by using (1.44) and operator
$$
\Gamma =y_{(1)}^I\frac \partial {\partial x^I}+2~y_{(2)}^I\frac
\partial {\partial y_{(1)}^I}+...+z~y_{(z)}^I\frac \partial
{\partial y_{(z-1)}^I},
$$
$$
M_{(1)J}^I=G_{(N)J}^I,
$$
$$
M_{(2)J}^I=\frac 12[\Gamma G_{(N)J}^I+G_{(N)K}^IM_{(1)J}^K],
$$
$$
..............
$$
$$
M_{(z)J}^I=\frac 1z[\Gamma M_{(z-1)J}^I+G_{(N)K}^IM_{(z-1)J}^K].
$$

The prolongations of FS--spaces can be generalized for Lagrange
s--spaces (on Lagrange spaces and theirs higher order extensions
see
 [160,161,162] and on supersymmetric extensions of Finsler geometry see
 [260]). Let $L^{n,m}=(\tilde M,g_{IJ})$ be a Lagrange s--space. The
Lagrangian $L:T\widetilde{M}\rightarrow \Lambda \,$ can be extended on $%
Osc^z\tilde M$ by using maps of the Lagrangian, $(L\circ \pi
_1^z)(u_{(z)})=L\left( u_{(1)}\right) ,$ and, as a consequence,
of the fundamental tensor (1.43), $(g_{IJ}\circ \pi
_1^z)(u_{(z)})=g_{IJ}\left( u_{(1)}\right) .$

\subsection{Higher order Lagrange s--spaces}

We introduce the notion of Lagrangian of z--order on a
differentiable s--manifold $\widetilde{M}$ as a map
$L^z:Osc^z\tilde M$ $\rightarrow \Lambda .$ In order to have
concordance with the definitions proposed by
 [162] we require the even part of the fundamental ds--tensor to be
of constant signature. Here we also note that questions to
considered in this subsection, being an supersymmetric approach,
are connected with the problem of elaboration of the so--called
higher order analytic mechanics (see, for instance,
 [64,150,149,226]).

A Lagrangian s--dif\-fe\-ren\-ti\-ab\-le of order $z$
($z=1,2,3,...)$ on s-dif\-fe\-ren\-ti\-ab\-le s--manifold
 $\widetilde{M}$ is an application $L^{(z)}:Osc^z%
\widetilde{M}\rightarrow \Lambda ,$ s--dif\-fe\-ren\-ti\-ab\-le
on $\widetilde{Osk^z\widetilde{M}}$ and smooth in the points of
 $Osc^z\widetilde{M}$ where $y_{(1)}^I=0.$

It is obvious that
$$
g_{IJ}(x,y_{(1)},...,y_{(z)})=\frac 12\frac{\partial
^2L^{(z)}}{\partial y_{(z)}^I\partial y_{(z)}^J}
$$
is a ds--tensor field because with respect to coordinate
transforms (1.3)
one holds transforms%
$$
K_I^{I^{\prime }}K_J^{J^{\prime }}g_{I^{\prime }J^{\prime
}}=g_{IJ.}
$$

A Lagrangian $L$ is regular if $srank|g_{IJ}|=(n,m).$

A Lagrange s--space of $z$--order is a pair $L^{(z,n,m)}=(\widetilde{M}%
,L^{(z)}),$ where $L^{(z)}$ is a s--differentiable regular Lagrangian of $z$%
--order, and with ds--tensor $g_{IJ}$ being of constant signature
on the even part of the basic s--manifold.

For details on nonsupersymmetric osculator bundles we cite 
 [162].

\section{Discussion}

We have explicitly constructed a new class of superspaces with
higher order anisotropy. The status of the results in this
Chapter and the relevant open questions are discussed as follows.

From the generally mathematical point of view it is possible a
definition of a supersymmetric differential geometric structure
imbedding both type of supersymmetric extensions of Finsler and
Lagrange geometry as well various Kaluza--Klein superspaces. The
first type of superspaces, considered as locally anisotropic,
are  characterized by nontrivial nonlinear connection structures
and corresponding distinguishing of geometric objects and basic
structure equations. The second type as a role is associated to
trivial nonlinear connections and higher order dimensions. A
substantial interest for further considerations presents the
investigations of physical consequences of models of field
interactions on higher and/or lower dimensional superspaces
provided with N--connection structure.

It worth noticing that higher order derivative theories are one
of currently central division in modern theoretical and
mathematical physics. It is necessary a rigorous formulation of
the geometric background for developing of higher order analytic
mechanics and corresponding extensions to classical quantum field
theories. Our results do not only contain a supersymmetric
extension of higher order fiber bundle geometry, but also propose
a general approach to the ''physics'' with local anisotropic
interactions. The elaborated in this Chapter formalism of
distinguished vector superbundles highlights a scheme by which
supergravitational and superstring theories with higher order
anisotropy can be constructed.



\chapter{HA--Supergravity}

In this Chapter we shall analyze three models of supergravity
with higher order anisotropy. We shall begin our considerations
with N--connection s--spaces in section 2.1. Such s--spaces are
generalizations of flat s--spaces containing a nontrivial
N--connection structure but with vanishing d--con\-nec\-ti\-on.
We shall introduce locally adapted s--vielbeins and define
s--fields and differential forms in N--connection s--spaces.
Sections 2.2 and 2.3 are correspondingly devoted to gauge s-field
and s--gravity theory in osculator s--bundles. In order to have
the possibility to compare our model with usual {\sf N}=1 ( one
dimensional supersymmetric extensions; see, for
instance, 
 286,170,215]) supergravitational models we develop a
supergavity theory on osculator s--bundle
$Osc^z\widetilde{M}_{(M)}$ where the even part of s--manifold
$\widetilde{M}_{(M)}$ has a local structure of Minkowski space
with action of Poincare group. In this case we do not have
problems connected with definition of spinors (Lorentz, Weyl or
Maiorana type) for spaces of arbitrary dimensions and can solve
Bianchi identities. As a matter of principle, by using our
results on higher dimensional and
locally anisotropic spaces, see 
 [256,255] we can introduce
distinguished spinor stuctures and develop variants of extended
supergravity with general higher order anisotropy. This approach
is based on global geometric constructions and allows us to avoid
tedious variational calculations and define the basic field
equations and conservation laws on s--spaces with local
anisotropy. That why, in section 2.5, we introduce
Einstein--Cartan equations on distinguished vector superbundles
(locally parametrized by arbitrary both type commuting and
anticommuting coordinates) in a geometric manner, in some line
following the geometric background for Einstein realtivity, but
in our case on dvs-bundels provided with arbitrary N--connection
and distinguished torsion and metric structures. We can consider
different models, for instance, with prescribed N--connection and
torsions, to develop a Einstein--Cartan like theory, or to follow
approaches from gauge gravity. In section 2.6 we propose a
variant of gauge like higher anisotropic supergravity being a
generalization to dvs--bundles of models of
locally anisotropic gauge gravity 
 [258,259,272] (see also Chapter 7
in this monograph).

\section{N--connection Superspaces}

Before we continue our analysis of the locally anisotropic
superspace, we shall first consider dvs--spaces provided with
N--connection structure, having trivial (flat like with respect
to la--frames) d--connection and d--metric. Such spaces will be
called as N--connection su\-per\-spa\-ces and denoted as
$\widetilde{{\cal E}}_N^{<z>}.$

\subsection{Supervielbeins in N--connection s--spaces}
Coordinates on s--space $\widetilde{{\cal E}}_N^{<z>}$ are denoted
$$
u^{<\alpha >}=(u^{\alpha
_p},y^{A_{p+1}},...,y^{A_z})=(x^I,y^{<A>})=
$$
$$
(\widehat{x}^i,\theta ^{\widehat{i}},\widehat{y}^{a_1},\zeta ^{\widehat{a}%
_1},...,\widehat{y}^{a_p},\zeta
^{\widehat{a}_p},...,\widehat{y}^{a_z},\zeta ^{\widehat{a}_z})
$$
We shall use la--frame decompositions of metric on $\widetilde{{\cal E}}%
_N^{<z>}:$%
$$
G_{<\alpha ><\beta >}(u)=l_{<\alpha >}^{<\underline{\alpha
}>}(u)l_{<\beta
>}^{<\underline{\beta }>}(u)G_{<\underline{\alpha }><\underline{\beta }>}.%
\eqno(2.1)
$$
Indices of type $<\underline{\alpha }>,<\underline{\beta }>,...$
are considered as abstract ones
 [180,181,\\ 182] s--vielbein indices. On
N-connection s--spaces we can fix la--frames and s--coordinates
when s--vielbein components $l_{<\alpha >}^{<\underline{\alpha
}>}$ and d--metrics $G_{<\underline{\alpha }><\underline{\beta
}>}$ and $G_{<\alpha
><\beta >}$ are constant.

We suppose that s--space $\widetilde{{\cal E}}_N^{<z>}$ is
provided with a set of $\sigma $--matrices
$$
\sigma _{\underleftarrow{i}\ \underrightarrow{i}}^{\underline{i}}=\sigma _{%
\underleftarrow{a_0}\ \underrightarrow{a_0}}^{\underline{a_0}}=l_i^{%
\underline{i}}\sigma _{\underleftarrow{i}\ \underrightarrow{i}}^i=l_{a_0}^{%
\underline{a_0}}\sigma _{\underleftarrow{a_0}\
\underrightarrow{a_0}}^{a_0},
$$
$$
\sigma _{\underleftarrow{a_1}\ \underrightarrow{a_1}}^{\underline{a_1}%
}=l_{a_1}^{\underline{a_1}}\sigma _{\underleftarrow{a_1}\ \underrightarrow{%
a_1}}^{a_1},...,\sigma _{\underleftarrow{a_p}\ \underrightarrow{a_p}}^{%
\underline{a_p}}=l_{a_p}^{\underline{a_p}}\sigma
_{\underleftarrow{a_p}\ \underrightarrow{a_p}}^{a_p},...,\sigma
_{\underleftarrow{a_z}\
\underrightarrow{a_z}}^{\underline{a_z}}=l_{a_z}^{\underline{a_z}}\sigma _{%
\underleftarrow{a_z}\ \underrightarrow{a_z}}^{a_z}
$$
necessary for spinor parametizations of anticommuting variables%
$$
\theta ^{\underline{i}}=\zeta ^{\underline{a}_0}=(\theta ^{\underleftarrow{i}%
}=\zeta ^{\underleftarrow{a_0}},\theta ^{\underrightarrow{i}}=\zeta ^{%
\underrightarrow{a_0}}),
$$
$$
\zeta ^{\underline{a_1}}=(\zeta ^{\underleftarrow{a_1}},\zeta ^{%
\underrightarrow{a_1}}),...,\zeta ^{\underline{a_p}}=(\zeta ^{%
\underleftarrow{a_p}},\zeta ^{\underrightarrow{a_p}}),...,\zeta ^{\underline{%
a_z}}=(\zeta ^{\underleftarrow{a_z}},\zeta
^{\underrightarrow{a_z}})
$$
(for simplicity, in sections 2.1--2.4 we consider Lorentz like spinor indices%
\\ $\left( ...,(\underleftarrow{a_p},\underrightarrow{a_p}),...\right) , $ $%
p=0,1,2,...,z$ for a 4--dimensional even component of a s--space,
but in our case provided with N--connection structure). For
symplicity, we shall omit dots before and after indices enabled
with subindex $p$ if this will not give rise to ambiguities.

The locally adapted to N--connection partial s--derivations are
introduced
in this manner:%
$$
\delta _{\underline{a_p}}=\frac \delta {\partial y^{\underline{a_p}%
}},~\delta _{\underleftarrow{a_p}}=\frac \delta {\partial \zeta ^{%
\underleftarrow{a_p}}}+i\sigma _{\underleftarrow{a_p}\ \underrightarrow{a_p}%
}^{a_p}\zeta ^{\underrightarrow{a_p}}\frac \delta {\partial
y^{a_p}},~\delta
_{\underrightarrow{a_p}}=-\frac \delta {\partial \zeta ^{\underleftarrow{a_p}%
}}-i\zeta ^{\underleftarrow{a_p}}\sigma _{\underleftarrow{a_p}\
\underrightarrow{a_p}}^{a_p}{}\frac \delta {\partial
y^{a_p}},\eqno(2.2)
$$
where $i$ is the imaginary unity and, for instance, $\delta _{\underline{a_0}%
}=\frac \delta {\partial y^{\underline{a_0}}}=\delta
_{\underline{i}}=\frac \delta {\partial x^{\underline{i}}}$.
La--derivations from (2.2) are consturcted by using operators
(1.12). In brief we denote (2.2) as $ \delta _{<\underline{\alpha
}>}=l_{<\underline{\alpha }>}^{<\alpha >}\delta _{<\alpha >}, $
where matrix $l_{<\underline{\alpha }>}^{<\alpha >}$ is
parametrized as
$$
l_{<\underline{\alpha }>}^{<\alpha >}=
$$
$$
\left(
\begin{array}{cccccccc}
\delta _{\underline{i}}^i & 0 & 0 & ... & 0 & 0 & 0 & ... \\
i\sigma _{\underleftarrow{\widehat{i}}\ \underrightarrow{i}}^i\theta ^{%
\underrightarrow{i}} & \delta _{\underleftarrow{\widehat{i}}}^{%
\underleftarrow{i}} & 0 & ... & 0 & 0 & 0 & ... \\
-i\theta ^{\underleftarrow{i}}\sigma _{\underleftarrow{i}\ \underrightarrow{%
\widehat{j}}}^i\varepsilon ^{\underrightarrow{\widehat{j}}\underrightarrow{%
\widehat{i}}} & 0 & -\delta _{\underrightarrow{i}}^{\underrightarrow{%
\widehat{i}}} & ... & 0 & 0 & 0 & ... \\
... & ... & ... & ... & ... & ... & ... & ... \\
0 & 0 & 0 & ... & \delta _{\underline{a_p}}^{a_p} & 0 & 0 & ... \\
0 & 0 & 0 & ... & i\sigma _{\underleftarrow{\widehat{a}_p}\ \underrightarrow{%
a_p}}^{a_p}\zeta ^{\underrightarrow{a_p}} & \delta _{\underleftarrow{%
\widehat{a}_p}}^{\underleftarrow{a_p}} & 0 & ... \\
0 & 0 & 0 & ... & -i\zeta ^{\underleftarrow{a_p}}\sigma _{\underleftarrow{a_p%
}\ \underrightarrow{\widehat{b}_p}}^{a_p}\varepsilon ^{\underrightarrow{%
\widehat{b}_p}\underrightarrow{\widehat{a}_p}} & 0 & -\delta _{%
\underrightarrow{a_p}}^{\underrightarrow{\widehat{a}_p}} & ... \\
... & ... & ... & ... & ... & ... & ... & ...
\end{array}
\right) ,
$$
$\varepsilon $--objects of type $\varepsilon ^{\underrightarrow{\widehat{b}_p%
}\underrightarrow{\widehat{a}_p}}$ are spinor metrics. The inverse matrix $%
l_{<\alpha >}^{<\underline{\alpha }>}$ satisfies conditions
$$
l_{<\underline{\alpha }>}^{<\alpha >}l_{<\beta >}^{<\underline{\alpha }%
>}=\delta _{<\beta >}^{<\alpha >},~l_{<\alpha >}^{<\underline{\alpha }>}l_{<%
\underline{\beta }>}^{<\alpha >}=\delta _{<\underline{\beta }>}^{<\underline{%
\alpha }>}
$$
and are parametrized as
$$
l_{<\alpha >}^{<\widehat{\alpha }>}=
$$
$$
\left(
\begin{array}{cccccccc}
\delta _i^{\underline{i}} & 0 & 0 & ... & 0 & 0 & 0 & ... \\
-i\sigma _{\underleftarrow{i}\ \underrightarrow{i}}^{\underline{i}}\theta ^{%
\underrightarrow{i}} & \delta _{\underleftarrow{i}}^{\underleftarrow{%
\widehat{i}}} & 0 & ... & 0 & 0 & 0 & ... \\
-i\theta ^{\underleftarrow{i}}\sigma _{\underleftarrow{i}\ \underrightarrow{j%
}}^i\varepsilon ^{\underrightarrow{j}\underrightarrow{i}} & 0 & -\delta _{%
\underrightarrow{\widehat{i}}}^{\underrightarrow{i}} & ... & 0 &
0 & 0 & ...
\\
... & ... & ... & ... & ... & ... & ... & ... \\
0 & 0 & 0 & ... & \delta _{a_p}^{\underline{a_p}} & 0 & 0 & ... \\
0 & 0 & 0 & ... & -i\sigma _{\underleftarrow{a_p}\ \underrightarrow{a_p}}^{%
\underline{a_p}}\zeta ^{\underrightarrow{a_p}} & \delta _{\underleftarrow{a_p%
}}^{\underleftarrow{\widehat{a}_p}} & 0 & ... \\
0 & 0 & 0 & ... & -i\zeta ^{\underleftarrow{a_p}}\sigma _{\underleftarrow{a_p%
}\ \underrightarrow{b_p}}^{\underline{a_p}}\varepsilon ^{\underrightarrow{b_p%
}\underrightarrow{a_p}} & 0 & -\delta _{\underrightarrow{\widehat{a}_p}}^{%
\underrightarrow{a_p}} & ... \\
... & ... & ... & ... & ... & ... & ... & ...
\end{array}
\right) .
$$
We call $l_{<\underline{\alpha }>}^{<\alpha >}~\left( l_{<\alpha >}^{<%
\underline{\alpha }>}\right) $ generalized supervielbien,
s--vielbein,
(inverse s--viel\-be\-in) of N--connection s--space.
Additionally to (2.2) we shall use differential operators:%
$$
P_{a_p}=\frac \delta {\partial y^{\widehat{a}_p}},~Q_{\underleftarrow{a_p}%
}=\frac \delta {\partial \theta ^{\underleftarrow{a_p}}}-i\sigma _{%
\underleftarrow{a_p}\ \underrightarrow{a_p}}^{a_p}\zeta ^{\underrightarrow{%
a_p}}\frac \delta {\partial
y^{a_p}},~Q_{\underrightarrow{a_p}}=-\frac
\delta {\partial \theta ^{\underleftarrow{a_p}}}+i\zeta ^{\underleftarrow{a_p%
}}\sigma _{\underleftarrow{a_p}\
\underrightarrow{a_p}}^{a_p}{}\frac \delta {\partial
y^{a_p}}.\eqno(2.3)
$$
For a trivial N--connection operators (2.2) and (2.3) are
transformed respectively into ''covariant'' and infinitesimal
generators on flat
s--spaces 
 [286,170,215].

\subsection{Locally anisotropic superfields}

Functions on $u^{<\alpha >}$ are called superfields (s--fields) in $%
\widetilde{{\cal E}}_N^{<z>}.$ For simplicity we consider a real
valued
s--field on dvs--bundle $\widetilde{{\cal E}}_N^{<1>}:$%
$$
V(u_{(p-1)},...,y_{(p)},\zeta _{(p)},...,y_{(z-1)},\zeta
_{(z-1)},y_{(z)},\zeta _{(z)})=
$$
$$
V^{+}(u_{(p-1)},...,y_{(p)},\zeta _{(p)},...,y_{(z-1)},\zeta
_{(z-1)},y_{(z)},\zeta _{(z)}),
$$
where by ''+'' is denoted the Hermitian conjugation. Every
s--field is a polynomial decomposition on variables $\left(
\theta ^i,...,\zeta
^{a_p},...\right) .$ For instance, with respect to $(\zeta ^{\underleftarrow{%
a_z}},\zeta ^{\underrightarrow{a_z}})=(\underleftarrow{\zeta _{(z)}},%
\underrightarrow{\zeta _{(z)}})$ we have a such type polynom (for
simplicity, here we omit spinor indices):%
$$
V(u_{(z-1)},y_{(z)},\zeta ^{\underleftarrow{a_z}},\zeta ^{\underrightarrow{%
a_z}})=C(u_{(z-1)},y_{(z)})+$$ $$i\underleftarrow{\zeta
_{(z)}}\chi
(u_{(z-1)},y_{(z)})-i\underrightarrow{\zeta _{(z)}}\overline{\chi }%
(u_{(z-1)},y_{(z)})+
$$
$$
\frac i2\underleftarrow{\zeta _{(z)}}\underleftarrow{\zeta
_{(z)}}(\mu (u_{(z-1)},y_{(z)})+i\nu (u_{(z-1)},y_{(z)}))-\frac
i2\underleftarrow{\zeta _{(z)}}\underleftarrow{\zeta _{(z)}}(\mu
(u_{(z-1)},y_{(z)})-
$$
$$
i\nu (u_{(z-1)},y_{(z)}))-\underleftarrow{\zeta _{(z)}}\sigma ^i%
\underrightarrow{\zeta
_{(z)}}v_i(u_{(z-1)},y_{(z)})+i\underleftarrow{\zeta
_{(z)}}\underleftarrow{\zeta _{(z)}}\underrightarrow{\zeta _{(z)}}\overline{%
\lambda }(u_{(z-1)},y_{(z)})-
$$
$$
i\underrightarrow{\zeta _{(z)}}\underrightarrow{\zeta _{(z)}}\underleftarrow{%
\zeta _{(z)}}\lambda (u_{(z-1)},y_{(z)})-\frac 12\underleftarrow{\zeta _{(z)}%
}\underleftarrow{\zeta _{(z)}}\delta _i\chi (u_{(z-1)},y_{(z)})\sigma ^i%
\underrightarrow{\zeta _{(z)}}+
$$
$$
\frac 12\underrightarrow{\zeta _{(z)}}\underrightarrow{\zeta _{(z)}}%
\underleftarrow{\zeta _{(z)}}\sigma ^i\delta _i\overline{\chi }%
(u_{(z-1)},y_{(z)})+
$$
$$
\underleftarrow{\zeta _{(z)}}\underleftarrow{\zeta _{(z)}}\underleftarrow{%
\zeta _{(z)}}\underleftarrow{\zeta _{(z)}}\left( \frac
12P(u_{(z-1)},y_{(z)})+\frac 14(\delta ^I\delta
_I)C(u_{(z-1)},y_{(z)})\right) ,...
$$
(in a similar manner we shall decompose on spinor variables\\ $(%
\underleftarrow{\zeta _{(z-1)}},\underrightarrow{\zeta _{(z-1)}}),...,(%
\underleftarrow{\zeta _{(1)}},\underrightarrow{\zeta _{(1)}}),(%
\underleftarrow{\theta },\underrightarrow{\theta })).$

\subsection{Differential forms in N--connection spaces}
For locally adapted differntials (1.13) we introduce the
supersymmetric
commutation rules%
$$
\delta u^{<\alpha >}\Lambda \delta u^{<\beta >}=-(-)^{|\alpha
\beta |}\delta u^{<\beta >}\Lambda \delta u^{<\alpha >}.
$$
For every integer $q$ we introduce the linear space generated by
basis elements\\ $\delta u^{<\alpha _1>},\delta u^{<\alpha
_2>},...,\delta u^{<\alpha _q>}$ where every multiple satisfies
the above presented
s--commutation rules. So, a function $\Phi (u)$ on $\widetilde{{\cal E}}%
_N^{<z>}$ is a 0--form, $\delta u^{<\alpha >}\Phi _{<\alpha >}$
is a 1--form, $\delta u^{<\alpha >}\Lambda \delta u^{<\beta
>}\Phi _{<\alpha
><\beta >}$ is a 2--form and so on. Forms can be multiplied by taking into
account that $u^{<\alpha >}\delta u^{<\beta >}=(-)^{|\alpha \beta
|}u^{<\beta >}\delta u^{<\alpha >}.$

Let
$$
\varpi =\delta u^{<\alpha _1>}\Lambda \delta u^{<\alpha
_2>}\Lambda ...\Lambda \delta u^{<\alpha _q>}
$$
be a q--form. The N--connection adapted differntial%
$$
\delta \varpi =\delta u^{<\alpha _1>}\Lambda \delta u^{<\alpha
_2>}\Lambda
...\Lambda \delta u^{<\alpha _q>}\Lambda \delta u^{<\alpha _{q+1}>}\frac{%
\delta \varpi }{\partial u^{<\alpha _{q+1}>}}
$$
is a (q+1)--form. One holds the Poincare lemma: $\delta \delta
=0.$ Under some topological restrictions
 [203,63,147] the inverse Poincare
lemma also holds: from $\delta \rho =0$ one follos that $\rho
=\delta \varpi $ (it should be noted here that on dvs--bundles we
must take into account the condition of existence of a
N--connection structure).

An arbitrary locally adapted basis $l^{<\widehat{\alpha }>}$ in
the space of 1--forms can be described by its s--vielbein matrix
(generally being
different from the (2.3)):%
$$
\delta u^{<\alpha >}l_{<\alpha >}^{<\underline{\alpha }>}(u_{(z)})=l^{<%
\underline{\alpha }>}.
$$
The inverse matrix $l_{<\underline{\alpha }>}^{<\alpha
>}(u_{(z)})$ is defined in a usual manner.

\section{HA--Ga\-u\-ge S--Fields}

This section is devoted to the geometric background of gauge
theory on curved s--spaces provided with N--connection structure.

\subsection{Gauge transforms in osculator s--bundles}

The structural group is a Lie group, acting on q--forms:%
$$
U=\exp \{i\Psi (u_{(z-1)},y_{(z)},\zeta ^{\underleftarrow{a_z}},\zeta ^{%
\underrightarrow{a_z}})\},
$$
$$
\Psi (u_{(z-1)},y_{(z)},\zeta ^{\underleftarrow{a_z}},\zeta ^{%
\underrightarrow{a_z}})=\sum\limits_{\widehat{e}}\gamma ^{\widehat{e}%
}(u_{(z-1)},y_{(z)},\zeta ^{\underleftarrow{a_z}},\zeta ^{\underrightarrow{%
a_z}})T_{\widehat{e}},
$$
where $T_{\widehat{e}}$ are generators of group :%
$$
\varsigma ^{\prime }=\varsigma U.
$$
The connection form%
$$
\varphi =\delta u^{<\alpha >}\varphi _{<\alpha >}(u_{(z-1)},y_{(z)},\zeta ^{%
\underleftarrow{a_z}},\zeta ^{\underrightarrow{a_z}})=l^{<\underline{\alpha }%
>}\varphi _{<\underline{\alpha }>}(u_{(z-1)},y_{(z)},\zeta ^{\underleftarrow{%
a_z}},\zeta ^{\underrightarrow{a_z}})\eqno(2.4)
$$
takes values in Lie algebra, i.e.%
$$
\varphi _{<\underline{\alpha }>}(u_{(z-1)},y_{(z)},\zeta ^{\underleftarrow{%
a_z}},\zeta ^{\underrightarrow{a_z}})=\sum\limits_{\widehat{e}}\varphi _{<%
\underline{\alpha }>}^{\widehat{e}}(u_{(z-1)},y_{(z)},\zeta ^{%
\underleftarrow{a_z}},\zeta
^{\underrightarrow{a_z}})T_{\widehat{e}}.
$$
This s--field is a higher order anisotropic generalization of
Yang--Mills
s--poten\-ti\-al. We have these transformation rules
$$
\varphi ^{\prime }=U^{-1}\varphi U+U^{-1}\delta U,
$$
$$
\varphi _{<\underline{\alpha }>}^{\prime }=U^{-1}\varphi _{<\underline{%
\alpha }>}U+U^{-1}\delta _{<\underline{\alpha }>}U,
$$
where $\delta _{<\underline{\alpha }>}=l_{<\underline{\alpha
}>}^{<\alpha
>}\delta _{<\alpha >}.$

The equation $K=\delta \varphi -\varphi \cdot \varphi $ shows
that we can construct ds--tensor values from connection
s--potential (2.4). Computing expressions
$$
\delta \varphi =l^{<\underline{\alpha }>}\delta \varphi
_{<\underline{\alpha }>}+\delta l^{<\underline{\alpha }>}~\varphi
_{<\underline{\alpha }>},
$$
$$
\delta \varphi _{<\underline{\alpha }>}=l^{<\underline{\beta }>}\delta _{<%
\underline{\beta }>}\varphi _{<\underline{\alpha }>},\delta l^{<\underline{%
\alpha }>}=$$ $$l^{<\underline{\beta }>}\delta _{<\underline{\beta }>}l^{<%
\underline{\alpha }>}\mbox{and}~\varphi \cdot \varphi
=l^{<\underline{\alpha
}>}l^{<\underline{\beta }>}\varphi _{<\underline{\alpha }>}\varphi _{<%
\underline{\beta }>},
$$
we get for the coefficients of $$K=l^{<\underline{\alpha }>}\Lambda l^{<%
\underline{\beta }>}K_{<\underline{\alpha }><\underline{\beta
}>},$$ where
$$K_{<\underline{\beta }><\underline{\alpha }>}=-(-)^{|\underline{\alpha }%
\underline{\beta }|}K_{<\underline{\alpha }><\underline{\beta
}>}$$ the
formula:%
$$
K_{<\underline{\beta }><\underline{\alpha }>}=\delta _{<\underline{\beta }%
>}\varphi _{<\underline{\alpha }>}-(-)^{|\underline{\alpha }\underline{\beta
}|}\delta _{<\underline{\alpha }>}\varphi _{<\underline{\beta }>}+
$$
$$
(-)^{|<\underline{\beta }>(<\underline{\alpha }>+<\underline{\mu }>)|}l_{<%
\underline{\alpha }>}^{<\alpha >}(\delta _{<\underline{\beta
}>}l_{<\alpha
>}^{<\underline{\gamma }>})\varphi _{<\underline{\gamma }>}-(-)^{|<%
\underline{\beta }><\underline{\mu }>|}l_{<\underline{\beta
}>}^{<\alpha
>}(\delta _{<\underline{\alpha }>}l_{<\alpha >}^{<\underline{\gamma }%
>})\varphi _{<\underline{\gamma }>}-
$$
$$
\varphi _{<\underline{\beta }>}\varphi _{<\underline{\alpha }>}+(-)^{|%
\underline{\alpha }\underline{\beta }|}\varphi _{<\underline{\alpha }%
>}\varphi _{<\underline{\beta }>}+w_{<\underline{\beta }><\underline{\alpha }%
>}^{<\gamma >}\varphi _{<\gamma >},
$$
where $w_{<\underline{\beta }><\underline{\alpha }>}^{<\gamma >}$
unholonomy coefficients are defined as in subsection 1.3.5. As an
example, we present
the structure of ds--components of the coefficients :%
$$
K_{a_pb_p}=\delta _{a_p}\varphi _{b_p}-\delta _{b_p}\varphi
_{a_p}+[\varphi _{a_p},\varphi _{b_p}]+w_{a_pb_p}^{<\alpha
>}\varphi _{<\alpha >},~
$$
$$
K_{a_p\underleftarrow{b_p}}=\delta _{a_p}\varphi _{\underleftarrow{b_p}%
}-\delta _{\underleftarrow{b_p}}\varphi _{a_p}+[\varphi _{a_p},\varphi _{%
\underleftarrow{b}}]+w_{a_p\underleftarrow{b_p}}^{<\alpha
>}\varphi _{<\alpha >},
$$
$$
K_{a_p\underrightarrow{b_p}}=\delta _{a_p}\varphi _{\underrightarrow{b_p}%
}-\delta _{\underrightarrow{b_p}}\varphi _{a_p}+[\varphi _{a_p},\varphi _{%
\underrightarrow{b_p}}]+w_{a_p\underrightarrow{b_p}}^{<\alpha
>}\varphi _{<\alpha >},~
$$
$$
K_{\underleftarrow{a_p}\underleftarrow{b_p}}=\delta _{\underleftarrow{a_p}%
}\varphi _{\underleftarrow{b_p}}+\delta _{\underleftarrow{b_p}}\varphi _{%
\underleftarrow{a_p}}+\{\varphi _{\underleftarrow{a_p}},\varphi _{%
\underleftarrow{b_p}}\}+w_{\underleftarrow{a_p\ }\underleftarrow{b_p}%
}^{<\alpha >}\varphi _{<\alpha >},
$$
$$
K_{\underrightarrow{a_p}\underrightarrow{b_p}}=\delta _{\underrightarrow{a_p}%
}\varphi _{\underrightarrow{b_p}}+\delta _{\underrightarrow{b_p}}\varphi _{%
\underrightarrow{a_p}}+\{\varphi _{\underrightarrow{a_p}},\varphi _{%
\underrightarrow{b_p}}\}+w_{\underrightarrow{a_p}\underrightarrow{b_p}%
}^{<\alpha >}\varphi _{<\alpha >},
$$
$$
K_{\underleftarrow{a_p}\ \underrightarrow{b_p}}=\delta _{\underleftarrow{a_p}%
}\varphi _{\underrightarrow{b_p}}+\delta _{\underrightarrow{b_p}}\varphi _{%
\underleftarrow{a_p}}+2i\sigma _{\underleftarrow{a_p}\ \underrightarrow{b_p}%
}^{\widehat{a}_p}\varphi _{\widehat{a}_p}+\{\varphi _{\underleftarrow{a_p}%
},\varphi _{\underrightarrow{b_p}}\}+w_{\underleftarrow{a_p}\
\underrightarrow{b_p}}^{<\alpha >}\varphi _{<\alpha >},
$$
where $p=0,1,2,...,z.$

In the next subsections we shall analyze constraints substantially
decreasing the number of independent fields containing components
of s--field $\varphi _{<\widehat{\alpha }>}$ without restrictions
on theirs dependence on coordinates $u^{<\alpha >}.$

\subsection{Abelian locally anisotropic s--fields}

This class of s--fields satisfy conditions $[\varphi _{<\alpha
>}\varphi _{<\beta >}\}=0.$ If constraints
$$
K_{\underleftarrow{a_p}\underleftarrow{b_p}}=\delta _{\underleftarrow{a_p}%
}\varphi _{\underleftarrow{b_p}}+\delta _{\underleftarrow{b_p}}\varphi _{%
\underleftarrow{a_p}}+w_{\underleftarrow{a_p\ }\underleftarrow{b_p}%
}^{<\alpha >}\varphi _{<\alpha >},K_{\underrightarrow{a_p}\underrightarrow{%
b_p}}=\delta _{\underrightarrow{a_p}}\varphi
_{\underrightarrow{b_p}}+\delta
_{\underrightarrow{b_p}}\varphi _{\underrightarrow{a_p}}+w_{\underrightarrow{%
a_p}\underrightarrow{b_p}}^{<\alpha >}\varphi _{<\alpha >}
$$
are imposed, there are such s--fields $A\left( u\right) ,B\left(
u\right) $ that
$$
\varphi _{\underleftarrow{a_p}}=-i\delta _{\underleftarrow{a_p}}A,~\varphi _{%
\underrightarrow{a_p}}=\delta _{\underrightarrow{a_p}}B,
$$
With resect to gauge shifts by a s--function $\kappa (u)$%
$$
A\rightarrow A+\kappa +S^{+},~S\mbox{ is a s--function satisfying}\ \delta _{%
\underrightarrow{i}}S=0,...,\delta _{\underrightarrow{a_p}}S=0,
$$
$$
B\rightarrow B-\kappa +T,\ T\mbox{ is a s--function satisfying}\ \delta _{%
\underrightarrow{i}}T=0,...,\delta _{\underrightarrow{a_p}}T=0
$$
one holds these transformation laws:%
$$
\varphi _{\underleftarrow{a_p}}\rightarrow \varphi _{\underleftarrow{a_p}%
}-i\delta _{\underleftarrow{a_p}}\kappa ,\varphi _{\underrightarrow{a_p}%
}=\varphi _{\underrightarrow{a_p}}-i\delta
_{\underrightarrow{a_p}}\kappa .
$$
If (additionally) equations%
$$
K_{\underleftarrow{a_p}\ \underrightarrow{b_p}}=\delta _{\underleftarrow{a_p}%
}\varphi _{\underrightarrow{b_p}}+\delta _{\underrightarrow{b_p}}\varphi _{%
\underleftarrow{a_p}}+2i\sigma _{\underleftarrow{a_p}\ \underrightarrow{b_p}%
}^{\widehat{a}_p}\varphi _{\widehat{a}_p}+w_{\underleftarrow{a_p}\
\underrightarrow{b_p}}^{<\alpha >}\varphi _{<\alpha >}=0\eqno(2.5)
$$
are satisfied, the s--functions $\varphi _{\widehat{i}},...,\varphi _{%
\widehat{a}_p},...$ can be expressed through $A$ and $B;$ gauge
transforms of these fields are parametrized as
$$
\varphi _{<\alpha >}\rightarrow \varphi _{<\alpha >}-i\delta
_{<\alpha
>}\kappa .
$$

So we can express s--field $\varphi _{<\alpha >},$ as well curvatures\\ $%
K_{ij},...,K_{a_pb_p},...,K_{i\underleftarrow{j}},K_{i\underrightarrow{j}%
},...,K_{a_p\underleftarrow{b_p}},K_{a_p\underrightarrow{b_p}},...$
as functions on $A,B.$ All invariants can be expressed through
values
$$
W_{\underleftarrow{a_p}}=\varepsilon ^{\underrightarrow{b_p}\
\underrightarrow{c_p}}\delta _{\underrightarrow{b_p}}\delta _{%
\underrightarrow{c_p}}\delta _{\underleftarrow{a_p}}(A+B),...,W_{%
\underrightarrow{a_p}}=\varepsilon ^{\underleftarrow{b_p}\underleftarrow{c_p}%
}\delta _{\underleftarrow{b_p}}\delta _{\underleftarrow{c_p}}\delta _{%
\underrightarrow{a_p}}(A+B),
$$
$$
K_{a_p\underleftarrow{b_p}}=\frac 18\sigma
_{a_p}^{\underrightarrow{c_p}\
\underleftarrow{d_p}}\varepsilon _{\underleftarrow{d_p}\ \underleftarrow{b_p}%
}W_{\underrightarrow{c_p}},K_{a_p\underrightarrow{b_p}}=-\frac
18\varepsilon
_{\underrightarrow{b_p}\ \underrightarrow{c_p}}\sigma _{a_p}^{%
\underrightarrow{c_p}\
\underleftarrow{d_p}}W_{\underleftarrow{d_p}},
$$
$$
K_{a_pb_p}=\frac i{64}\{(\varepsilon \sigma _{a_p}\sigma _{b_p})^{%
\underleftarrow{c_p}\ \underleftarrow{d_p}}(\delta _{\underleftarrow{d_p}}W_{%
\underleftarrow{c_p}}+
$$
$$
\delta
_{\underleftarrow{c_p}}W_{\underleftarrow{d_p}})+(\varepsilon
\sigma
_{a_p}\sigma _{b_p})^{\underrightarrow{c_p}\ \underrightarrow{d_p}}(\delta _{%
\underrightarrow{c_p}}W_{\underrightarrow{d_p}}+\delta _{\underrightarrow{d_p%
}}W_{\underrightarrow{c_p}})\}\ .
$$
From definition of $W_{\underleftarrow{i}},W_{\underrightarrow{i}},...,W_{%
\underleftarrow{a_p}},W_{\underrightarrow{a_p}},...$ one follows that%
$$
\delta _{\underrightarrow{d_p}}W_{\underleftarrow{c_p}}=0,\delta _{%
\underleftarrow{c_p}}W_{\underrightarrow{d_p}}=0,\delta ^{\underleftarrow{a_p%
}}W_{\underleftarrow{a_p}}-\delta _{\underrightarrow{b_p}}W^{%
\underrightarrow{b_p}}=0.
$$

The reality conditions for our theory are specified by conditions
$\varphi
=\varphi ^{+}$ within a gauge transform when $A+B=(A+B)^{+}$ $W_{%
\underrightarrow{i}}$ and $S=T.$ In this case the corresponding
Lagrangian is chosen as
$$
L\sim W^{\underleftarrow{i}}W_{\underleftarrow{i}}+W^{\underleftarrow{a_1}%
}W_{\underleftarrow{a_1}}+...+W^{\underleftarrow{a_z}}W_{\underleftarrow{a_z}%
}+W^{\underrightarrow{i}}W_{\underrightarrow{i}}+W^{\underrightarrow{a_1}}W_{%
\underrightarrow{a_p}}+...+W^{\underrightarrow{a_z}}W_{\underrightarrow{a_z}%
}.\eqno(2.6)
$$

In a similar manner we consider nonabelian gauge s--fields of
s--spaces with local anisotropy.

\subsection{Nonabelian locally anisotropic gauge s--fields}

Constraints are imposed as in the Abelian case:%
$$
K_{\underleftarrow{a_p}\
\underleftarrow{b_p}}=K_{\underrightarrow{a_p}\
\underrightarrow{b_p}}=K_{\underleftarrow{a_p}\
\underrightarrow{b_p}}=0.
$$
From $K_{\underleftarrow{i}\ \underleftarrow{j}}=...=K_{\underleftarrow{a_p}%
\ \underleftarrow{b_p}}=...0$ and $K_{\underrightarrow{i}\ \underrightarrow{j%
}}=...=K_{\underrightarrow{a_p}\ \underrightarrow{b_p}}=...0$ we
respectively obtain
$$
\varphi _{\underleftarrow{a_p}}=-e^{-A}\delta _{\underleftarrow{a_p}}e^A%
\eqno(2.7)
$$
and%
$$
\varphi _{\underrightarrow{a_p}}=-e^{-A}\delta
_{\underrightarrow{a_p}}e^A.
$$
Considering transforms
$$
e^A\rightarrow e^{S^{+}}e^Ae^\kappa ~\mbox{and}\ e^{-U}\rightarrow
e^{-T}e^{-A}e^\kappa ,\eqno(2.8)
$$
from which one follows transformation laws for $A$ and $B:$%
$$
A\rightarrow A+\kappa +S^{+}+...\ \mbox{and}\ B\rightarrow
B-\kappa +T+...,
$$
and imposing constraints of type (2.5) we can express, similarly
as in Abe\-li\-an case, the s--field $\varphi _{<\alpha >},$ and
curvatures $$
K_{ij},...,K_{a_pb_p},...,K_{i\underleftarrow{j}},K_{i\underrightarrow{j}%
},...,K_{a_p\underleftarrow{b_p}},K_{a_p\underrightarrow{b_p}},...$$
as
functions on $A,B,$ and, as a consequence, as functions of invariants $$W_{%
\underleftarrow{i}},W_{\underrightarrow{i}},...,W_{\underleftarrow{a_p}},W_{%
\underrightarrow{a_p}}.$$

Finally, in this subsection we remark that traces\\ $Tr(W^{\underleftarrow{i}%
}W_{\underleftarrow{i}}),Tr(W^{\underleftarrow{a_1}}W_{\underleftarrow{a_1}%
}),...,Tr(W^{\underleftarrow{a_z}}W_{\underleftarrow{a_z}})$ are
gauge
invariant and%
$$
\delta _{\underrightarrow{i}}Tr(W^{\underleftarrow{i}}W_{\underleftarrow{i}%
})=...=\delta _{\underrightarrow{a_p}}Tr(W^{\underleftarrow{a_p}}W_{%
\underleftarrow{a_p}})=...=0.
$$
So, we can use these traces and theirs complex conjugations in
order to define Lagrangians of type (2.6) for nonabelian gauge
theories. Reality conditions and gauge transforms can be
considered as in the Abelian case but by taking into account
changings (2.7) and (2.8).

\section{ Su\-per\-gra\-vi\-ty in Osculator S--Bundles}

The generalized s--vielbein $E_{<\alpha >}^{<\underline{\alpha
}>}$ and connection form $\Phi _{<\underline{\beta }><\mu
>}^{<\underline{\alpha }>},$ the last takes values in a Lie
algebra, are considered as basic variables on osculator bundle
$Osc^z\widetilde{M}$ with base $\widetilde{M}$ being of dimension
$\left( (3,1),1\right) $ where $\left( 3,1\right) $ denotes
respectively the dimension and signature of the even subspace and
$1$ is the dimention of the odd subspace. Our aim is to find
respectively a higher order extension of d--covarant equations
for the field of spin 2 and spin 3/2. As a posible structural
group we choose, for instance, the Lorentz subgroup (locally the
action of this group is split according to the fixed
N--connection structure). With respect to coordinate d--transforms
$$
\delta u^{<\mu ^{\prime }>}=\delta u^{<\nu >}\frac{\delta u^{<\mu
^{\prime }>}}{\partial u^{<\nu >}}
$$
one holds the transformation laws%
$$
E_{<\alpha ^{\prime }>}^{<\underline{\alpha }>}=E_{<\alpha >}^{<\underline{%
\alpha }>}\frac{\delta u^{<\alpha >}}{\partial u^{<\alpha
^{\prime }>}}\ \mbox{and}\ \Phi _{<\underline{\beta }><\mu
^{\prime }>}^{<\underline{\alpha
}>}=\frac{\delta u^{<\nu >}}{\partial u^{<\mu ^{\prime }>}}\Phi _{<%
\underline{\beta }><\nu >}^{<\underline{\alpha }>}.
$$

Let introduce 1--forms
$$
E^{<\underline{\alpha }>}=E_{<\mu >}^{<\underline{\alpha
}>}\delta u^{<\mu
>}\ \mbox{and}\ \Phi _{<\underline{\beta }>}^{<\underline{\alpha }>}=\Phi _{<%
\underline{\beta }><\mu >}^{<\underline{\alpha }>}\delta u^{<\mu
>}
$$
satisfying transformation laws of type:%
$$
E^{<\underline{\alpha ^{\prime }}>}=E^{<\underline{\alpha }>}X_{<\underline{%
\alpha }>}^{<\underline{\alpha }^{\prime }>}\ \mbox{and}\ \Phi _{<\underline{%
\beta }^{\prime }>}^{<\underline{\alpha }^{\prime }>}=X_{\quad <\underline{%
\beta }^{\prime }>}^{-1\ <\underline{\alpha }>}\Phi _{<\underline{\alpha }%
>}^{<\underline{\beta }>}X_{<\underline{\beta }>}^{<\underline{\alpha }%
^{\prime }>}+X_{\quad <\underline{\beta }^{\prime }>}^{-1\ <\underline{%
\alpha }>}\delta X_{<\underline{\alpha }>}^{<\underline{\alpha
}^{\prime }>}.
$$

The torsion and curvature are defined respectively by the first
and second
structure equations%
$$
\Omega ^{<\underline{\alpha }>}=\delta E^{<\underline{\alpha }>}-E^{<%
\underline{\beta }>}\Phi _{<\underline{\beta
}>}^{<\underline{\alpha }>},
$$
and
$$
R_{<\underline{\alpha }>}^{<\underline{\beta }>}=\delta \Phi _{<\underline{%
\alpha }>}^{<\underline{\beta }>}-\Phi _{<\underline{\alpha }>}^{<\underline{%
\gamma }>}\Phi _{<\underline{\gamma }>}^{<\underline{\beta }>}.
$$
The coefficients with values in Lie algebra are written in this form:%
$$
\Omega _{<\underline{\beta }><\underline{\gamma }>}^{<\underline{\alpha }%
>}=(-)^{|<\underline{\beta }>(<\underline{\gamma }>+<\underline{\mu }>)|}E_{<%
\underline{\gamma }>}^{<\mu >}E_{<\underline{\beta }>}^{<\nu
>}\delta _{<\nu
>}E_{<\mu >}^{<\underline{\alpha }>}-
$$
$$
(-)^{|<\underline{\gamma }><\underline{\mu }>)|}E_{<\underline{\beta }%
>}^{<\mu >}E_{<\underline{\gamma }>}^{<\nu >}\delta _{<\nu >}E_{<\mu >}^{<%
\underline{\alpha }>}-\Phi _{<\underline{\gamma }><\underline{\beta }>}^{<%
\underline{\alpha }>}+(-)^{|<\underline{\beta
}><\underline{\gamma }>|}\Phi
_{<\underline{\beta }><\underline{\gamma }>}^{<\underline{\alpha }>}%
\eqno(2.9)
$$
and
$$
R_{<\underline{\delta }><\underline{\varepsilon }><\underline{\alpha }>}^{<%
\underline{\beta }>}=(-)^{|<\underline{\delta }>(<\underline{\varepsilon }>+<%
\underline{\mu }>)|}E_{<\underline{\varepsilon }>}^{<\mu >}E_{<\underline{%
\delta }>}^{<\nu >}\delta _{<\nu >}\Phi _{<\underline{\alpha }><\mu >}^{<%
\underline{\beta }>}-
$$
$$
(-)^{|<\underline{\varepsilon }><\underline{\mu }>|}E_{<\underline{\delta }%
>}^{<\mu >}E_{<\underline{\varepsilon }>}^{<\nu >}\delta _{<\nu >}\Phi _{<%
\underline{\alpha }><\mu >}^{<\underline{\beta }>}-
$$
$$
(-)^{|<\underline{\delta }>(<\underline{\varepsilon }>+<\underline{\alpha }%
>+<\underline{\gamma }>)|}\Phi _{<\underline{\alpha }><\underline{%
\varepsilon }>}^{<\underline{\gamma }>}\Phi _{<\underline{\gamma }><%
\underline{\delta }>}^{<\underline{\beta }>}+$$
$$(-)^{|<\underline{\varepsilon }%
>(<\underline{\alpha }>+<\underline{\gamma }>)|}\Phi _{<\underline{\alpha }><%
\underline{\delta }>}^{<\underline{\gamma }>}\Phi _{<\underline{\gamma }><%
\underline{\varepsilon }>}^{<\underline{\beta }>}.
$$

Putting $E^{<\underline{\alpha }>}=l^{<\underline{\alpha }>},$ with $%
l_{<\alpha >}^{<\underline{\alpha }>}$ from (2.3), $\Phi
_{<\underline{\beta }><\underline{\gamma }>}^{<\underline{\alpha
}>}=0$ in (2.9) and for a vanishing N--connection we obtain that
torsion for a trivial osculator
s--space has components:%
$$
\Omega _{\underleftarrow{b_p}\ \underrightarrow{c_p}}^{(0)\underline{a_p}%
}=\Omega _{\underrightarrow{b_p}\ \underleftarrow{c_p}}^{(0)\underline{a_p}%
}=2i\sigma _{\underleftarrow{b_p}\ \underrightarrow{c_p}}^{\underline{a_p}%
}=0,\eqno(2.10)
$$
the rest of components are zero.

In order to consider the linearized osculator s--gravity we substitute $%
E_{<\alpha >}^{<\underline{\alpha }>}=l_{<\alpha >}^{<\underline{\alpha }%
>}+kh_{<\alpha >}^{<\underline{\alpha }>}$ in (2.3), where $k$ is the
interaction constant and $h_{<\alpha >}^{<\underline{\alpha }>}$
is linear perturbation of a ds--frame in N--connection s--space.
By straightfoward calculations we can verify that from equations
(2.10) and
$$
\Omega _{\underline{b_p}\underline{c_p}}^{\underline{a_p}}=\Omega _{%
\underleftarrow{b_p}\ \underleftarrow{c_p}}^{\underleftarrow{a_p}}=\Omega _{%
\underleftarrow{b_p}\ \underleftarrow{c_p}}^{\underrightarrow{a_p}}=\Omega _{%
\underrightarrow{b_p}\
\underrightarrow{c_p}}^{\underleftarrow{a_p}}=\Omega
_{\underleftarrow{b_p}\ \underleftarrow{c_p}}^{\underline{a_p}}=\Omega _{%
\underleftarrow{b_p}\ \underline{c_p}}^{\underline{a_p}}=0
$$
one follows only algebraic relations. In this case on the base
s--space of the osculator s--bundle we obtain exactly a dynamical
system of equations for spin 2 and spin 3/2 fields (for usual
supergravity see
 [70,81,93,47,288,215]). We can also solve nonlinear equations. It
is convenient to introduce the special gauge when for $\underleftarrow{%
\theta }=\underrightarrow{\theta }=...=\underleftarrow{\zeta _{(p)}}=%
\underrightarrow{\zeta _{(p)}}=...=0$ the s--vielbein and
ds--connection are prametrized
$$
E_i^{\underline{i}}=l_i^{\underline{i}}\left( u\right) ,...,E_{a_p}^{%
\underline{a_p}}=l_{a_p}^{\underline{a_p}}\left(
u_{(p-1)},x_{(p)}\right)
,...,E_{\underleftarrow{i}}^{\underleftarrow{\widehat{i}}}=\delta _{%
\underleftarrow{i}}^{\underleftarrow{\widehat{i}}},...,E_{\underleftarrow{a_p%
}}^{\underleftarrow{\widehat{a}_p}}=\delta _{\underleftarrow{a_p}}^{%
\underleftarrow{\widehat{a}_p}},...,...\eqno(2.11)
$$
$$
E_i^{\underleftarrow{i}}=\frac 12\psi _i^{\underleftarrow{i}%
}(x),...,E_{a_p}^{\underleftarrow{a_p}}=\frac 12\psi _{a_p}^{\underleftarrow{%
a_p}}(u_{(p-1)},y_{(p)}),...,
$$
$$
E_{i\underrightarrow{i}}=\frac 12\psi _{i\underrightarrow{j}%
}(u_{(p-1)},x_{(p)}),...,E_{a_p\underrightarrow{a_p}}=\frac 12\psi _{a_p%
\underrightarrow{a_p}}(u_{(p-1)},y_{(p)}),...,
$$
the rest of components of the s--vielbein are zero, and
$$
\Phi _{\underleftarrow{j}\ k}^{\underleftarrow{i}}=\varphi _{\underleftarrow{%
j}\ k}^{\underleftarrow{i}}(x),...,\Phi _{\underleftarrow{b_p}\ c_p}^{%
\underleftarrow{a_p}}=\varphi _{\underleftarrow{b_p}\ c_p}^{\underleftarrow{%
a_p}}(u_{(p-1)},y_{(p)}),...,\eqno(2.12)
$$
the rest of components of ds--connection are zero.

Fields $l_i^{\underline{i}}\left( u\right) ,...,l_{a_p}^{\underline{a_p}%
}\left( u_{(p-1)}\right) ,...,\psi
_i^{\underleftarrow{i}}(x),...,\psi
_{a_p}^{\underleftarrow{a_p}}(u_{(p-1)},y_{(p)}),...$ and\\ $\varphi _{%
\underleftarrow{j}\ k}^{\underleftarrow{i}}(x),...\varphi _{\underleftarrow{%
b_p}\ c_p}^{\underleftarrow{a_p}}(u_{(p-1)},y_{(p)}),...$ from
(2.11) and (2.12) are corresponding extensions of the tetrad,
Rarita--Schwinger and
connectiom fields on osculator bundle 
 [162].

We note that equations
$$
\Omega _{\underleftarrow{i}\ \underline{k}}^{\underrightarrow{j}}=\Omega _{%
\underleftarrow{i}\ \underline{k}}^{\underrightarrow{j}}=0\ \left(
...,\Omega _{\underleftarrow{b_p}\ \underline{c_p}}^{\underrightarrow{a_p}%
}=\Omega _{\underleftarrow{b_p}\ \underline{c_p}}^{\underrightarrow{a_p}%
}=0,...\right)
$$
defines the dynamics in $x$--space ( in $\left( u_{(p-1)},x_{(p)}\right) $%
--space).The rest of nonvanishing components of torsion are
computed by
putting components (2.11) and (2.12) into (2.9) :%
$$
\Omega _{ij\mid \underleftarrow{\theta }=\underrightarrow{\theta }=0}^{%
\widehat{k}}={\cal T}_{ij}^{\widehat{k}}=\frac i2(\psi _i\sigma ^{\widehat{k}%
}\overline{\psi }_j-\psi _j\sigma ^{\widehat{k}}\overline{\psi
}_i),...,
$$
$$
\Omega _{b_pc_p\mid \underleftarrow{\zeta
_{(p)}}=\underrightarrow{\zeta _{(p)}}=0}^{\widehat{a}_p}={\cal
T}_{b_pc_p}^{\widehat{a}_p}=\frac i2(\psi
_{b_p}\sigma ^{\widehat{k}}\overline{\psi }_{c_p}-\psi _{c_p}\sigma ^{%
\widehat{k}}\overline{\psi }_{b_p}),...,
$$
which shows that torsion in dvs--bundles can be generated by a
corresponding distribution of spin density, and
$$
\Omega _{ij\mid \underleftarrow{\theta }=\underrightarrow{\theta }=0}^{%
\underleftarrow{i}}={\cal T}_{ij}^{\underleftarrow{i}}=\frac 12(D_i\psi _j^{%
\underleftarrow{i}}-D_j\psi _i^{\underleftarrow{i}}),...,
$$
$$
\Omega _{b_pc_p\mid \underleftarrow{\zeta
_{(p)}}=\underrightarrow{\zeta
_{(p)}}=0}^{\underleftarrow{a_p}}={\cal T}_{b_pc_p}^{\underleftarrow{a_p}%
}=\frac 12(D_{b_p}\psi _{c_p}^{\underleftarrow{a_p}}-D_{c_p}\psi _{b_p}^{%
\underleftarrow{i}}),...,
$$
where $D_i,...,D_{b_p},...$ are usual d--covariant derivatives on
the base of osculator space.

We note that in this and next sections we shall omit tedious
calculations
being similar to those from 
 [288]; we shall present the final results
and emphasize that we can verify them straightforward manner by
taking into account the distinguished character of geometical
objects and the interactions with the N--connection fields).

The Bianchi identities in osculator s--bund\-le are written as
$$
\delta \Omega ^{<\underline{\alpha }>}+\Omega ^{<\underline{\beta }>}\Phi _{<%
\underline{\beta }>}^{<\underline{\alpha }>}-E^{<\underline{\beta }>}R_{~<%
\underline{\beta }>}^{<\underline{\alpha }>}=0,
$$
$$
\delta R_{~<\underline{\alpha }>}^{<\underline{\beta }>}+R_{~<\underline{%
\alpha }>}^{<\underline{\gamma }>}R_{~<\underline{\gamma }>}^{<\underline{%
\beta }>}-\Phi _{<\underline{\alpha }>}^{<\underline{\gamma }>}\Phi _{<%
\underline{\gamma }>}^{<\underline{\beta }>}=0,
$$
or, in coefficient form, as
$$
E^{<\underline{\gamma }>}E^{<\underline{\beta }>}E^{<\underline{\alpha }%
>}(E_{<\underline{\alpha }>}^{<\mu >}{\cal D}_{<\mu >}\Omega _{<\underline{%
\beta }><\underline{\gamma }>}^{<\underline{\delta }>}+
$$
$$
\Omega _{<\underline{\alpha }><\underline{\beta }>}^{<\underline{\tau }%
>}\Omega _{<\underline{\tau }><\underline{\gamma >}}^{<\underline{\delta }%
>}-R_{<\underline{\alpha }><\underline{\beta }><\underline{\gamma }>}^{<%
\underline{\delta }>}=0,\eqno(2.13)
$$
$$
E^{<\underline{\gamma }>}E^{<\underline{\beta }>}E^{<\underline{\alpha }%
>}\left( E_{<\underline{\alpha }>}^{<\mu >}{\cal D}_{<\mu >}R_{<\underline{%
\beta }><\underline{\gamma }><\underline{\delta }>}^{<\underline{\tau }%
>}+\Omega _{<\underline{\alpha }><\underline{\beta }>}^{<\underline{%
\varepsilon }>}R_{<\underline{\varepsilon }><\underline{\gamma }><\underline{%
\delta }>}^{<\underline{\tau }>}\right) =0,
$$
where the supersymmetric gauge d--covariant derivation ${\cal
D}_{<\mu >}$ acts, for instance, as
$$
{\cal D}_{<\mu >}X^{<\underline{\alpha }>}=\delta _{<\mu >}X^{<\underline{%
\alpha }>}+(-)^{|<\underline{\beta }><\mu >|}X^{<\underline{\beta }>}\Phi _{<%
\underline{\beta }><\mu >}^{<\underline{\alpha }>},
$$
and%
$$
{\cal D}_{<\mu >}X_{<\underline{\alpha }>}=\delta _{<\mu >}X_{<\underline{%
\alpha }>}-\Phi _{<\underline{\alpha }><\mu >}^{<\underline{\beta }>}X_{<%
\underline{\beta }>}.
$$

Introducing parametrizations (2.11) (the special gauge) in (2.13)
we find
equations:%
$$
\sigma _{\underleftarrow{b_p}\ \underrightarrow{c_p}}^{\underline{a_p}}l_{%
\underline{a_p}}^{d_p}l_{\underline{e_p}}^{e_p}\left( D_{d_p}\overline{\psi }%
_{e_p}^{\underrightarrow{c_p}}-D_{e_p}\overline{\psi }_{d_p}^{%
\underrightarrow{c_p}}\right) =0\eqno(2.14)
$$
(the Rarita--Shwinger equa\-ti\-ons on osculator bundle) and
$$
{\cal R}_{\underline{a_p}\ \underline{c_p}\ \underline{b_p}}^{\underline{b_p}%
}+il^{\underline{b_p}\ d_p}\psi _{d_p}^{\underleftarrow{e_p}}\sigma _{%
\underline{a_p}\ \underleftarrow{e_p}\ \underrightarrow{f_p}}{\cal T}_{%
\underline{b_p}\ \underline{c_p}}^{\underrightarrow{f_p}}+il^{\underline{b_p}%
d_p}\overline{\psi }_{d_p}^{\underrightarrow{f_p}}\sigma
_{\underline{a_p}\ \underleftarrow{e_p}\
\underrightarrow{f_p}}{\cal T}_{\underline{b_p}\
\underline{c_p}}^{\underleftarrow{e_p}}=0,\eqno(2.15)
$$
where components of sourse are
$$
{\cal T}_{\underline{a_p}\
\underline{b_p}}^{\underrightarrow{c_p}}=\frac
12l_{\underline{a_p}}^{a_p}l_{\underline{b_p}}^{b_p}(D_{a_p}\psi _{b_p}^{%
\underrightarrow{c_p}}-D_{b_p}\psi
_{a_p}^{\underrightarrow{c_p}}),
$$
and curvature (in the special gauge)\ is expressed as%
$$
{\cal R}_{\underline{a_p}\underline{b_p}c_pd_p}=l_{c_p}^{\underline{e_p}%
}l_{d_p}^{\underline{f_p}}R_{\underline{a_p}\underline{b_p}\underline{e_p}%
\underline{f_p}}-i(l_{c_p}^{\underline{f_p}}\psi _{d_p}^{\underleftarrow{s_p}%
}-l_{d_p}^{\underline{f_p}}\psi _{c_p}^{\underleftarrow{s_p}})\sigma _{%
\underline{f_p}\ \underleftarrow{s_p}\ \underrightarrow{t_p}}\Omega _{%
\underline{a_p}\ \underline{b_p}}^{\underrightarrow{t_p}}-
$$
$$
i(l_{d_p}^{\underline{f_p}}\overline{\psi }_{c_p\underrightarrow{s_p}%
}-l_{c_p}^{\underline{f_p}}\overline{\psi }_{d_p\underrightarrow{s_p}%
})\sigma _{\underline{f_p}}^{\underleftarrow{t_p}\ \underrightarrow{s_p}%
}\Omega _{\underline{a_p}\ \underline{b_p}\
\underleftarrow{t_p}\mid \underleftarrow{\zeta
_{(p)}}=\underrightarrow{\zeta _{(p)}}=0},
$$
$$
R_{\underline{b_p}c_pd_p}^{\underline{a_p}}=E_{c_p}^{\underline{c_p}%
}E_{d_p}^{\underline{d_p}}R_{\underline{b_p}\underline{c_p}\underline{d_p}}^{%
\underline{a_p}}+E_{c_p}^{\underleftarrow{s_p}}E_{d_p}^{\underline{d_p}}R_{%
\underline{b_p}\underleftarrow{s_p}\underline{d_p}}^{\underline{a_p}}+
$$
$$
E_{c_p}^{\underline{c_p}}E_{d_p}^{\underleftarrow{s_p}}R_{\underline{b_p}%
\underleftarrow{s_p}\underline{c_p}}^{\underline{a_p}}+E_{c_p%
\underrightarrow{t_p}}E_{d_p}^{\underline{d_p}}R_{.\underline{b_p}.%
\underline{d_p}}^{\underline{a_p}.\underrightarrow{t_p}}+E_{c_p}^{\underline{%
c_p}}E_{d_p\underrightarrow{t_p}}R_{\underline{b_p}\underline{c_p}}^{%
\underline{a_p}..\underrightarrow{t_p}}=
$$
$$
\delta _{d_p}\Phi _{\underline{b_p}c_p}^{\underline{a_p}}-\delta
_{c_p}\Phi
_{\underline{b_p}d_p}^{\underline{a_p}}+\Phi _{\underline{b_p}d_p}^{%
\underline{f_p}}\Phi _{\underline{f_p}c_p}^{\underline{a_p}}-\Phi _{%
\underline{b_p}c_p}^{\underline{f_p}}\Phi _{\underline{f_p}d_p}^{\underline{%
a_p}}+w_{c_pd_p}^{e_p}\Phi
_{e_p\underline{b_p}}^{\underline{a_p}}.
$$

Finally, in this section, we note that for trivial N--connection
sructures and on vector superbundles the equations (2.14) and
(2.15) are transformed into dynamical field equations for the
model of supergravity developed by
S.\ Deser and B. Zumino [70]. 

\section{Bianchi Identities: Osculat\-or S--Bun\-d\-les}

The purpose of this section is the analyse of Bianci identities
in the
framework of locally anisotropic supergraviry theory on osculator
s--bundle. We shall impose s--gravitational constraints and solve
these identities with respect to s--fields and theirs
ds--covariant derivations.

\subsection{Distinguiched Bianchi identities}

The Bianchi identities are written as (2.13). Constraints on
torsion are imposed in general form as (2.11) with the rest of
components being zero. By using the technique developed in
 [95] we shall solve in explicit form
the system (2.13) with the mentioned type of constraints on
osculator s--bundle in explicit form. We note that to do this we
shall not use an explicit form of ds--covariant derivation; the
necessary information is
contained in the s--symmetric commutator%
$$
[{\cal D}_{<\alpha >},{\cal D}_{<\beta >}\}=-R_{~\bullet <\alpha
><\beta
>}^{\bullet }-\Omega _{<\alpha ><\beta >}^{<\gamma >}{\cal D}_{<\gamma >}.
$$

In order to find solutions we distinguish identities (2.13) in this form:%
$$
R_{\underleftarrow{a_p}\underleftarrow{b_p}\underleftarrow{c_d}%
\underleftarrow{ep}}+R_{\underleftarrow{b_p}\underleftarrow{c_d}%
\underleftarrow{a_p}\underleftarrow{ep}}+R_{\underleftarrow{c_d}%
\underleftarrow{a_p}\underleftarrow{b_p}\underleftarrow{ep}}=0,\eqno(2.16)
$$

$$
R_{\underleftarrow{a_p}\underrightarrow{b_p}\underleftarrow{c_p}%
\underleftarrow{d_p}}+R_{\underrightarrow{b_p}\underleftarrow{c_p}%
\underleftarrow{a_p}\underleftarrow{d_p}}+2i\sigma _{\underleftarrow{c_p}%
\underrightarrow{b_p}}^{e_p}\Omega _{\underleftarrow{a_p}e_p\underleftarrow{%
d_p}}+2i\sigma _{\underleftarrow{a_p}\underrightarrow{b_p}}^{e_p}\Omega _{%
\underleftarrow{c_p}e_p\underleftarrow{d_p}}=0,\eqno(2.17)
$$

$$
R_{\underleftarrow{a_p}\underleftarrow{b_p}\underrightarrow{c_p}%
\underrightarrow{d_p}}=-2i\sigma _{\underleftarrow{b_p}\underrightarrow{c_p}%
}^{e_p}\Omega _{\underleftarrow{a_p}e_p\underrightarrow{d_p}}-2i\sigma _{%
\underleftarrow{a_p}\underrightarrow{c_p}}^{e_p}\Omega _{\underleftarrow{b_p}%
e_p\underrightarrow{d_p}},\eqno(2.18)
$$

$$
R_{\underrightarrow{a_p}\underrightarrow{b_p}\underline{c_p}\underline{d_p}%
}=-2i\sigma _{\underline{d_p}\underleftarrow{s_p}\underrightarrow{b_p}%
}\Omega _{\underrightarrow{a_p}\underline{c_p}}^{\underleftarrow{s_p}%
}-2i\sigma
_{\underline{d_p}\underleftarrow{s_p}\underrightarrow{a_p}}\Omega
_{\underrightarrow{b_p}\underline{c_p}}^{\underleftarrow{s_p}},\eqno(2.19)
$$

$$
R_{\underleftarrow{a_p}\underrightarrow{b_p}\underline{c_p}\underline{d_p}%
}=-2i\sigma _{\underline{d_p}\underleftarrow{s_p}\underrightarrow{b_p}%
}\Omega _{\underleftarrow{a_p}\underline{c_p}}^{\underleftarrow{s_p}%
}-2i\sigma
_{\underline{d_p}\underleftarrow{s_p}\underleftarrow{a_p}}\Omega
_{\underrightarrow{b_p}\underline{c_p}}^{\underleftarrow{s_p}},\eqno(2.20)
$$

$$
R_{\underline{a_p}\underline{b_p}\underline{c_p}\underline{d_p}}+R_{%
\underline{b_p}\underline{c_p}\underline{a_p}\underline{d_p}}+R_{\underline{%
c_p}\underline{a_p}\underline{b_p}\underline{d_p}}=0,\eqno(2.21)
$$
(linear equations without derivatives)%
$$
R_{\underleftarrow{a_p}\underline{b_p}\underleftarrow{c_p}\underleftarrow{d_p%
}}+R_{\underleftarrow{c}\underline{b_p}\underleftarrow{a_p}\underleftarrow{%
d_p}}+{\cal D}_{\underleftarrow{c_p}}\Omega _{\underleftarrow{a_p}\underline{%
b_p}\underleftarrow{d_p}}+{\cal D}_{\underleftarrow{a_p}}\Omega _{%
\underleftarrow{c_p}\underline{b_p}\underleftarrow{d_p}}=0,\eqno(2.22)
$$

$$
R_{\underline{a_p}\underleftarrow{b_p}\underrightarrow{c_p}\underrightarrow{%
d_p}}={\cal D}_{\underleftarrow{b_p}}\Omega _{\underrightarrow{c_p}%
\underline{a_p}\underrightarrow{d_p}}+{\cal
D}_{\underrightarrow{c_p}}\Omega
_{\underleftarrow{b_p}\underline{a_p}\underrightarrow{d_p}}+2i\sigma _{%
\underleftarrow{b_p}\underrightarrow{c_p}}^{m_p}\Omega _{m_p\underline{a_p}%
\underrightarrow{d_p}},\eqno(2.23)
$$

$$
{\cal D}_{\underrightarrow{k}}\Omega _{\underrightarrow{j}\underline{i}%
\underrightarrow{l}}+{\cal D}_{\underrightarrow{j}}\Omega _{\underrightarrow{%
k}\underline{i}\underrightarrow{l}}=0,\eqno(2.24)
$$
(linear identities containing derivations) and%
$$
R_{\underline{a_p}\underline{b_p}\underrightarrow{c_p}\underrightarrow{d_p}}=%
{\cal D}_{\underline{a_p}}\Omega _{\underline{b_p}\underrightarrow{c_p}%
\underrightarrow{d_p}}+{\cal D}_{\underline{b_p}}\Omega _{\underrightarrow{%
c_p}\underline{a_p}\underrightarrow{d_p}}+{\cal D}_{\underrightarrow{c_p}%
}\Omega
_{\underline{a_p}\underline{b_p}\underrightarrow{d_p}}+\eqno(2.25)
$$
$$
\Omega
_{\underline{b_p}\underrightarrow{c_p}}^{\underleftarrow{m_p}}\Omega
_{\underleftarrow{m_p}\underline{a_p}\underrightarrow{d_p}}+\Omega _{%
\underline{b_p}\underrightarrow{c_p}}^{\underrightarrow{m_p}}\Omega _{%
\underrightarrow{m_p}\underline{a_p}\underrightarrow{d_p}}+\Omega _{%
\underrightarrow{c_p}\underline{a_p}}^{\underleftarrow{m_p}}\Omega _{%
\underleftarrow{m_p}\underline{b_p}\underrightarrow{d_p}}+\Omega _{%
\underrightarrow{c_p}\underline{a_p}}^{\underrightarrow{m_p}}\Omega _{%
\underrightarrow{m_p}\underline{b_p}\underrightarrow{d_p}},...,
$$

$$
R_{\underline{a_p}\underline{b_p}\underrightarrow{c_p}\underrightarrow{d_p}}=%
{\cal D}_{\underline{a_p}}\Omega _{\underline{b_p}\underrightarrow{c_p}%
\underleftarrow{d_p}}+{\cal D}_{\underline{b_p}}\Omega _{\underrightarrow{c_p%
}\underline{a_p}\underleftarrow{d_p}}+{\cal
D}_{\underrightarrow{c_p}}\Omega
_{\underline{a_p}\underline{b_p}\underleftarrow{d_p}}+\eqno(2.26)
$$
$$
\Omega
_{\underline{b_p}\underrightarrow{c_p}}^{\underleftarrow{m_p}}\Omega
_{\underleftarrow{m_p}\underline{a_p}\underleftarrow{d_p}}+\Omega _{%
\underline{b_p}\underrightarrow{c_p}}^{\underrightarrow{m_p}}\Omega _{%
\underrightarrow{m_p}\underline{a_p}\underleftarrow{d_p}}+\Omega _{%
\underrightarrow{c_p}\underline{a_p}}^{\underleftarrow{m_p}}\Omega _{%
\underleftarrow{m_p}\underline{b_p}\underleftarrow{d_p}}+\Omega _{%
\underrightarrow{c_p}\underline{a_p}}^{\underrightarrow{m_p}}\Omega _{%
\underrightarrow{m_p}\underline{b_p}\underleftarrow{d_p}}=0,...,
$$

$$
{\cal D}_{\underline{a_p}}\Omega _{\underline{b_p}\underline{c_p}%
\underleftarrow{d_p}}+{\cal D}_{\underline{b_p}}\Omega _{\underline{c_p}%
\underline{a_p}\underleftarrow{d_p}}+{\cal D}_{\underline{c_p}}\Omega _{%
\underline{a_p}\underline{b_p}\underleftarrow{d_p}}+\Omega _{\underline{a_p}%
\underline{b_p}}^{\underleftarrow{m_p}}\Omega _{\underleftarrow{m_p}%
\underline{c_p}\underleftarrow{d_p}}+\Omega _{\underline{a_p}\underline{b_p}%
}^{\underrightarrow{m_p}}\Omega _{\underrightarrow{m_p}\underline{c_p}%
\underleftarrow{d_p}}+\eqno(2.27)
$$
$$
\Omega _{\underline{b_p}\underline{c_p}}^{\underleftarrow{m_p}}\Omega _{%
\underleftarrow{m_p}\underline{a_p}\underleftarrow{d_p}}+\Omega _{\underline{%
b_p}\underline{c_p}}^{\underrightarrow{m_p}}\Omega _{\underrightarrow{m_p}%
\underline{a_p}\underleftarrow{d_p}}+\Omega _{\underline{c_p}\underline{a_p}%
}^{\underleftarrow{m_p}}\Omega _{\underleftarrow{m_p}\underline{b_p}%
\underleftarrow{d_p}}+\Omega _{\underline{c_p}\underline{a_p}}^{%
\underrightarrow{m_p}}\Omega _{\underrightarrow{m_p}\underline{b_p}%
\underleftarrow{d_p}},...~
$$
(nonlinear identities).

For a trivial osculator s--bundle
$Osc^0\widetilde{M}=\widetilde{M},\dim
\widetilde{M}=(4,1),\widetilde{M}$ being a s--simmetric extension
of the Lorentz bundle formulas (2.16)--(2.27) are transformed
into the Bianchi identities for the locally isotropic s--gravity
model considered by
 [95].

\subsection{Solution of distinguished Bianchi identities}

It is convinient to use spinor decompositions of curvatures and
torsions
(see, for instance, 
 [180,181,182])
$$
R_{\underrightarrow{a_p}\underrightarrow{b_p}\underleftarrow{c_p}%
\underrightarrow{d_p}\underleftarrow{e_p}\underrightarrow{f_p}}=\sigma _{%
\underleftarrow{c_p}\underrightarrow{d_p}}^{d_p}\sigma _{\underleftarrow{e_p}%
\underrightarrow{f_p}}^{f_p}R_{\underrightarrow{a_p}\underrightarrow{b_p}%
d_pf_p},\quad \Omega _{\underrightarrow{a_p}\underleftarrow{b_p}%
\underrightarrow{c_p}\underleftarrow{d_p}}=\sigma _{\underleftarrow{b_p}%
\underrightarrow{c_p}}^{e_p}\Omega _{\underrightarrow{a_p}e_p\underleftarrow{%
d_p}}
$$
and
$$
R_{\underrightarrow{a_p}\underrightarrow{b_p}\underleftarrow{c_p}%
\underrightarrow{d_p}\underleftarrow{e_p}\underrightarrow{f_p}%
}=-2\varepsilon _{\underleftarrow{c_p}\underleftarrow{e_p}}R_{%
\underrightarrow{a_p}\underrightarrow{b_p}\underrightarrow{d_p}%
\underrightarrow{f_p}}+2\varepsilon _{\underrightarrow{d_p}\underrightarrow{%
f_p}}R_{\underrightarrow{a_p}\underrightarrow{b_p}\underleftarrow{c_p}%
\underleftarrow{e_p}},...\ .
$$

Let start with the solutions of linear equations without
derivatives. By straightforward calculations we can verify that
$$
\Omega _{\underrightarrow{i}\underleftarrow{j}\underrightarrow{k}%
\underleftarrow{l}}=-2i\varepsilon _{\underrightarrow{i}\underrightarrow{k}%
}\varepsilon _{\underleftarrow{j}\underleftarrow{l}}R_{(0)},...,\Omega _{%
\underrightarrow{a_p}\underleftarrow{b_p}\underrightarrow{c_p}%
\underleftarrow{d_p}}=-2i\varepsilon _{\underrightarrow{a_p}\underrightarrow{%
c_p}}\varepsilon _{\underleftarrow{b_p}\underleftarrow{d_p}}R_{(p)},...%
\eqno(2.28)
$$
and
$$
R_{\underrightarrow{i}\underrightarrow{j}\underrightarrow{k}\underrightarrow{%
l}}=4\left( \varepsilon
_{\underrightarrow{i}\underrightarrow{l}}\varepsilon
_{\underrightarrow{j}\underrightarrow{k}}+\varepsilon _{\underrightarrow{j}%
\underrightarrow{l}}\varepsilon _{\underrightarrow{i}\underrightarrow{k}%
}\right) R_{(0)},...,
$$
$$
R_{\underrightarrow{a_p}\underrightarrow{b_p}\underrightarrow{c_p}%
\underrightarrow{d_p}}=4\left( \varepsilon _{\underrightarrow{a_p}%
\underrightarrow{d_p}}\varepsilon _{\underrightarrow{b_p}\underrightarrow{c_p%
}}+\varepsilon _{\underrightarrow{b_p}\underrightarrow{d_p}}\varepsilon _{%
\underrightarrow{a_p}\underrightarrow{c_p}}\right) R_{(p)},...,
$$
where
$$
R_{(0)}=g^{kl}R_{kli}^i,...,R_{(p)}=g^{a_pc_p}R_{a_pc_pb_p}^{b_p},...,
$$
satisfy correspondingly identities (2.18) and (2.19).

Similarly we can check that spinor decompositions%
$$
\Omega _{\underleftarrow{a_p}\underleftarrow{b_p}\underrightarrow{c_p}%
\underleftarrow{d_p}}=\frac i4\left( \varepsilon _{\underleftarrow{b_p}%
\underleftarrow{d_p}}Q_{\underleftarrow{a_p}\underrightarrow{c_p}%
}-3\varepsilon _{\underleftarrow{a_p}\underleftarrow{b_p}}Q_{\underleftarrow{%
d_p}\underrightarrow{c_p}}-3\varepsilon _{\underleftarrow{a_p}%
\underleftarrow{d_p}}Q_{\underleftarrow{b_p}\underrightarrow{c_p}}\right)
,
$$
$$
R_{\underleftarrow{a_p}\underrightarrow{b_p}\underleftarrow{c_p}%
\underleftarrow{d_p}=}\varepsilon _{c_pa_p}Q_{\underleftarrow{d_p}%
\underrightarrow{b_p}}+\varepsilon _{d_pa_p}Q_{\underleftarrow{c_p}%
\underrightarrow{b_p}},
$$
$$
Q_{\underleftarrow{a_p}\underrightarrow{b_p}}^{*}=Q_{\underleftarrow{a_p}%
\underrightarrow{b_p}}
$$
solve identities (2.16),(2.17) and (2.20).

The identity (2.21) allows as to express a part of curvature
components through some components of torsion:
$$
R_{\underleftarrow{a_p}\underline{b_p}\underline{c_p}\underline{d_p}%
}=i(\sigma
_{\underline{b_p}\underleftarrow{a_p}\underrightarrow{e_p}}\Omega
_{\underline{c_p}\underline{d_p}}^{\underrightarrow{e_p}}-\sigma _{%
\underline{d_p}\underleftarrow{a_p}\underrightarrow{e_p}}\Omega _{\underline{%
b_p}\underline{c_p}}^{\underrightarrow{e_p}}-\sigma _{\underline{c_p}%
\underleftarrow{a_p}\underrightarrow{e_p}}\Omega _{\underline{d_p}\underline{%
b_p}}^{\underrightarrow{e_p}}).
$$

Now we consider these spinor decompositions of curvatures:%
$$
R_{\underleftarrow{a_p}\underrightarrow{a_p}\underleftarrow{b_p}%
\underrightarrow{b_p}\underleftarrow{c_p}\underrightarrow{c_p}%
\underleftarrow{d_p}\underrightarrow{d_p}}=\varepsilon _{\underleftarrow{a_p}%
\underleftarrow{b_p}}\varepsilon _{\underleftarrow{c_p}\underleftarrow{d_p}}%
\overline{\chi }_{\underrightarrow{a_p}\underrightarrow{b_p}\underrightarrow{%
c_p}\underrightarrow{d_p}}+\varepsilon _{\underrightarrow{a_p}%
\underrightarrow{b_p}}\varepsilon _{\underrightarrow{c_p}\underrightarrow{d_p%
}}\chi _{\underleftarrow{a_p}\underleftarrow{b_p}\underleftarrow{c_p}%
\underleftarrow{d_p}}-
$$
$$
\varepsilon _{\underleftarrow{a_p}\underleftarrow{b_p}}\varepsilon _{%
\underrightarrow{c_p}\underrightarrow{d_p}}\overline{\varphi }_{%
\underrightarrow{a_p}\underrightarrow{b_p}\underleftarrow{c_p}%
\underleftarrow{d_p}}-\varepsilon _{\underrightarrow{a_p}\underrightarrow{b_p%
}}\varepsilon _{\underleftarrow{c_p}\underleftarrow{d_p}}\varphi _{%
\underleftarrow{a_p}\underleftarrow{b_p}\underrightarrow{c_p}%
\underrightarrow{d_p}}.
$$
The identity (2.21) is satisfied if
$$
\varphi _{\underleftarrow{a_p}\underleftarrow{b_p}\underrightarrow{c_p}%
\underrightarrow{d_p}}=\overline{\varphi }_{\underrightarrow{c_p}%
\underrightarrow{d_p}\underleftarrow{a_p}\underleftarrow{b_p}},\chi
_{\quad
\underleftarrow{b_p}\underleftarrow{a_p}\underleftarrow{c_p}}^{%
\underleftarrow{a_p}}=\varepsilon _{\underleftarrow{b_p}\underleftarrow{c_p}%
}\Lambda ~(\Lambda \mbox{ is real}).
$$

The next step is the solution of linear identities (2.22),(2.23)
and (2.24) containing derivatives. Putting (2.28) into (2.24) we
find
$$
{\cal D}_{\underrightarrow{a_p}}R_{(p)}=0.
$$
The d--spinor%
$$
\Omega _{\underrightarrow{e_p}\underleftarrow{a_p}\underrightarrow{b_p}%
\underleftarrow{c_p}\underrightarrow{d_p}}=\sigma _{\underleftarrow{a_p}%
\underrightarrow{b_p}}^{\underline{a_p}}\sigma _{\underleftarrow{c_p}%
\underrightarrow{d_p}}^{\underline{b_p}}\Omega _{\underrightarrow{e_p}%
\underline{a_p}\underline{b_p}}
$$
can be decomposed into irreducible parts 
 [180,181]
$$
\Omega _{\underrightarrow{e_p}\underleftarrow{a_p}\underrightarrow{b_p}%
\underleftarrow{c_p}\underrightarrow{d_p}}=-\varepsilon _{\underleftarrow{a_p%
}\underleftarrow{c_p}}(W_{\underrightarrow{b_p}\underrightarrow{c_p}%
\underleftarrow{e_p}}+\varepsilon _{\underrightarrow{e_p}\underrightarrow{d_p%
}}\tau _{\underrightarrow{b_p}}+\varepsilon _{\underrightarrow{e_p}%
\underrightarrow{b_p}}\tau _{\underrightarrow{d_p}})+\varepsilon _{%
\underrightarrow{a_p}\underrightarrow{c_p}}\tau _{\underleftarrow{a_p}%
\underleftarrow{c_p}\underrightarrow{e_p}},
$$
where
$W_{\underrightarrow{b_p}\underrightarrow{c_p}\underleftarrow{e_p}}$
is an arbitrary d--spinor and $\tau _{\underrightarrow{b_p}}$ and $\tau _{%
\underleftarrow{a_p}\underleftarrow{c_p}\underrightarrow{e_p}}$
are
expressed through derivations of $Q_{\underleftarrow{a_p}\underrightarrow{b_p%
}}$ (see below ). A tedious but trivial calculus can convinse us
that the solution of (2.22) can be expressed as
$$
\Omega _{\underrightarrow{e_p}\underleftarrow{a_p}\underrightarrow{b_p}%
\underleftarrow{c_p}\underrightarrow{d_p}}=-\varepsilon _{\underleftarrow{a_p%
}\underleftarrow{c_p}}W_{\underrightarrow{b_p}\underrightarrow{c_p}%
\underleftarrow{e_p}}+\frac 12\varepsilon _{\underrightarrow{a_p}%
\underrightarrow{c_p}}({\cal D}_{\underleftarrow{a_p}}Q_{\underleftarrow{c_p}%
\underrightarrow{e_p}}+{\cal D}_{\underleftarrow{c_p}}Q_{\underleftarrow{b_p}%
\underrightarrow{e_p}})-
$$
$$
\frac 12\varepsilon
_{\underleftarrow{a_p}\underleftarrow{c_p}}(\varepsilon
_{\underrightarrow{e_p}\underrightarrow{d_p}}{\cal D}^{\underleftarrow{f_p}%
}Q_{\underleftarrow{f_p}\underrightarrow{b_p}}+\varepsilon _{%
\underrightarrow{e_p}\underrightarrow{b_p}}{\cal D}^{\underleftarrow{f_p}}Q_{%
\underleftarrow{f_p}\underrightarrow{d_p}}),
$$
$$
R_{\underleftarrow{a_p}\underleftarrow{b_p}\underrightarrow{c_p}%
\underleftarrow{d_p}\underleftarrow{e_p}}=\frac i2(\varepsilon _{%
\underleftarrow{a_p}\underleftarrow{b_p}}{\cal D}_{\underleftarrow{d_p}}Q_{%
\underleftarrow{e_p}\underrightarrow{c_p}}+\varepsilon _{\underleftarrow{a_p}%
\underleftarrow{d_p}}{\cal D}_{\underleftarrow{b_p}}Q_{\underleftarrow{e_p}%
\underrightarrow{c_p}}+
$$
$$
\varepsilon _{\underleftarrow{a_p}\underleftarrow{b_p}}{\cal D}_{%
\underleftarrow{e_p}}Q_{\underleftarrow{d_p}\underrightarrow{c_p}%
}+\varepsilon _{\underleftarrow{a_p}\underleftarrow{d_p}}{\cal D}_{%
\underleftarrow{e_p}}Q_{\underleftarrow{b_p}\underrightarrow{c_p}%
})+i(\varepsilon _{\underleftarrow{d_p}\underleftarrow{a_p}}\varepsilon _{%
\underleftarrow{b_p}\underleftarrow{e_p}}+\varepsilon _{\underleftarrow{e_p}%
\underleftarrow{a_p}}\varepsilon _{\underleftarrow{b_p}\underleftarrow{d_p}})%
{\cal D}^{\underleftarrow{f_p}}Q_{\underleftarrow{f_p}\underrightarrow{c_p}%
},
$$
$$
R_{\underleftarrow{a_p}\underleftarrow{b_p}\underrightarrow{c_p}%
\underrightarrow{d_p}\underrightarrow{e_p}}=2i\varepsilon _{\underleftarrow{%
a_p}\underleftarrow{b_p}}W_{\underrightarrow{d_p}\underrightarrow{e_p}%
\underrightarrow{c_p}}+\frac i2(\varepsilon _{\underrightarrow{c_p}%
\underrightarrow{d_p}}{\cal D}_{\underleftarrow{a_p}}Q_{\underleftarrow{b_p}%
\underrightarrow{e_p}}+\varepsilon _{\underrightarrow{c_p}\underrightarrow{%
e_p}}{\cal D}_{\underleftarrow{a_p}}Q_{\underleftarrow{b_p}\underrightarrow{%
d_p}}).
$$
This solution is compatible with (2.23) if
$$
{\cal D}^{c_p}Q_{\underleftarrow{c_p}\underrightarrow{e_p}}={\cal D}_{%
\underleftarrow{e_p}}R_{(p)}^{*}.
$$

So we have solved all linear identities.

Nonlinear relations (2.25),(2.26) and (2.27) can be transformed
into linear ones by using commutators of d--covariant
derivations. Omitting such algebraic transforms we present
expressions
$$
\overline{\chi }_{\underrightarrow{a_p}\underrightarrow{b_p}\underrightarrow{%
c_p}\underrightarrow{d_p}}=\frac 14({\cal D}_{\underrightarrow{a_p}}W_{%
\underrightarrow{b_p}\underrightarrow{c_p}\underrightarrow{d_p}}+{\cal D}_{%
\underrightarrow{b_p}}W_{\underrightarrow{c_p}\underrightarrow{d_p}%
\underrightarrow{a_p}}+{\cal D}_{\underrightarrow{c_p}}W_{\underrightarrow{%
d_p}\underrightarrow{a_p}\underrightarrow{b_p}}+{\cal D}_{\underrightarrow{%
d_p}}W_{\underrightarrow{a_p}\underrightarrow{b_p}\underrightarrow{c_p}})+
$$
$$
(\varepsilon _{\underrightarrow{a_p}\underrightarrow{d_p}}\varepsilon _{%
\underrightarrow{c_p}\underrightarrow{b_p}}+\varepsilon _{\underrightarrow{%
b_p}\underrightarrow{d_p}}\varepsilon _{\underrightarrow{c_p}%
\underrightarrow{a_p}})\times $$ $$
[\frac 1{16}(\overline{{\cal D}}_{\underrightarrow{e_p%
}}\overline{{\cal D}}^{\underrightarrow{e_p}}R_{(p)}^{*}+{\cal D}^{%
\underleftarrow{e_p}}{\cal D}_{\underleftarrow{e_p}}R_{(p)})+\frac 18Q_{%
\underleftarrow{e_p}\underrightarrow{e_p}}Q^{\underleftarrow{e_p}%
\underrightarrow{e_p}}-2RR^{*}]
$$
and
$$
\varphi _{\underleftarrow{a_p}\underleftarrow{b_p}\underrightarrow{c_p}%
\underrightarrow{d_p}}=\overline{\varphi }_{\underrightarrow{c_p}%
\underrightarrow{d_p}\underleftarrow{a_p}\underleftarrow{b_p}}=\frac 14(Q_{%
\underleftarrow{a_p}\underrightarrow{d_p}}Q_{\underleftarrow{b_p}%
\underrightarrow{c_p}}+Q_{\underleftarrow{b_p}\underrightarrow{d_p}}Q_{%
\underleftarrow{a_p}\underrightarrow{c_p}})+
$$
$$
\frac i8({\cal D}_{\underleftarrow{b_p}\underrightarrow{c_p}}Q_{%
\underleftarrow{a_p}\underrightarrow{d_p}}+{\cal D}_{\underleftarrow{a_p}%
\underrightarrow{c_p}}Q_{\underleftarrow{b_p}\underrightarrow{d_p}}+{\cal D}%
_{\underleftarrow{b_p}\underrightarrow{d_p}}Q_{\underleftarrow{a_p}%
\underrightarrow{c_p}}+{\cal D}_{\underleftarrow{a_p}\underrightarrow{d_p}%
}Q_{\underleftarrow{b_p}\underrightarrow{c_p}}+
$$
$$
{\cal D}_{\underrightarrow{c_p}}{\cal D}_{\underleftarrow{b_p}}Q_{%
\underleftarrow{a_p}\underrightarrow{d_p}}+{\cal D}_{\underrightarrow{c_p}}%
{\cal D}_{\underleftarrow{a_p}}Q_{\underleftarrow{b_p}\underrightarrow{d_p}}+%
{\cal D}_{\underrightarrow{d_p}}{\cal D}_{\underleftarrow{b_p}}Q_{%
\underleftarrow{a_p}\underrightarrow{c_p}}+{\cal D}_{\underrightarrow{d_p}}%
{\cal D}_{\underleftarrow{a_p}}Q_{\underleftarrow{b_p}\underrightarrow{c_p}%
})
$$
which solves (2.25); if conditions
$$
{\cal D}_{\underrightarrow{a_p}}W_{\underrightarrow{b_p}\underrightarrow{c_p}%
\underrightarrow{d_p}}=0
$$
are satisfied we obtaine solutions (2.26) and (2.27).

\section{Einstein--Cartan D--Structures}

In this section we shall introduce a set of Einstein like
gravitational equations, i.e. we shall formulate a variant of
higher order anisotropic
supergravity on dsv-bundle ${\cal E}^{<z>}$ over a supersmooth manifold $%
\widetilde{M}$. This model will contain as particular cases the
Miron and Anastasiei locally anisotropic gravity
 [9,160,161] on vector
bundles (they considered prescribed components of N-connection
and h(hh)-
and v(vv)--torsions; in our supesymmetric approach 
 [260] we used
algebraic equations for torsion and its source in order to close
the system of field equations). There are two ways in developing
supergravitational models. We can try to maintain similarity to
Einstein's general relativity (see in
 [16,172] an example of such type locally isotropic
supergravity) and to formulate a variant of Einstein--Cartan
theory on
dvs--bundles, this will be the aim of this section, or to
introduce into consideration generalized supervielbein variables
and to formulate a supersymmetric gauge like model of
la--supergravity (this approach is more accepted in the usual
locally isotropic supergravity, see as reviews
  [215,288,170]). The second variant will be analysed in the next section by
using the s-bundle of supersymmetric affine adapted frames on
la-superspaces.

Let consider a dvs--bundle ${\cal E}^{<z>}$ provided with some
compatible nonlinear connection N, d--connection $D$ and metric
$G$ structures (see detailes and conventions in section 1.3). For
a locally N-adapted frame we write
$$
D_{({\frac \delta {\delta u^{<\gamma >}}})}{\frac \delta {\delta
u^{<\beta
>}}}={{\Gamma }_{<\beta ><\gamma >}^{<\alpha >}}{\frac \delta {\delta
u^{<\alpha >}}},
$$
where the d--connection $D$ has the following coefficients:
$$
{{{\Gamma }^I}_{JK}}={{L^I}_{JK}},{{{\Gamma }^I}_{J<A>}}={{C^I}_{J<A>}},{{%
\Gamma }^I}_{<A>J}=0,{{\Gamma }^I}_{<A><B>}=0,\eqno(2.29)
$$
$$
{{\Gamma }^{<A>}}_{JK}=0,{{\Gamma }^{<A>}}_{J<B>}=0,{{\Gamma }^{<A>}}_{<B>K}=%
{L^{<A>}}_{<B>K},
$$
$$
{{{\Gamma }^{<A>}}_{<B><C>}}={{C^{<A>}}_{<B><C>}}.
$$
The nonholonomy coef\-fi\-ci\-ents ${{w^{<\gamma >}}_{<\alpha
><\beta >}}$ of
d--connection (2.29), defined as $[{\delta }_{<\alpha >},{\delta
}_{<\beta
>}\}={{w^{<\gamma >}}_{<\alpha ><\beta >}}{\delta }_{<\gamma >},$ are
computed as follows:
$$
{{w^K}_{IJ}}=0,{{w^K}_{<A>J}}=0,{{w^K}_{I<A>}}=0,{{w^K}_{<A><B>}}=0,{{w^{<A>}%
}_{IJ}}={R^{<A>}}_{IJ},
$$
$$
{{w^{<B>}}_{<A>I}}=-(-)^{|I<A>|}{\frac{\partial {N_I^B}}{\partial
y^{<A>}}},$$
$${{w^{<B>}}_{I<A>}}={\frac{\partial {N_I^{<B>}}}{\partial y^{<A>}}},{{w^{<C>}}%
_{<A><B>}}=0.
$$
By straightforward calculations we obtain respectively these
components of
torsion, $${\cal T}({\delta }_{<\gamma >},{\delta }_{<\beta >})={{\cal T}%
_{\cdot <\beta ><\gamma >}^{<\alpha >}}{\delta }_{<\alpha >},$$
and
curvature, $${\cal R}({\delta }_{<\beta >},{\delta }_{<\gamma >}){\delta }%
_{<\tau >}={{\cal R}_{.<\beta ><\gamma ><\tau >}^{<\alpha >}}{{\delta }%
_{<\alpha >}},$$ ds--tensors:
$$
{\cal T}_{\cdot JK}^I={T^I}_{JK},{\cal T}_{\cdot J<A>}^I={C^I}_{J<A>},{\cal T%
}_{\cdot J<A>}^I=-{C^I}_{J<A>},{\cal T}_{\cdot <A><B>}^I=0,
$$
$$
{\cal T}_{\cdot IJ}^{<A>}={R^{<A>}}_{IJ},{\cal T}_{\cdot I<B>}^{<A>}=-{%
P^{<A>}}_{<B>I},{\cal T}_{\cdot <B>I}^{<A>}={P^{<A>}}_{<B>I},
$$
$$
{\cal T}_{\cdot <B><C>}^{<A>}={S^{<A>}}_{<B><C>}
$$
and
$$
{\cal R}_{\cdot IKL}^J=R_{IKL}^J,{\cal R}_{\cdot BKL}^J=0,{\cal
R}_{\cdot JKL}^{<A>}=0,{\cal R}_{\cdot <B>KL}^{<A>}={R}_{\cdot
<B>KL}^{<A>},
$$
$$
{\cal R}_{J\cdot K<D>}^I={{P_J}^I}_{K<D>},{\cal R}_{<B>K<D>}^I=0,{\cal R}%
_{JK<D>}^{<A>}=0,
$$
$$
{\cal R}_{<B>K<D>}^{<A>}={P}_{<B>K<D>}^{<A>},{\cal R}_{J<D>K}^I=-(-)^{|<D>K|}%
{P}_{JK<D>}^I,
$$
$$
{\cal R}_{<B><D>K}^I=0,{\cal R}_{J<D>K}^{<A>}=0,
$$
$$
{\cal R}_{<B><D>K}^{<A>}=(-)^{|K<D>|}{P}_{<B>K<D>}^{<A>},{\cal R}%
_{J<C><D>}^I=S_{J<C><D>}^I,
$$
$$
{\cal R}_{<B><C><D>}^I=0,{\cal R}_{J<C><D>}^{<A>}=0,{\cal R}%
_{<B><C><D>}^{<A>}=S_{BCD}^A
$$
( see formulas (1.25) and (1.29)).

The locally adapted components ${\cal R}_{<\alpha ><\beta >}={\cal R}ic(D)({%
\delta }_{<\alpha >},{\delta }_{<\beta >})$ (we point that in
general on dvs-bundles ${\cal R}_{<\alpha ><\beta >}\ne
{(-)}^{\mid <\alpha ><\beta
>\mid }{\cal R}_{<\beta ><\alpha >})$ of the Ricci tensor are as follows:
$$
{\cal R}_{IJ}=R_{IJK}^K,{\cal R}_{I<A>}=-{}^{(2)}{P_{I<A>}}=-{P}_{IK<A>}^K%
$$
$$
{\cal R}_{<A>I}={}^{(1)}{P_{<A>I}}={P}_{<A>I<B>}^{<B>},{\cal R}%
_{<A><B>}=S_{<A><B><C>}^{<C>}=S_{<A><B>}.
$$
For the scalar curvature, ${\check {{\cal R}}}=Sc(D)=G^{<\alpha ><\beta >}%
{\cal R}_{<\alpha ><\beta >},$ we have
$$
Sc(D)=R+S,
$$
where $R=g^{IJ}R_{IJ}$ and $S=h^{<A><B>}S_{<A><B>}.$

The Einstein--Cartan equations on dvs--bund\-les are written as
$$
{\cal R}_{<\alpha ><\beta >}-{\frac 12}G_{<\alpha ><\beta >}{\check {{\cal R}%
}}+{\lambda }G_{<\alpha ><\beta >}={\kappa }_1{\cal J}_{<\alpha ><\beta >},%
\eqno(2.30)
$$
and
$$
T_{<\beta ><\gamma >}^{<\alpha >}+{{G_{<\beta >}}^{<\alpha >}}{{T^{<\tau >}}%
_{<\gamma ><\tau >}}-
$$
$$
{(-)}^{\mid <\beta ><\gamma >\mid }{{G_{<\gamma >}}^{<\alpha >}}{{T^{<\tau >}%
}_{<\beta ><\tau >}}={\kappa }_2{Q^{<\alpha >}}_{<\beta ><\gamma >},%
\eqno(2.31)
$$
where $${\cal J}_{<\alpha ><\beta >} \mbox{ and }
 {Q_{<\beta ><\gamma >}^{<\alpha >}}$$
are respectively components of energy--moment\-um and
spin--density of matter
ds--tensors on la--space, ${\kappa }_1$ and ${\kappa }_2$ are the
corresponding interaction constants and ${\lambda }$ is the
cosmological constant. To write in a explicit form the mentioned
matter sources of la--supergravity in (2.30) and (2.31) there are
necessary more detailed studies of models of interaction of
superfields on locally anisotropic superspaces (in previous
sections we presented details for a class of osculator
s--bundles; further generalizations with an explicit writing out
of terms higher order anisotropic interactions of s--fields on an
arbitrary dvs--bundle is connected with combersome calculations
and formulas; we omit such considerations in this monograph).

Equations (2.30), can be split into base- and fibre--components,
$$
R_{IJ}-{\frac 12}(R+S-{\lambda })g_{IJ}={\kappa }_1{\cal J}%
_{IJ},{}^{(1)}P_{<A>I}={\kappa }_1{}^{(1)}{\cal
J}_{<A>I},\eqno(2.32)
$$
$$
S_{<A><B>}-{\frac 12}(S+R-{\lambda })g_{<A><B>}={{\kappa }_2}{\tilde {{\cal J%
}}}_{<A><B>},{}^{(2)}P_{I<A>}=-{{\kappa }_2}{}^{(2)}{{\cal
J}_{I<A>}},
$$
are a supersymmetric higher order, with cosmological term,
generalization of the similar ones presented in
 [9,160,161], with prescribed
N-connection and h(hh)-- and v(vv)--torsions. We have added
algebraic equations (2.31) in order to close the system of
s--gravitational field equations (realy we have also to take into
account the system of constraints (1.34) if locally anisotropic
s--gravitational field is associated to a ds--metric (1.35)).

We point out that on la--superspaces the divergence $D_{<\alpha >}{\cal J}%
^{<\alpha >}$ does not vanish (this is a consequence of
generalized Bianchi and Ricci identities (1.30), (1.31) and
(1.39)). The d--covariant derivations of
the left and right parts of (2.30), equivalently of (2.32), are as follows:%
$$
D_{<\alpha >}[{\cal R}_{<\beta >}^{\cdot <\alpha >}-{\frac 12}({\check {%
{\cal R}}}-2{\lambda }){\delta }_{<\beta >}^{\cdot <\alpha >}]=
$$
$$
\left\{
\begin{array}{rl}
{\lbrack {{R_J}^I}-{\frac 12}{({R+S-2{\lambda })}}{{{\delta
}_J}^I}]}_{\mid
I}+{}^{(1)}{P^{<A>}}_{I\perp <A>}=0, &  \\
{\lbrack {{S_{<B>}}^{<A>}}-{\frac 12}{({R+S-2{\lambda })}}{{{\delta }_{<B>}}%
^{<A>}}]}_{\perp <A>}-{}^{(2)}{P^I}_{<B>\mid I}=0, &
\end{array}
\right.
$$
where%
$$
{}^{(1)}{P^{<A>}}_J={}^{(1)}{P_{<B>J}}g^{<A><B>},{}^{(2)}{P^I}_{<B>}={}^{(2)}%
{P_{J<B>}}g^{IJ},
$$
$$
{R^I}_J={R_{KJ}}g^{IK},{S^{<A>}}_{<B>}={S_{<C><B>}}h^{<A><C>},
$$
and
$$
{D_{<\alpha >}}{\cal J}_{\cdot <\beta >}^{<\alpha >}={\cal U}_{<\alpha >},%
\eqno(2.33)
$$
where
$$
D_{<\alpha >}{\cal J}_{\cdot <\beta >}^{<\alpha >}=\left\{
\begin{array}{rl}
{{\cal J}_{\cdot J\mid I}^I+{}^{(1)}{\cal J}_{\cdot J\perp <A>}^{<A>}}={%
\frac 1{{\kappa }_1}}{{\cal U}_J}, &  \\
{{}^{(2)}{\cal J}_{\cdot <A>\mid I}^I+{\cal J}_{\cdot <A>\perp <B>}^{<B>}}={%
\frac 1{{\kappa }_1}}{{\cal U}_{<A>}}, &
\end{array}
\right.
$$
and%
$$
{\cal U}_{<\alpha >}={\frac 12}(G^{<\beta ><\delta >}{{\cal
R}_{<\delta
><\varphi ><\beta >}^{<\gamma >}}{{\cal T}_{\cdot <\alpha ><\gamma
>}^{<\varphi >}}-\eqno(2.34)
$$
$$
{(-)}^{\mid <\alpha ><\beta >\mid }G^{<\beta ><\delta >}{{\cal
R}_{<\delta
><\varphi ><\alpha >}^{<\gamma >}}{{\cal T}_{\cdot <\beta ><\gamma
>}^{<\varphi >}}+{{\cal R}_{\cdot <\varphi >}^{<\beta >}}{{\cal T}_{\cdot
<\beta ><\alpha >}^{<\varphi >}}).
$$
So, it follows that ds-vector ${\cal U}_\alpha $ vanishes if
d-connection (2.29) is torsionless.

No wonder that conservation laws for values of energy--momentum
type, being a consequence of global automorphisms of spaces and
s--spaces, or, respectively, of theirs tangent spaces and
s--spaces (for models on curved spaces and s--spaces), on
la--superspaces are more sophisticate because, in general, such
automorphisms do not exist for a generic local anisotropy. We can
construct a higher order model of supergravity, in a way similar
to that for the Einstein theory if instead an arbitrary metric
d--connection the generalized Christoffel symbols ${\tilde \Gamma
}_{\cdot \beta \gamma }^\alpha $ (see (1.39)) are used. This is a
locally anisotropic supersymmetric model on the base s-manifold
$\widetilde{M}$ which looks like locally isotropic on the total
space of a dvs--bundle. More general supergravitational models
which are locally anisotropic on the both base and total spaces
can be generated by using deformations of d-connections of type
(1.40). In this case the vector ${\cal U}_\alpha $ from (2.34)
can be interpreted as a corresponding source of generic local
anisotropy satisfying generalized conservation laws of type
(2.33).

More completely the problem of formulation of conservation laws
for both locally isotropic and anisotropic higher order
supergravity can be solved in the frame of the theory of nearly
autoparallel maps of dvs-bundles (with specific deformations of
d-connections (1.40) and in consequence of torsion
  and curvature),
which have to generalize our constructions from
 [249,250,251,263,278,279], see section 3.4 and Chapter 8)

We end this section with the remark that field equations of type
(2.30), equivalently (2.32), for higher order supergravity can be
similarly introduced for the particular cases of higher order
anisotropic s--spaces provided with metric structure of type
(1.35) with coefficients parametrized as for higher order
prolongations of the Lagrange,  or Finsler, s--spaces (subsection
1.4.2).

\section{Gauge Like Locally Anisotropic Super\-gra\-vi\-ty}

The aim of this section is to introduce a set of gauge like
gravitational equations (wich are equivalent to Einstein
equations on dvs--bundels (2.30) if well defined conditions are
satisfied). This model will be a higher order anidotopic
supersymmetric extension of our constructions for gauge
la--gravity 
 [260,272,258,259] and of affine--gauge interpretation
of the Einstein gravity 
 [195,196,194,240,246].

The great part of theories of locally isotropic s--gravity are
formulated as gauge supersymmetric models based on supervielbein
formalism (see
 [174,215,286,288]). A similar model of supergravity on osculator s--bundles
have been considered in section 2.3. Here we shall analyse a
geometric background for such theories on dvs--bundles. Let
consider an arbitrary adapted to N-connection frame
 $l_{<\underline{\alpha }>}(u)=(l_{\underline{I}%
}(u),l_{<\underline{C}>}(u))$ on ${\cal E}^{<z>}$ and s-vielbein matrix%
$$
l_{{<\alpha >}}^{<\underline{\alpha }>}=\left(
\begin{array}{cccccc}
l_I^{\underline{I}} & 0 & ... & 0 & ... & 0 \\
0 & l_{A_1}^{\underline{A_1}} & ... & 0 & ... & 0 \\
... & ... & ... & 0 & ... & 0 \\
0 & 0 & 0 & l_{A_p}^{\underline{A_p}} & ... & 0 \\
0 & 0 & 0 & 0 & ... & 0 \\
0 & 0 & 0 & 0 & ... & l_{A_z}^{\underline{A_z}}
\end{array}
\right) \subset GL_{n,k}^{<m,l>}(\Lambda )=
$$
$$
GL(n,k,{\Lambda })\oplus GL(m_1,l_1,{\Lambda })\oplus ...\oplus GL(m_p,l_p,{%
\Lambda })\oplus ...\oplus GL(m_z,l_z,{\Lambda })
$$
for which
$$
{\frac \delta {\delta u^{<\alpha >}}}=l{_{<\alpha >}}^{<\underline{\alpha }%
>}(u)l_{<\underline{\alpha }>}(u),
$$
or, equivalently, ${\frac \delta {\partial x^I}}={l_I}^{\underline{I}}(u)l_{%
\underline{I}}(u)$ and ${\frac \delta {\partial y^{<C>}}}={l_{<C>}}^{<%
\underline{C}>}l_{<\underline{C}>}(u),$ and
$$
G_{<\alpha ><\beta >}(u)=l{_{<\alpha >}}^{<\underline{\alpha }>}(u)l{%
_{<\beta >}}^{<\underline{\beta }>}(u){\eta }_{<{\underline{\alpha }><}{%
\underline{\beta }>}},
$$
where, for simplicity, ${\eta }_{<\underline{\alpha
}><\underline{\beta }>}$ is a constant metric on vs--space
$V^{n,k}\oplus V^{<l,m>}.$

By $LN({\cal E}^{<z>})$ is denoted the set of all locally adapted
frames in
all points of sv--bundle ${\cal E}^{<z>} .$ For a surjective s--map ${%
\pi }_L$ from $LN({\cal E}^{<z>}{\cal )}$ to ${\cal E}^{<z>}$ and treating $%
GL_{n,k}^{<m,l>}(\Lambda )$ as the structural s--group we define
a principal s--bundle,
$$
{\cal L}N({\cal E}^{<z>})=(LN({\cal E}^{<z>}),{\pi }_L:LN({\cal
E}^{<z>})\to {\cal E}^{<z>},GL_{n,k}^{<m,l>}({\Lambda })),
$$
called as the s--bundle of linear adap\-ted fra\-mes on ${\cal
E}^{<z>}{\cal .}$

Let $I_{<\hat \alpha >}$ be the canonical basis of the sl--algebra ${\cal G}%
_{n,k}^{<m,l>}$ for a s--group\\ $GL_{n,k}^{<m,l>}({\Lambda })$
with a
cumulative index $<{\hat \alpha >}$. The structural coefficients\\ ${%
f_{<\hat \alpha ><\hat \beta >}}^{<\hat \gamma >}$ of ${\cal G}%
_{n,k}^{<m,l>} $ satisfy s--commutation rules
$$
[I_{<\hat \alpha >},I_{<\hat \beta >}\}={f_{<\hat \alpha ><\hat \beta >}}%
^{<\hat \gamma >}I_{<\hat \gamma >}.
$$
On ${\cal E}^{<z>}$ we consider the connection 1--form
$$
{\Gamma }={{\Gamma }^{<\underline{\alpha }>}}_{<{\underline{\beta
>}<}\gamma
>}(u)I_{<\underline{\alpha }>}^{<\underline{\beta }>}du^{<\gamma >},%
\eqno(2.35)
$$
where
$$
{{\Gamma }^{<\underline{\alpha }>}}_{<\underline{\beta }><\gamma >}(u)=l{^{<%
\underline{\alpha }>}}_{<\alpha >}l{^{<\beta >}}_{<\underline{\beta }>}{{%
\Gamma }^{<\alpha >}}_{<\beta ><\gamma >}+l{^{<\underline{\alpha
}>}}{\frac \delta {\partial u^\gamma }}l{^{<\alpha
>}}_{<\underline{\beta }>}(u),
$$
${{\Gamma }^{<\alpha >}}_{<\beta ><\gamma >}{\quad }$ are the
components of
the metric d--connection, s-matrix\\ $l{^{<\beta >}}_{<\underline{%
\beta }>}{\quad }$ is inverse to the s-vielbein matrix ${\quad }l{^{<%
\underline{\beta }>}}_{<\beta >},\quad $ and ${\quad }I_{<\underline{\beta >}%
}^{<\underline{\alpha }>}=\delta _{<\underline{\beta
>}}^{<\underline{\alpha
}>}$ is the standard distinguished basis in SL--algebra ${\cal G}%
_{n,k}^{<m,l>}.$

The curvature ${\cal B}$ of the connection (2.35),
$$
{\cal B}=d{\Gamma }+{\Gamma }\land {\Gamma }={\cal R}_{<\underline{\alpha }%
><\gamma ><\delta >}^{<\underline{\beta }>}I_{<\underline{\beta }>}^{<%
\underline{\alpha }>}{\delta u^{<\gamma >}}\land \delta u^{<\delta >}%
$$
has coefficients
$$
{\cal R}_{<\underline{\alpha }><\gamma ><\delta >}^{<\underline{\beta }>}=l{%
^{<\alpha >}}_{<\underline{\alpha }>}(u)l{^{<\underline{\beta
}>}}_{<\beta
>}(u){\cal R}_{<\alpha ><\gamma ><\delta >}^{<\beta >},
$$
where ${\cal R}_{<\alpha ><\gamma ><\delta >}^{<\beta >}$ are the
components of the ds--tensor (1.29).

In addition with ${\cal L}N({\cal E}^{<z>})$ we consider another
s--bundle, the s--bundle of locally adapted affine frames
$$
{\cal A}N({\cal E}^{<z>})=(AN({\cal E}^{<z>}),{{\pi }_A}:AN({\cal E}%
^{<z>})\to {\cal E}^{<z>},{{AF}_{n,k}^{<m,l>}}({\Lambda }))%
$$
with the structural s--group ${AN}_{n,k}^{<m,l>}(\Lambda
)=GL_{n,k}^{<m,l>}(\Lambda )\odot {\quad }{\Lambda }^{n,k}\oplus {\Lambda }%
^{<m,l>}$ being a semidirect product (denoted by $\odot $ ) of $%
GL_{n,k}^{<m,l>}(\Lambda )$ and ${\Lambda }^{n,k}\oplus {\Lambda
}^{<m,l>}.$
Because the LS--algebra ${{\cal A}f}_{n,k}^{<m,l>}$ of s--group $%
AF_{n,k}^{<m,l>}({\Lambda }),$ is a direct sum of ${\cal
G}_{n,k}^{<m,l>}$
and ${\Lambda }^{n,k}\oplus {\Lambda }^{<m,l>}$ we can write forms on ${\cal %
A}N({\cal E}^{<z>})$ as $\Theta =({\Theta }_1,{\Theta }_2),$ where ${\Theta }%
_1$ is the ${\cal G}_{n,k}^{<m,l>}$--component and ${\Theta }_2$ is the $({%
\Lambda }^{n,k}\oplus {\Lambda }^{<m,l>})$--component of the form
$\Theta .$ The connection (2.35) in ${\cal L}N({\cal E}^{<z>})$
induces a Cartan connection $\overline{\Gamma }$ in ${\cal
A}N({\cal E}^{<z>})$ (see the case
of usual affine frame bundles in
 [40,195,196,194] and generalizations for
locally anisotropic gauge gravity and supergravity in
 [272,260]).
This is the unique connection on dvs--bundle ${\cal A}N({\cal
E}^{<z>})$ represented as $i^{*}{\overline{\Gamma }}=({\Gamma
},{\chi }),$ where $\chi $
is the shifting form and $i:{\cal A}N({\cal E}^{<z>})\to {\cal L}N({\cal E}%
^{<z>})$ is the trivial reduction of dvs--bundles. If $l=(l_{<\underline{%
\alpha }>})$ is a local adapted frame in ${\cal L}N({\cal E}^{<z>})$ then ${%
\overline{l}}=i\circ l$ is a local section in ${\cal A}N({\cal
E}^{<z>})$ and
$$
{\overline{\Gamma }}=l\Gamma =(\Gamma ,\chi ),{\overline{{\cal B}}}={%
\overline{B}}{\cal B}=({\cal B},{\cal T}),\eqno(2.36)
$$
where ${\chi }=e_{<\underline{\alpha }>}\otimes {l^{<\underline{\alpha }>}}%
_{<\alpha >}du^{<\alpha >},{\quad }e_{<\underline{\alpha }>}$ is
the
standard basis in ${\Lambda }^{n,k}\oplus {\Lambda }^{<m,l>}$ and torsion $%
{\cal T}$ is introduced as
$$
{\cal T}=d{\chi }+[{\Gamma }\land \chi \}={\cal T}_{\cdot <\beta
><\gamma
>}^{<\underline{\alpha }>}e_{<\underline{\alpha }>}du^{<\beta >}\land
du^{<\gamma >},
$$
${\cal T}_{\cdot <\beta ><\gamma >}^{<\underline{\alpha }>}={l^{<\underline{%
\alpha }>}}_{<\alpha >}{T^{<\alpha >}}_{\cdot <\beta ><\gamma >}$
are defined by the components of the torsion ds--tensor (1.25).

By using metric $G$ (1.35) on dvs--bundle ${\cal E}^{<z>}$ we can
define the
dual ( Hodge ) operator ${*}_G:{\overline{\Lambda }}^{q,s}({\cal E}^{<z>})\to {%
\overline{\Lambda }}^{n-q,k-s}({\cal E}^{<z>})$ for forms with
values in
LS--algebras on ${\cal E}^{<z>}$ (see details, for instance, in 
 [288]), where ${\overline{\Lambda }}^{q,s}({\cal E}^{<z>})$ denotes the
s--algebra of exterior (q,s)--forms on ${\cal E}^{<z>}{\cal .}$

Let operator ${*}_G^{-1}$ be the inverse to operator $*$ and ${\hat \delta }%
_G$ be the adjoint to the absolute derivation d (associated to
the scalar product for s--forms) specified for (r,s)--forms as
$$
{\delta }_G={(-1)}^{r+s}{*}_G^{-1}\circ d\circ {*}_G.
$$
The both introduced operators act in the space of
LS--algebra--valued forms as
$$
{*}_G(I_{<\hat \alpha >}\otimes {\phi }^{<\hat \alpha
>})=I_{<\hat \alpha
>}\otimes ({*}_G{\phi }^{<\hat \alpha >})
$$
and
$$
{\delta }_G(I_{<\hat \alpha >}\otimes {\phi }^{<\hat \alpha
>})=I_{<\hat \alpha >}\otimes {\delta }_G{\phi }^{<\hat \alpha >}.
$$
If the supersymmetric variant of the Killing form for the
structural
s--group of a s--bundle into consideration is degenerate as a
s--matrix (for instance, this holds for s--bundle ${\cal
A}N({\cal E}^{<z>})$ ) we use an auxiliary nondegenerate bilinear
s--form in order to define formally a metric structure ${G_{{\cal
A}}}$ in the total space of the s--bundle. In this case we can
introduce operator ${\delta }_{{\cal E}}$ acting in the
total space and define operator ${\Delta }\doteq {\hat H}\circ {\delta }_{%
{\cal A}},$ where ${\hat H}$ is the operator of horizontal
projection. After $\hat H$--projection we shall not have
dependence on components of auxiliary bilinear forms.

Methods of abstract geometric calculus, by using operators ${{*}_G},{{*}_{%
{\cal A}}},{{\delta }_G},{{\delta }_{{\cal A}}}$ and ${\Delta },$
are
illustrated, for instance, in 
 [195,196] for locally isotropic spaces and
in 
 [260,272,258,259] for locally anisotropic, spaces. Because on
superspaces these operators act in a similar manner we omit
tedious
intermediate calculations and present the final necessary results. For ${%
\Delta }{\overline{B}}$ one computers
$$
{\Delta }{\overline{{\cal B}}}=({\Delta }{\cal B},{\cal {R\tau }}+{\cal R}%
i),
$$
where ${\cal {R\tau }}={\delta }_G{\cal J}+{*}_G^{-1}[{\Gamma },{*}{\cal J}%
\} $ and
$$
{\cal R}i={{*}_G^{-1}}[{\chi },{{*}_G}{\cal R}\}={(-1)}%
^{n+k+l_1+m_1+...+l_z+m_z}{{\cal R}_{<\alpha ><\mu >}}G^{<\alpha
><{\hat \alpha >}}{e_{<\hat \alpha >}}{\delta u^{<\mu
>}}.\eqno(2.37)
$$
Form ${\cal R}i$ from (2.37) is locally constructed by using the
components of the Ricci ds--tensor (see Einstein equations (2.30)
as one follows from the decomposition with respect to a locally
adapted basis ${\delta u^{<\alpha >}}$ (1.12)).

Equations
$$
{\Delta }{\overline{{\cal B}}}=0\eqno(2.38)
$$
are equivalent to the geometric form of Yang--Mills equations for
the connection ${\overline{\Gamma }}$ (see (2.36)). D.A. Popov
and L.I. Dikhin
proved 
 196,197] that such gauge equations coincide with the vacuum
Einstein equations if as components of connection form (2.35) the
usual Christoffel symbols are used. For spaces with local
anisotropy the torsion of a metric d--connection in general is
not vanishing and we have to introduce the source 1--form in the
right part of (2.38) even gravitational
interactions with matter fields are not considered 
 [272,258,259].

Let us consider the locally anisotropic supersymmetric matter
source
${\overline{{\cal J}}}$ constructed by using the same formulas as
for ${\Delta }{\overline{{\cal B}}}$ when instead of ${\cal
R}_{<\alpha ><\beta >}$ from
(2.37) is taken ${{\kappa }_1}({\cal J}_{<\alpha ><\beta >}-{\frac 12}%
G_{<\alpha ><\beta >}{\cal J})-{\lambda }(G_{<\alpha ><\beta >}-{\frac 12}%
G_{<\alpha ><\beta >}G_{<\tau >}^{\cdot <\tau >}).$ By
straightforward calculations we can verify that Yang--Mills
equations
$$
{\Delta }{\overline{{\cal B}}}=\overline{{\cal J}}\eqno(2.40)
$$
for torsionless connection ${\overline{\Gamma }}=({\Gamma },{\chi
})$ in s-bundle ${\cal A}N({\cal E}^{<z>})$ are equivalent to
Einstein equations (2.30) on dvs--bundle ${\cal E}^{<z>}{\cal .}$
But such types of gauge like la-super\-gra\-vi\-ta\-ti\-on\-al
equations, completed with algebraic equations for torsion and
s--spin source, are not variational in the total space of the
s--bundle ${\cal A}L{({\cal E}}^{<z>}{)}.$ This is a consequence
of the mentioned degeneration of the Killing form for the affine
structural group
 [40,195,196] which also holds for our la-supersymmetric
generalization. We point out that we have introduced equations
(2.39) in a ''pure'' geometric manner by using operators
${*},{\quad }{\delta }$ and horizontal projection ${\hat H}.$

We end this section by emphasizing that to construct a
variational gauge like supersymmetric la--gravitational model is
possible, for instance, by
considering a minimal extension of the gauge s--group $AF_{n,k}^{m,l}({%
\Lambda })$ to the de Sitter s--group $S_{n,k}^{m,l}({\Lambda }%
)=SO_{n,k}^{m,l}({\Lambda }),$ acting on space ${\Lambda
}_{n,k}^{m,l}\oplus {\cal R},$ and formulating a nonlinear
version of de Sitter gauge s--gravity
(see: 
 [240,194] for locally isotropic gauge gravity, 
 [272] for a
locally anisotropic variant and Chapter 7 in this monograph for
higher order anisotropic generalizations of gauge gravity).


\chapter{Supersymmetric NA--Maps}

The study of models of classical and quantum field interactions
in higher dimension superspaces with, or not, local anisotropy is
in order of the day. The development of this direction entails
great difficulties because of problematical character of the
possibility and manner of definition of conservation laws on
la--spaces. It will be recalled that conservation laws of
energy--momentum type are a consequence of existence of a global
group of automorphisms of the fundamental Mikowski spaces. As a
rule one considers the tangent space's automorphisms or
symmetries conditioned by the existence of Killing vectors on
curved (pseudo)Riemannian spaces. There are not any global or
local automorphisms on generic la--spaces and in result of this
fact, at first glance, there are a lot of substantial
difficulties with formulation of conservation laws and, in
general, of physical consistent field theories with local
anisotropy. R. Miron and M. Anastasiei investigated the nonzero
divergence of the matter energy--momentum d--tensor, the source
in Einstein equations on la--spaces, and considered an original
approach to the geometry of time--dependent Lagrangians
 [12,160,161]. In a series of papers
 [249,276,263,273,274,278,279,252,275,277] we attempt to solve
the problem of definition of energy-momentum values for locally
isotropic and anisotropic gravitational and matter fields
interactions and of conservation laws for basic physical values
on spaces with local anisotropy in the framework of the theory of
nearly geodesic and nearly autoparallel maps.

In this Chapter a necessary geometric background (the theory of
nearly autoparallel maps, in brief na-maps, and tensor integral
formalism) for formulation and investigation of conservation laws
on higher order isotropic and anisotropic superspaces is
developed. The class of na--maps contains as a particular case
the conformal transforms and is characterized by corresponding
 invariant conditions for generalized Weyl tensors and Thomas parameters
  [227,230]. We can connect the na--map theory with the formalism of
tensor integral and multitensors on distinguished vector
superbundles. This approaches based
 on generalized conformal transforms of superspaces with
 or not different types of higher order anisotropy consist a new division of
 differential supergeometry with applications in modern theoretical and
 mathematical physics.

We note that in most cases proofs of our theorems are mechanical
but rather tedious calculations similar to
those presented in 
 [230,252,263]. Some of them will be given in
detail, the rest will be sketched. We shall omit splitting of
formulas into even and odd components (see Chapter 8 on nearly
autoparallel maps and conservation laws for higher order (non
supersymmetric) anisotropic spaces).

Section 3.1 is devoted to the formulation of the theory of nearly
autoparallel maps of dvs--bundles. The classification of na--maps
and formulation of their invariant conditions are given in
section 3.2. In section 3.3 we define the nearly autoparallel
tensor--integral on locally anisotropic multispaces. The problem
of formulation of conservation laws on spaces with local
anisotropy is studied in section 3.4. Some conclusions are
 presented in section 3.5.

\section{NA--Maps of DVS--Bundles}

This section is devoted to an extension of the ng-- [230]
and na--map
 [249,\\ 250,252,273,274,278] theories by introducing into
consideration maps of dvs--bundles provided with compatible
N--connection, d--connection and metric structures.

We shall use pairs of open regions $(U,{\underline{U}})$ of
higher order
anisotropic s--spaces, $U{\subset }{\cal E}^{<z>},\,{\underline{U}}{\subset }%
{\underline{{\cal E}}}^{<z>}$, and 1--1 local maps $f:U{\to
}{\underline{U}}$ given by necessary class supersmooth functions
$f^{<a>}(u)$ and their inverse functions
$f^{<\underline{a}>}({\underline{u}})$ nondegenerated in every
point $u{\in }U$ and ${\underline{u}}{\in }{\underline{U}}.$

Two open regions $U$~ and ${\underline{U}}$~ are attributed to a
common for f--map coordinate system if this map is realized on
the principle of coordinate equality $q(u^\alpha ){\to
}{\underline{q}}(u^\alpha )$~ for
every point $q{\in }U$~ and its f--image ${\underline{q}}{\in }{\underline{U}%
}.$ We note that all calculations included in this work will be
local in nature and taken to refer to open subsets of mappings of
type
 ${\xi }{\supset }U {\longrightarrow} {\underline{U}}{\subset }{\underline{ \xi }}.$
 For simplicity,
we suppose that in a fixed common coordinate
system for $U$ and ${\underline{U}}$ spaces ${\cal E}^{<z>}$ and ${%
\underline{{\cal E}}}^{<z>}$ are characterized by a common
N--connection structure (in consequence of (1.34) by a
corresponding concordance of d--metric structure), i.e.
$$
N_{A_p}^{A_f}(u)={\underline{N}}_{A_p}^{A_f}(u)={\underline{N}}_{A_p}^{A_f}({%
\underline{u}}),
$$
which leads to the possibility to establish common local bases,
adapted to a given N--connection, on both regions $U$ and
${\underline{U}.}$ Let denote by ${\Gamma }_{{.<}{\beta
><}{\gamma >}}^{<\alpha >}$ a compatible with metric structure
d--connection on the dvs--bundle ${\cal E}^{<z>}.$ The linear
d--connection on the dvs--bundle $\underline{{\cal E}}^{<z>}$ is
considered to be a general one with torsion
$$
{\underline{T}}_{{.<}{\beta ><}{\gamma >}}^{<\alpha >}={\underline{\Gamma }}%
_{{.<}{\beta ><}{\gamma >}}^{<\alpha >}-{\underline{\Gamma }}_{{.<}{\gamma ><%
}{\beta >}}^{<\alpha >}+w_{{.<}{\beta ><}{\gamma >}}^{<\alpha >}.
$$
and nonmetricity
$$
{\underline{K}}_{<{\alpha ><}{\beta ><}{\gamma
>}}={{\underline{D}}_{<\alpha
>}}{\underline{G}}_{<{\beta ><}{\gamma >}}.\eqno(3.1)
$$

Geometrical objects on ${\underline{{\cal E}}}^{<z>}$ are
parametrized by
underlined symbols, for example, ${\underline{A}}^{<\alpha >},{\underline{B}}%
^{<{\alpha ><}{\beta >}},$~ or underlined indices, for example, $A^{%
\underline{a}},B^{{\underline{a}}{\underline{b}}}$ (in this
Chapter we shall
not underline indices for s--vielbein decompositions as in Chapters 1 and 2).%

It is convenient to use auxiliary s--sym\-met\-ric d--con\-nec\-ti\-ons, $${%
\gamma }_{{.<}{\beta ><}{\gamma >}}^{<\alpha >}=(-)^{|<\beta ><\gamma >|}{%
\gamma }_{{.<}{\gamma ><}{\beta >}}^{<\alpha >}~\mbox{ on } {\cal
E}^{<z>}$$
 and $${\underline{\gamma }}_{.<{\beta ><}{\gamma >}}^{<\alpha >}=(-)^{|<\beta
><\gamma >|}{\underline{\gamma }}_{{.<}{\gamma ><}{\beta >}}^{<\alpha >}
\mbox{ on } {\underline{{\cal E}}}^{<z>}$$ defined respectively as
$$
{\Gamma }_{{.<}{\beta ><}{\gamma >}}^{<\alpha >}={\gamma }_{{.<}{\beta ><}{%
\gamma >}}^{<\alpha >}+T_{{.<}{\beta ><}{\gamma >}}^{<\alpha
>}\quad {\rm \mbox{and}}\quad {\underline{\Gamma }}_{{.<}{\beta
><}{\gamma >}}^{<\alpha
>}={\underline{\gamma }}_{{.<}{\beta ><}{\gamma >}}^{<\alpha >}+{\underline{T%
}}_{{.<}{\beta ><}{\gamma >}}^{<\alpha >}.
$$

We are interested in definition of local 1--1 maps from $U$ to ${\underline{U%
}}$ characterized by s--symmetric, $P_{{.<}{\beta ><}{\gamma
>}}^{<\alpha
>}, $ and s--antisymmetric,
$$Q_{{.<}{\beta ><}{\gamma >}}^{<\alpha >}(
Q_{{.<}{\beta ><}{\gamma >}}^{<\alpha >}=-(-)^{|<\beta ><\gamma >|}Q_{{%
.<\gamma ><\beta >}}^{<\alpha >}$$ deformations:
$$
{\underline{\gamma }}_{{.<}{\beta ><}{\gamma >}}^{<\alpha >}={\gamma }_{{.<}{%
\beta ><}{\gamma >}}^{<\alpha >}+P_{{.<}{\beta ><}{\gamma >}}^{<\alpha >}%
\eqno(3.2)
$$
and
$$
{\underline{T}}_{{.<}{\beta ><}{\gamma >}}^{<\alpha >}=T_{{.<}{\beta ><}{%
\gamma >}}^{<\alpha >}+Q_{{.<}{\beta ><}{\gamma >}}^{<\alpha
>}.\eqno(3.3)
$$
The auxiliary linear covariant derivations induced by $${\gamma
}_{{.<}{\beta
><}{\gamma >}}^{<\alpha >} \mbox{ and }
{\underline{\gamma }}_{{.<}{\beta ><}{%
\gamma >}}^{<\alpha >}$$ are denoted respectively as $^{({\gamma })}D$~ and $%
^{({\gamma })}{\underline{D}}.$~

Locally adapted coordinate parametrizations of curves on $U$~are
written in form:
$$
u^{<\alpha >}=u^{<\alpha >}({\eta })=(x^{<I>}({\eta }),y^{<A>}({\eta })),~{%
\eta }_1<{\eta }<{\eta }_2,
$$
where corresponding tangent vector fields are defined as
$$
v^{<\alpha >}={\frac{{du^{<\alpha >}}}{d{\eta }}}=({\frac{{dx^I({\eta })}}{{d%
{\eta }}}},{\frac{{dy^{<A>}({\eta })}}{d{\eta }}}).
$$

\begin{definition}
\label{3.1d} A curve $l$~ is auto parallel, a--parallel, on
${\cal E}^{<z>}$ if its tangent ds--vector field $v^\alpha $~
satisfies a--parallel equations :
$$
vDv^{<\alpha >}=v^{<\beta >}{^{({\gamma })}D}_{<\beta >}v^{<\alpha >}={\rho }%
({\eta })v^{<\alpha >},\eqno(3.4)
$$
where ${\rho }({\eta })$~ is a scalar function on ${\cal
E}^{<z>}$.
\end{definition}

We consider a curve ${\underline{l}}{\subset \underline{{\cal
E}}}^{<z>}$ is
given in parametric form as\\ $u^{<\alpha >}=u^{<\alpha >}({\eta }),~{\eta }%
_1<{\eta }<{\eta }_2$ with tangent vector field $v^{<\alpha >}={\frac{{%
du^{<\alpha >}}}{{d{\eta }}}}{\ne }0$ and suppose that a
2--dimensional s--distribution $E_2({\underline{l}})$ is defined
along ${\underline{l}},$ i.e. in every point $u{\in
}{\underline{l}}$ is fixed a 2-dimensional ds--vector space
$E_2({\underline{l}}){\subset }{\underline{\xi }}.$ The
introduced distribution $E_2({\underline{l}})$~ is coplanar along ${%
\underline{l}}$~ if every ds--vector ${\underline{p}}^{<\alpha
>}(u_{(0)}^{<\beta >}){\subset }E_2({\underline{l}}),$ $u_{(0)}^{<\beta >}{%
\subset }{\underline{l}}$~ rests contained in the same
s--distribution after parallel transports along
${\underline{l}},$~ i.e. ${\underline{p}}^{<\alpha
>}(u^{<\beta >}({\eta })){\subset }E_2({\underline{l}}).$

\begin{definition}
\label{3.2d} A curve ${\underline{l}}$~ is nearly autoparallel,
or in brief
na--parallel, on space ~${\underline{{\cal E}}}^{<z>}$ if a coplanar along ${%
\underline{l}}$~ distribution $E_2({\underline{l}})$ containing tangent to ${%
\underline{l}}$~ vector field $v^{<\alpha >}({\eta })$,~ i.e. $v^{<\alpha >}(%
{\eta }){\subset }E_2({\underline{l}}),$~ is defined.
\end{definition}

We can define nearly autoparallel maps of la--spaces as an
anisotropic
generalization (see 
 [279,276], for ng-- 
  [230] and na--maps 
 [249,273,\\ 278,274,247]):

\begin{definition}
\label{3.3d} Nearly autoparallel maps, na--maps, of higher order
anisot\-rop\-ic s--spaces are de\-fined as lo\-cal 1--1
map\-pings of dvs---bund\-les, $${\cal E}^{<z>}{\to
\underline{{\cal E}}}^{<z>},$$ changing
every a--parallel on ${\cal E}^{<z>}$ into a na--parallel on ${\underline{%
{\cal E}}}^{<z>}.$
\end{definition}

Let formulate the general conditions when deformations (3.2) and
(3.3) charac\-ter\-ize na-maps : An a--parallel $l{\subset }U$~
is given by
func\-ti\-ons\\ $u^{<\alpha >}=u^{<{\alpha >}}({\eta }),v^{<\alpha >}={\frac{%
{du^{<\alpha >}}}{d{\eta }}}$, ${\eta }_1<{\eta }<{\eta }_2$,
satisfying equations (3.4). We consider that to this a--parallel
corresponds a na--parallel ${\underline{l}}\subset
{\underline{U}}$ given by the same parameterization in a common
for a chosen na--map coordinate system on $U$~
and ${\underline{U}}.$ This condition holds for vectors ${\underline{v}}%
_{(1)}^{<\alpha >}=v{\underline{D}}v^{<\alpha >}$~ and
${\underline v}_{(2)}^{<\alpha >}=v%
{\underline{D}}v_{(1)}^{<\alpha >}$ satisfying equality
$$
{\underline{v}}_{(2)}^{<\alpha >}={\underline{a}}({\eta })v^{<\alpha >}+{%
\underline{b}}({\eta }){\underline{v}}_{(1)}^{<\alpha >}\eqno(3.5)
$$
for some scalar s--functions ${\underline{a}}({\eta })$ and ${\underline{b}}(%
{\eta }),$ see definitions 3.4 and 3.5. Introducing (3.2) and
(3.3) into
expressions for ${\underline{v}}_{(1)}^{<\alpha >}$~ and ${\underline{v}}%
_{(2)}^{<\alpha >}$~ in (3.5) we obtain:
$$
v^{<\beta >}v^{<\gamma >}v^{<\delta >}(D_{<\beta >}P_{{.<}{\gamma
><}{\delta
>}}^{<\alpha >}+P_{{.<}{\beta ><}{\tau >}}^{<\alpha >}P_{{.<}{\gamma ><}{%
\delta >}}^{<\tau >}+Q_{{.<}{\beta ><}{\tau >}}^{<\alpha >}P_{{.<}{\gamma ><}%
{\delta >}}^{<\tau >})=\eqno(3.6)
$$
$$
bv^{<\gamma >}v^{<\delta >}P_{{.<}{\gamma ><}{\delta >}}^{<\alpha
>}+av^{<\alpha >},
$$
where
$$
b({\eta },v)={\underline{b}}-3{\rho },\qquad \mbox{and}\qquad a({\eta },v)={%
\underline{a}}+{\underline{b}}{\rho }-v^{<\alpha >}{\delta }_{<\alpha >}{%
\rho }-{\rho }^2\eqno(3.7)
$$
are called the deformation parameters of na--maps.

The deformation of torsion $Q_{{.<}{\beta ><}{\tau >}}^{<\alpha
>}$ must
satisfy algebraic compatibility conditions for a given nonmetricity tensor ${%
\underline{K}}_{<{\alpha ><}{\beta ><}{\gamma >}}$~ (see (3.1))
 on ${\underline{{\cal E}}%
}^{<z>}$\\ ( or metricity conditions if d--connection
${\underline{D}}_\alpha $
 on ${\underline{{\cal E}}}^{<z>}$ is required to be metric) :
$$
D_{<\alpha >}G_{<{\beta ><}{\gamma >}}-P_{{.<}{\alpha >}(<{\beta >}%
}^{<\delta >}G_{<{\gamma >}\}{<}{\delta >}}-{\underline{K}}_{<{\alpha ><}{%
\beta ><}{\gamma >}}=Q_{{.<}{\alpha >}(<{\beta >}}^{<\delta >}G_{<{\gamma >}%
\}<{\delta >}},\eqno(3.8)
$$
where $\{{\quad }]$ denotes the s--symmetric alternation
satisfying, for instance, conditions
$$
Q_{{.<}{\alpha >}(<{\beta >}}^{<\delta >}G_{<{\gamma >}\}<{\delta >}}=Q_{{.<}%
{\alpha >}<{\beta >}}^{<\delta >}G_{<{\gamma >}<{\delta
>}}+(-)^{|<\beta
><\gamma >|}Q_{{.<}{\alpha >}<\gamma {>}}^{<\delta >}G_{<\beta {>}<{\delta >}%
};
$$
we shall use also operation $[...\}$ defined, for instance, as%
$$
A_{[<\alpha >}B_{<\beta >\}}=A_{<\alpha >}B_{<\beta
>}-(-)^{|<\alpha ><\beta
>|}A_{<\beta >}B_{<\alpha >}.
$$

So, we have proved the

\begin{theorem}
\label{3.1t} The na--maps from a dvs--bundle ${\cal E}^{<z>}$ to a
dvs--bundle ${\underline{{\cal E}}}^{<z>}$ with a fixed common
nonlinear connection structure
$$
N_{A_p}^{A_f}(u)={\underline{N}}_{A_p}^{A_f}(u)
$$
and given d--connections,
$$
{\Gamma }_{{.<}{\beta ><}{\gamma >}}^{<\alpha >}\mbox{ on }{\cal E}^{<z>}%
\mbox{ and }{\underline{\Gamma }}_{{.}{\beta }{\gamma }}^\alpha \mbox{ on }{%
\underline{{\cal E}}}^{<z>},
$$
are locally parametrized by the solutions of equations (3.6) and
(3.8) for every point $u^{<\alpha >}$ and direction $v^{<\alpha
>}$ on $U{\subset \underline{{\cal E}}}^{<z>}.$
\end{theorem}

We call (3.6) and (3.8) the basic equations for na--maps of
higher order
anisotropic s--spaces. They are a generalization for such
s--spaces of
corresponding Sinyukov's equations 
 [230] for isotropic spaces provided
with symmetric affine connection structure.

\section{ Classification of Na--Maps}

We can classify na--maps by considering possible polynomial
parametrizations on variables $v^{<\alpha >}$~ of deformations
parameters $a$ and $b$ from (3.6) and (3.7) ).
 \begin{theorem}
\label{3.2t} There are four classes of na--maps characterized by
corresponding deformation parameters and deformation ds--tensors
and basic
equations:

\begin{enumerate}
\item  for $na_{(0)}$--maps, ${\pi }_{(0)}$--maps,
$$
P_{<{\beta ><}{\gamma >}}^{<\alpha >}(u)={\psi }_{({<\beta >}}{\delta }_{<{%
\gamma >}\}}^{<\alpha >}
$$
(${\delta }_{<\beta >}^{<\alpha >}$~ is the Kronecker symbol and ${\psi }%
_{<\beta >}={\psi }_{<\beta >}(u)$~ is a covariant ds--vector
field);

\item  for $na_{(1)}$--maps
$$
a(u,v)=a_{<{\alpha ><}{\beta >}}(u)v^{<\alpha >}v^{<\beta >},\quad
b(u,v)=b_{<\alpha >}(u)v^{<\alpha >}
$$
and $P_{{.<}{\beta ><}{\gamma >}}^{<\alpha >}(u)$~ is the
solution of equations
$$
D_{(<{\alpha >}}P_{{.<}{\beta ><}{\gamma >}\}}^{<\delta >}+P_{(<{\alpha ><}{%
\beta >}}^{<\tau >}P_{{.<}{\gamma >}\}<{\tau >}}^{<\delta >}-P_{(<{\alpha ><}%
{\beta >}}^{<\tau >}Q_{{.<}{\gamma >}\}<{\tau >}}^{<\delta
>}=\eqno(3.9)
$$
$$
b_{(<{\alpha >}}P_{{.<}{\beta ><}{\gamma >}\}}^{<{\delta >}}+a_{(<{\alpha ><}%
{\beta >}}{\delta }_{<{\gamma >}\}}^{<\delta >};
$$

\item  for $na_{(2)}$--maps
$$
a(u,v)=a_{<\beta >}(u)v^{<\beta >},\quad b(u,v)={\frac{{b_{<{\alpha ><}{%
\beta >}}v^{<\alpha >}v^{<\beta >}}}{{{\sigma }_{<\alpha >}(u)v^{<\alpha >}}}%
}, $$ $${\sigma }_{<\alpha >}v^{<\alpha >}{\neq }0,
$$
$$
P_{{.<}{\alpha ><}{\beta >}}^{<\tau >}(u)={{\psi }_{(<{\alpha >}}}{\delta }%
_{<{\beta >}\}}^{<\tau >}+{\sigma }_{(<{\alpha >}}F_{<{\beta
>}\}}^{<\tau >}
$$
and $F_{<\beta >}^{<\alpha >}(u)$~ is the solution of equations
$$
{D}_{(<{\gamma >}}F_{<{\beta >}\}}^{<\alpha >}+F_{<\delta >}^{<\alpha >}F_{(<%
{\gamma >}}^{<\delta >}{\sigma }_{<{\beta >}\}}-Q_{{.<}{\tau >}(<{\beta >}%
}^{<\alpha >}F_{<{\gamma >}\}}^{<\tau >}=\eqno(3.10)
$$
$$
{\mu }_{(<{\beta >}}F_{<{\gamma >}\}}^{<\alpha >}+{\nu
}_{(<{\beta >}}{\delta }_{<{\gamma >}\}}^{<\alpha >}
$$
$({\mu }_{<\beta >}(u),{\nu }_{<\beta >}(u),{\psi }_{<\alpha >}(u),{\sigma }%
_{<\alpha >}(u)$~ are covariant ds--vectors) ;

\item  for $na_{(3)}$--maps
$$
b(u,v)={\frac{{{\alpha }_{<{\beta ><}{\gamma ><}{\delta
>}}v^{<\beta
>}v^{<\gamma >}v^{<\delta >}}}{{{\sigma }_{<{\alpha ><}{\beta >}}v^{<\alpha
>}v^{<\gamma >}}}},
$$
$$
P_{{.<}{\beta ><}{\gamma >}}^{<\alpha >}(u)={\psi }_{(<{\beta >}}{\delta }_{<%
{\gamma >}\}}^{<\alpha >}+{\sigma }_{<{\beta ><}{\gamma >}}{\varphi }%
^{<\alpha >},
$$
where ${\varphi }^{<\alpha >}$~ is the solution of equations

$$
D_{<\beta >}{\varphi }^{<\alpha >}={\nu }{\delta }_{<\beta >}^{<\alpha >}+{%
\mu }_{<\beta >}{\varphi }^{<\alpha >}+{\varphi }^{<\gamma
>}Q_{{.<}{\gamma
><}{\delta >}}^{<\alpha >},\eqno(3.11)
$$
where ${\alpha }_{<{\beta ><}{\gamma ><}{\delta >}}(u),{\sigma }_{<{\alpha ><%
}{\beta >}}(u),{\psi }_{<\beta >}(u),{\nu }(u)$~ and ${\mu
}_{<\beta >}(u)$ are ds---tensors.
\end{enumerate}
\end{theorem}

{\bf Proof.} We sketch respectively:

\begin{enumerate}
\item  Using a ds--tensor ${P^{<\alpha >}}_{<\beta ><\gamma >}(u)={\psi }%
_{(<\beta >}{\delta }_{<\gamma >\}}^{<\alpha >}$ one
 can show that a--paral\-lel
equations (3.4) on ${\cal E}^{<z>}$ transform into similar ones on
\underline{${\cal E}$}$^{<z>}$ if and only if deformations of
type (3.2) are considered.

\item  From corresponding to $na_{(1)}$--maps parametrizations of $a(u,v)$
and $b(u,v)$ (see conditions of the theorem) for a $v^\alpha \neq 0$ on $%
U\in {\cal E}^{<z>}$ and after a redefinition of deformation
parameters we obtain that equations (3.6) hold if and only if
${P^{<\alpha >}}_{<\beta
><\gamma >}$ satisfies (3.3).

\item  In a similar manner we obtain basic $na_{(2)}$--map equations (3.10)
from (3.6) by considering $na_{(2)}$--parametrizations of
deformation parameters and d--tensor.

\item  For $na_{(3)}$--maps we must take into consideration deformations of
torsion (3.3) and introduce $na_{(3)}$--parametrizations for $b(u,v)$ and\\ ${%
P^{<\alpha >}}_{<\beta ><\gamma >}$ into the basic na--equations
(3.6). The last ones, for $na_{(3)}$--maps, are equivalent to
equations (3.11) (with a corresponding redefinition of
deformation parameters). \qquad $\Box $
\end{enumerate}

We point out that for ${\pi }_{(0)}$-maps we have not
differential equations on\\ $P_{{.<}{\beta ><}{\gamma
>}}^{<\alpha >}$ (in the isotropic case one considers a first
order system of differential equations on metric
 [230]; we omit constructions with deformation of metric in this section).\

To formulate invariant conditions for reciprocal na--maps (when
every
a-parallel on ${\underline{{\cal E}}}^{<z>}$~ is also transformed
into na--parallel on ${\cal E}^{<z>}$ ) we introduce into
consideration the
curvature and Ricci tensors defined for auxiliary connection ${\gamma }_{{.<}%
{\beta ><}{\gamma >}}^{<\alpha >}$~ :
$$
r_{<{\alpha ><\beta ><}{\tau >}}^{{<}{\delta >}}={\delta }_{[<{\beta >}}{%
\gamma }_{{.<}{\tau >}\}<{\alpha >}}^{<\delta >}+$$ $${\gamma }_{{.<}{\rho >}[<{%
\beta >}}^{<\delta >}{\gamma }_{{.<}{\tau >}\}<{\alpha >}}^{<\rho >}+{{%
\gamma }^{<\delta >}}_{<\alpha ><\phi >}{w^{<\phi >}}_{<\beta
><\tau >}
$$
and, respectively, $r_{<{\alpha ><}{\tau >}}=r_{<\alpha ><\tau
><\gamma
>}^{<\gamma >}$, where $[\quad \}$ denotes antisymmetric s--alternation of
indices. We define values:
$$
^{(0)}T_{{.<}{\alpha ><}{\beta >}}^{<\mu >}={\Gamma
}_{{.<}{\alpha ><}{\beta
>}}^{<\mu >}-T_{{.<}{\alpha ><}{\beta >}}^{<\mu >}-\frac 1{n_E}({\delta }_{(<%
{\alpha >}}^{<\mu >}{\Gamma }_{{.<}{\beta >}\}<{\delta >}}^{<\delta >}-{%
\delta }_{(<{\alpha >}}^{<\mu >}T_{{.<}{\beta >}\}<{\gamma
>}}^{<\delta >}),
$$
$$
{}^{(0)}{W}_{<\alpha ><\beta ><\gamma >}^{<\tau >}={r}_{<\alpha
><\beta
><\gamma >}^{<\tau >}+\frac 1{n_E}[{\gamma }_{\cdot <\varphi ><\tau
>}^{<\tau >}{\delta }_{(<\alpha >}^{<\tau >}{w^{<\varphi >}}_{<\beta
>\}<\gamma >}-
$$
$$
({\delta }_{<\alpha >}^{<\tau >}{r}_{[<\gamma ><\beta >\}}+{\delta }%
_{<\gamma >}^{<\tau >}{r}_{[<\alpha ><\beta >\}}-{\delta
}_{<\beta >}^{<\tau
>}{r}_{[<\alpha ><\gamma >\}})]-$$
$$\frac 1{(n_E)^2}[{\delta }_{<\alpha
>}^{<\tau >}(2{\gamma }_{\cdot <\varphi ><\tau >}^{<\tau >}{w^{<\varphi >}}%
_{[<\gamma ><\beta >\}}- {\gamma }_{\cdot <\tau >[<\gamma
>}^{<\tau >}{w^{<\varphi >}}_{<\beta
>\}<\varphi >})+$$
$${\delta }_{<\gamma >}^{<\tau >}(2{\gamma }_{\cdot <\varphi
><\tau >}^{<\tau >}{w^{<\varphi >}}_{<\alpha ><\beta >}-{\gamma }_{\cdot
<\alpha ><\tau >}^{<\tau >}{w^{<\varphi >}}_{<\beta ><\varphi >})-
$$
$$
{\delta }_{<\beta >}^{<\tau >}(2{\gamma }_{\cdot <\varphi ><\tau >}^{<\tau >}%
{w^{<\varphi >}}_{<\alpha ><\gamma >}-{\gamma }_{\cdot <\alpha
><\tau
>}^{<\tau >}{w^{<\varphi >}}_{<\gamma ><\varphi >})],
$$
$$
{^{(3)}T}_{{.<}{\alpha ><}{\beta >}}^{<\delta >}={\gamma }_{{.<}{\alpha ><}{%
\beta >}}^{<\delta >}+{\epsilon }{\varphi }^{<\tau >}{^{({\gamma })}D}%
_{<\beta >}q_{<\tau >}+\frac 1{n_E}({\delta }_{<\alpha
>}^{<\gamma >}-$$
$${\epsilon }{\varphi }^{<\delta >}q_{<\alpha >})[{\gamma }_{{.<}{\beta ><}{%
\tau >}}^{<\tau >}+ {\epsilon }{\varphi }^{<\tau >}{^{({\gamma
})}D}_{<\beta >}q_{<\tau >}+
$$
$$\frac
1{n_E-1}q_{<\beta >}({\epsilon }{\varphi }^{<\tau >}{\gamma }_{{.<}{\tau ><}{%
\lambda >}}^{<\lambda >}+{\varphi }^{<\lambda >}{\varphi }^{<\tau >}{^{({%
\gamma })}D}_{<\tau >}q_{<\lambda >})]-
$$
$$
\frac 1{n_E}({\delta }_{<\beta >}^{<\delta >}-{\epsilon }{\varphi
}^{<\delta
>}q_{<\beta >})[{\gamma }_{{.<}{\alpha ><}{\tau >}}^{<\tau >}+{\epsilon }{%
\varphi }^{<\tau >}{^{({\gamma })}D}_{<\alpha >}q_{<\tau >}+
$$
$$
\frac 1{n_E-1}q_{<\alpha >}({\epsilon }{\varphi }^{<\tau >}{\gamma }_{{.<}{%
\tau ><}{\lambda >}}^{<\lambda >}+{\varphi }^{<\lambda >}{\varphi }^{<\tau >}%
{^{(\gamma )}D}_{<\tau >}q_{<\lambda >})],
$$

$$
{^{(3)}W}_{<\beta ><\gamma ><\delta >}^{<\alpha >}={\rho }_{<{\beta >}{.<}{%
\gamma ><}{\delta >}}^{{.<}{\alpha >}}+{\epsilon }{\varphi
}^{<\alpha
>}q_{<\tau >}{\rho }_{<{\beta >}{.<}{\gamma ><}{\delta >}}^{{.<}{\tau >}}+$$
$$({\delta }_{<\delta >}^{<\alpha >}-
{\epsilon }{\varphi }^{<\alpha >}q_{<\delta >})p_{<{\beta
><}{\gamma >}}-
$$
$$({%
\delta }_{<\gamma >}^{<\alpha >}-{\epsilon }{\varphi }^{<\alpha
>}q_{<\gamma
>})p_{<{\beta ><}{\delta >}}-({\delta }_{<\beta >}^{<\alpha >}-{\epsilon }{%
\varphi }^{<\alpha >}q_{<\beta >})p_{[<{\gamma ><}{\delta >}\}},
$$
$$
(n_E-2)p_{<{\alpha ><}{\beta >}}=-{\rho }_{<{\alpha ><}{\beta >}}-{\epsilon }%
q_{<\tau >}{\varphi }^{<\gamma >}{\rho }_{<{\alpha >}{.<}{\beta ><}{\gamma >}%
}^{{.<}{\tau >}}+\frac 1{n_E}[{\rho }_{<{\tau >}{.<}{\beta ><}{\alpha >}}^{{%
.<}{\tau >}}+$$
$${\epsilon }q_{<\beta >}{\varphi }^{<\tau >}{\rho }_{<{\alpha ><}%
{\tau >}}-
{\epsilon }q_{<\tau >}{\varphi }^{<\gamma >}{\rho }_{<{\gamma >}{.<}{\beta ><%
}{\alpha >}}^{{.<}{\tau >}}+$$
$${\epsilon }q_{<\alpha >}(-{\varphi }^{<\gamma >}{%
\rho }_{<{\tau >}{.<}{\beta ><}{\gamma >}}^{{.<}{\tau >}}+{\epsilon }%
q_{<\tau >}{\varphi }^{<\gamma >}{\varphi }^{<\delta >}{\rho }_{<{\gamma >}{%
.<}{\beta ><}{\vec \delta >}}^{{.<}{\tau >}})],
$$
where $q_{<\alpha >}{\varphi }^{<\alpha >}={\epsilon }=\pm 1,$
$n_E$ is the
dimension of dvs--bundle,%
$$
{{\rho }^{<\alpha >}}_{<\beta ><\gamma ><\delta >}=r_{<\beta
>\cdot <\gamma
><\delta >}^{\cdot <\alpha >}+$$ $${\frac 12}({\psi }_{(<\beta >}{\delta }%
_{<\varphi >\}}^{<\alpha >}+{\sigma }_{<\beta ><\varphi
>}{\varphi }^{<\tau
>}){w^{<\varphi >}}_{<\gamma ><\delta >}
$$
( we write $${\rho }_{\cdot <\beta ><\gamma ><\delta >}^{<\alpha >}={%
\underline{r}}_{<\beta >\cdot <\gamma ><\delta >}^{\cdot <\alpha >}-{\frac 12%
}({\psi }_{(<\beta >}{\delta }_{<{\varphi >}\}}^{<\alpha >}-{\sigma }%
_{<\beta ><\varphi >}{\varphi }^{<\tau >}){w^{<\varphi
>}}_{<\gamma ><\delta
>}$$ for a corresponding value on \underline{${\cal E}$}$^{<z>})$ and ${\rho }%
_{<\alpha ><\beta >}={\rho }_{\cdot <\alpha ><\beta ><\tau
>}^{<\tau >}.$

Similar values,
$$
^{(0)}{\underline{T}}_{{.<}{\beta ><}{\gamma >}}^{<\alpha >},^{(0)}{%
\underline{W}}_{{.<}{\alpha ><}{\beta ><}{\gamma >}}^{<\nu >},{\hat T}_{{.<}{%
\beta ><}{\gamma >}}^{<\alpha >},{\check T}_{{.<}{\beta ><}{\tau >}%
}^{<\alpha >},
$$
$$
{\hat W}_{{.<}{\alpha ><}{\beta ><}{\gamma >}}^{<\delta >},{\check W}_{{.<}{%
\alpha ><}{\beta ><}{\gamma >}}^{<\delta >},^{(3)}{\underline{T}}_{{.<}{%
\alpha ><}{\beta >}}^{<\delta >},^{(3)}{\underline{W}}_{{.<}{\beta ><}{%
\gamma ><}{\delta >}}^{<\alpha >},
$$
are given, correspondingly, by auxiliary connections ${\underline{\Gamma }}_{%
{.<}{\alpha ><}{\beta >}}^{<\mu >},$~
$$
{\star {\gamma }}_{{.<}{\beta ><}{\lambda >}}^{<\alpha >}={\gamma }_{{.<}{%
\beta ><}{\lambda >}}^{<\alpha >}+{\epsilon }F_{<\tau >}^{<\alpha >}{^{({%
\gamma })}D}_{(<{\beta >}}F_{<{\lambda >}\}}^{<\tau >},$$ $$ {\check \gamma }%
_{{.<}{\beta ><}{\lambda >}}^{<\alpha >}={\widetilde{\gamma }}_{{.<}{\beta ><%
}{\lambda >}}^{<\alpha >}+{\epsilon }F_{<\tau >}^{<\lambda >}{\widetilde{D}}%
_{(<{\beta >}}F_{<{\lambda >}\}}^{<\tau >},
$$
$$
{\widetilde{\gamma }}_{{.<}{\beta ><}{\tau >}}^{<\alpha >}={\gamma }_{{.<}{%
\beta ><}{\tau >}}^{<\alpha >}+{\sigma }_{(<{\beta >}}F_{<{\tau >}%
\}}^{<\alpha >},$$ $$ {\hat \gamma }_{{.<}{\beta ><}{\lambda >}}^{<\alpha >}=%
{\star {\gamma }}_{{.<}{\beta ><}{\lambda >}}^{<\alpha
>}+{\widetilde{\sigma }}_{(<{\beta >}}{\delta }_{<{\lambda
>}\}}^{<\alpha >},
$$
where ${\widetilde{\sigma }}_{<\beta >}={\sigma }_{<\alpha
>}F_{<\beta
>}^{<\alpha >}.$

\begin{theorem} \label{3.3t} Four classes of reciprocal
 na--maps of higher order an\-isot\-rop\-ic
s--spaces are characterized by corresponding invariant criterions:
\begin{enumerate}
\item  for a--maps $^{(0)}T_{{.<}{\alpha ><}{\beta >}}^{<\mu >}=^{(0)}{%
\underline{T}}_{{.<}{\alpha ><}{\beta >}}^{<\mu >},$
$$
{}^{(0)}W_{{.<}{\alpha ><}{\beta ><}{\gamma >}}^{<\delta >}=^{(0)}{%
\underline{W}}_{{.<}{\alpha ><}{\beta ><}{\gamma >}}^{<\delta
>};\eqno(3.12)
$$

\item  for $na_{(1)}$--maps
$$
3({^{({\gamma })}D}_{<\lambda >}P_{{.<}{\alpha ><}{\beta >}}^{<\delta >}+P_{{%
.<}{\tau ><}{\lambda >}}^{<\delta >}P_{{.<}{\alpha ><}{\beta
>}}^{<\tau
>})= $$
$$r_{(<{\alpha >}{.<}{\beta >}\}<{\lambda >}}^{{.<}{\delta >}}-{\underline{%
r}}_{(<{\alpha >}{.<}{\beta >}\}<{\lambda >}}^{{.<}{\delta >}}+
[T_{{.<}{\tau >}(<{\alpha >}}^{<\delta >}P_{{.<}{\beta ><}{\lambda >}%
\}}^{<\tau >}+$$ $$Q_{{.<}{\tau >}(<{\alpha >}}^{<\delta >}P_{{.<}{\beta ><}{%
\lambda >}\}}^{<\tau >}+b_{(<{\alpha >}}P_{{.<}{\beta ><}{\lambda >}%
\}}^{<\delta >}+{\delta }_{(<{\alpha >}}^{<\delta >}a_{<{\beta ><}{\lambda >}%
\}}];
$$

\item  for $na_{(2)}$--maps ${\hat T}_{{.<}{\beta ><}{\tau >}}^{<\alpha >}={%
\star T}_{{.<}{\beta ><}{\tau >}}^{<\alpha >},$
$$
{\hat W}_{{.<}{\alpha ><}{\beta ><}{\gamma >}}^{<\delta >}={\star W}_{{.<}{%
\alpha ><}{\beta ><}{\gamma >}}^{<\delta >};
$$

\item  for $na_{(3)}$--maps $^{(3)}T_{{.<}{\beta ><}{\gamma >}}^{<\alpha
>}=^{(3)}{\underline{T}}_{{.<}{\beta ><}{\gamma >}}^{<\alpha >},$
$$
{}^{(3)}W_{{.<}{\beta ><}{\gamma ><}{\delta >}}^{<\alpha >}=^{(3)}{%
\underline{W}}_{{.<}{\beta ><}{\gamma ><}{\delta >}}^{<\alpha
>}.\eqno(3.13)
$$
\end{enumerate}
\end{theorem}

{\bf Proof. }

\begin{enumerate}
\item  First we prove that a--invariant conditions (3.12) hold. Deformations
of d---connections of type
$$
{}^{(0)}{\underline{\gamma }}_{\cdot <\alpha ><\beta >}^{<\mu >}={{\gamma }%
^{<\mu >}}_{<\alpha ><\beta >}+{\psi }_{(<\alpha >}{\delta
}_{<\beta
>\}}^{<\mu >}\eqno(3.14)
$$
define a--maps. Contracting indices $<\mu >$ and $<\beta >$ we
can write
$$
{\psi }_{<\alpha >}=\frac 1{n_E}({{\underline{\gamma }}^{<\beta
>}}_{<\alpha
><\beta >}-{{\gamma }^{<\beta >}}_{<\alpha ><\beta >}).\eqno(3.15)
$$
Introducing d--vector ${\psi }_{<\alpha >}$ into previous
relation and expressing
$$
{{\gamma }^{<\alpha >}}_{<\beta ><\tau >}=-{T^{<\alpha >}}_{<\beta ><\tau >}+%
{{\Gamma }^{<\alpha >}}_{<\beta ><\tau >}
$$
and similarly for underlined values we obtain the first invariant
conditions from (3.12).

Putting deformation (3.14) into the formula for
$$
{\underline{r}}_{<\alpha >\cdot <\beta ><\gamma >}^{\cdot <\tau
>}\quad \mbox{and}\quad {\underline{r}}_{<\alpha ><\beta
>}={\underline{r}}_{<\alpha
><\beta ><\tau >}^{\cdot <\tau >}
$$
we obtain respectively relations
$$
{\underline{r}}_{<\alpha >\cdot <\beta ><\gamma >}^{\cdot <\tau
>}-r_{<\alpha >\cdot <\beta ><\gamma >}^{\cdot <\tau >}={\delta }_{<\alpha
>}^{<\tau >}{\psi }_{[<\gamma ><\beta >\}}+$$
$${\psi }_{<\alpha >[<\beta >}{%
\delta }_{<\gamma >\}}^{<\tau >}+{\delta }_{(<\alpha >}^{<\tau >}{\psi }%
_{<\varphi >\}}{w^{<\varphi >}}_{<\beta ><\gamma >}
$$
and
$$
{\underline{r}}_{<\alpha ><\beta >}-r_{<\alpha ><\beta >}=$$
$${\psi }_{[<\alpha
><\beta >\}}+(n_E-1){\psi }_{<\alpha ><\beta >}+{\psi }_{<\varphi >}{%
w^{<\varphi >}}_{<\beta ><\alpha >}+{\psi }_{<\alpha >}{w^{<\varphi >}}%
_{<\beta ><\varphi >},\eqno(3.16)
$$
where
$$
{\psi }_{<\alpha ><\beta >}={}^{({\gamma })}D_{<\beta >}{\psi }_{<\alpha >}-{%
\psi }_{<\alpha >}{\psi }_{<\beta >}.
$$
Putting (3.14) into (3.16) we can express ${\psi }_{[<\alpha
><\beta >\}}$ as

$$
{\psi }_{[\alpha \beta \}}=\frac
1{n_E+1}[{\underline{r}}_{[<\alpha ><\beta
>\}}+\frac 2{n_E+1}{\underline{\gamma }}_{\cdot <\varphi ><\tau >}^{<\tau >}{%
w^{<\varphi >}}_{[<\alpha ><\beta >\}}- \eqno(3.17)
$$
$$
\frac 1{n_E+1}{\underline{\gamma }}_{\cdot <\tau >[<\alpha >}^{<\tau >}{%
w^{<\varphi >}}_{<\beta >\}<\varphi >}]-\frac
1{n_E+1}[r_{[<\alpha ><\beta
>\}}+
$$
$$
\frac 2{n_E+1}{{\gamma }^{<\tau >}}_{<\varphi ><\tau >}{w^{<\varphi >}}%
_{[<\alpha ><\beta >\}}-$$ $$\frac 1{n_E+1}{{\gamma }^{<\tau
>}}_{<\tau >[<\alpha
>}{w^{<\varphi >}}_{<\beta >\}<\varphi >}].
$$
To simplify our consideration we can choose an a--transform,
pa\-ra\-met\-ri\-zed by corresponding $\psi $--vector from
(3.14), (or fix a local coordinate cart) the antisymmetrized
relations (3.17) to be satisfied by the ds--tensor
$$
{\psi }_{<\alpha ><\beta >}=\frac
1{n_E+1}[{\underline{r}}_{<\alpha ><\beta
>}+\frac 2{n_E+1}{\underline{\gamma }}_{\cdot <\varphi ><\tau >}^{<\tau >}{%
w^{<\varphi >}}_{<\alpha ><\beta >}-$$
$$\frac 1{n_E+1}{\underline{\gamma }}%
_{\cdot <\alpha ><\tau >}^{<\tau >}{w^{<\varphi >}}_{<\beta
><\varphi >}- r_{<\alpha ><\beta >}-$$
$$\frac 2{n_E+1}{{\gamma }^{<\tau >}}_{<\varphi ><\tau >}%
{w^{<\varphi >}}_{<\alpha ><\beta >}+\frac 1{n_E+1}{{\gamma }^{<\tau >}}%
_{<\alpha ><\tau >}{w^{<\varphi >}}_{<\beta ><\varphi
>}]\eqno(3.18)
$$
Introducing expressions (3.14),(3.17) and (3.18) into deformation
of curvature (3.15) we obtain the second conditions (3.12) of
a-map invariance:
$$
^{(0)}W_{<\alpha >\cdot <\beta ><\gamma >}^{\cdot <\delta >}={}^{(0)}{%
\underline{W}}_{<\alpha >\cdot <\beta ><\gamma >}^{\cdot <\delta
>},
$$
where the Weyl d--tensor on $\underline{{\cal E}}^{<z>}$ (the
extension of
the usual one for geodesic maps on (pseudo)--Riemannian spaces to
the case of dvs---bundles provided with N--connection structure)
is defined as
$$
{}^{(0)}{\underline{W}}_{<\alpha >\cdot <\beta ><\gamma >}^{\cdot <\tau >}={%
\underline{r}}_{<\alpha >\cdot <\beta ><\gamma >}^{\cdot <\tau
>}+\frac
1{n_E}[{\underline{\gamma }}_{\cdot <\varphi ><\tau >}^{<\tau >}{\delta }%
_{(<\alpha >}^{<\tau >}{w^{<\varphi >}}_{<\beta >\}<\gamma >}-$$
$$({\delta }_{<\alpha >}^{<\tau >}{\underline{r}}_{[<\gamma ><\beta >\}}+
{\delta }_{<\gamma >}^{<\tau >}{\underline{r}}_{[<\alpha ><\beta >\}}-{%
\delta }_{<\beta >}^{<\tau >}{\underline{r}}_{[<\alpha ><\gamma
>\}})]-$$
$$\frac1{(n_E)^2}[{\delta }_{<\alpha >}^{<\tau >}(2{\underline{\gamma }}_{\cdot
<\varphi ><\tau >}^{<\tau >}{w^{<\varphi >}}_{[<\gamma ><\beta
>\}}-
{\underline{\gamma }}_{\cdot <\tau >[<\gamma >}^{<\tau >}{w^{<\varphi >}}%
_{<\beta >\}<\psi >})+$$
$${\delta }_{<\gamma >}^{<\tau >}(2{\underline{\gamma }}%
_{\cdot <\varphi ><\tau >}^{<\tau >}{w^{<\varphi >}}_{<\alpha
><\beta >}-
{\underline{\gamma }}_{\cdot <\alpha ><\tau >}^{<\tau >}{w^{<\varphi >}}%
_{<\beta ><\varphi >})-$$ $${\delta }_{<\beta >}^{<\tau >}(2{\underline{\gamma }}%
_{\cdot <\varphi ><\tau >}^{<\tau >}{w^{<\varphi >}}_{<\alpha ><\gamma >}-{%
\underline{\gamma }}_{\cdot <\alpha ><\tau >}^{<\tau >}{w^{<\varphi >}}%
_{<\gamma ><\varphi >})]
$$
The formula for $^{(0)}W_{<\alpha >\cdot <\beta ><\gamma
>}^{\cdot <\tau >}$ written similarly with respect to
non--underlined values is presented in section 1.2.

\item  To obtain $na_{(1)}$--invariant conditions we rewrite $na_{(1)}$%
--equations (3.9) as to consider in explicit form covariant derivation $^{({%
\gamma })}D$ and deformations (3.2) and (3.3):
$$
2({}^{({\gamma })}D_{<\alpha >}{P^{<\delta >}}_{<\beta ><\gamma >}+{}^{({%
\gamma })}D_{<\beta >}{P^{<\delta >}}_{<\alpha ><\gamma >}+{}^{({\gamma }%
)}D_{<\gamma >}{P^{<\delta >}}_{<\alpha ><\beta >}+$$
$${P^{<\delta >}}_{<\tau><\alpha >}{P^{<\tau >}}_{<\beta ><\gamma >}+
{P^{<\delta >}}_{<\tau ><\beta >}{P^{<\tau >}}_{<\alpha ><\gamma
>}+$$
$${P^{<\delta >}}_{<\tau ><\gamma >}{P^{<\tau >}}_{<\alpha ><\beta >})=
{T^{<\delta >}}_{<\tau >(<\alpha >}{P^{<\tau >}}_{<\beta ><\gamma
>\}}+$$
$${H^{<\delta >}}_{<\tau >(<\alpha >}{P^{<\tau >}}_{<\beta ><\gamma
>\}}+$$
$$b_{(<\alpha >}{P^{<\delta >}}_{<\beta ><\gamma >\}}+a_{(<\alpha ><\beta
>}{\delta }_{<\gamma >\}}^{<\delta >}.\eqno(3.19)
$$
Alternating the first two indices in (3.19) we have
$$
2({\underline{r}}_{(<\alpha >\cdot <\beta >\}<\gamma >}^{\cdot
<\delta
>}-r_{(<\alpha >\cdot <\beta >\}<\gamma >}^{\cdot <\delta >})=2({}^{(\gamma
)}D_{<\alpha >}{P^{<\delta >}}_{<\beta ><\gamma >}+
$$
$$
{}^{(\gamma )}D_{<\beta >}{P^{<\delta >}}_{<\alpha ><\gamma
>}-2{}^{(\gamma )}D_{<\gamma >}{P^{<\delta >}}_{<\alpha ><\beta
>}+$$
 $${P^{<\delta >}}_{<\tau><\alpha >}{P^{<\tau >}}_{<\beta ><\gamma >}+
$$
$$
{P^{<\delta >}}_{<\tau ><\beta >}{P^{<\tau >}}_{<\alpha ><\gamma >}-2{%
P^{<\delta >}}_{<\tau ><\gamma >}{P^{<\tau >}}_{<\alpha ><\beta
>}).
$$
Substituting the last expression from (3.19) and rescalling the
deformation parameters and d--tensors we obtain the conditions
(3.9).

\item  We prove the invariant conditions for $na_{(0)}$--maps satisfying
conditions
$$
\epsilon \neq 0\quad \mbox{and}\quad \epsilon -F_{<\beta
>}^{<\alpha
>}F_{<\alpha >}^{<\beta >}\neq 0
$$
Let define the auxiliary d--connection
$$
{\tilde \gamma }_{\cdot <\beta ><\tau >}^{<\alpha >}={\underline{\gamma }}%
_{\cdot <\beta ><\tau >}^{<\alpha >}-{\psi }_{(<\beta >}{\delta
}_{<\tau
>)}^{<\alpha >}={{\gamma }^{<\alpha >}}_{<\beta ><\tau >}+{\sigma }_{(<\beta
>}F_{<\tau >\}}^{<\alpha >}\eqno(3.20)
$$
and write
$$
{\tilde D}_{<\gamma >}={}^{({\gamma })}D_{<\gamma >}F_{<\beta >}^{<\alpha >}+%
{\tilde \sigma }_{<\gamma >}F_{<\beta >}^{<\alpha >}-{\epsilon }{\sigma }%
_{<\beta >}{\delta }_{<\gamma >}^{<\alpha >},
$$
where ${\tilde \sigma }_{<\beta >}={\sigma }_{<\alpha >}F_{<\beta
>}^{<\alpha >},$ or, as a consequence from the last equality,
$$
{\sigma }_{(<\alpha >}F_{<\beta >\}}^{<\tau >}={\epsilon
}F_{<\lambda
>}^{<\tau >}({}^{({\gamma })}D_{(<\alpha >}F_{<\beta >\}}^{<\alpha >}-{%
\tilde D}_{(<\alpha >}F_{<\beta >\}}^{<\lambda >})+{\tilde \sigma }%
_{(<\alpha >}{\delta }_{<\beta >\}}^{<\tau >}.
$$
Introducing auxiliary connections
$$
{\star {\gamma }}_{\cdot <\beta ><\lambda >}^{<\alpha >}={\gamma
}_{\cdot
<\beta ><\lambda >}^{<\alpha >}+{\epsilon }F_{<\tau >}^{<\alpha >}{}^{({%
\gamma })}D_{(<\beta >}F_{<\lambda >\}}^{<\tau >}
$$
and
$$
{\check \gamma }_{\cdot <\beta ><\lambda >}^{<\alpha >}={\tilde \gamma }%
_{\cdot <\beta ><\lambda >}^{<\alpha >}+{\epsilon }F_{<\tau >}^{<\alpha >}{%
\tilde D}_{(<\beta >}F_{<\lambda >\}}^{<\tau >}
$$
we can express deformation (3.20) in a form characteristic for
a--maps:
$$
{\hat \gamma }_{\cdot <\beta ><\gamma >}^{<\alpha >}={\star {\gamma }}%
_{\cdot <\beta ><\gamma >}^{<\alpha >}+{\tilde \sigma }_{(<\beta >}{\delta }%
_{<\lambda >\}}^{<\alpha >}.\eqno(3.21)
$$
Now it's obvious that $na_{(2)}$--invariant conditions (3.21) are
equivalent with a--invariant conditions (3.12) written for
d--connection (3.21).

\item  Finally, we prove the last statement, for $na_{(3)}$--maps, of the
theorem 3.3. Let
$$
q_{<\alpha >}{\varphi }^{<\alpha >}=e=\pm 1,\eqno(3.22)
$$
where ${\varphi }^{<\alpha >}$ is contained in
$$
{\underline{\gamma }}_{\cdot <\beta ><\gamma >}^{<\alpha >}={{\gamma }%
^{<\alpha >}}_{<\beta ><\gamma >}+{\psi }_{(<\beta >}{\delta
}_{<\gamma
>\}}^{<\alpha >}+{\sigma }_{<\beta ><\gamma >}{\varphi }^{<\alpha >}.%
\eqno(3.23)
$$
Acting with operator $^{({\gamma })}{\underline{D}}_{<\beta >}$
on (3.23) we write
$$
{}^{({\gamma })}{\underline{D}}_{<\beta >}q_{<\alpha >}={}^{({\gamma }%
)}D_{<\beta >}q_{<\alpha >}-{\psi }_{(<\alpha >}q_{<\beta >\}}-e{\sigma }%
_{<\alpha ><\beta >}.\eqno(3.24)
$$
Contracting (3.24) with ${\varphi }^{<\alpha >}$ we can express
$$
e{\varphi }^{<\alpha >}{\sigma }_{<\alpha ><\beta >}=$$
$${\varphi }^{<\alpha
>}({}^{({\gamma })}D_{<\beta >}q_{<\alpha >}-{}^{({\gamma })}{\underline{D}}%
_{<\beta >}q_{<\alpha >})-{\varphi }_{<\alpha >}q^{<\alpha >}q_{<\beta >}-e{%
\psi }_{<\beta >}.$$ Introducing the last formula in (3.23),
contracted on indices $<\alpha >$ and $<\gamma >,$ we obtain
$$
n_E{\psi }_{<\beta >}={\underline{\gamma }}_{\cdot <\alpha ><\beta
>}^{<\alpha >}-{{\gamma }^{<\alpha >}}_{<\alpha ><\beta >}+e{\psi }_{<\alpha
>}{\varphi }^{<\alpha >}q_{<\beta >}+$$
$$e{\varphi }^{<\alpha >}{\varphi }%
^{<\beta >}({}^{({\gamma })}{\underline{D}}_{<\beta >}-{}^{({\gamma }%
)}D_{<\beta >}).\eqno(3.25)$$
From these relations and (3.22), we have%
$$
n_E{\psi }_{<\alpha >}{\varphi }^{<\alpha >}={\varphi }^{<\alpha >}({%
\underline{\gamma }}_{\cdot <\alpha ><\beta >}^{<\alpha >}-{{\gamma }%
^{<\alpha >}}_{<\alpha ><\beta >})+
$$
$$
e{\varphi }^{<\alpha >}{\varphi }^{<\beta >}({}^{({\gamma })}{\underline{D}}%
_{<\beta >}q_{<\alpha >}-{}^{({\gamma })}D_{<\beta >}q_{<\alpha
>}).
$$

Using the equalities and identities (3.24) and (3.25) we can
express deformations (3.23) as the first $na_{(3)}$--invariant
conditions from (3.13).

To prove the second class of $na_{(3)}$--invariant conditions we
introduce two additional ds--tensors:
$$
{{\rho }^{<\alpha >}}_{<\beta ><\gamma ><\delta >}=r_{<\beta
>\cdot <\gamma
><\delta >}^{\cdot <\alpha >}+$$ $${\frac 12}({\psi }_{(<\beta >}{\delta }%
_{<\varphi >\}}^{<\alpha >}+{\sigma }_{<\beta ><\varphi
>}{\varphi }^{<\tau
>}){w^{<\varphi >}}_{<\gamma ><\delta >}{\quad }
$$
and
$$
{\underline{\rho }}_{\cdot <\beta ><\gamma ><\delta >}^{<\alpha >}={%
\underline{r}}_{<\beta ><\gamma ><\delta >}^{\cdot <\alpha >}-$$ $${\frac 12}({%
\psi }_{(<\beta >}{\delta }_{<{\varphi >}\}}^{<\alpha >}-{\sigma
}_{<\beta
><\varphi >}{\varphi }^{<\tau >}){w^{<\varphi >}}_{<\gamma ><\delta >}.%
\eqno(3.26)
$$
Considering deformation (3.23) and (3.26) we write relation
$$
{\tilde \sigma }_{\cdot <\beta ><\gamma ><\delta >}^{<\alpha >}={\underline{%
\rho }}_{\cdot <\beta ><\gamma ><\delta >}^{<\alpha >}-{\rho
}_{\cdot <\beta
><\gamma ><\delta >}^{<\alpha >}=\eqno(3.27)
$$
$$
{\psi }_{<\beta >[<\delta >}{\delta }_{<\gamma >\}}^{<\alpha >}-{\psi }_{[<{%
\gamma ><}{\delta >}\}}{\delta }_{<\beta >}^{<\alpha >}-{\sigma
}_{<\beta
><\gamma ><\delta >}{\varphi }^{<\alpha >},
$$
where
$$
{\psi }_{<\alpha ><\beta >}={}^{({\gamma })}D_{<\beta >}{\psi }_{<\alpha >}+{%
\psi }_{<\alpha >}{\psi }_{<\beta >}-({\nu }+{\varphi }^{<\tau >}{\psi }%
_{<\tau >}){\sigma }_{<\alpha ><\beta >},
$$
and
$$
{\sigma }_{<\alpha ><\beta ><\gamma >}={}^{({\gamma
})}D_{[<\gamma >}{\sigma
}_{<{\beta >}\}<{\alpha >}}+$$ $${\mu }_{[<\gamma >}{\sigma }_{<{\beta >}\}<{%
\alpha >}}-{\sigma }_{<{\alpha >}[<{\gamma >}}{\sigma }_{<{\beta >}\}<{\tau >%
}}{\varphi }^{<\tau >}.
$$
Multiplying (3.27) on $q_{<\alpha >}$ we can write (taking into
account relations (3.22)) the relation
$$
e{\sigma }_{<\alpha ><\beta ><\gamma >}=-q_{<\tau >}{\tilde
\sigma }_{\cdot <\alpha ><\beta ><\delta >}^{<\tau >}+{\psi
}_{<\alpha >[<\beta >}q_{<\gamma
>\}}-{\psi }_{[<\beta ><\gamma >\}}q_{<\alpha >}.\eqno(3.28)
$$
The next step is to express ${\psi }_{<\alpha ><\beta >}$ in
terms of ds--objects on ${\cal E}^{<z>}.$ To do this we contract
indices $<\alpha >$ and $<\beta >$ in (3.27) and obtain
$$
n_E{\psi }_{[<\alpha ><\beta >\}}=-{\sigma }_{\cdot <\tau
><\alpha ><\beta
>}^{<\tau >}+$$ $$eq_{<\tau >}{\varphi }^{<\lambda >}{\sigma }_{\cdot <\lambda
><\alpha ><\beta >}^{<\tau >}-e{\tilde \psi }_{[<\alpha >}{\tilde \psi }%
_{<\beta >\}}.
$$
Then contracting indices $<\alpha >$ and $<\delta >$ in (3.27)
and using
(3.28) we write%
$$
(n_E-2){\psi }_{<\alpha ><\beta >}={\tilde \sigma }_{\cdot
<\alpha ><\beta
><\tau >}^{<\tau >}-
$$
$$
eq_{<\tau >}{\varphi }^{<\lambda >}{\tilde \sigma }_{\cdot
<\alpha ><\beta
><\lambda >}^{<\tau >}+{\psi }_{[<\beta ><\alpha >\}}+e({\tilde \psi }%
_{<\beta >}q_{<\alpha >}-{\hat \psi }_{(<\alpha >}q_{<\beta
>\}},\eqno(3.29)
$$
where ${\hat \psi }_{<\alpha >}={\varphi }^{<\tau >}{\psi
}_{<\alpha ><\tau
>}.$ If the both parts of (3.29) are contracted with ${\varphi }^{<\alpha
>}, $ one follows that
$$
(n_E-2){\tilde \psi }_{<\alpha >}={\varphi }^{<\tau >}{\sigma
}_{\cdot <\tau
><\alpha ><\lambda >}^{<\lambda >}-$$ $$eq_{<\tau >}{\varphi }^{<\lambda >}{%
\varphi }^{<\delta >}{\sigma }_{<\lambda ><\alpha ><\delta
>}^{<\tau
>}-eq_{<\alpha >},
$$
and, in consequence of ${\sigma }_{<\beta >(<\gamma ><\delta
>\}}^{<\alpha
>}=0,$ we have
$$
(n_E-1){\varphi }={\varphi }^{<\beta >}{\varphi }^{<\gamma >}{\sigma }%
_{\cdot <\beta ><\gamma ><\alpha >}^{<\alpha >}.
$$
By using the last expressions we can write
$$
(n_E-2){\underline{\psi }}_{<\alpha >}={\varphi }^{<\tau
>}{\sigma }_{\cdot
<\tau ><\alpha ><\lambda >}^{<\lambda >}-eq_{<\tau >}{\varphi }^{<\lambda >}{%
\varphi }^{<\delta >}{\sigma }_{\cdot <\lambda ><\alpha ><\delta
>}^{<\tau
>}-\eqno(3.30)
$$
$$
e{(n}_E{-1)}^{-1}q_{<\alpha >}{\varphi }^{<\tau >}{\varphi }^{<\lambda >}{%
\sigma }_{\cdot <\tau ><\lambda ><\delta >}^{<\delta >}.
$$
Contracting (3.29) with ${\varphi }^{<\beta >}$ we have
$$
(n_E){\hat \psi }_{<\alpha >}={\varphi }^{<\tau >}{\sigma
}_{\cdot <\alpha
><\tau ><\lambda >}^{<\lambda >}+{\tilde \psi }_{<\alpha >}
$$
and taking into consideration (3.30) we can express ${\hat \psi
}_{<\alpha
>} $ through\\ ${\sigma }_{\cdot <\beta ><\gamma ><\delta >}^{<\alpha >}.$

As a consequence of (3.28)--(3.30) we obtain this formulas for d--tensor ${%
\psi }_{<\alpha ><\beta >}:$
$$
(n_E-2){\psi }_{<\alpha ><\beta >}={\sigma }_{\cdot <\alpha
><\beta ><\tau
>}^{<\tau >}-eq_{<\tau >}{\varphi }^{<\lambda >}{\sigma }_{\cdot <\alpha
><\beta ><\lambda >}^{<\tau >}+
$$
$$
\frac 1{n_E}\{-{\sigma }_{\cdot <\tau ><\beta ><\alpha >}^{<\tau
>}+eq_{<\tau >}{\varphi }^{<\lambda >}{\sigma }_{\cdot <\lambda ><\beta
><\alpha >}^{<\tau >}-$$ $$q_{<\beta >}(e{\varphi }^{<\tau >}{\sigma }_{\cdot
<\alpha ><\tau ><\lambda >}^{<\lambda >}-
$$
$$
q_{<\tau >}{\varphi }^{<\lambda >}{\varphi }^{<\delta >}{\sigma
}_{\cdot <\alpha ><\lambda ><\delta >}^{<\tau >})+eq_{<\alpha
>}[{\varphi }^{<\lambda
>}{\sigma }_{\cdot <\tau ><\beta ><\lambda >}^{<\tau >}-$$
$$eq_{<\tau >}{\varphi
}^{<\lambda >}{\varphi }^{<\delta >}{\sigma }_{\cdot <\lambda
><\beta
><\delta >}^{<\tau >}-
\frac e{n_E-1}q_{<\beta >}({\varphi }^{<\tau >}{\varphi }^{<\lambda >}{%
\sigma }_{\cdot <\tau ><\gamma ><\delta >}^{<\delta >}-$$
$$eq_{<\tau >}{\varphi }%
^{<\lambda >}{\varphi }^{<\delta >}{\varphi }^{<\varepsilon >}{\sigma }%
_{\cdot <\lambda ><\delta ><\varepsilon >}^{<\tau >})]\}.
$$

Finally, putting the last formula and (3.28) into (3.27) and
after a rearrangement of terms we obtain the second group of
$na_{(3)}$--invariant conditions (3.13). If necessary we can
rewrite these conditions in terms of geometrical objects on
${\cal E}^{<z>}$ and $\underline{{\cal E}}^{<z>}.$ To do this we
mast introduce splittings (3.26) into (3.13). \qquad $\Box $
\end{enumerate}

For the particular case of $na_{(3)}$--maps when
$$
{\psi }_{<\alpha >}=0,{\varphi }_{<\alpha >}=g_{<\alpha ><\beta >}{\varphi }%
^{<\beta >}={\frac \delta {\delta u^{<\alpha >}}}(\ln {\Omega }),{\Omega }%
(u)>0
$$
and
$$
{\sigma }_{<\alpha ><\beta >}=g_{<\alpha ><\beta >}
$$
we define a subclass of conformal transforms
${\underline{g}}_{<\alpha
><\beta >}(u)={\Omega }^2(u)g_{<\alpha ><\beta >}$ which, in consequence of
the fact that d--vector ${\varphi }_{<\alpha >}$ must satisfy
equations
(3.11), generalizes the class of concircular transforms (see 
 [230] for
references and details on concircular mappings of Riemannaian
spaces).

The basic na--equations (3.9)--(3.11) are systems of first order
partial
differential equations. The study 
  [278] of their geometrical properties and
definition of integral varieties, general and particular
solutions are possible by using the formalism of Pffaf systems
 [57]. We point out
that by using algebraic methods we can always verify if systems of
na--equations of type (3.9)--(3.11) are, or not, involute, even
to find their explicit solutions it is a difficult task (see more
detailed
considerations for isotropic ng--maps in 
 [230] and, on language of
Pffaf systems for na--maps, in 
 [247]). We can also formulate the
Cauchy problem for na--equations on ${\cal E}^{<z>}$~ and choose
deformation parameters (3.7) as to make involute mentioned
equations for the case of maps to a given background space
${\underline{{\cal E}}}^{<z>}$. If a solution, for example, of
$na_{(1)}$--map equations exists, we say that
space ${\cal E}^{<z>}$ is $na_{(1)}$--projective to space ${\underline{{\cal %
E}}}^{<z>}.$ In general, we have to introduce chains of na--maps
in order to obtain involute systems of equations for maps
(superpositions of na-maps) from ${\cal E}^{<z>}$ to
${\underline{{\cal E}}}^{<z>}:$
$$U \buildrel {ng<i_{1}>} \over \longrightarrow  {U_{\underline 1}}
\buildrel ng<i_2> \over \longrightarrow \cdots \buildrel
ng<i_{k-1}> \over \longrightarrow U_{\underline {k-1}} \buildrel
ng<i_k> \over \longrightarrow {\underline U} $$
where $U\subset {\cal E}^{<z>},U_{\underline{1}}\subset {\cal E}_{\underline{%
1}}^{<z>},\ldots ,U_{k-1}\subset {\cal
E}_{k-1}^{<z>},{\underline{U}}\subset {\cal E}_k^{<z>}$ with
corresponding splittings of auxiliary symmetric connections
$$
{\underline{\gamma }}_{.<{\beta ><}{\gamma >}}^{<\alpha >}=_{<i_1>}P_{.<{%
\beta ><}{\gamma >}}^{<\alpha >}+_{<i_2>}P_{.<{\beta ><}{\gamma
>}}^{<\alpha
>}+\cdots +_{<i_k>}P_{.<{\beta ><}{\gamma >}}^{<\alpha >}
$$
and torsion
$$
{\underline{T}}_{.<{\beta ><}{\gamma >}}^{<\alpha >}=T_{.<{\beta ><}{\gamma >%
}}^{<\alpha >}+_{<i_1>}Q_{.<{\beta ><}{\gamma >}}^{<\alpha >}+_{<i_2>}Q_{.<{%
\beta ><}{\gamma >}}^{<\alpha >}+\cdots +_{<i_k>}Q_{.<{\beta ><}{\gamma >}%
}^{<\alpha >}
$$
where cumulative indices $<i_1>=0,1,2,3,$ denote possible types
of na--maps.

\begin{definition}
\label{3.4d} A dvs--bundle ${\cal E}^{<z>}$~ is nearly
conformally projective to
 dvs--bundle ${\underline{{\cal E}}}^{<z>},$ $$ nc:{\cal E}^{<z>}{\to }{{%
\underline{{\cal E}}}^{<z>}},$$ if there is a finite chain of na--maps from $%
{\cal E}^{<z>}$ to ${\underline{{\cal E}}}^{<z>}.$
\end{definition}

For nearly conformal maps we formulate :

\begin{theorem}
\label{3.4t} For every fixed triples $(N_{A_p}^{A_f},{\Gamma }_{{.<}{\beta ><%
}{\gamma >}}^{<\alpha >},U\subset {\cal E}^{<z>})$ and\\ $(N_{A_p}^{A_f},{%
\underline{\Gamma }}_{{.<}{\beta ><}{\gamma >}}^{<\alpha >}$, ${\underline{U}%
}\subset {\underline{{\cal E}}}^{<z>})$, of components of
nonlinear
connection, d--connec\-ti\-on and d--metric being of class $C^r(U),C^r({%
\underline{U}})$, $r>3,$ there is a finite chain of na--maps $nc:U\to {%
\underline{U}}.$
\end{theorem}

Proof is similar to that for isotropic maps 
 [249,273,252] (we have to
introduce a finite number of na-maps with corresponding
components of deformation parameters and deformation tensors in
order to transform step by step coefficients of d-connection
${\Gamma }_{<\gamma ><\delta >}^{<\alpha
>} $ into ${\underline{\Gamma }}_{<\beta ><\gamma >}^{<\alpha >}).$

We introduce the concept of the Category of la--spaces, ${\cal C}({\cal E}%
^{<z>}).$ The elements of ${\cal C}({\cal E}^{<z>})$ consist from $Ob{\cal C}%
({\cal E}^{<z>})=\{{\cal E}^{<z>},{\cal E}_{<i_1>}^{<z>},{\cal E}%
_{<i_2>}^{<z>},{\ldots },\}$ being dvs--bundles, for simplicity
in this
work, having common N--connection structures, and $Mor{\cal C}({\cal E}%
^{<z>})=\{nc({\cal E}_{<i_1>}^{<z>},{\cal E}_{<i_2>}^{<z>})\}$
being chains of na--maps interrelating higher order anisotropic
s--spaces. We point out that we can consider equivalent models of
physical theories on every object of ${\cal C}({\cal E}^{<z>})$
(see details for isotropic gravitational
models in 
 [249,252,278,273,274] and anisotropic gravity in
 [263,276,279]). One of the main purposes of this chapter is to
develop a ds---tensor and variational formalism on ${\cal C}({\cal E}%
^{<z>}), $ i.e. on higher order anisotropic multispaces,
interrelated with nc--maps. Taking into account the distinguished
character of geometrical objects on dvs---spaces we call tensors
on ${\cal C}({\cal E}^{<z>})$ as distinguished tensors on
dvs--bundle Category, or dc--tensors.

Finally, we emphasize that presented in that section definitions
and theorems can be generalized for dvs---bundles with arbitrary
given structures of nonlinear connection, linear d--connection
and metric
structures. Proofs are similar to those from 
 [251,230].

\section{Na--Maps and Tensor--Integral}

Our aim in this section is to define ds--tensor integration not
only for ds--bitensors, objects defined on the same dvs--bundle,
but for dc--tensors,
defined on two dvs--bundles, ${\cal E}^{<z>}$ and \underline{${\cal E}$}$%
^{<z>}$, even it is necessary on higher order anisotropic
multispaces. A. Mo\'or tensor--integral formalism having a lot of
applications in classical and quantum gravity
 [234,289,100] was extended for locally isotropic
multispaces in 
 [278,273]. The unispacial locally anisotropic version
is given in
 [258,102].

Let $T_u{\cal E}^{<z>}$~ and $T_{\underline{u}}{\underline{{\cal
E}}}^{<z>}$ be tangent spaces in corresponding points\\ $u{\in
}U{\subset {\cal E}}^{<z>}$
and ${\underline{u}}{\in }{\underline{U}}{\subset \underline{{\cal E}}}%
^{<z>} $ and, respectively, $T_u^{*}{\cal E}^{<z>}$ and $T_{\underline{u}%
}^{*}{\underline{{\cal E}}}^{<z>}$ be their duals (in general, in
this section we shall not consider that a common coordinatization
is introduced for open regions $U$ and ${\underline{U}}$ ). We
call as the dc--tensors on the pair of dvs--bundles (${\cal
E}^{<z>},{\underline{{\cal E}}}^{<z>}$ ) the elements of
distinguished tensor algebra
$$
({\otimes }_\alpha T_u{\cal E}^{<z>}){\otimes }({\otimes }_\beta T_u^{*}%
{\cal E}^{<z>}){\otimes }({\otimes }_\gamma T_{\underline{u}}{\underline{%
{\cal E}}}^{<z>}){\otimes }({\otimes }_\delta T_{\underline{u}}^{*}{%
\underline{{\cal E}}}^{<z>})
$$
defined over the dvs--bundle ${\cal E}^{<z>}{\otimes \underline{{\cal E}}}%
^{<z>},$ for a given $nc:{\cal E}^{<z>}{\to \underline{{\cal
E}}}^{<z>}$.

We admit the convention that underlined and non--underlined
indices refer,
respectively, to the points ${\underline{u}}$ and $u$. Thus $Q_{{.<}{%
\underline{\alpha }>}}^{<\beta >},$ for instance, are the
components of
dc--tensor $Q{\in }T_u{\cal E}^{<z>}{\otimes }T_{\underline{u}}{\underline{%
{\cal E}}}^{<z>}.$

Let open regions $U$ and ${\underline{U}}$ be homeomorphic to a
superspace sphere ${\cal R}^{2n_E}$ and introduce an isomorphism
${\mu }_{{u},{\underline{u}}}$ between $%
T_u{\cal E}^{<z>}$ and $T_{\underline{u}}{\underline{{\cal
E}}}^{<z>}$ (given by a map $nc:U{\to }{\underline{U}}).$ We
consider that for every
ds--vector $v^\alpha {\in }T_u{\cal E}^{<z>}$ corresponds the vector ${\mu }%
_{{u},{\underline{u}}}(v^{<\alpha >})=v^{<\underline{\alpha }>}{\in }T_{%
\underline{u}}{\underline{{\cal E}}}^{<z>},$ with components $v^{<\underline{%
\alpha }>}$ being linear functions of $v^{<\alpha >}$:
$$
v^{<\underline{\alpha }>}=h_{<\alpha >}^{<\underline{\alpha }>}(u,{%
\underline{u}})v^{<\alpha >},\quad v_{<\underline{\alpha }>}=h_{<\underline{%
\alpha }>}^{<\alpha >}({\underline{u}},u)v_{<\alpha >},
$$
where $h_{<\underline{\alpha }>}^{<\alpha >}({\underline{u}},u)$
are the components of dc--tensor associated with ${\mu
}_{u,{\underline{u}}}^{-1}$. In a similar manner we have
$$
v^{<\alpha >}=h_{<\underline{\alpha }>}^{<\alpha >}({\underline{u}},u)v^{<%
\underline{\alpha }>},\quad v_{<\alpha >}=h_{<\alpha >}^{<\underline{\alpha }%
>}(u,{\underline{u}})v_{<\underline{\alpha }>}.
$$

In order to reconcile just presented definitions and to assure
the identity for trivial maps ${\cal E}^{<z>}{\to }{\cal
E}^{<z>},u={\underline{u}},$ the transport dc--tensors must
satisfy conditions :
$$
h_{<\alpha >}^{<\underline{\alpha }>}(u,{\underline{u}})h_{<\underline{%
\alpha >}}^{<\beta >}({\underline{u}},u)={\delta }_{<\alpha
>}^{<\beta
>},h_{<\alpha >}^{<\underline{\alpha }>}(u,{\underline{u}})h_{<\underline{%
\beta }>}^{<\alpha >}({\underline{u}},u)={\delta }_{<\underline{\beta }>}^{<%
\underline{\alpha }>}
$$
and
$$
{\lim }_{{({\underline{u}}{\to }u})}h_{<\alpha >}^{<\underline{\alpha }>}(u,{%
\underline{u}})={\delta }_{<\alpha >}^{<\underline{\alpha }>},\quad {\lim }_{%
{({\underline{u}}{\to }u})}h_{<\underline{\alpha }>}^{<\alpha >}({\underline{%
u}},u)={\delta }_{<\underline{\alpha }>}^{<\alpha >}.
$$

Let ${\overline{S}}_p{\subset }{\overline{U}}{\subset }{\overline{{\cal E}}}%
^{<z>}$ is a homeomorphic to $p$-dimensional sphere and suggest
that chains of na--maps are used to connect regions:
$$ U \buildrel nc_{(1)} \over \longrightarrow {\overline S}_p
     \buildrel nc_{(2)} \over \longrightarrow {\underline U}.$$

\begin{definition}
\label{3.5d} The tensor integral in ${\overline{u}}{\in
}{\overline{S}}_p$
of a dc--tensor\\ $N_{<{\varphi >}{.<}{\underline{\tau }>}{.<}{\overline{%
\alpha }}_1>{\cdots <}{\overline{\alpha }}_p>}^{{.<}{\gamma >}{.<}{%
\underline{\kappa }>}}$ $({\overline{u}},u),$ completely
antisymmetric on the indices\\ $<{{\overline{\alpha
}}_1>},{\ldots },<{\overline{\alpha }}_p>,$ over domain
${\overline{S}}_p,$ is defined as
$$
N_{<{\varphi >}{.<}{\underline{\tau }>}}^{{.<}{\gamma >}{.<}{\underline{%
\kappa }>}}({\underline{u}},u)=I_{({\overline{S}}_p)}^{\underline{U}}N_{<{%
\varphi >}{.<}{\overline{\tau }>}{.<}{\overline{\alpha }}_1>{\ldots <}{%
\overline{\alpha }}_p>}^{{.<}{\gamma >}{.<}{\overline{\kappa }>}}({\overline{%
u}},{\underline{u}})dS^{<{\overline{\alpha }}_1>{\ldots <}{\overline{\alpha }%
}_p>}=
$$
$$
{\int }_{({\overline{S}}_p)}h_{<\underline{\tau }>}^{<\overline{\tau }>}({%
\underline{u}},{\overline{u}})h_{<\overline{\kappa }>}^{<\underline{\kappa }%
>}({\overline{u}},{\underline{u}})N_{<{\varphi >}{.<}{\overline{\tau }>}{.<}{%
\overline{\alpha }}_1>{\cdots <}{\overline{\alpha }}_p>}^{{.<}{\gamma >}{.<}{%
\overline{\kappa }>}}({\overline{u}},u)d{\overline{S}}^{<{\overline{\alpha }}%
_1>{\cdots <}{\overline{\alpha }}_p>},\eqno(3.31)
$$
where $dS^{<{\overline{\alpha }}_1>{\cdots <}{\overline{\alpha }}_p>}={%
\delta }u^{<{\overline{\alpha }}_1>}{\land }{\cdots }{\land }{\delta }u_p^{<%
\overline{\alpha }>}$.
\end{definition}

We note that in this work we a dealing only with locally trivial
geometric constructions so ambiguities connected with global
definition of integration
and Stoke formulas on different types of s--superspaces 
 [203,204]
 are avoided. Let
suppose that transport dc--tensors $h_{<\alpha
>}^{<\underline{\alpha }>}$~ and $h_{<\underline{\alpha
}>}^{<\alpha >}$~ admit covariant derivations of or\-der two and
pos\-tu\-la\-te ex\-is\-ten\-ce of de\-for\-ma\-ti\-on
dc--ten\-sor\\ $B_{<{\alpha ><}{\beta >}}^{{..<}{\gamma >}}(u,{\underline{u}}%
)$~ satisfying relations
$$
D_{<\alpha >}h_{<\beta >}^{<\underline{\beta }>}(u,{\underline{u}})=B_{<{%
\alpha ><}{\beta >}}^{{..<}{\gamma >}}(u,{\underline{u}})h_{<\gamma >}^{<%
\underline{\beta }>}(u,{\underline{u}})\eqno(3.32)
$$
and, taking into account that $D_{<\alpha >}{\delta }_{<\gamma
>}^{<\beta
 >}=0,$
 $$
D_{<\alpha >}h_{<\underline{\beta >}}^{<\beta >}({\underline{u}},u)=-B_{<{%
\alpha ><}{\gamma >}}^{{..<}{\beta >}}(u,{\underline{u}})h_{<\underline{%
\beta }>}^{<\gamma >}({\underline{u}},u).
$$
By using formulas (1.25) and (1.29) for torsion and curvature of
d--con\-nec\-ti\-on ${\Gamma }_{<{\beta ><}{\gamma >}}^{<\alpha
>}$~ we calculate s--commutators:
$$
D_{[<{\alpha >}}D_{<{\beta >}\}}h_{<\gamma >}^{<\underline{\gamma >}}=-(R_{<{%
\gamma ><\alpha ><}{\beta >}}^{{<\lambda >}}+T_{{.<}{\alpha ><}{\beta >}%
}^{<\tau >}B_{<{\tau ><}{\gamma >}}^{{..<}{\lambda >}})h_{<\lambda >}^{<%
\underline{\gamma >}}.\eqno(3.33)
$$
On the other hand from (3.32) one follows that
$$
D_{[<{\alpha >}}D_{<{\beta >}\}}h_{<\gamma >}^{<\underline{\gamma >}}=(D_{[<{%
\alpha >}}B_{<{\beta >}\}<{\gamma >}}^{{..<}{\lambda >}}+B_{[<{\alpha >}{|<}{%
\tau >}{|}{.}}^{{..<}{\lambda >}}B_{<{\beta >}\}<{\gamma >}{.}}^{{..<}{\tau >%
}})h_{<\lambda >}^{<\underline{\gamma >}},\eqno(3.34)
$$
where ${|<}{\tau >}{|}$~ denotes that index $<{\tau >}$~ is
excluded from the action of s--antisymmetrization $[{\quad }\}$.
From (3.33) and (3.34) we obtain
$$
D_{[<{\alpha >}}B_{<{\beta >}\}<{\gamma >}}^{{..<}{\lambda >}}+B_{[<{\beta >}%
{|<}{\gamma >}{|}}B_{<{\alpha >}\}<{\tau >}}^{{..<}{\lambda >}}=$$
$$(R_{<{\gamma
>}{.<}{\alpha ><}{\beta >}}^{{.<}{\lambda >}}+T_{{.<}{\alpha ><}{\beta >}%
}^{<\tau >}B_{<{\tau ><}{\gamma >}}^{{..<}{\lambda >}}).
$$

Let ${\overline{S}}_p$~ be the boundary of
${\overline{S}}_{p-1}$. The Stoke's formula for tensor--integral
(3.31) is defined as
$$
I_{{\overline{S}}_p}N_{<{\varphi >}{.<}{\overline{\tau }>}{.<}{\overline{%
\alpha }}_1>{\ldots <}{\overline{\alpha }}_p>}^{{.<}{\gamma >}{.<}{\overline{%
\kappa }>}}dS^{<{\overline{\alpha }}_1>{\ldots
<}{\overline{\alpha }}_p>}=
$$
$$
I_{{\overline{S}}_{p+1}}{^{{\star }{(p)}}{\overline{D}}}_{[<{\overline{%
\gamma }>}{|}}N_{<{\varphi >}{.<}{\overline{\tau
}>}{.}{|<}{\overline{\alpha
}}_1>{\ldots <}{{\overline{\alpha }}_p>\}}}^{{.<}{\gamma >}{.<}{\overline{%
\kappa }>}}dS^{<{\overline{\gamma }><}{\overline{\alpha }}_1>{\ldots <}{%
\overline{\alpha }}_p>},
$$
where
$$
{^{{\star }{(p)}}D}_{[<{\overline{\gamma }>}{|}}N_{<{\varphi >}{.<}{%
\overline{\tau }>}{.}{|<}{\overline{\alpha }}_1>{\ldots <}{\overline{\alpha }%
}_p>\}}^{{.<}{\gamma >}{.<}{\overline{\kappa }>}}=D_{[<{\overline{\gamma }>}{%
|}}N_{<{\varphi >}{.<}{\overline{\tau }>}{.}{|<}{\overline{\alpha }}_1>{%
\ldots <}{\overline{\alpha }}_p>\}}^{{.<}{\gamma >}{.<}{\overline{\kappa }>}%
}-$$ $$B_{[<{\overline{\gamma }>}{|<}{\overline{\tau }>}}^{{..<}{\overline{%
\epsilon }>}}N_{<{\varphi >}{.<}{\overline{\epsilon }>}{.}{|<}{\overline{%
\alpha }}_1>{\ldots <}{\overline{\alpha }}_p>\}}^{{.<}{\gamma >}{.<}{%
\overline{\kappa }>}}+
pT_{{.}[<{\overline{\gamma }><}{\overline{\alpha }}_1>{|}}^{<\underline{%
\epsilon }>}N_{<{\varphi >}{.<}{\overline{\tau }>}{.<}{\overline{\epsilon }>}%
{|<}{\overline{\alpha }}_2>{\ldots <}{\overline{\alpha
}}_p>\}}^{{.<}{\gamma
>}{.<}{\overline{\kappa }>}}+$$
$$B_{[<{\overline{\gamma }>}{|<}{\overline{%
\epsilon }>}}^{..<{\overline{\kappa }>}}N_{<{\varphi >}{.<}{\overline{\tau }>%
}{.}{|<}{\overline{\alpha }}_1>{\ldots <}{\overline{\alpha }}_p>\}}^{{.<}{%
\gamma >}{.<}{\overline{\epsilon }>}}.
$$
We define the dual element of the hypersurfaces element $dS^{<{\alpha }_1>{%
\ldots <}{\alpha }_p>}$ as
$$
d{\cal S}_{<{\beta }_1>{\ldots <}{\beta }_{q-p}>}={\frac 1{{p!}}}{\epsilon }%
_{<{\beta }_1>{\ldots <}{\beta }_{k-p}><{\alpha }_1>{\ldots <}{\alpha }%
_p>}dS^{<{\alpha }_1>{\ldots <}{\alpha }_p>},\eqno(3.35)
$$
where ${\epsilon }_{<{\gamma }_1>{\ldots <}{\gamma }_q>}$ is
completely s--antisymmetric on its indices and
$$
{\epsilon }_{12{\ldots }(n_E)}=\sqrt{{|}G{|}},G=s\det {|}G_{<{\alpha ><}{%
\beta >}}|,
$$
$G_{<{\alpha ><}{\beta >}}$ is taken from (1.35). The dual of dc--tensor $N_{<{%
\varphi >}{.<}{\overline{\tau }>}{.<}{\overline{\alpha }}_1>{\ldots <}{%
\overline{\alpha }}_p>}^{{.<}{\gamma ><}{\overline{\kappa }>}}$
is defined
as the dc--tensor  ${\cal N}_{<{\varphi >}{.<}{\overline{\tau }>}}^{{.<}{%
\gamma >}{.<}{\overline{\kappa }><}{\overline{\beta }}_1>{\ldots <}{%
\overline{\beta }}_{n_E-p}>}$ satisfying
$$
N_{<{\varphi >}{.<}{\overline{\tau }>}{.<}{\overline{\alpha }}_1>{\ldots <}{%
\overline{\alpha }}_p>}^{{.<}{\gamma >}{.<}{\overline{\kappa }>}}={\frac 1{{%
p!}}}{\cal N}_{<{\varphi >}{.<}{\overline{\tau }>}}^{{.<}{\gamma >}{.<}{%
\overline{\kappa }><}{\overline{\beta }}_1>{\ldots <}{\overline{\beta }}%
_{n_E-p}>}{\epsilon }_{<{\overline{\beta }}_1>{\ldots <}{\overline{\beta }}%
_{n_E-p}><{\overline{\alpha }}_1>{\ldots <}{\overline{\alpha }}_p>}.%
\eqno(3.36)
$$
Using (3.14), (3.35) and (3.36) we can write
$$
I_{{\overline{S}}_p}N_{<{\varphi >}{.<}{\overline{\tau }>}{.<}{\overline{%
\alpha }}_1>{\ldots <}{\overline{\alpha }}_p>}^{{.<}{\gamma >}{.<}{\overline{%
\kappa }>}}dS^{<{\overline{\alpha }}_1>{\ldots <}{\overline{\alpha }}_p>}=%
\eqno(3.37)
$$
$$
{\int }_{{\overline{S}}_{p+1}}{^{\overline{p}}D}_{<\overline{\gamma }>}{\cal %
N}_{<{\varphi >}{.<}{\overline{\tau }>}}^{{.<}{\gamma >}{.<}{\overline{%
\kappa }><}{\overline{\beta }}_1>{\ldots <}{\overline{\beta }}_{n_E-p-1}><{%
\overline{\gamma }>}}d{\cal S}_{<{\overline{\beta }}_1>{\ldots <}{\overline{%
\beta }}_{n_E-p-1}>},
$$
where
$$
{^{\overline{p}}D}_{<\overline{\gamma }>}{\cal N}_{<{\varphi >}{.<}{%
\overline{\tau }>}}^{{.<}{\gamma >}{.<}{\overline{\kappa }><}{\overline{%
\beta }}_1>{\ldots <}{\overline{\beta }}_{n_E-p-1}><{\overline{\gamma }>}}=$$ $${%
\overline{D}}_{<\overline{\gamma }>}{\cal N}_{{\varphi }{.}{\overline{\tau }}%
}^{{.<}{\gamma >}{.<}{\overline{\kappa }><}{\overline{\beta }}_1>{\ldots <}{%
\overline{\beta }}_{n_E-p-1}><{\overline{\gamma }>}}+
$$
$$
(-1)^{(n_E-p)}(n_E-p+1)T_{{.<}{\overline{\gamma }><}{\overline{\epsilon }>}%
}^{[<{\overline{\epsilon }>}}{\cal N}_{<{\varphi >}{.<}{\overline{\tau }>}}^{%
{.}{|<}{\gamma >}{.<}{\overline{\kappa }>}{|<}{\overline{\beta }}_1>{\ldots <%
}{\overline{\beta }}_{n_E-p-1}>\}<{\overline{\gamma }>}}-
$$
$$
B_{<{\overline{\gamma }><}{\overline{\tau }>}}^{{..<}{\overline{\epsilon }>}}%
{\cal N}_{<{\varphi >}{.<}{\overline{\epsilon }>}}^{{.<}{\gamma >}{.<}{%
\overline{\kappa }><}{\overline{\beta }}_1>{\ldots <}{\overline{\beta }}%
_{n_E-p-1}><{\overline{\gamma }>}}+$$ $$B_{<{\overline{\gamma }><}{\overline{%
\epsilon }>}}^{{..<}{\overline{\kappa }>}}{\cal N}_{<{\varphi >}{.<}{%
\overline{\tau }>}}^{{.<}{\gamma >}{.<}{\overline{\epsilon }><}{\overline{%
\beta }}_1>{\ldots <}{\overline{\beta
}}_{n_E-p-1}><{\overline{\gamma }>}}.
$$
The equivalence of (3.36) and (3.37) is a consequence of equalities%
$$
D_{<\gamma >}{\epsilon }_{<{\alpha }_1>{\ldots <}{\alpha }_k>}=0\
$$
and
$$
\ {\epsilon }_{<{\beta }_1>{\ldots <}{\beta }_{n_E-p}><{\alpha }_1>{\ldots <}%
{\alpha }_p>}{\epsilon }^{<{\beta }_1>{\ldots <}{\beta }_{n_E-p}><{\gamma }%
_1>{\ldots <}{\gamma }_p>}=$$ $$p!(n_E-p)!{\delta }_{<{\alpha }_1>}^{[<{\gamma }%
_1>}{\cdots }{\delta }_{<{\alpha }_p>}^{<{\gamma }_p>\}}.
$$
The developed in this section tensor integration formalism will
be used in the next section for definition of a class of
conservation laws on s--spaces with higher order anisotropy.

\section{Ds--Tensor Integral Conservation Laws }

There are not  global and local groups of automorphisms on
generic higher order anisotropic s--spaces and, in consequence,
the definition of conservation laws on such s--spaces is a
challenging task. Our main idea is
to use chains of na--maps from a given, called hereafter as the
fundamental higher order anisotropic s--space, dvs--bundle, to an
auxiliary one with trivial curvatures and torsions admitting a
global group of automorphisms. The aim of this section is to
formulate conservation laws for higher order anisotropic
s--gravitational fields by using dc--objects and tensor--integral
values, na--maps and variational calculus on the Category of
dvs--bundle.

\subsection{Divergence of energy--momentum ds--ten\-sor}

R. Miron and M. Anastasiei 
 [160,161] pointed to this specific form of
conservation laws of matter on la--spaces.  Calculating  the
divergence of
the energy--momentum ds---tensor from equations\ (2.31) on dvs--bundle ${\cal E}%
^{<z>}$ we find%
$$
D_{<\alpha >}{E}_{<\beta >}^{<\alpha >}={\frac 1{\ {\kappa
}_1}}U_{<\beta >}, \eqno(3.38)$$ and concluded that ds--vector
$$
U_{<\alpha >}={\frac 12}(G^{<\beta ><\delta >}{R}_{<\delta ><\phi
><\beta
>}^{<\gamma >}{\ T}_{\cdot <\alpha ><\gamma >}^{<\phi >}-$$
$$G^{<\beta ><\delta
>}{R}_{<\delta ><\phi ><\alpha >}^{<\gamma >}{\ T}_{\cdot <\beta ><\gamma
>}^{<\phi >}+{R_{<\phi >}^{<\beta >}}{\ T}_{\cdot <\beta ><\alpha >}^{<\phi
>})
$$
vanishes if and only if d--connection $D$ is without torsion.

Here we note the multiconnection character of higher order
anisotropic s--spaces. For example, for a ds--metric (1.35) on
${\cal E}^{<z>}$ we can equivalently introduce another (see
(1.39)) metric linear connection $\tilde D.$ The Einstein
equations
$$
{\tilde R}_{<\alpha ><\beta >}-{\frac 12}G_{<\alpha ><\beta >}{\tilde R}={%
\kappa }_1{\tilde E}_{<\alpha ><\beta >}\eqno(3.39)
$$
constructed by using connection (3.20) have vanishing divergences
$$
{\tilde D}^{<\alpha >}({{\tilde R}_{<\alpha ><\beta >}}-{\frac
12}G_{<\alpha
><\beta >}{\tilde R})=0\mbox{ and }{\tilde D}^{<\alpha >}{\tilde E}_{<\alpha
><\beta >}=0,
$$
similarly as those on (pseudo)Riemannian spaces. We conclude that
by using the connection (1.39) we construct a model of higher
order anisotropic
s--gravity which looks like locally isotropic on the total space of ${\cal E}%
^{<z>}.$ More general s--gravitational models with higher order
anisotropy can be obtained by using deformations of connection
${\tilde \Gamma }_{\cdot <\beta ><\gamma >}^{<\alpha >},$
$$
{{\Gamma }^{<\alpha >}}_{<\beta ><\gamma >}={\tilde \Gamma
}_{\cdot <\beta
><\gamma >}^{<\alpha >}+{P^{<\alpha >}}_{<\beta ><\gamma >}+{Q^{<\alpha >}}%
_{<\beta ><\gamma >},
$$
were, for simplicity, ${{\Gamma }^{<\alpha >}}_{<\beta ><\gamma
>}$ is chosen to be also metric and satisfy Einstein equations
(2.31). We can consider deformation d--tensors ${P^{<\alpha
>}}_{<\beta ><\gamma >}$ generated (or not) by deformations of
type (3.9),(3.10) and (3.11) for na--maps. In this case
ds---vector $U_{<\alpha >}$ can be interpreted as a generic
source of local anisotropy on ${\cal E}^{<z>}$ satisfying
generalized conservation laws (3.38).

\subsection{D--conservation laws}

From (3.31) we obtain a tensor integral on ${\cal C}({\cal
E}^{<z>})$ of a d--tensor :
$$
N_{<{\underline{\tau }>}}^{{.<}{\underline{\kappa }>}}(\underline{u})=I_{{%
\overline{S}}_p}N_{<{\overline{\tau }>}{..<}{\overline{\alpha }}_1>{\ldots <}%
{\overline{\alpha }}_p>}^{{..<}{\overline{\kappa }>}}({\overline{u}})h_{<{%
\underline{\tau }>}}^{<{\overline{\tau }>}}({\underline{u}},{\overline{u}}%
)h_{<{\overline{\kappa }>}}^{<{\underline{\kappa }>}}({\overline{u}},{%
\underline{u}})dS^{<{\overline{\alpha }}_1>{\ldots <}{\overline{\alpha }}%
_p>}.
$$

We note that tensor--integral can be defined not only for
dc--tensors but
and for d--tensors on ${\cal E}^{<z>}$. Really, suppressing indices ${%
\varphi }$~ and ${\gamma }$~ in (3.36) and (3.37), considering
instead of a deformation dc--tensor a deformation tensor
$$
B_{<{\alpha ><}{\beta >}}^{{..<}{\gamma >}}(u,{\underline{u}})=B_{<{\alpha ><%
}{\beta >}}^{{..<}{\gamma >}}(u)=P_{{.<}{\alpha ><}{\beta >}}^{<\gamma >}(u)%
\eqno(3.40)
$$
(we consider deformations induced by a nc--transform) and integration\\ $%
I_{S_p}{\ldots }dS^{<{\alpha }_1>{\ldots <}{\alpha }_p>}$ in la--space $%
{\cal E}^{<z>}$ we obtain from (3.31) a tensor--integral on ${\cal C}({\cal E}%
^{<z>})$~ of a ds---tensor:
$$
N_{<{\underline{\tau }>}}^{{.<}{\underline{\kappa }>}}({\underline{u}}%
)=I_{S_p}N_{<{\tau >}{.<}{\alpha }_1>{\ldots <}{\alpha }_p>}^{.<{\kappa >}%
}(u)h_{<{\underline{\tau }>}}^{<\tau >}({\underline{u}},u)h_{<\kappa >}^{<%
\underline{\kappa }>}(u,{\underline{u}})dS^{<{\alpha }_1>{\ldots <}{\alpha }%
_p>}.
$$
Taking into account (3.39) and using formulas (1.25),(1.29) and
(3.3) we can calculate that curvature
$$
{\underline{R}}_{<{\gamma ><\alpha ><}{\beta >}}^{<{\lambda >}}=D_{[<{\beta >%
}}B_{<{\alpha >}\}<{\gamma >}}^{{..<}{\lambda >}}+B_{[<{\alpha
>}{|<}{\gamma
>}{|}}^{{..<}{\tau >}}B_{<{\beta >}\}<{\tau >}}^{{..<}{\lambda >}}+T_{{.<}{%
\alpha ><}{\beta >}}^{<{\tau >}{..}}B_{<{\tau ><}{\gamma >}}^{{..<}{\lambda >%
}}
$$
of connection ${\underline{\Gamma }}_{{.<}{\alpha ><}{\beta
>}}^{<\gamma
>}(u)={\Gamma }_{{.<}{\alpha ><}{\beta >}}^{<\gamma >}(u)+B_{<{\alpha ><}{%
\beta >}{.}}^{{..<}{\gamma >}}(u),$ with $B_{<{\alpha ><}{\beta >}}^{{..<}{%
\gamma >}}(u)$\\ taken from (3.40), vanishes, ${\underline{R}}_{<{\gamma >}{.<%
}{\alpha ><}{\beta >}}^{{.<}{\lambda >}}=0.$ So, we can conclude
that
la--space ${\cal E}^{<z>}$ admits a tensor integral structure on ${\cal {C}}(%
{\cal E}^{<z>})$ for ds---tensors associated to deformation tensor $B_{<{%
\alpha ><}{\beta >}}^{{..<}{\gamma >}}(u)$ if the nc--image
~${\cal E}^{<z>}$ is locally parallelizable. That way we
generalize the one space tensor integral constructions in
 [100,102,258] were the possibility to
introduce tensor integral structure on a curved space was
restricted by the condition that this space is locally
parallelizable. For $q=n_E$ relations
(3.37), written for ds---tensor ${\cal N}_{<\underline{\alpha }>}^{{.<}{%
\underline{\beta }><}{\underline{\gamma }>}}$ (we change indices $<{%
\overline{\alpha }>},<{\overline{\beta }>},{\ldots }$ into $<{\underline{%
\alpha }>},<{\underline{\beta }>},{\ldots })$ extend the Gauss formula on $%
{\cal {C}}({\cal E}^{<z>})$:
$$
I_{S_{q-1}}{\cal N}_{<\underline{\alpha }>}^{{.<}{\underline{\beta }><}{%
\underline{\gamma }>}}d{\cal S}_{<\underline{\gamma }>}=I_{{\underline{S}}_q}%
{^{\underline{q-1}}D}_{<{\underline{\tau }>}}{\cal N}_{<{\underline{\alpha }>%
}}^{{.<}{\underline{\beta }><}{\underline{\tau }>}}d{\underline{V}},%
\eqno(3.41)
$$
where $d{\underline{V}}={\sqrt{{|}{\underline{G}}_{<{\alpha ><}{\beta >}}{|}}%
}d{\underline{u}}^1{\ldots }d{\underline{u}}^q$ and
$$
{^{\underline{q-1}}D}_{<{\underline{\tau }>}}{\cal N}_{<\underline{\alpha }%
>}^{{.<}{\underline{\beta }><}{\underline{\tau }>}}=D_{<{\underline{\tau }>}}%
{\cal N}_{<\underline{\alpha }>}^{{.<}{\underline{\beta
}><}{\underline{\tau
}>}}-T_{{.<}{\underline{\tau }><}{\underline{\epsilon }>}}^{<{\underline{%
\epsilon }>}}{\cal N}_{<{\underline{\alpha }>}}^{<{\underline{\beta }><}{%
\underline{\tau }>}}-$$
$$B_{<{\underline{\tau }><}{\underline{\alpha }>}}^{{..<}{%
\underline{\epsilon }>}}{\cal N}_{<{\underline{\epsilon }>}}^{{.<}{%
\underline{\beta }><}{\underline{\tau }>}}+B_{<{\underline{\tau }><}{%
\underline{\epsilon }>}}^{{..<}{\underline{\beta }>}}{\cal N}_{<{\underline{%
\alpha }>}}^{{.<}{\underline{\epsilon }><}{\underline{\tau
}>}}.\eqno(3.42)
$$

Let consider physical values $N_{<{\underline{\alpha }>}}^{{.<}{\underline{%
\beta }>}}$ on \underline{${\cal E}$}$^{<z>}$~ defined on its density ${\cal %
N}_{{\underline{\alpha }}}^{{.<}{\underline{\beta }><}{\underline{\gamma }>}%
},$ i. e.
$$
N_{<{\underline{\alpha }>}}^{{.<}{\underline{\beta }>}}=I_{{\underline{S}}%
_{q-1}}{\cal N}_{<{\underline{\alpha }>}}^{{.<}{\underline{\beta >}<}{%
\underline{\gamma }>}}d{\cal S}_{<{\underline{\gamma
}>}}\eqno(3.43)
$$
with this conservation law (due to (3.41) and (3.42)):
$$
{^{\underline{q-1}}D}_{<{\underline{\gamma }>}}{\cal
N}_{<{\underline{\alpha }>}}^{{.<}{\underline{\beta
}><}{\underline{\gamma }>}}=0.\eqno(3.44)
$$
We note that these conservation laws differ from covariant
conservation laws for well known physical values such as density
of electric current or of
energy-- momentum tensor. For example, taking density ${E}_{<\beta >}^{{.<}{%
\gamma >}},$ with corresponding to (3.42) and (3.44) conservation
law,
$$
{^{\underline{q-1}}D}_{<{\underline{\gamma }>}}{E}_{<{\underline{\beta }>}%
}^{<{\underline{\gamma }>}}=D_{<{\underline{\gamma }>}}{E}_{<{\underline{%
\beta }>}}^{<{\underline{\gamma }>}}-T_{{.<}{\underline{\epsilon }><}{%
\underline{\tau }>}}^{<{\underline{\tau }>}}{E}_{<{\underline{\beta }>}}^{{.<%
}{\underline{\epsilon }>}}-B_{<{\underline{\tau }><}{\underline{\beta }>}}^{{%
..<}{\underline{\epsilon }>}}{E}_{<\underline{\epsilon }>}^{<{\underline{%
\tau }>}}=0,
$$
we can define values (see (3.42) and (3.43))
$$
{\cal P}_{<\underline{\alpha }>}=I_{{\underline{S}}_{q-1}}{E}_{<{\underline{%
\alpha }>}}^{{.<}{\underline{\gamma }>}}d{\cal
S}_{<{\underline{\gamma }>}}.
$$
Defined conservation laws (3.42) for ${E}_{<{\underline{\beta }>}}^{{.<}{%
\underline{\epsilon }>}}$ have nothing to do with those for
energy--momentum tensor $E_{<\alpha >}^{{.<}{\gamma >}}$ from
Einstein equations for
la--gravity 
 [160,161] or with ${\tilde E}_{<\alpha ><\beta >}$ from
(3.44), (3.42) and (3.43) with vanishing divergence
 $D_{<\gamma >}{\tilde E}_{<\alpha >}^{{.<}{%
\gamma >}}=0.$ So ${\tilde E}_{<\alpha >}^{{.<}{\gamma >}}{\neq }{E}%
_{<\alpha >}^{{.<}{\gamma >}}.$ A similar conclusion was made in
 [100]
for unispacial locally isotropic tensor integral. In the case of
multispatial tensor integration we have another possibility
(firstly pointed
in 
 [278,258] for Einstein-Cartan spaces), namely, to identify ${E}_{<{%
\underline{\beta }>}}^{{.<}{\underline{\gamma }>}}$ from (3.43)
with the
na-image of ${\ {E}}_{<\beta >}^{{.<}{\gamma >}}$ on dvs--bundle ${\cal E}%
^{<z>}$ (we shall consider this construction for a
nonsupersymmetric case in
 Chapter 6, subsection 6.7.6).

\section{Concluding Remarks}

We  defined a new class (a generalization of conformal
transforms) of maps of
 distinguished vector superbundles with deformation of linear d--connection.
 This class consists from nearly autoparallel maps transforming every
 autoparallel on the first s--space into nearly autoparallel on the second
 s--space. There are four types of na--maps characterized by corresponding
 basic equations and invariant  conditions (see Theorems 3.2 and 3.3). We
 proved that by using chains of na--maps ( nearly conformal transforms ) every
 d--connection on a given dvs--bundle can be transformed into another given
 d--connection on a second dvs--bundle.  In consequence one follows that
 we can equivalently modelate physical processes on all types of higher order
 anisotropic s--spaces being connected by chains of nonsingular na--maps.
 This result can be applied for definition of conservation laws on
 s--spaces where such laws and equations of motion takes a more simplified
 form.

 The problem of formulation of conservation laws for supersymmertic
 interactions with higher order anisotropy have been also considered
 by using the formalism of tensor integral. We introduced Stoke's
 and Gauss type formulas on distinguished  vector s--bundles and found that
 tensor integral conservation laws can be formulated for values on auxiliary
 s--spaces connected with the fundamental one with a chain of na--maps.

 Finally, we note that Chapter 8 is devoted to a detailed  investigation of
 na--maps and conservation laws on (non supersymmetric)  higher  order
 anisotropic spaces of energy--momentum type values for gravitational models
 with local anisotropy.


\chapter{HA--Superstrings}

The superstring theory holds the greatest promise as the
unification theory of all fundamental interactions. The
superstring models contains a lot a characteristic features of
Kaluza--Klein approaches, supersymmetry and supergravity, local
field theory and dual models. We note that in the string theories
the nonlocal one dimensional quantum objects (strings) mutually
interacting by linking and separating together are considered as
fundamental values. Perturbations of the quantized string are
identified with quantum particles. Symmetry and conservation laws
in the string and superstring theory can be considered as
sweeping generalizations of gauge principles which consists the
basis of quantum field models. The new physical concepts are
formulated in the framework a ''new'' for physicists mathematical
formalism of the algebraic geometry and topology 
 [106].

The relationship between two dimensional $\sigma $-models and
strings has been considered
 [153,80,53,229,7] in order to discuss the
effective low energy field equations for the massless models of
strings. Nonlinear $\sigma $-models makes up a class of quantum
field systems for which the fields are also treated as
coordinates of some manifolds. Interactions are introduced in a
geometric manner and admit a lot of applications and
generalizations in classical and quantum field and string
theories. The geometric structure of nonlinear sigma models
manifests the existence of topological nontrivial configuration,
admits a geometric interpretation of conterterms and points to a
substantial interrelation between extended supersymmetry and
differential supergeometry. In connection to this a new approach
based on nonlocal, in general, higher order
anisotropic constructions seem to be emerging 
 [254,269,261]. We
consider the reader to be familiar with basic results from
supergeometry (see, for instance,
 [63,147,290,203]), supergravity theories
 [86,215,288,286,287] and superstrings 
 [115,289,139,140].

In this Chapter we shall present an introduction into the theory
of higher order anisotropic superstrings being a natural
generalization to locally anisotropic (la) backgrounds (we shall
write in brief la-backgrounds, la-spaces and la-geometry) of the
Polyakov's covariant functional-integral
approach to string theory 
 [193]. Our aim is to show that a
corresponding low-energy string dynamics contains the motion
equations for field equations on higher order anisotropic
superspaces and to analyze the geometry of the perturbation
theory of the locally anisotropic supersymmetric sigma models. We
note that this Chapter is devoted to supersymmetric models of
locally anisotropic superstrings (details on the so called
bosonic higher order anisotropic strings are given in the Chapter
9).

The plan of presentation in the Chapter is as follows. Section
4.1 contains an introduction into the geometry of two dimensional
higher order anisotropic sigma models and an locally anisotropic
approach to heterotic strings. In section 4.2 the background
field method for $\sigma $-models is generalized for a
distinguished calculus locally adapted to the N--connection
structure in higher order anisotropic superspaces. Section 4.3 is
devoted to a study of Green--Schwartz action in distinguished
vector superbundles. Fermi strings in higher order anisotropic
spaces are considered in section 4.4. An example of one--loop and
two--loop calculus for anomalies of locally anisotropic strings
is presented in section 4.5. Conclusions are drawn in section
4.6.\

\section{Superstrings in HA--Spaces}

This section considers the basic formalism for superstrings in
dvs--bundles. We shall begin our study with nonsupersymmetric two
dimensional higher order anisotropic sigma models. Then we shall
analyze supersimmetric extensions and locally anisotropic
generalizations of the Green--Schvarz action.

\subsection{Two dimensional ha--sigma s--models}

Let $\widehat{{\cal E}}^{<z>}$ be a higher order anisotropic
space (not
superspace) with coordinates $\widehat{u}^{<\alpha >}=\widehat{u}(z)=(%
\widehat{x}^i=\widehat{x}(z),\widehat{y}^{<a>}=\widehat{y}%
(z))=(x^i,y^{a_1},....,y^{a_p},...,y^{a_z)},$ d--metric $\widehat{g}%
_{<\alpha ><\beta >};$ we use denotation $\left( N_2,\gamma
_{\ddot a\ddot e}\right) $ for a two dimensional world sheet with
metric $\gamma _{\ddot a\ddot e}(z^{\ddot u})$ of signature (+,-)
and local coordinates $z=z^{\ddot u},$ where $\ddot a,\ddot
e,\ddot u,...=1,2.$

The action of a bosonic string in a dv--bundle $\widehat{{\cal
E}}^{<z>}$ is
postulated as
$$
I_\sigma =\frac 1{\lambda ^2}\int d^2z\{\frac 12\sqrt{\gamma
}\gamma ^{\ddot
a\ddot e}\partial _{\ddot a}\widehat{u}^{<\alpha >}(z)\partial _{\ddot e}%
\widehat{u}^{<\beta >}(z)\},\eqno(4.1)
$$
which defines the so called two dimensional sigma model ($\sigma
$--model) with d--metric $\widehat{g}_{<\alpha ><\beta >}$ in
higher order anisotropic spaces (dv--bundles with
N--connec\-ti\-on) and $\lambda $ being constant. We shall give a
detailed study of different modifications of the
model (4.1) in Chapter 9, see also 
 [269,261]. Here we shall
consider a supersymmetric generalization of the string action by
applying the techniques of two dimensional (1,1)--supersymmetry
by changing of scalar fields $u(z)$ into real {\sf N=1} s--fields
(without constraints; for
locally isotropic constructions see 
 [24,45,68,87,111,176])
 $\widehat{u}(z,\theta )$ which are polynoms with respect to Maiorana
anticommuting spinor coordinate $\theta :$%
$$
\widehat{u}^{<\alpha >}(z,\theta )=\widehat{u}^{<\alpha >}(z)+\overline{%
\theta }\lambda ^{<\alpha >}(z)+\frac 12\overline{\theta }\theta
F^{<\alpha
>}(z).\eqno(4.2)
$$

We adopt next conventions and denotations with respect to
2--di\-men\-si\-on \index{Dirac matrices!two--dimension}
Dirac matrices $\gamma _{\ddot a}$ and matrix of charge conjugation $C:$%
$$
\{\gamma _{\ddot a},\gamma _{\ddot e}\}=2\eta _{\ddot a\ddot e}{\bf 1,~}%
tr(\gamma _{\ddot e}\gamma _{\ddot a})=2\eta _{\ddot e\ddot
a},~(\gamma _5)^2=1;
$$
$$
\widehat{\partial }=\gamma ^{\ddot e}\partial _{\ddot e};~C\gamma
_{\ddot a}^TC^{-1}=-\gamma _{\ddot a},~C=-C^T=C^{-1}.
$$
For Maiorana spinors $\theta _{\tilde a},\chi _{\tilde n},...$
one holds \index{Maiorana!spinors} relations
$$
\overline{\theta }=\theta ^{+}\gamma ^0,~\overline{\theta ^{\tilde a}}%
=C^{\tilde a\tilde n}\theta _{\tilde n},
$$
$$
\overline{\theta }\chi =\overline{\chi }\theta ,\overline{\theta
}\gamma
_{\ddot a}\chi =-\overline{\chi }\gamma _{\ddot a}\theta ,\overline{\theta }%
\gamma _5\chi =-\overline{\chi }\gamma _5\theta .
$$

Let introduce in the two dimensional (1,1)--superspace the
covariant
derivations%
$$
D_{\tilde n}=\frac \partial {\partial \overline{\theta }^{\tilde n}}-i(%
\widehat{\partial }\theta )_{\tilde n},
$$
satisfying algebra%
$$
\{D_{\tilde n},D_{\tilde o}\}=2i(\widehat{\partial }C)_{\tilde
n\tilde o}\equiv 2i\partial _{\tilde n\tilde o}
$$
and the integration measure on anticommuting variables with properties%
$$
\int d\theta _{\tilde n}=0,\int d\theta _{\tilde n}\theta
^{\tilde a}=\delta _{\tilde n}^{\tilde a},\frac 1{2i}\int
d^2\theta (\overline{\theta }\theta )=1.
$$

The (1,1)--supersymmetric generalization of (4.1) in terms of
s--fields
(4.2) is written as%
$$
I_{\sigma S}=\frac 1{8i\pi \alpha ^{\prime }}\int d^2z\int d^2\theta \{%
\widehat{g}_{<\alpha ><\beta >}(\widehat{u})-\widehat{b}_{<\alpha ><\beta >}(%
\widehat{u})\}\overline{D}\widehat{u}^{<\alpha >}(1+\gamma _5)D\widehat{u}%
^{<\beta >},
$$
where $\lambda ^2=2\pi \alpha ^{\prime }$ which is a higher order
anisotropic generalization of the Curtright--Zachos
 [68] nonlinear
\index{Nonlinear sigma model!generalization}
sigma model. Integrating on $\theta \,$ and excluding auxiliary fields $%
F^{<\alpha >}$ according to theirs algebraic equations we obtain
from the
last expression:%
$$
I_{\sigma S}=\frac 12\int d^2z\{\widehat{g}_{<\alpha ><\beta
>}\partial
^{\ddot e}\widehat{u}^{<\alpha >}\partial _{\ddot e}\widehat{u}^{<\beta >}+%
\eqno(4.3)
$$
$$
\varepsilon ^{\ddot e\ddot \imath }\widehat{b}_{<\alpha ><\beta
>}\partial
_{\ddot e}\widehat{u}^{<\alpha >}\partial _{\ddot \imath }\widehat{u}%
^{<\beta >}+i\widehat{g}_{<\alpha ><\beta >}\overline{\lambda
}^{<\alpha
>}\gamma ^{\ddot e}\widehat{D}_{\ddot e}^{(-)}\lambda ^{<\beta >}+
$$
$$
\frac 18\widehat{R}_{<\beta ><\alpha ><\gamma ><\delta >}^{(-)}\overline{%
\lambda }^{<\alpha >}(1+\gamma _5)\lambda ^{<\gamma >}\overline{\lambda }%
^{<\beta >}(1+\gamma _5)\lambda ^{<\delta >}\},
$$
where
$$
\widehat{D}_{\ddot e}^{(\pm )}\lambda ^{<\beta >}=[\delta
_{<\alpha
>}^{<\beta >}\partial _{\ddot e}+\widehat{\widetilde{\Gamma }}_{<\alpha
><\gamma >}^{<\beta >}\partial _{\ddot e}\widehat{u}^{<\gamma >}\pm \widehat{%
B}_{<\alpha ><\gamma >}^{<\beta >}\varepsilon _{\ddot e\ddot
o}\partial ^{\ddot o}\widehat{u}^{<\gamma >}]\lambda ^{<\gamma >},
$$
$\widehat{\widetilde{\Gamma }}_{<\alpha ><\gamma >}^{<\beta >}$
are Christoffel d--symbols\ (1.39) on dv--bundle. In order to
have compatible with the N--connection structure motions of
la--strings we consider \index{Locally anisotropic
strings!la--strings} \index{la--strings}
 [269,261] these relations between ds--tensor $b_{<\alpha ><\beta >},$
strength $$\widehat{B}_{<\alpha ><\beta ><\gamma >}=\delta _{[<\alpha >}%
\widehat{b}_{<\beta ><\gamma >\}}$$ and torsion $$T_{<\alpha
><\gamma
>}^{<\beta >}$$ (see (1.29)):%
$$
\delta _{<\alpha >}\widehat{b}_{<\beta ><\gamma
>}=\widehat{g}_{<\alpha
><\delta >}\widehat{T}_{<\beta ><\gamma >}^{<\delta >},\eqno(4.4)
$$
with s--integrability conditions
$$
\Omega _{a_pa_s}^{a_f}\delta _{a_f}\widehat{b}_{<\beta ><\gamma
>}=\delta _{[a_h}\widehat{T}_{a_s\}<\beta ><\gamma
>},~(f<p,s;p,s=0,1,...,z),\eqno(4.5)
$$
where $\Omega _{a_pa_s}^{a_f}$ are the coefficients of the
N--connection curvature (1.11). In this case we can express
$\widehat{B}_{<\alpha ><\beta
><\gamma >}=\widehat{T}_{[<\alpha ><\beta ><\gamma >\}}.$ Conditions (4.4)
and (4.5) define a model of higher order anisotropic superstrings when the $%
\sigma $--modes s--antisymmetric strength is introduced from the
higher order anisotropic background torsion. More general
constructions are possible by using normal coordinates locally
adapted to both N--connection and torsion structures on
background s--spaces. For simplicity, we omit such considerations
in this work.

Ds--tensor $\widehat{R}_{<\alpha ><\beta ><\gamma ><\delta >}$
from (4.3)
denotes the curvature with torsion $B:$%
$$
\widehat{R}_{<\beta ><\alpha ><\gamma ><\delta >}^{(\pm )}[\widehat{\Gamma }%
^{(\pm )}]=\widetilde{R}_{<\beta ><\alpha ><\gamma ><\delta >}\mp
$$
$$
D_{<\gamma >}\widehat{B}_{<\alpha ><\beta ><\delta >}\pm D_{<\delta >}%
\widehat{B}_{<\alpha ><\beta ><\gamma >}+
$$
$$
\widehat{B}_{<\tau ><\alpha ><\gamma >}\widehat{B}_{<\delta
><\beta
>}^{<\tau >}-\widehat{B}_{<\tau ><\alpha ><\delta >}\widehat{B}_{<\gamma
><\beta >}^{<\tau >},
$$
where $\widetilde{R}_{<\beta ><\alpha ><\gamma ><\delta >}$ is
the curvature
of the torsionless Christoffel d--symbols (1.39),%
$$
\widehat{\Gamma }_{<\beta ><\gamma >}^{<\alpha >(\pm
)}=\widehat{g}^{<\alpha
><\tau >}\widehat{\Gamma }_{<\tau ><\beta ><\gamma >}^{(\pm )},\ \widehat{%
\Gamma }_{<\tau ><\beta ><\gamma >}^{(\pm )}=$$ $$\widehat{\widetilde{\Gamma }}%
_{<\tau ><\beta ><\gamma >}\pm B_{<\tau ><\beta ><\gamma >}^{(\pm
)},
$$
$$
\widetilde{\Gamma }_{<\tau ><\beta ><\gamma >}=\frac 12(\delta _{<\beta >}%
\widetilde{g}_{<\tau ><\gamma >}+\delta _{<\gamma
>}\widetilde{g}_{<\beta
><\tau >}-\delta _{<\tau >}\widetilde{g}_{<\beta ><\gamma >}).
$$

In order to define a locally supersymmetric generalization of the
model (4.3) we consider a supersymmetric calculus of the set of
(1,1)--multiplets
of higher order anisotropic matter ($\widehat{\varphi }^{<\alpha >},\widehat{%
\lambda }^{<\alpha >}=\lambda ^{<\alpha >})$ with the multiplet of
(1,1)--supergravity $\left( e_{\ddot e}^{\underline{\ddot
e}},\psi _{\ddot
e}\right) $ in two dimensions (see 
 [137,139,140] for locally
isotropic constructions).

The global supersymmetric variant of action (4.3), for $2\pi
\alpha ^{\prime }=1,$ is written as
$$
I_0[\varphi ,\lambda ]=\frac 12\int d^2z[\widehat{g}_{<\alpha
><\beta
>}\partial ^{\underline{\ddot e}}\widehat{u}^{<\alpha >}\partial _{%
\underline{\ddot e}}\widehat{u}^{<\beta >}+\widehat{b}_{<\alpha
><\beta
>}\varepsilon ^{\underline{\ddot e}\ \underline{\ddot a}}\partial _{%
\underline{\ddot e}}\widehat{u}^{<\alpha >}\partial _{\underline{\ddot a}}%
\widehat{u}^{<\beta >}+\eqno(4.6)
$$
$$
i\widehat{g}_{<\alpha ><\beta >}\overline{\lambda }^{<\alpha >}\gamma ^{%
\underline{\ddot e}}(D_{\underline{\ddot e}}\lambda )^{<\beta >}+i\widehat{B}%
_{<\alpha ><\beta ><\gamma >}\overline{\lambda }^{<\alpha
>}\gamma _5\gamma ^{\underline{\ddot e}}(\partial
_{\underline{\ddot e}}\widehat{u}^{<\beta
>})\lambda ^{<\gamma >}+
$$
$$
\frac 16\widehat{\widetilde{R}}_{<\beta ><\alpha ><\gamma ><\delta >}(%
\overline{\lambda }^{<\alpha >}\lambda ^{<\gamma >})(\overline{\lambda }%
^{<\beta >}\lambda ^{<\delta >})-
$$
$$
\frac 14\widehat{\widetilde{D}}_{<\varepsilon >}B_{<\alpha ><\beta ><\tau >}(%
\overline{\lambda }^{<\alpha >}\gamma _5\lambda ^{<\beta >})(\overline{%
\lambda }^{<\tau >}\lambda ^{<\varepsilon >})-
$$
$$
\frac 14\widetilde{B}_{<\alpha ><\beta ><\tau
>}\widetilde{B}_{<\gamma
><\delta >}^{<\tau >}(\overline{\lambda }^{<\alpha >}\gamma _5\lambda
^{<\beta >})(\overline{\lambda }^{<\gamma >}\gamma _5\lambda
^{<\delta >})],
$$
where covariant derivation $\widehat{\widetilde{D}}_{<\varepsilon
>}$ is defined by torsionless Christoffel d--symbols.

The action (4.6) is invariant under global s--transforms with
Maiorana
spinor parameter $\varepsilon :$%
$$
\bigtriangleup \widehat{\varphi }^{<\alpha
>}=\overline{\varepsilon }\ \lambda ^{<\alpha >},
$$
$$
\bigtriangleup \lambda ^{<\alpha >}=-i(\widehat{\partial }\widehat{\varphi }%
^{<\alpha >})\varepsilon +\frac \varepsilon 2(\widehat{\widetilde{\Gamma }}%
_{<\beta ><\gamma >}^{<\alpha >}\overline{\lambda }^{<\beta
>}\lambda
^{<\gamma >}-\widetilde{B}_{<\beta ><\gamma >}^{<\alpha >}\overline{\lambda }%
^{<\beta >}\gamma _5\lambda ^{<\gamma >}).
$$

Defining Maiorana--Weyl spinors $\lambda _{\pm }^{<\alpha >}$
(MW--spinors)\ \index{Maiorana--Weyl spinors} instead of $\lambda
^{<\alpha >}$ we can rewrite the action (4.6) in a more
convenient form:%
$$
I_0[\varphi ,\lambda _{\pm }]=\frac 12\int
d^2z[\widetilde{g}_{<\alpha
><\beta >}\partial ^{\underline{\ddot e}}\widetilde{u}^{<\alpha >}\partial _{%
\underline{\ddot e}}\widetilde{u}^{<\beta >}+\widehat{b}_{<\alpha
><\beta
>}\varepsilon ^{\underline{\ddot e}\ \underline{\ddot a}}\partial _{%
\underline{\ddot e}}\widehat{u}^{<\alpha >}\partial _{\underline{\ddot a}}%
\widehat{u}^{<\beta >}+\eqno(4.6a)
$$
$$
i\widehat{g}_{<\alpha ><\beta >}\overline{\lambda }_{+}^{<\alpha >}(\widehat{%
D}^{+}\lambda _{+})^{<\beta >}+i\widehat{g}_{<\alpha ><\beta >}\overline{%
\lambda }_{-}^{<\alpha >}(\widehat{D}^{-}\lambda _{-})^{<\beta >}+
$$
$$
\frac 14\widehat{\widetilde{R}}_{<\beta ><\alpha ><\gamma ><\delta >}^{+}(%
\overline{\lambda }^{<\alpha >}\gamma _5\lambda ^{<\beta >})(\overline{%
\lambda }^{<\gamma >}\gamma _5\lambda ^{<\delta >})],
$$
with s--symmetric transformation law%
$$
\bigtriangleup \varphi ^{<\alpha >}=\bigtriangleup _{+}\varphi
^{<\alpha
>}+\bigtriangleup _{-}\varphi ^{<\alpha >}=\overline{\varepsilon }_{+}\
\lambda _{-}^{<\alpha >}+\overline{\varepsilon }_{-}\ \lambda
_{+}^{<\alpha
>},
$$
$$
\bigtriangleup \lambda _{\pm }^{<\alpha >}=-i(\widehat{\partial
}u^{<\alpha
>})\varepsilon _{\mp }-\widetilde{\Gamma }_{<\beta ><\gamma >}^{<\alpha
>(\pm )}\lambda _{\pm }^{<\beta >}\bigtriangleup _{\pm }\varphi ^{<\gamma
>}.
$$
For simplicity, in the rest of this Chapter we shall omit
''hats'' on geometrical objects if ambiguities connected with
indices for manifolds and supermanifolds will not arise.

In string theories one considers variations of actions of type
(4.6) with respect to s--symmetric transformation laws and
decompositions with respect to powers of $\lambda .$ Coefficients
proportional to $\lambda ^5$ vanishes because they do not contain
derivations of $\varepsilon (z)$--parameters. In order to
compensate the therms proportional to $\lambda $ and $\lambda ^3$
one adds the so--called Nether term
$$
I^{(N)}=\frac 12\int d^2z[2g_{<\alpha ><\beta >}(\partial
_{\underline{\ddot
e}}\varphi ^{<\alpha >})(\overline{\lambda }^{<\beta >}\gamma ^{\underline{%
\ddot o}}\gamma ^{\underline{\ddot e}}\psi _{\underline{\ddot
o}})-
$$
$$
\frac i3B_{<\alpha ><\beta ><\gamma >}(\overline{\lambda
}^{<\alpha >}\gamma
_5\gamma ^{\underline{\ddot e}}\lambda ^{<\beta >})(\overline{\lambda }%
^{<\gamma >}\psi _{\underline{\ddot e}})-%
$$
$$
\frac i3B_{<\alpha ><\beta ><\gamma >}(\overline{\lambda
}^{<\alpha >}\gamma ^{\underline{\ddot e}}\lambda ^{<\beta
>})(\overline{\lambda }^{<\gamma
>}\gamma _5\psi _{\underline{\ddot e}}),
$$
where $\psi _{\underline{\ddot e}}$ is the higher order
an\-isot\-rop\-ic
generalization of Maiorana gravitino with s--symmetric transformation law%
\index{Maiorana gravitino!higher order an\-isot\-rop\-ic}
$$
\bigtriangleup \psi _{\underline{\ddot e}}=-\partial _{\underline{\ddot e}%
}\varepsilon +...
$$

From the standard variation, but locally adapted to the
N--connection, of the $I^{(N)}$ with a next covariantization
(with respect to $\left( e_{\ddot e}^{\underline{\ddot e}},\psi
_{\ddot e}\right) )$ of the theory. In result
(it's convenient to use MW--spinors) we introduce this action:%
\index{Maiorana--Weyl spinors!MW--spinors}
$$
I=\frac 12\int d^2z\ e\ [\gamma ^{\ddot a\ddot u}g_{<\alpha
><\beta
>}\partial _{\ddot a}u^{<\alpha >}\partial _{\ddot u}u^{<\beta
>}+e^{-1}\varepsilon ^{\ddot a\ddot u}b_{<\alpha ><\beta >}\partial _{\ddot
a}u^{<\alpha >}\partial _{\ddot u}u^{<\beta >}+\eqno(4.7)
$$
$$
ig_{<\alpha ><\beta >}\overline{\lambda }_{+}^{<\alpha >}(\widehat{D}%
^{+}\lambda _{+})^{<\beta >}+ig_{<\alpha ><\beta >}\overline{\lambda }%
_{-}^{<\alpha >}(\widehat{D}^{-}\lambda _{-})^{<\beta >}+
$$
$$
\frac 14\widetilde{R}_{<\beta ><\alpha ><\gamma ><\delta >}^{+}(\overline{%
\lambda }^{<\alpha >}\gamma _5\lambda ^{<\beta >})(\overline{\lambda }%
^{<\gamma >}\gamma _5\lambda ^{<\delta >})]+
$$
$$
g_{<\alpha ><\beta >}(2\partial _{\ddot a}u^{<\alpha >}+\overline{\psi }%
_{\ddot a}\lambda _{+}^{<\alpha >}+\overline{\psi }_{\ddot
a}\lambda _{-}^{<\alpha >})\times
$$
$$
(\overline{\lambda }_{+}^{<\beta >}\{\gamma ^{\ddot a\ddot
u}+e^{-1}\varepsilon ^{\ddot a\ddot u}\}\psi _{\ddot u}+(\overline{\lambda }%
_{-}^{<\beta >}\{\gamma ^{\ddot a\ddot u}-e^{-1}\varepsilon
^{\ddot a\ddot u}\}\psi _{\ddot u})+
$$
$$
\frac{2i}3B_{<\alpha ><\beta ><\gamma >}\{(\overline{\lambda
}_{+}^{<\alpha
>}\gamma ^{\ddot a}\overline{\lambda }_{+}^{<\beta >})(\overline{\lambda }%
_{+}^{<\gamma >}\psi _{\ddot a})-(\overline{\lambda
}_{-}^{<\alpha >}\gamma
^{\ddot a}\overline{\lambda }_{-}^{<\beta >})(\overline{\lambda }%
_{-}^{<\gamma >}\psi _{\ddot a})\}],
$$
where $e=\det |e_{\ddot e}^{\underline{\ddot e}}|,$ for which the
higher
order anisotropic laws of supersymmetric transforms holds:%
$$
\bigtriangleup e_{\ddot e}^{\underline{\ddot e}}=2i\overline{\varepsilon }%
\gamma ^{\underline{\ddot e}}\psi _{\ddot e},\bigtriangleup \psi
_{\ddot e}=-D_{\ddot e}\varepsilon ,\eqno(4.8)
$$
$$
\bigtriangleup \varphi ^{<\alpha >}=\bigtriangleup _{+}\varphi
^{<\alpha
>}+\bigtriangleup _{-}\varphi ^{<\alpha >}=\overline{\varepsilon }_{+}\
\lambda _{-}^{<\alpha >}+\overline{\varepsilon }_{-}\ \lambda
_{+}^{<\alpha
>},
$$
$$
\bigtriangleup \lambda _{\pm }^{<\alpha >}=-i(\widehat{\partial
}u^{<\alpha
>}+\{\overline{\lambda }_{+}^{<\alpha >}\psi _{\ddot a}+\overline{\lambda }%
_{-}^{<\alpha >}\psi _{\ddot a}\}\gamma ^{\ddot a})\varepsilon _{\mp }-%
\widetilde{\Gamma }_{<\beta ><\gamma >}^{<\alpha >(\pm )}\lambda
_{\pm }^{<\beta >}\bigtriangleup _{\pm }\varphi ^{<\gamma >}.
$$

Restricting our considerations in (4.7) and (4.8) only with
$\lambda
_{+}^{<\alpha >}$--spinors and $\varepsilon _{-}$--parameters, when $%
\varepsilon _{+},\lambda _{-}^{<\alpha >}=0,$ we obtain the
action for the heterotic higher order anisotropic string on
background $\left( g_{<\alpha
><\beta >},b_{<\alpha ><\beta >}\right) $ with (1,0)--local supersymmetry $%
\left( \psi _{\ddot a}\rightarrow \psi _{\ddot a(-)}\right) .$
This action can be interpreted as the ''minimal'' interaction of
the higher order anisotropic (1,0)--matter $\left( \varphi
^{<\alpha >},\lambda _{+}^{<\alpha
>}\right) $ with (1,0)--supergravity $\left( e_{\ddot e}^{\underline{\ddot e}%
},\psi _{\ddot e}\right) .$

\subsection{Locally anisotropic heterotic strings}

\index{Locally anisotropic!heterotic strings} \index{Hetrotic
strings!locally anisotropic} As an illustration of application of
s--field methods in locally anisotropic s--spaces we shall
construct the action for a model of higher order anisotropic
s--string.

The (1,0)--superspaces can be parametrized by two Bose coordinates $%
(z^{\ddagger },z^{=})$ and one Fermy coordinate $\theta ^{+};\ddot
u=(z^{\ddagger },z^{=},\theta ^{+}).$ One represents vector indices as $%
(++,--)\equiv (\ddagger ,=)$ taking into account that by $(+,-)$
there are denoted spirality $\pm 1/2.$ \index{Coordinate!Bose}
\index{Coordinate!Fermi}

The standard derivations
$$
D_{\ddot A}=\{D_{+},\partial _{\ddagger },\partial
_{=}\};D_{+}=\frac
\partial {\partial \theta ^{+}}+i\theta ^{+}\partial _{+},
$$
in the flat (1,0)--superspace 
 [214] satisfy algebra%
$$
\{D_{+},D_{+}\}=2i\partial _{\ddagger },~\partial _{\ddagger
}\partial
_{=}=\Box ,~[\partial _{\underline{a}},D_{+}]=[\partial _{\underline{a}%
},\partial _{\underline{b}}]=0,
$$
and s--space integration measure%
$$
\int d\theta _{+}=\frac \partial {\partial \theta ^{+}},d^3\ddot
u^{-}=d^2zd\theta _{+}.
$$

In the flat (1,0)--superspace one defines scalar and spinor s--fields%
$$
\varphi (z,\theta )=A(z)+\theta ^{+}\lambda _{+}(z),\psi
_{-}(z,\theta )=\eta _{-}(z)+\theta ^{+}F(z)
$$
and action
$$
I=\int d^3\ddot uL=\int d^2z(D_{+}L)_{|\theta =0}
$$
with a charged Lagrangian, $L=L_{-},$ in order to have the Lorentz
invariance.

The (1,0)--multiplet of supergravity is described by a set of
covariant \index{Supergravity!(1,0)--multiplet}
derivations%
$$
\bigtriangledown _{\ddot A}=E_{\ddot A}^{\underline{\ddot U}}D_{\underline{%
\ddot U}}+\omega _{\ddot A}^{(M)}\equiv E_{\ddot A}+\Omega _{\ddot A},%
\eqno(4.9)
$$
where $E_{\ddot A}^{\underline{\ddot U}}$ is a s--vielbein and
$\omega
_{\ddot A}^{(M)}$ is the Lorentz connection with L--generator $M:$%
$$
[M,\lambda _{\pm }]=\pm \frac 12\lambda _{\pm }.
$$

The covariant constraints in s--space (for (1,0)--supergravity
 [46,85]) are given by relations:%
$$
\{\nabla _{+},\nabla _{+}\}=2i\nabla _{\ddagger },~[\nabla
_{+},\nabla _{=}]=-2i\Sigma ^{+}M,\eqno(4.10)
$$
$$
[\nabla _{+},\nabla _{\ddagger }]=0,~[\nabla _{\ddagger },\nabla
_{=}]=-\Sigma ^{+}\nabla _{+}-R^{(2)}M,
$$
where $R^{(2)}=2\nabla _{+}\Sigma ^{+}$ and $\Sigma ^{+}$ defines
the covariant strength of (1,0)--su\-per\-gra\-vi\-ty in s--space.

As a consequence of (4.10) only $E_{+}^{=},E_{=}^{\ddagger
},E_{+}^{\ddagger },E_{+}^{+}$ and $E_{=}^{=}$ are independent
s--fields; the rest of components of vielbein and connection can
be expressed through them. The conditions of covariance of
derivations (4.9) lead to these transformation laws with respect
to d--coordinate and local Lorentz transforms with
corresponding parameters $K^{\underline{\ddot U}}$ and $\Lambda _l:$%
$$
\bigtriangledown _{\ddot A}^{\prime }=e^K\bigtriangledown _{\ddot
A}e^{-K},~K=K^{\underline{\ddot U}}D_{\underline{\ddot
U}}+\Lambda _lM;
$$
for vielbeins we have%
$$
\bigtriangleup E_{\ddot A}^{\underline{\ddot U}}=-\nabla _{\ddot A}K^{%
\underline{\ddot U}}+K^{\underline{\ddot I}}D_{\underline{\ddot
I}}E_{\ddot
A}^{\underline{\ddot U}}+E_{\ddot A}^{\underline{\ddot O}}K^{\underline{%
\ddot I}}[D_{\underline{\ddot I}},D_{\underline{\ddot O}}\}^{\underline{%
\ddot U}}+\Lambda _l[M,E_{\ddot A}^{\underline{\ddot U}}],
$$
from which one follows that%
$$
\bigtriangleup E_{+}^{\ddagger }=K^{\underline{\ddot U}}D_{\underline{\ddot U%
}}E_{+}^{\ddagger }+2iE_{+}^{+}K^{+}-E_{+}^{\underline{\ddot U}}D_{%
\underline{\ddot U}}K^{\ddagger }+\frac 12\Lambda
_lE_{+}^{\ddagger }.
$$

It is convenient to use the s--symmetric gauge (when $E_{+}^{\ddagger }=0),$%
$$
K^{+}=-\frac i2\left( E_{+}^{+}\right) ^{-1}\nabla
_{+}K^{\ddagger },
$$
and to introduce the Lorentz--invariant scalar s--field $S$ and
Lorentz
compensator $L$ satisfying correspondingly conditions%
$$
E_{+}^{+}(E_{=}^{=})^{1/2}=e^{-S}~\mbox{ and }%
~E_{+}^{+}(E_{=}^{=})^{-1/2}=e^L.
$$
In a Lorentz invariant theory we can always choose the gauge
$L=0;$ in this gauge the s--conformal transforms are accompanied
by a corresponding
compensating Lorentz transform with parameter%
$$
\Lambda _l=(E_{+}^{+})^{-1}\nabla _{+}K^{+}-\frac
12(E_{=}^{=})^{-1}\nabla _{=}K^{=}=\frac 12(\nabla _{\ddagger
}K^{\ddagger }-\nabla _{=}K^{=})+...
$$

The solution of constraints (4.10) in the s--symmetric gauge $%
E_{+}^{\ddagger }=L=0$ and in the linear approximation is 
 [85]
$$
\nabla _{+}=(1-\frac S2)D_{+}+H_{+}^{=}\partial
_{=}-(D_{+}S+\partial _{=}H_{+}^{=})M,
$$
$$
\nabla _{\ddagger }=(1-S)\partial _{\ddagger }+i[\frac 12(\partial
_{=}H_{+}^{=})+(D_{+}S)]D_{+}-
$$
$$
i[D_{+}H_{+}^{=}]\partial _{=}-(\partial _{\ddagger
}S-iD_{+}\partial _{=}H_{+}^{=})M,
$$
$$
\nabla _{=}=(1-S)\partial _{=}-\frac i2[(D_{+}H_{=}^{\ddagger
})]D_{+}+H_{=}^{\ddagger }\partial _{\ddagger }+(\partial
_{=}S+\partial _{\ddagger }H_{=}^{\ddagger })M,
$$
$$
\Sigma ^{+}=\frac i2[D_{+}(\partial _{\ddagger }H_{=}^{\ddagger
}+2\partial _{=}S)+\partial _{=}^2H_{+}^{=}]+...,
$$
where s--fields $\left( H_{+}^{=},H_{=}^{\ddagger }\right) $ and
$S$ are prepotentials of the system. These s--potentials have to
be used in the quantum field theory.

The linearized expression for s--field density $E^{-1}=Ber(E_{\ddot A}^{%
\underline{\ddot U}})$ is computed as
$$
E^{-1}=s\det (E_{\ddot A}^{\underline{\ddot
U}})=e^{3S/2}[1+iH_{=}^{\ddagger
}(D_{+}H_{+}^{=}+H_{+}^{=}\partial _{=}H_{+}^{=})]^{-1}.
$$

The action for heterotic string in higher order an\-isot\-rop\-ic
\index{Higher order an\-isot\-rop\-ic!(1,0)--superspace}
(1,0)--su\-per\-spa\-ce accounting for the background of massless
modes of locally anisotropic graviton, antisymmetric d--tensor,
dilaton and gauge bosons is introduced in a manner similar to
locally isotropic models
 [105,116,139,140] but with corresponding extension to distinguished and
locally adapted to N--connection geometric objects:%
$$
I_{HS}=\frac 1{4\pi \alpha ^{\prime }}\int d^3\ddot
u^{-}E^{-1}\{i\nabla _{+}u^{<\alpha >}\nabla _{=}u^{<\beta
>}[g_{<\alpha ><\beta >}(u)+b_{<\alpha
><\beta >}(u)]+\eqno(4.11)
$$
$$
\Psi _{(-)}^{|I|}[\delta _{|I||J|}\nabla
_{+}+A_{|I||J|}^{+}(u)]\Psi _{(-)}^{|J|}+\alpha ^{\prime }\Phi
(u)\Sigma ^{+}\},
$$
where $A_{|I||J|}^{+}(u)=A_{|I||J|<\alpha >}\nabla _{+}u^{<\alpha
>}$ is the gauge boson background, $\Phi (u)$ is the dilaton
field and $\Psi _{(-)}^{|I|}$ are hetrotic fermions,
$|I|,|J|,...=1,2,...,,{\sf N.}$

\section{Background D--Field Methods }

The background--quantum decomposition of superfields of (1,0)
higher order \index{Background!d--field methods} anisotropic
supergravity considered in previous section can be performed in
a standard manner 
 [85] by taking into account the distinguished
character of geometrical objects on locally anisotropic s--spaces;
constraints (4.10) should be solved in terms of
background--covariant derivations and quantum s--fields $\left(
H_{+}^{=},H_{=}^{\ddagger },S\right) ,$ quantum s--fields $L$ and
$E_{+}^{\ddagger }$ are gauged in an algebraic manner (not
introducing into considerations ghosts) and note that by using
quantum scale transforms we can impose gauge $S=0$ (also without
Faddeev--Popov ghosts). Finally, after a background--quantum
decomposition of superfields of (1.0) higher order anisotropic
supergravity in the just pointed out manner we can fix the
quantum gauge invariance, putting zero values for quantum fields
(in absences of supergravitational and conformal anomalies and
for topological trivial background configurations). In this case
all (1,0) supergravity fields can be considered as background
ones.

The fixing of gauge symmetry as a vanishing of quantum s--fields
induces the
ghost action%
$$
I_{FP}=\int d^3z^{-}E\{b_{=}^{\ddagger }\nabla
_{+}c^{=}+b_{+}^{=}\nabla _{=}c^{\ddagger }\}.\eqno(4.12)
$$

The aim of this section is to compute the renormalized effective
action, \index{Action!renormalized effective} more exactly, it
anomaly part on the background of s--fields (1,0) higher order
anisotropic supergravity for the model defined by the action
(4.11).

In order to integrate on quantum fields in (4.11) in a
d---covariant manner we use background--quantum decompositions of
the action with respect to normal locally adapted coordinates
along autoparallels (see details in Chapter III ) defined by
Christoffel d--symbols (1.39). We emphasize the multiconnection
character of locally anisotropic spaces; every geometric
construction with a fixed d--connection structure (from some
purposes considered as a simple or more convenient one) can be
transformed, at least locally, into a another similar one for a
corresponding d--connection by using deformations of connections
(1.40) (or (3.2) and (3.3) if we are interested in na--map
deformations). Let
$$
\frac{\delta ^2X^{<\alpha >}}{\partial s^2}+\{\frac{<\alpha
>}{<\beta
><\gamma >}\}\frac{\delta X^{<\beta >}}{\partial s}\frac{\delta X^{<\gamma >}%
}{\partial s}=0,
$$
where $X^{<\alpha >}(s=0)=X^{<\alpha >},X^{<\alpha
>}(s=1)=X^{<\alpha >}+\pi ^{<\alpha >}$ and $\pi ^{<\alpha >}$
are quantum fluctuations with respect to background $X^{<\alpha
>}.\,$Covariant quantum fields $\zeta ^{<\alpha >}$
(''normal fields'') are defined as%
$$
\zeta ^{<\alpha >}=\frac{\delta X^{<\alpha >}}{\partial s}\mid
_{s=0}\equiv \zeta ^{<\alpha >}(s)\mid _{s=0}.
$$
We shall use covariantized in a $\sigma $--model manner
derivations in (1,0) higher order anisotropic s--space
$$
{\cal D}_{\ddot A}\equiv ({\cal D}_{+},{\cal D}_{=})=\nabla
_{\ddot
A}+\Gamma _{\ddot A},\Gamma _{\ddot A<\beta >}^{<\alpha >}\equiv \{\frac{%
<\alpha >}{<\gamma ><\beta >}\}\nabla _{\ddot A}X^{<\gamma >},
$$
on properties of derivation $\nabla _{\ddot A}$ see (4.3) and
(4.4), into distinguished autoparallel (in a $\sigma $--model
manner) d--covariant derivation with properties
$$
D\left( s\right) T_{<\alpha >...}=\zeta ^{<\beta >}{\cal
D}_{<\beta
>}T_{<\alpha >...},D(s)\zeta ^{<\alpha >}(s)=0.
$$

The derivation ${\cal D}_{\ddagger }$ is defined as
$$
2i{\cal D}_{\ddagger }\equiv \{{\cal D}_{+},{\cal D}_{+}\};
$$
we note that ${\cal D}_{\ddagger }\neq \nabla _{\ddagger
}^{cov}.$ One holds
the next relations:%
$$
{\cal D}_{+}\zeta ^{<\beta >}=\nabla _{+}\zeta ^{<\beta >}+\{\frac{<\beta >}{%
<\tau ><\sigma >}\}\nabla _{+}X^{<\sigma >}\zeta ^{<\tau >},
$$
$$
{\cal D}_{=}\zeta ^{<\beta >}=\nabla _{=}\zeta ^{<\beta >}+\{\frac{<\beta >}{%
<\tau ><\sigma >}\}\nabla _{=}X^{<\sigma >}\zeta ^{<\tau >},
$$
$$
{\cal D}_{\ddagger }\zeta ^{<\beta >}=\nabla _{\ddagger
}^{cov}\zeta ^{<\beta >}-\frac i2R_{~<\tau ><\sigma ><\nu
>}^{<\beta >}\nabla _{+}X^{<\sigma >}\nabla _{+}X^{<\nu >}\zeta
^{<\tau >},
$$
$$
D(s){\cal D}_{\ddot A}\zeta ^{<\beta >}(s)=\zeta ^{<\nu
>}(s)\zeta ^{<\tau
>}(s)R_{~<\tau ><\sigma ><\nu >}^{<\beta >}\nabla _{\ddot A}X^{<\sigma >},
$$
where
$$
\nabla _{\ddagger }^{cov}\zeta ^{<\beta >}=\nabla _{\ddagger
}\zeta ^{<\beta
>}+\{\frac{<\beta >}{<\tau ><\sigma >}\}\nabla _{\ddagger }X^{<\tau >}\zeta
^{<\sigma >}.
$$

For heterotic fermions the d--covariant and gauge--covariant
formalism of \index{Heterotic fermions} computation of
quantum--background decomposition of action can be performed
by using the prescription%
$$
\Psi _{(-)}^{|I|}(s): \Psi _{(-)}^{|I|}(0)=\Psi _{(-)}^{|I|}, \Psi
_{(-)}^{|I|}(s=1)=\Psi _{(-)}^{|I|}+\Delta _{(-)}^{|I|},
D_{(-)}^2\Psi _{-}^{|I|}(s)=0,
$$
where $\Delta _{-}^{|I|}$ are quantum fluctuations with respect to
background $\Psi ;$ functions $\Psi _{(-)}^{|I|}(s)$ interpolate
in a gauge--covariant manner with $\Psi _{(-)}^{|I|}+\Delta
_{-}^{|I|}$ because
of definition of the operator $D(s):$%
$$
D(s)\Psi _{(-)}^{|I|}=[\delta ^{|I||J|}\frac \delta {\partial
s}+A_{<\alpha
>}^{|I||J|}\frac{\delta X^{<\alpha >}}{\partial s}]\Psi _{(-)}^{|J|}=\frac{%
\delta \Psi _{(-)}^{|I|}}{\partial s}+A_{<\alpha >}^{|I||J|}\zeta
^{<\alpha
>}\Psi _{(-)}^{|J|}.
$$

As d--covariant and gauge--covariant quantum s--fields we use
spinors
$$
\chi _{(-)}^{|J|}\equiv D(s)\Psi _{(-)}^{|J|}(s)\mid _{s=0}
$$
satisfying conditions%
$$
D(s)\chi _{(-)}^{|J|}(s)=0.
$$
We also define the next derivation in (1,0) higher order
anisotropic s--superspace
$$
({\cal D}_{+}\Psi _{(-)})^{|I|}\equiv (\delta _{|J|}^{|I|}\nabla
_{+}+A_{|J|<\alpha >}^{|I|}\nabla _{+}X^{<\alpha >})\Psi
_{(-)}{}^{|J|}
$$
for which one holds identities%
$$
D(s){\cal D}_{+}\Psi _{(-)}{}^{|I|}={\cal D}_{+}\chi
_{(-)}{}^{|I|}+F_{|J|<\alpha ><\beta >}^{|I|}\zeta ^{<\alpha
>}\nabla _{+}X^{<\beta >}\Psi _{(-)}{}^{|I|},
$$
$$
D(s)F_{|J|<\alpha ><\beta >}^{|I|}=\zeta ^{<\gamma >}{\cal
D}_{<\gamma
>}F_{|J|<\alpha ><\beta >}^{|I|},
$$
where ${\cal D}_{<\gamma >}$ is both d-covariant and gauge
invariant derivation and the strength d--tensor of gauge fields
is defined by using
la--derivation operators,%
$$
F_{|J|<\alpha ><\beta >}^{|I|}=\delta _{<\alpha >}A_{|J|<\beta
>}^{|I|}-\delta _{<\beta >}A_{|J|<\alpha >}^{|I|}+A_{|K|<\alpha
>}^{|I|}A_{|J|<\beta >}^{|K|}-A_{|K|<\beta >}^{|I|}A_{|J|<\alpha >}^{|K|}.
$$

The action (4.11) consist from four groups of terms of different nature:%
$$
I_{HS}=I^{(1)}+I^{(2)}+I^{(3)}+I^{(4)}.\eqno(4.13)
$$

The first term
$$
I^{(1)}=-\frac i{4\pi \alpha ^{\prime }}\int d^3z^{-}Eg_{<\alpha
><\beta
>}(X)\nabla _{+}X^{<\alpha >}\nabla _{=}X^{<\beta >}
$$
is associated to the higher order anisotropic gravitational
sector and has the next background--quantum decomposition
$$
I^{(1)}[X+\pi (\zeta )]=I_0^{(1)}+I_1^{(1)}+I_2^{(1)}+...,
$$
where
$$
I_b^{(1)}=\frac 1{b!}\frac{d^bI^{(1)}}{ds^b}\mid
_{s=0};b=0,1,2,....
$$
Thes terms are computed in a usual (but distinguished to the
N--connection
structure) manner:%
$$
I_1^{(1)}=-\frac i{2\pi \alpha ^{\prime }}\int
d^3z^{-}Eg_{<\alpha ><\beta
>}({\cal D}_{+}\zeta ^{<\alpha >})\nabla _{=}X^{<\beta >},\eqno(4.14)
$$
$$
I_2^{(1)}=-\frac i{4\pi \alpha ^{\prime }}\int
d^3z^{-}E\{g_{<\alpha ><\beta
>}{\cal D}_{+}\zeta ^{<\alpha >}{\cal D}_{=}\zeta ^{<\beta >}+
$$
$$
R_{<\lambda ><\sigma ><\tau ><\nu >}\nabla _{+}X^{<\sigma >}\nabla
_{=}X^{<\nu >}\zeta ^{<\lambda >}\zeta ^{<\tau >}\},
$$
$$
I_3^{(1)}=-\frac i{4\pi \alpha ^{\prime }}\int d^3z^{-}E\{\frac
23(R_{<\lambda ><\sigma ><\tau ><\nu >}\nabla _{+}X^{<\sigma >}{\cal D}%
_{=}\zeta ^{<\nu >}+
$$
$$
R_{<\lambda ><\sigma ><\tau ><\nu >}\nabla _{=}X^{<\sigma >}{\cal D}%
_{+}\zeta ^{<\nu >})\zeta ^{<\lambda >}\zeta ^{<\tau >}+
$$
$$
\frac 13{\cal D}_{<\mu >}R_{<\lambda ><\sigma ><\tau ><\nu
>}\nabla _{+}X^{<\sigma >}\nabla _{=}X^{<\nu >}\zeta ^{<\mu
>}\zeta ^{<\lambda
>}\zeta ^{<\tau >},
$$
$$
I_4^{(1)}=-\frac i{4\pi \alpha ^{\prime }}\int d^3z^{-}E\{\frac 14({\cal D}%
_{<\mu >}R_{<\lambda ><\sigma ><\tau ><\nu >}\nabla _{+}X^{<\sigma >}{\cal D}%
_{=}\zeta ^{<\nu >}+
$$
$$
{\cal D}_{<\mu >}R_{<\lambda ><\sigma ><\tau ><\nu >}\nabla _{=}X^{<\sigma >}%
{\cal D}_{+}\zeta ^{<\nu >})\zeta ^{<\mu >}\zeta ^{<\lambda
>}\zeta ^{<\tau
>}+
$$
$$
\frac 13R_{<\lambda ><\sigma ><\tau ><\nu >}\zeta ^{<\lambda
>}\zeta ^{<\tau
>}{\cal D}_{+}\zeta ^{<\sigma >}{\cal D}_{=}\zeta ^{<\nu >}+
$$
$$
[\frac 13R_{<\lambda ><\sigma ><\tau ><\nu >}R_{<\delta
><\varepsilon >\ \cdot \ <\gamma >}^{\qquad \quad <\nu >}+
$$
$$
\frac 1{12}{\cal D}_{<\delta >}{\cal D}_{<\gamma >}R_{<\lambda
><\sigma
><\varepsilon ><\tau >}]\nabla _{+}X^{<\sigma >}\nabla _{=}X^{<\varepsilon
>}\zeta ^{<\lambda >}\zeta ^{<\tau >}\zeta ^{<\delta >}\zeta ^{<\gamma >},
$$
$$
I_5^{(1)}=-\frac i{4\pi \alpha ^{\prime }}\int d^3z^{-}E\times
$$
$$
\{\frac 16({\cal D}_{<\mu >}R_{<\lambda ><\sigma ><\nu ><\tau >}{\cal D}%
_{+}\zeta ^{<\sigma >}{\cal D}_{=}\zeta ^{<\nu >}\zeta ^{<\mu
>}\zeta ^{<\lambda >}\zeta ^{<\tau >}+...\},
$$
$$
I_6^{(1)}=-\frac i{4\pi \alpha ^{\prime }}\int d^3z^{-}E\{\frac 1{20}{\cal D}%
_{<\alpha >}{\cal D}_{<\mu >}R_{<\lambda ><\sigma ><\nu ><\tau >}+
$$
$$
\frac 2{45}R_{<\lambda ><\gamma ><\nu ><\tau >}R_{<\alpha
><\sigma >\ \cdot
\ <\mu >}^{\qquad \quad <\gamma >}{\cal D}_{+}\zeta ^{<\sigma >}{\cal D}%
_{=}\zeta ^{<\nu >}\zeta ^{<\alpha >}\zeta ^{<\mu >}\zeta
^{<\lambda >}\zeta ^{<\tau >}+...\},
$$
where by dots, in this section, are denoted those terms
(containing multiples $\left( \nabla X\right) )$ which are not
important for calculation of anomalies, see section 4.5.

The second term in (4.13)
$$
I^{(2)}=-\frac i{4\pi \alpha ^{\prime }}\int d^3z^{-}Eb_{<\alpha
><\beta
>}(X)\nabla _{+}X^{<\alpha >}\nabla _{=}X^{<\beta >}
$$
having the next background--quantum decomposition
\index{Decomposition!background--quantum}
$$
I^{(2)}[X+\pi (\zeta )]=I_0^{(2)}+I_1^{(2)}+I_2^{(2)}+...,
$$
where
$$
I_b^{(2)}=\frac 1{b!}\frac{d^bI^{(2)}}{ds^b}\mid
_{s=0};b=0,1,2,....
$$
describes a Wess--Zumino--Witten like model of interactions (in a
higher
order anisotropic variant). The diagram vertexes depends only on intensity $%
H $ of antisymmetric d--tensor $b:$%
$$
H_{<\tau ><\mu ><\nu >}=\frac 32\delta _{[<\tau >}b_{<\mu ><\nu
>]}=
$$
$$
\frac 12(\delta _{<\tau >}b_{<\mu ><\nu >}+\delta _{<\mu
>}b_{<\nu ><\tau
>}-\delta _{<\nu >}b_{<\mu ><\tau >}).
$$
By straightforward calculations we find the coefficients:%
$$
I_1^{(2)}=-\frac i{2\pi \alpha ^{\prime }}\int d^3z^{-}E\zeta
^{<\tau
>}\nabla _{+}X^{<\alpha >}\nabla _{=}X^{<\beta >}H_{<\tau ><\alpha ><\beta
>},
$$
$$
I_2^{(2)}=-\frac i{4\pi \alpha ^{\prime }}\int d^3z^{-}E\{\zeta ^{<\tau >}%
{\cal D}_{+}\zeta ^{<\alpha >}\nabla _{=}X^{<\beta >}H_{<\tau
><\alpha
><\beta >}+
$$
$$
\zeta ^{<\tau >}\nabla _{+}X^{<\alpha >}{\cal D}_{=}\zeta ^{<\beta
>}H_{<\tau ><\alpha ><\beta >}+
$$
$$
\zeta ^{<\lambda >}\zeta ^{<\tau >}\nabla _{+}X^{<\alpha >}\nabla
_{=}X^{<\beta >}{\cal D}_{<\lambda >}H_{<\tau ><\alpha ><\beta
>}\},
$$
$$
I_3^{(2)}=-\frac i{4\pi \alpha ^{\prime }}\int d^3z^{-}E\{\frac
23\zeta ^{<\tau >}{\cal D}_{+}\zeta ^{<\mu >}{\cal D}_{=}\zeta
^{<\nu >}H_{<\tau
><\mu ><\nu >}+
$$
$$
\frac 23\zeta ^{<\tau >}\zeta ^{<\rho >}\zeta ^{<\gamma
>}R_{~\cdot <\gamma
><\rho >[<\delta >}^{<\mu >}H_{<\nu >]<\tau ><\mu >}\nabla _{+}X^{<\delta
>}\nabla _{=}X^{<\nu >}+
$$
$$
\frac 23\zeta ^{<\tau >}\zeta ^{<\lambda >}({\cal D}_{+}\zeta
^{<\mu
>}\nabla _{=}X^{<\nu >}+{\cal D}_{=}\zeta ^{<\nu >}\nabla _{+}X^{<\mu >})%
{\cal D}_{<\lambda >}H_{<\tau ><\mu ><\nu >}+
$$
$$
\frac 13\zeta ^{<\sigma >}\zeta ^{<\lambda >}\zeta ^{<\tau
>}\nabla
_{+}X^{<\alpha >}\nabla _{=}X^{<\beta >}{\cal D}_{<\sigma >}{\cal D}%
_{<\lambda >}H_{<\tau ><\alpha ><\beta >}\},
$$
$$
I_4^{(2)}=-\frac i{4\pi \alpha ^{\prime }}\int d^3z^{-}E\{\frac
12\zeta ^{<\tau >}\zeta ^{<\lambda >}{\cal D}_{+}\zeta ^{<\mu
>}{\cal D}_{=}\zeta ^{<\nu >}H_{<\tau ><\mu ><\nu >}+
$$
$$
\frac 16\zeta ^{<\lambda >}\zeta ^{<\tau >}\zeta ^{<\rho >}\zeta
^{<\gamma
>}\nabla _{+}X^{<\delta >}\nabla _{=}X^{<\nu >}{\cal D}_{<\lambda
>}(R_{~\cdot <\gamma ><\rho >[<\delta >}^{<\mu >}H_{<\nu >]<\tau ><\mu >})+
$$
$$
\frac 23\zeta ^{<\tau >}\zeta ^{<\rho >}\zeta ^{<\gamma
>}R_{~\cdot <\gamma
><\rho >[<\delta >}^{<\mu >}H_{<\nu >]<\tau ><\mu >}\times (\nabla
_{+}X^{<\delta >}{\cal D}_{=}X^{<\nu >}+
$$
$$
{\cal D}_{+}X^{<\nu >}\nabla _{=}X^{<\delta >})+\zeta ^{<\lambda
>}\zeta ^{<\tau >}\zeta ^{<\alpha >}\zeta ^{<\beta >}\nabla
_{+}X^{<\nu >}\nabla _{=}X^{<\rho >}\times
$$
$$
(\frac 13{\cal D}_{<\lambda >}H_{<\tau ><\mu >[<\rho >}R_{<\nu
><\alpha >\
\cdot \ <\beta >}^{\qquad \quad <\mu >}-\frac 1{12}{\cal D}_{<\alpha >}{\cal %
D}_{<\beta >}{\cal D}_{<\gamma >}H_{<\tau ><\nu ><\rho >})+
$$
$$
\frac 14\zeta ^{<\tau >}\zeta ^{<\lambda >}\zeta ^{<\gamma >}({\cal D}%
_{+}\zeta ^{<\mu >}\nabla _{=}X^{<\nu >}+
$$
$$
{\cal D}_{=}\zeta ^{<\nu >}\nabla _{+}X^{<\mu >}){\cal D}_{<\gamma >}{\cal D}%
_{<\lambda >}H_{<\tau ><\mu ><\nu >}\},
$$
$$
I_5^{(2)}=-\frac i{4\pi \alpha ^{\prime }}\int d^3z^{-}E\{\frac
12\zeta
^{<\tau >}\zeta ^{<\lambda >}\zeta ^{<\gamma >}{\cal D}_{+}\zeta ^{<\mu >}%
{\cal D}_{=}\zeta ^{<\nu >}\times
$$
$$
[\frac 15{\cal D}_{<\gamma >}{\cal D}_{<\lambda >}H_{<\tau ><\mu
><\nu
>}+\frac 2{15}R_{~\cdot <\gamma ><\lambda >[<\mu >}^{<\rho >}H_{<\nu >]<\tau
><\rho >}\}+...,
$$
$$
I_6^{(2)}=-\frac i{4\pi \alpha ^{\prime }}\int d^3z^{-}E\zeta
^{<\tau
>}\zeta ^{<\lambda >}\zeta ^{<\alpha >}\zeta ^{<\beta >}{\cal D}_{+}\zeta
^{<\rho >}{\cal D}_{=}\zeta ^{<\nu >}\times
$$
$$
[\frac 19{\cal D}_{<\lambda >}H_{<\tau ><\mu >[<\rho >}R_{<\nu
>]<\alpha >\ \cdot \ <\beta >}^{\qquad \quad <\mu >}+\frac
1{18}{\cal D}_{<\lambda
>}R_{~\cdot <\gamma ><\rho >[<\delta >}^{<\mu >}H_{<\nu >]<\tau ><\mu >}+
$$
$$
\frac 1{18}{\cal D}_{<\alpha >}{\cal D}_{<\beta >}{\cal
D}_{<\lambda
>}H_{<\tau ><\rho ><\nu >}]+...,
$$
where by dots are denoted terms not being important for
calculation of anomalies and operations of symmetrization ( ) and
antisymmetrization are taken without coefficients.

The third term in (4.13)
$$
I^{(3)}=I_0^{(3)}+I_1^{(3)}+I_2^{(3)}+...+I_s^{(3)}+...
$$
is of Fradkin--Tseitlin dilaton type with coefficients%
\index{Dilaton}
$$
I_0^{(3)}=-\frac 1{4\pi \alpha ^{\prime }}\int d^3z^{-}E\alpha
^{\prime }\Sigma ^{+}\Phi ,
$$
$$
I_1^{(3)}=-\frac 1{4\pi \alpha ^{\prime }}\int d^3z^{-}E\alpha
^{\prime }\Sigma ^{+}\zeta ^{<\alpha >}\delta _{<\alpha >}\Phi ,
$$
$$
I_2^{(3)}=-\frac 1{4\pi \alpha ^{\prime }}\int d^3z^{-}E\alpha
^{\prime }\Sigma ^{+}\frac 1{2!}\zeta ^{<\beta >}\zeta ^{<\alpha
>}{\cal D}_{<\alpha
>}{\cal D}_{<\beta >}\Phi ,...,
$$
$$
I_s^{(3)}=-\frac 1{4\pi \alpha ^{\prime }}\int d^3z^{-}E\alpha
^{\prime
}\Sigma ^{+}\frac 1{s!}\zeta ^{<\alpha _1>}...\zeta ^{<\alpha _s>}{\cal D}%
_{<\alpha _1>}...{\cal D}_{<\alpha _s>}\Phi .
$$

The forth term in (4.13)%
$$
I^{(4)}=-\frac 1{4\pi \alpha ^{\prime }}\int d^3z^{-}E~\Psi _{(-)}{}^{|I|}%
{\cal D}_{+}\Psi _{(-)}{}^{|I|}
$$
has the next background--quantum decomposition
$$
I^{(4)}[\Psi +\Delta (\chi ),X+\pi (\zeta
)]=I_0^{(4)}+I_1^{(4)}+I_2^{(4)}+...,
$$
with coefficients%
$$
I_1^{(4)}=-\frac 1{4\pi \alpha ^{\prime }}\int d^3z^{-}E~\{\chi
_{(-)}{}^{|I|}{\cal D}_{+}\chi _{(-)}{}^{|I|}+\Psi _{(-)}{}^{|I|}{\cal D}%
_{+}\Psi _{(-)}{}^{|I|}+\eqno(4.15)
$$
$$
\Psi _{(-)}{}^{|I|}F_{|I||J|<\mu ><\nu >}\zeta ^{<\mu >}\nabla
_{+}X^{<\nu
>}\Psi _{(-)}{}^{|J|}\},
$$
$$
I_2^{(4)}=-\frac 1{4\pi \alpha ^{\prime }}\int d^3z^{-}E~\{\chi
_{(-)}{}^{|I|}{\cal D}_{+}\chi _{(-)}{}^{|I|}+
$$
$$
2\chi _{(-)}{}^{|I|}F_{|I||J|<\mu ><\nu >}\zeta ^{<\mu >}\nabla
_{+}X^{<\nu
>}\Psi _{(-)}{}^{|J|}-
$$
$$
\frac 12\zeta ^{<\nu >}\zeta ^{<\mu >}\Psi _{(-)}{}^{|I|}\Psi _{(-)}{}^{|J|}%
{\cal D}_{<\lambda >}F_{|I||J|<\mu ><\nu >}\nabla _{+}X^{<\nu >}-
$$
$$
\frac 12\Psi _{(-)}{}^{|I|}\Psi _{(-)}{}^{|J|}F_{|I||J|<\mu ><\nu
>}\zeta ^{<\mu >}{\cal D}_{+}\zeta ^{<\nu >}\},
$$
$$
I_3^{(4)}=-\frac 1{4\pi \alpha ^{\prime }}\int d^3z^{-}E~\{\chi
_{(-)}{}^{|I|}F_{|I||J|<\mu ><\nu >}\zeta ^{<\mu >}\nabla
_{+}X^{<\nu >}\chi _{(-)}{}^{|J|}+
$$
$$
\zeta ^{<\lambda >}\zeta ^{<\mu >}\chi _{(-)}{}^{|I|}{\cal
D}_{<\lambda
>}F_{|I||J|<\mu ><\nu >}\nabla _{+}X^{<\nu >}\Psi _{(-)}{}^{|J|}+
$$
$$
\chi _{(-)}{}^{|I|}F_{|I||J|<\mu ><\nu >}\zeta ^{<\mu >}{\cal
D}_{+}\zeta ^{<\nu >}\Psi _{(-)}{}^{|J|}-
$$
$$
\frac 13\zeta ^{<\lambda >}\zeta ^{<\nu >}\Psi _{(-)}{}^{|I|}{\cal D}%
_{<\lambda >}F_{|I||J|<\mu ><\nu >}\Psi _{(-)}{}^{|J|}{\cal
D}_{+}\zeta ^{<\mu >}-
$$
$$
\frac 16\zeta ^{<\lambda >}\zeta ^{<\tau >}\zeta ^{<\mu >}\Psi
_{(-)}{}^{|I|}\Psi _{(-)}{}^{|J|}\nabla _{+}X^{<\nu >}\times
$$
$$
({\cal D}_{<\tau >}{\cal D}_{<\lambda >}F_{|I||J|<\mu ><\nu
>}-R_{<\lambda
>]<\nu >\ \cdot \ <\tau >}^{\qquad \quad <\gamma >}F_{|I||J|<\mu ><\nu >}),
$$
$$
I_4^{(4)}=-\frac 1{4\pi \alpha ^{\prime }}\int d^3z^{-}E~\{\frac
12\chi
_{(-)}{}^{|I|}F_{|I||J|<\mu ><\nu >}\chi _{(-)}{}^{|J|}\zeta ^{<\nu >}{\cal D%
}_{+}\zeta ^{<\mu >}-
$$
$$
\frac 12\zeta ^{<\lambda >}\zeta ^{<\mu >}\chi _{(-)}{}^{|I|}({\cal D}%
_{<\lambda >}F_{|I||J|<\mu ><\nu >})\chi _{(-)}{}^{|J|}\nabla
_{+}X^{<\nu
>}+
$$
$$
\frac 23\zeta ^{<\lambda >}\zeta ^{<\mu >}\chi _{(-)}{}^{|I|}({\cal D}%
_{<\lambda >}F_{|I||J|<\mu ><\nu >})\Psi _{(-)}{}^{|J|}{\cal
D}_{+}\zeta ^{<\nu >}+$$ $$\frac 13\zeta ^{<\gamma >}\zeta
^{<\lambda >}\zeta ^{<\mu >}\chi _{(-)}{}^{|I|}\times
$$
$$
({\cal D}_{<\gamma >}{\cal D}_{<\lambda >}F_{|I||J|<\mu ><\nu
>}+R_{<\gamma
><\nu >\ \cdot \ <\lambda >}^{\qquad \quad <\rho >}F_{|I||J|<\mu ><\rho
>})\nabla _{+}X^{<\nu >}\Psi _{(-)}{}^{|J|}-
$$
$$
\frac 1{24}\zeta ^{<\lambda >}\zeta ^{<\mu >}\zeta ^{<\gamma
>}\Psi _{(-)}{}^{|I|}\Psi _{(-)}{}^{|J|}(3{\cal D}_{<\mu >}{\cal
D}_{<\lambda
>}F_{|I||J|<\gamma ><\nu >}+
$$
$$
R_{<\lambda ><\nu >\ \cdot \ <\gamma >}^{\qquad \quad <\rho
>}F_{|I||J|<\tau
><\rho >}){\cal D}_{+}\zeta ^{<\nu >}-\frac 1{24}\zeta ^{<\tau >}\zeta
^{<\lambda >}\zeta ^{<\mu >}\zeta ^{<\gamma >}\Psi
_{(-)}{}^{|I|}\Psi _{(-)}{}^{|J|}\times
$$
$$
[{\cal D}_{<\gamma >}{\cal D}_{<\tau >}{\cal D}_{<\lambda
>}F_{|I||J|<\mu
><\nu >}+({\cal D}_{<\gamma >}R_{<\lambda ><\nu >\ \cdot \ <\mu >}^{\qquad
\quad <\rho >})F_{|I||J|<\tau ><\rho >}+
$$
$$
3({\cal D}_{<\gamma >}F_{|I||J|<\tau ><\rho >})R_{<\lambda ><\nu
>\ \cdot \ <\mu >}^{\qquad \quad <\rho >}]\nabla _{+}X^{<\nu >}\},
$$
$$
I_5^{(4)}=-\frac 1{4\pi \alpha ^{\prime }}\int d^3z^{-}E~\{-\frac
13\zeta ^{<\mu >}\zeta ^{<\lambda >}\chi _{(-)}{}^{|I|}
$$
$$
({\cal D}_{<\lambda >}F_{|I||J|<\mu ><\nu >})\chi _{(-)}{}^{|J|}{\cal D}%
_{+}\zeta ^{<\mu >}+...\},
$$
$$
I_6^{(4)}=-\frac 1{4\pi \alpha ^{\prime }}\int d^3z^{-}E~\{-\frac
1{24}\zeta ^{<\gamma >}\zeta ^{<\mu >}\zeta ^{<\lambda >}\chi
_{(-)}{}^{|I|}\chi _{(-)}{}^{|J|}\times
$$
$$
(3{\cal D}_{<\gamma >}{\cal D}_{<\lambda >}F_{|I||J|<\mu ><\nu
>}+F_{|I||J|<\gamma ><\rho >}R_{<\lambda ><\nu >\ \cdot \ <\mu >}^{\qquad
\quad <\rho >}){\cal D}_{+}\zeta ^{<\nu >}\}+...
$$

The kinetic terms for quantum fields $\zeta ^{<\mu >}$ and $\chi
_{(-)}{}^{|J|}$ in the decompositions (2.14) and (2.15) define the
propagators $\left( 2\pi \alpha ^{\prime }=1\right) $%
$$
<\zeta ^{<\mu >}(\ddot u)\zeta ^{<\nu >}(\ddot u^{\prime })>=g^{<\mu ><\nu >}%
\frac{D_{+}}{\Box }\delta _{(-)}^3(\ddot u,\ddot u^{\prime })=
$$
$$
g^{<\mu ><\nu >}\frac 1{(2\pi )^2}\int d^{d_2}p\frac
1{(-p^2)}D_{+}[e^{ip(z-z^{\prime })}\delta _{(-)}(\theta -\theta
^{\prime })],
$$
$$
<\chi _{(-)}{}^{|I|}(\ddot u)\chi _{(-)}{}^{|J|}(\ddot u^{\prime
})>=i\delta ^{|I||J|}\frac{\partial _{=}D_{+}}{\Box }\delta
_{(-)}^3(\ddot u,\ddot u^{\prime })=
$$
$$
\frac{\delta ^{|I||J|}}{(2\pi )^2}\int d^{d_2}p\frac{p_{=}}{p^2}%
D_{+}[e^{ip(z-z^{\prime })}\delta _{(-)}^3(\ddot u,\ddot
u^{\prime })].
$$
Finally, we remark that background--quantum decompositions of the
action (4.11) for heterotic string define the Feynman rules
(vertixes and \index{Feynman rules!for heterotic string}
\index{Heterotic string!Feynman rules} propagators) for the
corresponding generalization of the two--dimensional sigma model
which are basic for a perturbation quantum formalism in higher
order anisotropic spaces.

\section{Green--Schwarz DVS--Action }

The Green--Scwarz covariant action ( GS--action ) for
superstrings can be \index{Action!Green--Scwarz}
\index{Green--Scwarz!covariant action} \index{GS--action}
considered as a two dimensional $\sigma $--model with
Wess--Zumino--Wit\-ten \index{Wess--Zumino--Wit\-ten term} term
and flat, dimension $d=10,$ s--space as the tangent space
 [91,109,110,158]. The GS--action was generalized for the curved
background {\sf N=1}, $d=10$ of the superspace under the
condition that
motion equations hold 
 [293] and under similar conditions for {\sf N=2}%
, $d=10$ supergravity
 [96].

The GS--action in dimensions $d=3,4,6,10$ can be represented as
$$
I=\frac 12\int d^2z\sqrt{-\gamma }\gamma ^{\ddot e\ddot \imath
}\partial
_{\ddot e}u^{<\alpha >}\partial _{\ddot \imath }u^{<\beta >}(l_{<\alpha >}^{<%
\underline{\alpha }>}l_{<\beta >}^{<\underline{\beta }>})\widehat{\eta }_{<%
\underline{\alpha }><\underline{\beta }>}+\eqno(4.16)
$$
$$
\frac 12\int d^2z\varepsilon ^{\ddot e\ddot \imath }\partial
_{\ddot e}u^{<\alpha >}\partial _{\ddot \imath }u^{<\beta
>}B_{<\alpha ><\beta >}
$$
by using of the flat  vielbein $l_{<\alpha >}^{<\underline{\alpha
}>}$ and 2--form
$$
B=\frac 12\delta u^{<\alpha >}\Lambda \delta u^{<\beta
>}B_{<\alpha ><\beta
>}
$$
in the flat $d=10$ s--space with coordinates $u^{<\alpha >}.$ An
important role in formulation of the GS--action plays the fact
that 3--form $H=dB$ is closed, $dH=0.$ Because $H$ is
s--invariant the 2--form $B$ changes on
complete derivation under s--transforms%
$$
\delta ^{\star }H=0,\delta ^{\star }dB=d\delta ^{\star
}B=0,\delta ^{\star }B=d\Lambda ,
$$
where $\delta ^{\star }$ is dual to $d,$ which ensures the
s--invariance of the GS--action.

We generalize the action (4.16) for higher order anisotropic
s--spaces by changing the flat vielbein $l_{<\alpha
>}^{<\underline{\alpha }>}$ into the locally anisotropic,
$\widehat{E}_{<\alpha >}^{<\underline{\alpha }>},$ with a
possible dependence of the Lagrangian on scalar fields (see
 [202] for locally isotropic spaces),%
$$
I=\frac 12\int d^2z[\sqrt{-\gamma }\gamma ^{\ddot e\ddot \imath
}\partial
_{\ddot e}u^{<\alpha >}\partial _{\ddot \imath }u^{<\beta >}(E_{<\alpha >}^{<%
\underline{\alpha }>}E_{<\beta >}^{<\underline{\beta }>})\widehat{\eta }_{<%
\underline{\alpha }><\underline{\beta }>}+\eqno(4.17)
$$
$$
\varepsilon ^{\ddot e\ddot \imath }\partial _{\ddot e}u^{<\alpha
>}\partial _{\ddot \imath }u^{<\beta >}E_{<\alpha
>}^{<\underline{\alpha }>}E_{<\beta
>}^{<\underline{\beta }>}B_{<\underline{\alpha }><\underline{\beta }>}+
$$
$$
VV_{\underline{\ddot e}}^{\ddot e}\overline{\Psi }^{|\underline{I}|}\gamma ^{%
\underline{\ddot e}}(\partial _{\ddot e}\delta _{|\underline{I}||\underline{J%
}|}+E_{\ddot e}^{<\alpha
>}A_{|\underline{I}||\underline{J}|<\alpha >})\Psi
^{|\underline{J}|}],
$$
where $P$ is a scalar function, $\gamma _{\ddot e\ddot \imath }=V_{\ddot e}^{%
\underline{\ddot e}}V_{\ddot \imath }^{\underline{i}}\gamma _{\underline{%
\ddot e}\underline{\ddot \imath }},~V=\det (V_{\ddot e}^{\underline{\ddot e}%
}),u^{<\alpha >}$ are coordinates of the higher order anisotropic
s--space and $\Psi ^{|\underline{J}|}$ are two dimensional
(heterotic) MW--fermions in the fundamental representation of the
interior symmetry group $Gr$.

As background s--fields we shall consider
$$
E^{<\underline{\alpha }>}=d\widehat{u}^{<\beta >}E_{<\beta >}^{<\underline{%
\alpha }>}(z),~B=\frac 12E^{<\underline{\alpha }>}E^{<\underline{\beta }%
>}B_{<\underline{\alpha }><\underline{\beta }>}(z),
$$
$$
A_{|\underline{I}||\underline{J}|}=A_{|\underline{I}||\underline{J}|<%
\underline{\alpha }>}E^{<\underline{\alpha }>},
$$
where%
$$
E_{\ddot a}^{<\underline{\alpha }>}\equiv \partial _{\ddot
a}u^{<\alpha
>}E_{<\alpha >}^{<\underline{\alpha }>}=(\widehat{E}_{\ddot a}^{<\underline{%
\alpha }>},E_{\ddot a}^{<\underline{\alpha }>}=E_{\ddot
a}^{\underline{\ddot a}}).
$$
S--fields $A_{|\underline{I}||\underline{J}|}$ belong to the
adjoint representation of the interior symmetry group $Gr$.

The action (4.17) is invariant under transforms\\ $\left( \triangle E^{<%
\underline{\alpha }>}\equiv \triangle u^{<\alpha >}E_{<\alpha >}^{<%
\underline{\alpha }>}\right) :$%
$$
\triangle \widehat{E}^{<\underline{\alpha }>}=0,\triangle E^{\ddot
e}=2(\Gamma _{<\underline{\alpha }>})^{\ddot e\ddot
o}\widehat{E}_{\ddot
e}^{<\underline{\alpha }>}V_{\underline{\ddot e}}^{\ddot e}k_{\ddot o}^{%
\underline{\ddot o}},\eqno(4.18)
$$
$$
\triangle \Psi ^{|\underline{I}|}=-(\triangle E^{<\underline{\alpha }>})A_{<%
\underline{\alpha }>}^{|\underline{I}||\underline{J}|}\Psi ^{|\underline{J}%
|},\triangle V_{\ddot e}^{\underline{\ddot e}}=-\frac 12(\gamma
_{\ddot
e\ddot \imath }+\varepsilon _{\ddot e\ddot \imath })M^{\underline{\ddot o}%
\ddot \imath }k_{\underline{\ddot o}}^{\underline{\ddot e}},
$$
where $\varepsilon _{\ddot e\ddot \imath }$ is defined as a
two--dimensional tensor, parameter $k_{<\alpha
>}^{\underline{\ddot e}}$ is anti--self--dual as a two--vector,
juggling of indices of $d$ --dimensional Dirac matrices is
realized by using the $d$--dimensional matrix of charge
conjugation and ,
for simplicity, we can consider matrices $(\Gamma _{<\underline{\alpha }%
>})^{\ddot e\ddot o}$ as symmetric; we shall define below the value $%
M^{<\alpha >\ddot \imath }.$

The variation of action (4.17) under transforms (4.18) can be written as%
$$
\triangle I=\int d^2z\frac 12[e^P\triangle (V\gamma ^{\ddot e\ddot \imath })%
\widehat{E}_{\ddot e}^{<\underline{\alpha }>}\widehat{E}_{\ddot \imath }^{<%
\underline{\beta }>}\eta _{<\underline{\alpha }><\underline{\beta }>}+%
\eqno(4.19)
$$
$$
\triangle E^{\ddot o}(e^P(V\gamma ^{\ddot e\ddot \imath
})\widehat{E}_{\ddot
e}^{<\underline{\alpha }>}\widehat{E}_{\ddot \imath }^{<\underline{\beta }%
>}\eta _{<\underline{\alpha }><\underline{\beta }>}D_{\ddot
o}P-2e^P\triangle (V\gamma ^{\ddot e\ddot \imath })E_{\ddot e}^{<\underline{%
\alpha }>}E_{\ddot \imath }^{<\underline{\beta }>}T_{<\underline{\alpha }%
>\ddot o<\underline{\beta }>}+
$$
$$
\varepsilon ^{\ddot e\ddot \imath }E_{\ddot e}^{<\underline{\alpha }%
>}E_{\ddot \imath }^{<\underline{\beta }>}H_{<\underline{\alpha }><%
\underline{\beta }>\ddot o}-V\overline{\Psi
}^{|\underline{I}|}\gamma
^{\ddot e}\Psi ^{|\underline{J}|}E_{\ddot e}^{<\underline{\alpha }>}F_{<%
\underline{\alpha }>\ddot o|\underline{I}||\underline{J}|})+$$ $$\triangle (VV_{%
\underline{\ddot e}}^{\ddot e})\overline{\Psi }^{|\underline{I}|}\gamma ^{%
\underline{\ddot e}}({\cal D}_{\ddot e}\Psi )^{|\underline{J}|},
$$
where ${\cal D}_{\ddot e}$ is the $Gr$--covariant derivation and
the torsion 2--form $T^{<\underline{\alpha }>}$, the strength
3--form $H$ and the supersymmetric Yang--Mills strength 2--form
\index{Yang--Mills strength!supersymmetric}
$F^{|\underline{I}||\underline{J}%
|}$ are respectively defined by relations%
$$
T^{<\underline{\alpha }>}=dE^{<\underline{\alpha }>}+E^{<\underline{\beta }%
>}\Omega _{<\underline{\beta }>}^{<\underline{\alpha }>}=\frac 12E^{<%
\underline{\beta }>}E^{<\underline{\gamma }>}T_{<\underline{\beta }><%
\underline{\gamma }>}^{<\underline{\alpha }>},
$$
$$
H=dB=\frac 16E^{<\underline{\gamma }>}E^{<\underline{\beta }>}E^{<\underline{%
\alpha }>}H_{<\underline{\alpha }><\underline{\beta }><\underline{\gamma }%
>},
$$
$$
F^{|\underline{I}||\underline{J}|}=dA^{|\underline{I}||\underline{J}|}+A^{|%
\underline{I}||\underline{K}|}A^{|\underline{K}||\underline{J}|}=\frac 12E^{<%
\underline{\beta }>}E^{<\underline{\gamma }>}F_{<\underline{\beta }><%
\underline{\gamma }>}^{|\underline{I}||\underline{J}|}.
$$

The variation of action (4.19) under transforms (4.18) vanishes
if and only if there are satisfied the next conditions:

1) 3--form $H$ is closed under condition
$$
(\Gamma ^{<\underline{\alpha }>})_{\ddot e\ddot \imath }(\Gamma _{<%
\underline{\alpha }>})_{\ddot o\ddot u}+(\Gamma ^{<\underline{\alpha }%
>})_{\ddot \imath \ddot o}(\Gamma _{<\underline{\alpha }>})_{\ddot e\ddot
u}+(\Gamma ^{<\underline{\alpha }>})_{\ddot o\ddot e}(\Gamma _{<\underline{%
\alpha }>})_{\ddot \imath \ddot u}=0,
$$
which holds for dimensions $d=3,4,6,10;$

2) there are imposed constraints
$$
\widehat{T}_{\ddot e\ddot \imath }^{<\underline{\alpha }>}=-i(\widehat{%
\Gamma }^{<\underline{\alpha }>})_{\ddot e\ddot \imath },~\widehat{\eta }%
_{<\gamma >(<\alpha >}\widehat{T}_{<\beta >)\ddot a}^{<\gamma >}=\widehat{%
\eta }_{<\alpha ><\beta >}B_{\ddot a},~F_{\ddot e\ddot \imath }^{|\underline{%
I}||\underline{J}|}=0,\eqno(4.20)
$$
$$
F_{<\underline{\alpha }>\ddot \imath }^{|\underline{I}||\underline{J}|}=(%
\widehat{\Gamma }_{<\underline{\alpha }>})_{\ddot e\ddot \imath }w^{\ddot e|%
\underline{I}||\underline{J}|},~H_{\ddot e\ddot \imath \ddot
o}=0,~H_{\ddot
e\ddot \imath <\underline{\alpha }>}=-ie^P(\widehat{\Gamma }_{<\underline{%
\alpha }>})_{\ddot e\ddot \imath },~
$$
$$
\widehat{H}_{\ddot u<\alpha ><\beta >}=2e^P(\Gamma _{<\underline{\alpha }%
>})_{\ddot o}^{\ddot e}(\Gamma _{<\underline{\beta }>})_{\ddot u}^{\ddot
o}H_{\ddot e};
$$

3) The coefficient $M^{\underline{\ddot o}\ddot \imath }$ from
(4.18) is
taken in the form%
$$
M^{\underline{\ddot o}\ddot \imath }=4iE^{\underline{\ddot o}\ddot \imath }-4%
\widehat{E}_{<\underline{\alpha }>}^{\ddot \imath }(\widehat{\Gamma }^{<%
\underline{\alpha }>})^{\underline{\ddot o}\underline{\ddot u}}H_{\underline{%
\ddot u}}-\Psi ^{|\underline{I}|}\gamma ^{\ddot \imath }\Psi ^{|\underline{J}%
|}w_{|\underline{I}||\underline{J}|}^{\underline{\ddot o}}e^{-P},
$$
where
$$
D_{\underline{\ddot u}}P+2H_{\underline{\ddot u}}-2B_{\underline{\ddot u}%
}=0;
$$

4) The last term in (4.19) vanishes because of conditions of
chirality, \index{Conditions of chirality}
$$
\Psi ^{|\underline{I}|}=-\gamma _5\Psi ^{|\underline{I}|}.
$$

In the locally isotropic s--gravity it is known
 [86,175,21]
that s--field equations of type (4.20) are compatible with Bianchi
identities and can be interpreted as standard constraints defining
supergravity in the superspace. Considering locally adapted to
N--connections geometric objects end equations (4.20) we obtain a
variant of higher order anisotropic supergravity (see sections
2.3 and 2.6
 in this monograph
 [260,265,266,267] for details on locally anisotropic
supergravity) which for dimensions $d=n+m=10$ contain,
distinguished by the N--connection structure, motion equations of
{\sf N=1} of higher order anisotropic supergravity and
super--Yang--Mills matter.

The above presented constructions can be generalized in order to
obtain a variant of higher order anisotropic {\sf N=2, }$d=10$
supergravity from so--called IIB--superstrings
 [96] (which, in our case, will be
modified to be locally anisotropic). To formulate the model we
use a locally adapted s--vielbein 1--form $E^{<\underline{\alpha
}>}=\delta u^{<\alpha
>}E_{<\alpha >}^{<\underline{\alpha }>},$ a $SO(1,9)\otimes U(1)$ connection
1--form $\Omega _{<\underline{\alpha }>}^{<\underline{\beta }>},$
a 2--form of complex potential $A$ and one real 4--form $B.$
Strengths are defined in
a standard manner:%
$$
T^{<\underline{\alpha }>}=DE^{<\underline{\alpha }>}=\delta E^{<\underline{%
\alpha }>}+E^{<\underline{\beta }>}\Omega _{<\underline{\beta }>}^{<%
\underline{\alpha }>},\eqno(4.21)
$$
$$
R_{<\underline{\alpha }>}^{<\underline{\beta }>}=\delta \Omega _{<\underline{%
\alpha }>}^{<\underline{\beta }>}+\Omega _{<\underline{\alpha }>}^{<%
\underline{\gamma }>}\Omega _{<\underline{\gamma
}>}^{<\underline{\beta }>},
$$
$$
F=\delta A,~G=\delta B+A\overline{F}-\overline{A}F.
$$

On the mass shell (on locally anisotropic spaces we shall consider
distinguished metrics) ds--tensors (4.21) are expressed in terms
of one
scalar s--field $V\in SU(1,1):$%
$$
V=\left(
\begin{array}{cc}
q & s \\
\overline{u} & \overline{v}
\end{array}
\right) ,q\overline{q}-s\overline{s}=1.
$$
Excluding a scalar by using the local U(1)--invariance we can use
the first components of complex s--fields $\left( q,s\right) $ as
physical scalar fields of the theory.

The constraints defining IIB supergravity in $d=n+m=10$ higher
order \index{Constraints!defining IIB supergravity} anisotropic
s--space contain equations (on every anisotropic ''shell'', in
locally adapted frames, they generalize constraints of IIB
supergavity
 [113]; see sections 2.2 and 2.3 for denotations on higher order anisotropic
s--spaces):%
$$
T_{\underleftarrow{b_p}\underleftarrow{c_p}}^{a_p}=T_{\underrightarrow{b_p}%
\underrightarrow{c_p}}^{a_p}=0,T_{\underleftarrow{b_p}\underrightarrow{c_p}%
}^{a_p}=-i\sigma _{\underleftarrow{b_p}\underrightarrow{c_p}}^{a_p},T_{%
\underleftarrow{b_p}c_p}^{a_p}=T_{\underrightarrow{b_p}c_p}^{a_p}=0,%
\eqno(4.22)
$$
$$
F_{\underleftarrow{a_p}\underleftarrow{b_p}\underleftarrow{c_p}}=F_{%
\underleftarrow{a_p}\underleftarrow{b_p}\underrightarrow{c_p}}=F_{%
\underleftarrow{a_p}\underrightarrow{b_p}\underrightarrow{c_p}}=F_{%
\underrightarrow{a_p}\underrightarrow{b_p}\underrightarrow{c_p}}=F_{a_p%
\underleftarrow{b_p}\underrightarrow{c_p}}=0,
$$
$$
F_{a_p\underleftarrow{b_p}\underleftarrow{c_p}}=-iq(\sigma _{a_p})_{%
\underleftarrow{b_p}\underleftarrow{c_p}},~F_{a_pb_p\underrightarrow{c_p}%
}=-q(\sigma _{a_pb_p})_{\underrightarrow{c_p}}^{\underrightarrow{d_p}%
}\Lambda _{\underrightarrow{d_p}},
$$
$$
F_{a_p\underrightarrow{b_p}\underrightarrow{c_p}}=-is(\sigma _{a_p})_{%
\underrightarrow{b_p}\underrightarrow{c_p}},~F_{a_pb_p\underleftarrow{c_p}%
}=s(\sigma _{a_pb_p})_{\underleftarrow{c_p}}^{\underleftarrow{d_p}}\overline{%
\Lambda }_{\underleftarrow{d_p}},
$$
$$
H_{a_p\underleftarrow{b_p}\underleftarrow{c_p}}=-i(q-\overline{s})(\sigma
_{a_p})_{\underleftarrow{b_p}\underleftarrow{c_p}},~H_{a_p\underrightarrow{%
b_p}\underrightarrow{c_p}}=-i(q-\overline{s})(\sigma _{a_p})_{%
\underrightarrow{b_p}\underrightarrow{c_p}},
$$
$$
H_{a_pb_p\underrightarrow{c_p}}=-(q-\overline{s})(\sigma _{a_pb_p})_{%
\underrightarrow{c_p}}^{\underrightarrow{d_p}}\Lambda _{\underrightarrow{d_p}%
},~H_{a_p\underrightarrow{b_p}\underrightarrow{c_p}}=-i(\overline{q}%
-s)(\sigma _{a_pb_p})_{\underleftarrow{c_p}}^{\underleftarrow{d_p}}\overline{%
\Lambda }_{\underleftarrow{d_p}},
$$
where
$$
(\sigma _{a_p}\sigma _{b_p})_{\underrightarrow{c_p}}^{\underrightarrow{d_p}%
}=(\sigma
_{a_pb_p})_{\underrightarrow{c_p}}^{\underrightarrow{d_p}}+\eta
_{a_pb_p}\delta
_{\underrightarrow{c_p}}^{\underrightarrow{d_p}},\eqno(4.23)
$$
the same formula holds for ''$\underleftarrow{}"$--underlined
spinors,
spinor $\Lambda _{\underrightarrow{d_p}}$ will be used below for fixing of $%
U(1)$ gauge and 3--form $H\equiv F+\overline{F}=\delta
\widetilde{B}$ is real and closed (this condition is crucial in
the construction of the GS--action on the background of
IIB--supergravity, with respect to usual isotropic string model
see
 [96]).

The action (4.16) can be generalized for {\sf N=2} higher order
anisotropic
s--spaces in this manner:%
$$
I_S=\int d^2z\{\sqrt{-\gamma }\gamma ^{\ddot \imath \ddot u}P(q,s)\widehat{E}%
_{\ddot \imath }^{<\underline{\alpha }>}\widehat{E}_{\ddot u}^{<\underline{%
\beta }>}\widehat{\eta }_{<\underline{\alpha }><\underline{\beta
}>}+\frac 12\varepsilon ^{\ddot \imath \ddot u}E_{\ddot \imath
}^{<\alpha >}E_{\ddot u}^{<\beta >}\widetilde{B}_{<\alpha ><\beta
>},\eqno(4.24)
$$
where $P(q,s)$ is a function of scalar fields $q$ and $s,$ and
$$
E_{\ddot \imath }^{<\underline{\alpha }>}\equiv \partial _{\ddot
\imath }u^{<\alpha >}E_{<\alpha >}^{<\underline{\alpha
}>}=(\widehat{E}_{\ddot
\imath }^{<\underline{\alpha }>},...,E_{\ddot \imath }^{\underleftarrow{a_p}%
},...,\overline{E}_{\ddot \imath }^{\underrightarrow{b_p}},...).
$$

The variation of the Lagrangian in (4.23) under respective
k--transforms of type (4.18) can be written as
$$
\triangle L=\sqrt{-\gamma }\gamma ^{\ddot \imath \ddot
u}PE_{\ddot \imath }^{<\gamma >}\triangle E^{<\beta >}T_{<\beta
><\gamma >}^{<\underline{\delta
}>}E_{\ddot u}^{<\underline{\beta }>}\widehat{\eta }_{<\underline{\delta }><%
\underline{\beta }>}+\eqno(4.25)
$$
$$
\frac 12\varepsilon ^{\ddot \imath \ddot u}E_{\ddot \imath
}^{<\gamma
>}E_{\ddot u}^{<\beta >}\triangle E^{<\tau >}H_{<\tau ><\beta ><\gamma >}+
$$
$$
\frac 12[\triangle (\sqrt{-\gamma }\gamma ^{\ddot \imath \ddot u}P)+\sqrt{%
-\gamma }\gamma ^{\ddot \imath \ddot u}\triangle
P]\widehat{E}_{\ddot \imath
}^{<\underline{\alpha }>}\widehat{E}_{\ddot u}^{<\underline{\beta }>}%
\widehat{\eta }_{<\underline{\alpha }><\underline{\beta }>}.
$$

Taking into account constraints (4.22) we can express variation
(4.25) as
$$
\triangle L=(\{-i(\varepsilon ^{\ddot \imath \ddot u}\gamma
^{\ddot e\ddot a}(q-\overline{s})+\gamma ^{\ddot \imath \ddot
u}\varepsilon ^{\ddot e\ddot a}P)\widehat{E}_{\ddot e}^{<\gamma
>}\widehat{E}_{\ddot \imath }^{<\beta
>}(\sigma _{<\beta >}\sigma _{<\gamma >})_{\underleftarrow{\tau }}^{%
\underleftarrow{\delta }}E_{\ddot u}^{\underleftarrow{\tau }}k_{\ddot a%
\underleftarrow{\delta }}-
$$
$$
i[(-\gamma )^{-1/2}\varepsilon ^{\ddot \imath \ddot u}\varepsilon
^{\ddot e\ddot a}(\overline{q}-s)+\sqrt{-\gamma }\gamma ^{\ddot
\imath \ddot
u}\gamma ^{\ddot e\ddot a}P]\widehat{E}_{\ddot e}^{<\gamma >}\widehat{E}%
_{\ddot \imath }^{<\beta >}(\sigma _{<\beta >}\sigma _{<\gamma >})_{%
\underleftarrow{\tau }}^{\underleftarrow{\delta }}\overline{E}_{\ddot u}^{%
\underleftarrow{\tau }}k_{\ddot a\underleftarrow{\delta }}+
$$
$$
\widehat{E}_{\ddot \imath }^{<\gamma >}(\widehat{E}_{\ddot u}^{<\beta >}%
\widehat{E}_{\ddot e}^{<\alpha >}\widehat{\eta }_{<\beta ><\alpha
>})(\sigma _{<\gamma >})^{\underleftarrow{\alpha
}\underleftarrow{\beta }}[\varepsilon
^{\ddot \imath \ddot u}\gamma ^{\ddot e\ddot a}(\overline{q}-s)\overline{%
\Lambda }_{\underleftarrow{\alpha }}+
$$
$$
(-\gamma )^{-1/2}\varepsilon ^{\ddot \imath \ddot u}\varepsilon
^{\ddot
e\ddot a}(q-\overline{s})\Lambda _{\underleftarrow{\alpha }}]k_{\ddot a%
\underleftarrow{\beta }}\}+h.c.)+
$$
$$
\frac 12\triangle (\sqrt{-\gamma }\gamma ^{\ddot \imath \ddot u})\widehat{E}%
_{\ddot \imath }^{<\gamma >}\widehat{E}_{\ddot u}^{<\beta >}\widehat{\eta }%
_{<\gamma ><\beta >}P+\frac 12\sqrt{-\gamma }\gamma ^{\ddot \imath \ddot u}%
\widehat{E}_{\ddot \imath }^{<\gamma >}\widehat{E}_{\ddot u}^{<\beta >}%
\widehat{\eta }_{<\gamma ><\beta >}\triangle P,
$$
where $h.c.$ denotes Hermitian conjugation.

Using relation (4.23) and fixing the $U(1)$--gauge as to have
$$
P=q-\overline{s}=\overline{q}-s~\mbox{ and }~\triangle P=(q-\overline{s}%
)(\triangle E^{\underleftarrow{\alpha }}\Lambda _{\underleftarrow{\alpha }%
}-\triangle \overline{E}^{\underleftarrow{\alpha }}\overline{\Lambda }_{%
\underleftarrow{\alpha }})
$$
we can obtain zero values of the coefficients before $(\sigma _{a_pb_p})$%
--terms. The rest of terms in $\triangle L$ vanish for a
corresponding fixing of the variation $\triangle (\sqrt{-\gamma
}\gamma ^{\ddot \imath \ddot u}).$

So, in this section we have constructed a model of higher order
an\-isot\-rop\-ic
IIB--superstring on the background of IIB supergravity with broken chiral $%
U(1)$--subgroup of the supersymmetry $SU(1,1)$--group of automorphisms of $%
N=2,d=n+m=10$ supergravity. We omit in this Chapter calculus for
supersymmetric $\beta$--functions; we shall present similar
details in \index{B!$\beta$--functions} sections 9.1 and 9.2
 for higher order anisotropic nonsupersymmetric $\sigma$--models.

\section{Fermi Strings in HA--Spaces}

There are some types of Fermi strings in dependence of the number {\sf %
N=0, 1, 2, 4} of supersymmetry generators (see, for instance,
 [115,139,140] for reviews and basic references on this classification for
\index{Fermi strings} locally isotropic strings). The aim of this
section is to present basic results on Fermi and heterotic
strings on higher order anisotropic backgrounds: the construction
of actions and calculation of superconformal anomalies.

We note that there are two possible interpretations of models
considered in this Chapter. On one hand they can be considered as
locally anisotropic supersymmetric two dimensional supersymmetric
nonlinear sigma models connected with supergravity. Under
quantization of such type theories the superconformal invariance
is broken; the two point functions of graviton and gravitino,
computed from the quantum effective action can became nontrivial
in result (this conclusion was made [1]
 for locally
isotropic sigma models and, in general, holds good for locally
anisotropic generalizations). On the other hand our models can be
interpreted as Fermi strings on higher order anisotropic
background. We shall follow the second treatment.

The effective ''off--shell'' action $\Gamma $ for a (infinite)
set of locally anisotropic fields is introduced (in a manner
similar to
 [80]) as%
$$
\Gamma [G,H,...]=\sum_\chi e^{\widetilde{\sigma }\chi }\int
[D\gamma _{\ddot e\ddot \imath }^{(e)}][Du^{<\alpha >}]\exp (-I),
$$
$$
I=\frac 1{2\pi \alpha ^{\prime }}\int d^2z\{\frac 12\sqrt{\gamma ^{(e)}}%
\gamma _{(e)}^{\ddot a\ddot \imath }\partial _{\ddot
a}\widehat{u}^{<\alpha
>}\partial _{\ddot \imath }\widehat{u}^{<\beta >}G_{<\alpha ><\beta >}(u)+
$$
$$
\varepsilon ^{\ddot a\ddot \imath }\partial _{\ddot
a}\widehat{u}^{<\alpha
>}\partial _{\ddot \imath }\widehat{u}^{<\beta >}H_{<\alpha ><\beta
>}(u)+...\},
$$
where dots are used instead of possible sources (with higher order
derivations), compatible with the reparametrization invariance,
of another types of perturbations and $\gamma _{\ddot e\ddot
\imath }^{(e)}$ is the Euclid two dimensional metric. In (4.26)
we consider in explicit form the components of locally
anisotropic graviton and antisymmetric d--tensor and, for
simplicity, omit the dilaton field and topological considerations.
\begin{figure}[htbp]
\begin{picture}(380,180) \setlength{\unitlength}{1pt}
\thicklines

\put(70,35){\circle{30}} \put(85,35){\line(1,0){30}}
\put(25,35){\line(1,0){30}} \put(85,2){\makebox(30,20){$a)$}}
\put(95,40){\makebox(30,20){$\sigma$}}
\put(10,40){\makebox(30,20){$\sigma$}}

\put(190,35){\circle{30}} \put(145,35){\line(1,0){30}}
\put(222,35){\circle{30}} \put(237,35){\line(1,0){30}}
\put(230,2){\makebox(30,20){$b)$}}
\put(240,40){\makebox(30,20){$\sigma$}}
\put(125,40){\makebox(30,20){$\sigma$}}
\put(200,51){\makebox(35,20){${\alpha}' R$}}

\put(330,35){\circle{30}} \put(285,35){\line(1,0){30}}
\put(345,35){\line(1,0){30}} \put(330,61){\oval(12,20)}
\put(345,2){\makebox(30,20){$c)$}}
\put(360,40){\makebox(30,20){$\sigma$}}
\put(275,40){\makebox(30,20){$\sigma$}}
\put(335,50){\makebox(30,20){${\alpha}' R$}}

\put(70,135){\circle{30}} \put(85,135){\line(1,0){30}}
\put(25,135){\line(1,0){30}} \put(85,102){\makebox(30,20){$d)$}}
\put(95,140){\makebox(30,20){$\sigma$}}
\put(10,140){\makebox(30,20){$\sigma$}} \put(64,135){\oval(20,12)}
\put(27,115){\makebox(30,20){${\alpha}' R$}}

\put(200,135){\circle{30}} \put(155,135){\line(1,0){30}}
\put(215,135){\line(1,0){30}} \put(210,102){\makebox(30,20){$e)$}}
\put(230,115){\makebox(30,20){$\sigma$}}
\put(145,115){\makebox(30,20){$\sigma$}}
\put(200,147){\oval(20,10)[b]}
\put(210,145){\makebox(40,20){$\sqrt{{\alpha}'} S$}}
\put(153,145){\makebox(40,20){$\sqrt{{\alpha}'} S$}}

\put(330,135){\circle{30}} \put(285,135){\line(1,0){30}}
\put(345,135){\line(1,0){30}} \put(330,119){\line(0,1){32}}
\put(345,102){\makebox(30,20){$f)$}}
\put(360,140){\makebox(30,20){$\sigma$}}
\put(275,140){\makebox(30,20){$\sigma$}}
\put(310,150){\makebox(30,20){$\sqrt{{\alpha}'} S$}}
\put(310,98){\makebox(30,20){$\sqrt{{\alpha}'} S$}}

\end{picture}
\caption{\it The  diagrams de\-fin\-ing the con\-for\-m\-al
a\-no\-ma\-ly of a clos\-ed bo\-son string  in a ha--space}
\end{figure}
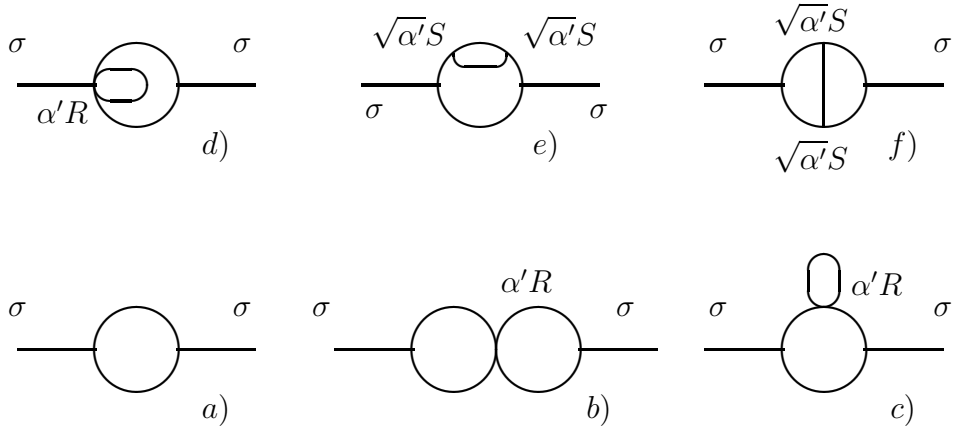

\index{Con\-form\-al anomaly} \index{Boson string!con\-form\-al
an\-o\-ma\-ly} The problem of calculation of $\Gamma $ is split
into two steps: the first is the calculation of the effective
action on an higher order anisotropic with fixed Euler
characteristic $\chi $ then the averaging on all metrics
and topologies. In order to solve the first task we shall compute%
$$
\exp \{-W[G,H,g]\}=\int [D\eta ]\exp \{-I[u+\upsilon (z),g]\}.
$$

The general structure $W$ is constructed from dimension and
symmetry
considerations 
 [193]
$$
W=-\frac{\beta (u)}\varepsilon \int R(z)\sqrt{\gamma
_{(e)}}d^2z+\gamma
(u)\int (R(z)\sqrt{\gamma _{(e)}})_z\Box _{zz^{\prime }}^{-1}(R(z)\sqrt{%
\gamma _{(e)}})_{z^{\prime }}d^2zd^2z^{\prime },\eqno(4.27)
$$
where the dimensional regularization $(\varepsilon =2-d_2$ on
world sheet),\thinspace $\gamma _{(e)}$ is the determinant of the
two dimensional metric, $R=R_{\quad \ddot a\ddot \imath }^{\ddot
a\ddot \imath },$ one holds the relation $\beta (u)=4\gamma (u)$
for dimensionless functions (we can computer them as
perturbations on $\alpha ^{\prime })$ and note that the second
term in (4.27) is the Weyl anomaly which shall be computed by
using normal locally adapted to N--connection coordinates on
higher order anisotropic space.

For a two--dimensional conformal flat two--dimensional, $d_2=2,$ metric $%
\gamma _{\ddot e\ddot \imath }^{(e)}$ we find
$$
\gamma _{\ddot e\ddot \imath }^{(e)}=e^{2\sigma }\delta _{\ddot
e\ddot \imath },~R\left( u\right) =-2e^{-2\sigma }\Box \sigma
,~\gamma
_{(e)}^{\ddot a\ddot \imath }=e^{-2\sigma }\delta ^{\ddot a\ddot \imath },~%
\sqrt{\gamma _{(e)}}=e^{2\sigma }.
$$

From decomposition $\gamma _{\ddot e\ddot \imath }^{(e)}=\delta
_{\ddot e\ddot \imath }+h_{\ddot e\ddot \imath }$ with respect to
first and second order terms on $h$ we have
$$
\sqrt{\gamma ^{(e)}}\gamma _{(e)}^{\ddot a\ddot \imath
}=[1+\varepsilon \sigma (z)+\frac \varepsilon 2(d-4)\sigma
^2(z)]\delta ^{\ddot a\ddot \imath }
$$
when $h_{\ddot e\ddot \imath }=2\sigma h_{\ddot e\ddot \imath
},h\equiv h_{\ddot a}^{\ddot a}=2\sigma d_{(2)}.$ We give similar
formulas for the
frame decomposition of two metric%
$$
e^{1/2}e_{\underline{\ddot e}}^{\ddot e}=(1+\frac 12\varepsilon
\sigma
)\delta _{\underline{\ddot e}}^{\ddot e},e^{-1/2}e_{\ddot e\underline{\ddot e%
}}=(1-\frac 12\varepsilon \sigma )\delta _{\ddot
e\underline{\ddot e}},
$$
which are necessary for dealing with spinors in curved spaces.

Transforming the quantum field $\zeta \rightarrow \sqrt{2\pi
\alpha ^{\prime
}}\zeta ,\,$where $\zeta $ is the tangent d--vector in the point $u\in {\cal %
E}^{<z>}$ of the higher order anisotropic space and using the
conformal--flat part of the two dimensional metric we obtain this
effective
action necessary for further calculations%
$$
I_{int}^{eff}=\int d^2z[\frac 13(2\pi \alpha ^{\prime
})^{1/2}\varepsilon ^{\ddot e\ddot \imath }H_{<\alpha ><\beta
><\gamma >}(u)\partial _{\ddot e}\zeta ^{<\alpha >}\partial
_{\ddot \imath }\zeta ^{<\beta >}\zeta ^{<\gamma >}+\eqno(4.28)
$$
$$
\frac{2\pi \alpha ^{\prime }}6\varepsilon \sigma (z)R_{<\alpha
><\beta
><\gamma ><\delta >}(u)\partial _{\ddot e}\zeta ^{<\alpha >}\partial _{\ddot
e}\zeta ^{<\delta >}\zeta ^{<\beta >}\zeta ^{<\gamma >}+
$$
$$
\frac 12\varepsilon \sigma (z)\partial _{\ddot e}\zeta ^{<\alpha
>}\partial _{\ddot e}\zeta ^{<\alpha >}+\frac{2\pi \alpha
^{\prime }}6R_{<\alpha
><\beta ><\gamma ><\delta >}(u)\partial _{\ddot e}\zeta ^{<\alpha >}\partial
_{\ddot e}\zeta ^{<\delta >}\zeta ^{<\beta >}\zeta ^{<\gamma >}+
$$
$$
\frac{2\pi \alpha ^{\prime }}4D_{<\alpha >}H_{<\alpha ><\beta
><\gamma
>}(u)\varepsilon ^{\ddot e\ddot \imath }\partial _{\ddot e}\zeta ^{<\beta
>}\partial _{\ddot \imath }\zeta ^{<\gamma >}\zeta ^{<\alpha >}\zeta
^{<\delta >},
$$
where $(u)$ higher order anisotropic space coordinates not
depending on two
coordinates $z.$ So, the anomaly in (4.27) takes the form%
$$
4\vartheta \int d^2z~\sigma (z)\Box \sigma (z)\eqno(4.29)
$$

If $\widetilde{\sigma }$ in (4.26)\ is the connection on
topologies constant the first term in (4.27) can be absorbed by
the renormalization of this connection constant. The set of
two--loop diagrams defining (4.29) is illustrated in the figure
4.1. We note that we must take into account tedpoles because of
the compactness of the string world sheet there are not infrared
divergences.

In the one--loop approximation (figure 4.1) we find%
\index{One--loop approximation} \index{Approximation!one--loop}
$$
\beta ^{(1)}=\frac 1{24\pi }(n+m_1+...+m_z),~\gamma ^{(1)}=\frac
1{96\pi }(n+m_1+...+m_z);
$$
there is correspondence with classical results 
 [193] if we consider a
trivial distinguishing of the space--time dimension
$n_E=n+m_1+...+m_z$.

The two--loop terms (figure 4.1) from (4.29) are computed as
\index{approximation!two--loop}
$$
b)=-\frac{\pi \alpha ^{\prime }\varepsilon ^2R}{3(2\pi )^6}\int
d^2k\sigma (k)\sigma (-k)\int d^{d_2}pd^{d_2}q\times
$$
$$
\frac{(p\cdot k-p^2)^2(q\cdot k-q^2)+(p\cdot k-p^2)^2(p\cdot q)(q\cdot k-q^2)%
}{p^2(k-p)^2q^2(k-q)^2},
$$
$$
c)=\frac{\pi \alpha ^{\prime }\varepsilon ^2R}{3(2\pi )^4}G(0)\int
d^2k\sigma (k)\sigma (-k)\int d^{d_2}p\frac{(p\cdot
k-p^2)^2}{p^2(k-p)^2},
$$
$$
d)=-\frac{\pi \alpha ^{\prime }\varepsilon ^2R}{3(2\pi
)^4}G(0)\int d^2k\sigma (k)\sigma (-k)\int d^{d_2}p\frac{(p\cdot
k-p^2)^2}{p^2(k-p)^2},
$$
$$
e)+f)=\frac{2\varepsilon ^2\pi \alpha ^{\prime }H_{<\alpha
><\beta ><\gamma
>}^2}{(2\pi )^6}\varepsilon ^{\ddot e\ddot \imath }\varepsilon ^{\ddot
a\ddot u}\int d^2k\sigma (k)\sigma (-k)\int d^{d_2}pd^{d_2}q\times
$$
$$
\frac 1{p^2(k-p)^2q^2(k-p-q)^2}\times
$$
$$
\{\frac{(k\cdot p-p^2)(k\cdot q-q^2)p_{\ddot e}(k-q)_{\ddot \imath
}(k-q)_{\ddot a}q_{\ddot u}}{(k-q)^2}+
$$
$$
\frac{(k\cdot p-p^2)^2q_{\ddot e}(k-q)_{\ddot \imath }q_{\ddot
a}(k-p)_{\ddot u}}{(k-p)^2}\},
$$
where
$$
\sigma (z)\equiv \frac 1{(2\pi )^2}\int d^2p\sigma (p)\exp (-ipx).
$$

The contributions of tedpole nonvanishing diagrams mutually
compensate. The
sum of the rest of contributions results in the anomaly%
\index{Anomaly}
$$
\gamma ^{(2)}=\frac{\alpha ^{\prime }}{64\pi }(-R+\frac
13H^2)=\frac{\alpha ^{\prime }}{64\pi }(-\widehat{R}-\frac 23H^2),
$$
where $H^2=H_{<\alpha ><\beta ><\gamma >}H^{<\alpha ><\beta
><\gamma >}$ and $\widehat{R}$ is the scalar curvature with
torsion.

Computing $W$ in the leading order on $\alpha ^{\prime }$ for the
closed boson string, it is not difficult to find the effective
action $\Gamma $ for massless perturbations of the string (of the
metric $G_{<\alpha ><\beta >}$ and field $H_{<\alpha ><\beta >})$
on the tree $\left( \chi =2\right) $
level. Taking into account the identity%
$$
\int (R(z)\sqrt{\gamma _{(e)}})_z\Box _{zz^{\prime
}}^{-1}(R(z)\sqrt{\gamma _{(e)}})_{z^{\prime }}d^2zd^2z^{\prime
}=16\pi
$$
for the metric on sphere we find%
$$
\Gamma ^{(0)}[G,H]\sim \int \frac{\delta ^{n_E}u}{(2\pi \alpha
^{\prime
})^{n_E/2}}\sqrt{G(u)}[1+\frac{\alpha ^{\prime }}4(-R+\frac 13H^2)],%
\eqno(4.30)
$$
where $n_E=n+m_1+...+m_z$ is the dimension of higher order
anisotropic space.
 Formula (4.30) generalizes for such type of spaces (scalar curvature $R$
and torsion $H$ are for, distinguished by N--connection, on
la--space) of that presented in
 [53,54]. The cosmological constant in (4.30)
arises due to the taxion modes in the spectrum of boson strings
and is
absent for superstrings. From vanishing of $\beta $--functions for $\widehat{%
R}$ and  tacking into account the contributions of
reparametrization
ghosts 
 [193] into the anomaly of boson string we obtain into the
leading approximation
$$
\beta =\frac{n_E-8}{24\pi }+\frac{\alpha ^{\prime }}{16\pi }(-\widehat{R}%
-\frac 23H^2)+...,
$$
$$
\gamma =\frac{n_E-26}{96\pi }+\frac{\alpha ^{\prime }}{64\pi }(-\widehat{R}%
-\frac 23H^2)+....
$$

In consequence, the correction to the critical dimension is
$$
D_c=26+\alpha ^{\prime }H^2+{\it O}([\alpha ^{\prime
}]^2).\eqno(4.31)
$$
We emphasize that torsion in (4.31) can be interpreted in a
different manner that in the case of locally isotropic theories
where $H_{...}$ is considered as an antisymmetric strength of a
specific gauge field (see the
Wess--Zumino--Witten model 
 [287,293]). For locally anisotropic
spaces we suggested the idea that the $H_{...}$--terms are
induced by the distinguished components of torsions of, in our
case, higher order
anisotropic spaces. 
\begin{figure}[htbp]
\begin{picture}(380,70) \setlength{\unitlength}{1pt}
\thicklines

\put(210,35){\circle{30}} \put(225,35){\line(1,0){30}}
\put(165,35){\line(1,0){30}} \put(235,40){\makebox(30,20){$H$}}
\put(150,40){\makebox(30,20){$H$}} \put(199,24){\line(1,1){22}}
\put(203,21){\line(1,1){21}} \put(210,20){\line(1,1){16}}
\put(196,28){\line(1,1){21}} \put(194,34){\line(1,1){16}}
\end{picture}
\caption{\it The  diagrams defining
su\-per\-gra\-vi\-ta\-ti\-on\-al and su\-per\-con\-form\-al
ano\-ma\-li\-es in the the\-ory of high\-er or\-der
an\-isot\-rop\-ic su\-per\-strings}
\end{figure}
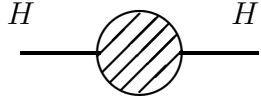

The presented in this section constructions can be generalized
for the case of {\sf N=1 }and {\sf N=2} higher order anisotropic
Fermi strings. Let decompose action $I_S[e_{\ddot
e}^{\underline{\ddot e}},u+\zeta (z),\psi
(z)] $ to within forth order on quantum fields $\zeta ^{<\alpha >}$ and $%
\psi _{\ddot e}^{<\alpha >}(z).$ After redefinition $\zeta \rightarrow \sqrt{%
2\pi \alpha ^{\prime }}\zeta ,\psi \rightarrow e^{-1/4}\sqrt{2\pi
\alpha
^{\prime }}\psi $ we obtain the next additional to (4.28) term:%
$$
I_{F(int)}=\int d^2z[\frac i4\varepsilon \sigma \overline{\psi }^{<%
\underline{\alpha }>}\gamma ^{\ddot e}\partial _{\ddot e}\psi ^{<\underline{%
\alpha }>}-
$$
$$
\frac i2\sqrt{2\pi \alpha ^{\prime }}(1-\frac \varepsilon 2\sigma
)H_{<\alpha ><\beta ><\gamma >}\overline{\psi }^{<\alpha >}\gamma
^{\ddot e}(\partial _{\ddot e}\zeta ^{<\gamma >})\psi ^{<\beta >}+
$$
$$
i\frac{\pi \alpha ^{\prime }}2(1+\frac \varepsilon 2\sigma
)R_{<\alpha
><\beta ><\gamma ><\delta >}\overline{\psi }^{<\alpha >}\gamma ^{\ddot
e}(\partial _{\ddot e}\zeta ^{<\delta >})\psi ^{<\beta >}\zeta
^{<\gamma >}+
$$
$$
\frac{\pi \alpha ^{\prime }}{16}\widehat{R}_{<\alpha ><\beta
><\gamma
><\delta >}\overline{\psi }^{<\alpha >}(1+\gamma _5)\psi ^{<\gamma >}%
\overline{\psi }^{<\beta >}(1+\gamma _5)\psi ^{<\delta >}].
$$

Fixing the gauges
$$
{\sf N=1:~}e_{\ddot e}^{\underline{\ddot e}}=e^\sigma \delta _{\ddot e}^{%
\underline{\ddot e}}~\psi _{\ddot e}=\frac 12\gamma _{\ddot
e}\lambda ,
$$
$$
{\sf N=2:~}e_{\ddot e}^{\underline{\ddot e}}=e^\sigma \delta _{\ddot e}^{%
\underline{\ddot e}}~\psi _{\ddot e}=\frac 12\gamma _{\ddot
e}\lambda ,~A^{\ddot e}=\frac 12\varepsilon ^{\ddot e\ddot
u}\partial _{\ddot u}\rho ,
$$
where $\lambda $ is the Maiorana (${\sf N=1}$ ) or Dirac (${\sf
N=2}$ ) spinor. Because of supersymmetry it is enough
[80] to compute only the coefficient before Weyl anomaly in order
to get the superconformal \index{Anomaly!Weyl}
\index{Weyl!anomaly} anomalies. The one--loop results are
\index{Superconformal anomalies}
$$
{\sf N=1:}~\beta ^{(1)}=\frac{n_E}{16\pi },~\gamma ^{(1)}=\frac{n_E}{64\pi }%
,
$$
$$
{\sf N=2:}~\beta ^{(1)}=\frac{n_E}{8\pi },~\gamma
^{(1)}=\frac{n_E}{32\pi },
$$
from which, taking into account the reparametrization and
superconformal
ghosts we obtain these values of critical dimension ($\gamma =0):$%
$$
{\sf N=1:}~\beta _t^{(1)}=\frac{n_E-2}{16\pi },~\gamma ^{(1)}=\frac{n_E-10}{%
64\pi },
$$
$$
{\sf N=2:}~\beta ^{(1)}=\frac{n_E}{8\pi },~\gamma ^{(1)}=\frac{n_E-2}{32\pi }%
.
$$

From formal point of view in the two--loop approximation we must
consider diagrams b)--m) from fig. 4.2. By straightforward
calculations by using methods similar to those presented in
 [80] we conclude that all
two--loop contributions b)--m) vanish. In result we conclude that
for Fermi
strings one holds the next formulas:%
$$
{\sf N=1:}~\beta =\frac{n_E-2}{16\pi }-\frac{\alpha ^{\prime }H^2}{24\pi }%
+...,~\gamma =\frac{n_E-10}{64\pi }-\frac{\alpha ^{\prime
}H^2}{96\pi },
$$
$$
{\sf N=2:}~\beta =\frac{n_E}{8\pi }-\frac{\alpha ^{\prime }H^2}{24\pi }%
+...,~\gamma =\frac{n_E-2}{32\pi }-\frac{\alpha ^{\prime
}H^2}{96\pi }
$$
or%
$$
{\sf N=1:}~D_c=10+\frac 23\alpha ^{\prime }H^2+...,
$$
$$
{\sf N=2:}~D_c=2+\frac 13\alpha ^{\prime }H^2...
$$
in the leading order on $\alpha ^{\prime }.$

Finally, we note that because $H^2$ contains components of
N--connection and torsion of d--connection on higher order
anisotropic space we conclude that a possible local anisotropy of
space--time can change the critical dimension of Fermi strings.

 \section{Anomalies in LAS--Mo\-dels}

Anomalies in quantum field theories are considered beginning with
works
 [2] and
 [35]. Conformal and gravitational anomalies have been
analyzed in
 [61,72,6] (see also reviews
 [298,169,139,140,92]). The aim of this section is to investigate
anomalies in higher order anisotropic (1,0)--superspaces.
\index{Anomalies}

\subsection{One--loop calculus}

Supergravitational and conform anomalies of heterotic locally anisotropic $%
\sigma $--models connected with (1,0) higher order anisotropic
supergravity are defined by the finite (anomaly) parts of
diagrams of self--energy type (fig. 4.2). In order to compute
anomalies it is necessary to consider all the vertexes of the
theory with no more than the linear dependence on \index{Vertexes}
potentials $H_{+}^{=},H_{=}^{\ddagger }$ (see subsection 4.1.2
and section 4.4 for denotations on higher order anisotropic
heterotic superstrings). We
consider this ''effective'' action (without ghosts):%
$$
I=I_0+I_{int},~I_{int}=I_0^{\prime }+I_1,\eqno(4.32)
$$
where
$$
I_0=-\int d^3z^{-}[iD_{+}\zeta ^{<\alpha >}\partial _{=}\zeta ^{<\beta >}%
\widehat{g}_{<\alpha ><\beta >}+\chi _{-}^{|I|}D_{+}\chi
_{-}^{|I|}],
$$
$$
I_0^{\prime }=-\frac 12\int d^3z^{-}\{iD_{+}\zeta ^{<\alpha
>}\partial _{=}\zeta ^{<\beta >}(\zeta ^{<\gamma
>}\widehat{A}_{<\gamma ><\alpha
><\beta >}+
$$
$$
\zeta ^{<\delta >}\zeta ^{<\gamma >}\widehat{B}_{<\delta ><\gamma
><\alpha
><\beta >}+\zeta ^{<\varepsilon >}\zeta ^{<\delta >}\zeta ^{<\gamma >}%
\widehat{{\cal D}}_{<\varepsilon ><\delta ><\gamma ><\alpha
><\beta >}+
$$
$$
\zeta ^{<\tau >}\zeta ^{<\varepsilon >}\zeta ^{<\delta >}\zeta ^{<\gamma >}%
\widehat{M}_{<\varepsilon ><\delta ><\gamma ><\tau ><\alpha
><\beta >})+\chi _{(-)}^{|I|}\chi _{(-)}^{|J|}D_{+}\zeta ^{<\beta
>}\times
$$
$$
(\zeta ^{<\alpha >}\widehat{C}_{<\alpha ><\beta >}^{|I||J|}+\zeta
^{<\gamma
>}\zeta ^{<\alpha >}\widehat{E}_{<\gamma ><\alpha ><\beta >}^{|I||J|}+\zeta
^{<\varepsilon >}\zeta ^{<\gamma >}\zeta ^{<\alpha >}\widehat{K}%
_{<\varepsilon ><\gamma ><\alpha ><\beta >}^{|I||J|}),
$$
$$
I_1=-\frac 12\int d^3z^{-}\{(\widehat{g}_{<\alpha ><\beta
>}+\zeta ^{<\gamma
>}\widehat{A}_{<\gamma ><\alpha ><\beta >}+
$$
$$
\zeta ^{<\delta >}\zeta ^{<\gamma >}\widehat{B}_{<\delta ><\gamma
><\alpha
><\beta >}+\zeta ^{<\varepsilon >}\zeta ^{<\delta >}\zeta ^{<\gamma >}%
\widehat{{\cal D}}_{<\varepsilon ><\delta ><\gamma ><\alpha
><\beta >}+
$$
$$
\zeta ^{<\tau >}\zeta ^{<\varepsilon >}\zeta ^{<\delta >}\zeta ^{<\gamma >}%
\widehat{M}_{<\varepsilon ><\delta ><\gamma ><\tau ><\alpha
><\beta
>})+[iH_{+}^{=}\partial _{=}\zeta ^{<\alpha >}\partial _{=}\zeta ^{<\beta
>}+
$$
$$
\frac 12D_{+}\zeta ^{<\alpha >}(D_{+}H_{=}^{\ddagger })D_{+}\zeta
^{<\beta
>}+i(D_{+}\zeta ^{<\alpha >})H_{=}^{\ddagger }(\partial _{\ddagger }\zeta
^{<\beta >})]+
$$
$$
\chi _{(-)}^{|I|}H_{+}^{=}\partial _{=}\chi _{(-)}^{|J|}+\chi
_{(-)}^{|I|}\chi _{(-)}^{|J|}H_{+}^{=}\partial _{=}\zeta ^{<\beta
>}\times
$$
$$
(\zeta ^{<\alpha >}\widehat{C}_{<\alpha ><\beta >}^{|I||J|}+\zeta
^{<\gamma
>}\zeta ^{<\alpha >}\widehat{E}_{<\gamma ><\alpha ><\beta >}^{|I||J|}+\zeta
^{<\varepsilon >}\zeta ^{<\gamma >}\zeta ^{<\alpha >}\widehat{K}%
_{<\varepsilon ><\gamma ><\alpha ><\beta >}^{|I||J|})+
$$
$$
\frac i{4\pi }\sum\limits_p^6\frac 1{p!}\zeta ^{<\alpha
_1>}...\zeta ^{<\alpha _p>}\widehat{{\cal D}}_{<\alpha
_1>}...\widehat{{\cal D}}_{<\alpha _p>}\Phi (D_{+}\partial
_{\ddagger }H_{=}^{\ddagger }+\partial _{=}^2H_{+}^{=})\},
$$
where%
$$
\widehat{A}_{<\gamma ><\alpha ><\beta >}=\frac
23\widehat{H}_{<\gamma
><\alpha ><\beta >},
$$
$$
\widehat{B}_{<\delta ><\gamma ><\alpha ><\beta >}=\frac 13\widehat{R}%
_{<\delta ><\alpha ><\beta ><\gamma >}+\frac 12\widehat{{\cal D}}_{<\delta >}%
\widehat{H}_{<\gamma ><\alpha ><\beta >},
$$
$$
\widehat{{\cal D}}_{<\varepsilon ><\delta ><\gamma ><\alpha
><\beta >}=\frac
15\widehat{{\cal D}}_{<\varepsilon >}\widehat{{\cal D}}_{<\delta >}\widehat{H%
}_{<\gamma ><\alpha ><\beta >}+
$$
$$
\frac 16\widehat{{\cal D}}_{<\varepsilon >}\widehat{R}_{<\delta
><\alpha
><\beta ><\gamma >}+\frac 2{15}R_{<\delta ><\varepsilon >[<\alpha >}^{<\tau
>}\widehat{H}_{<\beta >]<\gamma ><\tau >},
$$
$$
\widehat{M}_{<\varepsilon ><\delta ><\gamma ><\tau ><\alpha
><\beta >}=\frac
1{20}\widehat{{\cal D}}_{<\varepsilon >}\widehat{{\cal D}}_{<\delta >}%
\widehat{R}_{<\gamma ><\alpha ><\beta ><\tau >}+
$$
$$
\frac 2{45}\widehat{R}_{<\beta ><\tau ><\gamma ><\vartheta
>}R_{<\delta
><\varepsilon ><\alpha >}^{<\vartheta >}+\widehat{{\cal D}}_{<\varepsilon >}%
\widehat{{\cal D}}_{<\delta >}\widehat{{\cal D}}_{<\gamma >}\widehat{H}%
_{<\tau ><\alpha ><\beta >}+
$$
$$
\frac 1{18}\widehat{{\cal D}}_{<\varepsilon
>}\widehat{R}_{<\gamma ><\delta
>[<\alpha >}^{<\vartheta >}\widehat{H}_{<\beta >]<\tau ><\vartheta >}+\frac
19\widehat{{\cal D}}_{<\gamma >}\widehat{H}_{<\tau ><\vartheta >[<\alpha >}%
\widehat{R}_{|<\delta ><\varepsilon >|<\beta >]}^{<\vartheta >},
$$
$$
\widehat{C}_{<\alpha ><\beta >}^{|I||J|}=-\frac
12\widehat{F}_{<\alpha
><\beta >}^{|I||J|},\widehat{E}_{<\gamma ><\alpha ><\beta >}^{|I||J|}=-\frac
13\widehat{{\cal D}}_{<\gamma >}\widehat{F}_{<\alpha ><\beta
>}^{|I||J|},
$$
$$
\widehat{K}_{<\varepsilon ><\gamma ><\alpha ><\beta
>}^{|I||J|}=-\frac
1{24}(3\widehat{{\cal D}}_{<\varepsilon >}\widehat{{\cal D}}_{<\gamma >}%
\widehat{F}_{<\alpha ><\beta
>}^{|I||J|}+\widehat{F}_{<\varepsilon ><\tau
>}^{|I||J|}R_{<\alpha ><\gamma ><\beta >}^{<\tau >}.
$$
For simplicity, in this section we shall omit tilde
''$\widetilde{}"$ over geometric objects (such as curvatures and
torsions computed for Christoffel distinguished symbols (1.39))
but maintain hats ''$\widehat{}"$ in order to point out even
components on the s--space).

We write the supergravitational anomaly in this general form (see 
 [287] for locally isotropic models):%
$$
\frac 1{32\pi }\int d^3z^{-}\{\gamma _1D_{+}H_{+}^{=}\frac{\partial _{=}^4}{%
\Box }H_{+}^{=}-i\gamma _2D_{+}H_{=}^{\ddagger }\frac{\partial _{=}^3}{\Box }%
H_{=}^{\ddagger }\},
$$
where background depending coefficients will be defined from a
perturbation
calculus on $\alpha ^{\prime }$ by using (4.32). 
\begin{figure}[htbp]
\begin{picture}(380,70) \setlength{\unitlength}{1pt}
\thicklines

\put(210,35){\circle{30}} \put(225,35){\line(1,0){30}}
\put(165,35){\line(1,0){30}} \put(235,35){\makebox(30,20){$H$}}
\put(150,35){\makebox(30,20){$H$}}
\put(235,10){\makebox(30,20){$p$}}
\put(150,10){\makebox(30,20){$p$}}
\put(175,45){\makebox(30,20){${\partial}_{=}$}}
\put(215,45){\makebox(30,20){${\partial}_{=}$}}
\put(175,5){\makebox(30,20){${\partial}_{=}$}}
\put(215,5){\makebox(30,20){${\partial}_{=}$}}
\end{picture}
\caption{\it $(1,0)$ su\-per\-graf de\-fin\-ing the one--lo\-op
a\-no\-ma\-ly}
\end{figure}

For computation of supergrafs we use a standard techniques
 [85]  of
\begin{figure}[htbp]
\begin{picture}(380,70) \setlength{\unitlength}{1pt}
\thicklines

\put(70,35){\circle{30}} \put(85,35){\line(1,0){30}}
\put(25,35){\line(1,0){30}}
\put(95,10){\makebox(30,20){$H^{\ddagger}_{=}$}}
\put(10,10){\makebox(30,20){$H^{\ddagger}_{=}$}}
\put(35,45){\makebox(30,20){${\partial}_{\ddagger}$}}
\put(75,45){\makebox(30,20){${\partial}_{\ddagger}$}}
\put(35,5){\makebox(30,20){${D}_{+}$}}

\put(190,35){\circle{30}} \put(205,35){\line(1,0){30}}
\put(145,35){\line(1,0){30}}
\put(215,35){\makebox(30,20){$H^{=}_{+}$}}
\put(130,35){\makebox(30,20){$H^{=}_{+}$}}
\put(155,45){\makebox(30,20){${\partial}_{=}^2$}}
\put(195,45){\makebox(30,20){${\partial}_{=}$}}
\put(195,5){\makebox(30,20){${\partial}_{=}$}}

\put(300,35){\circle{30}} \put(315,35){\line(1,0){30}}
\put(255,35){\line(1,0){30}}
\put(325,10){\makebox(30,20){$H^{=}_{+}$}}
\put(240,10){\makebox(30,20){$H^{=}_{+}$}}
\put(265,45){\makebox(30,20){${\partial}_{=}$}}
\put(305,45){\makebox(30,20){${\partial}_{=}$}}
\put(265,5){\makebox(30,20){${\partial}_{=}$}}
\put(305,5){\makebox(30,20){${\partial}_{=}$}}

\end{picture}
\caption{\it $(1,0)$ supergrafs de\-fin\-ing the 1--lo\-op
di\-la\-ton con\-tri\-bu\-ti\-on to the anomaly}
\end{figure}
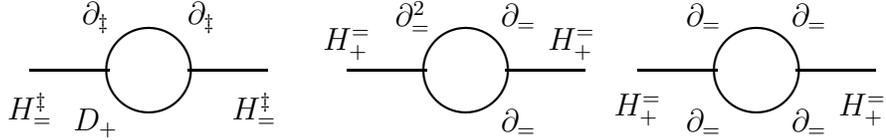
reducing to integrals in momentum space which is standard
practice in quantum field theory. For instance, the one loop
diagram corresponding to
figure 4.3 is computed%
\begin{figure}[htbp]
\begin{picture}(380,70) \setlength{\unitlength}{1pt}
\thicklines

\put(210,35){\circle{30}} \put(225,35){\line(1,0){30}}
\put(165,35){\line(1,0){30}}
\put(235,10){\makebox(30,20){$H_{=}^{\ddagger}$}}
\put(150,10){\makebox(30,20){$H^{=}_{+}$}}
\put(199,24){\line(1,1){22}} \put(203,21){\line(1,1){21}}
\put(210,20){\line(1,1){16}} \put(196,28){\line(1,1){21}}
\put(194,34){\line(1,1){16}}
\end{picture}
\caption{\it The  diagrams  being un\-es\-sen\-ti\-al for
 cal\-cu\-la\-ti\-on of a\-no\-ma\-li\-es} \end{figure}

$$
\frac 1{(2\pi )^2}\int
d^2k\frac{k_{=}^2(k_{=}+p_{=})^2}{k^2(k+p)^2}=-\frac i{24\pi
}\frac{p_{=}^4}{p^2}.
$$
Diagrams of type illustrated on fig. 4.4 give rise only to local
contributions in the anomaly and are not introduced because of
dimensional
considerations. 
\begin{figure}[htbp]
\begin{picture}(380,70) \setlength{\unitlength}{1pt}
\thicklines

\put(90,35){\circle{30}} \put(122,35){\circle{30}}
\put(137,35){\line(1,0){30}} \put(45,35){\line(1,0){30}}
\put(55,45){\makebox(30,20){$D_{+}$}}
\put(92,50){\makebox(30,20){$B$}}
\put(125,45){\makebox(30,20){${\partial}_{=}$}}
\put(92,2){\makebox(30,20){$a)$}}

\put(240,35){\circle{30}} \put(272,35){\circle{30}}
\put(287,35){\line(1,0){30}} \put(195,35){\line(1,0){30}}
\put(200,45){\makebox(30,20){$D_{+}$}}
\put(242,50){\makebox(30,20){$B$}}
\put(200,10){\makebox(30,20){${\partial}_{=}$}}
\put(242,2){\makebox(30,20){$b)$}}
\end{picture}
\caption{\it $B$--depending 2--lo\-op cor\-rec\-ti\-ons to the
a\-no\-ma\-ly}
\end{figure}

We note that comparing with similar locally isotropic results 
 [140]
the torsions $H_{\cdot }^{\cdot }$ are generated by components of
the distinguished torsion of the higher order anisotropic
background.

\subsection{Two--loop calculus}

The two--loop $B$--depending corrections to the anomaly are
defined by diagrams illustrated in fig. 4.6. The one--loop
results including ghosts (4.12) are similar to locally
isotropic ones 
 [85]. Thus we present a brief summary (we must take
into account the splitting of dimensions in higher order anisotropic spaces):%
$$
W_{eff}^{1-loop}=\frac 1{96\pi }\int d^3z^{-}\{(n+m_1+...+m_z-26+\frac{N_E}%
2)D_{+}H_{+}^{=}\frac{\partial _{=}^4}{\Box }H_{+}^{=}-
$$
$$
\frac{3i}2(n+m_1+...+m_z-10)D_{+}H_{=}^{\ddagger }\frac{\partial
_{\ddagger }^3}{\Box }H_{=}^{\ddagger }\}.\eqno(4.33)
$$
If the action (4.33) is completed by local conterterms
$$
H_{+}^{=}\Box H_{=}^{\ddagger },~S\partial
_{=}^2H_{+}^{=},~S\partial _{\ddagger }D_{+}H_{=}^{\ddagger
},~S\partial _{=}D_{+}S
$$
and conditions $N_E-(n+m_1+...+m_z)=22$ and $\gamma _1=\gamma
_2\equiv
\gamma $ are satisfied, we obtain from (4.33) a gauge invariant action:%
$$
W_{eff}+W_{loc}=\frac 1{16\pi }(n+m_1+...+m_z-10)\int d^3z^{-}\Sigma ^{+}(%
\frac{D_{+}}{\Box })\Sigma ^{+},\eqno(4.34)
$$
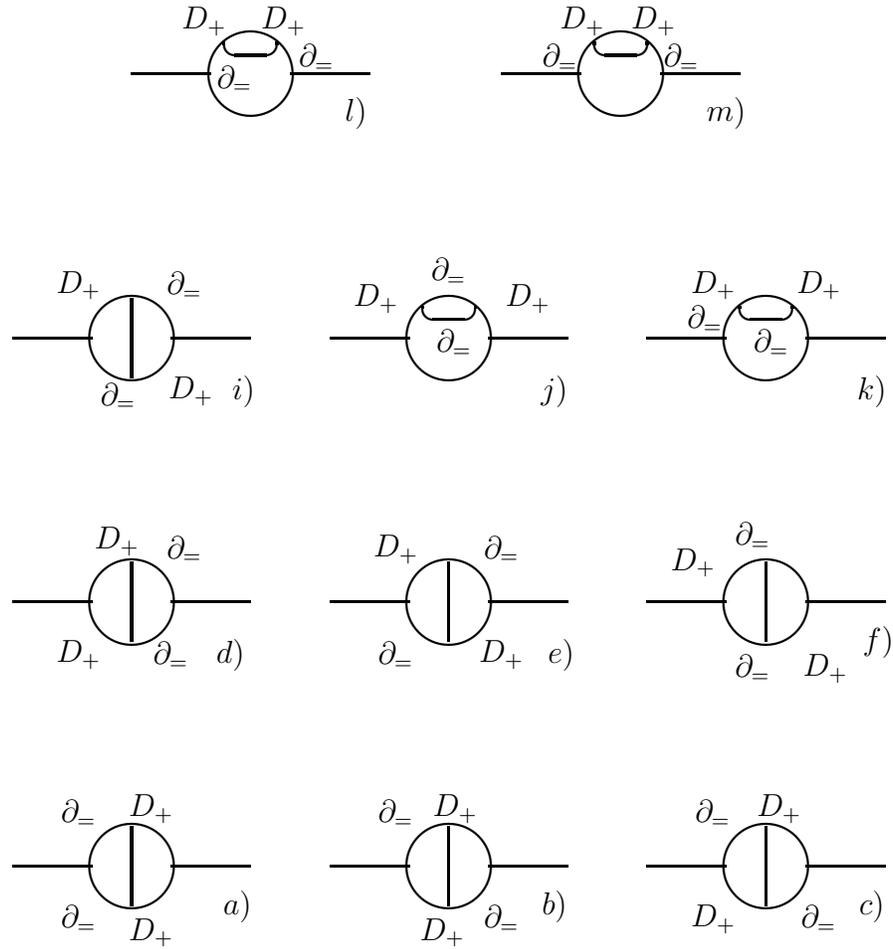
\begin{figure}[htbp]
\begin{picture}(380,400) \setlength{\unitlength}{1pt}
\thicklines

\put(70,35){\circle{30}} \put(85,35){\line(1,0){30}}
\put(25,35){\line(1,0){30}} \put(70,20){\line(0,1){30}}
\put(70,5){\makebox(15,10){$D_{+}$}}
\put(95,10){\makebox(30,20){$a)$}}
\put(35,45){\makebox(30,20){${\partial}_{=}$}}
\put(70,52){\makebox(15,10){$D_{+}$}}
\put(35,5){\makebox(30,20){${\partial}_{=}$}}

\put(190,35){\circle{30}} \put(205,35){\line(1,0){30}}
\put(145,35){\line(1,0){30}} \put(190,20){\line(0,1){30}}
\put(215,10){\makebox(30,20){$b)$}}
\put(155,45){\makebox(30,20){${\partial}_{=}$}}
\put(195,5){\makebox(30,20){${\partial}_{=}$}}
\put(180,5){\makebox(15,10){$D_{+}$}}
\put(185,52){\makebox(15,10){$D_{+}$}}

\put(310,35){\circle{30}} \put(325,35){\line(1,0){30}}
\put(265,35){\line(1,0){30}} \put(310,20){\line(0,1){30}}
\put(335,10){\makebox(30,20){$c)$}}
\put(275,45){\makebox(30,20){${\partial}_{=}$}}
\put(300,47){\makebox(30,20){$D_{+}$}}
\put(275,5){\makebox(30,20){$D_{+}$}}
\put(315,5){\makebox(30,20){${\partial}_{=}$}}
\put(70,135){\circle{30}} \put(85,135){\line(1,0){30}}
\put(25,135){\line(1,0){30}} \put(70,120){\line(0,1){30}}
\put(100,110){\makebox(15,10){$d)$}}
\put(57,153){\makebox(15,10){${D}_{+}$}}
\put(75,145){\makebox(30,20){${\partial}_{=}$}}
\put(35,105){\makebox(30,20){${D}_{+}$}}
\put(70,105){\makebox(30,20){${\partial}_{=}$}}
\put(190,135){\circle{30}} \put(205,135){\line(1,0){30}}
\put(145,135){\line(1,0){30}} \put(190,120){\line(0,1){30}}
\put(225,110){\makebox(15,10){$e)$}}
\put(155,145){\makebox(30,20){${D}_{+}$}}
\put(195,145){\makebox(30,20){${\partial}_{=}$}}
\put(195,105){\makebox(30,20){${D}_{+}$}}
\put(155,105){\makebox(30,20){${\partial}_{=}$}}
\put(310,135){\circle{30}} \put(325,135){\line(1,0){30}}
\put(265,135){\line(1,0){30}} \put(310,120){\line(0,1){30}}
\put(345,115){\makebox(15,10){$f)$}}
\put(275,145){\makebox(15,10){${D}_{+}$}}
\put(290,150){\makebox(30,20){${\partial}_{=}$}}
\put(290,100){\makebox(30,20){${\partial}_{=}$}}
\put(325,105){\makebox(15,10){$D_{+}$}}
\put(70,235){\circle{30}} \put(85,235){\line(1,0){30}}
\put(25,235){\line(1,0){30}} \put(70,220){\line(0,1){30}}
\put(105,210){\makebox(15,10){$i)$}}
\put(35,245){\makebox(30,20){${D}_{+}$}}
\put(75,245){\makebox(30,20){${\partial}_{=}$}}
\put(50,203){\makebox(30,20){${\partial}_{=}$}}
\put(85,210){\makebox(15,10){$D_{+}$}}
\put(190,235){\circle{30}} \put(205,235){\line(1,0){30}}
\put(145,235){\line(1,0){30}} \put(190,247){\oval(20,10)[b]}
\put(215,205){\makebox(30,20){$j)$}}
\put(175,250){\makebox(30,20){${\partial}_{=}$}}
\put(155,245){\makebox(15,10){$D_{+}$}}
\put(177,223){\makebox(30,20){${\partial}_{=}$}}
\put(205,240){\makebox(30,20){$D_{+}$}}
\put(310,235){\circle{30}} \put(325,235){\line(1,0){30}}
\put(265,235){\line(1,0){30}} \put(310,247){\oval(20,10)[b]}
\put(335,205){\makebox(30,20){$k)$}}
\put(275,245){\makebox(30,20){$D_{+}$}}
\put(315,245){\makebox(30,20){$D_{+}$}}
\put(272,232){\makebox(30,20){${\partial}_{=}$}}
\put(297,223){\makebox(30,20){${\partial}_{=}$}}
\put(115,335){\circle{30}} \put(130,335){\line(1,0){30}}
\put(70,335){\line(1,0){30}} \put(115,347){\oval(20,10)[b]}
\put(140,310){\makebox(30,20){$l)$}}
\put(90,350){\makebox(15,10){$D_{+}$}}
\put(99,328){\makebox(20,10){${\partial}_{=}$}}
\put(120,350){\makebox(15,10){$D_{+}$}}
\put(130,336){\makebox(20,10){${\partial}_{=}$}}
\put(255,335){\circle{30}} \put(270,335){\line(1,0){30}}
\put(210,335){\line(1,0){30}} \put(255,347){\oval(20,10)[b]}
\put(280,310){\makebox(30,20){$m)$}}
\put(233,350){\makebox(15,10){${D}_{+}$}}
\put(260,350){\makebox(15,10){${D}_{+}$}}
\put(217,331){\makebox(30,20){${\partial}_{=}$}}
\put(263,331){\makebox(30,20){${\partial}_{=}$}}

\end{picture}
\caption{\it $A^2$--depending 2--loop cor\-rec\-ti\-ons to the
anomaly}
\end{figure}

The action (4.34) is the conformal anomaly of our model ($\Sigma
^{+}$ \index{Conformal anomaly} depends on $S).$ The critical
parameters of the higher order anisotropic heterotic string
$$
n+m_1+...+m_z=10~\mbox{and } ~N_E=32
$$
make up the conditions of cancelation of it anomalies (the
original locally
isotropic result was obtained in 
 [97]). We can also compute and add
the one--loop dilaton (see fig. 4.5) contribution to (4.34):
$$
W_{eff}^{(1,\Phi )}=\frac{{\cal D}^2\Phi }{128\pi ^2}\int
d^3z^{-}[iD_{+}H_{=}^{\ddagger }\frac{\partial _{\ddagger }^3}{\Box }%
H_{=}^{\ddagger }-D_{+}H_{+}^{=}\frac{\partial _{=}^4}{\Box
}H_{+}^{=}],
$$
$$
\gamma ^{(1,\Phi )}=-\frac 1{4\pi }{\cal D}^2\Phi .
$$
computed
$$
a)=I_1(p)=\int \frac{d^2kd^2q}{16\pi ^4}\frac{%
q_{=}^2(q_{=}+p_{=})(k_{=}+p_{=})}{(q^2-\mu ^2)[(q+p)^2-\mu
^2]}\times
$$
$$
\frac{k^2}{(k^2-\mu ^2)[(k+p)^2-\mu ^2]}=\frac 1{64\pi ^2}\frac{p_{=}^4}{p^2}%
+{\it O}\left( \mu ^2\right) ,
$$
$$
b)=I_2(p)=\int \frac{d^2kd^2q}{16\pi
^4}\frac{q_{=}(q_{=}+p_{=})}{(q^2-\mu ^2)[(q+p)^2-\mu ^2]}\times
$$
$$
\frac{k^2(k_{=}+p_{=})^2}{(k^2-\mu ^2)[(k+p)^2-\mu ^2]}=-\frac 1{32\pi ^2}%
\frac{p_{=}^4}{p^2}+{\it O}\left( \mu ^2\right) ,
$$
where the mass parameter $\mu ^2$ is used as a infrared regulator.

Two--loop A$^2$--dependent diagrams are illustrated in fig.4.7.
The supergrafs f),i) and j) are given by the momentum integral
$I_f=I_i=I_j=I(p) :$
$$
I(p)=\int \frac{d^2kd^2q}{(2\pi )^4}\frac{k_{=}q_{=}(k_{=}+q_{=}+p_{=})^2}{%
k^2q^2(k+q+p)^2}=\frac{p_{=}^4}{96\pi ^2p^2};%
$$
to this integral there are also proportional the anomaly parts of
diagrams a)-e),k),l) and m). The rest of possible $A^2$--type
diagrams do not contribute to the anomaly part of the effective
action. After a straightforward computation of two loop diagrams
we have
$$
\gamma ^{2-loop}=\frac 1{16\pi }(-\widetilde{R}+\frac 13H^2);
$$
there are not dilaton contributions in the two--loop
approximation.

The anomaly coefficient $\gamma $ is connected with the central
charge of Virasoro superalgebra of heterotic string on the
background of massless \index{Virasoro superalgebra} modes (under
the conditions of vanishing of $\beta $-functions or,
equivalently, if the motion equations are satisfied, see
 [241,243,231,178,67,173,48]):%
$$
2\gamma +\alpha ^{\prime }\left( {\cal D}_{<\beta >}\Phi \right) ^2=%
\widetilde{\beta }^\Phi \equiv \beta ^\Phi -\frac 14\beta
_{<\alpha ><\beta
>}^gg^{<\alpha ><\beta >}=
$$
$$
\frac{n+m_1+...+m_z-10}2+\frac{\alpha ^{\prime }}2L_{eff},
$$
$$
<T>_{-}=\frac 1{4\pi }\widetilde{\beta }^\Phi \Sigma ^{+}+...,
$$
where $<T>_{-}$is the averaged supertrace, $\beta _{<\alpha
><\beta >}^g$ is
the metric $\beta $-function, $\widetilde{\beta }^\Phi $ is the dilaton $%
\beta $-function and by dots there are denoted the terms
vanishing on the motion equations. We note that the $\beta
$-functions and effective Lagrangian are defined in the string
theory only with the exactness of
redefinition of fields 
 [48]. From a standard calculus according the
perturbation theory on $\alpha ^{\prime }$ we have
\index{Perturbation theory}
$$
\beta _{<\alpha ><\beta >}^g=\alpha ^{\prime
}(\widetilde{R}_{<\alpha
><\beta >}-H_{<\alpha ><\beta >}^2)
$$
and
$$
S_{eff}^{(0)}\equiv \int d^{10}X\sqrt{|g|}L_{eff}^{(0)}=
$$
$$
\frac 12\int d^{10}X\sqrt{|g|}\{-\widetilde{R}-4{\cal D}^2\Phi +4({\cal D}%
_{<\alpha >}\Phi )^2+\frac 13H^2\},
$$
where%
$$
H_{<\alpha ><\beta >}^2\equiv H_{<\alpha ><\gamma ><\delta
>}H_{<\beta
>}^{\quad <\gamma ><\delta >},
$$
which is a higher order anisotropic generalization of models
developed in
 [38,58,36].

\section{Conclusions}

To develop in a straightforward manner self--consistent physical
theories,
 define local conservation laws,give a corresponding treatment of geometrical
objects and so on, on different extensions on Finsler spaces with
nonlinear structure of metric form and of connections, torsions
and curvatures is a highly conjectural task. Only the approach on
modeling of geometric models of the mentioned type (super)spaces
on vector (super)bundles provided with compatible nonlinear and
distinguished connections and metric structures make ''visible''
the possibility (see, for instance,
 [160,161,256,272,255,258,\\ 259,264,267]), manner of elaboration,
as well common features and differences of models of fundamental
physical fields with generic locally an\-isot\-rop\-ic
interactions. From viewpoint of the string theory fundamental
ideas only some primarily changes in established material have
been introduced in this Chapter. But we did not try to a simple
straightforward repetition of standard material in context of some
sophisticate geometries. Our main purposes were to illustrate
that the higher order  anisotropic supergravity is also naturally
contained in the framework of low energy superstring dynamics and
to develop a corresponding geometric and computational technique
for supersymmetric sigma models in locally anisotropic
backgrounds.

The above elaborated methods of perturbative calculus of
anomalies of hetrotic sigma models in higher order anisotropic
superspaces, as a matter of principle, can be used in every
finite order on $\alpha ^{\prime }$ (for instance, by using the
decomposition (4.32) for effective action we can, in a similar
manner as for one- and two--loop calculations presented in section
4.5, find corrections of anomalies up to fifth order inclusive)
and are compatible with the well known results for locally
isotropic strings and sigma models. We omit such considerations
in this work (which could make up a background for a monograph on
locally anisotropic string theory).



\chapter{Stochastics in LAS--Spaces}

We shall describe the analytic results which combine the
fermionic Brownian motion with stochastic integration in higher
order anisotropic spaces. It will be shown that a wide class of
stochastic differential equations in locally anisotropic
superspaces have solutions. Such solutions will be than used to
derive a Feynman--Kac formula for higher order anisotropic
systems. We shall achieve this by introducing locally anisotropic
superpaths parametrized by a commuting and an anticommuting time
variable. The supersymmetric stochastic techniques employed in
this Chapter was developed
by A. Rogers in a series of works 
 [206,207,205,209,210]
(superpaths have been also considered in papers
 [103,82]
 and 
 [198]). One of
the main our purposes is to extend this formalism in order to
formulate the theory of higher order anisotropic processes in
distinguished vector
superbundles 
 [260,262,265,266,267,253,268].
Stochastic calculus for bosonic and fermionic Brownian paths will
provide a geometric approach to Brownian motion in locally
anisotropic superspaces.

Sections 5.1 and 5.2 of this Chapter contain correspondingly a
brief introduction  into the subject and a brief review of
fermionic Brownian motion and path  integration. Section 5.3
considers distinguished stochastic integrals in the  presence of
fermionic paths. Some results on calculus on a (1,1)--dimensional
superspace and two supersymmetric formulae for superpaths are
described in section 5.4 and than, in section 5.5,  the theorem
on the existence of unique  solutions to a useful class of
distinguished stochastic differential  equations is proved and
the distinguished supersymmetric Feynman--Kac formula  is
established. Section 5.6 defines some higher order anisotropic
manifolds  which can be constructed from a vector bundle over a
vector  bundle provided with compatible nonlinear and
distinguished connection and metric structures. In section 5.7 a
geometric formulation of Brownian paths on  higher order
anisotropic manifolds is contained; these paths are used to give
a Feynman--Kac formula for the Laplace--Beltrami operator for
twisted  differential forms. This formula is used to give a proof
of the index theorem  using supersymmetry of the higher order
anisotropic superspaces in section 5.8  (we shall apply the
methods developed in
 [4,82]
  and 
 [209,210].

\section{Introduction}

The stochastic calculus have been recently generalized to various
extensions of Finsler geometry and has many applications in
modern theoretical and mathematical physics and biology
 [13,14,131,141,189,253,262]
 (see
also constructions connected with path integration techniques for
Euclidean spaces and Riemannian manifolds
 [71,74,75,76,90,123,124,125,132]). We intend to present a
manifestly supersymmetric formalism for investigation of
diffusion processes in superspaces with higher order anisotropy.
Stochastic calculus is characterized by many uses in modern
physics (see, for instance,
 131,141,282,73,74,75,76]), biology
 [14,13] and economy 
 [189] and became more familiar to theoretical physics. There are a lot
monographs and textbooks where the stochastic processes and
diffusion are considered from a rigorous mathematical view point
or with the aim to make the presentation more accessible for
applications in different branches of science and economy. In
section 10.1 a brief introduction into the theory of stochastic
differential equations in Euclidean spaces and necessary basic
definitions are presented (one can consult the Chapter 10 and
some of the just cited monographs if this is necessary). Here we
do not assume any prior knowledge of stochastic calculus and
consider the reader oriented to differential geometry and
supergravity theories.

In their simplest form, path integrals can be defined on Euclidean space $%
{\cal R}^q,$ parametrized by coordinates $x,$ with respect to
Hamiltonians
of the form%
$$
H=-\frac 12\sum\limits_{i=s}^q\partial _i\partial _i+V(x),
$$
where $V(x)$ is the interaction potential, which leads to the
Feynman--Kac \index{Feynman--Kac!for\-mu\-la}
for\-mu\-la%
$$
\exp (Ht)\varphi (x)=\int d\mu \exp
(-\int\limits_0^tV(x+b_s)ds)\varphi (x+b_s),
$$
where $b_s$ denotes a Brownian paths and the Wiener measure is
often written \index{Brownian paths} \index{Wiener measure} in
the physics literature as
$$
d\mu \cong {\cal D}x\exp (-\frac 12\int\limits_0^t\left( \frac{dx}{ds}%
\right) ^2ds.
$$
The Feynman--Kac formula can be constructed for every Hamiltonian
which is a second--order elliptic differential operator. The
range of possible Hamiltonians used in the definition of
stochastic calculus can be extended
by including terms of the form $\exp \left( -\int\limits_0^t\sum\limits_{%
\overline{a}=1}^q\varphi
_{\overline{a}}(s)db_s^{\overline{a}}\right) $ in the integrand
or by replacing the simple Brownian paths $b_s$ by paths $x_{t}$
which satisfy the stochastic differential equation%
$$
dx^j=\varphi ^j(t)dt+\sum\limits_{\overline{a}=1}^qc_{\overline{a}%
}^j(s)db_s^{\overline{a}}.
$$
In this case the Feynman--Kac formula can be defined for
Hamiltonians which are arbitrary second--order elliptic operators
having the second--order part
of form%
$$
-\frac 12\sum\limits_{\overline{a},j,k=1}^qc_{\overline{a}}^j(x)c_{\overline{%
a}}^k(x)\frac{\partial ^2}{\partial x^j\partial x^k}.
$$

In our considerations we shall use the main results of the paper
 [209] where a supersymmetric Feynman--Kac formula for Hamiltonians $H$
which are the square of a supercharge $Q$ being a Dirac--like
operator. We \index{supercharge} shall generalize this approach
(see also applications of a companion paper
 [210]) in order to obtain Feynman--Kac like formulas for the heat
kernels of the Laplace--Beltrami operators in locally anisotropic
superspaces). Here we also note that because the canonical
anticommutation relations are also the defining relations of a
Clifford algebra, the methods developed in this Chapter are also
applicable to various geometrical operators on Clifford and
spinor bundles and various bundles of differential forms on
distinguished vector (super)bundles
 [160,161,162,256,255,258,259,264,267].

\section{Fermionic Brownian LA--Motion}

The geometry of higher order anisotropic (super)spaces is
considered in detail in Chapters 1 and 2 of this book. This
section introduces notation, and summarizes some necessary
aspects in order to develop an approach to higher order
an\-isot\-rop\-ic Brownian motion. \index{Brownian motoin!higher
order an\-isot\-rop\-ic}

For each positive integer $l,$ ${\cal B}_l$ denotes the Grassmann algebra on $%
{\cal R}^l.$ ${\cal B}_l$ is defined over reals with generators ${\bf 1,}%
\beta _{\left( 1\right) },...,\beta _{\left( l\right) }$
satisfying
anticommuting relations%
$$
\beta _{\left( i\right) }\beta _{\left( j\right) }=-\beta
_{\left( j\right) }\beta _{\left( i\right) },~{\bf 1}\beta
_{\left( i\right) }=\beta _{\left( i\right) }{\bf
1,}~(i),(j)=1,...,l.
$$
The Grassmann algebra ${\cal B}_l$ is $Z_2$--graduated, ${\cal B}_l={\cal B}%
_{l,0}\oplus {\cal B}_{l,1}$ (the elements of ${\cal B}_{l,0}$ and ${\cal B}%
_{l,1}$ are respectively even and odd elements which in turn can
be represented as sums of terms containing the product of an even
and, correspondingly, odd numbers of anticommuting generators.
Any two odd elements anticommute, while even elements commute
both with one another and
with odd elements. We introduce denotation ${\cal R}_l^{n,k}\doteq {\cal R}%
^n\times \left( {\cal B}_{l,1}\right) ^k.\,$ For every set we write $%
{\cal R}_l^{n,k[J]}=\prod\limits_{j\in J}{\cal R}_{|j|l}^{n,k},$ where $%
{\cal R}_{|j|)l}^{n,k}\doteq {\cal R}^n\times \left( {\cal B}%
_{|j|l,1}\right) ^k,$ with ${\cal B}_{|j|l}$ being the Grassmann
algebra with $l$ anticommuting generators $\beta _{|j|\left(
1\right) },...,\beta _{|j|\left( l\right) }$ in addition to the
commuting generator ${\bf 1.}$ We note that it is necessary to
use different anticommuting generators for every $j\in J.$

We parametrize elements of ${\cal R}_l^{n,k}$ by coordinates
$$
{\bf x=}\left( x,\theta \right) =x^I=\left( x^i=\widehat{x}^i,\theta ^{%
\widehat{i}}\right) =\left( x^1,...,x^n,\theta ^{\widehat{1}},...,\theta ^{%
\widehat{k}}\right) .
$$
For our further considerations we shall also use s--space ${\cal R}%
_l^{n,2k[J]}$ (where $J$ is a finite set containing $K$ elements)
provided
with coordinates of type%
$$
(\underline{x}^1,...,\underline{x}^K,\underline{\theta }^1,...,\underline{%
\theta }^K,\underline{\rho }^1,...,\underline{\rho }^K)=
$$
$$
(x^{1,1},...,x^{K,n},\theta ^{1,1},...,\theta ^{1,k},...,\theta
^{K,k},\rho ^{1,k},...,\rho ^{K,k}).
$$

Having defined classes of functions ${\it C}^\infty \left( {\cal R}^n,{\cal R%
}\right) $ or $L^2\left( {\cal R}^n,{\cal C}\right) $ on ${\cal
R}^n$ we can in a similar manner introduce corresponding
analogous of classes of functions on ${\cal R}_l^{n,k}$, denoted
${\it C}^{\prime \infty }\left(
{\cal R}^n,{\cal R}\right) $ or $L^{\prime 2}\left( {\cal R}^n,{\cal C}%
\right) .$ For a Banach space ${\cal S}$ by ${\it C}^{\prime
\infty }\left(
{\cal R}^n,{\cal S}\right) $ we denote the class of functions%
$$
f:{\cal R}_l^{n,k}\rightarrow {\cal B}_l\otimes {\cal S}\eqno(5.1)
$$
such that
$$
f(x^1,...,x^n,\theta ^{\widehat{1}},...,\theta ^{\widehat{k}%
})=\sum\limits_{\{\mu \}\in M_k}f_{\{i\}}(x^1,...,x^n)\theta ^{\widehat{i}%
_1}...\theta ^{\widehat{l}_w},
$$
where $f_{\{\mu \}}\in {\it C}^\infty \left( {\cal R}^n,{\cal S}\right) ,$ $%
\{i\}=\widehat{i}_1,...\widehat{i}_w$ is a multi--index with $1\leq \widehat{%
i}_1<...<\widehat{i}_w<k$ and $M_k$ is the set of multi--indices,
including the empty one. A function on ${\cal R}_l^{n,k}$ is
bounded if each of the coefficients functions $f_{\{\mu \}}$ are
bounded. Differentiation of a function of ${\cal R}_l^{n,k}$ with
respect $i$ the even element will be denoted as $\partial _i$ and
differentiation with respect to the $\widehat{j} $th odd element
will be denoted as $\partial _{\widehat{j}}.$ Details on analysis
of functions of Grassmann variables may be found in
 [203,205].

The class of functions (5.1) can be extended on trivial
distinguished vector bundles (dvs--bundles) if we consider
distinguished sets of indices; in this
case we write%
$$
f:({\cal R}_l^{n,k},{\cal R}_l^{m_1,l_1},...,{\cal R}_l^{m_p,l_p},...,{\cal R%
}_l^{m_z,l_z})\rightarrow {\cal B}_l\otimes {\cal S,}
$$
$$
f(u^{<\alpha >})=f(u^{\alpha
_p})=\sum\limits_{\{<\widehat{a}_p>\}\in
M_k}f_{\{\widehat{a}_p\}}(\widehat{u}^1,...,\widehat{u}^{n+m_1+...+m_z})%
\zeta ^{\widehat{a_p}_1}...\zeta ^{\widehat{a_p}_w},
$$
where $p$ labels the number of distinguished ''shells'', $$f_{\{\widehat{a}%
_p\}}\in C^\infty ({\cal R}_l^{n_E},{\cal S),}(n_E=n+m_1+...+m_p)),\{<%
\widehat{a}_p>\}=\{\widehat{a}_{p1},\widehat{a}_{p2},...,\widehat{a}_{pw}\}$$
are multi--indices with $1\leq
\widehat{a}_{p1}<...<\widehat{a}_{pw}\leq l_p$ and
$M_k=\{M_{k1},...,M_{kp}\}.$ If all coefficients
$f_{\{\widehat{a}_p\}}$
are bounded the function $f$ on $$({\cal R}_l^{n,k},{\cal R}_l^{m_1,l_1},...,%
{\cal R}_l^{m_p,l_p},...,{\cal R}_l^{m_z,l_z})$$ is said to be
bounded.

We shall also use distinguishings of indices and denote
corresponding spaces
and local coordinates respectively by ${\cal R}_{l_E}^{n_E,k_E}=(...,{\cal R}%
_{l_p}^{n_p,k_p},...)$ and
$$
(\underline{x}^1,...,\underline{x}^K,\underline{\theta }^1,...,\underline{%
\theta }^K,\underline{\rho }^1,...,\underline{\rho }^K,...,\underline{y}%
_{(p)}^1,...,\underline{y}_{(p)}^{K_p},\underline{\zeta }_{(p)}^1,...,%
\underline{\zeta }_{(p)}^{K_p},\underline{\rho
}_{(p)}^1,...,\underline{\rho }_{(p)}^{K_p},...)=
$$
$$
(x^{1,1},...,x^{K,n},\theta ^{1,1},...,\theta ^{1,k},...,\theta
^{K,k},\rho ^{1,k},...,\rho ^{K,k},...,
$$
$$
y_{(p)}^{1,1},...,y_{(p)}^{K_p,m_p},\zeta _{(p)}^{1,1},...,\zeta
_{(p)}^{1,k},...,\zeta _{(p)}^{K_p,k},\rho _{(p)}^{1,k},...,\rho
_{(p)}^{K_p,k},...).
$$

The usual partial derivations with respect to coordinates
$u^{<\alpha >}$ will be denoted $\partial _{<\alpha >}=\frac
\partial {\partial u^{<\alpha
>}}$ and the locally adapted to the N--connection derivations will be
denoted $\delta _{<\alpha >}=\frac \delta {\partial u^{<\alpha
>}}.$

The integration of a function of $w$ commuting variables $\zeta ^{\widehat{%
a_p}_1},...,\zeta ^{\widehat{a_p}_w}$ is defined according the
rule
 [37]
$$
\int d^wf(\zeta )=f_{1...w},
$$
if $f(\zeta )=f_{\widehat{a}_{p1}...\widehat{a}_{pw}}\zeta ^{\widehat{a_p}%
_1},...,\zeta ^{\widehat{a_p}_w}+$ terms of lower order in $\zeta
.$ The necessary details on the analysis of functions of
Grassmann variables may
be found, for instance, in 
 [203,205,146,147].

Because ${\cal R}_l^{n,2k[J]}$ is not a simply product of copies of ${\cal R}%
_l^{n,2k}$ we have a quite sophisticate definition of
distinguished Grassmann random variables:

\begin{definition}
Let be finite subsets of $I_p,\,$ with $J_{M_p}\subset J_{M_p+1}$ for each $$%
M_p=1,2,...,(p=0,1,...,z),M_E = \{M_p\},J_E=\{J_{M_p}\},$$
$$Gr_{M_p}\in
L^{2^{\prime }}\left( {\cal R}_l^{n,2k[J_{M_p}]},{\cal C}\right) ,$$  $$%
Gr_{M_E}\in L^{2^{\prime }}\left( {\cal R}_{l_E}^{n_E,2k_E[J_{M_E}]},{\cal C}%
\right) ,{\cal C}$$ is the complex number field, and
$$
I_{M_E}(Gr)=\int d^{\sharp (J_{M_E})}\zeta d^{\sharp
(J_{M_E})}\rho ~\times \eqno(5.2)
$$
$$
f_{J_E}(\underline{x}^1,...,\underline{x}^K,\underline{\theta }^1,...,%
\underline{\theta }^K,\underline{\rho }^1,...,\underline{\rho }^K,...,%
\underline{y}_{(p)}^1,...,\underline{y}_{(p)}^{K_p},\underline{\zeta }%
_{(p)}^1,...,\underline{\zeta }_{(p)}^{K_p},$$
 $$\underline{\rho }_{(p)}^1,...,%
\underline{\rho }_{(p)}^{K_p},...)\times
Gr_{M_E}(\underline{\theta }^1,...,\underline{\theta }^K,\underline{\rho }%
^1,...,\underline{\rho }^K,...,\underline{\zeta }_{(p)}^1,...,\underline{%
\zeta }_{(p)}^{K_p},\underline{\rho }_{(p)}^1,...,\underline{\rho }%
_{(p)}^{K_p},...);
$$

a) a collection $\left( J_{M_E},G_{M_E}\right) $ is called a
distinguished
random variable on Grassmann Wiener space of the sequences $%
\{...,I_{M_p},...\}$ tends to limits (denoted as$\int d\mu _{f(p)}GrM_p$ or $%
{\cal E}_F(Gr_{M_p})$ as $M_p$ tends to infinity for every
$p=0,1,...,z;$

b) a sequence of pairs $\left( J_{M_E},G_{M_E}\right) $ for which
distinguished integrals (5.2) do not necessarily satisfy the
divergence condition is called a generalized random variable.
\end{definition}

For finite sets $I_{M_E}$ the defined distinguished Grassmann
random \index{Random variables!Grassmann} \index{Grassman!random
variables} variables (in brief dr--variables) are said to be
finitely defined. There \index{dr--variables} are $2n_E$ finitely
defined dr--variables $\zeta _r^{<\alpha >},\rho
_r^{<\alpha >}$ for every $r\in I_E$ and corresponding $J=\{r\}:$%
$$
\zeta _r^{<\alpha >}(...,\zeta _{(p)}^1,...,\zeta
_{(p)}^{m_p},\rho _{(p)}^1,...,\rho _{(p)}^{m_p},...)=\zeta
^{<\alpha >},
$$
$$
\rho _r^{<\alpha >}(...,\zeta _{(p)}^1,...,\zeta _{(p)}^{m_p},\rho
_{(p)}^1,...,\rho _{(p)}^{m_p},...)=\rho ^{<\alpha >}.
$$

\begin{definition}
The $\left( 0,2n_E\right) $--dimensional process
$$
(...,\zeta _{r(p)}^1,...,\zeta _{r(p)}^{m_p},\rho
_{r(p)}^1,...,\rho _{r(p)}^{m_p},...)
$$
is called distinguished fermionic Brownian motion.
\end{definition}
\index{Brownian motion!distinguished fermionic}

For some applications it is necessary to combine distinguished
fermionic path integrals with ordinary bosonic path integrals
using the usual Wiener measure and Brownian motion. We can obtain
a distinguished super Wiener measure by defining a measure (in
generalized sense) on the $\left(
n_E,k_E\right) $--dimensional super Wiener space ${\cal R}%
_{l_E}^{n_E,2k_E[I_E]}$ of paths in superspace by defining finite
\index{super Wiener space} distributions on ${\cal
R}_{l_E}^{n_E,2k_E[J_E]}$ (where $J_E$ is a finite subset of
$I_E$ containing $K_E$ elements) to be
$$
F_{J_E}(\underline{x}^1,...,\underline{x}^K,\underline{\theta }^1,...,%
\underline{\theta }^K,\underline{\rho }^1,...,\underline{\rho }^K,...,$$ $$%
\underline{y}_{(p)}^1,...,\underline{y}_{(p)}^{K_p},\underline{\zeta }%
_{(p)}^1,...,\underline{\zeta }_{(p)}^{K_p},\underline{\rho }_{(p)}^1,...,%
\underline{\rho }_{(p)}^{K_p},...)=
$$
$$
f_{J_E}(\underline{x}^1,...,\underline{x}^K,\underline{y}_{(p)}^1,...,%
\underline{y}_{(p)}^{K_p},...)\times
$$
$$
\phi (\underline{\theta }^1,...,\underline{\theta }^K,\underline{\rho }%
^1,...,\underline{\rho }^K,...,\underline{\zeta }_{(p)}^1,...,\underline{%
\zeta }_{(p)}^{K_p},\underline{\rho }_{(p)}^1,...,\underline{\rho }%
_{(p)}^{K_p},...),
$$
where
$$
f_{J_E}(\underline{x}^1,...,\underline{x}^K,...,\underline{y}_{(p)}^1,...,%
\underline{y}_{(p)}^{K_p},...)=f_{J_E}\left( \underline{u}\right)
=
$$
$$
P_{t^1}(\underline{0},\underline{u}^1)...P_{t^K-t^{K-1}}(\underline{u}^{K-1},%
\underline{u}^K)
$$
with%
$$
P_t(\underline{u}_1,\underline{u}_2)=\left( \frac{2\pi }t\right)
^{n_E/2}\exp \left(
-\frac{(\underline{u}_1-\underline{u}_2)^2}{2t}\right)
$$
and
$$
d\mu _E^{(r)}=\phi _{J_E}(\underline{\theta }^1,...,\underline{\theta }^K,%
\underline{\rho }^1,...,\underline{\rho }^K,...,\underline{\zeta }%
_{(p)}^1,...,\underline{\zeta }_{(p)}^{K_p},\underline{\rho }_{(p)}^1,...,%
\underline{\rho }_{(p)}^{K_p},...)=\eqno(5.3)
$$
$$
\exp [-i(\underline{\rho }^1\cdot \underline{\theta }^1+\underline{\rho }%
^2\cdot (\underline{\theta }^2-\underline{\theta }^1)+...+\underline{\rho }%
^K\cdot (\underline{\theta }^K-\underline{\theta }^{K-1})+...+
$$
$$
\underline{\rho }_{(p)}^1\cdot \underline{\zeta }^1+\underline{\rho }%
_{(p)}^2\cdot (\underline{\zeta }_{(p)}^2-\underline{\zeta }_{(p)}^1)+...+%
\underline{\rho }_{(p)}^{K_p}\cdot (\underline{\zeta }_{(p)}^{K_p}-%
\underline{\zeta }_{(p)}^{K_p-1})+...)]
$$
is the distinguished super Wiener measure. \index{Measure!super
Wiener} \index{Distinguished!super Wiener measure}

Our aim is to relate fermionic Wiener integrals to purely bosonic
integrals. In order to do this we introduce definitions:

\begin{definition}
a) Let $\{F_s:0\leq s\leq t\},$ where $t$ is a positive number,
is a collection of random variables on ${\cal
R}_{l_E}^{n_E,2k_E[0,t]}.$ Than $F_s$ is said to be a
distinguished stochastic pro\-cess on this space.

\index{distinguished stochastic pro\-cess}

b) If to the set of random variables $F_s,$ for each $s\in
I=[0,t],$ one corresponds a sequence of pairs $\left(
J_{s,A},F_{s,A}:A=1,2,...\right) ,$ and $J_{s,A}\subset [0,s],\,$
the stochastic process $F_s$ is adapted.
\end{definition}

We note that a usual definition of adapted stochastic processes
was not \index{Stochastic process!adapted} \index{Stochastic
process!distinguished} possible because on super Wiener spaces we
are not dealing directly with sets of $\sigma $ algebras of
measurable sets.

\begin{definition}
Let $J_E$ be a finite subset of $I_E$ containing $K_E$ elements
and function
$$
f ( \underline{x}^1,...,\underline{x}^K,\underline{\theta }^1,...,%
\underline{\theta }^K,\underline{\rho }^1,...,\underline{\rho
}^K,...,$$ $$
\underline{y}_{(p)}^1,...,\underline{y}_{(p)}^{K_p},\underline{\zeta }%
_{(p)}^1,...,\underline{\zeta }_{(p)}^{K_p},\underline{\rho }_{(p)}^1,...,%
\underline{\rho }_{(p)}^{K_p},...) =
$$
$$
\sum\limits_{<\mu >\in M_{n_{K_E}}}\sum\limits_{<\nu >\in
M_{n_{K_E}}}f_{<\mu >}^{<\nu >}\left( \underline{x}^1,...,\underline{x}%
^K,...,\underline{y}_{(p)}^1,...,\underline{y}_{(p)}^{K_p},...\right)
\zeta ^{<\mu >}\rho ^{<\nu >}
$$
$$
\in L^{2^{\prime }}\left( {\cal
R}_{l_E}^{n_E,2k_E[J_{M_E}]},{\cal C}\right) .
$$
The functions $\left| f\right| _1\in L^2\left( {\cal R}^{n_EK_E},{\cal R}%
\right) $ and $\left| f\right| _2\in L^1\left( {\cal R}^{n_EK_E},{\cal R}%
\right) $ are defined by
$$
|f|_1\left( \underline{x}^1,...,\underline{x}^K,...,\underline{y}%
_{(p)}^1,...,\underline{y}_{(p)}^{K_p},...\right) =
$$
$$
\sum\limits_{<\mu >\in M_{n_{K_E}}}\sum\limits_{<\nu >\in
M_{n_{K_E}}}|f_{<\mu >}^{<\nu >}\left( \underline{x}^1,...,\underline{x}%
^K,...,\underline{y}_{(p)}^1,...,\underline{y}_{(p)}^{K_p},...\right)
|
$$
and%
$$
(|f|_2\left( \underline{x}^1,...,\underline{x}^K,...,\underline{y}%
_{(p)}^1,...,\underline{y}_{(p)}^{K_p},...\right) )^2=
$$
$$
\sum\limits_{<\mu >\in M_{n_{K_E}}}\sum\limits_{<\nu >\in
M_{n_{K_E}}}|f_{<\mu >}^{<\nu >}\left( \underline{x}^1,...,\underline{x}%
^K,...,\underline{y}_{(p)}^1,...,\underline{y}_{(p)}^{K_p},...\right)
|^2.
$$
\end{definition}

For instance, if a distinguished random variable $Gr$ on super
Wiener space, defined finitely on $J_E\subset I_E,\,$ is written
as
$$
Gr ( \underline{x}^1,...,\underline{x}^K,\underline{\theta }^1,...,%
\underline{\theta }^K,\underline{\rho }^1,...,\underline{\rho }^K,...,$$ $$%
\underline{y}_{(p)}^1,...,\underline{y}_{(p)}^{K_p},\underline{\zeta }%
_{(p)}^1,...,\underline{\zeta }_{(p)}^{K_p},\underline{\rho }_{(p)}^1,...,%
\underline{\rho }_{(p)}^{K_p},...) =\eqno(5.4)
$$
$$
\sum\limits_{<\mu >\in M_{n_{K_E}}}\sum\limits_{<\nu >\in
M_{n_{K_E}}}Gr_{<\mu >}^{<\nu >}\left( \underline{x}^1,...,\underline{x}%
^K,...,\underline{y}_{(p)}^1,...,\underline{y}_{(p)}^{K_p},...\right)
\zeta ^{<\mu >}\rho _{<\nu >}
$$
the expectation ${\cal E}_b(|Gr|_1)$ with respect to bosonic
Wiener measure
is computed by using formula%
$$
{\cal E}_b(|Gr|_1)=\sum\limits_{<\mu >\in M_{n_{K_E}}}\int
du^{n_EK_E}\sum\limits_{<\nu >\in M_{n_{K_E}}}f_{J_E} ( \underline{x}%
^1,...,\underline{x}^K,...,\underline{y}_{(p)}^1,...,\underline{y}%
_{(p)}^{K_p},... )
$$
$$   \times
|Gr_{<\mu >}^{<\nu >}\left( \underline{x}^1,...,\underline{x}^K,...,%
\underline{y}_{(p)}^1,...,\underline{y}_{(p)}^{K_p},...\right) |.
\eqno(5.5)
$$

\begin{lemma}
Let ${\cal E}_s(Gr)$ be the expectation of variable (3.4) with
respect to distinguished super Wiener measure (3.3) and ${\cal
E}_b(|Gr|_1)\,$ is the bosonic expectation (3.5). Then $|{\cal
E}_s(Gr)|\leq {\cal E}_b(|Gr|_1).$
\end{lemma}

\index{Expectation}
{\it Proof.} One holds the inequality%
$$
{\cal E}_b(|Gr|)\leq \sum\limits_{<\mu >\in
M_{n_{K_E}}}\sum\limits_{<\nu
>\in M_{n_{K_E}}}|\int d\widehat{u}^{n_EK_E}d\zeta ^{n_EK_E}d\rho
^{n_EK_E}\times
$$
$$
f_{J_E}\left( \underline{x}^1,...,\underline{x}^K,...,\underline{y}%
_{(p)}^1,...,\underline{y}_{(p)}^{K_p},...\right) \times
$$
$$
Gr_{<\mu >}^{<\nu >}\left( \underline{x}^1,...,\underline{x}^K,...,%
\underline{y}_{(p)}^1,...,\underline{y}_{(p)}^{K_p},...\right)
\zeta ^{<\mu
>}\rho _{<\nu >}d\mu _E^{(r)}|\leq
$$
$$
\sum\limits_{<\mu >\in M_{n_{K_E}}}\sum\limits_{<\nu >\in M_{n_{K_E}}}|\int d%
\widehat{u}^{n_EK_E}f_{J_E}\left( \underline{x}^1,...,\underline{x}^K,...,%
\underline{y}_{(p)}^1,...,\underline{y}_{(p)}^{K_p},...\right)
\times
$$
$$
Gr_{<\mu >}^{<\nu >}\left( \underline{x}^1,...,\underline{x}^K,...,%
\underline{y}_{(p)}^1,...,\underline{y}_{(p)}^{K_p},...\right)
$$
because for even Grassmann elements on every distinguished
''shell'', whose squares are zero, theirs Taylor expansions
contain each non--zero element
with coefficient exactly 1. Thus%
$$
|{\cal E}_s(Gr)|\leq \sum\limits_{<\mu >\in M_{n_{K_E}}} \int d\widehat{u}%
^{n_EK_E}\sum\limits_{<\nu >\in M_{n_{K_E}}}f_{J_E}\left( \underline{x}%
^1,...,\underline{x}^K,...,\underline{y}_{(p)}^1,...,\underline{y}%
_{(p)}^{K_p},...\right)
$$
$$               \times
|Gr_{<\mu >}^{<\nu >}\left( \underline{x}^1,...,\underline{x}^K,...,%
\underline{y}_{(p)}^1,...,\underline{y}_{(p)}^{K_p},...\right) |={\cal E}%
_b(|Gr|_1).
$$
$\Box $

We conclude this section by emphasizing that on distinguished
vector superbundles with trivial N--connection structures we can
apply the Rogers'
supersymmetric stochastic calculus 
 [209,210,206,207] in a
\index{Supersymmetric!stochastic calculus} \index{Stochastic
calculus!supersymmetric} straightforward manner, step by step, on
every ''shell'' of local anisotropy.

\section{ Stochastic D--In\-teg\-ra\-tion}

The aim of this section is to generalize the It$\widehat{o}$
theorem
 [123,124,125] in a manner as to obtain the chain rule for the
distinguished supersymmetric stochastic differentiation.
\index{Stochastic differentiation!supersymmetric}
\index{Supersymmetric!stochastic differentiation}

The It$\widehat{o}$ formula in a special case of $k$--dimensional
Brownian motion can be written as
$$
f(b_t,t)-f(b_0,0)=\int\limits_0^t\sum\limits_{i=1}^k\partial
_if(b_s,s)db_s^i+\int_0^t\partial _sf(b_s,s)ds+\frac
12\int\limits_0^t\sum\limits_{i=1}^k\partial _i\partial
_if(b_s)ds,
$$
where $f$ is a suitable well--behaved function of ${\cal R}^k\times {\cal R}%
^{+}.$ In the case of purely fermionic Brownian motion one writes
$$
f(\theta _t,t)-f(\theta _0,0)=_I\int\limits_0^t\partial
_sf(\theta _s,s)ds,
$$
where ''=$_I"$ means that there are two random variables on ${\cal R}%
^{(0,2n_E)[(0,t)]}$ with equal expectations. For, simplicity we
shall consider $k=2n_E.$

Let $F_s$ be a [0,t]--adapted process on the distinguished super
Wiener space ${\cal R}_{l_E}^{n_E,2k_E[[0,t]]}.$ We suppose that
for $\forall s\in
I_E~F_s$ corresponds to the sequence of pairs of subsets and functions $%
\left( J_{s,A},F_{s,A};A=1,2,...\right)
,~J_A=\bigcup\limits_{r=1}^{2^A-1}J_{t_{r,}A},$ where for $r=1,...,2^A-1,t_r=%
\frac{rt}{2^A},$ and consider functions $K_A=\left(
\sum\limits_{r=1}^{2^A-1}F_{t_r,A}\frac t{2^A}\right) $ on ${\cal R}%
_{l_E}^{n_E,2k_E[J_A]}.$

\begin{definition}
a) One says that the sequence $\left( J_A,K_A\right) $ defines a
distinguished Grassmann random variable which is denoted
$\int_0^tF_sds,$ i.e. $F_s$ has a time integral if ${\cal
E}_s(K_A)$ tends to a limit as $A$ tends to infinity.
\index{Random variable!distinguished Grassmann}

b) Let $0<u<t,$ $p(A,u)$ is the greatest integer such that
$p(A,u)t/2^A$ and
$L_A^u=\sum\limits_{r=1}^{p(A,u)}F_{t_r,A}\frac t{2^A}.$ Than, if ${\cal E}%
_s(L_A^u)$ tends to a limit as $A\rightarrow \infty ,$ the
sequence of pairs $\left( J_A,L_A^u:A=1,2,...\right) $ defines a
distinguished Grassmann random variable which is denoted
$\int_0^uF_sds.$
\end{definition}

The definition 5.5 holds also good for distinguished generalized
random variables.

We introduce the next denotations: $M_\Omega ^{1^{\prime }}[0,t]$
is the set of all $[0,t]$--a\-dapt\-ed processes $F_s$ on the
distinguished super Wiener space ${\cal
R}_{l_E}^{n_E,2k_E[[0,t]]}$ such that $\int_0^t|F_s|_1ds$ exists
and is finite and ${\cal E}_b(|F_s|_1)$ is bounded on $[0,t];$ $
M_\Omega^{2^{\prime }}[0,t]$ denotes the set of all
[0,t]--adapted processes $F_s$ on the distinguished super Wiener
space ${\cal R}_{l_E}^{n_E,2k_E[[0,t]]}$
such that $$\int_0^t|F_s|_2^2 ds$$ exists and is finite and ${\cal E}%
_b(|F_s|_2^2)$ is bounded on $[0,t]$ (we use correspondingly sequences $$%
\left( J_{s,A},|F_{s,A}|_1:A=1,2,...\right) $$ and $$\left(
J_{s,A},|F_{s,A}|_2^2:A=1,2,...\right) $$ as in the definition
5.3 and note that following definition 5.5 the time integrals are
Riemann integrals).

\begin{theorem}
Let $F^{<\alpha >}\in M_\Omega ^{2^{\prime }}[0,t]$ for $<\alpha
>=1,2,...k,$
for each $A=1,2,...$ let $J_A=\{t_1,...,t_{2^A-1}\}$ with $t_r=rt/2^A$ for $%
r=1,...,2^A-1$ and consider
$$
G_A\left( \underline{\widehat{u}}^1,...,\underline{\widehat{u}}^{2^A-1},%
\underline{\zeta }^1,...,\underline{\zeta }^{2^A-1},\underline{\rho }^1,...,%
\underline{\rho }^{2^A-1}\right) =
$$
$$
\sum\limits_{<\alpha
>=1}^{n_E}\sum\limits_{r=1}^{2^A-1}F^{A,<\alpha
>}\left( \underline{\widehat{u}}^1,...,\underline{\widehat{u}}^r,\underline{%
\zeta }^1,...,\underline{\zeta }^r,\underline{\rho }^1,...,\underline{\rho }%
^r\right) \left( \underline{\widehat{u}}^{r+1,<\alpha >}-\underline{\widehat{%
u}}^{r,<\alpha >}\right)
$$
$$
\in L^{2^{\prime }}\left( {\cal
R}_{l_E}^{n_E,2k_E[J_{M_E}]},{\cal C}\right) .
$$
Then the sequence $(J_A,G_A)$ defines a distinguished Grassmann
random variable denoted $\int_0^t\sum_{<\alpha
>=1}^{n_E}F_s^{<\alpha
>}db_s^{<\alpha >},$ where $|\int_0^t\sum_{<\alpha >=1}^{n_E}F_s^{<\alpha
>}db_s^{<\alpha >}|_2^2$ is also a distinguished random variable, and one
holds the equality
$$
{\cal E}\left( |\int_0^t\sum_{<\alpha >=1}^{n_E}F_s^{<\alpha
>}db_s^{<\alpha
>}|_2^2\right) ={\cal E}\left( \int_0^t\sum_{<\alpha >=1}^{n_E}|F_s^{<\alpha
>}|_2^2ds\right) .
$$
\end{theorem}

{\it Proof.} We note that $\int d\mu _f\left( G_A\right) $
converges to a random variable on $n_E$--dimensional
distinguished Wiener space and, as a consequence $(J_A,G_A)$
defines a distinguished super random variable. From the standard
results
$$
{\cal E}_b\left( (b_t-b_s)(b_t-b_s)\right) =|s-t|
$$
and using that, if $q<r,$
$$
F_{t_r,A<\mu ><\nu >}\left( \underline{\widehat{u}}^1,...,\underline{%
\widehat{u}}^r\right) F_{t_q,A<\mu ><\nu >}\left( \underline{\widehat{u}}%
^1,...,\underline{\widehat{u}}^q\right) (u^{q+1}-u^q)
$$
is independent of $(u^{r+1}-u^r)$ we obtain
$$
{\cal E}_b\left( \left| G_A\right| _2^2\right) ={\cal E}_b\left(
\sum\limits_{r=1}^{2^A-1}|F_{t_r,A}|_2^2(t_{r+1}-t_r)\right)
$$
which tends to a correct limit as $A\rightarrow \infty \,$ since $F$ is in $%
M_\Omega ^{2^{\prime }}[0,t].\Box $

Combining both type of integrations we obtain the following
definition of distinguished stochastic integral.

\begin{definition}
Let $t$ be a positive real number and consider a stochastic
process on distinguished super Wiener space ${\cal
R}_{l_E}^{n_E,2k_E[[0,t]]}$ such that
$$
Z_{t_1}-Z_{t_2}=\int\limits_{t_1}^{t_2}A_sds+\int\limits_{t_1}^{t_2}\sum%
\limits_{<\alpha >=1}^{n_E}C_s^{<\alpha >}db_s^{<\alpha >},
$$
where $0\leq t_1\leq t_2\leq t$ and $A_s\in M_\Omega ^{1^{\prime
}}[0,t]$ and $C_s^{<\alpha >}\in M_\Omega ^{2^{\prime }}[0,t]$
are [0,t]--adapted processes on the distinguished Wiener
space.Then $Z_s$ is said to be a distinguished stochastic
in\-teg\-ral.
\end{definition}
\index{Distinguished!stochastic in\-teg\-ral}

Now we formulate the key result of this section (the generalized
for dis\-ting\-uish\-ed su\-per\-spa\-ces It\^o theorem):
\index{It\^o theorem}
\begin{theorem}
Consider distinguished stochastic integrals $Z_s^{<\beta >}$\\
$(<\beta
>=1,2,...,n_E)$ on the distinguished super Wiener space ${\cal R}%
_{l_E}^{n_E,2k_E[[0,t]]}$ with
$$
Z_s^{<\underline{\beta }>}-Z_0^{<\underline{\beta }>}=\int\limits_0^tA_s^{<%
\underline{\beta }>}dt+\int\limits_{t_1}^{t_2}\sum\limits_{<\alpha
>=1}^{n_E}C_s^{<\alpha ><\underline{\beta }>}db_s^{<\alpha >}(s)
$$
and suppose that ${\cal E}\left( A_s^{<\underline{\beta
}>}\right) ^2$ is bounded for $\forall s\in [0,t].$ Then, if\\
$H\in C^{5^{\prime }}\left( {\cal R}^{(n_E,k_E)},{\cal C}\right)
$ and $Z_{s,A}^{<\beta >}$ is the $A$th
term in the sequence defining the random variable $Z_s^{<\underline{\beta }%
>},$ the sequence $H(Z_{s,A}^{<\underline{\beta }>})$ defines a
distinguished stochastic process according the stochastic integral%
$$
H_s-H_0=_I\int\limits_0^t\sum\limits_{<\underline{\beta }>}^{n_E+k_E}\sum%
\limits_{<\alpha >}^{n_E}\delta _{<\underline{\beta }>}H\left(
Z_s\right) C_s^{<\alpha ><\underline{\beta }>}db_s^{<\alpha >}+
$$
$$
\int\limits_0^t(\sum\limits_{<\underline{\beta }>}^{n_E+k_E}A_s^{<\underline{%
\beta }>}\delta _{<\underline{\beta }>}H\left( Z_s\right) +
$$
$$
\frac 12\sum\limits_{<\alpha >}^{n_E}\sum\limits_{<\underline{\beta }%
>}^{n_E+k_E}\sum\limits_{<\underline{\gamma }>}^{n_E+k_E}C_s^{<\alpha ><%
\underline{\beta }>}C_s^{<\alpha ><\underline{\gamma }>}\delta _{<\underline{%
\gamma }>}\delta _{<\underline{\beta }>}H\left( Z_s\right) )ds.
$$
\end{theorem}

{\it Proof.} From%
$$
H\left( Z_{t,A}\right) =H\left( Z_{0,A}\right)
+\sum\limits_{r=1}^{2^A-1}\left( H\left( Z_{t_{r+1},A}\right)
-H\left( Z_{t_r,A}\right) \right)
$$
we have
$$
H\left( Z_{t_{r+1},A}\right) -H\left( Z_{t_r,A}\right) =\sum\limits_{<%
\underline{\beta }>}^{n_E+k_E}\Delta ^{rA}\left( Z^{<\underline{\beta }%
>}\right) \delta _{<\underline{\beta }>}H\left( Z_{t_r,A}\right) +
$$
$$
\sum\limits_{<\underline{\beta }>}^{n_E+k_E}\sum\limits_{<\underline{\gamma }%
>}^{n_E+k_E}\Delta ^{rA}\left( Z^{<\underline{\beta }>}\right) \Delta
^{rA}\left( Z^{<\underline{\gamma }>}\right) \delta _{<\underline{\beta }%
>}\delta _{<\underline{\gamma }>}H\left( Z_{t_r,A}\right) +R_r,
$$
where $\Delta ^{rA}\left( Z^{<\underline{\beta }>}\right) =Z_{t_{r+1},A}^{<%
\underline{\beta }>}-Z_{t_r,A}^{<\underline{\beta }>}\,$ and
$R_r$ is the remainder term to be analyzed. As in the classical
case (see a similar
considerations in 
 [209]) by using truncated Taylor series) one finds
that $\left| R_r\right| \leq const/2^{2A}$, where $const$ is independent of $%
A,$ and using lemma 5.1 we obtain%
$$
{\cal E}\left( H\left( Z_{t,A}\right) -H\left( Z_{0,A}\right) \right) ={\cal %
E(}\sum\limits_{r=1}^{2^A-1}(\sum\limits_{<\underline{\beta }%
>}^{n_E+k_E}\Delta ^{rA}\left( Z^{<\underline{\beta }>}\right) \delta _{<%
\underline{\beta }>}H\left( Z_{t_r,A}\right) +
$$
$$
\frac 12\sum\limits_{<\underline{\beta }>}^{n_E+k_E}\sum\limits_{<\underline{%
\gamma }>}^{n_E+k_E}\Delta ^{rA}\left( Z^{<\underline{\beta
}>}\right)
\Delta ^{rA}\left( Z^{<\underline{\gamma }>}\right) \delta _{<\underline{%
\beta }>}\delta _{<\underline{\gamma }>}H\left( Z_{t_r,A}\right)
))+R_A^{\prime },
$$
where $\left| R_A^{\prime }\right| <C/2^A$ for some $C$
independent of $A.$ So the result follows on taking limits as
$A\rightarrow \infty .\Box $

\section{ Su\-per\-sym\-met\-ric It\^{o} D--For\-mu\-las}

The purpose of the section is to derive some It$\widehat{o}$
formulas for random variables on $\left( m,m\right)
$--dimensional distinguished super Wiener space (in brief, we
call It\^o d--formulas). We shall develop a stochastic version of
calculus on a $\left( 1,1\right) $--dimensional superspace
(parametized by a real variable $t$ and an odd Grassmann variable
$\tau ).$ One defines the superderivative $D_T$ by the operator%
$$
D_T=\frac \partial {\partial \tau }+\tau \frac \partial {\partial
t}
$$
which acts on functions of the form
$$
F(t,\tau )=A(t)+\tau B(t),\eqno(5.6)
$$
where $A(t)$ has a time derivative ($A$ and $B$ may have either
Grassmann parity and $A(t)$ has a time derivative). If $B(t)$ is
also differentiable,
we have $D_T^2=\partial /\partial t$ and for a system where the Hamiltonian $%
H$ is the square of a supercharge $Q,$ which is an odd operator,
the imaginary--time Schr\"odinger equation $\frac{\partial
f}{\partial t}=-Hf$ has a square root $D_Tf=Qf.$ In our further
considerations we shall use the formula
$$
D_T\left( \exp (-Ht-Q\tau )\right) =\left( \exp (-Ht-Q\tau
)\right) (Q-2\tau H).\eqno(5.7)
$$

Now we define an integral with respect to a $\left( 1,1\right) $%
--dimensional variable $S=(s,\sigma )$ including both odd and
even limits
 [208]:%
$$
\int\limits_0^\tau d\sigma \int\limits_0^tdsF(s,\sigma )\doteq
\int d\sigma \int\limits_0^{t+\sigma \tau }dsF(s,\sigma ),
$$
where the even integral on the right--hand side os evaluated by regarding\\ $%
\int_0^udsF(s,\sigma )$ as a function of $u$ (evaluating this
expression when $u=t+\sigma \tau $ by Taylor expansion about
$t=u)$ and the integration with respect to the odd variable
$\sigma $ is then carried out as Berezin integration. For a
function of type (5.6) we have
$$
\int\limits_0^\tau d\sigma \int\limits_0^tdsF(s,\sigma )=\tau
A(t)+\int\limits_0^tB(s)ds.
$$
From the above definition of integration one follows the next
important for
calculations result:%
$$
\int\limits_0^\tau d\sigma \int\limits_0^tdsD_SF(s,\sigma
)=F(t,\tau )-F(0,0)
$$
and (the rules for the superderivatives of such integral with
respect to its
upper limits)%
$$
D_T\int\limits_0^\tau d\sigma \int\limits_0^tdsG(s,\sigma
)=G(t,\tau ).
$$

\begin{definition}
Let $G_s$ and $H_s$ are adapted stochastic processes on
distinguished super
Wiener spaces and $H_s\in M_\Omega ^{1^{\prime }}[0,t].$ Then, if $%
F_s=G_s+\sigma H_s,$%
$$
\int\limits_0^\tau d\sigma \int\limits_0^tdsF_S\doteq
\int\limits_0^tdsH_s+\tau G_t.
$$
\end{definition}

We note that in this monograph we follow the convention that
repeated indices are to be summed over their range. The
superpaths $u_S$ and $\zeta _{S}$ defined below are the
stochastic version of the standard superfields used in
supersymmetric quantum mechanics.

\begin{theorem}
Let $u_s^{<\beta >}$ and $f_{(s)<\underline{\alpha }>}^{<\beta
>}$ be adapted stochastic processes on $\left( k,k\right)
$--dimensional
distinguished super Wiener space (in this section indices takes values $%
<\beta >=1,2,...,n_E$ and $<\underline{\alpha }>=1,2,...,k)$ and
$\phi $ be a function in $C^{5^{\prime }}\left( {\cal
R}^{n_E,k}\right) $ which is
linear in the odd argument and let $\left( \underline{u},\underline{\zeta }%
\right) \in {\cal R}_l^{(n_E,k)}.$ For any $q\in C^{5^{\prime }}\left( {\cal %
R}\right) $ let
$$
q_{t,\tau }\left( \widehat{u},\zeta \right) \doteq q\left(
\int\limits_0^\tau d\sigma \int\limits_0^tds~\phi \left( u+u_S+\frac i{\sqrt{%
2}}\sigma \eta _s,\zeta +\zeta _S\right) \right)
$$
where
$$
u_S^{<\beta >}=u_s^{<\beta >}+\frac i{\sqrt{2}}\sigma \xi _s^{<\underline{%
\alpha }>}f_{(s)<\underline{\alpha }>}^{<\beta >},~\zeta _S^{<\underline{%
\alpha }>}=\xi _s^{<\underline{\alpha }>}+\sqrt{2}i\sigma \ddot a_s^{<%
\underline{\alpha }>},
$$
$$
\eta _s^{<\beta >}=\zeta ^{<\underline{\alpha }>}f_{(s)<\underline{\alpha }%
>}^{<\beta >}, $$
$$\xi _s^{<\underline{\alpha }>}=\zeta _s^{<\underline{\alpha }%
>}-i\rho _s^{<\underline{\alpha }>}
$$
and the notation $\ddot a_s^{<\underline{\alpha }>}$ is to be
interpreted in combination with $ds$ as \\ $\ddot
a_s^{<\underline{\alpha }>}ds=da_s.$ Then
$$
q_{t,\tau }\left( \widehat{u},\zeta \right)
-q_{0,0}=_I\int\limits_0^\tau d\sigma
\int\limits_0^tds~\{q_{s,\sigma }^{\prime }\left(
\widehat{u},\zeta \right) \phi \left( u+u_S+\frac
i{\sqrt{2}}\sigma \eta _s,\zeta +\zeta _S\right) -
$$
$$
[2\sigma q_{s,\sigma }^{\prime \prime }\left( \widehat{u},\zeta
\right) \delta _{<\underline{\alpha }>}\phi \left( u+u_S+\frac
i{\sqrt{2}}\sigma \eta _s,\zeta +\zeta _S\right)
$$
$$
\delta _{<\underline{\alpha }>}\phi \left( u+u_S+\frac
i{\sqrt{2}}\sigma \eta _s,\zeta +\zeta _S\right) ]\}.
$$
\end{theorem}

{\it Proof. }We have%
$$
q_{t,\tau }\left( \widehat{u},\zeta \right) -q_{0,0}\left(
\widehat{u},\zeta \right) =
$$
$$
q_{t,0}\left( \widehat{u},\zeta \right) +\tau \phi (u+u_t,\zeta
+\xi _t)q_{t,0}^{\prime }-q_{0,0}=\tau \phi (u+u_t,\zeta +\xi
_t)q_{t,0}^{\prime }+
$$
$$
q(\int\limits_0^tds\left( \frac i{\sqrt{2}}(\zeta ^{<\underline{\alpha }%
>}+\xi _s^{<\underline{\alpha }>})f_{(s)<\underline{\alpha }>}^{<\beta
>}\right) \delta _{<\beta >}\phi (u+u_s,\zeta +\xi _s))+
$$
$$
\sqrt{2}i\int\limits_0^tda_s^{<\underline{\alpha }>}\delta _{<\underline{%
\alpha }>}\phi (u+u_s,\zeta +\xi _s)-q_{0,0}\left(
\widehat{u},\zeta \right) =_I
$$
$$
\tau \phi (u+u_t,\zeta +\xi _t)q_{t,0}^{\prime }\left(
\widehat{u},\zeta \right)
+\sqrt{2}i\int\limits_0^tda_s^{<\underline{\alpha
}>}q_{0,s}^{\prime }\left( \widehat{u},\zeta \right) \delta
_{<\underline{\alpha }>}\phi (u+u_s,\zeta +\xi _s)+
$$
$$
\int\limits_0^tds(\frac i{\sqrt{2}}q_{s,0}^{\prime }(\zeta ^{<\underline{%
\alpha }>}+\xi _s^{<\underline{\alpha }>})f_{(s)<\underline{\alpha }%
>}^{<\beta >}\delta _{<\beta >}\phi (u+u_s,\zeta +\xi _s)-
$$
$$
q_{s,0}^{\prime \prime }\delta _{<\underline{\alpha }>}\phi
(u+u_s,\zeta +\xi _s)\delta _{<\underline{\alpha }>}\phi
(u+u_s,\zeta +\xi _s))
$$
and (using theorem 5.2)%
$$
\int\limits_0^\tau d\sigma \int\limits_0^tds~q_{s,\sigma
}^{\prime }\left( \widehat{u},\zeta \right) \phi \left(
u+u_S+\frac i{\sqrt{2}}\sigma \eta _s,\zeta +\zeta _S\right) =_I
$$
$$
\tau \phi (u+u_t,\zeta +\xi _t)q_{t,0}^{\prime }\left(
\widehat{u},\zeta \right)
+\sqrt{2}i\int\limits_0^tda_s^{<\underline{\alpha
}>}q_{0,s}^{\prime }\left( \widehat{u},\zeta \right) \delta
_{<\underline{\alpha }>}\phi (u+u_s,\zeta +\xi _s)+
$$
$$
\frac i{\sqrt{2}}\int\limits_0^tdsq_{s,0}^{\prime }(\zeta ^{<\underline{%
\alpha }>}+\xi _s^{<\underline{\alpha }>})f_{(s)<\underline{\alpha }%
>}^{<\beta >}\delta _{<\beta >}\phi (u+u_s,\zeta +\xi _s)+
$$
$$
\int\limits_0^tds~\phi (u+u_s,\zeta +\xi _s)q_{s,0}^{\prime
\prime }\left( \widehat{u},\zeta \right) \phi (u+u_s,\zeta +\xi
_s).
$$
The final result follows from the relation%
$$
\int\limits_0^tds~\phi (u+u_s,\zeta +\xi _s)q_{s,0}^{\prime
\prime }\left( \widehat{u},\zeta \right) \phi (u+u_s,\zeta +\xi
_s)=_I
$$
$$
\int\limits_0^tdsq_{s,0}^{\prime \prime }\left( \widehat{u},\zeta
\right) \sum\limits_{<\underline{\alpha }>}^{n_E}\delta
_{<\underline{\alpha }>}\phi (u+u_s,\zeta +\xi _s)\delta
_{<\underline{\alpha }>}\phi (u+u_s,\zeta +\xi _s).
$$
$\Box $

The second basic result in this section is a distinguished
supersymmetric version of the restricted It\^o theorem.
\index{Restricted It\^o theorem}

\begin{theorem}
Let $\widehat{u}_s^{<\alpha >}$ be an adapted distinguished
stochastic
process such that%
$$
\widehat{u}_s^{<\alpha >}-\widehat{u}_0^{<\alpha >}=\int\limits_0^tl_{<%
\underline{\beta }>}^{<\alpha >}(\widehat{u}_s)da_s^{<\underline{\beta }%
>}+\int\limits_0^tL_{<\underline{\beta }>}^{<\alpha >}ds,
$$
where for $<\alpha >=1,...,n_E$ and $<\underline{\beta
}>=1,...,k$ (we can
consider, for instance, $k=n_E),$ $l_{<\underline{\beta }>}^{<\alpha >}(%
\widehat{u}_s)$ is a function on ${\cal R}^{n_E}$ and $L_{<\underline{\beta }%
>}^{<\alpha >}$ is an adapted distinguished stochastic process on $\left(
k,k\right) $--dimensional distinguished super Wiener space. Then,
using notations of theorem 5.3, there is a function $f\in {\it
C}^{5^{\prime
}}\left( {\cal R}^{n_E,k}\right) $ satisfying conditions%
$$
f\left( \widehat{u}+\widehat{u}_T+\frac i{\sqrt{2}}\tau \eta
,\zeta +\zeta _s\right) -f(\widehat{u},\zeta )=_I
$$
$$
\int\limits_0^\tau d\sigma \int\limits_0^tds[\psi ^{<\underline{\beta }>}l_{<%
\underline{\beta }>}^{<\alpha >}(\widehat{u}+\widehat{u}_s)\delta
_{<\alpha
>}f\left( \widehat{u}+\widehat{u}_S+\frac i{\sqrt{2}}\tau \eta ,\zeta +\zeta
_s\right) +
$$
$$
\sigma l_{<\underline{\beta }>}^{<\alpha >}(\widehat{u}+\widehat{u}_s)l_{<%
\underline{\beta }>}^{<\beta >}(\widehat{u}+\widehat{u}_s)\delta
_{<\alpha
>}\delta _{<\beta >}f\left( \widehat{u}+\widehat{u}_S+\frac i{\sqrt{2}}\tau
\eta ,\zeta +\zeta _s\right) +
$$
$$
\int\limits_0^\tau d\sigma \int\limits_0^tdu_s^{<\alpha >}~\sigma
~\delta _{<\alpha >}f\left( \widehat{u}+\widehat{u}_S+\frac
i{\sqrt{2}}\tau \eta ,\zeta +\zeta _s\right) .
$$
\end{theorem}

{\it Proof.} Applying the It\^o\thinspace formula from the
theorem 5.3 we
write%
$$
f\left( \widehat{u}+\widehat{u}_T+\frac i{\sqrt{2}}\tau \eta
,\zeta +\zeta _s\right) -f(\widehat{u},\zeta )=_I
$$
$$
\int\limits_0^\tau d\sigma \int\limits_0^tdu_s^{<\alpha >}~\sigma
~\delta _{<\alpha >}f\left( \widehat{u}+\widehat{u}_s,\zeta
+\zeta _s\right) +
$$
$$
\frac i{\sqrt{2}}\tau (\zeta ^{<\underline{\alpha }>}+\xi _t^{<\underline{%
\alpha }>})l_{<\underline{\alpha }>}^{<\alpha >}(\widehat{u}+\widehat{u}%
_s)\delta _{<\alpha >}f\left( \widehat{u}+\widehat{u}_t,\zeta
+\zeta _t\right) +
$$
$$
\frac 12\int\limits_0^\tau d\sigma \int\limits_0^tdu_s^{<\alpha
>}~\sigma
~l_{<\underline{\alpha }>}^{<\alpha >}(\widehat{u}+\widehat{u}_s)l_{<%
\underline{\alpha }>}^{<\beta >}(\widehat{u}+\widehat{u}_s)\delta
_{<\alpha
>}\delta _{<\beta >}f\left( \widehat{u}+\widehat{u}_s,\zeta +\zeta _s\right)
.
$$
The result follows from a simple integration on $s$ and a
corresponding regroupation of terms.$\Box $

The theorems 5.3 and 5.4 (which are extensions for distinguished
classes of
indices of similar results proved by Alice Rogers 
 [209]) will be
used in section 5.5 for a proof of a variant of Feynman--Kac
formula for locally anisotropic stochastic processes.

\section{ Feynman--Kac D--Formula}

The subject of this section is to show that a reasonable class of
stochastic \index{Stochastic d--equations} differential equations
on locally anisotropic superspaces can be solved. The main
theorem will establish that distinguished differential equations
of the
form%
$$
dU_s^{<\alpha >}=A_{<\underline{\alpha }>}^{<\alpha >}(U_s,\zeta
_s,\rho
_s,s)da_s^{<\underline{\alpha }>}+B^{<\alpha >}(U_s,\zeta _s,\rho _s,s)ds%
\eqno(5.8)
$$
have unique solutions if coefficients $A_{<\underline{\alpha
}>}^{<\alpha >}$ and $B^{<\alpha >}$ satisfy some regularity
conditions.

We explain precisely what we mean by a distinguished stochastic
differential equation (in brief, stochastic d--equation) of form
(5.8):

\begin{definition}
Let coefficients $B^{<\alpha >}(U_s,\zeta _s,\rho _s,s)$ and $A_{<\underline{%
\alpha }>}^{<\alpha >}(U_s,\zeta _s,\rho _s,s)$ take
correspondingly values
in $M_\Omega ^{1^{\prime }}[0,t]$ and $M_\Omega ^{2^{\prime }}[0,t]$ and $%
A\in {\cal C}^{n_E}.\,$ Than if $U_s^{<\alpha >}$ is a
distinguished stochastic process in $M_\Omega ^{2^{\prime
}}[0,t]$ such that
$$
U_s^{<\alpha >}-U_0^{<\alpha >}=\int\limits_0^tA_{<\underline{\alpha }%
>}^{<\alpha >}(U_s,\zeta _s,\rho _s,s)da_s^{<\underline{\alpha }%
>}+\int\limits_0^tB^{<\alpha >}(U_s,\zeta _s,\rho _s,s)ds,\eqno(5.9a)
$$
$$
U_0=A,\eqno(5.9b)
$$
where $\left( a_s,\zeta _s,\rho _s\right) $ are Brownian paths in
$\left(
k,k_E\right) $--dimensional distinguished super Wiener space one says that $%
U_s$ satisfies the stochastic differential d--equation (5.9a)
with initial conditions\ (5.9b).
\end{definition}

We note that two random variables $F$ and $G$ are said to be equal if ${\cal %
E}(|F - G|_2 )=0. $

The basic result of this section is formulated as

\begin{theorem}
Suppose that under conditions of definition 5.8 and all $s\in
[0,t]$ and all $\left( <\mu >,<\nu >\right) \in \left(
A_{k_E}\times A_{k_E}\right) $the
coefficient functions of $A_{<\underline{\alpha }>}^{<\alpha >}$ and $%
B^{<\alpha >}$ satisfy conditions%
$$
\left| A_{<\underline{\alpha }><\nu >}^{<\alpha ><\mu >}(\widehat{u}%
_1,s)-A_{<\underline{\alpha }><\nu >}^{<\alpha ><\mu >}(\widehat{u}%
_2,s)\right| \leq C\frac{\left| \widehat{u}_1-\widehat{u}_2\right| }{%
n_Ek\times 2^{2k_E}},
$$
$$
\left| A_{<\underline{\alpha }><\nu >}^{<\alpha ><\mu >}(\widehat{u}%
_1,s)\right| \leq C\frac{1+\left| \widehat{u}_1\right| }{n_Ek\times 2^{2k_E}}%
,
$$
$$
\left| B_{<\nu >}^{<\alpha ><\mu >}(\widehat{u}_1,s)-B_{<\nu
>}^{<\alpha
><\mu >}(\widehat{u}_2,s)\right| \leq C\frac{\left| \widehat{u}_1-\widehat{u}%
_2\right| }{n_E\times 2^{2k_E}},
$$
$$
\left| B_{<\nu >}^{<\alpha ><\mu >}(\widehat{u}_1,s)\right| \leq C\frac{%
1+\left| \widehat{u}_1\right| }{n_E\times 2^{2k_E}},
$$
for some fixed positive number $C.$ Then there exists a unique
solution to
the stochastic differential d--equation%
$$
dU_s^{<\alpha >}=A_{<\underline{\alpha }>}^{<\alpha >}(U_s,\zeta
_s,\rho
_s,s)da_s^{<\underline{\alpha }>}+B^{<\alpha >}(U_s,\zeta _s,\rho _s,s)ds,%
\eqno(5.10)
$$
$$
U_0=A.
$$
\end{theorem}

{\it Proof. }Supposing that $\widehat{u}_s$ and $\widehat{v}_s$
are two
solutions of (5.10) and using theorem 5.1 we can write%
$$
{\cal E}|\widehat{u}_s-\widehat{v}_s|_2^2\leq
2\int\limits_0^s\left|
B^{<\alpha >}(\widehat{u}_z,\zeta _z,\rho _z,z)-B^{<\alpha >}(\widehat{v}%
_z,\zeta _z,\rho _z,z)\right| _2^2dz+
$$
$$
2\int\limits_0^s\left| A_{<\underline{\beta }>}^{<\alpha >}(\widehat{u}%
_z,\zeta _z,\rho _z,z)-A_{<\underline{\beta }>}^{<\alpha >}(\widehat{v}%
_z,\zeta _z,\rho _z,z)\right| _2^2dz\leq
$$
$$
2C^2(1+s)\int\limits_0^s\left| \widehat{u}_s-\widehat{v}_s\right|
_2^2dz,
$$
i.e. ${\cal E}|\widehat{u}_s-\widehat{v}_s|_2^2=0~\forall s\in
[0,t],$ and thus any solution which exists is unique.

The existence of solutions is established by considering the sequence $%
U_{s(r)}^{<\alpha >}$ $(r=1,2,...)$ of stochastic processes on the
distinguished super Wiener space\\ ${\cal R}_l^{(n_E,2k_E)[[0,t]]}$ with $%
U_{s(0)}^{<\alpha >}=0$ and, for $r>0,$%
$$
U_{s(r)}^{<\alpha >}=A+\int\limits_0^sB^{<\alpha
>}(U_{s(r-1)},\zeta _z,\rho
_z,z)dz+\int\limits_0^sA_{<\underline{\beta }>}^{<\alpha
>}(U_{s(r-1)},\zeta _z,\rho _z,z)da_s^{<\underline{\beta }>}.
$$

The next step is the following inductive hyposes, for all
integers $q$ up to and including some fixed integer $r:$

a) $U_{s(r)}^{<\alpha >}$ take values in $M_\Omega ^{2^{\prime
}}[0,t];$

b) ${\cal E(}|U_{s(q)}^{<\alpha >}-U_{s(q-1)}^{<\alpha >}|_2^2)\leq \frac{%
(C_1s)^q}{q!},\,$where $C_1$ is a positive constant.

So, $U_{s(1)}^{<\alpha >}$ is well defined and%
$$
{\cal E}|U_{s(1)}^{<\alpha >}-U_{s(0)}^{<\alpha >}|_2^2\leq
2|\int\limits_0^sB^{<\alpha >}(u_0,\zeta _z,\rho
_z,z)dz|_2^2+2\int\limits_0^s|A_{<\underline{\beta }>}^{<\alpha
>}(u_0,\zeta _z,\rho _z,z)|_2^2dz,
$$
from which one follows%
$$
{\cal E}|U_{s(1)}^{<\alpha >}-U_{s(0)}^{<\alpha >}|_2^2\leq
2C^2s^2+C^2s(1+|A|^2)\leq C_1\mbox{ if }C_1\geq
2C^2(t+1)(1+|A|^2).
$$
The inductive step
$$
{\cal E}|U_{s(r+1)}^{<\alpha >}-U_{s(r)}^{<\alpha >}|_2^2\leq
C_1\int\limits_0^t{\cal E}|U_{s(r)}^{<\alpha >}- $$
$$U_{s(r-1)}^{<\alpha
>}|_2^2ds\leq C_1\int\limits_0^t\frac{(C_1s)^r}{r!}ds=\frac{(C_1t)^{r+1}}{%
(r+1)!}
$$
implies that $U_{s(r+1)}^{<\alpha >}$ takes values in $M_\Omega
^{2^{\prime
}}[0,t],$ i.e. the inductive hypothesis is satisfied and the sequence $%
U_{s(r)}^{<\alpha >}$ converges in the manner required by the
theorem.$\Box $

Now we formulate the distinguished variant of the main theorem
from
 [209] establishing a Feynman--Kac for\-mu\-la for a Hamiltonian $H$ which in
\index{Feynman--Kac!for\-mu\-la} our case is considered for
locally anisotropic interactions.

Let $\left( a_s,\zeta _s,\rho _s\right) $ denotes Brownian motion
in $\left( k,k\right) $--dimensional distinguished super Wiener
space, $g$ is a $C^5$ Riemannian metric on ${\cal R}^k$ which has
components $g_{<\alpha ><\beta
>}$ and consider an orthonormal basis $e^{<\underline{\alpha }>}$ of
distinguished 1--forms with components $e_{<\alpha
>}^{<\underline{\alpha }>}
$ satisfying conditions%
$$
\widehat{g}_{<\alpha ><\beta >}(\widehat{u})=e_{<\alpha >}^{<\underline{%
\alpha }>}(\widehat{u})e_{<\beta >}^{<\underline{\alpha
}>}(\widehat{u}).
$$
We consider that all derivatives of $e_{<\alpha
>}^{<\underline{\alpha }>}$ up to fifth order are required to be
uniformaly bounded on ${\cal R}^k.$ and consider $\widehat{u}_s$
to be the unique solution to the distinguished
stochastic differential equation%
$$
d\widehat{u}_s^{<\alpha >}=e_{<\underline{\alpha }>}^{<\alpha >}(\widehat{u}%
+u_s)da_s^{<\underline{\alpha }>}+
$$
$$
\frac 12(\delta ^{<\underline{\alpha }><\underline{\beta }>}+\zeta _s^{<%
\underline{\alpha }>}\zeta _s^{<\underline{\beta }>})e_{<\underline{\alpha }%
>}^{<\beta >}(\widehat{u}+u_s)\delta _{<\beta >}e_{<\underline{\alpha }%
>}^{<\alpha >}(\widehat{u}+u_s)ds,
$$
$$
\widehat{u}_0^{<\alpha >}=0,
$$
where $e_{<\underline{\alpha }>}^{<\alpha >}$ and $e_{<\underline{\alpha }%
>}^{<\beta >}\delta _{<\beta >}e_{<\underline{\beta }>}^{<\alpha >}$ are to
be extended to even Grassmann elements by Taylor expansion
truncated at the
first--derivative term, and introduce variables%
$$
\widehat{u}_S^{<\alpha >}=\widehat{u}_s^{<\alpha >}+\frac
i{\sqrt{2}}\sigma \zeta _s^{<\underline{\alpha
}>}e_{<\underline{\alpha }>}^{<\alpha >}\left(
\widehat{u}+\widehat{u}_s\right) ,
$$
$$
~\zeta _S^{<\underline{\alpha }>}=\zeta _s^{<\underline{\alpha }>}+\sqrt{2}%
i\sigma a_s^{<\underline{a}>}~\mbox{ and }~\eta _s^{<\alpha >}=\zeta ^{<%
\underline{\alpha }>}e_{<\underline{\alpha }>}^{<\alpha >}\left( \widehat{u}+%
\widehat{u}_s\right) .
$$

\begin{theorem}
Consider the operator $\psi ^{<\underline{\alpha }>}=\zeta ^{<\underline{%
\alpha }>}+\frac \delta {\partial \zeta ^{<\underline{\alpha
}>}}$ acting on
the spaces $L^{2^{\prime }}({\cal R}_l^{(k,k)})$ and $\varphi $ being a $%
{\it C}^5$ function on ${\cal R}^{(k,k)}$ which is linear in the
odd argument and has all derivatives up to fifth order uniformly
bounded. Then if $Q=\psi ^{<\underline{\alpha }>}\left(
e_{<\underline{\alpha }>}^{<\alpha
>}\left( \widehat{u}\right) \delta _{<\beta >}\right) +\varphi \left(
\widehat{u},\psi \right) $ acts on $L^{2^{\prime }}({\cal
R}_l^{(k,k)}),$
where $H=Q^2,F\in C^{5^{\prime }}\left( {\cal R}_l^{(k,k)}\right) $ and $%
\left( \widehat{u},\zeta \right) \in {\cal R}_l^{(n_E,k)},\,$ one
holds the formula
$$
\exp (-Ht-Q\tau )F\left( \widehat{u},\zeta \right) =
$$
$$
{\cal E}\{\exp [\int\limits_0^\tau d\sigma
\int\limits_0^tds~\varphi \left( \widehat{u}+\widehat{u}_S+\frac
i{\sqrt{2}}\sigma \eta _s,\zeta +\zeta _S\right) ]\times
$$
$$
F\left( \widehat{u}+\widehat{u}_T+\frac i{\sqrt{2}}\sigma \eta
_t,\zeta +\zeta _t\right) \}.
$$
\end{theorem}

{\it Proof. }We introduce on $L^{2^{\prime }}({\cal
R}_l^{(k,k)})$ the
operator $\overline{V}_{t,\tau }$ defined by completion of $V_{t,\tau }:{\it %
C}_0^{\infty ^{\prime }}({\cal R}_l^{(k,k)})\rightarrow L^{2^{\prime }}(%
{\cal R}_l^{(k,k)})$ with%
$$
V_{t,\tau }P(\widehat{u},\zeta )={\cal E}\{\exp
[\int\limits_0^\tau d\sigma
\int\limits_0^tds~\varphi \left( \widehat{u}+\widehat{u}_S+\frac i{\sqrt{2}%
}\sigma \eta _s,\zeta +\zeta _S\right) ]\times
$$
$$
P\left( \widehat{u}+\widehat{u}_T+\frac i{\sqrt{2}}\sigma \eta
_t,\zeta +\zeta _t\right) \},
$$
where $P\in {\it C}_0^{\infty ^{\prime }}({\cal R}_l^{(k,k)}).$
Applying the
It\^o formula for products to the integrand in $V_{t,\tau }f(\widehat{u}%
,\zeta ),$ taking expectations and after some algebraic transforms we obtain%
$$
V_{t,\tau }P(\widehat{u},\zeta )-P(\widehat{u},\zeta )=
$$
$$
\int\limits_0^\tau d\sigma \int\limits_0^tds~V_{s,\sigma }QP(\widehat{u}%
,\zeta )-2\sigma V_{s,\sigma }HP(\widehat{u},\zeta ).
$$
As a consequence,
$$
D_TV_{t,\tau }P(\widehat{u},\zeta )=V_{t,\tau }\left( Q-2\tau H\right) P(%
\widehat{u},\zeta ),
$$
using equation (5.7) and taking into account the uniqueness of
solutions to such differential equations we deduce as required by
the conditions of the
theorem that%
$$
V_{t,\tau }P(\widehat{u},\zeta )=\exp (-Ht-Q\tau
)P(\widehat{u},\zeta ).
$$
$\Box $

\section{S--Differential D--Forms}

The section is devoted to a geometric approach to supermanifolds
which will provide the appropriate arena for the Brownian paths
(see sections 5.3 and 5.4) in locally anisotropic superspaces. We
shall consider a class of superspaces which can be constructed in
a natural way from vector bundles over higher order anisotropic
space provided with a Riemannian metric structure.
\index{Distinguished s--manifolds} \index{Differential forms}

Let $\widetilde{{\cal E}}_N$ be a smooth, compact
$n_E$--dimensional real distinguished vector bundle $\left(
n_E=n+m_1+...+m_z\right) $ provided with N- and d--connection and
Riemannian metric structures and let $E_{H}$
be a smooth $m_H$--dimensional Hermitian vector bundle over $\widetilde{%
{\cal E}}_N.\,$ Suppose that $\{U_{\underbrace{\alpha }}|\underbrace{\alpha }%
\in \Lambda \}$ by sets which are both coordinate neighborhoods of $%
\widetilde{{\cal E}}_N$ and local trivializations of $E_H.$ For each $%
\underbrace{\alpha },\underbrace{\beta },...\in \Lambda $ we
consider
coordinate maps on $\widetilde{{\cal E}}_N$ of type $\phi _{\underbrace{%
\alpha }}:U_{\underbrace{\alpha }}\rightarrow {\cal R}^n$ and
transition
functions of the bundle $E_{H}$ of type $h_{\underbrace{\alpha }%
\underbrace{\beta }}:U_{\underbrace{\alpha }}\cap U_{\underbrace{\beta }%
}\rightarrow U(m_H)$ (for every point $q\in U_{\underbrace{\alpha }}\cap U_{%
\underbrace{\beta }}$ the transition functions are parametrized by unitary $%
m_H\times m_H$ matrices distinguished correspondingly to shells of
anisotropy). We denote by
$$
\{\tau _{\underbrace{\alpha }\underbrace{\beta }}:\phi _{\underbrace{\beta }%
}\left( U_{\underbrace{\alpha }}\cap U_{\underbrace{\beta
}}\right) \rightarrow \phi _{\underbrace{\alpha }}\left(
U_{\underbrace{\alpha }}\cap U_{\underbrace{\beta }}\right) \}
$$
the coordinate transition functions on $\widetilde{M}$ with $\tau _{%
\underbrace{\alpha }\underbrace{\beta }}=\phi
_{\underbrace{\alpha }}\circ
\phi _{\underbrace{\beta }}^{-1}$ and by $(m_{\underbrace{\alpha }%
\underbrace{\beta }})_{\quad <\varepsilon >}^{<\gamma
>}(\widehat{u})=\left(
\partial \widehat{u}_{\underbrace{\alpha }}^{<\gamma >}/\partial \widehat{u}%
_{\underbrace{\beta }}^{<\varepsilon >}\right) $ be the
corresponding Jacobian matrix.

In our further considerations we shall use the $\left( n_E,m_H+n_E\right) $%
--di\-men\-si\-on\-al distinguished supermanifold
 $S\left( E_H\right) $ built over $%
\widetilde{{\cal E}}_N$ with local coordinates%
$$
(x_{\underbrace{\alpha }}^1,...,x_{\underbrace{\alpha }}^n,\theta _{%
\underbrace{\alpha }}^1,...,\theta _{\underbrace{\alpha }}^n,\eta _{%
\underbrace{\alpha }}^1,...,\eta _{\underbrace{\alpha
}}^{m_H},...,
$$
$$
y_{(p)\underbrace{\alpha }}^1,...,y_{(p)\underbrace{\alpha
}}^{m_p},\zeta _{(p)\underbrace{\alpha }}^1,...,\zeta
_{(p)\underbrace{\alpha }}^{m_p},\eta _{(p)\underbrace{\alpha
}}^1,...,\eta _{(p)\underbrace{\alpha }}^{m_H},...),
$$
where $p=1,...,z,$ and transition functions%
$$
T_{\underbrace{\alpha }\underbrace{\beta }}:(x_{\underbrace{\alpha }%
}^1,...,x_{\underbrace{\alpha }}^n,\theta _{\underbrace{\alpha }%
}^1,...,\theta _{\underbrace{\alpha }}^n,\eta _{\underbrace{\alpha }%
}^1,...,\eta _{\underbrace{\alpha }}^{m_H},...,\eqno(5.11)
$$
$$
y_{(p)\underbrace{\alpha }}^1,...,y_{(p)\underbrace{\alpha
}}^{m_p},\zeta _{(p)\underbrace{\alpha }}^1,...,\zeta
_{(p)\underbrace{\alpha }}^{m_p},\eta
_{(p)\underbrace{\alpha }}^1,...,\eta _{(p)\underbrace{\alpha }%
}^{m_H},...)\rightarrow
$$
$$
(\tau _{\underbrace{\alpha }\underbrace{\beta }}^1(\widehat{u}_{\underbrace{%
\beta }}),...,\tau _{\underbrace{\alpha }\underbrace{\beta }}^n(\widehat{u}_{%
\underbrace{\beta }}),
$$
$$
(m_{\underbrace{\alpha }\underbrace{\beta }})_{\quad <i>}^{<1>}(\widehat{u}_{%
\underbrace{\beta }})\theta _{\underbrace{\alpha }}^{<i>},...,(m_{%
\underbrace{\alpha }\underbrace{\beta }})_{\quad <i>}^{<n>}(\widehat{u}_{%
\underbrace{\beta }})\theta _{\underbrace{\alpha }}^{<i>},
$$
$$
(h_{\underbrace{\alpha }\underbrace{\beta }})_r^1\left( \phi _{\underbrace{%
\beta }}^{-1}(\widehat{u}_{\underbrace{\beta }})\right) \eta _{\underbrace{%
\beta }}^r,...,(h_{\underbrace{\alpha }\underbrace{\beta
}})_r^{m_H}\left( \phi _{\underbrace{\beta
}}^{-1}(\widehat{u}_{\underbrace{\beta }})\right) \eta
_{\underbrace{\beta }}^r,...,
$$
$$
(\tau _{(p)\underbrace{\alpha }\underbrace{\beta }}^1(\widehat{u}_{%
\underbrace{\beta }}),...,\tau _{(p)\underbrace{\alpha }\underbrace{\beta }%
}^{m_p}(\widehat{u}_{\underbrace{\beta }}),
$$
$$
(m_{\underbrace{\alpha }\underbrace{\beta }})_{\quad <a_p>}^{<1>}(\widehat{u}%
_{\underbrace{\beta }})\theta _{\underbrace{\alpha }}^{<a_p>},...,(m_{%
\underbrace{\alpha }\underbrace{\beta }})_{\quad <a_p>}^{<m_p>}(\widehat{u}_{%
\underbrace{\beta }})\theta _{\underbrace{\alpha }}^{<a_p>},
$$
$$
(h_{\underbrace{\alpha }\underbrace{\beta }})_r^1\left( \phi _{\underbrace{%
\beta }}^{-1}(\widehat{u}_{\underbrace{\beta }})\right) \eta _{(p)%
\underbrace{\beta }}^r,...,
$$
$$
(h_{\underbrace{\alpha }\underbrace{\beta }})_r^{m_H}\left( \phi _{%
\underbrace{\beta }}^{-1}(\widehat{u}_{\underbrace{\beta
}})\right) \eta _{(p)\underbrace{\beta }}^r,...).
$$
(We shall omit indices of type $\underbrace{\alpha
},\underbrace{\beta }$ in our further considerations if this will
not give rise to ambiguities).

We denote by ${\it C}^{\infty ^{\prime }}\left( S\left(
E_H\right) \right) $
the space of functions $f$ which locally take the form%
$$
f\left( \widehat{u},...\zeta _{(p)}...,\eta \right)
=\sum\limits_{<\mu
^p>\in \widehat{M}_{n_E}}\sum\limits_{r=1}^{m_H}f_{<\widehat{\mu }^p>r}(%
\widehat{u})\zeta ^{<\widehat{\mu }^p>}\eta ^r,
$$
where $<\widehat{\mu }^p>=\widehat{\mu }_1^p...\widehat{\mu
}_{k^p}^p$ is a multi--index with $1\leq \widehat{\mu
}_1^p<...<\widehat{\mu }_{k^p}^p\leq m_p,$ the symbol
$\widehat{M}_{n_E}$ is used for the set of all such
multi--indices (including the empty one),\thinspace $\zeta ^{<\widehat{\mu }%
^p>}=\zeta ^{\widehat{\mu }_1^p}...\zeta ^{\widehat{\mu
}_{k_p}^p}$ and each
$f_{<\widehat{\mu }^p>r}$ take values in ${\it C}^\infty \left( {\cal R}%
^{n_E},{\cal C}\right) ,$ such functions are linear in the $\eta
^r$ but multilinear in the $\zeta ^{\widehat{\mu }_n^p}.$ As a
result of choise of
the transition functions (5.11), there is a globally defined map%
$$
I:\Gamma (\Omega (\widetilde{M})\otimes E_H)\rightarrow C^{\infty
^{\prime }}(S(E_H),{\cal C)}
$$
which may be obtained from the local prescription%
$$
I(s)(\widehat{u},\zeta ,\eta )=\sum\limits_{<\mu >\in \widehat{M}%
_{n_E}}\sum\limits_{r=1}^{m_H}f_{<\widehat{\mu }^p>r}(\widehat{u})\zeta ^{<%
\widehat{\mu }^p>}\eta ^r,
$$
if%
$$
s(\widehat{u})=\sum\limits_{<\widehat{\mu }>\in \widehat{M}%
_{n_E}}\sum\limits_{r=1}^{m_H}f_{<\widehat{\mu
}^p>r}(\widehat{u})\delta \widehat{u}^{<\widehat{\mu }^p>}\eta ^r,
$$
where $\Omega (\widetilde{M})$ is the bundle of smooth forms on $\widetilde{M%
}$ and $\left( e^1,...,e^{m_h}\right) $ is the appropriate basis
of the fibre of the bundle $E_H.$ This map is also an isomorphism
of vector spaces and of sheaves. For simplicity we shall omit
''hats'' on indices and consider $<\widehat{\alpha }>=<\alpha >$
in this section.

The aim of our constructions in this section is to allow one to
express the Hodge--de Rham operator on twisted differential forms
on $\widetilde{M}$ as a differential operator on the extended
space of functions $C^{\infty ^{\prime }}(S(E_H),{\cal C).}$ In
explicit form we introduce the Hodge--de
Rham operator in the form%
\index{Hodge--de Rham operator}
$$
d+\delta =\psi ^{<\alpha >}(\delta _{<\alpha >}-\widetilde{\Gamma
}_{<\alpha
><\gamma >}^{<\beta >}\zeta ^{<\gamma >}\delta _{<\beta >}-A_{<\alpha
>r}^s\eta ^r\frac \partial {\partial \eta ^s}),\eqno(5.12)
$$
where $\psi ^{<\alpha >}=\zeta ^{<\alpha >}-g^{<\alpha ><\beta
>}\delta _{<\beta >},$ $\widetilde{\Gamma }_{<\alpha ><\gamma
>}^{<\beta >}$ are the
Christoffel d--symbols (1.39) of the Riemannian metric on a dv--bundle $%
{\cal E}_N.$ For simplicity, in this chapter we shall analyze
locally anisotropic supersymmetric stochastic processes on higher
order anisotropic spaces provided with d--connections of type; a
further generalization is possible by introducing deformations of
connections of type (1.40), for instance in (5.12) in order to
define a deformed variant of
 the Hodge--de Ram operator.
\begin{lemma}
Let $L=\frac 12(d+\delta )^2$ be the Laplace--Beltrami operator
on the space of functions $C^{\infty ^{\prime }}(S(E_H),{\cal
C).}$ Then\thinspace we can write this operator as
$$
L=-\frac 12(B-\widetilde{R}_{<\alpha >}^{<\beta
>}(\widehat{u})\zeta ^{<\alpha >}\delta _{<\alpha >}-\frac
12\widetilde{R}_{<\alpha ><\beta
>}^{\qquad <\widehat{\gamma }><\widehat{\delta }>}\zeta ^{<\beta >}\zeta
^{<\alpha >}\delta _{<\widehat{\gamma }>}\delta _{<\widehat{\delta }>}+%
\eqno(5.13)
$$
$$
\frac 14[\psi ^{<\alpha >},\psi ^{<\beta >}]F_{<\alpha ><\beta
>r}^s\eta ^r\frac \partial {\partial \eta ^s}),
$$
where $$\widetilde{R}_{<\alpha ><\beta >}^{\qquad <\widehat{\gamma }><%
\widehat{\delta }>}$$ are the components of the curvature of
$$\left( \widetilde{M},g\right) ,F_{<\alpha ><\beta >r}^s$$ is
the curvature of the
connection on $E_H,$ and $B$ is the twisted Bochner Laplacian,%
\index{Bochner Laplacian}
$$
B=g^{<\alpha ><\beta >}\left( D_{<\alpha >}D_{<\beta >}-\widetilde{\Gamma }%
_{<\alpha ><\beta >}^{<\gamma >}D_{<\gamma >}\right)
$$
with
$$
D_{<\alpha >}=\delta _{<\alpha >}-\widetilde{\Gamma }_{<\alpha
><\beta
>}^{<\gamma >}\zeta ^{<\beta >}\frac \delta {\delta \zeta ^{<\gamma
>}}-A_{<\alpha >r}^s\eta ^r\frac \partial {\partial \eta ^s}.
$$
\end{lemma}

{\it Proof.} We have
$$
L=\frac 12(\psi ^{<\alpha >}D_{<\alpha >}\psi ^{<\beta >}D_{<\beta >})=%
\eqno(5.14)
$$
$$
\frac 12\left( (\frac 12\{\psi ^{<\alpha >}\psi ^{<\beta
>}\}+\frac 12[\psi ^{<\alpha >}\psi ^{<\beta >}])\right)
D_{<\alpha >}D_{<\beta >}+\psi ^{<\alpha >}[D_{<\alpha >}\psi
^{<\beta >}]D_{<\beta >}
$$
and
$$
\frac 12\{\psi ^{<\alpha >}\psi ^{<\beta >}\}=-g^{<\alpha ><\beta
>}.
$$
Calculating in explicit form we obtain%
$$
\frac 12[\psi ^{<\alpha >}\psi ^{<\beta >}]D_{<\alpha >}D_{<\beta
>}=
$$
$$
-\frac 12[\psi ^{<\alpha >}\psi ^{<\beta >}]((\delta _{<\alpha >}\widetilde{%
\Gamma }_{<\beta ><\delta >}^{<\gamma >}-\widetilde{\Gamma
}_{<\alpha
><\delta >}^{<\varepsilon >}\widetilde{\Gamma }_{<\beta ><\varepsilon
>}^{<\gamma >})\zeta ^{<\delta >}\frac \delta {\partial \zeta ^{<\gamma >}}+
$$
$$
\delta _{<\alpha >}A_{<\beta >r}^s-A_{<\alpha >r}^tA_{<\beta
>t}^s)\eta
^r\frac \partial {\partial \eta ^s}=\widetilde{R}_{<\beta >}^{<\alpha >}(%
\widehat{u})\zeta ^{<\beta >}\frac \delta {\partial \zeta
^{<\alpha >}}+
$$
$$
\frac 12\widetilde{R}_{<\gamma ><\delta >}^{\qquad <\alpha ><\beta >}(%
\widehat{u})\zeta ^{<\delta >}\zeta ^{<\gamma >}\frac \delta
{\partial \zeta ^{<\alpha >}}\frac \delta {\partial \zeta
^{<\beta >}}-\frac 14[\psi ^{<\alpha >}\psi ^{<\beta
>}]F_{<\alpha ><\beta >r}^s\eta ^r\frac \partial {\partial \eta
^s}.
$$
Taking into account that
$$
[D_{<\alpha >},\psi ^{<\beta >}]=-\widetilde{\Gamma }_{<\alpha
><\delta
>}^{<\beta >}\psi ^{<\delta >}
$$
and
$$
\psi ^{<\alpha >}[D_{<\alpha >},\psi ^{<\beta >}]D_{<\beta
>}=-\psi ^{<\alpha >}\psi ^{<\beta >}\widetilde{\Gamma }_{<\alpha
><\beta >}^{<\delta
>}D_{<\delta >}=\widehat{g}^{<\alpha ><\beta >}\widetilde{\Gamma }_{<\alpha
><\beta >}^{<\delta >}.
$$
we obtain from (5.14) the result (5.13) required by the
theorem.$\Box $

In our stochastic considerations we shall also apply is the super extension $%
S(O(\widetilde{{\cal E}}_N),E)$ of the bundle of orthonormal
distinguished
frames of the higher order anisotropic base space (see 
 [262,268]
and Chapter 10 for details on nonsupersymmetric locally
anisotropic stochastic processes) defined similarly as the
Hermitian vector bundle,of course, with corresponding coordinate
and local trivialization
neighborhoods. Let $(u_{\underbrace{\beta }}^{<\alpha >},b_{\underbrace{%
\beta }}^{<\mu |}),$ where indices run as $<\alpha >,<\beta
>,...=1,...,n_E$
and $<\mu |,<\nu |,...=1,...,\frac 12n_E(n_E-1),$ be local coordinates on $O(%
\widetilde{{\cal E}}_N)|_{U_{\underbrace{\alpha }}}.$ Then $S(O(\widetilde{%
{\cal E}}_N),E)$ is the $\left[ \frac 12n_E(n_E+1),n_E+m_H\right] $%
--dimensional distinguished supermanifold with local coordinates $(u_{%
\underbrace{\beta }}^{<\alpha >},b_{\underbrace{\beta }}^{<\mu |},\zeta _{%
\underbrace{\beta }}^{<\alpha >},\eta _{\underbrace{\beta }}^r).$
On
overlapping neighborhoods $U_{\underbrace{\alpha }}$ and $U_{\underbrace{%
\beta }}$ the coordinates $(u_{\underbrace{\beta }}^{<\alpha >},\zeta _{%
\underbrace{\beta }}^{<\alpha >},\eta _{\underbrace{\beta }}^r)$
have the
transition functions defined by (5.11) while the coordinates $b_{\underbrace{%
\beta }}^{<\mu |}$ transform as on the bundle of orthonormalized
distinguished frames $O(\widetilde{{\cal E}}_N).$

\section{Locally Anisotropic Brownian S--Paths}

In this section we construct Brownian paths on the higher order
anisotropic \index{Brownian s--paths!locally anisotropic}
superspaces. We shall use two approaches developed for Riemannian
spaces and correspondingly generalize them for locally
anisotropic curved spaces. The first one has been used in
conjunction with fermionic Brownian motion
 [207] in order to analyse the Hodge--de Rham operator on a Riemann
manifold and to prove the Gauss--Bonnet--Cern formula. The second
approach to Brownian paths on manifolds
 [74,117] is to use paths which are
solutions of stochastic differential equations. An elegant way to
do this is to introduce the symmetric product for the
Stratonovich integral, which in our case is defined in a
distinguished manner:

\begin{definition}
Let $X_s$ and $Y_s$ are stochastic integrals with%
$$
dX_s=f_{<\underline{\alpha }>}da_s^{<\underline{\alpha }>}+f_{0,s}ds~%
\mbox{ and } dY_s=g_{<\underline{\alpha
}>,s}da_s^{<\underline{\alpha }>}+g_{0,s}ds,
$$
where $a_s^{<\underline{\alpha }>}(<\underline{\alpha
}>=1,...,n_E$ denotes a $n_E$--dimensional Brownian motion and
$da_s^{<\underline{\alpha }>}$ denotes the It\^o differential.
Then
$$
Y_s\circ dX_s\doteq Y_s(f_{<\alpha >,s}da_s^{<\underline{\alpha }%
>}+f_{0,s}ds)+\frac 12f_{<\alpha >,s}g_{<\underline{\alpha }>,s}ds.
$$
\end{definition}

Let $A_{<\underline{\alpha }>}$ be a system of $n_E$--dimensional
d--vector field on a dvs--bundle $\widetilde{{\cal E}}_N.$ In a
locally adapted to
N-connection coordinate system we can write $A_{<\underline{\alpha }>}=A_{<%
\underline{\alpha }>}^{<\alpha >}\frac \delta {\partial
u^{<\alpha >}}$ and
consider the distinguished stochastic differential equations%
$$
du_s^{<\alpha >}=A_{<\underline{\alpha }>}^{<\alpha >}(u_s)\circ da_s^{<%
\underline{\alpha }>}+A_0^{<\alpha >}(u_s)ds.\eqno(5.15)
$$
It is known that the solution to such an equation exists globally
on usual manifolds
 [74,117]; such equations enable one to define a notion of
stochastic flow on a manifold and to construct functions on $\widetilde{%
{\cal E}}_N\times {\cal R}^{+}$ which satisfy the differential
equation
$$
\frac{\partial f}{\partial t}=\frac 12A_{<\underline{\alpha }>}A_{<%
\underline{\alpha }>}f~\mbox{ with }~f(u,0)=h(u),\eqno(5.16)
$$
where $h$ is a smooth function on $\widetilde{{\cal E}}_N$ and $A_{<%
\underline{\alpha }>}$ has suitable properties as in the
following theorem:

\begin{theorem}
Let $u_s^{<\alpha >}$ be a solution of (5.15) with initial conditions $%
x_0=x\in \widetilde{{\cal E}}_N.\,$ Then $f(u,t)={\cal
E}(h(u_t))$ satisfies the distinguished differential equation
(5.16).
\end{theorem}

{\it Proof.} We schetch the proof by using the It\^o formula and
omit details of the patching of solutions over different
coordinate neighborhoods. Having
$$
{\cal E}\left( h(u_t)\right) -h(u)={\cal E}(\int\limits_0^t\delta
_{<\alpha
>}h\left( u_s\right) du_s^{<\alpha >}+
$$
$$
\frac 12\int\limits_0^tA_{<\underline{\alpha }>}^{<\alpha >}(u_s)A_{<%
\underline{\alpha }>}^{<\beta >}(u_s)\delta _{<\alpha >}\delta
_{<\beta
>}h\left( u_s\right) ds)
$$
we write
$$
du_s^{<\alpha >}=A_{<\underline{\alpha }>}^{<\alpha >}(u_s)da^{<\underline{%
\alpha }>}+\frac 12A_{<\underline{\alpha }>}^{<\beta
>}(u_s)\left( \delta _{<\beta >}A_{<\underline{\alpha
}>}^{<\alpha >}(u_s)\right) ds.
$$
Because $da_s^{<\underline{\alpha }>}$ have zero expectation and
are
independent of $a^{<\underline{\alpha }>}$ when $u\leq s$ we compute%
$$
{\cal E}\left( h(u_t)\right) -h(u)=\frac 12{\cal E}(\int\limits_0^tA_{<%
\underline{\alpha }>}^{<\alpha >}(u_s)(\delta _{<\alpha >}A_{<\underline{%
\alpha }>}^{<\beta >}(u_s))\delta _{<\beta >}h\left( u_s\right) +
$$
$$
\int\limits_0^tA_{<\underline{\alpha }>}^{<\alpha >}(u_s)A_{<\underline{%
\alpha }>}^{<\beta >}(u_s)\delta _{<\alpha >}\delta _{<\beta
>}h\left(
u_s\right) ds)=\frac 12\int\limits_0^tA_{<\underline{\alpha }>}A_{<%
\underline{\alpha }>}h(u_s)ds.
$$
So, for a suitable $A_{<\underline{\alpha }>},$ we have
$$
{\cal E}\left( h(u_t)\right) =e^{\frac 12\left( A_{<\underline{\alpha }>}A_{<%
\underline{\alpha }>}\right) }h(u)
$$
and thus $f(u,t)={\cal E}\left( h(u_t)\right) $ solves
(5.15).$\Box $

No we shall construct paths on dv--bundles provided with
distinguished Riemannian metric structure, which help in the
study of the Laplacian and related operators. We shall use
$O\left( \widetilde{{\cal E}}_N\right) $, the bundle of
orthonarmal distinguished frames on $\widetilde{{\cal E}}_N,\,$
and consider canonical d--vector fields on this bundle as the
d--vector
fields $A_{<\underline{\alpha }>}.\,$ Let $\left( u,l_{<\underline{\alpha }%
>}\right) $ be a point in $O\left( \widetilde{{\cal E}}_N\right) ,u\in
\widetilde{{\cal E}}_N$ and $l_{<\underline{\alpha }>}$ is a
orthonormal basis of the tangent spaces at $\dot u.\,$ With
respect to locally adapted bases $l_{<\underline{\alpha
}>}=l_{<\underline{\alpha }>}^{<\alpha >}\frac
\delta {\partial ^{<\alpha >}}$ we introduce the canonical vector fields on $%
O\left( \widetilde{{\cal E}}_N\right) :$%
$$
V_{<\underline{a}>}=l_{<\underline{\alpha }>}-l_{<\underline{\alpha }%
>}^{<\alpha >}l_{<\underline{\beta }>}^{<\beta >}\widetilde{\Gamma }%
_{<\alpha ><\beta >}^{<\gamma >}(u)\frac \partial {\partial l_{<\underline{%
\beta }>}^{<\gamma >}},
$$
where $\widetilde{\Gamma }_{<\alpha ><\beta }^{<\gamma >}$ are the
Christoffel d--symbols (1.39). In this case from (5.15), one
follows the
stochastic differential equations%
$$
du_s^{<\alpha >}=l_{<\underline{\alpha }>,s}^{<\alpha >}\circ da_s^{<%
\underline{\alpha }>},~dl_{<\underline{\alpha }>,s}^{<\alpha >}=-l_{<%
\underline{\alpha }>,s}^{<\tau >}l_{<\underline{\beta }>,s}^{<\varepsilon >}%
\widetilde{\Gamma }_{<\tau ><\varepsilon >}^{<\alpha >}(u_s)\circ da_s^{<%
\underline{\beta }>}.
$$
If $f$ is a smooth function on $O\left( \widetilde{{\cal E}}_N\right) ,$%
$$
{\cal E}\left( f\left( u_t,l_t\right) \right) -f\left( u,l\right)
=\frac
12\int\limits_0^tV_{<\underline{a}>}V_{<\underline{a}>}f\left(
u_s,l_s\right) ds.
$$
When $f\,$ depends on $u$ only, i.e. $f=\varphi \circ \pi $ where
$\pi :O\left( \widetilde{{\cal E}}_N\right) \rightarrow
\widetilde{{\cal E}}_N$
is the projection map, and $\varphi \in {\it C}^\infty \left( \widetilde{%
{\cal E}}_N\right) ),$%
$$
{\cal E}\left( \varphi \left( u_t,\right) \right) -\varphi \left(
u\right) =-\int\limits_0^tL_{scal}~\varphi \left( u_s\right) ds,
$$
where $L_{scal}=-\frac 12\left( g^{<\alpha ><\beta >}\delta
_{<\alpha
>}\delta _{<\beta >}+g^{<\alpha ><\beta >}\widetilde{\Gamma }_{<\alpha
><\beta >}^{<\tau >}\delta _{<\tau >}\right) $ is the scalar Laplacian. This
\index{Scalar Laplacian} holds since
$$
\frac 12V_{<\underline{a}>}V_{<\underline{a}>}f\left( u\right)
=-L_{scal}f\left( u\right)
$$
and
$$
l_{<\underline{\alpha }>,s}^{<\alpha >}l_{<\underline{\alpha
}>,s}^{<\beta
>}=\widehat{g}^{<\alpha ><\beta >}\left( u_s\right)
$$
almost certainly (a.c.) (in Chapter 10 we shall use also the term
 almost sure, a.s.). Hence one finds that $f\left( u,t\right) ={\cal E}%
\left( \varphi \left( u_t,\right) \right) $ satisfies
$$
\frac{\partial f}{\partial t}=-L_{scal}f.
$$

We shall define the Brownian paths in superspace which lead to\\
Feynman--Kac formula for the Laplace--Beltrami operator
$$L=\left( d+\delta \right) ^2$$ \index{Laplace--Beltrami
operator} acting on $C^{\infty ^{\prime }}(S(E_H),{\cal C).}$
Denoting by $\left( \zeta _t^{<\underline{\alpha }>},\rho
_t^{<\underline{\alpha }>}\right) $ the $n_E$--dimensional
fermionic Brownian paths and by $\left( u^{<\alpha
>},l_{<\underline{\alpha }>}^{<\alpha >},\zeta ^{<\alpha >},\eta ^r\right) $
the coordinates of a pont in the extended bundle of orthogonal
distinguished frames $S(O(\widetilde{{\cal E}}_N),E)$ we consider
$n_E+n_E^2+m_H$
stochastic differential equations%
$$
u_t^{<\alpha >}=u^{<\alpha >}+\int\limits_0^tl_{<\underline{\alpha }%
>,s}^{<\alpha >}\circ da_s^{<\underline{\alpha }>},\eqno(5.17)
$$
$$
l_{<\underline{\alpha }>,t}^{<\alpha >}=l_{<\underline{\alpha
}>}^{<\alpha
>}-\int\limits_0^tl_{<\underline{\alpha }>,s}^{<\gamma >}\widetilde{\Gamma }%
_{<\tau ><\beta >}^{<\alpha >}\left( u_s\right) l_{<\underline{\beta }%
>,s}^{<\beta >}\circ da_s^{<\underline{\beta }>},
$$
$$
\xi _t^{<\alpha >}=\zeta ^{<\alpha >}+\zeta _t^{<\alpha >}l_{<\underline{%
\alpha }>,t}^{<\alpha >}+\int\limits_0^t(-\zeta _t^{<\alpha >}dl_{<%
\underline{\alpha }>,s}^{<\alpha >}-
$$
$$
-\xi _s^{<\tau >}\widetilde{\Gamma }_{<\tau ><\beta >}^{<\alpha
>}\left(
u_s\right) l_{<\underline{\beta }>,s}^{<\beta >}\circ da_s^{<\underline{%
\beta }>}+\frac i4\xi _s^{<\beta >}\widetilde{R}_{<\beta ><\tau
><\varepsilon >}^{<\alpha >}\left( u_s\right) \xi _s^{<\tau >}\pi
_s^{<\varepsilon >}ds),
$$
$$
\eta _t^p=\eta ^p+\int\limits_0^t(-l_{<\underline{\alpha
}>,s}^{<\gamma
>}\eta _s^qA_{<\gamma >q}^p\left( u_s\right) \circ da_s^{<\underline{\alpha }%
>}+
$$
$$
\frac 14\eta _s^q\left( \xi _s^{<\alpha >}+i\pi _s^{<\alpha
>}\right) \left( \xi _s^{<\beta >}+i\pi _s^{<\beta >}\right)
F_{<\alpha ><\beta >q}^p\left( u_s\right) ds),
$$
where $\pi _s^{<\alpha >}=l_{<\underline{\alpha }>,s}^{<\gamma >}\rho _s^{<%
\underline{\alpha }>}.$

The existence of local solutions of distinguished stochastic
differential equations of type (5.17) is a consequence of theorem
5.5; a usual patching techniques allow a global solution to be
constructed since they transforms d--covariantly under
coordinates changes.

For our further purposes it is convenient to introduce into
consideration
d--vector fields $W_{<\underline{\alpha }>}\,$on $S(O(\widetilde{{\cal E}}%
_N),E),$ written in the form%
$$
W_{<\underline{\alpha }>}=l_{<\underline{\alpha }>}^{<\gamma
>}\frac \delta
{\partial u^{<\gamma >}}-l_{<\underline{\alpha }>}^{<\tau >}l_{<\underline{%
\alpha }>}^{<\beta >}\widetilde{\Gamma }_{<\tau ><\beta
>}^{<\gamma >}\frac
\partial {\partial l_{<\underline{\alpha }>}^{<\gamma >}}-
$$
$$
l_{<\underline{\alpha }>}^{<\tau >}\zeta ^{<\beta >}\widetilde{\Gamma }%
_{<\tau ><\beta >}^{<\gamma >}\frac \delta {\partial \zeta ^{<\gamma >}}-l_{<%
\underline{\alpha }>}^{<\tau >}\eta ^rA_{<\gamma >r}^s\frac
\partial {\partial \eta ^s},
$$
which are the canonical vector fields on $S(O(\widetilde{{\cal
E}}_N),E).\,$
The d--vectors $W_{<\underline{\alpha }>}$ when acting on functions on $S(O(%
\widetilde{{\cal E}}_N),E)$ which are independent of the $l_{<\underline{%
\alpha }>}^{<\gamma >},$ i.e. on functions of the form $f=\varphi
\circ \pi
_s$ where $\pi _s$ is the canonical projection of $S(O(\widetilde{{\cal E}}%
_N),E)$ onto $S\left( E\right) ,$ are related to the
Laplace--Beltrami
operator $L=\frac 12\left( d+\delta \right) ^2$ by formula%
$$
L=-\frac 12(W_{<\underline{\alpha }>}W_{<\underline{\alpha }>}-\widetilde{R}%
_{<\beta >}^{<\alpha >}\zeta ^{<\beta >}\frac \delta {\partial
\zeta ^{<\alpha >}}-
$$
$$
\frac 12\widetilde{R}_{<\gamma ><\delta >}^{\qquad <\alpha ><\beta >}(%
\widehat{u})\zeta ^{<\delta >}\zeta ^{<\gamma >}\frac \delta
{\partial \zeta ^{<\alpha >}}\frac \delta {\partial \zeta
^{<\beta >}}+\frac 14[\psi ^{<\alpha >}\psi ^{<\beta
>}]F_{<\alpha ><\beta >r}^s\eta ^r\frac \partial {\partial \eta
^s})
$$
(a consequence from lemma 5.2).

Now, applying the theorem 5.7 we establish the Feynman--Kac
for\-mu\-la: \index{Feynman--Kac!for\-mu\-la}

\begin{theorem}
For a distinguished stochastic process $\left( u_t^{<\alpha >},l_{<%
\underline{\alpha }>,t}^{<\alpha >},\eta _t^r\right) $ satisfying
equations (5.17) one holds the formula
$$
\exp (-Lt)\varphi (u,\zeta ,\eta )={\cal E}\left( \varphi
(u_t,\xi _t,\eta _t)\right)
$$
where $L=\frac 12\left( d+\delta \right) ^2$ is the
Laplace--Beltrami operator acting on\\  $C^{\infty ^{\prime
}}(S(E_H),{\cal C)}.$
\end{theorem}

{\it Proof. }From the distinguished superspace It\^o formula (see
theorem 5.2) and using properties of fermionic paths (see section
5.2) we have
$$
{\cal E}\left( \varphi (u_t,\xi _t,\eta _t)\right) -\varphi (u,\zeta ,\eta )=%
{\cal E}(\frac 12\int\limits_0^tW_{<\underline{\alpha }>}W_{<\underline{%
\alpha }>}\varphi (u_s,\xi _s,\eta _s)-
$$
$$
\frac i4\widetilde{R}_{<\beta ><\gamma ><\delta >}^{<\alpha
>}(u_s)\xi _s^{<\gamma >}\pi _s^{<\delta >}\xi _s^{<\beta >}\frac
\delta {\partial \zeta ^{<\alpha >}}\varphi (u_s,\xi _s,\eta _s)+
$$
$$
\frac 14(\xi _s^{<\alpha >}+i\pi _s^{<\alpha >})(\xi _s^{<\beta
>}+i\pi _s^{<\beta >})F_{<\alpha ><\beta >p}^q\eta _t^p\frac
\partial {\partial \eta ^q}\varphi (u_s,\xi _s,\eta _s)ds)=
$$
$$
{\cal E}(\frac 12\int\limits_0^t(W_{<\underline{\alpha }>}W_{<\underline{%
\alpha }>}-\widetilde{R}_{<\beta >}^{<\alpha >}\zeta ^{<\beta
>}\frac \delta {\partial \zeta ^{<\alpha >}}-
$$
$$
\frac 12\widetilde{R}_{<\gamma ><\delta >}^{\qquad <\alpha ><\beta >}(%
\widehat{u})\zeta ^{<\delta >}\zeta ^{<\gamma >}\frac \delta
{\partial \zeta ^{<\alpha >}}\frac \delta {\partial \zeta
^{<\beta >}}+
$$
$$
\frac 14[\psi ^{<\alpha >}\psi ^{<\beta >}]F_{<\alpha ><\beta
>r}^s\eta ^r\frac \partial {\partial \eta ^s}.
$$
Putting
$$
f(u,\zeta ,\eta ,t)={\cal E}\left( \varphi (u_t,\xi _t,\eta
_t)\right) ,
$$
and writing
$$
f(u,\zeta ,\eta ,t)-f(u,\zeta ,\eta
,0)=-\int\limits_0^tLf(u,\zeta ,\eta ,s)ds
$$
we obtain the result of the theorem.$\Box $

\section{Atiyah--Singer D--Index Theorem}

Our aim is to extend the Atiyah--Singer index theorem
 [23,135] in order
\index{Atiyah--Singer index theorem} to include contributions of
the N--connection structure. For simplicity, the basic formulas
will be written with respect to the components of curvatures of
Christoffel d--symbols (1.39) in dvs--bundles. Different
generalizations can be obtained by using deformations of
d--connections (1.40), but we shall omit them in this monograph.
Although such formulas in the simplest case of torsionless
connections defined by Christoffel d--symbols contain
contributions from components of N--connection the geometric
constructions connected with the proof of the Atiyah--Singer
theorem hold good for higher order anisotropic s--spaces having
the even part provided with a distinguished Riemannian metric and
corresponding Christoffel like (containing locally adapted
partial derivations instead of usual derivation)
d--connection. For instance, the formula of McKean 
 [135] and Witten 
 [291] expressing the index of the complex in terms of the heat kernel
of the Laplacian.

Let consider $\left( \widetilde{{\cal E}}_N,g_{<\alpha ><\beta
>}\right) $ a compact distinguished Riemannian manifold of
dimension $n_E=2n_E^{\prime }\,$
and a $m_H$--dimensional Hermitian vector bundle over $\widetilde{{\cal E}}%
_N.$ By using corresponding extensions on $\widetilde{{\cal
E}}_N$ of the
Hodge--de Rham operator $d+\delta $ and the Laplace--Beltrami operator $%
L=\frac 12\left( d+\delta \right) ^2$ can write the McKean and
Singer \index{McKean and Singer formula} formula
$$
Index\left( d+\delta \right) =Str\exp (-Lt),\eqno(5.18)
$$
where $Str$ denotes the supertrace. Identifying the space of
twisted forms on $\widetilde{{\cal E}}_N$ with the space of
functions $C^{\infty ^{\prime }}(S(E_H),{\cal C)}$ on the
supermanifold $S(E_H)$ we can define the supertrace in this
manner: The standard involution $\tau $ may be defined on
$C^{\infty ^{\prime }}(S(E_H),{\cal C)}$ by the formula
$$
\tau \left( \sum\limits_{<\mu ^p>\in \widehat{M}_{n_E}}\sum%
\limits_{r=1}^{m_H}f_{<\widehat{\mu }^p>r}(\widehat{u})\zeta
^{<\widehat{\mu }^p>}\eta ^r\right) =
$$
$$
\sum\limits_{<\mu ^p>\in \widehat{M}_{n_E}}\sum\limits_{r=1}^{m_H}\int_{%
{\cal B}}\frac{d^{n_E}\rho }{\sqrt{\det \left( g_{<\alpha ><\beta >}\right) }%
}\times $$ $$ \exp \left( i\rho ^{<\alpha >}\zeta ^{<\beta
>}\widehat{g}_{<\alpha ><\beta
>}\left( u\right) f_{<\widehat{\mu }^p>r}\rho ^{<\mu ^p>}\eta ^r\right) ,
$$
where $\rho ^1,...,\rho ^{n_E}$ are anticommuting variables and $%
\int_{{\cal B}}$ denotes the Berezin integration. The supertrace
can be
defined for a suitable operator $O$ on\\ $C^{\infty ^{\prime }}(S(E_H),{\cal %
C)}$ by the formula $StrO=Tr\tau O,$ where $Tr$ denotes the
conventional trace.

The extension to higher order anisotropic spaces of the
Atiyah--Singer index theorem for the twisted Hirzebruch signature
complex is formulated in this form:

\begin{theorem}
One holds the formula%
$$
Index\left( d+\delta \right) =\int\limits_{\widetilde{{\cal
E}}_N}\left[ tr\exp \left( -\frac F{2\pi }\right) \det \left(
\frac{i\Omega /2\pi }{\tanh (i\Omega /2\pi )}\right)
^{1/2}\right] ,
$$
where $F$ is the curvature 2--form of a connection on the bundle
$E_H,\Omega
$ is the Riemann d--curvature form of $\left( \widetilde{{\cal E}}%
_N,g_{<\alpha ><\beta >}\right) $ and the square brackets indicate
projection onto the $n_E$--form component of the integrand. An
equivalent result (combining with (5.18)) is
$$
Str\exp \left( -Ht\right) =\int\limits_{\widetilde{{\cal
E}}_N}\left[ tr\exp \left( -\frac F{2\pi }\right) \det \left(
\frac{i\Omega /2\pi }{\tanh (i\Omega /2\pi )}\right)
^{1/2}\right] ,
$$
which holds for all $t.$
\end{theorem}

One also can present a stronger local version of the just
formulated theorem:

\begin{theorem}
With the notation of the theorem 5.9, if $p\in \widetilde{{\cal
E}}_N,$
there is the limit%
$$
\lim _{t\rightarrow 0}~str\exp \left( -Ht\right) (p,p)dvol=\left[
tr\exp \left( -\frac F{2\pi }\right) \det \left( \frac{i\Omega
/2\pi }{\tanh (i\Omega /2\pi )}\right) ^{1/2}\right] \mid _p,
$$
where $str$ denotes the $2^{n_E}m_H\times 2^{n_E}m_H\,$ matrix
supertrace, as opposed to the full operator supertrace $Str,$ so
that $str\exp (-Ht)(p,q)
$ is then the kernel of the operator on ${\it C}^\infty \left( \widetilde{%
{\cal E}}_N\right) $ obtained by this partial supertrace.
\end{theorem}

{\it Proof. }To prove this theorem we shall use the distinguished
Feynman--Kac formula (see theorem 5.8), to analyse the operator
$\exp (-Ht)$ and by using Duhamel's formula, to investigate the
kernel and show that in the limit as $t\rightarrow 0$ only the
required terms survive (in general we shall follow the methods
developed in
 [89] and 
 [210] but a
considerable attention will be given to the contribution of
N--connection structure and of distinguishing of indices).

The matrix supertrace of an operator can be expressed in terms of
a Berezin
integral of its kernel. Let denote by $P\left( q\right) $ the 2$^p$%
--dimensional space of polynomial functions of $q$ anticommuting
variables. A linear operator $O$ on this space has a kernel
$O\left( \xi ,\zeta \right) =O\left( \xi _1,...,\xi _q,\zeta
_1,...,\zeta _q\right) $ being a function of $2p$ anticommuting
variables and satisfying
$$
Of\left( \xi \right) =\int_{{\cal B}}d^qO\left( \xi ,\zeta
\right) f\left( \zeta \right)
$$
for every functions $f$ in $P\left( q\right) .$ By direct
calculation we can obtain
$$
trO=\int_{{\cal B}}d^q\xi O\left( \xi ,-\xi \right) .
$$
In a similar manner for $Q$ being an operator on $C^{\infty
^{\prime }}(S(E_H),{\cal C)}$ one has the local coordinate
expression
$$
strQ(u,v)=trQ(u,v)=\eqno(5.19)
$$
$$
\int_{{\cal B}}d^{n_E}\rho d^{n_E}\zeta d^{m_H}\eta Q(u,v,\rho
,-\zeta ,\eta ,-\eta )\frac{\exp \left( i\rho ^{<\alpha >}\zeta
^{<\beta >}g_{<\alpha
><\beta >}(u)\right) }{\sqrt{\det \left( g_{<\alpha ><\beta >}(u)\right) }}.
$$

As a next step we construct a locally equivalent d--metric and
d--con\-nec\-ti\-on on the distinguished superextension of ${\cal
R}^{n_E}{\cal \times C.}$ Here we note that this theorem has a
local character and is sufficient to replace the dv--bundle
$\widetilde{{\cal E}}_N$ by ${\cal R}^{n_E}$ and the Hermitian
bundle $E_H\,\,$ over $\widetilde{{\cal E}}_N\,$ by the trivial
bundle ${\cal R}^{n_E}{\cal \times C}^{m_H}$ over ${\cal
R}^{n_E}$ with metric and connection satisfying certain
conditions. We shall construct a
suitable metric and connection. Consider an open subset $W\subset \widetilde{%
{\cal E}}_N,\,\,$containing $w$ $\in W,$ which has compact
closure and is both a coordinate neighborhood of\thinspace $M$
and local trivialization
neighborhood of the bundle $E_H$ and that $U$ is also an open subset of $%
\widetilde{{\cal E}}_N$ containing $w$ with $\overline{U}\subset W.$ Let $%
\phi :W\rightarrow {\cal R}^{n_E}$ be a system of normal
coordinates centered in $w,$ which satisfy conditions $\det
\left( g_{<\alpha ><\beta
>}\right) \mid _U=1,\phi (w)=0,$ and choose a local trivialization of the
bundle $E_H$ such that $A_{<\gamma >r}^s(0)=0.$ The required
metric and connection on ${\cal R}^{n_E}$ are introduced
respectively as satisfying
conditions%
$$
(\overbrace{g})_{<\alpha ><\beta >}\left( u^1,...,u^{n_E}\right)
=g_{<\alpha
><\beta >}(\phi ^{-1}\left( u^1,...,u^{n_E}\right) )~\mbox{when}~u\in \phi
\left( U\right) ,
$$
$$
\overbrace{g}_{<\alpha ><\beta >}\left( u^1,...,u^{n_E}\right)
=\delta _{<\alpha ><\beta >}~\mbox{when}~u\notin \phi \left(
U\right)
$$
and $\det \left( (\overbrace{g})_{<\alpha ><\beta >}\right) =1$ through $%
{\cal R}^{n_E}$ and
$$
(\overbrace{A})_{<\gamma >r}^s(u^1,...,u^{n_E})=A_{<\gamma
>r}^s(\phi ^{-1}(u^1,...,u^{n_E}))~\mbox{when}~u\in \phi \left(
U\right) ,
$$
$$
(\overbrace{A})_{<\gamma
>r}^s(u^1,...,u^{n_E})=0~\mbox{when}~u\notin \phi \left( U\right)
$$
We note with $"\overbrace{}"$ geometric objects on ${\cal R}^{n_E}$ and $%
{\cal R}^{n_E}{\cal \times C}^{m_H},$%
$$
(\overbrace{R})_{\quad <\beta ><\gamma ><\delta >}^{<\alpha
>}\left( \widehat{u}\right) =R_{\quad <\beta ><\gamma ><\delta
>}^{<\alpha >}\left( \phi ^{-1}(\widehat{u})\right)
$$
and%
$$
(\overbrace{F})_{<\alpha ><\beta >q}^p\left( \widehat{u}\right)
=F_{<\alpha
><\beta >q}^p\left( \phi ^{-1}(\widehat{u})\right)
$$
on $\phi \left( U\right) .$ From standard Taylor expansions in
normal
coordinates%
$$
(\overbrace{g})_{<\alpha ><\beta >}\left( \widehat{u}\right)
=\delta
_{<\alpha ><\beta >}-\frac 13\widehat{u}^{<\gamma >}\widehat{u}^{<\delta >}%
\widetilde{R}_{<\gamma ><\alpha ><\delta ><\beta >}\left(
0\right) +...
$$
$$
(\overbrace{\Gamma })_{<\alpha ><\beta >}^{<\gamma
>}(\widehat{u})=$$
$$\frac 13%
\widehat{u}^{<\delta >}((\overbrace{R})_{\quad <\alpha ><\delta
><\beta
>}^{<\gamma >}\left( 0\right) +(\overbrace{R})_{\quad <\beta ><\delta
><\alpha >}^{<\gamma >}\left( 0\right) )+...$$
$$
(\overbrace{A})_{<\gamma >r}^s(\widehat{u})=-\frac
12\widehat{u}^{<\delta
>}F_{<\gamma ><\delta >r}^s(0)+....,
$$
cutting and pasting arguments we obtain
$$
\lim _{t\rightarrow 0}\left( str\exp (-Ht)\right) (w,w)-str\exp (-\overbrace{%
H}t(0,0))=0,\eqno(5.20)
$$
i.e. in the rest of this section it will be sufficient to
consider objects on \\ $S\left( {\cal R}^{n_E}{\cal \times
C}^{m_H}\right) .$

Let denote by $\left( \overbrace{H}\right) ^0=-\frac 12\delta
_{<\alpha
>}\delta _{<\beta >}$ the trivial N--connection flat Laplacian in $S\left(
{\cal R}^{n_E}{\cal \times C}^{m_H}\right) $ and $\left( \overbrace{K}%
\right) _t\left( u,u^{\prime },\zeta ,\zeta ^{\prime },\eta ,\eta
^{\prime }\right) $ and\\ $\left( \overbrace{K}\right) _t^0\left(
u,u^{\prime },\zeta ,\zeta ^{\prime },\eta ,\eta ^{\prime
}\right) $ be the heat kernels
$$
\exp \left( -\left( \overbrace{H}\right) t\left( u,u^{\prime
},\zeta ,\zeta ^{\prime },\eta ,\eta ^{\prime }\right) \right)
\mbox{ and }\exp \left( -\left( \overbrace{H}\right) ^0t\left(
u,u^{\prime },\zeta ,\zeta ^{\prime },\eta ,\eta ^{\prime
}\right) \right)
$$
satisfying the Duhamel's formula (see
 [89])%
$$
\left( \overbrace{K}\right) _t\left( u,u^{\prime },\zeta ,\zeta
^{\prime },\eta ,\eta ^{\prime }\right) -\left(
\overbrace{K}\right) _t^0\left( u,u^{\prime },\zeta ,\zeta
^{\prime },\eta ,\eta ^{\prime }\right) =
$$
$$
\int\limits_0^tdse^{-(t-s)\left( \overbrace{H}\right) }\left(
\left( \overbrace{H}\right) -\left( \overbrace{H}\right)
^0\right) \left( \overbrace{K}\right) _s^0\left( u,u^{\prime
},\zeta ,\zeta ^{\prime },\eta ,\eta ^{\prime }\right)
$$
where all differential operators act with respect to variables
$u,\zeta $ and $\eta .$ Expressing
$$
\left( \overbrace{K}\right) _s^0\left( u,u^{\prime },\zeta ,\zeta
^{\prime },\eta ,\eta ^{\prime }\right) =\int_{{\cal
B}}d^{n_E}\zeta d^{n_E}\zeta
^{\prime }d^{m_H}\eta \int\limits_0^tds\{(e^{-(t-s)\left( \overbrace{H}%
\right) }\times
$$
$$
\left( \left( \overbrace{H}\right) -\left( \overbrace{H}\right)
^0\right) \left( \overbrace{K}\right) _s^0\left( 0,0,\zeta ,\zeta
^{\prime },\eta ,\eta ^{\prime }\right) )\mid _{\eta ^{\prime
}=-\eta }\exp [-i\zeta \zeta ]\}.
$$
Applying the Feynman--Kac formula (see theorem 5.8)
$$
e^{-(t-s)\left( \overbrace{H}\right) }\left( \left(
\overbrace{H}\right) -\left( \overbrace{H}\right) ^0\right)
\left( \overbrace{K}\right) _s^0\left( 0,0,\zeta ,\zeta ^{\prime
},\eta ,\eta ^{\prime }\right) =
$$
$$
{\cal E}\int_{{\cal B}}d^{n_E}\rho d^{m_H}k\left( 2\pi s\right)
^{-n_E/2}F_s\left( (\overbrace{u})_{t-s},(\overbrace{\xi })_{t-s},\rho ,(%
\overbrace{\eta })_{t-s},k\right)
$$
$$
\exp (-i\rho \zeta ^{\prime })\exp \left( -ik\eta ^{\prime
}\right) ,
$$
where
$$  F_s(u,\zeta ,\rho ,\eta ,k)=\left( \left( \overbrace{H}\right) -\left(
\overbrace{H}\right) ^0\right) \times $$
$$\left[ \exp (-(u^2/2s))\exp (-i\rho \zeta)\exp \left( -ik\eta \right)
 \right]
$$
and $\left( \overbrace{u}\right) _s,\left( \overbrace{\zeta
}\right) _s$ and $\left( \overbrace{\eta }\right) _s$ satisfy the
distinguished stochastic equations (5.17) (in normal coordinates
with initial conditions $\left(
\overbrace{u}\right) _0=0,$ $\left( \overbrace{l}\right) _{<\underline{%
\alpha }>0}^{<\alpha >}=\delta _{<\underline{\alpha }>}^{<\alpha
>},\left( \overbrace{\xi }\right) _0=\zeta $ and $\left(
\overbrace{\eta }\right)
_0=\eta )$ we can write%
$$
str\left( \overbrace{K}\right) _t(0,0)={\cal E[}\int_{{\cal
B}}d^{n_E}\zeta d^{n_E}\zeta ^{\prime }d^{n_E}\rho
d^{m_H}k\int\limits_0^tds\left( 2\pi s\right) ^{-n_E/2}\times
$$
$$
F_s\left( (\overbrace{u})_{t-s},(\overbrace{\xi })_{t-s},\rho ,(\overbrace{%
\eta })_{t-s},k\right) \exp (-i\zeta \zeta ^{\prime })\exp
(-i\rho \zeta ^{\prime })\exp \left( -ik\eta \right) ].
$$
Integrating the last expression on $\zeta ^{\prime }$ and $\rho
^{\prime }$ we obtain
$$
str\left( \overbrace{K}\right) _t(0,0)={\cal E[}\int_{{\cal
B}}d^{n_E}\zeta d^{m_H}\eta d^{m_H}k\int\limits_0^tds\left( 2\pi
s\right) ^{-n_E/2}\times
$$
$$
F_s\left( (\overbrace{u})_{t-s},(\overbrace{\xi })_{t-s},\zeta ,(\overbrace{%
\eta })_{t-s},k\right) \exp \left( -ik\eta \right) ].
$$

Now we construct a simpler Hamiltonian $\left( \overbrace{H}\right) ^1$ on $%
S\left( {\cal R}^{n_E}{\cal \times C}^{m_H}\right) $ with heat kernel $%
\left( \overbrace{K}\right) _t^1\left( u,u^{\prime },\zeta ,\zeta
^{\prime },\eta ,\eta ^{\prime }\right) $ such that
$$
\lim _{t\rightarrow 0}str\left( \overbrace{K}\right) _t^1\left(
0,0\right) =\lim _{t\rightarrow 0}str\left( \overbrace{K}\right)
_t\left( 0,0\right)
$$
which allows us to calculate the required supertrace. Considering
a simplification of distinguished stochastic differential
equations (5.17)
$$
\widehat{u}_t^{1<\alpha >}=a_t^{<\alpha >},
$$
$$
(\overbrace{\xi })_{t-s}^{1<\alpha >}=\zeta ^{<\alpha >}+\zeta _t^{<%
\underline{\alpha }>}\delta _{<\underline{\alpha }>}^{<\alpha >}+\frac 13(%
\overbrace{\xi })_s^{1<\beta >}\widetilde{R}_{<\beta >}^{<\alpha
>}ds-
$$
$$
\frac 13\int\limits_0^t((\overbrace{\xi })_s^{1<\alpha >}\widehat{u}%
_s^{1<\gamma >}(\widetilde{R}_{<\gamma ><\delta ><\beta
>}^{\qquad \quad \quad \quad <\alpha >}+\widetilde{R}_{<\gamma
><\beta ><\delta >}^{\qquad \quad \quad \quad <\alpha
>})da_s^{<\beta >}+
$$
$$
\frac i4(\overbrace{\xi })_s^{1<\beta >}(\overbrace{\xi })_s^{1<\gamma >}(%
\overbrace{\pi })_s^{1<\delta >}\widetilde{R}_{<\beta ><\gamma
>~~<\delta
>}^{\qquad \quad <\alpha >}ds),
$$
$$
(\overbrace{\eta })_t^{1r}=\eta ^r+\frac 14\int\limits_0^t(\overbrace{\eta }%
)_t^{1q}\left( \zeta _s^{<\alpha >}-i\pi _s^{<\alpha >}\right)
\left( \zeta _s^{<\beta >}-i\pi _s^{<\beta >}\right) F_{<\alpha
><\beta >q}^rds,
$$
where $\pi _s^{1<\beta >}=\rho _s^{<\underline{\alpha }>}\delta _{<%
\underline{\alpha }>}^{<\beta >},R_{<\gamma ><\alpha ><\delta ><\beta >}=(%
\overbrace{R})_{<\gamma ><\alpha ><\delta ><\beta >}^1(0)$ and\\
$F_{<\alpha
><\beta >q}^r=(\overbrace{F})_{<\alpha ><\beta >q}^r(0),\,$ with indices
raised and lowered by\\ $\left( \overbrace{g}\right) _{<\alpha
><\beta
>}^1(0)=\delta _{<\alpha ><\beta >},$ we introduce the modified Hamiltonian\\
 $\left( \overbrace{H}\right) ^1$ satisfying conditions%
$$
{\cal E}\left( f\left( \widehat{u}_t^1,(\overbrace{\xi })_t^1,(\overbrace{%
\eta })_t^1\right) \right) -f\left( 0,\zeta ,\eta \right)
=\eqno(5.21)
$$
$$
{\cal E}\left( -\int\limits_0^t\left( \overbrace{H}\right)
^1f\left( \widehat{u}_s^1,(\overbrace{\xi })_s^1,(\overbrace{\eta
})_s^1\right) ds\right)
$$
where $f\in {\it C}^{\infty ^{\prime }}(S\left( {\cal R}^{n_E}{\cal \times C}%
^{m_H}\right) )$ and
$$
\left( \overbrace{H}\right) ^1=-\frac 12\delta _{<\alpha >}\delta
_{<\alpha
>}+\frac 13\widetilde{R}_{<\beta >}^{<\alpha >}(\widehat{u})\widehat{\zeta }%
^{<\beta >}\frac \delta {\partial \zeta ^{<\alpha >}}+
$$
$$
\frac 14\widetilde{R}_{<\delta ><\gamma >}^{\qquad <\alpha ><\beta >}(%
\widehat{u})\widehat{\zeta }^{<\delta >}\widehat{\zeta }^{<\gamma
>}\frac \delta {\partial \zeta ^{<\alpha >}}\frac \delta
{\partial \zeta ^{<\beta
>}}-
$$
$$
+\frac 12\widetilde{R}_{<\beta >}^{<\alpha >}(\widehat{u})\widehat{\zeta }%
^{<\beta >}\frac \delta {\partial \zeta ^{<\alpha >}}-\frac
14\psi ^{<\alpha
>}\psi ^{<\beta >}F_{<\alpha ><\beta >r}^s\eta ^r\frac \partial {\partial
\eta ^s}+
$$
$$
\frac 13\widehat{\zeta }^{<\delta >}\widehat{u}^{<\gamma >}\left( \widetilde{%
R}_{<\gamma >\quad <\delta >}^{\quad <\alpha >\quad <\beta >}+\widetilde{R}%
_{<\gamma ><\delta >}^{\qquad <\alpha ><\beta >}\right) \frac
\delta {\partial \zeta ^{<\beta >}}\delta _{<\alpha >}+
$$
$$
\frac 1{18}\widehat{\zeta }^{<\varepsilon >}\widehat{\zeta }^{<\delta >}%
\widehat{u}^{<\tau >}\widehat{u}^{<\gamma >}(\widetilde{R}_{<\tau
><\delta
><\varepsilon >}^{\qquad \quad \quad \quad <\alpha >}+\widetilde{R}_{<\tau
><\varepsilon ><\delta >}^{\qquad \quad \quad \quad <\alpha >})\times
$$
$$
\left( \widetilde{R}_{<\gamma >\quad <\delta >}^{\quad <\beta
>\quad <\gamma
>}+\widetilde{R}_{<\gamma ><\delta >}^{\qquad <\beta ><\gamma >}\right)
\frac \delta {\partial \zeta ^{<\alpha >}}\frac \delta {\partial
\zeta ^{<\beta >}}
$$
(in this formulas the objects enabled with ''hats'' are
operators).

Applying again the Duhamel's formula and introducing variable%
$$
G_s(u,\xi ,\zeta ,\eta ,k)=\exp (-(\frac{u^2}{2s}+i\zeta \xi
+ik\eta )
$$
we find that
$$
\widetilde{K}_t(0,0)-\widetilde{K}_t^1(0,0)={\cal E}\int_{{\cal B}%
}d^{n_E}\zeta d^{m_H}\eta d^{m_H}k\int\limits_0^tds\left( 2\pi
s\right) ^{-n_E/2}\times
$$
$$
[\left( \left( \overbrace{H}\right) -\left( \overbrace{H}\right)
^0\right)
G_s(u,\xi ,\zeta ,\eta ,k)\mid _{u=\widehat{u}_{t-s},\xi =(\overbrace{\xi }%
)_{t-s},\eta =(\overbrace{\eta })_{t-s}}-
$$
$$
\left( \left( \overbrace{H}^1\right) -\left( \overbrace{H}\right)
^0\right)
G_s(u,\xi ,\zeta ,\eta ,k)\mid _{u=\widehat{u}_{t-s}^1,\xi =(\overbrace{\xi }%
)_{t-s}^1,\eta =(\overbrace{\eta })_{t-s}^1}].
$$
It is convenient to introduce scaled variables $\omega
=\sqrt{2\pi t}\zeta $
and then after the Berezin integration we find%
$$
\widetilde{K}_t(0,0)-\widetilde{K}_t^1(0,0)=A(t)+B(t),
$$
where%
$$
A(t)={\cal E}\int_{{\cal B}}d^{n_E}\omega d^{m_H}\eta
d^{m_H}k\left( 2\pi t\right) ^{-n_E/2}\int\limits_0^tds\left(
2\pi s\right) ^{-n_E/2}\times
$$
$$
[(\left( \overbrace{H}\right) -\left( \overbrace{H}\right)
^1)G_s(u,\xi
,\frac \omega {\sqrt{2\pi t}},\eta ,k)\mid _{u=\widehat{u}_{t-s}^1,\xi =(%
\overbrace{\xi })_{t-s}^1,\eta =(\overbrace{\eta })_{t-s}^1}]
$$
and%
$$
B(t)={\cal E}\int_{{\cal B}}d^{n_E}\omega d^{m_H}\eta
d^{m_H}k\left( 2\pi t\right) ^{-n_E/2}\int\limits_0^tds\left(
2\pi s\right) ^{-n_E/2}\times
$$
$$
[(\left( \overbrace{H}\right) -\left( \overbrace{H}\right)
^0)G_s(u,\xi
,\frac \omega {\sqrt{2\pi t}},\eta ,k)\mid _{u=\widehat{u}_{t-s},\xi =(%
\overbrace{\xi })_{t-s},\eta =(\overbrace{\eta })_{t-s}}-
$$
$$
(\left( \overbrace{H}\right) -\left( \overbrace{H}\right)
^0)G_s(u,\xi
,\frac \omega {\sqrt{2\pi t}},\eta ,k)\mid _{u=\widehat{u}_{t-s}^1,\xi =(%
\overbrace{\xi })_{t-s}^1,\eta =(\overbrace{\eta })_{t-s}^1}].
$$

Using the standard flat space Brownian motion techniques we can
show that for any suitable regular function $f$ on $S\left( {\cal
R}^{n_E}{\cal \times
C}^{m_H}\right) $ we write%
$$
{\cal E}f\left( \widehat{u}_{t-s}^1,(\overbrace{\xi })_{t-s}^1,(\overbrace{%
\eta })_{t-s}^1\right) =\exp \left[ -\left( \overbrace{H}\right)
^1(t-s)\right] f\left( u,\frac \omega {\sqrt{2\pi t}},\eta \right) =%
\eqno(5.22)
$$
$$
\mbox{ (taking into account (5.21) ) }=
$$
$$
{\cal E}[\exp (\int\limits_0^{t-s}(-\frac i4\left( \zeta
_s^{<\alpha >}-i\pi _s^{<\alpha >}\right) \left( \zeta _s^{<\beta
>}-i\pi _s^{<\beta >}\right) F_{<\alpha ><\beta >q}^r\eta
_s^{0q}k_{rs}^0ds-
$$
$$
\frac 14\widetilde{R}_{<\alpha ><\beta ><\gamma ><\delta >}\zeta
_s^{<\alpha
>}\zeta _s^{<\beta >}\rho _s^{<\gamma >}\rho _s^{<\delta >}+\frac i3%
\widetilde{R}_{<\alpha ><\beta >}\zeta _s^{<\alpha >}\rho
_s^{<\beta >}ds-
$$
$$
\frac i3\zeta _s^{<\varepsilon >}\rho _s^{<\delta >}a_s^{<\tau >}(\widetilde{%
R}_{<\tau ><\delta ><\varepsilon ><\alpha >}+\widetilde{R}_{<\tau
><\varepsilon ><\delta ><\alpha >})da_s^{<\alpha >}))\times
$$
$$
f\left( b_{t-s},\frac \omega {\sqrt{2\pi t}}+\zeta _{t-s},\eta
-\eta _{t-s}^0\right) ].
$$

Estimations of solutions in a manner similar to that presented in
 [209,\\ 210] enable us to show that $A(t)\rightarrow 0$ as $t\rightarrow
0 $ and $B\left( t\right) \rightarrow 0$ as $t\rightarrow 0.\,$
Thus
$$
\lim _{t\rightarrow 0}str\exp \left[ -\left( \overbrace{H}\right)
t(0,0)\right] =\lim _{t\rightarrow 0}str\exp \left[ -\left( \overbrace{H}%
\right) ^1t(0,0)\right] .\eqno(5.23)
$$

Now we begin the final step in the proof of this theorem by using
the standard flat--space path--integral techniques. Applying once
again the
Duhamel's formula and using (5.22) we obtain%
$$
str\exp [-\left( \overbrace{H}\right) ^1t(0,0)]={\cal E[}\int_{{\cal B}%
}d^{n_E}\omega d^{m_H}\eta d^{m_H}k\left( 2\pi t\right)
^{-n_E/2}\times
$$
$$
\int\limits_0^tds\left( 2\pi s\right) ^{-n_E/2}d^{n_E}\zeta
^{\prime }\{\exp
\left[ -\left( \overbrace{H}\right) ^1(t-s)\right] (\left( \overbrace{H}%
\right) ^1-\left( \overbrace{H}\right) ^0)
$$
$$
G_s(0,\frac \omega {\sqrt{2\pi t}},\zeta ^{\prime },\eta ,-\eta
)\exp (-(ik\eta )\exp \left( -i\frac \omega {\sqrt{2\pi t}}\zeta
^{\prime }\right) \}]=
$$
$$
{\cal E[}\int_{{\cal B}}d^{n_E}\omega d^{m_H}\eta
d^{m_H}k\int\limits_0^tds\left( 2\pi s\right) ^{-n_E/2}\times
$$
$$
[\exp (\int\limits_0^{t-s}(\frac 14\left( \zeta _s^{<\alpha
>}-i\pi _s^{<\alpha >}\right) \left( \zeta _s^{<\beta >}-i\pi
_s^{<\beta >}\right) F_{<\alpha ><\beta >q}^r\eta
_s^{0q}k_{rs}^0ds-
$$
$$
\frac 14\widetilde{R}_{<\alpha ><\beta ><\gamma ><\delta >}\zeta
_s^{<\alpha
>}\zeta _s^{<\beta >}\rho _s^{<\gamma >}\rho _s^{<\delta >}+\frac i3%
\widetilde{R}_{<\alpha ><\beta >}\zeta _s^{<\alpha >}\rho
_s^{<\beta >}ds+
$$
$$
\frac i3\zeta _s^{<\varepsilon >}\rho _s^{<\delta >}a_s^{<\tau >}(\widetilde{%
R}_{<\tau ><\delta ><\varepsilon ><\alpha >}+\widetilde{R}_{<\tau
><\varepsilon ><\delta ><\alpha >})da_s^{<\alpha >}))\times
$$
$$
(\left( \overbrace{H}\right) ^1-\left( \overbrace{H}\right)
^0)G_s(u,\xi ,\frac \omega {\sqrt{2\pi t}},\alpha ,k)\mid
_{u=\widehat{a}_{t-s},\xi =\xi _{t-s},\alpha =\eta _{t-s}^0}\exp
(ik\eta )]].
$$
After some usual estimates for flat space Brownian paths it can be seen that%
$$
\lim _{t\rightarrow 0}str\exp \left[ -\left( \overbrace{H}\right)
^1t(0,0)\right] ={\cal E\{}\lim _{t\rightarrow 0}\int_{{\cal B}%
}d^{n_E}\omega d^{m_H}\eta d^{m_H}k\left( 2\pi t\right)
^{-n_E/2}\times
$$
$$
\int\limits_0^tds\left( 2\pi s\right) ^{-n_E/2}[(\left(
\overbrace{H}\right) ^2-\left( \overbrace{H}\right) ^0)
$$
$$
G_s(u,\xi ,\frac \omega {\sqrt{2\pi t}},\alpha ,k)\mid _{u=\widehat{a}%
_{t-s},\xi =\xi _{t-s},\alpha =\eta _{t-s}^0}\exp (ik\eta )]\},
$$
where%
$$
\left( \overbrace{H}\right) ^2=-(\frac 12\delta _{<\alpha
>}\delta _{<\alpha
>}-
$$
$$
\frac 13\frac{\omega ^{<\alpha >}}{\sqrt{2\pi t}}\frac{\omega ^{<\beta >}}{%
\sqrt{2\pi t}}\left( \widetilde{R}_{<\gamma >\quad <\beta
><\alpha >}^{\quad <\delta >}+\widetilde{R}_{<\gamma ><\beta
>\qquad <\alpha >}^{\qquad \qquad <\delta >}\right)
\widehat{u}^{<\gamma >}\delta _{<\delta >}-
$$
$$
\frac 1{18}\frac{\omega ^{<\varepsilon >}}{\sqrt{2\pi
t}}\frac{\omega
^{<\beta >}}{\sqrt{2\pi t}}\frac{\omega ^{<\alpha >}}{\sqrt{2\pi t}}\frac{%
\omega ^{<\nu >}}{\sqrt{2\pi t}}(\widetilde{R}_{<\tau ><\mu
><\varepsilon
><\alpha >}+\widetilde{R}_{<\tau ><\varepsilon ><\mu ><\alpha >})\times
$$
$$
\left( \widetilde{R}_{<\gamma >\quad <\beta ><\nu >}^{\quad <\mu >}+%
\widetilde{R}_{<\gamma ><\beta >\qquad <\nu >}^{\qquad \qquad
<\mu >}\right) \widehat{u}^{<\tau >}\widehat{u}^{<\gamma >}+
$$
$$
\frac 14\frac{\omega ^{<\alpha >}}{\sqrt{2\pi t}}\frac{\omega ^{<\beta >}}{%
\sqrt{2\pi t}}\widetilde{R}_{<\alpha ><\beta ><\varepsilon ><\delta >}%
\widehat{\chi }^{<\varepsilon >}\widehat{\chi }^{<\delta
>}+\frac{\omega
^{<\alpha >}}{\sqrt{2\pi t}}\frac{\omega ^{<\beta >}}{\sqrt{2\pi t}}%
F_{<\alpha ><\beta >r}^s\widehat{\eta }^r\frac \partial {\partial
\eta ^s},
$$
where $\widehat{\chi }^{<\varepsilon >}=\zeta ^{<\varepsilon
>}+\frac \delta
{\partial \zeta ^{<\varepsilon >}}.\,$ Thus%
$$
\lim _{t\rightarrow 0}str\exp \left[ -\left( \overbrace{H}\right)
^1t(0,0)\right] =\lim _{t\rightarrow 0}str\exp \left[ -\left( \overbrace{H}%
\right) ^2t(0,0)\right] .\eqno(5.24)
$$
The operator $\left( \overbrace{H}\right) ^2$ decouples into
operators
acting separately on the $u$- $\zeta $- and $\eta $--variables:%
$$
\left( \overbrace{H}\right) ^2=\left( \overbrace{H}\right)
_u^2+\left( \overbrace{H}\right) _\zeta ^2+\left(
\overbrace{H}\right) _\eta ^2,
$$
where%
$$
\left( \overbrace{H}\right) _u^2=-(\frac 12\delta _{<\alpha
>}\delta _{<\alpha >}+\frac 12\frac{\omega ^{<\alpha
>}}{\sqrt{2\pi t}}\frac{\omega ^{<\beta >}}{\sqrt{2\pi
t}}\widetilde{R}_{<\alpha ><\beta ><\nu >}^{\qquad \qquad <\mu
>}\widehat{u}^{<\nu >}\delta _{<\mu >}+
$$
$$
\frac 18\frac{\omega ^{<\nu >}}{\sqrt{2\pi t}}\frac{\omega ^{<\beta >}}{%
\sqrt{2\pi t}}\frac{\omega ^{<\tau >}}{\sqrt{2\pi t}}\frac{\omega
^{<\alpha
>}}{\sqrt{2\pi t}}\widetilde{R}_{<\tau ><\nu ><\varepsilon ><\mu >}%
\widetilde{R}_{<\alpha ><\beta ><\gamma >}^{\qquad \qquad <\mu >}\widehat{u}%
^{<\varepsilon >}\widehat{u}^{<\gamma >}),
$$
$$
\left( \overbrace{H}\right) _\zeta ^2=-\frac 14\frac{\omega ^{<\alpha >}}{%
\sqrt{2\pi t}}\frac{\omega ^{<\beta >}}{\sqrt{2\pi
t}}\widetilde{R}_{<\alpha
><\beta ><\varepsilon ><\delta >}\widehat{\chi }^{<\varepsilon >}\widehat{%
\chi }^{<\delta >}
$$
and%
$$
\left( \overbrace{H}\right) _\eta ^2=-\frac{\omega ^{<\alpha >}}{\sqrt{2\pi t%
}}\frac{\omega ^{<\beta >}}{\sqrt{2\pi t}}F_{<\alpha ><\beta >r}^s\widehat{%
\eta }^r\frac \partial {\partial \eta ^s}.
$$

In order to evaluate $\exp \left[ -\left( \overbrace{H}\right)
_u^2t(0,0)\right] \,\,$ we use result given in
 [69] for ${\cal R}^2:$%
$$
L=-\frac 12\delta _{<\alpha >}\delta _{<\alpha >}+\frac{iB}2\left(
x^1\partial _2-x^2\partial _1\right) +\frac 18B^2\left[
(x^1)^2+(x^2)^2\right]
$$
then\ $ \exp [-Lt(0,0)]=\frac B{4\pi \sinh \left( \frac
12Bt\right) }. $ Let introduce $\Omega _{<\alpha >}^{\quad <\beta
>}=\frac 12\omega ^{<\varepsilon >}\omega ^{<\tau >}$\\
$\widetilde{R}_{<\varepsilon ><\tau
><\alpha >}^{\qquad \qquad <\beta >}$ a distinguished $n_E\times n_E$
matrix, skew--diagonalized as
$$
\Omega =\left( \Omega _{<\alpha >}^{\quad <\beta >}\right) =$$
{\footnotesize $$ \left(
\begin{array}{cccccccccccc}
0 & \Omega _1^{(0)} & ... & 0 & 0 & ... & 0 & 0 & ... & 0 & 0 & ... \\
-\Omega _1^{(0)} & 0 & ... & 0 & 0 & ... & 0 & 0 & ... & 0 & 0 & ... \\
... & ... & ... & ... & ... & ... & ... & ... & ... & ... & ... & ... \\
0 & 0 & ... & 0 & \Omega _{n/2}^{(0)} & ... & 0 & 0 & ... & 0 & 0 & ... \\
0 & 0 & ... & -\Omega _{n/2}^{(0)} & 0 & ... & 0 & 0 & ... & 0 & 0 & ... \\
... & ... & ... & ... & ... & ... & ... & ... & ... & ... & ... & ... \\
0 & 0 & ... & 0 & 0 & ... & 0 & \Omega _1^{(p)} & ... & 0 & 0 & ... \\
0 & 0 & ... & 0 & 0 & ... & -\Omega _1^{(p)} & 0 & ... & 0 & 0 & ... \\
... & ... & ... & ... & ... & ... & ... & ... & ... & ... & ... & ... \\
0 & 0 & ... & 0 & 0 & ... & 0 & 0 & ... & 0 & \Omega _{m_p/2}^{(p)} & ... \\
0 & 0 & ... & 0 & 0 & ... & 0 & 0 & ... & -\Omega _{m_p/2}^{(p)}
& 0 & ...
\\
... & ... & ... & ... & ... & ... & ... & ... & ... & ... & ... &
...
\end{array}
\right)
$$}

Now we can compute%
$$
\exp \left[ -\left( \overbrace{H}\right) _u^2t(0,0)\right]
=\prod\limits_{k_{(0)}=1}^{n/2}...\prod\limits_{k_{(p)}=1}^{m_{(p)}/2}...%
\prod\limits_{k_{(z)}=1}^{m_{(z)}/2}\times
$$
$$
\frac{i\Omega _{k_{(0)}}^{(0)}}{2\pi t}\frac 1{\sinh (i\Omega
_{k_{(0)}}^{(0)}/2\pi )}....\frac{i\Omega _{k_{(p)}}^{(p)}}{2\pi
t}\frac
1{\sinh (i\Omega _{k_{(p)}}^{(p)}/2\pi )}...\frac{i\Omega _{k_{(z)}}^{(z)}}{%
2\pi t}\frac 1{\sinh (i\Omega _{k_{(z)}}^{(z)}/2\pi )}.
$$
Using fermion path 
 [206] by direct calculation we find
$$
\exp \left[ -\left( \overbrace{H}\right) _\zeta ^2(\zeta ,\zeta
^{\prime })\right] =\int_{{\cal B}}d^n\rho _{(0)}...d^{m_p}\rho
_{(p)}...d^{m_z}\rho _{(z)}
$$
$$
\{\exp [-i\rho _{(0)}(\theta -\theta ^{\prime })]...\exp [-i\rho
_{(p)}(\zeta _{(p)}-\zeta _{(p)}^{\prime })]...\exp [-i\rho
_{(z)}(\zeta _{(z)}-\zeta _{(z)}^{\prime })]\times
$$
$$
\prod\limits_{k_{(0)}=1}^{n/2}[ \cosh (\frac{i\Omega _{k_{(0)}}}{2\pi }%
)+\left( \theta ^{2k_{(0)}-1}+i\rho ^{2k_{(0)}-1}\right) \left(
\theta
^{2k_{(0)}}+i\rho ^{2k_{(0)}}\right) \sinh (\frac{i\Omega _{k_{(0)}}}{2\pi }%
)] \times ...
$$
$$ \times
\prod\limits_{k_{(p)}=1}^{m_p/2}[ \cosh (\frac{i\Omega
_{k_{(p)}}}{2\pi })+\left( \zeta ^{2k_{(p)}-1}+i\rho
^{2k_{(p)}-1}\right) \left( \zeta
^{2k_{(p)}}+i\rho ^{2k_{(p)}}\right) \sinh (\frac{i\Omega _{k_{(p)}}}{2\pi }%
)] ...
$$
$$\times
\prod\limits_{k_{(z)}=1}^{m_z/2}[ \cosh (\frac{i\Omega
_{k_{(z)}}}{2\pi })+\left( \zeta ^{2k_{(z)}-1}+i\rho
^{2k_{(z)}-1}\right) \left( \zeta
^{2k_{(z)}}+i\rho ^{2k_{(z)}}\right) \sinh (\frac{i\Omega _{k_{(z)}}}{2\pi }%
)]
$$
and%
$$
str\exp \left[ -\left( \overbrace{H}\right) ^2t(0,0)\right] =
$$
$$
\int_{{\cal B}}d^n\omega _{(0)}...d^{m_p}\omega
_{(p)}...d^{m_z}\omega
_{(z)}\prod\limits_{k_{(0)}=1}^{n/2}...\prod%
\limits_{k_{(p)}=1}^{m_{(p)}/2}...\prod\limits_{k_{(z)}=1}^{m_{(z)}/2}\times
$$
$$
\frac{i\Omega _{k_{(0)}}^{(0)}}{2\pi t}\frac 1{\sinh (i\Omega
_{k_{(0)}}^{(0)}/2\pi )}....\frac{i\Omega _{k_{(p)}}^{(p)}}{2\pi
t}\frac
1{\sinh (i\Omega _{k_{(p)}}^{(p)}/2\pi )}...\frac{i\Omega _{k_{(z)}}^{(z)}}{%
2\pi t}\frac 1{\sinh (i\Omega _{k_{(z)}}^{(z)}/2\pi )}
$$
$$
\cosh (\frac{i\Omega _{k_{(0)}}}{4\pi })...\cosh (\frac{i\Omega _{k_{(p)}}}{%
4\pi })...\cosh (\frac{i\Omega _{k_{(z)}}}{4\pi })tr\left[ \exp
\left(
-\omega ^{<\alpha >}\omega ^{<\beta >}\right) \frac{F_{<\alpha ><\beta >}}{%
2\pi }\right] .
$$
Using (5.20),(5.23) and (5.24) we obtain the required result of the theorem%
$$
str\exp \left[ -H(w,w)\right] dvol=
$$
$$
\left[ tr\exp \left( -\frac F{2\pi }\right) \det \left(
\frac{i\Omega /2\pi }{\tanh (i\Omega /2\pi )}\right)
^{1/2}\right] \mid _w.
$$
$\Box $

\part{ Higher Order Anisotropic Interac\-ti\-ons}


\chapter{HA--Spinors}

Some of fundamental problems in physics advocate the extension to
locally anisotropic and higher order anisotropic backgrounds of
physical theories
 [159,161,13,29,18,162,272,265,266,267]. In order to
construct physical models on higher order anisotropic spaces it
is necessary a corresponding generalization of the spinor theory.
Spinor variables and interactions of spinor fields on Finsler
spaces were used in a heuristic manner, for instance, in works
 [18,177], where the problem of a
rigorous definition of la-spinors for la-spaces was not
considered. Here we note that, in general, the nontrivial
nonlinear connection and torsion structures and possible
incompatibility of metric and connections makes the solution of
the mentioned problem very sophisticate. The geometric definition
of la-spinors and a detailed study of the relationship between
Clifford, spinor and nonlinear and distinguished connections
structures in vector bundles, generalized Lagrange and Finsler
spaces are presented in
refs. 
 [256,255,264].

The purpose of the Chapter is to summarize and extend our
investigations
 [256,255,264,272,260] on formulation of the theory of classical
and quantum field interactions on higher order anisotropic
spaces. We receive primary attention to the development of the
necessary geometric framework: to propose an abstract spinor
formalism and formulate the differential geometry of higher order
anisotropic spaces . The next step is the investigation of higher
order anisotropic interactions of fundamental fields on generic
higher order anisotropic spaces (in brief we shall use instead of
higher order anisotropic the abbreviation ha-, for instance,
ha--spaces, ha--interactions and ha--spinors). \index{Ha--spaces}
\index{Ha--interactions} \index{Ha--spinors}

In order to develop the higher order anisotropic spinor theory it
will be convenient to extend the Penrose and Rindler abstract
index formalism
 [180,181,182] (see also the Luehr and Rosenbaum index free methods
 [154]) proposed for spinors on locally isotropic spaces. We note that in
order to formulate the locally anisotropic physics usually we have
dimensions $d>4$ for the fundamental, in general higher order
anisotropic space--time and to take into account physical effects
of the nonlinear connection structure. In this case the 2-spinor
calculus does not play a
 preferential role.

Section 6.1 of this Chapter contains an introduction into the
geometry of higher order anisotropic spaces, the  distinguishing
of geometric objects by N--connection structures  in such spaces
is analyzed, explicit formulas for coefficients of torsions and
curvatures of N- and d--connections are presented and the field
equations for gravitational interactions with higher order
anisotropy are formulated. The distinguished Clifford algebras
are introduced in section 6.2 and higher order anisotropic
Clifford bundles are defined in
 section 6.3. We present a study of almost complex structure for the case of
 locally anisotropic spaces modeled in the framework of the almost Hermitian
 model of generalized Lagrange spaces in section 6.4. The d--spinor
 techniques is analyzed in section 6.5 and the differential
 geometry of higher order anisotropic spinors is formulated in section 6.6.
 The section 6.7 is devoted to geometric aspects of the theory of field
 interactions with higher order anisotropy (the d--tensor and d--spinor form
 of basic field equations for gravitational, gauge and d--spinor fields are
 introduced). Finally, an outlook and conclusions on ha--spinors are given in
 section 6.8.

\section{ Basic Geometric Objects in Ha--Spaces}

We review some results and methods of the differential geometry
of vector bundles provided with nonlinear and distinguished
connections and metric
structures 
 [160,161,162,265,266,267].\ This section
serves the twofold purpose of establishing of abstract index
denotations and starting the geometric backgrounds which are used
in the next sections of the Chapter. We note that a number of
formulas can be obtained as even components of similar geometric
objects and equations introduced in Chapters 1 and 2 of the first
part of the monograph (for locally trivial considerations we can
formally omit ''tilde'' on denotations of s-bundles, change
s-indices into even ones and supersimmetric commutation rules into
usual ones).

\subsection{ N--connections: distinguish\-ing of ge\-o\-met\-ric ob\-jects}

Let ${\cal E}^{<z>}{\cal =}$ $\left(
E^{<z>},p,M,Gr,F^{<z>}\right) $ be a
locally trivial distinguished vector bundle, dv-bundle, where $F^{<z>}={\cal %
R}^{m_1}\oplus ...\oplus {\cal R}^{m_z}$ (a real vector space of dimension $%
m=m_1+...+m_z,\dim F=m,$ ${\cal R\ }$ denotes the real number
field) is the typical fibre, the structural group is chosen to be
the group of automorphisms of ${\cal R}^m$ , i.e. $Gr=GL\left(
m,{\cal R}\right) ,\,$ and
$p:E^{<z>}\rightarrow M$ (defined by intermediar projections $%
p_{<z,z-1>}:E^{<z>}\rightarrow
E^{<z-1>},p_{<z-1,z-2>}:E^{<z-1>}\rightarrow
E^{<z-2>},...p:E^{<1>}\rightarrow M)$  is a differentiable
surjection of a differentiable manifold $E$ (total space, $\dim
E=n+m)$ to a differentiable manifold $M$ (base space, $\dim M=n
).$ Local coordinates on ${\cal E}^{<z>}$ are denoted as
$$
u^{<{\bf \alpha >}}=\left( x^{{\bf i}},y^{<{\bf a>\ }}\right) =\left( x^{%
{\bf i}}\doteq y^{{\bf a}_0},y^{{\bf a}_1},....,y^{{\bf a}_z}\right) =%
$$
$$
(...,y^{{\bf a}_{(p)}},...)=\{y_{(p)}^{{\bf a}_{(p)}}\}=\{y^{{\bf a}%
_{(p)}}\},
$$
or in brief ${\bf u=u_{<z>}=(x,y}_{(1)},...,{\bf
y}_{(p)},...,{\bf y}_{(z)})$ where boldfaced indices will be
considered as coordinate ones for which the
Einstein summation rule holds (Latin indices ${\bf i,j,k,...=a}_0{\bf ,b}_0%
{\bf ,c}_0{\bf ,...}=1,2,...,n$ will parametrize coordinates of
geometrical objects with respect to a base space $M,$ Latin
indices ${\bf a}_p , {\bf b}_p,$ ${\bf c}_p,...=$
$1,2,...,m_{(p)}$ will parametrize fibre coordinates of
geometrical objects and Greek indices ${\bf \alpha ,\beta ,\gamma
,...}$ are considered as cumulative ones for coordinates of
objects defined on the
total space of a v-bundle). We shall correspondingly use abstract indices $%
\alpha =(i,a),$ $\beta =(j,b),\gamma =(k,c),...$ in the Penrose
manner
 [180,181,182] in order to mark geometical objects and theirs (base,
fibre)-components or, if it will be convenient, we shall consider
boldfaced letters (in the main for pointing to the operator
character of tensors and spinors into consideration) of type
${\bf A\equiv }A=\left( A^{(h)},A^{(v_1)},...,A^{(v_z)}\right)
{\bf ,b=}\left( b^{(h)},b^{(v_1)},...,b^{(v_z)}\right) ,...,{\bf
R},$ ${\bf \omega },$ ${\bf \Gamma},...$ for geometrical objects
on ${\cal E}$ and theirs splitting into horizontal (h), or base,
and vertical (v), or fibre, components. For simplicity, we shall
prefer writing out of abstract indices instead of boldface ones
if this will not give rise to ambiguities.

Coordinate trans\-forms $u^{<{\bf \alpha ^{\prime }>\ }}=u^{<{\bf
\alpha ^{\prime }>\ }}\left( u^{<{\bf \alpha >}}\right) $ on
${\cal E}^{<z>}$ are writ\-ten as
$$
\{u^{<\alpha >}=\left( x^{{\bf i}},y^{<{\bf a>}}\right)
\}\rightarrow \{u^{<\alpha ^{\prime }>}=\left( x^{{\bf i^{\prime
}\ }},y^{<{\bf a^{\prime }>}}\right) \}
$$
and written as recurrent maps
$$
x^{{\bf i^{\prime }\ }}=x^{{\bf i^{\prime }\ }}(x^{{\bf
i}}),~rank\left(
\frac{\partial x^{{\bf i^{\prime }\ }}}{\partial x^{{\bf i}}}\right) =n,%
\eqno(6.1)
$$
$$
y_{(1)}^{{\bf a_1^{\prime }\ }}=K_{{\bf a}_1{\bf \ }}^{{\bf a_1^{\prime }}%
}(x^{{\bf i\ }})y_{(1)}^{{\bf a}_1},K_{{\bf a}_1{\bf \ }}^{{\bf a_1^{\prime }%
}}(x^{{\bf i\ }})\in GL\left( m_1,{\cal R}\right) ,
$$
$$
............
$$
$$
y_{(p)}^{{\bf a_p^{\prime }\ }}=K_{{\bf a}_p{\bf \ }}^{{\bf a_p^{\prime }}%
}(u_{(p-1)})y_{(p)}^{{\bf a}_p},K_{{\bf a}_p{\bf \ }}^{{\bf a_p^{\prime }}%
}(u_{(p-1)})\in GL\left( m_p,{\cal R}\right) ,
$$
$$
.............
$$
$$
y_{(z)}^{{\bf a_z^{\prime }\ }}=K_{{\bf a}_z{\bf \ }}^{{\bf a_z^{\prime }}%
}(u_{(z-1)})y_{(z-1)}^{{\bf a}_z},K_{{\bf a}_z{\bf \ }}^{{\bf a_z^{\prime }}%
}(u_{(z-1)})\in GL\left( m_z,{\cal R}\right)
$$
where matrices $K_{{\bf a}_1{\bf \ }}^{{\bf a_1^{\prime }}}(x^{{\bf i\ }%
}),...,K_{{\bf a}_p{\bf \ }}^{{\bf a_p^{\prime }}}(u_{(p-1)}),...,K_{{\bf a}%
_z{\bf \ }}^{{\bf a_z^{\prime }}}(u_{(z-1)})$ are functions of
necessary
smoothness class. In brief we write transforms (6.1) in the form%
$$
x^{{\bf i^{\prime }\ }}=x^{{\bf i^{\prime }\ }}(x^{{\bf i}}),y^{<{\bf %
a^{\prime }>\ }}=K_{<{\bf a>}}^{<{\bf a^{\prime }>}}y^{<{\bf a>}}.
$$
In general form we shall write $K$--matrices $K_{<\alpha {\bf
>}}^{<\alpha
{\bf ^{\prime }>}}=\left( K_{{\bf i}}^{{\bf i^{\prime }}},K_{<{\bf a>}}^{<%
{\bf a^{\prime }>}}\right) ,$ where $K_{{\bf i}}^{{\bf i^{\prime }}}=\frac{%
\partial x^{{\bf i^{\prime }\ }}}{\partial x^{{\bf i}}}.$

A local coordinate parametrization of ${\cal E}^{<z>}$ naturally
defines a \index{Basis!coordinate} coordinate basis of the module
of d--vector fi\-elds \index{module!of d--vector fi\-elds}
 $\Xi \left( {\cal E}%
^{<z>}\right) ,$%
$$
\partial _{<\alpha >}=(\partial _i,\partial _{<a>})=(\partial _i,\partial
_{a_1},...,\partial _{a_p},...,\partial _{a_z})=\eqno(6.2)
$$
$$
\frac \partial {\partial u^{<\alpha >}}=\left( \frac \partial
{\partial x^i},\frac \partial {\partial y^{<a>}}\right) =\left(
\frac \partial {\partial x^i},\frac \partial {\partial
y^{a_1}},...\frac \partial {\partial y^{a_p}},...,\frac \partial
{\partial y^{a_z}}\right) ,
$$
and the reciprocal to (6.2) coordinate basis
\index{Basis!reciprocal}
$$
d^{<\alpha
>}=(d^i,d^{<a>})=(d^i,d^{a_1},...,d^{a_p},...,d^{a_z})=\eqno(6.3)
$$
$$
du^{<\alpha
>}=(dx^i,dy^{<a>})=(dx^i,dy^{a_1},...,dy^{a_p},...,dy^{a_z}),
$$
which is uniquely defined from the equations%
$$
d^{<\alpha >}\circ \partial _{<\beta >}=\delta _{<\beta
>}^{<\alpha >},
$$
where $\delta _{<\beta >}^{<\alpha >}$ is the Kronecher symbol and by ''$%
\circ $$"$ we denote the inner (scalar) product in the tangent bundle ${\cal %
TE}^{<z>}.$

The concept of {\bf nonlinear connection,} in brief,
N--connection, is \index{Nonlinear connection!N--connection}
fundamental in the geometry of locally anisotropic and higher
order anisotropic spaces (see a detailed study and basic
references in
 [160,161,162] and Chapter 1 of this monograph for a supersymmetric
definition of N--connection). In a dv--bundle ${\cal E}^{<z>}$ it
is defined as a distribution $\{N:E_u\rightarrow
H_uE,T_uE=H_uE\oplus V_u^{(1)}E\oplus ...\oplus
V_u^{(p)}E...\oplus V_u^{(z)}E\}$ on $E^{<z>}$ being a global
decomposition, as a Whitney sum, into horizontal,${\cal HE,\ }$
and vertical, ${\cal VE}^{<p>}{\cal ,}p=1,2,...,z$ subbundles of
the tangent bundle ${\cal TE:}$ \index{Whitney sum}
$$
{\cal TE}=H{\cal E}\oplus V{\cal E}^{<1>}\oplus ...\oplus V{\cal E}%
^{<p>}\oplus ...\oplus V{\cal E}^{<z>}.\eqno(6.4)
$$

Locally a N-connection in ${\cal E}^{<z>}$ is given by it components $N_{<%
{\bf a}_f>}^{<{\bf a}_p>}({\bf u}),z\geq p>f\geq 0$ (in brief we
shall write $N_{<a_f>}^{<a_p>}(u)$ ) with respect to bases (6.2)
and (6.3)):

$$
{\bf N}=N_{<a_f>}^{<a_p>}(u)\delta ^{<a_f>}\otimes \delta
_{<a_p>},(z\geq p>f\geq 0),
$$

We note that a linear connection in a dv-bundle ${\cal E}^{<z>}$
can be
considered as a particular case of a N-connection when $%
N_i^{<a>}(u)=K_{<b>i}^{<a>}\left( x\right) y^{<b>},$ where functions $%
K_{<a>i}^{<b>}\left( x\right) $ on the base $M$ are called the
Christoffel coefficients. \index{Christoffel coefficients}

To coordinate locally geometric constructions with the global
splitting of ${\cal E}^{<z>}$ defined by a N-connection
structure, we have to introduce a
lo\-cal\-ly adap\-ted bas\-is ( la--basis, la--frame ):%
\index{La--basis} \index{La--frame}
$$
\delta _{<\alpha >}=(\delta _i,\delta _{<a>})=(\delta _i,\delta
_{a_1},...,\delta _{a_p},...,\delta _{a_z}),\eqno(6.5)
$$
with components parametrized as

$$
\delta _i=\partial _i-N_i^{a_1}\partial
_{a_1}-...-N_i^{a_z}\partial _{a_z},
$$
$$
\delta _{a_1}=\partial _{a_1}-N_{a_1}^{a_2}\partial
_{a_2}-...-N_{a_1}^{a_z}\partial _{a_z},
$$
$$
................
$$
$$
\delta _{a_p}=\partial _{a_p}-N_{a_p}^{a_{p+1}}\partial
_{a_{p+1}}-...-N_{a_p}^{a_z}\partial _{a_z},
$$
$$
...............
$$
$$
\delta _{a_z}=\partial _{a_z}
$$
and it dual la--basis%
\index{La--basis!dual}
$$
\delta ^{<\alpha >}=(\delta ^i,\delta ^{<a>})=\left( \delta
^i,\delta ^{a_1},...,\delta ^{a_p},...,\delta ^{a_z}\right)
,\eqno(6.6)
$$
$$
\delta x^i=dx^i,
$$
$$
\delta y^{a_1}=dy^{a_1}+M_i^{a_1}dx^i,
$$
$$
\delta y^{a_2}=dy^{a_2}+M_{a_1}^{a_2}dy^{a_1}+M_i^{a_2}dx^i,
$$
$$
.................
$$
$$
\delta
y^{a_p}=dy^{a_p}+M_{a_{p-1}}^{a_p}dy^{p-1}+M_{a_{p-2}}^{a_p}dy^{a_{p-2}}+...+M_i^{a_p}dx^i,
$$
$$
...................
$$
$$
\delta
y^{a_z}=dy^{a_z}+M_{a_{z-1}}^{a_z}dy^{z-1}+M_{a_{z-2}}^{a_z}dy^{a_{z-2}}+...+M_i^{a_z}dx^i
$$
(for details on expressing of coefficients $M_{a_f}^{a_p},$ the
dual coefficients of a N--connections, by recurrent formulas
through the components $N_{a_f}^{a_p},$ see subsection 1.22; we
have to consider the even part for the calculus presented there).

The {\bf \ nonholonomic coefficients } $${\bf w}=\{w_{<\beta
><\gamma
>}^{<\alpha >}\left( u\right) \}$$ of locally adapted to the N--connection
structure frames  are defined as%
\index{Nonholonomic coefficients}
$$
\left[ \delta _{<\alpha >},\delta _{<\beta >}\right] =\delta
_{<\alpha
>}\delta _{<\beta >}-\delta _{<\beta >}\delta _{<\alpha >}= $$ $$w_{<\beta
><\gamma >}^{<\alpha >}\left( u\right) \delta _{<\alpha >}.
$$

The {\bf algebra of tensorial distinguished fields} $DT\left( {\cal E}%
^{<z>}\right) $ (d--fields, d--tensors, d--objects) on ${\cal
E}^{<z>}$ is \index{Algebra!of d--tensors} \index{D--fields}
\index{D--tensors} \index{D--objects}
introduced as the tensor algebra\\ ${\cal T}=\{{\cal T}%
_{qs_1...s_p...s_z}^{pr_1...r_p...r_z}\}$ of the dv-bundle ${\cal
E}_{\left( d\right) }^{<z>},$
$$
p_d:{\cal HE}^{<z>}{\cal \oplus V}^1{\cal E}^{<z>}{\cal \oplus }...{\cal %
\oplus V}^p{\cal E}^{<z>}{\cal \oplus }...{\cal \oplus V}^z{\cal E}^{<z>}%
{\cal \rightarrow E}^{<z>}{\cal .\,}
$$
${\cal \ }$ An element ${\bf t}\in {\cal
T}_{qs_1...s_z}^{pr_1...r_z},$ d-tensor field of type $\left(
\begin{array}{cccccc}
p & r_1 & ... & r_p & ... & r_z \\
q & s_1 & ... & s_p & ... & s_z
\end{array}
\right) ,$ can be written in local form as%
$$
{\bf t}%
=t_{j_1...j_qb_1^{(1)}...b_{r_1}^{(1)}...b_1^{(p)}...b_{r_p}^{(p)}...b_1^{(z)}...b_{r_z}^{(z)}}^{i_1...i_pa_1^{(1)}...a_{r_1}^{(1)}...a_1^{(p)}...a_{r_p}^{(p)}...a_1^{(z)}...a_{r_z}^{(z)}}\left(
u\right) \delta _{i_1}\otimes ...\otimes \delta _{i_p}\otimes
d^{j_1}\otimes ...\otimes d^{j_q}\otimes
$$
$$
\delta _{a_1^{(1)}}\otimes ...\otimes \delta
_{a_{r_1}^{(1)}}\otimes \delta ^{b_1^{(1)}}...\otimes \delta
^{b_{s_1}^{(1)}}\otimes ...\otimes \delta _{a_1^{(p)}}\otimes
...\otimes \delta _{a_{r_p}^{(p)}}\otimes ...\otimes
$$
$$
\delta ^{b_1^{(p)}}...\otimes \delta ^{b_{s_p}^{(p)}}\otimes
\delta _{a_1^{(z)}}\otimes ...\otimes \delta
_{a_{rz}^{(z)}}\otimes \delta ^{b_1^{(z)}}...\otimes \delta
^{b_{s_z}^{(z)}}.
$$

We shall respectively use denotations $X{\cal (E}^{<z>})$ (or $X{%
\left( M\right) ),\ }\Lambda ^p\left( {\cal E}^{<z>}\right) $ (or
$\Lambda ^p\left( M\right) ) $ and ${\cal F(E}^{<z>})$ (or $%
{\cal F}$ $\left( M\right) $) for the module of d-vector fields on ${\cal E}%
^{<z>}$ (or $M$ ), the exterior algebra of p-forms on ${\cal
E}^{<z>}$
(or $M)$ and the set of real functions on ${\cal E}^{<z>}$(or $%
M). $

In general, d--objects on ${\cal E}^{<z>}$ are introduced as
geometric objects with various group and coordinate transforms
coordinated with the N--connection structure on ${\cal E}^{<z>}.$
For example, a d--connection $D$ on ${\cal E}^{<z>}$ is defined
as a linear connection $D$ on $E^{<z>}$ conserving under a
parallelism the global
decomposition (6.4) into horizontal and vertical subbundles of ${\cal TE}%
^{<z>}$ .

A N-connection in ${\cal E}^{<z>}$ induces a corresponding
decomposition of d-tensors into sums of horizontal and vertical
parts, for example, for every d-vector $X\in {\cal X(E}^{<z>})$
and 1-form $\widetilde{X}\in \Lambda ^1\left( {\cal
E}^{<z>}\right) $ we have respectively
$$
X=hX+v_1X+...+v_zX{\bf \ \quad }\mbox{and \quad }\widetilde{X}=h\widetilde{X}%
+v_1\widetilde{X}+...~v_z\widetilde{X}.\eqno(6.7)
$$
In consequence, we can associate to every d-covariant derivation
along the d-vector (6.7), $D_X=X\circ D,$ two new operators of h-
and v-covariant derivations defined respectively as
$$
D_X^{(h)}Y=D_{hX}Y%
$$
and
$$
D_X^{\left( v_1\right) }Y=D_{v_1X}Y{\bf ,...,D_X^{\left(
v_z\right) }Y=D_{v_zX}Y\quad }\forall Y{\bf \in }{\cal
X(E}^{<z>}) ,
$$
for which the following conditions hold:%
$$
D_XY{\bf =}D_X^{(h)}Y + D_X^{(v_1)}Y+...+D_X^{(v_z)}Y, \eqno(6.8)
$$
$$
D_X^{(h)}f=(hX{\bf )}f
$$
and
$$
D_X^{(v_p)}f=(v_pX{\bf )}f,\quad X,Y{\bf \in }{\cal X\left(
E\right) ,}f\in {\cal F}\left( M\right) ,p=1,2,...z.
$$

We define a {\bf metric structure }${\bf G\ }$in the total space
$E^{<z>}$ \index{Metric structure} of dv-bundle ${\cal
E}^{<z>}{\cal =}$ $\left( E^{<z>},p,M\right) $ over a connected
and paracompact base $M$ as a symmetric covariant tensor field of
type $\left( 0,2\right) $, $G_{<\alpha ><\beta >,}$ being
nondegenerate and of constant signature on $E^{<z>}.$

Nonlinear connection ${\bf N}$ and metric ${\bf G}$ structures on ${\cal E}%
^{<z>}$ are mutually compatible it there are satisfied the
conditions:
$$
{\bf G}\left( \delta _{a_f},\delta _{a_p}\right) =0,\mbox{or equivalently, }%
G_{a_fa_p}\left( u\right) -N_{a_f}^{<b>}\left( u\right)
h_{a_f<b>}\left( u\right) =0,\eqno(6.9)
$$
where $h_{a_pb_p}={\bf G}\left( \partial _{a_p},\partial
_{b_p}\right) $ and $G_{b_fa_p}={\bf G}\left( \partial
_{b_f},\partial _{a_p}\right) ,0\leq
f<p\leq z,\,$ which gives%
$$
N_{c_f}^{b_p}\left( u\right) =h^{<a>b_p}\left( u\right)
G_{c_f<a>}\left( u\right) \eqno(6.10)
$$
(the matrix $h^{a_pb_p}$ is inverse to $h_{a_pb_p}).$ In
consequence one
obtains the following decomposition of metric :%
$$
{\bf G}(X,Y){\bf =hG}(X,Y)+{\bf v}_1{\bf G}(X,Y)+...+{\bf v}_z{\bf G}(X,Y)%
{\bf ,}\eqno(6.11)
$$
where the d-tensor ${\bf hG}(X,Y){\bf =G}(hX,hY)$ is of type
$\left(
\begin{array}{cc}
0 & 0 \\
2 & 0
\end{array}
\right) $ and the d-tensor ${\bf v}_p{\bf G}(X,Y)={\bf
G}(v_pX,v_pY)$ is of type $\left(
\begin{array}{ccccc}
0 & ... & 0(p) & ... & 0 \\
0 & ... & 2 & ... & z
\end{array}
\right) .$ With respect to la--basis (6.6) the d--metric (6.11) is written as%
$$
{\bf G}=g_{<\alpha ><\beta >}\left( u\right) \delta ^{<\alpha
>}\otimes \delta ^{<\beta >}=g_{ij}\left( u\right) d^i\otimes
d^j+h_{<a><b>}\left( u\right) \delta ^{<a>}\otimes \delta
^{<b>},\eqno(6.12)
$$
where $g_{ij}={\bf G}\left( \delta _i,\delta _j\right) .$

A metric structure of type (6.11) (equivalently, of type (6.12))
or a metric on $E^{<z>}$ with components satisfying constraints
(6.9), equivalently (6.10)) defines an adapted to the given
N-connection inner (d--scalar) \index{Inner product} product on
the tangent bundle ${\cal TE}^{<z>}{\cal .}$

We shall say that a d-connection $\widehat{D}_X$ is compatible
with the d-scalar product on ${\cal TE}^{<z>}{\cal \ }$ (i.e. is
a standard d-connection) if
$$
\widehat{D}_X\left( {\bf X\cdot Y}\right) =\left( \widehat{D}_X{\bf Y}%
\right) \cdot {\bf Z+Y\cdot }\left( \widehat{D}_X{\bf Z}\right)
,\forall {\bf X,Y,Z}{\bf \in }{\cal X(E}^{<z>}){\cal .}
$$
An arbitrary d--connection $D_X$ differs from the standard one
$\widehat{D}_X$ \index{D--connection}
by an operator $\widehat{P}_X\left( u\right) =\{X^{<\alpha >}\widehat{P}%
_{<\alpha ><\beta >}^{<\gamma >}\left( u\right) \},$ called the
deformation d-tensor with respect to $\widehat{D}_X,$ which is
just a d-linear transform of ${\cal E}_u^{<z>},$ $\forall u\in
{\cal E}^{<z>}{\cal .}$ The explicit form of $\widehat{P}_X$ can
be found by using the corresponding axiom
defining linear connections 
 [154]
$$
\left( D_X-\widehat{D}_X\right) fZ=f\left(
D_X-\widehat{D}_X\right) Z{\bf ,}
$$
written with respect to la-bases (6.5) and (6.6). From the last
expression we obtain
$$
\widehat{P}_X\left( u\right) =\left[ (D_X-\widehat{D}_X)\delta
_{<\alpha
>}\left( u\right) \right] \delta ^{<\alpha >}\left( u\right) ,
$$
therefore
$$
D_XZ{\bf \ }=\widehat{D}_XZ{\bf \ +}\widehat{P}_XZ.\eqno(6.13)
$$

A d-connection $D_X$ is {\bf metric (}or  compatible with met\-ric ${\bf G%
}$) on ${\cal E}^{<z>}$ if%
$$
D_X{\bf G}=0,\forall X{\bf \in }{\cal X(E}^{<z>}).\eqno(6.14)
$$
\index{D--connection!compatible with met\-ric}

Locally adapted components $\Gamma _{<\beta ><\gamma >}^{<\alpha
>}$ of a d-connection $D_{<\alpha >}=(\delta _{<\alpha >}\circ
D)$ are defined by the
equations%
$$
D_{<\alpha >}\delta _{<\beta >}=\Gamma _{<\alpha ><\beta
>}^{<\gamma
>}\delta _{<\gamma >},
$$
from which one immediately follows%
$$
\Gamma _{<\alpha ><\beta >}^{<\gamma >}\left( u\right) =\left(
D_{<\alpha
>}\delta _{<\beta >}\right) \circ \delta ^{<\gamma >}.\eqno(6.15)
$$

The operations of h- and v$_{(p)}$-covariant derivations, $%
D_k^{(h)}=\{L_{jk}^i,L_{<b>k\;}^{<a>}\}$ and $D_{c_p}^{(v_p)}=%
\{C_{jc_p}^i,C_{<b>c_p}^i,C_{jc_p}^{<a>},C_{<b>c_p}^{<a>}\}$ (see
(6.8)), are introduced as corresponding h- and
v$_{(p)}$-paramet\-ri\-za\-ti\-ons of
(6.15):%
$$
L_{jk}^i=\left( D_k\delta _j\right) \circ d^i,\quad
L_{<b>k}^{<a>}=\left( D_k\delta _{<b>}\right) \circ \delta
^{<a>}\eqno(6.16)
$$
and%
$$
C_{jc_p}^i=\left( D_{c_p}\delta _j\right) \circ \delta ^i,\quad
C_{<b>c_p}^{<a>}=\left( D_{c_p}\delta _{<b>}\right) \circ \delta ^{<a>}%
\eqno(6.17)
$$
$$
C_{<b>c_p}^i=\left( D_{c_p}\delta _{<b>}\right) \circ \delta
^i,\quad C_{jc_p}^{<a>}=\left( D_{c_p}\delta _j\right) \circ
\delta ^{<a>}.
$$
A set of components (6.16) and (6.17), $D\Gamma =\left(
L_{jk}^i,L_{<b>k}^{<a>},C_{j<c>}^i,C_{<b><c>}^{<a>}\right) ,\,$
completely defines the local action of a d-connection $D$ in
${\cal E}^{<z>}.$ For instance, taken a d-tensor field of type
$\left(
\begin{array}{cccc}
1 & ... & 1(p) & ... \\
1 & ... & 1(p) & ...
\end{array}
\right) ,$ ${\bf t}=t_{jb_p}^{ia_p}\delta _i\otimes \delta
_{a_p}\otimes \delta ^j\otimes \delta ^{b_p},$ and a d-vector
${\bf X}=X^i\delta
_i+X^{<a>}\delta _{<a>}$ we have%
$$
D_X{\bf t=}D_X^{(h)}{\bf t+}D_X^{(v_1)}{\bf %
t+..+.D_X^{(v_p)}t+...+D_X^{(v_z)}t=}
$$
$$
\left( X^kt_{jb_p|k}^{ia_p}+X^{<c>}t_{jb_p\perp
<c>}^{ia_p}\right) \delta _i\otimes \delta _{a_p}\otimes
d^j\otimes \delta ^{b_p},
$$
where the h--covariant derivative is written as%
\index{Derivative!h--covariant}
$$
t_{jb_p|k}^{ia_p}=\frac{\delta t_{jb_p}^{ia_p}}{\delta x^k}%
+L_{hk}^it_{jb_p}^{ha_p}+L_{c_pk}^{a_p}t_{jb_p}^{ic_p}-L_{jk}^ht_{hb_p}^{ia_p}-L_{b_pk}^{c_p}t_{jc_p}^{ia_p}
$$
and the v--covariant derivatives are written as%
\index{Derivatives!v--covariant}
$$
t_{jb_p\perp <c>}^{ia_p}=\frac{\partial t_{jb_p}^{ia_p}}{\partial y^{<c>}}%
+C_{h<c>}^it_{jb_p}^{ha_p}+C_{d_p<c>}^{a_p}t_{jb_p}^{id_p}-C_{j<c>}^ht_{hb_p}^{ia_p}-C_{b_p<c>}^{d_p}t_{jd_p}^{ia_p}.
$$
For a scalar function $f\in {\cal F(E}^{<z>})$ we have%
$$
D_i^{(h)}=\frac{\delta f}{\delta x^i}=\frac{\partial f}{\partial x^i}%
-N_i^{<a>}\frac{\partial f}{\partial y^{<a>}},
$$
$$
D_{a_f}^{(v_f)}=\frac{\delta f}{\delta x^{a_f}}=\frac{\partial
f}{\partial x^{a_f}}-N_{a_f}^{a_p}\frac{\partial f}{\partial
y^{a_p}},1\leq f<p\leq z-1,
$$
$$
\mbox{ and }D_{c_z}^{(v_z)}f=\frac{\partial f}{\partial y^{c_z}}.
$$

We emphasize that the geometry of connections in a dv-bundle
${\cal E}^{<z>}$ is very reach. If a triple of fundamental
geometric objects
$$(N_{a_f}^{a_p}\left( u\right) ,
 \Gamma _{<\beta ><\gamma >}^{<\alpha >}\left(u\right) ,
 G_{<\alpha ><\beta >}\left( u\right) ) $$ is fixed on ${\cal E}^{<z>},$
a multiconnection structure (with corresponding different
\index{Multiconnection structure} rules of covariant derivation,
which are, or not, mutually compatible and with the same, or not,
induced d-scalar products in ${\cal TE}^{<z>}{\cal )}$ is defined
on this dv-bundle. For instance, we enumerate some of connections
and covariant derivations which can present interest in
investigation of locally anisotropic gravitational and matter
field interactions:

\begin{enumerate}
\item  Every N-connection in ${\cal E}^{<z>}{\cal ,}$ with coefficients $%
N_{a_f}^{a_p}\left( u\right) $ being differentiable on
y-variables, induces a structure of linear connection \\
$\widetilde{N}_{<\beta ><\gamma >}^{<\alpha
>},$ where $$\widetilde{N}_{b_pc_f}^{a_p}=\frac{\partial N_{c_f}^{a_p}}{%
\partial y^{b_p}}$$ and $$\widetilde{N}_{b_pc_p}^{a_p}\left( u\right) =0.$$ For
some $$Y\left( u\right) =Y^i\left( u\right) \partial
_i+Y^{<a>}\left( u\right) \partial _{<a>}$$ and $$B\left(
u\right) =B^{<a>}\left( u\right)
\partial _{<a>}$$ one writes%
$$
D_Y^{(\widetilde{N})}B=\left[ Y^{c_f}\left( \frac{\partial
B^{a_p}}{\partial
y^{c_f}}+\widetilde{N}_{b_pi}^{a_p}B^{b_p}\right)
+Y^{b_p}\frac{\partial B^{a_p}}{\partial y^{b_p}}\right] \frac
\partial {\partial y^{a_p}}~(0\leq f<p\leq z).
$$

\item  The d--connection of Berwald type 
 [39]
$$
\Gamma _{<\beta ><\gamma >}^{(B)<\alpha >}=\left(
L_{jk}^i,\frac{\partial N_k^{<a>}}{\partial
y^{<b>}},0,C_{<b><c>}^{<a>}\right) ,\eqno(6.18)
$$
where
$$
L_{.jk}^i\left( u\right) =\frac 12g^{ir}\left( \frac{\delta
g_{jk}}{\partial
x^k}+\frac{\delta g_{kr}}{\partial x^j}-\frac{\delta g_{jk}}{\partial x^r}%
\right) ,
$$

$$
C_{.<b><c>}^{<a>}\left( u\right) =\frac 12h^{<a><d>}\left(
\frac{\delta
h_{<b><d>}}{\partial y^{<c>}}+\frac{\delta h_{<c><d>}}{\partial y^{<b>}}-%
\frac{\delta h_{<b><c>}}{\partial y^{<d>}}\right) ,\eqno(6.19)
$$

which is hv-metric, i.e. $D_k^{(B)}g_{ij}=0$ and
$D_{<c>}^{(B)}h_{<a><b>}=0.$

\item  The canonical d--connection ${\bf \Gamma ^{(c)}}$ associated to a
\index{Canonical d--connection} \index{D--connection!canonical}
metric ${\bf G}$ of type (6.12) $\Gamma _{<\beta ><\gamma
>}^{(c)<\alpha
>}=\left(
L_{jk}^{(c)i},L_{<b>k}^{(c)<a>},C_{j<c>}^{(c)i},C_{<b><c>}^{(c)<a>}\right)
,$
with coefficients%
$$
L_{jk}^{(c)i}=L_{.jk}^i,C_{<b><c>}^{(c)<a>}=C_{.<b><c>}^{<a>}%
\mbox{ (see (6.19)}
$$
$$
L_{<b>i}^{(c)<a>}=\widetilde{N}_{<b>i}^{<a>}+
$$
$$
\frac 12h^{<a><c>}\left( \frac{\delta h_{<b><c>}}{\delta x^i}-\widetilde{N}%
_{<b>i}^{<d>}h_{<d><c>}-\widetilde{N}_{<c>i}^{<d>}h_{<d><b>}\right)
,
$$
$$
C_{j<c>}^{(c)i}=\frac 12g^{ik}\frac{\partial g_{jk}}{\partial y^{<c>}}.%
\eqno(6.20)
$$
This is a metric d--connection which satisfies conditions
\index{D--connection!metric}
$$
D_k^{(c)}g_{ij}=0,D_{<c>}^{(c)}g_{ij}=0,D_k^{(c)}h_{<a><b>}=0,D_{<c>}^{(c)}h_{<a><b>}=0.
$$

\item  We can consider N-adapted Christoffel d--symbols%
\index{Christoffel!d--symbols}
$$
\widetilde{\Gamma }_{<\beta ><\gamma >}^{<\alpha >}=$$ $$\frac
12G^{<\alpha
><\tau >}\left( \delta _{<\gamma >}G_{<\tau ><\beta >}+\delta _{<\beta
>}G_{<\tau ><\gamma >}-\delta _{<\tau >}G_{<\beta ><\gamma >}\right) ,%
\eqno(6.21)
$$
which have the components of d-connection $$\widetilde{\Gamma
}_{<\beta
><\gamma >}^{<\alpha >}=\left( L_{jk}^i,0,0,C_{<b><c>}^{<a>}\right) $$ with $%
L_{jk}^i$ and $C_{<b><c>}^{<a>}$ as in (6.19) if $G_{<\alpha
><\beta >}$ is taken in the form (6.12).
\end{enumerate}

Arbitrary linear connections on a dv--bundle ${\cal E}^{<z>}$ can
be also characterized by theirs deformation tensors (see (6.13))
with respect, for
instance, to d--connect\-i\-on (6.21):%
$$
\Gamma _{<\beta ><\gamma >}^{(B)<\alpha >}=\widetilde{\Gamma
}_{<\beta
><\gamma >}^{<\alpha >}+P_{<\beta ><\gamma >}^{(B)<\alpha >},\Gamma _{<\beta
><\gamma >}^{(c)<\alpha >}=\widetilde{\Gamma }_{<\beta ><\gamma >}^{<\alpha
>}+P_{<\beta ><\gamma >}^{(c)<\alpha >}
$$
or, in general,%
$$
\Gamma _{<\beta ><\gamma >}^{<\alpha >}=\widetilde{\Gamma
}_{<\beta ><\gamma
>}^{<\alpha >}+P_{<\beta ><\gamma >}^{<\alpha >},
$$
where $P_{<\beta ><\gamma >}^{(B)<\alpha >},P_{<\beta ><\gamma
>}^{(c)<\alpha >}$ and $P_{<\beta ><\gamma >}^{<\alpha >}$ are respectively
the deformation d--ten\-sors of d-connect\-i\-ons (6.18),\
(6.20), or of a general one.

\subsection{ D--Torsions and d--curvatures }

The curvature ${\bf \Omega }$$\,$ of a nonlinear connection ${\bf
N}$ in a \index{Curvature of N--connection}
dv-bundle ${\cal E}^{<z>}$ can be defined as the Nijenhuis tensor field $%
N_v\left( X,Y\right) $ associated to ${\bf N\ }$ 
 [160,161]:
$$
{\bf \Omega }=N_v={\bf \left[ vX,vY\right] +v\left[ X,Y\right]
-v\left[ vX,Y\right] -v\left[ X,vY\right] ,X,Y}\in {\cal
X(E}^{<z>}){\cal ,}
$$
where $v=v_1\oplus ...\oplus v_z.$ In local form one has%
$$
{\bf \Omega }=\frac 12\Omega _{b_fc_f}^{a_p}\delta
^{b_f}\bigwedge \delta ^{c_f}\otimes \delta _{a_p},(0\leq f<p\leq
z),
$$
where%
$$
\Omega _{b_fc_f}^{a_p}=\frac{\delta N_{c_f}^{a_p}}{\partial y^{b_f}}-\frac{%
\partial N_{b_f}^{a_p}}{\partial y^{c_f}}+N_{b_f}^{<b>}\widetilde{N}%
_{<b>c_f}^{a_p}-N_{c_f}^{<b>}\widetilde{N}_{<b>b_f}^{a_p}.\eqno(6.22)
$$

The torsion ${\bf T}$ of d--connection ${\bf D\ }$ in ${\cal
E}^{<z>}$ is \index{Torsion!of d--connection}
defined by the equation%
$$
{\bf T\left( X,Y\right) =XY_{\circ }^{\circ }T\doteq }D_X{\bf
Y-}D_Y{\bf X\ -\left[ X,Y\right] .}\eqno(6.23)
$$
One holds the following h- and v$_{(p)}-$--decompositions%
$$
{\bf T\left( X,Y\right) =T\left( hX,hY\right) +T\left(
hX,vY\right) +T\left( vX,hY\right) +T\left( vX,vY\right)
.}\eqno(6.24)
$$
We consider the projections: ${\bf hT\left( X,Y\right)
,v}_{(p)}{\bf T\left(
hX,hY\right) ,hT\left( hX,hY\right) ,...}$ and say that, for instance, ${\bf %
hT\left( hX,hY\right) }$ is the h(hh)-torsion of\\ ${\bf D}$ , ${\bf v}_{(p)}%
{\bf T\left( hX,hY\right) \ }$ is the v$_p$(hh)-torsion of ${\bf
D}$ and so on.

The torsion (6.23) is locally determined by five d-tensor fields,
torsions, defined as
$$
T_{jk}^i={\bf hT}\left( \delta _k,\delta _j\right) \cdot d^i,\quad
T_{jk}^{a_p}={\bf v}_{(p)}{\bf T}\left( \delta _k,\delta
_j\right) \cdot \delta ^{a_p},
$$
$$
P_{jb_p}^i={\bf hT}\left( \delta _{b_p},\delta _j\right) \cdot
d^i,\quad P_{jb_f}^{a_p}={\bf v}_{(p)}{\bf T}\left( \delta
_{b_f},\delta _j\right) \cdot \delta ^{a_p},
$$
$$
S_{b_fc_f}^{a_p}={\bf v}_{(p)}{\bf T}\left( \delta _{c_f},\delta
_{b_f}\right) \cdot \delta ^{a_p}.
$$
Using formulas (6.5),(6.6),(6.22) and (6.23) we can computer in
explicit form the components of torsions (6.24) for a
d--connection of type (6.16) and (6.17):
$$
T_{.jk}^i=T_{jk}^i=L_{jk}^i-L_{kj}^i,\quad
T_{j<a>}^i=C_{.j<a>}^i,T_{<a>j}^i=-C_{j<a>}^i,\eqno(6.25)
$$
$$
T_{.j<a>}^i=0,T_{.<b><c>}^{<a>}=S_{.<b><c>}^{<a>}=C_{<b><c>}^{<a>}-C_{<c><b>}^{<a>},
$$
$$
T_{.b_fc_f}^{a_p}=\frac{\delta N_{c_f}^{a_p}}{\partial
y^{b_f}}-\frac{\delta
N_{b_f}^{a_p}}{\partial y^{c_f}},T_{.<b>i}^{<a>}=P_{.<b>i}^{<a>}=\frac{%
\delta N_i^{<a>}}{\partial y^{<b>}}%
-L_{.<b>j}^{<a>},T_{.i<b>}^{<a>}=-P_{.<b>i}^{<a>}.
$$

The curvature ${\bf R}$ of d--connection in ${\cal E}^{<z>}$ is
defined by \index{Curvature!of d--connection} the equation
$$
{\bf R\left( X,Y\right) Z=XY_{\bullet }^{\bullet }R\bullet Z}=D_XD_Y{\bf Z}%
-D_YD_X{\bf Z-}D_{[X,Y]}{\bf Z.}\eqno(6.26)
$$
One holds the next properties for the h- and v-decompositions of curvature:%
$$
{\bf v}_{(p)}{\bf R\left( X,Y\right) hZ=0,\ hR\left( X,Y\right) v}_{(p)}{\bf %
Z=0,~v}_{(f)}{\bf R\left( X,Y\right) v}_{(p)}{\bf Z=0,}
$$
$$
{\bf R\left( X,Y\right) Z=hR\left( X,Y\right) hZ+vR\left(
X,Y\right) vZ,}
$$
where ${\bf v=v}_1+...+{\bf v}_z.$ From (6.26) and the equation ${\bf %
R\left( X,Y\right) =-R\left( Y,X\right) }$ we get that the
curvature of a d-con\-nec\-ti\-on ${\bf D}$ in ${\cal E}^{<z>}$
is completely determined by
the following d-tensor fields:%
$$
R_{h.jk}^{.i}=\delta ^i\cdot {\bf R}\left( \delta _k,\delta
_j\right) \delta _h,~R_{<b>.jk}^{.<a>}=\delta ^{<a>}\cdot {\bf
R}\left( \delta _k,\delta _j\right) \delta _{<b>},\eqno(6.27)
$$
$$
P_{j.k<c>}^{.i}=d^i\cdot {\bf R}\left( \delta _{<c>},\delta
_{<k>}\right) \delta _j,~P_{<b>.<k><c>}^{.<a>}=\delta ^{<a>}\cdot
{\bf R}\left( \delta _{<c>},\delta _{<k>}\right) \delta _{<b>},
$$
$$
S_{j.<b><c>}^{.i}=d^i\cdot {\bf R}\left( \delta _{<c>},\delta
_{<b>}\right) \delta _j,~S_{<b>.<c><d>}^{.<a>}=\delta ^{<a>}\cdot
{\bf R}\left( \delta _{<d>},\delta _{<c>}\right) \delta _{<b>}.
$$
By a direct computation, using (6.5),(6.6),(6.16),(6.17) and
(6.27) we get :
$$
R_{h.jk}^{.i}=\frac{\delta L_{.hj}^i}{\delta x^h}-\frac{\delta L_{.hk}^i}{%
\delta x^j}+L_{.hj}^mL_{mk}^i-L_{.hk}^mL_{mj}^i+C_{.h<a>}^iR_{.jk}^{<a>},%
\eqno(6.28)
$$
$$
R_{<b>.jk}^{.<a>}=\frac{\delta L_{.<b>j}^{<a>}}{\delta
x^k}-\frac{\delta
L_{.<b>k}^{<a>}}{\delta x^j}%
+L_{.<b>j}^{<c>}L_{.<c>k}^{<a>}-L_{.<b>k}^{<c>}L_{.<c>j}^{<a>}+C_{.<b><c>}^{<a>}R_{.jk}^{<c>},
$$
$$
P_{j.k<a>}^{.i}=\frac{\delta L_{.jk}^i}{\partial y^{<a>}}%
+C_{.j<b>}^iP_{.k<a>}^{<b>}-
$$
$$
\left( \frac{\partial C_{.j<a>}^i}{\partial x^k}%
+L_{.lk}^iC_{.j<a>}^l-L_{.jk}^lC_{.l<a>}^i-L_{.<a>k}^{<c>}C_{.j<c>}^i\right)
,
$$
$$
P_{<b>.k<a>}^{.<c>}=\frac{\delta L_{.<b>k}^{<c>}}{\partial y^{<a>}}%
+C_{.<b><d>}^{<c>}P_{.k<a>}^{<d>}-
$$
$$
\left( \frac{\partial C_{.<b><a>}^{<c>}}{\partial x^k}+L_{.<d>k}^{<c>%
\,}C_{.<b><a>}^{<d>}-L_{.<b>k}^{<d>}C_{.<d><a>}^{<c>}-L_{.<a>k}^{<d>}C_{.<b><d>}^{<c>}\right)
,
$$
$$
S_{j.<b><c>}^{.i}=\frac{\delta C_{.j<b>}^i}{\partial
y^{<c>}}-\frac{\delta C_{.j<c>}^i}{\partial
y^{<b>}}+C_{.j<b>}^hC_{.h<c>}^i-C_{.j<c>}^hC_{h<b>}^i,
$$
$$
S_{<b>.<c><d>}^{.<a>}=\frac{\delta C_{.<b><c>}^{<a>}}{\partial y^{<d>}}-%
\frac{\delta C_{.<b><d>}^{<a>}}{\partial y^{<c>}} $$ $$
+C_{.<b><c>}^{<e>}C_{.<e><d>}^{<a>}-C_{.<b><d>}^{<e>}C_{.<e><c>}^{<a>}.
$$

We note that torsions (6.25) and curvatures (6.28) can be
computed by particular cases of d-connections when d-connections
(6.17), (6.20) or (6.22) are used instead of (6.16) and (6.17).

The components of the Ricci d--tensor \index{Ricci!d--tensor}
$$
R_{<\alpha ><\beta >}=R_{<\alpha >.<\beta ><\tau >}^{.<\tau >}
$$
with respect to locally adapted frame (6.6) are as follows:%
$$
R_{ij}=R_{i.jk}^{.k},\quad
R_{i<a>}=-^2P_{i<a>}=-P_{i.k<a>}^{.k},\eqno(6.29)
$$
$$
R_{<a>i}=^1P_{<a>i}=P_{<a>.i<b>}^{.<b>},\quad
R_{<a><b>}=S_{<a>.<b><c>}^{.<c>}.
$$
We point out that because, in general, $^1P_{<a>i}\neq
~^2P_{i<a>}$ the Ricci d--tensor is non symmetric.

Having defined a d--metric of type (6.12) in ${\cal E}^{<z>}$ we
can introduce the scalar curvature of d--connection ${\bf D}$:
$$
{\overleftarrow{R}}=G^{<\alpha ><\beta >}R_{<\alpha ><\beta >}=R+S,%
$$
where $R=g^{ij}R_{ij}$ and $S=h^{<a><b>}S_{<a><b>}.$

For our further considerations it will be also useful to use an
alternative way of definition torsion (6.23) and curvature (6.26)
by using the commutator
$$
\Delta _{<\alpha ><\beta >}\doteq \nabla _{<\alpha >}\nabla
_{<\beta
>}-\nabla _{<\beta >}\nabla _{<\alpha >}=2\nabla _{[<\alpha >}\nabla
_{<\beta >]}.$$ For components (6.25) of d--torsion we have
$$
\Delta _{<\alpha ><\beta >}f=T_{.<\alpha ><\beta >}^{<\gamma
>}\nabla _{<\gamma >}f\eqno(6.30)
$$
for every scalar function $f\,\,$ on ${\cal E}^{<z>}{\cal .}$
Curvature can be introduced as an operator acting on arbitrary
d--vector $V^{<\delta >}:$

$$
(\Delta _{<\alpha ><\beta >}-T_{.<\alpha ><\beta >}^{<\gamma
>}\nabla _{<\gamma >})V^{<\delta >}=R_{~<\gamma >.<\alpha ><\beta
>}^{.<\delta
>}V^{<\gamma >}\eqno(6.31)
$$
(we note that in this Chapter we shall follow conventions of
Miron and
Anastasiei 
 [160,161] on d--tensors; we can obtain corresponding
Penrose and Rindler abstract index formulas
 [181,182] just for a
trivial N-connection structure and by changing denotations for
components of torsion and curvature in this manner:\ $T_{.\alpha
\beta }^\gamma \rightarrow T_{\alpha \beta }^{\quad \gamma }$ and
$R_{~\gamma .\alpha \beta }^{.\delta }\rightarrow R_{\alpha \beta
\gamma }^{\qquad \delta }).$

Here we also note that torsion and curvature of a d--connection on ${\cal E}%
^{<z>}$ satisfy generalized for ha--spaces Ricci and Bianchi
identities which
in terms of components (6.30) and (6.31) are written respectively as%
$$
R_{~[<\gamma >.<\alpha ><\beta >]}^{.<\delta >}+\nabla _{[<\alpha
>}T_{.<\beta ><\gamma >]}^{<\delta >}+T_{.[<\alpha ><\beta >}^{<\nu
>}T_{.<\gamma >]<\nu >}^{<\delta >}=0\eqno(6.32)
$$
and%
\index{Ricci identities} \index{Bianchi identities}
$$
\nabla _{[<\alpha >}R_{|<\nu >|<\beta ><\gamma >]}^{\cdot <\sigma
>}+T_{\cdot [<\alpha ><\beta >}^{<\delta >}R_{|<\nu >|.<\gamma >]<\delta
>}^{\cdot <\sigma >}=0.
$$
Identities (6.32) can be proved similarly as in 
 [181] by taking into
account that indices play a distinguished character.

We can also consider a ha-generalization of the so-called
conformal Weyl \index{Conformal Weyl tensor}
tensor (see, for instance, 
 [181]) which can be written as a d-tensor
in this form:%
$$
C_{\quad <\alpha ><\beta >}^{<\gamma ><\delta >}=R_{\quad <\alpha
><\beta
>}^{<\gamma ><\delta >}-\frac 4{n+m_1+...+m_z-2}R_{\quad [<\alpha
>}^{[<\gamma >}~\delta _{\quad <\beta >]}^{<\delta >]}+\eqno(6.33)
$$
$$
\frac
2{(n+m_1+...m_z-1)(n+m_1+...+m_z-2)}{\overleftarrow{R}~\delta
_{\quad [<\alpha >}^{[<\gamma >}~\delta _{\quad <\beta
>]}^{<\delta >]}.}
$$
This object is conformally invariant on ha--spaces provided with
   d--con\-nec\-ti\-on generated by d--metric structures.

\subsection{ Field equations for ha--gravity}

The Einstein equations in some models of higher order anisotropic
supergravity have been considered in 
 [267]. Here we note that the
Einstein equations and conservation laws on v--bundles provided
with N-connection structures were studied in detail in
 [160,161,9,10,279,276,263].
 In ref. 
 [272] we proved that the
la-gravity can be formulated in a gauge like manner and analyzed
the conditions when the Einstein la-gravitational field equations
are equivalent to a corresponding form of Yang-Mills equations.
In this subsection we shall write the higher order anisotropic
gravitational field equations in a form more convenient for
theirs equivalent reformulation in ha-spinor variables.

We define d-tensor $\Phi _{<\alpha ><\beta >}$ as to satisfy
conditions
$$
-2\Phi _{<\alpha ><\beta >}\doteq R_{<\alpha ><\beta >}-\frac
1{n+m_1+...+m_z}\overleftarrow{R}g_{<\alpha ><\beta >}
$$
which is the torsionless part of the Ricci tensor for locally
isotropic
spaces 
 [181,182], i.e. $\Phi _{<\alpha >}^{~~<\alpha >}\doteq 0$.\
The Einstein equations on ha--spaces \index{Einstein equations!on
ha--spaces}
$$
\overleftarrow{G}_{<\alpha ><\beta >}+\lambda g_{<\alpha ><\beta
>}=\kappa E_{<\alpha ><\beta >},\eqno(6.34)
$$
where%
$$
\overleftarrow{G}_{<\alpha ><\beta >}=R_{<\alpha ><\beta >}-\frac 12%
\overleftarrow{R}g_{<\alpha ><\beta >}
$$
is the Einstein d--tensor, $\lambda $ and $\kappa $ are
correspondingly the \index{Einstein d--tensor} cosmological and
gravitational constants and by $E_{<\alpha ><\beta >}$ is denoted
the locally anisotropic energy--momentum d--tensor, can be
rewritten
in equivalent form:%
$$
\Phi _{<\alpha ><\beta >}=-\frac \kappa 2(E_{<\alpha ><\beta
>}-\frac 1{n+m_1+...+m_z}E_{<\tau >}^{~<\tau >}~g_{<\alpha
><\beta >}).\eqno(6.35)
$$

Because ha--spaces generally have nonzero torsions we shall add
to (6.35) (equivalently to (6.34)) a system of algebraic d--field
equations with the source $S_{~<\beta ><\gamma >}^{<\alpha >}$
being the locally anisotropic spin density of matter (if we
consider a variant of higher order anisotropic Einstein--Cartan
theory ): \index{Einstein--Cartan theory}
$$
T_{~<\alpha ><\beta >}^{<\gamma >}+2\delta _{~[<\alpha >}^{<\gamma
>}T_{~<\beta >]<\delta >}^{<\delta >}=\kappa S_{~<\alpha ><\beta
>.}^{<\gamma >}\eqno(6.36)
$$
From (6.32 ) and (6.36) one follows the conservation law of
higher order
anisot\-rop\-ic spin matter:%
$$
\nabla _{<\gamma >}S_{~<\alpha ><\beta >}^{<\gamma >}-T_{~<\delta
><\gamma
>}^{<\delta >}S_{~<\alpha ><\beta >}^{<\gamma >}=E_{<\beta ><\alpha
>}-E_{<\alpha ><\beta >}.
$$

Finally, in this section, we remark that all presented geometric
constructions contain those elaborated for generalized Lagrange
spaces
 [160,161] (for which a tangent bundle $TM$ is considered instead of a
v-bundle ${\cal E}^{<z>}$ ) and for constructions on the so
called osculator bundles with different prolongations and
extensions of Finsler and Lagrange
metrics 
 [162]. We also note that the Lagrange (Finsler) geometry is
characterized by a metric of type (6.12) with components parametized as $%
g_{ij}=\frac 12\frac{\partial ^2{\cal L}}{\partial y^i\partial
y^j}$ $\left( g_{ij}=\frac 12\frac{\partial ^2\Lambda
^2}{\partial y^i\partial y^j}\right) $ and $h_{ij}=g_{ij},$ where
${\cal L=L}$ $(x,y)$ $\left( \Lambda =\Lambda
\left( x,y\right) \right) $ is a Lagrangian $\left( \mbox{Finsler metric}%
\right) $ on $TM$ (see details in
 [160,161,159,29].

\section{Distinguished Clifford Algebras}

The typical fiber of dv-bundle $\xi _d\ ,\ \pi _d:\ HE\oplus
V_1E\oplus ...\oplus V_zE\rightarrow E$ is a d-vector space,
${\cal F}=h{\cal F}\oplus v_1{\cal F\oplus ...}\oplus v_z{\cal
F},$ split into horizontal $h{\cal F}$ and verticals $v_p{\cal
F,}p=1,...,z$ subspaces, with metric $G(g,h)$ induced by v-bundle
metric (6.12). Clifford algebras (see, for example,
\index{Clifford algebras!distinguished}
Refs. 
 [133,245,182]) formulated for d-vector spaces will be called
Clifford d-algebras 
 [256,255,275]. In this section we shall consider
the main properties of Clifford d--algebras. The proof of
theorems will be \index{Clifford d--algebras} based on the
technique developed in ref.
 [133] correspondingly adapted
to the distinguished character of spaces in consideration.

Let $k$ be a number field (for our purposes $k={\cal R}$ or $k={\cal C},%
{\cal R}$ and ${\cal C},$ are, respectively real and complex
number fields) and define ${\cal F},$ as a d-vector space on $k$
provided with nondegenerate symmetric quadratic form (metric)\
$G.$ Let $C$ be an algebra on $k$ (not necessarily commutative)
and $j\ :\ {\cal F}$ $\rightarrow C$ a homomorphism of underlying
vector spaces such that $j(u)^2=\;G(u)\cdot 1\ (1$ is the unity
in algebra $C$ and d-vector $u\in {\cal F}).$ We are interested
in definition of the pair $\left( C,j\right) $ satisfying the next
universitality conditions. For every $k$-algebra $A$ and arbitrary
homomorphism $\varphi :{\cal F}\rightarrow A$ of the underlying
d-vector spaces, such that $\left( \varphi (u)\right)
^2\rightarrow G\left( u\right) \cdot 1,$ there is a unique
homomorphism of algebras $\psi \ :\ C\rightarrow A$ transforming
the diagram 1
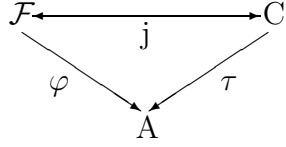
\begin{figure}[htbp]
\begin{center}
\begin{picture}(100,50) \setlength{\unitlength}{1pt}
\thinlines \put(0,45){${\cal F}$} \put(96,45){C} \put(48,2){A}
\put(50,38){j} \put(50,48){ \vector(-1,0){45}}
\put(50,48){\vector(1,0){45}} \put(5,42){\vector(3,-2){45}}
\put(98,42){\vector(-3,-2){45}} \put(15,20){$\varphi$}
\put(80,20){$\tau$}
\end{picture}
\end{center}
\caption{Diagram 1}
\end{figure}
into a commutative one.\ The algebra solving this problem will be
denoted as
$C\left( {\cal F},A\right) $ [equivalently as $C\left( G\right) $ or $%
C\left( {\cal F}\right) ]$ and called as Clifford d--algebra
associated with pair $\left( {\cal F},G\right) .$

\begin{theorem}
The above-presented diagram has a unique solution $\left(
C,j\right) $ up to isomorphism.
\end{theorem}

{\bf Proof:} (We adapt for d-algebras that of ref.
 [133], p. 127.) For
a universal problem the uniqueness is obvious if we prove the
existence of
solution $C\left( G\right) $ . To do this we use tensor algebra ${\cal L}%
^{(F)}=\oplus {\cal L}_{qs}^{pr}\left( {\cal F}\right) $ =$\oplus
_{i=0}^\infty T^i\left( {\cal F}\right) ,$ where $T^0\left( {\cal
F}\right)
=k$ and $T^i\left( {\cal F}\right) =k$ and $T^i\left( {\cal F}\right) ={\cal %
F}\otimes ...\otimes {\cal F}$ for $i>0.$ Let $I\left( G\right) $
be the bilateral ideal generated by elements of form $\epsilon
\left( u\right) =u\otimes u-G\left( u\right) \cdot 1$ where $u\in
{\cal F}$ and $1$ is the unity element of algebra ${\cal L}\left(
{\cal F}\right) .$ Every element from $I\left( G\right) $ can be
written as $\sum\nolimits_i\lambda
_i\epsilon \left( u_i\right) \mu _i,$ where $\lambda _{i},\mu _i\in {\cal L}(%
{\cal F})$ and $u_i\in {\cal F}.$ Let $C\left( G\right) $ =${\cal L}({\cal F}%
)/I\left( G\right) $ and define $j:{\cal F}\rightarrow C\left(
G\right) $ as
the composition of monomorphism $i:{{\cal F}\rightarrow L}^1 ({\cal F}%
)\subset {\cal L}({\cal F})$ and projection $p:{\cal L}\left( {\cal F}%
\right) \rightarrow C\left( G\right) .$ In this case pair $\left(
C\left( G\right) ,j\right) $ is the solution of our problem. From
the general properties of tensor algebras the homomorphism
$\varphi :{\cal F}\rightarrow A$ can be extended to ${\cal
L}({\cal F})$ , i.e., the diagram 2
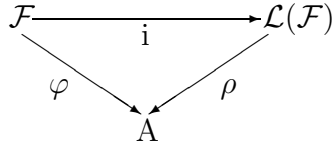
\begin{figure}[htbp]
\begin{center}
\begin{picture}(100,50) \setlength{\unitlength}{1pt}
\thinlines \put(0,45){${\cal F}$} \put(96,45){${\cal L}({\cal
F})$} \put(48,2){A} \put(50,38){i} \put(50,48){ \line(-1,0){45}}
\put(50,48){\vector(1,0){45}} \put(5,42){\vector(3,-2){45}}
\put(98,42){\vector(-3,-2){45}} \put(15,20){$\varphi$}
\put(80,20){$\rho$}
\end{picture}
\end{center}
\caption{Diagram 2}
\end{figure}
is commutative, where $\rho $ is a monomorphism of algebras.
Because $\left( \varphi \left( u\right) \right) ^2=G\left(
u\right) \cdot 1,$ then $\rho $ vanishes on ideal $I\left(
G\right) $ and in this case the necessary homomorphism $\tau $ is
defined. As a consequence of uniqueness of $\rho ,$ the
homomorphism $\tau $ is unique.

Tensor d-algebra ${\cal L}({\cal F )}$ can be considered as a
${\cal Z}/2$ graded algebra. Really, let us in\-tro\-duce ${\cal
L}^{(0)}({\cal F}) =
\sum_{i=1}^\infty T^{2i}\left( {\cal F}\right) $ and ${\cal L}^{(1)}({\cal F}%
) =\sum_{i=1}^\infty T^{2i+1}\left( {\cal F}\right) .$ Setting
$I^{(\alpha )}\left( G\right) =I\left( G\right) \cap {\cal
L}^{(\alpha )}({\cal F}).$
Define $C^{(\alpha )}\left( G\right) $ as $p\left( {\cal L}^{(\alpha )} (%
{\cal F})\right) ,$ where $p:{\cal L}\left( {\cal F}\right)
\rightarrow C\left( G\right) $ is the canonical projection. Then
$C\left( G\right) =C^{(0)}\left( G\right) \oplus C^{(1)}\left(
G\right) $ and in consequence we obtain that the Clifford
d-algebra is ${\cal Z}/2$ graded.

It is obvious that Clifford d-algebra functorially depends on
pair $\left( {\cal F},G\right) .$ If $f:{\cal F}\rightarrow{\cal
F}^{\prime }$ is a homomorphism of k-vector spaces, such that
$G^{\prime }\left( f(u)\right) =G\left( u\right) ,$ where $G$ and
$G^{\prime }$ are, respectively, metrics on ${\cal F}$ and ${\cal
F}^{\prime },$ then $f$ induces an homomorphism of
d-algebras%
$$
C\left( f\right) :C\left( G\right) \rightarrow C\left( G^{\prime
}\right)
$$
with identities $C\left( \varphi \cdot f\right) =C\left( \varphi
\right) C\left( f\right) $ and $C\left( Id_{{\cal F}}\right)
=Id_{C({\cal F)}}.$

If ${\cal A}^{\alpha}$ and ${\cal B}^{\beta}$ are ${\cal
Z}/2$--graded
 d--algebras,
then their graded tensorial product $%
{\cal A}^\alpha \otimes {\cal B}^\beta $ is defined as a
d-algebra for k-vector d-space ${\cal A}^\alpha \otimes {\cal
B}^\beta $ with the graded product induced as $\left( a\otimes
b\right) \left( c\otimes d\right) =\left( -1\right) ^{\alpha
\beta }ac\otimes bd,$ where $b\in {\cal B}^\alpha $ and $c\in
{\cal A}^\alpha \quad \left( \alpha ,\beta =0,1\right) .$

Now we reformulate for d--algebras the Chevalley theorem 
 [60]:

\begin{theorem}
\label{6.2t} The Clifford d-algebra
$$
C\left( h{\cal F}\oplus v_1{\cal F\oplus ...}\oplus v_z{\cal F}%
,g+h_1+...+h_z\right)
$$
is naturally isomorphic to $C(g)\otimes C\left( h_1\right) \otimes
...\otimes C\left( h_z\right) .$
\end{theorem}

{\bf Proof. }Let $n:h{\cal F}\rightarrow C\left( g\right) $ and $%
n_{(p)}^{\prime }:v_{(p)}{\cal F}\rightarrow C\left(
h_{(p)}\right) $ be canonical maps and map
$$
m:h{\cal F}\oplus v_1{\cal F...}\oplus v_z{\cal F}\rightarrow
C(g)\otimes C\left( h_1\right) \otimes ...\otimes C\left(
h_z\right)
$$
is defined as%
$$
m(x,y_{(1)},...,y_{(z)})=
$$
$$
n(x)\otimes 1\otimes ...\otimes 1+1\otimes n^{\prime
}(y_{(1)})\otimes ...\otimes 1+1\otimes ...\otimes 1\otimes
n^{\prime }(y_{(z)}),
$$
$x\in h{\cal F},y_{(1)}\in v_{(1)}{\cal F,...},y_{(z)}\in
v_{(z)}{\cal F.}$ We have
$$
\left( m(x,y_{(1)},...,y_{(z)})\right) ^2=\left[ \left( n\left(
x\right) \right) ^2+\left( n^{\prime }\left( y_{(1)}\right)
\right) ^2+...+\left( n^{\prime }\left( y_{(z)}\right) \right)
^2\right] \cdot 1=
$$
$$
[g\left( x\right) +h\left( y_{(1)}\right) +...+h\left(
y_{(z)}\right) ].
$$
\ Taking into account the universality property of Clifford
d-algebras we
conclude that $m_1+...+m_z$ induces the homomorphism%
$$
\zeta :C\left( h{\cal F}\oplus v_1{\cal F}\oplus ...\oplus v_z{\cal F}%
,g+h_1+...+h_z\right) \rightarrow
$$
$$
C\left( h{\cal F},g\right) \widehat{\otimes }C\left( v_1{\cal
F},h_1\right) \widehat{\otimes }...C\left( v_z{\cal F},h_z\right)
.
$$
We also can define a homomorphism%
$$
\upsilon :C\left( h{\cal F},g\right) \widehat{\otimes }C\left( v_1{\cal F}%
,h_{(1)}\right) \widehat{\otimes }...\widehat{\otimes }C\left( v_z{\cal F}%
,h_{(z)}\right) \rightarrow
$$
$$
C\left( h{\cal F}\oplus v_1{\cal F\oplus ...}\oplus v_z{\cal F}%
,g+h_{(1)}+...+h_{(z)}\right)
$$
by using formula $$\upsilon \left( x\otimes y_{(1)}\otimes
...\otimes y_{(z)}\right) =\delta \left( x\right) \delta
_{(1)}^{\prime }\left( y_{(1)}\right) ...\delta _{(z)}^{\prime
}\left( y_{(z)}\right) ,$$ where homomorphysms $\delta $ and
$\delta _{(1)}^{\prime },...,\delta _{(z)}^{\prime }$ are,
respectively, induced by im\-bed\-dings of $h{\cal F}$
and $v_1{\cal F}$ into $$h{\cal F}\oplus v_1{\cal F\oplus ...}\oplus v_z{\cal %
F}:$$
$$
\delta :C\left( h{\cal F},g\right) \rightarrow C\left( h{\cal F}\oplus v_1%
{\cal F\oplus ...}\oplus v_z{\cal F},g+h_{(1)}+...+h_{(z)}\right)
,
$$
$$
\delta _{(1)}^{\prime }:C\left( v_1{\cal F},h_{(1)}\right)
\rightarrow
C\left( h{\cal F}\oplus v_1{\cal F\oplus ...}\oplus v_z{\cal F}%
,g+h_{(1)}+...+h_{(z)}\right) ,
$$
$$
...................................
$$
$$
\delta _{(z)}^{\prime }:C\left( v_z{\cal F},h_{(z)}\right)
\rightarrow
C\left( h{\cal F}\oplus v_1{\cal F\oplus ...}\oplus v_z{\cal F}%
,g+h_{(1)}+...+h_{(z)}\right) .
$$

Superpositions of homomorphisms $\zeta $ and $\upsilon $ lead to identities%
$$
\upsilon \zeta =Id_{C\left( h{\cal F},g\right) \widehat{\otimes }C\left( v_1%
{\cal F},h_{(1)}\right) \widehat{\otimes }...\widehat{\otimes }C\left( v_z%
{\cal F},h_{(z)}\right) },\eqno(6.37)
$$
$$
\zeta \upsilon =Id_{C\left( h{\cal F},g\right) \widehat{\otimes }C\left( v_1%
{\cal F},h_{(1)}\right) \widehat{\otimes }...\widehat{\otimes }C\left( v_z%
{\cal F},h_{(z)}\right) }.
$$
Really, d--algebra $$C\left( h{\cal F}\oplus v_1{\cal F\oplus ...}\oplus v_z%
{\cal F},g+h_{(1)}+...+h_{(z)}\right) $$ is generated by elements of type $%
m(x,y_{(1)},...y_{(z)}).$ Calculating
$$
\upsilon \zeta \left( m\left( x,y_{(1)},...y_{(z)}\right) \right)
=\upsilon (n\left( x\right) \otimes 1\otimes ...\otimes
1+1\otimes n_{(1)}^{\prime }\left( y_{(1)}\right) \otimes
...\otimes 1+...+
$$
$$
1\otimes ....\otimes n_{(z)}^{\prime }\left( y_{(z)}\right)
)=\delta \left( n\left( x\right) \right) \delta \left(
n_{(1)}^{\prime }\left( y_{(1)}\right) \right) ...\delta \left(
n_{(z)}^{\prime }\left( y_{(z)}\right) \right) =
$$
$$
m\left( x,0,...,0\right)
+m(0,y_{(1)},...,0)+...+m(0,0,...,y_{(z)})=m\left(
x,y_{(1)},...,y_{(z)}\right) ,
$$
we prove the first identity in (6.37).

On the other hand, d-algebra
$$
C\left( h{\cal F},g\right) \widehat{\otimes }C\left( v_1{\cal F}%
,h_{(1)}\right) \widehat{\otimes }...\widehat{\otimes }C\left( v_z{\cal F}%
,h_{(z)}\right)
$$
is generated by elements of type
$$
n\left( x\right) \otimes 1\otimes ...\otimes ,1\otimes
n_{(1)}^{\prime }\left( y_{(1)}\right) \otimes ...\otimes
1,...1\otimes ....\otimes n_{(z)}^{\prime }\left( y_{(z)}\right) ,
$$
we prove the second identity in (6.37).

Following from the above--mentioned properties of homomorphisms
$\zeta $ and $\upsilon $ we can assert that the natural
isomorphism is explicitly constructed.$\Box $

In consequence of theorem 6.2 we conclude that all operations
with Clifford d--algebras can be reduced to calculations for
$$C\left( h{\cal F},g\right) \mbox{ and } C\left( v_{(p)}{\cal
F},h_{(p)}\right) $$
 which are usual Clifford
algebras of dimension $2^n$ and, respectively, $2^{m_p}$
 [133,22].

Of special interest is the case when $k={\cal R}$ and ${\cal F}$
is isomorphic to vector space ${\cal R}^{p+q,a+b}$ provided with
quadratic form
$-x_1^2-...-x_p^2+x_{p+q}^2-y_1^2-...-y_a^2+...+y_{a+b}^2.$ In
this case, the Clifford algebra, denoted as $\left(
C^{p,q},C^{a,b}\right) ,\,$ is
generated by symbols $%
e_1^{(x)},e_2^{(x)},...,e_{p+q}^{(x)},e_1^{(y)},e_2^{(y)},...,e_{a+b}^{(y)}$
satisfying properties $\left( e_i\right) ^2=-1~\left( 1\leq i\leq
p\right) ,\left( e_j\right) ^2=-1~\left( 1\leq j\leq a\right)
,\left( e_k\right) ^2=1~(p+1\leq k\leq p+q),$

$\left( e_j\right) ^2=1~(n+1\leq s\leq
a+b),~e_ie_j=-e_je_i,~i\neq j.\,$ Explicit calculations of
$C^{p,q}$ and $C^{a,b}$ are possible by using
isomorphisms 
 [133,182]
$$
C^{p+n,q+n}\simeq C^{p,q}\otimes M_2\left( {\cal R}\right) \otimes
...\otimes M_2\left( {\cal R}\right) \cong C^{p,q}\otimes
M_{2^n}\left( {\cal R}\right) \cong M_{2^n}\left( C^{p,q}\right) ,
$$
where $M_s\left( A\right) $ denotes the ring of quadratic matrices of order $%
s$ with coefficients in ring $A.$ Here we write the simplest isomorphisms $%
C^{1,0}\simeq {\cal C}, C^{0,1}\simeq {\cal R}\oplus {\cal R ,}$ and $%
C^{2,0}={\cal H},$ where by ${\cal H}$ is denoted the body of
quaternions.
We summarize this calculus as (as in ref. 
 [22])%
$$
C^{0,0}={\cal R}, C^{1,0}={\cal C}, C^{0,1}={\cal R}\oplus {\cal R}, C^{2,0}=%
{\cal H}, C^{0,2}= M_2\left( {\cal R}\right) ,
$$
$$
C^{3,0}={\cal H}\oplus {\cal H} , C^{0,3} = M_2\left( {\cal
R}\right), C^{4,0}=M_2\left( {\cal H}\right) , C^{0,4}=M_2\left(
{\cal H}\right) ,
$$
$$
C^{5,0}=M_4\left( {\cal C}\right) ,~C^{0,5}=M_2\left( {\cal
H}\right) \oplus M_2\left( {\cal H}\right) ,~C^{6,0}=M_8\left(
{\cal R}\right) ,~C^{0,6}=M_4\left( {\cal H}\right) ,
$$
$$
C^{7,0}=M_8\left( {\cal R}\right) \oplus M_8\left( {\cal R}\right)
,~C^{0,7}=M_8\left( {\cal C}\right) ,~C^{8,0}=M_{16}\left( {\cal
R}\right) ,~C^{0,8}=M_{16}\left( {\cal R}\right) .
$$
One of the most important properties of real algebras
$C^{0,p}~\left( C^{0,a}\right) $ and $C^{p,0}~\left(
C^{a,0}\right) $ is eightfold periodicity of $p(a).$

Now, we emphasize that $H^{2n}$-spaces  admit locally a structure
of Clifford algebra on complex vector spaces. Really, by using
almost \ Hermitian structure $J_\alpha ^{\quad \beta }$ and
considering
complex space ${\cal C}^n$ with nondegenarate quadratic form $%
\sum_{a=1}^n\left| z_a\right| ^2,~z_a\in {\cal C}^2$ induced
locally by metric (2.12) (rewritten in complex coordinates
$z_a=x_a+iy_a)$ we define Clifford algebra
$\overleftarrow{C}^n=\underbrace{\overleftarrow{C}^1\otimes
...\otimes \overleftarrow{C}^1}_n,$ where $\overleftarrow{C}^1={\cal %
C\otimes }_R{\cal C=C\oplus C}$ or in consequence, $\overleftarrow{C}%
^n\simeq C^{n,0}\otimes _{{\cal R}}{\cal C}\approx C^{0,n}\otimes _{{\cal R}}%
{\cal C}.$ Explicit calculations lead to isomorphisms $\overleftarrow{C}%
^2=C^{0,2}\otimes _{{\cal R}}{\cal C}\approx M_2\left( {\cal
R}\right) \otimes _{{\cal R}}{\cal C}\approx M_2\left(
\overleftarrow{C}^n\right)
,~C^{2p}\approx M_{2^p}\left( {\cal C}\right) $ and $\overleftarrow{C}%
^{2p+1}\approx M_{2^p}\left( {\cal C}\right) \oplus M_{2^p}\left( {\cal C}%
\right) ,$ which show that complex Clifford algebras, defined locally for $%
H^{2n}$-spaces, have periodicity 2 on $p.$

Considerations presented in the proof of theorem 6.2 show that map $j:{\cal F%
}\rightarrow C\left( {\cal F}\right) $ is monomorphic, so we can
identify
space ${\cal F}$ with its image in $C\left( {\cal F},G\right) ,$ denoted as $%
u\rightarrow \overline{u},$ if $u\in C^{(0)}\left( {\cal
F},G\right) ~\left( u\in C^{(1)}\left( {\cal F},G\right) \right)
;$ then $u=\overline{u}$ ( respectively, $\overline{u}=-u).$

\begin{definition}
The set of elements $u\in C\left( G\right) ^{*},$ where $C\left(
G\right) ^{*}$ denotes the multiplicative group of invertible
elements of $C\left( {\cal F},G\right) $ satisfying
$\overline{u}{\cal F}u^{-1}\in {\cal F},$ is called the twisted
Clifford d-group, denoted as $\widetilde{\Gamma }\left( {\cal
F}\right) .$
\end{definition}

Let $\widetilde{\rho }:\widetilde{\Gamma }\left( {\cal F}\right)
\rightarrow GL\left( {\cal F}\right) $ be the homorphism given by
$u\rightarrow \rho
\widetilde{u},$ where $\widetilde{\rho }_u\left( w\right) =\overline{u}%
wu^{-1}.$ We can verify that $\ker \widetilde{\rho }={\cal
R}^{*}$is a subgroup in $\widetilde{\Gamma }\left( {\cal
F}\right) .$

Canonical map $j:{\cal F}\rightarrow C\left( {\cal F}\right) $
can be interpreted as the linear map ${\cal F}\rightarrow C\left(
{\cal F}\right) ^0 $ satisfying the universal property of
Clifford d-algebras. This leads to a homomorphism of algebras,
$C\left( {\cal F}\right) \rightarrow C\left(
{\cal F}\right) ^t,$ considered by an anti-involution of $C\left( {\cal F}%
\right) $ and denoted as $u\rightarrow ~^tu.$ More exactly, if
$u_1...u_n\in {\cal F,}$ then $t_u=u_n...u_1$ and
$^t\overline{u}=\overline{^tu}=\left( -1\right) ^nu_n...u_1.$

\begin{definition}
The spinor norm of arbitrary $u\in C\left( {\cal F}\right) $ is defined as\\
$S\left( u\right) =~^t\overline{u}\cdot u\in C\left( {\cal
F}\right) .$
\end{definition}

It is obvious that if $u,u^{\prime },u^{\prime \prime }\in
\widetilde{\Gamma }\left( {\cal F}\right) ,$ then $S(u,u^{\prime
})=S\left( u\right) S\left( u^{\prime }\right) $ and \\ $S\left(
uu^{\prime }u^{\prime \prime }\right) =S\left( u\right) S\left(
u^{\prime }\right) S\left( u^{\prime \prime }\right) .$ For
$u,u^{\prime }\in {\cal F} S\left( u\right) =-G\left( u\right) $
and $S\left( u,u^{\prime }\right) =S\left( u\right) S\left(
u^{\prime }\right) =S\left( uu^{\prime }\right) .$

Let us introduce the orthogonal group $O\left( G\right) \subset
GL\left( G\right) $ defined by metric $G$ on ${\cal F}$ and
denote sets $SO\left( G\right) =\{u\in O\left( G\right) ,\det
\left| u\right| =1\},~Pin\left( G\right) =\{u\in
\widetilde{\Gamma }\left( {\cal F}\right) ,S\left( u\right)
=1\}$ and $Spin\left( G\right) =Pin\left( G\right) \cap C^0\left( {\cal F}%
\right) .$ For ${{\cal F}\cong {\cal R}}^{n+m}$ we write
$Spin\left( n_E\right) .$ By straightforward calculations (see
similar considerations in ref.
 [133]) we can verify the exactness of these sequences:%
$$
1\rightarrow {\cal Z}/2\rightarrow Pin\left( G\right) \rightarrow
O\left( G\right) \rightarrow 1,
$$
$$
1\rightarrow {\cal Z}/2\rightarrow Spin\left( G\right)
\rightarrow SO\left( G\right) \rightarrow 0,
$$
$$
1\rightarrow {\cal Z}/2\rightarrow Spin\left( n_E\right)
\rightarrow SO\left( n_E\right) \rightarrow 1.
$$
We conclude this section by emphasizing that the spinor norm was
defined
with respect to a quadratic form induced by a metric in dv-bundle ${\cal E}%
^{<z>}$. This approach differs from those presented in Refs.
 [18] and 
 [177].

\section{Clifford HA--Bundles}

We shall consider two variants of generalization of spinor
constructions defined for d-vector spaces to the case of
distinguished vector bundle \index{Clifford bundles!higher order
an\-i\-sot\-rop\-ic} spaces enabled with the structure of
N-connection. The first is to use the extension to the category
of vector bundles. The second is to define the Clifford fibration
associated with compatible linear d-connection and metric $G$ on
a vector bundle. We shall analyze both variants.

\subsection{Clifford d--module structure in dv--bundles}

Because functor ${\cal F}\to C({\cal F})$ is smooth we can extend
it to the category of vector bundles of type $\xi ^{<z>}=\{\pi
_d:HE^{<z>}\oplus
V_1E^{<z>}\oplus ...\oplus V_zE^{<z>}\rightarrow E^{<z>}\}.$ Recall that by $%
{\cal F}$ we denote the typical fiber of such bundles. For $\xi
^{<z>}$ we \index{Clifford d--module} obtain a bundle of
algebras, denoted as $C\left( \xi ^{<z>}\right) ,\,$ such that
$C\left( \xi ^{<z>}\right) _u=C\left( {\cal F}_u\right) .$
Multiplication in every fibre defines a continuous map $C\left(
\xi ^{<z>}\right) \times C\left( \xi ^{<z>}\right) \rightarrow
C\left( \xi
^{<z>}\right) .$ If $\xi ^{<z>}$ is a vector bundle on number field $k$%
,\thinspace \thinspace the structure of the $C\left( \xi ^{<z>}\right) $%
-module, the d-module, the d-module, on $\xi ^{<z>}$ is given by
the continuous map $C\left( \xi ^{<z>}\right) \times _E\xi
^{<z>}\rightarrow \xi
^{<z>}$ with every fiber ${\cal F}_u$ provided with the structure of the $%
C\left( {\cal F}_u\right) -$module, correlated with its $k$-module
structure, Because ${\cal F}\subset C\left( {\cal F}\right) ,$ we
have a fiber to fiber map ${\cal F}\times _E\xi ^{<z>}\rightarrow
\xi ^{<z>},$ inducing on every fiber the map ${\cal F}_u\times
_E\xi _{(u)}^{<z>}\rightarrow \xi _{(u)}^{<z>}$ (${\cal
R}$-linear on the first factor and $k$-linear on the second one
). Inversely, every such bilinear
map defines on $\xi ^{<z>}$ the structure of the $C\left( \xi ^{<z>}\right) $%
-module by virtue of universal properties of Clifford d-algebras.
Equivalently, the above-mentioned bilinear map defines a morphism
of v-bundles $m:\xi ^{<z>}\rightarrow HOM\left( \xi ^{<z>},\xi
^{<z>}\right) \quad [HOM\left( \xi ^{<z>},\xi ^{<z>}\right) $
denotes the bundles of homomorphisms] when $\left( m\left(
u\right) \right) ^2=G\left( u\right) $ on every point.

Vector bundles $\xi ^{<z>}$ provided with $C\left( \xi ^{<z>}\right) $%
-structures are objects of the category with morphisms being
morphisms of
dv-bundles, which induce on every point $u\in \xi ^{<z>}$ morphisms of $%
C\left( {\cal F}_u\right) -$modules. This is a Banach category
contained in the category of finite-dimensional d-vector spaces
on filed $k.$ We shall not use category formalism in this work,
but point to its advantages in further formulation of new
directions of K-theory (see , for example, an
introduction in Ref. 
 [133]) concerned with la-spaces.

Let us denote by $H^s\left( {\cal E}^{<z>},GL_{n_E}\left( {\cal
R}\right) \right) ,\,$ where $n_E=n+m_1+...+m_z,\,$ the
s-dimensional cohomology group
of the algebraic sheaf of germs of continuous maps of dv-bundle ${\cal E}%
^{<z>}$ with group $GL_{n_E}\left( {\cal R}\right) $ the group of
automorphisms of ${\cal R}^{n_E}\,$ (for the language of
algebraic topology
see, for example, Refs. 
 [133] and
 [98]. We shall also use the
group $SL_{n_E}\left( {\cal R}\right) =\{A\subset GL_{n_E}\left( {\cal R}%
\right) ,\det A=1\}.\,$ Here we point out that cohomologies\\
$H^s(M,Gr)$ characterize the class of a principal bundle $\pi
:P\rightarrow M$ on $M$ with structural group $Gr.$ Taking into
account that we deal with bundles distinguished by an
N-connection we introduce into consideration cohomologies
$H^s\left( {\cal E}^{<z>},GL_{n_E}\left( {\cal R}\right) \right)
$ as distinguished classes (d-classes) of bundles ${\cal E}^{<z>}$
provided with a global N-connection structure.

For a real vector bundle $\xi ^{<z>}$ on compact base ${\cal
E}^{<z>}$ we can define the orientation on $\xi ^{<z>}$ as an
element $\alpha _d\in H^1\left( {\cal E}^{<z>},GL_{n_E}\left(
{\cal R}\right) \right) $ whose
image on map%
$$
H^1\left( {\cal E}^{<z>},SL_{n_E}\left( {\cal R}\right) \right)
\rightarrow H^1\left( {\cal E}^{<z>},GL_{n_E}\left( {\cal
R}\right) \right)
$$
is the d-class of bundle ${\cal E}^{<z>}.$

\begin{definition}
The spinor structure on $\xi ^{<z>}$ is defined as an element\\
$\beta _d\in H^1\left( {\cal E}^{<z>},Spin\left( n_E\right)
\right) $ whose image in the
composition%
$$
H^1\left( {\cal E}^{<z>},Spin\left( n_E\right) \right)
\rightarrow H^1\left(
{\cal E}^{<z>},SO\left( n_E\right) \right) \rightarrow H^1\left( {\cal E}%
^{<z>},GL_{n_E}\left( {\cal R}\right) \right)
$$
is the d-class of ${\cal E}^{<z>}.$
\end{definition}

The above definition of spinor structures can be reformulated in
terms of \index{Spinor structures} principal bundles. Let $\xi
^{<z>}$ be a real vector bundle of rank n+m on a compact base
${\cal E}^{<z>}.$ If there is a principal bundle $P_d$ with
structural group $SO( n_E ) $ \newline  [ or $Spin( n_E ) ],$ this bundle $%
\xi ^{<z>}$ can be provided with orientation (or spinor)
structure. The
bundle $P_d$ is associated with element\\ $\alpha _d\in H^1\left( {\cal E}%
^{<z>},SO(n_{<z>})\right) $ [or $\beta _d\in H^1\left( {\cal E}%
^{<z>},Spin\left( n_E\right) \right) .$

We remark that a real bundle is oriented if and only if its first
Stiefel-Whitney d--class vanishes,
$$
w_1\left( \xi _d\right) \in H^1\left( \xi ,{\cal Z}/2\right) =0,
$$
where $H^1\left( {\cal E}^{<z>},{\cal Z}/2\right) $ is the first
group of Chech cohomology with coefficients in ${\cal Z}/2,$
Considering the second \index{Stiefel--Whitney class}
Stiefel--Whitney class $w_2\left( \xi ^{<z>}\right) \in H^2\left( {\cal E}%
^{<z>},{\cal Z}/2\right) $ it is well known that vector bundle
$\xi ^{<z>}$ admits the spinor structure if and only if
$w_2\left( \xi ^{<z>}\right) =0.$ Finally, in this subsection, we
emphasize that taking into account that base space ${\cal
E}^{<z>}$ is also a v-bundle, $p:E^{<z>}\rightarrow M,$ we have
to make explicit calculations in order to express cohomologies
$H^s\left( {\cal E}^{<z>},GL_{n+m}\right) \,$ and $H^s\left(
{\cal E}^{<z>},SO\left( n+m\right) \right) $ through cohomologies
$H^s\left( M,GL_n\right) ,H^s\left( M,SO\left( m_1\right) \right)
,$ $...H^s\left( M,SO\left( m_z\right) \right) ,$ , which depends
on global topological structures of spaces $M$ and ${\cal
E}^{<z>}$ $.$ For general bundle and base spaces this requires a
cumbersome cohomological calculus.

\subsection{Clifford fibration}

Another way of defining the spinor structure is to use Clifford
fibrations. Consider the principal bundle with the structural
group $Gr$ being a subgroup of orthogonal group $O\left( G\right)
,$ where $G$ is a quadratic nondegenerate form (see(2.12))
defined on the base (also being a bundle space) space ${\cal
E}^{<z>}.$ The fibration associated to principal fibration
$P\left( {\cal E}^{<z>},Gr\right) $ with a typical fiber having
Clifford algebra $C\left( G\right) $ is, by definition, the
Clifford fibration $PC\left( {\cal E}^{<z>},Gr\right) .$ We can
always define a \index{Clifford!fibration} metric on the Clifford
fibration if every fiber is isometric to $PC\left( {\cal
E}^{<z>},G\right) $ (this result is proved for arbitrary quadratic
forms $G$ on pseudo-Riemannian bases 
 [245]). If, additionally, $%
Gr\subset SO\left( G\right) $ a global section can be defined on
$PC\left( G\right) .$

Let ${\cal P}\left( {\cal E}^{<z>},Gr\right) $ be the set of
principal bundles with differentiable base ${\cal E}^{<z>}$ and
structural group $Gr.$
If $g:Gr\rightarrow Gr^{\prime }$ is an homomorphism of Lie groups and $%
P\left( {\cal E}^{<z>},Gr\right) \subset {\cal P}\left( {\cal E}%
^{<z>},Gr\right) $ (for simplicity in this section we shall
denote mentioned
bundles and sets of bundles as $P,P^{\prime }$ and respectively, ${\cal P},%
{\cal P}^{\prime }),$ we can always construct a principal bundle
with the property that there is as homomorphism $f:P^{\prime
}\rightarrow P$ of
principal bundles which can be projected to the identity map of ${\cal E}%
^{<z>}$ and corresponds to isomorphism $g:Gr\rightarrow
Gr^{\prime }.$ If the inverse statement also holds, the bundle
$P^{\prime }$ is called as the extension of $P$ associated to $g$
and $f$ is called the extension homomorphism denoted as
$\widetilde{g.}$

Now we can define distinguished spinor structures on bundle
spaces (compare with definition 6.3 ).

\begin{definition}
Let $P\in {\cal P}\left( {\cal E}^{<z>},O\left( G\right) \right)
$ be a principal bundle. A distinguished spinor structure of $P,$
equivalently a ds-structure of ${\cal E}^{<z>}$ is an extension
$\widetilde{P}$ of $P$
associated to homomorphism $h:PinG\rightarrow O\left( G\right) $ where $%
O\left( G\right) $ is the group of orthogonal rotations,
generated by metric $G,$ in bundle ${\cal E}^{<z>}.$
\end{definition}

So, if $\widetilde{P}$ is a spinor structure of the space ${\cal
E}^{<z>},$ then $$\widetilde{P}\in {\cal P}\left( {\cal
E}^{<z>},PinG\right) .$$

The definition of spinor structures on varieties was given in ref.
 [65]. In Refs.
 [65] 
and 
 [66] it is proved that a necessary and
sufficient condition for a space time to be orientable is to
admit a global field of orthonormalized frames. We mention that
spinor structures can be also defined on varieties modeled on
Banach spaces
 [8]. As we have
shown in this subsection, similar constructions are possible for
the cases when space time has the structure of a v-bundle with an
N-connection.

\begin{definition}
A special distinguished spinor structure, ds-structure, of principal bundle $$%
P=P\left( {\cal E}^{<z>},SO\left( G\right) \right) $$ is a principal bundle $$%
\widetilde{P}=\widetilde{P}\left( {\cal E}^{<z>},SpinG\right) $$
for which a homomorphism of principal bundles
$$\widetilde{p}:\widetilde{P}\rightarrow P,$$ projected on the
identity map of ${\cal E}^{<z>}$ and corresponding to
representation%
$$
R:SpinG\rightarrow SO\left( G\right) ,
$$
is defined.
\end{definition}

In the case when the base space variety is oriented, there is a
natural bijection between tangent spinor structures with a common
base. For special ds--structures we can define, as for any spinor
structure, the concepts of \index{Ds--structures} spin tensors,
spinor connections, and spinor covariant derivations (see
subsection 6.6.1 and Refs. 
 [255,275,264]).

\section{Almost Complex Spinor Structures}

Almost complex structures are an important characteristic of
$H^{2n}$-spaces \index{Almost complex!structures} \index{Almost
complex!spinor structures} and of osculator bundles
$Osc^{k=2k_1}(M),$ where $k_1=1,2,...$ . For
simplicity in this section we restrict our analysis to the case of $H^{2n}$%
-spaces. We can rewrite the almost Hermitian metric 
 [160,161], $%
H^{2n} $-metric ( see considerations from subsection 6.1.1 with
respect to \index{Almost Hermitian metric} metrics and conditions
of type (6.12) and correspondingly (6.14) ), in
complex form 
 [256]:

$$
G=H_{ab}\left( z,\xi \right) dz^a\otimes dz^b,\eqno(6.38)
$$
where
$$
z^a=x^a+iy^a,~\overline{z^a}=x^a-iy^a,~H_{ab}\left(
z,\overline{z}\right) =g_{ab}\left( x,y\right) \mid _{y=y\left(
z,\overline{z}\right) }^{x=x\left( z,\overline{z}\right) },
$$
and define almost complex spinor structures. For given metric (6.38) on $%
H^{2n}$-space there is always a principal bundle $P^U$ with
unitary structural group U(n) which allows us to transform
$H^{2n}$-space into v-bundle $\xi ^U\approx P^U\times _{U\left(
n\right) }{\cal R}^{2n}.$ This statement will be proved after we
introduce complex
\begin{figure}[htbp]
\begin{center}
\begin{picture}(255,50) \setlength{\unitlength}{1pt}
\thinlines \put(0,45){$U(n)$} \put(116,45){$SO(2n)$}
\put(53,2){${Spin}^c (2n)$} \put(70,38){i} \put(70,48){
\line(-1,0){45}} \put(70,48){\vector(1,0){45}}
\put(25,42){\vector(3,-2){45}} \put(78,13) {\vector(3,2){45}}
\put(245,10){(6.39)}
\put(35,20){$\sigma$} \put(100,20){${\rho}^c$}
\end{picture}
\end{center}
\caption{Diagram 3}
\end{figure}
spinor structures on oriented real vector bundles 
 [133].

Let us consider momentarily $k={\cal C}$ and introduce into
consideration
[instead of the group $Spin(n)]$ the group $Spin^c\times _{{\cal Z}%
/2}U\left( 1\right) $ being the factor group of the product
$Spin(n)\times
U\left( 1\right) $ with the respect to equivalence%
$$
\left( y,z\right) \sim \left( -y,-a\right) ,\quad y\in Spin(m).
$$
This way we define the short exact sequence%
$$
1\rightarrow U\left( 1\right) \rightarrow Spin^c\left( n\right) \stackrel{S^c%
}{\to }SO\left( n\right) \rightarrow 1,
$$
where $\rho ^c\left( y,a\right) =\rho ^c\left( y\right) .$ If
$\lambda $ is oriented , real, and rank $n,$ $\gamma $-bundle
$\pi :E_\lambda \rightarrow
M^n,$ with base $M^n,$ the complex spinor structure, spin structure, on $%
\lambda $ is given by the principal bundle $P$ with structural group $%
Spin^c\left( m\right) $ and isomorphism $\lambda \approx P\times
_{Spin^c\left( n\right) }{\cal R}^n.$ For such bundles the
categorial equivalence can be defined as
$$
\epsilon ^c:{\cal E}_{{\cal C}}^T\left( M^n\right) \rightarrow {\cal E}_{%
{\cal C}}^\lambda \left( M^n\right) ,\eqno(6.40)
$$
where $\epsilon ^c\left( E^c\right) =P\bigtriangleup
_{Spin^c\left( n\right)
}E^c$ is the category of trivial complex bundles on $M^n,{\cal E}_{{\cal C}%
}^\lambda \left( M^n\right) $ is the category of complex
v-bundles on $M^n$ with action of Clifford bundle $C\left(
\lambda \right) ,P\bigtriangleup _{Spin^c(n)}$ and $E^c$ is the
factor space of the bundle product $P\times
_ME^c$ with respect to the equivalence $\left( p,e\right) \sim \left( p%
\widehat{g}^{-1},\widehat{g}e\right) ,p\in P,e\in E^c,$ where $\widehat{g}%
\in Spin^c\left( n\right) $ acts on $E$ by via the imbedding
$Spin\left( n\right) \subset C^{0,n}$ and the natural action
$U\left( 1\right) \subset {\cal C}$ on complex v-bundle $\xi
^c,E^c=tot\xi ^c,$ for bundle $\pi ^c:E^c\rightarrow M^n.$

Now we return to the bundle $\xi ={\cal E}^{<1>}.$ A real
v-bundle (not being a spinor bundle) admits a complex spinor
structure if and only if there exist a homomorphism $\sigma
:U\left( n\right) \rightarrow Spin^c\left( 2n\right) $ making the
diagram 3 commutative. The explicit construction of $\sigma $ for
arbitrary $\gamma $-bundle is given in refs.
 [133] and
 [22]. For $H^{2n}$-spaces it is obvious that a diagram
similar to (6.39) can be defined for the tangent bundle $TM^n,$
which
directly points to the possibility of defining the $^cSpin$-structure on $%
H^{2n}$-spaces.

Let $\lambda $ be a complex, rank\thinspace $n,$ spinor bundle
with
$$
\tau :Spin^c\left( n\right) \times _{{\cal Z}/2}U\left( 1\right)
\rightarrow U\left( 1\right) \eqno(6.41)
$$
the homomorphism defined by formula $\tau \left( \lambda ,\delta
\right) =\delta ^2.$ For $P_s$ being the principal bundle with
fiber $Spin^c\left( n\right) $ we introduce the complex linear
bundle $L\left( \lambda ^c\right) =P_S\times _{Spin^c(n)}{\cal
C}$ defined as the factor space of $P_S\times
{\cal C}$ on equivalence relation%
$$
\left( pt,z\right) \sim \left( p,l\left( t\right) ^{-1}z\right) ,
$$
where $t\in Spin^c\left( n\right) .$ This linear bundle is
associated to complex spinor structure on $\lambda ^c.$

If $\lambda ^c$ and $\lambda ^{c^{\prime }}$ are complex spinor
bundles, the Whitney sum $\lambda ^c\oplus \lambda ^{c^{\prime
}}$ is naturally provided with the structure of the complex
spinor bundle. This follows from the \index{Complex spinor bundle}
holomorphism%
$$
\omega ^{\prime }:Spin^c\left( n\right) \times Spin^c\left(
n^{\prime }\right) \rightarrow Spin^c\left( n+n^{\prime }\right)
,\eqno(6.42)
$$
given by formula $\left[ \left( \beta ,z\right) ,\left( \beta
^{\prime },z^{\prime }\right) \right] \rightarrow \left[ \omega
\left( \beta ,\beta ^{\prime }\right) ,zz^{\prime }\right] ,$
where $\omega $ is the
homomorphism making the following diagram 4 commutative.%
\begin{figure}[htbp]
\begin{center}
\begin{picture}(190,50) \setlength{\unitlength}{1pt}
\thinlines \put(0,45){$Spin(n)\times Spin(n')$}
\put(160,45){$Spin(n+n')$} \put(15,2){$O(n)\times O(n')$}
\put(168,02){$O(n+n')$} \put(100,45){ \vector(1,0){55}}
\put(48,40){\vector(0,-1){25}} \put(90,5){\vector(1,0){70}}
\put(185,40) {\vector(0,-1){25}}
\end{picture}
\end{center}
\caption{Diagram 4}
\end{figure}
Here, $z,z^{\prime }\in U\left( 1\right) .$ It is obvious that
$L\left( \lambda ^c\oplus \lambda ^{c^{\prime }}\right) $ is
isomorphic to $L\left( \lambda ^c\right) \otimes L\left( \lambda
^{c^{\prime }}\right) .$

We conclude this section by formulating our main result on
complex spinor
structures for $H^{2n}$-spaces: 

\begin{theorem}
Let $\lambda ^c$ be a complex spinor bundle of rank $n$ and
$H^{2n}$-space considered as a real vector bundle $\lambda
^c\oplus \lambda ^{c^{\prime }}$ provided with almost complex
structure $J_{\quad \beta }^\alpha ;$ multiplication on $i$ is
given by $\left(
\begin{array}{cc}
0 & -\delta _j^i \\
\delta _j^i & 0
\end{array}
\right) $. Then, the diagram 5 is commutative up to isomorphisms
$\epsilon ^c
$ and $\widetilde{\epsilon }^c$ defined as in (6.40), ${\cal H}$ is functor $%
E^c\rightarrow E^c\otimes L\left( \lambda ^c\right) $ and ${\cal E}_{{\cal C}%
}^{0,2n}\left( M^n\right) $ is defined by functor ${\cal
E}_{{\cal C}}\left( M^n\right) \rightarrow {\cal E}_{{\cal
C}}^{0,2n}\left( M^n\right) $ given as correspondence
$E^c\rightarrow \Lambda \left( {\cal C}^n\right) \otimes E^c$
(which is a categorial equivalence), $\Lambda \left( {\cal
C}^n\right) $ is the exterior algebra on ${\cal C}^n.$ $W$ is the
real bundle $\lambda ^c\oplus \lambda ^{c^{\prime }}$ provided
with complex structure.
\end{theorem}

\begin{figure}[htbp]
\begin{center}
\begin{picture}(150,50) \setlength{\unitlength}{1pt}
\thinlines \put(0,45){${\cal E}_{\cal C}^{0,2n} (M^{2n})$}
\put(145,45){${\cal E}_{\cal C}^{{\lambda}^c \oplus {\lambda}^c }
(M^n )$} \put(81,0){${\cal E}_{\cal C}^W (M^n )$}
\put(99,38){${\epsilon}^c$} \put(99,48){ \line(-1,0){45}}
\put(99,48){\vector(1,0){45}} \put(54,42){\vector(3,-2){45}}
\put(147,42){\vector(-3,-2){45}}
\put(64,20){${\tilde{\varepsilon}}^c$} \put(129,20){$\cal H$}
\end{picture}
\end{center}
\caption{Diagram 5}
\end{figure}
{\bf Proof: }We use composition of homomorphisms%
$$
\mu :Spin^c\left( 2n\right) \stackrel{\pi }{\to }SO\left( n\right) \stackrel{%
r}{\to }U\left( n\right) \stackrel{\sigma }{\to }Spin^c\left(
2n\right) \times _{{\cal Z}/2}U\left( 1\right) ,
$$
commutative diagram 6 and introduce composition of homomorphisms%
$$
\mu :Spin^c\left( n\right) \stackrel{\Delta }{\to }Spin^c\left(
n\right) \times Spin^c\left( n\right) \stackrel{{\omega }^c}{\to
}Spin^c\left( n\right) ,
$$
where $\Delta $ is the diagonal homomorphism and $\omega ^c$ is
defined as in (6.42). Using homomorphisms (6.41) and (6.42) we
obtain formula $\mu \left( t\right) =\mu \left( t\right) r\left(
t\right) .$

Now consider bundle $P\times _{Spin^c\left( n\right)
}Spin^c\left( 2n\right) $ as the principal $Spin^c\left(
2n\right) $-bundle, associated to $M\oplus M $ being the factor
space of the product $P\times Spin^c\left( 2n\right) $ on the
equivalence relation $\left( p,t,h\right) \sim \left( p,\mu \left(
t\right) ^{-1}h\right) .$ In this case the categorial equivalence
(6.40) can be rewritten as
$$
\epsilon ^c\left( E^c\right) =P\times _{Spin^c\left( n\right)
}Spin^c\left( 2n\right) \Delta _{Spin^c\left( 2n\right) }E^c
$$
and seen as factor space of $P\times Spin^c\left( 2n\right)
\times _ME^c$ on equivalence relation
$$
\left( pt,h,e\right) \sim \left( p,\mu \left( t\right)
^{-1}h,e\right) \mbox{and}\left( p,h_1,h_2,e\right) \sim \left(
p,h_1,h_2^{-1}e\right)
$$
(projections of elements $p$ and $e$ coincides on base $M).$
Every element of $\epsilon ^c\left( E^c\right) $ can be
represented as $P\Delta _{Spin^c\left( n\right) }E^c,$ i.e., as a
factor space $P\Delta E^c$ on equivalence relation $\left(
pt,e\right) \sim \left( p,\mu ^c\left( t\right) ,e\right) ,$ when
$t\in Spin^c\left( n\right) .$
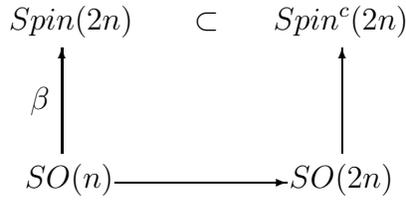
\begin{figure}[htbp]
\begin{center}
\begin{picture}(140,70) \setlength{\unitlength}{1pt}
\thinlines \put(0,60){$Spin(2n)$} \put(100,60){${Spin}^c (2n)$}
\put(6,0){$SO(n)$} \put(106,0){$SO(2n)$} \put(36,2){
\vector(1,0){65}} \put(20,13){\vector(0,1){40}}
\put(126,13) {\vector(0,1){40}}
\put(8,30){$\beta$} \put(70,60){$\subset$}
\end{picture}
\end{center}
\caption{Diagram 6}
\end{figure}
The complex line bundle $L\left( \lambda ^c\right) $ can be
interpreted as the factor space of\\ $P\times _{Spin^c\left(
n\right) }{\cal C}$ on equivalence relation $\left( pt,\delta
\right) \sim \left( p,r\left( t\right) ^{-1}\delta \right) .$
\index{Complex line bundle}

Putting $\left( p,e\right) \otimes \left( p,\delta \right) \left(
p,\delta
e\right) $ we introduce morphism%
$$
\epsilon ^c\left( E\right) \times L\left( \lambda ^c\right)
\rightarrow \epsilon ^c\left( \lambda ^c\right)
$$
with properties $\left( pt,e\right) \otimes \left( pt,\delta
\right) \rightarrow \left( pt,\delta e\right) =\left( p,\mu
^c\left( t\right) ^{-1}\delta e\right) ,$

$\left( p,\mu ^c\left( t\right) ^{-1}e\right) \otimes \left(
p,l\left( t\right) ^{-1}e\right) \rightarrow \left( p,\mu
^c\left( t\right) r\left( t\right) ^{-1}\delta e\right) $
pointing to the fact that we have defined
the isomorphism correctly and that it is an isomorphism on every fiber. $%
\Box $

\section{ D--Spinor Techniques}

The purpose of this section is to show how a corresponding
abstract spinor technique entailing notational and calculations
advantages can be developed
for arbitrary splits of dimensions of a d-vector space ${\cal F}=h{\cal F}%
\oplus v_1{\cal F\oplus ...}\oplus v_z{\cal F}$, where $\dim
h{\cal F}=n$ and $\dim v_p{\cal F}=m_p.$ For convenience we shall
also present some necessary coordinate expressions.

The problem of a rigorous definition of spinors on la-spaces
(la-spinors,
d-spinors) was posed and solved 
 [256,255,275] (see previous sections
6.2--6.4 for generalizations on higher order anisotropic
superspaces) in the framework of the formalism of Clifford and
spinor structures on v-bundles provided with compatible nonlinear
and distinguished connections and metric.
We introduced d-spinors as corresponding objects of the Clifford d-algebra $%
{\cal C}\left( {\cal F},G\right) $, defined for a d-vector space
${\cal F}$
in a standard manner (see, for instance, 
 [22]) and proved that
operations with ${\cal C}\left( {\cal F},G\right) \ $ can be
reduced to
calculations for ${\cal C}\left( h{\cal F},g\right) ,{\cal C}\left( v_1{\cal %
F},h_1\right) ,...$ and ${\cal C}\left( v_z{\cal F},h_z\right) ,$
which are
usual Clifford algebras of respective dimensions 2$^n,2^{m_1},...$ and 2$%
^{m_z}$ (if it is necessary we can use quadratic forms $g$ and
$h_p$ correspondingly induced on $h{\cal F}$ and $v_p{\cal F}$ by
a metric ${\bf G}
$ (6.12)). Considering the orthogonal subgroup $O{\bf \left( G\right) }%
\subset GL{\bf \left( G\right) }$ defined by a metric ${\bf G}$
we can define the d-spinor norm and parametrize d-spinors by
ordered pairs of
elements of Clifford algebras ${\cal C}\left( h{\cal F},g\right) $ and $%
{\cal C}\left( v_p{\cal F},h_p\right) ,p=1,2,...z.$ We emphasize
that the splitting of a Clifford d-algebra associated to a
dv-bundle ${\cal E}^{<z>}$ is a straightforward consequence of
the global decomposition (6.3) defining a N-connection structure
in ${\cal E}^{<z>}.$

In this section we shall omit detailed proofs which in most cases
are mechanical but rather tedious. We can apply the methods
developed in
 [180,181,182,154] in a straightforward manner on h- and v-subbundles in
order to verify the correctness of affirmations.

\subsection{Clifford d--algebra, d--spinors and d--twistors}

In order to relate the succeeding constructions with Clifford
d-algebras \index{Clifford d--algebra} \index{D--spinors}
\index{D--twistrs}
 [256,255] we consider a la-frame decomposition of the metric (6.12):%
$$
G_{<\alpha ><\beta >}\left( u\right) =l_{<\alpha >}^{<\widehat{\alpha }%
>}\left( u\right) l_{<\beta >}^{<\widehat{\beta }>}\left( u\right) G_{<%
\widehat{\alpha }><\widehat{\beta }>},
$$
where the frame d-vectors and constant metric matrices are
distinguished as

$$
l_{<\alpha >}^{<\widehat{\alpha }>}\left( u\right) =\left(
\begin{array}{cccc}
l_j^{\widehat{j}}\left( u\right) & 0 & ... & 0 \\
0 & l_{a_1}^{\widehat{a}_1}\left( u\right) & ... & 0 \\
... & ... & ... & ... \\
0 & 0 & .. & l_{a_z}^{\widehat{a}_z}\left( u\right)
\end{array}
\right) ,
$$
$$
G_{<\widehat{\alpha }><\widehat{\beta }>}=\left(
\begin{array}{cccc}
g_{\widehat{i}\widehat{j}} & 0 & ... & 0 \\
0 & h_{\widehat{a}_1\widehat{b}_1} & ... & 0 \\
... & ... & ... & ... \\
0 & 0 & 0 & h_{\widehat{a}_z\widehat{b}_z}
\end{array}
\right) ,
$$
$g_{\widehat{i}\widehat{j}}$ and $h_{\widehat{a}_1\widehat{b}_1},...,h_{%
\widehat{a}_z\widehat{b}_z}$ are diagonal matrices with $g_{\widehat{i}%
\widehat{i}}=$ $h_{\widehat{a}_1\widehat{a}_1}=...=h_{\widehat{a}_z\widehat{b%
}_z}=\pm 1.$

To generate Clifford d-algebras we start with matrix equations%
$$
\sigma _{<\widehat{\alpha }>}\sigma _{<\widehat{\beta }>}+\sigma _{<\widehat{%
\beta }>}\sigma _{<\widehat{\alpha }>}=-G_{<\widehat{\alpha }><\widehat{%
\beta }>}I,\eqno(6.43)
$$
where $I$ is the identity matrix, matrices $\sigma _{<\widehat{\alpha }%
>}\,(\sigma $-objects) act on a d-vector space ${\cal F}=h{\cal F}\oplus v_1%
{\cal F}\oplus ...\oplus v_z{\cal F}$ and theirs components are
distinguished as
$$
\sigma _{<\widehat{\alpha }>}\,=\left\{ (\sigma _{<\widehat{\alpha }>})_{%
\underline{\beta }}^{\cdot \underline{\gamma }}=\left(
\begin{array}{cccc}
(\sigma _{\widehat{i}})_{\underline{j}}^{\cdot \underline{k}} & 0
& ... & 0
\\
0 & (\sigma _{\widehat{a}_1})_{\underline{b}_1}^{\cdot
\underline{c}_1} &
... & 0 \\
... & ... & ... & ... \\
0 & 0 & ... & (\sigma _{\widehat{a}_z})_{\underline{b}_z}^{\cdot \underline{c%
}_z}
\end{array}
\right) \right\} ,\eqno(6.44)
$$
indices \underline{$\beta $},\underline{$\gamma $},... refer to
spin spaces of type ${\cal S}=S_{(h)}\oplus S_{(v_1)}\oplus
...\oplus S_{(v_z)}$ and
underlined Latin indices \underline{$j$},$\underline{k},...$ and $\underline{%
b}_1,\underline{c}_1,...,\underline{b}_z,\underline{c}_z...$ refer
respectively to h-spin space ${\cal S}_{(h)}$ and v$_p$-spin space ${\cal S}%
_{(v_p)},(p=1,2,...,z)\ $which are correspondingly associated to a h- and v$%
_p$-decomposition of a dv-bundle ${\cal E}^{<z>}.$ The
irreducible algebra of matrices $\sigma _{<\widehat{\alpha }>}$
of minimal dimension $N\times N,$ where
$N=N_{(n)}+N_{(m_1)}+...+N_{(m_z)},$ $\dim {\cal
S}_{(h)}$=$N_{(n)}$
and $\dim {\cal S}_{(v_p)}$=$N_{(m_p)},$ has these dimensions%
$$
{N_{(n)}=\left\{
\begin{array}{rl}
{\ 2^{(n-1)/2},} & n=2k+1 \\
{2^{n/2},\ } & n=2k;
\end{array}
\right. }$$ and $$ {N}_{(m_p)}{=} \left\{ \begin{array}{rl}
2^{(m_p-1)/2}, & m_p=2k_p+1 \\
2^{m_p}, & m_p=2k_p
\end{array}\right.
 ,
$$
where $k=1,2,...,k_p=1,2,....$

The Clifford d-algebra is generated by sums on $n+1$ elements of form%
$$
A_1I+B^{\widehat{i}}\sigma _{\widehat{i}}+C^{\widehat{i}\widehat{j}}\sigma _{%
\widehat{i}\widehat{j}}+D^{\widehat{i}\widehat{j}\widehat{k}}\sigma _{%
\widehat{i}\widehat{j}\widehat{k}}+...\eqno(6.45)
$$
and sums of $m_p+1$ elements of form%
$$
A_{2(p)}I+B^{\widehat{a}_p}\sigma _{\widehat{a}_p}+C^{\widehat{a}_p\widehat{b%
}_p}\sigma _{\widehat{a}_p\widehat{b}_p}+D^{\widehat{a}_p\widehat{b}_p%
\widehat{c}_p}\sigma
_{\widehat{a}_p\widehat{b}_p\widehat{c}_p}+...
$$
with antisymmetric coefficients\\ $C^{\widehat{i}\widehat{j}}=C^{[\widehat{i}%
\widehat{j}]},C^{\widehat{a}_p\widehat{b}_p}=C^{[\widehat{a}_p\widehat{b}%
_p]},D^{\widehat{i}\widehat{j}\widehat{k}}=D^{[\widehat{i}\widehat{j}%
\widehat{k}]},D^{\widehat{a}_p\widehat{b}_p\widehat{c}_p}=D^{[\widehat{a}_p%
\widehat{b}_p\widehat{c}_p]},...$ and matrices $\sigma _{\widehat{i}\widehat{%
j}}=\sigma _{[\widehat{i}}\sigma _{\widehat{j}]},\sigma _{\widehat{a}_p%
\widehat{b}_p}=\sigma _{[\widehat{a}_p}\sigma _{\widehat{b}_p]},\sigma _{%
\widehat{i}\widehat{j}\widehat{k}}=\sigma _{[\widehat{i}}\sigma _{\widehat{j}%
}\sigma _{\widehat{k}]},...$ . Really, we have 2$^{n+1}$ coefficients $%
\left( A_1,C^{\widehat{i}\widehat{j}},D^{\widehat{i}\widehat{j}\widehat{k}%
},...\right) $ and 2$^{m_p+1}$ coefficients $( A_{2(p)},C^{\widehat{a}%
_p\widehat{b}_p},D^{\widehat{a}_p\widehat{b}_p\widehat{c}_p},...)
$ of the Clifford algebra on ${\cal F.}$

For simplicity, in this subsection, we shall present the
necessary geometric constructions only for h-spin spaces ${\cal
S}_{(h)}$ of dimension $N_{(n)}.$ Considerations for a v-spin
space ${\cal S}_{(v)}$ are similar but with proper
characteristics for a dimension $N_{(m)}.$

In order to define the scalar (spinor) product on ${\cal
S}_{(h)}$ we introduce into consideration this finite sum
(because of a finite number of elements $\sigma
_{[\widehat{i}\widehat{j}...\widehat{k}]}):$
$$
^{(\pm
)}E_{\underline{k}\underline{m}}^{\underline{i}\underline{j}}=\delta
_{\underline{k}}^{\underline{i}}\delta _{\underline{m}}^{\underline{j}%
}+\frac 2{1!}(\sigma
_{\widehat{i}})_{\underline{k}}^{.\underline{i}}(\sigma
^{\widehat{i}})_{\underline{m}}^{.\underline{j}}+\frac{2^2}{2!}(\sigma _{%
\widehat{i}\widehat{j}})_{\underline{k}}^{.\underline{i}}(\sigma ^{\widehat{i%
}\widehat{j}})_{\underline{m}}^{.\underline{j}}+\frac{2^3}{3!}(\sigma _{%
\widehat{i}\widehat{j}\widehat{k}})_{\underline{k}}^{.\underline{i}}(\sigma
^{\widehat{i}\widehat{j}\widehat{k}})_{\underline{m}}^{.\underline{j}}+...%
\eqno(6.46)
$$
which can be factorized as
$$
^{(\pm )}E_{\underline{k}\underline{m}}^{\underline{i}\underline{j}}=N_{(n)}{%
\ }^{(\pm )}\epsilon _{\underline{k}\underline{m}}{\ }^{(\pm )}\epsilon ^{%
\underline{i}\underline{j}}\mbox{ for }n=2k\eqno(6.47)
$$
and%
$$
^{(+)}E_{\underline{k}\underline{m}}^{\underline{i}\underline{j}%
}=2N_{(n)}\epsilon _{\underline{k}\underline{m}}\epsilon ^{\underline{i}%
\underline{j}},{\ }^{(-)}E_{\underline{k}\underline{m}}^{\underline{i}%
\underline{j}}=0\mbox{ for }n=3(mod4),\eqno(6.48)
$$
$$
^{(+)}E_{\underline{k}\underline{m}}^{\underline{i}\underline{j}}=0,{\ }%
^{(-)}E_{\underline{k}\underline{m}}^{\underline{i}\underline{j}%
}=2N_{(n)}\epsilon _{\underline{k}\underline{m}}\epsilon ^{\underline{i}%
\underline{j}}\mbox{ for }n=1(mod4).
$$

Antisymmetry of $\sigma _{\widehat{i}\widehat{j}\widehat{k}...}$
and the construction of the objects (6.45),(6.46),\\ (6.47) and
(6.48) define the
properties of $\epsilon $-objects $^{(\pm )}\epsilon _{\underline{k}%
\underline{m}}$ and $\epsilon _{\underline{k}\underline{m}}$
which have an
eight-fold periodicity on $n$ (see details in 
 [182] and, with respect
to la-spaces, 
 [256]).

For even values of $n$ it is possible the decomposition of every
h-spin
space ${\cal S}_{(h)}$into irreducible h-spin spaces ${\bf S}_{(h)}$ and $%
{\bf S}_{(h)}^{\prime }$ (one considers splitting of h-indices,
for instance, \underline{$l$}$=L\oplus L^{\prime
},\underline{m}=M\oplus
M^{\prime },...;$ for v$_p$-indices we shall write $\underline{a}%
_p=A_p\oplus A_p^{\prime },\underline{b}_p=B_p\oplus B_p^{\prime
},...)$ and defines new $\epsilon $-objects
$$
\epsilon ^{\underline{l}\underline{m}}=\frac 12\left( ^{(+)}\epsilon ^{%
\underline{l}\underline{m}}+^{(-)}\epsilon ^{\underline{l}\underline{m}%
}\right) \mbox{ and }\widetilde{\epsilon }^{\underline{l}\underline{m}%
}=\frac 12\left( ^{(+)}\epsilon
^{\underline{l}\underline{m}}-^{(-)}\epsilon
^{\underline{l}\underline{m}}\right) \eqno(6.49)
$$
We shall omit similar formulas for $\epsilon $-objects with lower
indices.

We can verify, by using expressions (6.48) and straightforward
calculations,
these para\-met\-ri\-za\-ti\-ons on symmetry properties of $\epsilon $%
-objects (6.49)
$$
\epsilon ^{\underline{l}\underline{m}}=\left(
\begin{array}{cc}
\epsilon ^{LM}=\epsilon ^{ML} & 0 \\
0 & 0
\end{array}
\right) \mbox{ and }\widetilde{\epsilon }^{\underline{l}\underline{m}%
}=\left(
\begin{array}{cc}
0 & 0 \\
0 & \widetilde{\epsilon }^{LM}=\widetilde{\epsilon }^{ML}
\end{array}
\right) \mbox{ for }n=0(mod8);\eqno(6.50)
$$
$$
\epsilon ^{\underline{l}\underline{m}}=-\frac 12{}^{(-)}\epsilon ^{%
\underline{l}\underline{m}}=\epsilon ^{\underline{m}\underline{l}},%
\mbox{ where }^{(+)}\epsilon
^{\underline{l}\underline{m}}=0,\mbox{ and }
$$
$$
\widetilde{\epsilon }^{\underline{l}\underline{m}}=-\frac
12{}^{(-)}\epsilon
^{\underline{l}\underline{m}}=\widetilde{\epsilon }^{\underline{m}\underline{%
l}}\mbox{ for }n=1(mod8);
$$
$$
\epsilon ^{\underline{l}\underline{m}}=\left(
\begin{array}{cc}
0 & 0 \\
\epsilon ^{L^{\prime }M} & 0
\end{array}
\right) \mbox{ and }\widetilde{\epsilon }^{\underline{l}\underline{m}%
}=\left(
\begin{array}{cc}
0 & \widetilde{\epsilon }^{LM^{\prime }}=-\epsilon ^{M^{\prime
}L} \\ 0 & 0
\end{array}
\right) \mbox{ for }n=2(mod8);
$$
$$
\epsilon ^{\underline{l}\underline{m}}=-\frac 12{}^{(+)}\epsilon ^{%
\underline{l}\underline{m}}=-\epsilon ^{\underline{m}\underline{l}},%
\mbox{ where }^{(-)}\epsilon
^{\underline{l}\underline{m}}=0,\mbox{ and }
$$
$$
\widetilde{\epsilon }^{\underline{l}\underline{m}}=\frac
12{}^{(+)}\epsilon
^{\underline{l}\underline{m}}=-\widetilde{\epsilon }^{\underline{m}%
\underline{l}}\mbox{ for }n=3(mod8);
$$
$$
\epsilon ^{\underline{l}\underline{m}}=\left(
\begin{array}{cc}
\epsilon ^{LM}=-\epsilon ^{ML} & 0 \\
0 & 0
\end{array}
\right) $$ and $$\widetilde{\epsilon }^{\underline{l}\underline{m}%
}=\left(
\begin{array}{cc}
0 & 0 \\
0 & \widetilde{\epsilon }^{LM}=-\widetilde{\epsilon }^{ML}
\end{array}
\right) \mbox{ for }n=4(mod8);
$$
$$
\epsilon ^{\underline{l}\underline{m}}=-\frac 12{}^{(-)}\epsilon ^{%
\underline{l}\underline{m}}=-\epsilon ^{\underline{m}\underline{l}},%
\mbox{ where }^{(+)}\epsilon
^{\underline{l}\underline{m}}=0,\mbox{ and }
$$
$$
\widetilde{\epsilon }^{\underline{l}\underline{m}}=-\frac
12{}^{(-)}\epsilon
^{\underline{l}\underline{m}}=-\widetilde{\epsilon }^{\underline{m}%
\underline{l}}\mbox{ for }n=5(mod8);
$$
$$
\epsilon ^{\underline{l}\underline{m}}=\left(
\begin{array}{cc}
0 & 0 \\
\epsilon ^{L^{\prime }M} & 0
\end{array}
\right) \mbox{ and }\widetilde{\epsilon }^{\underline{l}\underline{m}%
}=\left(
\begin{array}{cc}
0 & \widetilde{\epsilon }^{LM^{\prime }}=\epsilon ^{M^{\prime }L}
\\ 0 & 0
\end{array}
\right) \mbox{ for }n=6(mod8);
$$
$$
\epsilon ^{\underline{l}\underline{m}}=\frac 12{}^{(-)}\epsilon ^{\underline{%
l}\underline{m}}=\epsilon ^{\underline{m}\underline{l}},\mbox{ where }%
{}^{(+)}\epsilon ^{\underline{l}\underline{m}}=0,\mbox{ and }
$$
$$
\widetilde{\epsilon }^{\underline{l}\underline{m}}=-\frac
12{}^{(-)}\epsilon
^{\underline{l}\underline{m}}=\widetilde{\epsilon }^{\underline{m}\underline{%
l}}\mbox{ for }n=7(mod8).
$$

Let denote reduced and irreducible h-spinor spaces in a form
pointing to the
symmetry of spinor inner products in dependence of values $n=8k+l$ ($%
k=0,1,2,...;l=1,2,...7)$ of the dimension of the horizontal
subbundle (we shall write respectively $\bigtriangleup $ and
$\circ $ for antisymmetric and symmetric inner products of
reduced spinors and $\diamondsuit =(\bigtriangleup ,\circ )$ and
$\widetilde{\diamondsuit }=(\circ ,\bigtriangleup )$ for
corresponding parametrizations of inner products, in brief {\it
i.p.}, of irreducible spinors; properties of scalar products of
spinors are defined by $\epsilon $-objects (6.50); we shall use
$\Diamond $
for a general {\it i.p.} when the symmetry is not pointed out):%
$$
{\cal S}_{(h)}{\ }\left( 8k\right) ={\bf S}_{\circ }\oplus {\bf
S}_{\circ }^{\prime};\quad \eqno(6.51)
$$
$$
{\cal S}_{(h)}{\ }\left( 8k+1\right) ={\cal S}_{\circ }^{(-)}\
\mbox{({\it i.p.} is defined by an }^{(-)}\epsilon
\mbox{-object);}
$$
$$
{\cal S}_{(h)}{\ }\left( 8k+2\right) =\{
\begin{array}{c}
{\cal S}_{\Diamond }=({\bf S}_{\Diamond },{\bf S}_{\Diamond }),\mbox{ or} \\
{\cal S}_{\Diamond }^{\prime }=({\bf S}_{\widetilde{\Diamond }}^{\prime },%
{\bf S}_{\widetilde{\Diamond }}^{\prime });
\end{array}
\qquad
$$
$$
{\cal S}_{(h)}\left( 8k+3\right) ={\cal S}_{\bigtriangleup
}^{(+)}\ \mbox{({\it i.p.} is defined by an }^{(+)}\epsilon
\mbox{-object);}
$$
$$
{\cal S}_{(h)}\left( 8k+4\right) ={\bf S}_{\bigtriangleup }\oplus {\bf S}%
_{\bigtriangleup }^{\prime };\quad
$$
$$
{\cal S}_{(h)}\left( 8k+5\right) ={\cal S}_{\bigtriangleup
}^{(-)}\ \mbox{({\it i.p. }is defined by an }^{(-)}\epsilon
\mbox{-object),}
$$
$$
{\cal S}_{(h)}\left( 8k+6\right) =\{
\begin{array}{c}
{\cal S}_{\Diamond }=({\bf S}_{\Diamond },{\bf S}_{\Diamond }),\mbox{ or} \\
{\cal S}_{\Diamond }^{\prime }=({\bf S}_{\widetilde{\Diamond }}^{\prime },%
{\bf S}_{\widetilde{\Diamond }}^{\prime });
\end{array}
$$
\qquad
$$
{\cal S}_{(h)}\left( 8k+7\right) ={\cal S}_{\circ }^{(+)}\
\mbox{({\it i.p. } is defined by an }^{(+)}\epsilon
\mbox{-object)}.
$$
We note that by using corresponding $\epsilon $-objects we can
lower and rise indices of reduced and irreducible spinors (for
$n=2,6(mod4)$ we can
exclude primed indices, or inversely, see details in 
 [180,181,182]).

The similar v-spinor spaces are denoted by the same symbols as in
(6.51) provided with a left lower mark ''$|"$ and parametrized
with respect to the values $m=8k^{\prime }+l$ (k'=0,1,...;
l=1,2,...,7) of the dimension of the vertical subbundle, for
example, as
$$
{\cal S}_{(v_p)}(8k^{\prime })={\bf S}_{|\circ }\oplus {\bf
S}_{|\circ
}^{\prime },{\cal S}_{(v_p)}\left( 8k+1\right) ={\cal S}_{|\circ }^{(-)},...%
\eqno(6.52)
$$
We use '' $\widetilde{}$ ''-overlined symbols,
$$
{\widetilde{{\cal S}}}_{(h)}\left( 8k\right) ={\widetilde{{\bf
S}}}_{\circ }\oplus \widetilde{S}_{\circ }^{\prime
},{\widetilde{{\cal S}}}_{(h)}\left( 8k+1\right)
={\widetilde{{\cal S}}}_{\circ }^{(-)},...\eqno(6.53)
$$
and
$$
{\widetilde{{\cal S}}}_{(v_p)}(8k^{\prime })={\widetilde{{\bf
S}}}_{|\circ
}\oplus {\widetilde{S}}_{|\circ }^{\prime },{\widetilde{{\cal S}}}%
_{(v_p)}\left( 8k^{\prime }+1\right) ={\widetilde{{\cal
S}}}_{|\circ }^{(-)},...\eqno(6.54)
$$
respectively for the dual to (6.50) and (6.51) spinor spaces.

The spinor spaces (6.50)-(6.54) are called the prime spinor
spaces, in brief p-spinors. They are considered as building
blocks of distinguished, for simplicity we consider $\left(
n,m_1\right) $--spinor spaces constructed in
this manner:%
$$
{\cal S}(_{\circ \circ ,\circ \circ })={\bf S_{\circ }\oplus
S_{\circ }^{\prime }\oplus S_{|\circ }\oplus S_{|\circ }^{\prime
},}{\cal S}(_{\circ \circ ,\circ }\mid ^{\circ })={\bf S_{\circ
}\oplus S_{\circ }^{\prime }\oplus S_{|\circ }\oplus
\widetilde{S}_{|\circ }^{\prime },}\eqno(6.55)
$$
$$
{\cal S}(_{\circ \circ ,}\mid ^{\circ \circ })={\bf S_{\circ
}\oplus
S_{\circ }^{\prime }\oplus \widetilde{S}_{|\circ }\oplus \widetilde{S}%
_{|\circ }^{\prime },}{\cal S}(_{\circ }\mid ^{\circ \circ \circ })={\bf %
S_{\circ }\oplus \widetilde{S}_{\circ }^{\prime }\oplus \widetilde{S}%
_{|\circ }\oplus \widetilde{S}_{|\circ }^{\prime },}
$$
$$
...............................................
$$
$$
{\cal S}(_{\triangle },_{\triangle })={\cal S}_{\triangle
}^{(+)}\oplus
S_{|\bigtriangleup }^{(+)},S(_{\triangle },^{\triangle })={\cal S}%
_{\triangle }^{(+)}\oplus \widetilde{S}_{|\triangle }^{(+)},
$$
$$
................................
$$
$$
{\cal S}(_{\triangle }|^{\circ },_ \diamondsuit )={\bf
S}_{\triangle }\oplus \widetilde{S_{\circ }}^{\prime }\oplus
{\cal S}_{|
\diamondsuit },{\cal S}%
(_{\triangle }|^{\circ },^\diamondsuit )={\bf S}_{\triangle
}\oplus \widetilde{S_{\circ }}^{\prime }\oplus {\cal
\widetilde{S}}_{|}^\diamondsuit ,
$$
$$
................................
$$
Considering the operation of dualization of prime components in
(6.55) we can generate different isomorphic variants of
distinguished $\left(
n,m_1\right) $-spinor spaces. If we add anisotropic ''shalls'' with $%
m_2,...,m_z,$ we have to extend correspondingly spaces (6.55), for instance,%
$$
{\cal S}(_{\circ \circ ,\circ \circ (1)},...,_{\infty
(p)},...,_{\infty (z)})={\bf S_{\circ }\oplus S_{\circ }^{\prime
}\oplus S_{|(1)\circ }\oplus S_{|(1)\circ }^{\prime }\oplus ...}
$$
$$
{\bf \oplus S_{|(p)\circ }\oplus S_{|(p)\circ }^{\prime }\oplus
...\oplus S_{|(z)\circ }\oplus S_{|(z)\circ ,}^{\prime }}
$$
and so on.

We define a d-spinor space ${\cal S}_{(n,m_1)}\ $ as a direct sum
of a horizontal and a vertical spinor spaces of type (6.55), for
instance,
$$
{\cal S}_{(8k,8k^{\prime })}={\bf S}_{\circ }\oplus {\bf
S}_{\circ }^{\prime
}\oplus {\bf S}_{|\circ }\oplus {\bf S}_{|\circ }^{\prime },{\cal S}%
_{(8k,8k^{\prime }+1)}\ ={\bf S}_{\circ }\oplus {\bf S}_{\circ
}^{\prime }\oplus {\cal S}_{|\circ }^{(-)},...,
$$
$$
{\cal S}_{(8k+4,8k^{\prime }+5)}={\bf S}_{\triangle }\oplus {\bf S}%
_{\triangle }^{\prime }\oplus {\cal S}_{|\triangle }^{(-)},...
$$
The scalar product on a ${\cal S}_{(n,m_1)}\ $ is induced by
(corresponding to fixed values of $n$ and $m_1$ ) $\epsilon
$-objects (6.50) considered for
h- and v$_1$-components. We present also an example for ${\cal S}%
_{(n,m_1+...+m_z)}:$%
$$
{\cal S}_{(8k+4,8k_{(1)}+5,...,8k_{(p)}+4,...8k_{(z)})}=
$$
$$
{\bf S}_{\triangle }\oplus {\bf S}_{\triangle }^{\prime }\oplus {\cal S}%
_{|(1)\triangle }^{(-)}\oplus ...\oplus {\bf S}_{|(p)\triangle }\oplus {\bf S%
}_{|(p)\triangle }^{\prime }\oplus ...\oplus {\bf S}_{|(z)\circ }\oplus {\bf %
S}_{|(z)\circ }^{\prime }.
$$

Having introduced d-spinors for dimensions $\left(
n,m_1+...+m_z\right) $ we can write out the generalization for
ha--spaces of twistor equations
 [181] by using the distinguished $\sigma $-objects (6.44):%
\index{Twistor equations!for ha--spaces}
$$
(\sigma _{(<\widehat{\alpha }>})_{|\underline{\beta
}|}^{..\underline{\gamma }}\quad \frac{\delta \omega
^{\underline{\beta }}}{\delta u^{<\widehat{\beta
}>)}}=\frac 1{n+m_1+...+m_z}\quad G_{<\widehat{\alpha }><\widehat{\beta }%
>}(\sigma ^{\widehat{\epsilon }})_{\underline{\beta }}^{..\underline{\gamma }%
}\quad \frac{\delta \omega ^{\underline{\beta }}}{\delta u^{\widehat{%
\epsilon }}},\eqno(6.56)
$$
where $\left| \underline{\beta }\right| $ denotes that we do not
consider symmetrization on this index. The general solution of
(6.56) on the d-vector space ${\cal F}$ looks like as
$$
\omega ^{\underline{\beta }}=\Omega ^{\underline{\beta }}+u^{<\widehat{%
\alpha }>}(\sigma _{<\widehat{\alpha }>})_{\underline{\epsilon }}^{..%
\underline{\beta }}\Pi ^{\underline{\epsilon }},\eqno(6.57)
$$
where $\Omega ^{\underline{\beta }}$ and $\Pi
^{\underline{\epsilon }}$ are constant d-spinors. For fixed
values of dimensions $n$ and $m=m_1+...m_z$ we mast analyze the
reduced and irreducible components of h- and v$_p$-parts of
equations (6.56) and their solutions (6.57) in order to find the
symmetry
properties of a d-twistor ${\bf Z^\alpha \ }$ defined as a pair of d-spinors%
$$
{\bf Z}^\alpha =(\omega ^{\underline{\alpha }},\pi _{\underline{\beta }%
}^{\prime }),
$$
where $\pi _{\underline{\beta }^{\prime }}=\pi _{\underline{\beta
}^{\prime }}^{(0)}\in {\widetilde{{\cal S}}}_{(n,m_1,...,m_z)}$
is a constant dual d-spinor. The problem of definition of spinors
and twistors on ha-spaces was
firstly considered in 
 [275] (see also
 [246,250]) in connection
with the possibility to extend the equations (6.57) and theirs
solutions (6.58), by using nearly autoparallel maps, on curved,
locally isotropic or anisotropic, spaces. In this subsection the
definition of twistors have been extended to higher order
anisotropic spaces with trivial N-- and d--connections.

\subsection{ Mutual transforms of d-tensors and d-spinors}

The spinor algebra for spaces of higher dimensions can not be
considered as a real alternative to the tensor algebra as for
locally isotropic spaces of
dimensions $n=3,4$ 
 [180,181,182]. The same holds true for ha-spaces
and we emphasize that it is not quite convenient to perform a
spinor calculus for dimensions $n,m>>4$. Nevertheless, the
concept of spinors is important for every type of spaces, we can
deeply understand the fundamental properties of geometical
objects on ha-spaces, and we shall consider in this subsection
some questions concerning transforms of d-tensor objects into
d-spinor ones.

\subsubsection{ Transformation of d-tensors into d-spinors}

In order to pass from d-tensors to d-spinors we must use $\sigma
$-objects (6.44) written in reduced or irreduced form \quad (in
dependence of fixed values of dimensions $n$ and $m$ ):

$$
(\sigma _{<\widehat{\alpha }>})_{\underline{\beta }}^{\cdot \underline{%
\gamma }},~(\sigma ^{<\widehat{\alpha }>})^{\underline{\beta }\underline{%
\gamma }},~(\sigma ^{<\widehat{\alpha }>})_{\underline{\beta }\underline{%
\gamma }},...,(\sigma _{<\widehat{a}>})^{\underline{b}\underline{c}%
},...,$$ $$(\sigma _{\widehat{i}})_{\underline{j}\underline{k}},...,(\sigma _{<%
\widehat{a}>})^{AA^{\prime }},...,(\sigma ^{\widehat{i}})_{II^{\prime }},....%
\eqno(6.58)
$$
It is obvious that contracting with corresponding $\sigma
$-objects (6.58) we can introduce instead of d-tensors indices
the d-spinor ones, for
instance,%
$$
\omega ^{\underline{\beta }\underline{\gamma }}=(\sigma ^{<\widehat{\alpha }%
>})^{\underline{\beta }\underline{\gamma }}\omega _{<\widehat{\alpha }%
>},\quad \omega _{AB^{\prime }}=(\sigma ^{<\widehat{a}>})_{AB^{\prime
}}\omega _{<\widehat{a}>},\quad ...,\zeta _{\cdot \underline{j}}^{\underline{%
i}}=(\sigma ^{\widehat{k}})_{\cdot \underline{j}}^{\underline{i}}\zeta _{%
\widehat{k}},....
$$
For d-tensors containing groups of antisymmetric indices there is
a more simple procedure of theirs transforming into d-spinors
because the objects
$$
(\sigma _{\widehat{\alpha }\widehat{\beta }...\widehat{\gamma }})^{%
\underline{\delta }\underline{\nu }},\quad (\sigma ^{\widehat{a}\widehat{b}%
...\widehat{c}})^{\underline{d}\underline{e}},\quad ...,(\sigma ^{\widehat{i}%
\widehat{j}...\widehat{k}})_{II^{\prime }},\quad ...\eqno(6.59)
$$
can be used for sets of such indices into pairs of d-spinor
indices. Let us enumerate some properties of $\sigma $-objects of
type (6.59) (for
simplicity we consider only h-components having q indices $\widehat{i},%
\widehat{j},\widehat{k},...$ taking values from 1 to $n;$ the properties of v%
$_p$-components can be written in a similar manner with respect to indices $%
\widehat{a}_p,\widehat{b}_p,\widehat{c}_p...$ taking values from 1 to $m$):%
$$
(\sigma _{\widehat{i}...\widehat{j}})^{\underline{k}\underline{l}}%
\mbox{
 is\ }\left\{ \
\begin{array}{c}
\mbox{symmetric on }\underline{k},\underline{l}\mbox{ for
}n-2q\equiv
1,7~(mod~8); \\ \mbox{antisymmetric on }\underline{k},\underline{l}%
\mbox{ for }n-2q\equiv 3,5~(mod~8)
\end{array}
\right\} \eqno(6.60)
$$
for odd values of $n,$ and an object
$$
(\sigma _{\widehat{i}...\widehat{j}})^{IJ}~\left( (\sigma _{\widehat{i}...%
\widehat{j}})^{I^{\prime }J^{\prime }}\right)
$$
$$
\mbox{ is\ }\left\{
\begin{array}{c}
\mbox{symmetric on }I,J~(I^{\prime },J^{\prime })\mbox{ for
}n-2q\equiv
0~(mod~8); \\ \mbox{antisymmetric on }I,J~(I^{\prime },J^{\prime })%
\mbox{ for }n-2q\equiv 4~(mod~8)
\end{array}
\right\} \eqno(6.61)
$$
or%
$$
(\sigma _{\widehat{i}...\widehat{j}})^{IJ^{\prime }}=\pm (\sigma _{\widehat{i%
}...\widehat{j}})^{J^{\prime }I}\{
\begin{array}{c}
n+2q\equiv 6(mod8); \\
n+2q\equiv 2(mod8),
\end{array}
\eqno(6.62)
$$
with vanishing of the rest of reduced components of the d-tensor $(\sigma _{%
\widehat{i}...\widehat{j}})^{\underline{k}\underline{l}}$ with
prime/unprime sets of indices.

\subsubsection{ Transformation of d-spinors into d-tensors; fundamental
d-spinors}

We can transform every d-spinor $\xi ^{\underline{\alpha }}=\left( \xi ^{%
\underline{i}},\xi ^{\underline{a}_1},...,\xi
^{\underline{a}_z}\right) $ into a corresponding d-tensor. For
simplicity, we consider this construction only for a h-component
$\xi ^{\underline{i}}$ on a h-space being of
dimension $n$. The values%
$$
\xi ^{\underline{\alpha }}\xi ^{\underline{\beta }}(\sigma ^{\widehat{i}...%
\widehat{j}})_{\underline{\alpha }\underline{\beta }}\quad \left( n%
\mbox{ is odd}\right) \eqno(6.63)
$$
or
$$
\xi ^I\xi ^J(\sigma ^{\widehat{i}...\widehat{j}})_{IJ}~\left(
\mbox{or }\xi
^{I^{\prime }}\xi ^{J^{\prime }}(\sigma ^{\widehat{i}...\widehat{j}%
})_{I^{\prime }J^{\prime }}\right) ~\left( n\mbox{ is even}\right)
\eqno(6.64)
$$
with a different number of indices $\widehat{i}...\widehat{j},$
taken together, defines the h-spinor $\xi ^{\underline{i}}\,$ to
an accuracy to the sign. We emphasize that it is necessary to
choose only those h-components of d-tensors (6.63) (or (6.64))
which are symmetric on pairs of indices $\underline{\alpha
}\underline{\beta }$ (or $IJ\,$ (or $I^{\prime }J^{\prime }$ ))
and the number $q$ of indices $\widehat{i}...\widehat{j}$
satisfies the condition (as a respective consequence of the
properties
(6.60) and/or (6.61), (6.62))%
$$
n-2q\equiv 0,1,7~(mod~8).\eqno(6.65)
$$
Of special interest is the case when
$$
q=\frac 12\left( n\pm 1\right) ~\left( n\mbox{ is odd}\right)
\eqno(6.66)
$$
or
$$
q=\frac 12n~\left( n\mbox{ is even}\right) .\eqno(6.67)
$$
If all expressions (6.63) and/or (6.64) are zero for all values
of $q\,$ with the exception of one or two ones defined by the
conditions (6.65), (6.66) (or (6.67)), the value $\xi
^{\widehat{i}}$ (or $\xi ^I$ ($\xi ^{I^{\prime }}))$ is called a
fundamental h-spinor. Defining in a similar manner the
fundamental v-spinors we can introduce fundamental d-spinors as
pairs of
fundamental h- and v-spinors. Here we remark that a h(v$_p$)-spinor $\xi ^{%
\widehat{i}}~(\xi ^{\widehat{a}_p})\,$ (we can also consider
reduced components) is always a fundamental one for $n(m)<7,$
which is a consequence of (6.67)).

Finally, in this section, we note that the geometry of
fundamental h- and v-spinors is similar to that of usual
fundamental spinors (see Appendix to
the monograph 
 [182]). We omit such details in this work, but
emphasize that constructions with fundamental d-spinors, for a
la-space, must be adapted to the corresponding global splitting
by N-connection of the space.



\section{ Differential Geometry of D--Spi\-nors}

This section is devoted to the differential geometry of
d--spinors in higher order anisotropic spaces.
\index{Differential geometry!of d--spinors} We shall use
denotations of type
$$
v^{<\alpha >}=(v^i,v^{<a>})\in \sigma ^{<\alpha >}=(\sigma
^i,\sigma ^{<a>})
$$
and%
$$
\zeta ^{\underline{\alpha }_p}=(\zeta ^{\underline{i}_p},\zeta ^{\underline{a%
}_p})\in \sigma ^{\alpha _p}=(\sigma ^{i_p},\sigma ^{a_p})\,
$$
for, respectively, elements of modules of d-vector and irreduced
d-spinor
fields (see details in 
 [256]). D-tensors and d-spinor tensors
(irreduced or reduced) will be interpreted as elements of corresponding $%
{\cal \sigma }$--modules, for instance,
$$
q_{~<\beta >...}^{<\alpha >}\in {\cal \sigma ^{<\alpha >}~_{<\beta >....}}%
,\psi _{~\underline{\beta }_p\quad ...}^{\underline{\alpha
}_p\quad
\underline{\gamma }_p}\in {\cal \sigma }_{~\underline{\beta _p}\quad ...}^{%
\underline{\alpha }_p\quad \underline{\gamma }_p}~,\xi _{\quad
J_pK_p^{\prime }N_p^{\prime }}^{I_pI_p^{\prime }}\in {\cal \sigma
}_{\quad J_pK_p^{\prime }N_p^{\prime }}^{I_pI_p^{\prime }}~,...
$$

We can establish a correspondence between the la-adapted metric
$g_{\alpha
\beta }$ (6.12) and d-spinor metric $\epsilon _{\underline{\alpha }%
\underline{\beta }}$ ( $\epsilon $-objects (6.50) for both h- and v$_p$%
-subspaces of ${\cal E}^{<z>}{\cal \,}$ ) of a ha-space ${\cal
E}^{<z>}$ by
using the relation%
$$
g_{<\alpha ><\beta >}=-\frac 1{N(n)+N(m_1)+...+N(m_z)}\times
\eqno(6.68)
$$
$$
((\sigma _{(<\alpha >}(u))^{\underline{\alpha }\underline{\beta
}}(\sigma
_{<\beta >)}(u))^{\underline{\delta }\underline{\gamma }})\epsilon _{%
\underline{\alpha }\underline{\gamma }}\epsilon _{\underline{\beta }%
\underline{\delta }},
$$
where%
$$
(\sigma _{<\alpha >}(u))^{\underline{\nu }\underline{\gamma
}}=l_{<\alpha
>}^{<\widehat{\alpha }>}(u)(\sigma _{<\widehat{\alpha }>})^{<\underline{\nu }%
><\underline{\gamma }>},\eqno(6.69)
$$
which is a consequence of formulas (6.43)-(6.50). In brief we can
write (6.68) as
$$
g_{<\alpha ><\beta >}=\epsilon _{\underline{\alpha }_1\underline{\alpha }%
_2}\epsilon _{\underline{\beta }_1\underline{\beta }_2}\eqno(6.70)
$$
if the $\sigma $-objects are considered as a fixed structure, whereas $%
\epsilon $-objects are treated as caring the metric ''dynamics ''
, on la-space. This variant is used, for instance, in the
so-called 2-spinor
geometry 
 [181,182] and should be preferred if we have to make
explicit the algebraic symmetry properties of d-spinor objects by
using
 metric decomposition (6.70). An
alternative way is to consider as fixed the algebraic structure
of $\epsilon $-objects and to use variable components of $\sigma
$-objects of type (6.69) for developing a variational d-spinor
approach to gravitational and matter field interactions on
ha-spaces ( the spinor Ashtekar variables
 [20]
are introduced in this manner).

We note that a d--spinor metric \index{D--spinor metric}
$$
\epsilon _{\underline{\nu }\underline{\tau }}=\left(
\begin{array}{cccc}
\epsilon _{\underline{i}\underline{j}} & 0 & ... & 0 \\
0 & \epsilon _{\underline{a}_1\underline{b}_1} & ... & 0 \\
... & ... & ... & ... \\
0 & 0 & ... & \epsilon _{\underline{a}_z\underline{b}_z}
\end{array}
\right)
$$
on the d-spinor space ${\cal S}=({\cal S}_{(h)},{\cal S}_{(v_1)},...,{\cal S}%
_{(v_z)})$ can have symmetric or antisymmetric h (v$_p$) -components $%
\epsilon _{\underline{i}\underline{j}}$ ($\epsilon _{\underline{a}_p%
\underline{b}_p})$ , see $\epsilon $-objects (6.50). For
simplicity, in this section (in order to avoid cumbersome
calculations connected with eight-fold periodicity on dimensions
$n$ and $m_p$ of a ha-space ${\cal E}^{<z>}$)
 we shall develop a general d-spinor formalism only by using irreduced
spinor spaces ${\cal S}_{(h)}$ and ${\cal S}_{(v_p)}.$

\subsection{ D-covariant derivation on ha--spaces}

Let ${\cal E}^{<z>}$ be a ha-space. We define the action on a
d-spinor of a
d-covariant operator%
\index{D--covariant operator!for d--spinors}
$$
\nabla _{<\alpha >}=\left( \nabla _i,\nabla _{<a>}\right)
=(\sigma _{<\alpha
>})^{\underline{\alpha }_1\underline{\alpha }_2}\nabla _{^{\underline{\alpha
}_1\underline{\alpha }_2}}=$$ $$\left( (\sigma _i)^{\underline{i}_1\underline{i}%
_2}\nabla _{^{\underline{i}_1\underline{i}_2}},~(\sigma _{<a>})^{\underline{a%
}_1\underline{a}_2}\nabla
_{^{\underline{a}_1\underline{a}_2}}\right) =
$$
$$
\left( (\sigma _i)^{\underline{i}_1\underline{i}_2}\nabla _{^{\underline{i}_1%
\underline{i}_2}},~(\sigma
_{a_1})^{\underline{a}_1\underline{a}_2}\nabla
_{(1)^{\underline{a}_1\underline{a}_2}},...,(\sigma _{a_p})^{\underline{a}_1%
\underline{a}_2}\nabla
_{(p)^{\underline{a}_1\underline{a}_2}},...,(\sigma
_{a_z})^{\underline{a}_1\underline{a}_2}\nabla _{(z)^{\underline{a}_1%
\underline{a}_2}}\right)
$$
(in brief, we shall write
$$
\nabla _{<\alpha >}=\nabla _{^{\underline{\alpha }_1\underline{\alpha }%
_2}}=\left( \nabla _{^{\underline{i}_1\underline{i}_2}},~\nabla _{(1)^{%
\underline{a}_1\underline{a}_2}},...,\nabla _{(p)^{\underline{a}_1\underline{%
a}_2}},...,\nabla _{(z)^{\underline{a}_1\underline{a}_2}}\right) )
$$
as maps
$$
\nabla _{{\underline{\alpha }}_1{\underline{\alpha }}_2}\ :\ {\cal \sigma }^{%
\underline{\beta }}\rightarrow \sigma _{<\alpha >}^{\underline{\beta }%
}=\sigma _{{\underline{\alpha }}_1{\underline{\alpha
}}_2}^{\underline{\beta }}=
$$
$$
\left( \sigma _i^{\underline{\beta }}=\sigma _{{\underline{i}}_1{\underline{i%
}}_2}^{\underline{\beta }},\sigma _{(1)a_1}^{\underline{\beta }}=\sigma _{(1)%
{\underline{\alpha }}_1{\underline{\alpha }}_2}^{\underline{\beta }%
},...,\sigma _{(p)a_p}^{\underline{\beta }}=\sigma _{(p){\underline{\alpha }}%
_1{\underline{\alpha }}_2}^{\underline{\beta }},...,\sigma _{(z)a_z}^{%
\underline{\beta }}=\sigma _{(z){\underline{\alpha }}_1{\underline{\alpha }}%
_2}^{\underline{\beta }}\right)
$$
satisfying conditions%
$$
\nabla _{<\alpha >}(\xi ^{\underline{\beta }}+\eta ^{\underline{\beta }%
})=\nabla _{<\alpha >}\xi ^{\underline{\beta }}+\nabla _{<\alpha >}\eta ^{%
\underline{\beta }},
$$
and%
$$
\nabla _{<\alpha >}(f\xi ^{\underline{\beta }})=f\nabla _{<\alpha >}\xi ^{%
\underline{\beta }}+\xi ^{\underline{\beta }}\nabla _{<\alpha >}f
$$
for every $\xi ^{\underline{\beta }},\eta ^{\underline{\beta }}\in {\cal %
\sigma ^{\underline{\beta }}}$ and $f$ being a scalar field on ${\cal E}%
^{<z>}.$ It is also required that one holds the Leibnitz rule%
$$
(\nabla _{<\alpha >}\zeta _{\underline{\beta }})\eta ^{\underline{\beta }%
}=\nabla _{<\alpha >}(\zeta _{\underline{\beta }}\eta ^{\underline{\beta }%
})-\zeta _{\underline{\beta }}\nabla _{<\alpha >}\eta
^{\underline{\beta }}
$$
and that $\nabla _{<\alpha >}\,$ is a real operator, i.e. it
commuters with
the operation of complex conjugation:%
$$
\overline{\nabla _{<\alpha >}\psi _{\underline{\alpha }\underline{\beta }%
\underline{\gamma }...}}=\nabla _{<\alpha >}(\overline{\psi }_{\underline{%
\alpha }\underline{\beta }\underline{\gamma }...}).
$$

Let now analyze the question on uniqueness of action on d-spinors
of an operator $\nabla _{<\alpha >}$ satisfying necessary
conditions . Denoting by $\nabla _{<\alpha >}^{(1)}$ and $\nabla
_{<\alpha >}$ two such d-covariant
operators we consider the map%
$$
(\nabla _{<\alpha >}^{(1)}-\nabla _{<\alpha >}):{\cal \sigma ^{\underline{%
\beta }}\rightarrow \sigma _{\underline{\alpha }_1\underline{\alpha }_2}^{%
\underline{\beta }}}.\eqno(6.71)
$$
Because the action on a scalar $f$ of both operators $\nabla
_\alpha ^{(1)}$
and $\nabla _\alpha $ must be identical, i.e.%
$$
\nabla _{<\alpha >}^{(1)}f=\nabla _{<\alpha >}f,
$$
the action (6.71) on $f=\omega _{\underline{\beta }}\xi ^{\underline{\beta }%
} $ must be written as
$$
(\nabla _{<\alpha >}^{(1)}-\nabla _{<\alpha >})(\omega _{\underline{\beta }%
}\xi ^{\underline{\beta }})=0.
$$
In consequence we conclude that there is an element $\Theta _{\underline{%
\alpha }_1\underline{\alpha }_2\underline{\beta }}^{\quad \quad \underline{%
\gamma }}\in {\cal \sigma }_{\underline{\alpha }_1\underline{\alpha }_2%
\underline{\beta }}^{\quad \quad \underline{\gamma }}$ for which%
$$
\nabla _{\underline{\alpha }_1\underline{\alpha }_2}^{(1)}\xi ^{\underline{%
\gamma }}=\nabla _{\underline{\alpha }_1\underline{\alpha }_2}\xi ^{%
\underline{\gamma }}+\Theta _{\underline{\alpha }_1\underline{\alpha }_2%
\underline{\beta }}^{\quad \quad \underline{\gamma }}\xi ^{\underline{\beta }%
}\eqno(6.72)
$$
and%
$$
\nabla _{\underline{\alpha }_1\underline{\alpha }_2}^{(1)}\omega _{%
\underline{\beta }}=\nabla _{\underline{\alpha }_1\underline{\alpha }%
_2}\omega _{\underline{\beta }}-\Theta _{\underline{\alpha }_1\underline{%
\alpha }_2\underline{\beta }}^{\quad \quad \underline{\gamma }}\omega _{%
\underline{\gamma }}~.
$$
The action of the operator (6.71) on a d-vector $v^{<\beta >}=v^{\underline{%
\beta }_1\underline{\beta }_2}$ can be written by using formula
(6.72) for
both indices $\underline{\beta }_1$ and $\underline{\beta }_2$ :%
$$
(\nabla _{<\alpha >}^{(1)}-\nabla _{<\alpha >})v^{\underline{\beta }_1%
\underline{\beta }_2}=\Theta _{<\alpha >\underline{\gamma
}}^{\quad \underline{\beta }_1}v^{\underline{\gamma
}\underline{\beta }_2}+\Theta
_{<\alpha >\underline{\gamma }}^{\quad \underline{\beta }_2}v^{\underline{%
\beta }_1\underline{\gamma }}=
$$
$$
(\Theta _{<\alpha >\underline{\gamma }_1}^{\quad \underline{\beta
}_1}\delta
_{\underline{\gamma }_2}^{\quad \underline{\beta }_2}+\Theta _{<\alpha >%
\underline{\gamma }_1}^{\quad \underline{\beta }_2}\delta _{\underline{%
\gamma }_2}^{\quad \underline{\beta }_1})v^{\underline{\gamma }_1\underline{%
\gamma }_2}=Q_{\ <\alpha ><\gamma >}^{<\beta >}v^{<\gamma >},
$$
where%
$$
Q_{\ <\alpha ><\gamma >}^{<\beta >}=Q_{\qquad \underline{\alpha }_1%
\underline{\alpha }_2~\underline{\gamma }_1\underline{\gamma }_2}^{%
\underline{\beta }_1\underline{\beta }_2}=\Theta _{<\alpha >\underline{%
\gamma }_1}^{\quad \underline{\beta }_1}\delta _{\underline{\gamma }%
_2}^{\quad \underline{\beta }_2}+\Theta _{<\alpha >\underline{\gamma }%
_1}^{\quad \underline{\beta }_2}\delta _{\underline{\gamma
}_2}^{\quad \underline{\beta }_1}.\eqno(6.73)
$$
The d-commutator $\nabla _{[<\alpha >}\nabla _{<\beta >]}$
defines the d-torsion (see (6.23)-(6.25) and (6.30)). So,
applying operators $\nabla _{[<\alpha >}^{(1)}\nabla _{<\beta
>]}^{(1)}$ and $\nabla _{[<\alpha
>}\nabla _{<\beta >]}$ on $f=\omega _{\underline{\beta }}\xi ^{\underline{%
\beta }}$ we can write
$$
T_{\quad <\alpha ><\beta >}^{(1)<\gamma >}-T_{~<\alpha ><\beta
>}^{<\gamma
>}=Q_{~<\beta ><\alpha >}^{<\gamma >}-Q_{~<\alpha ><\beta >}^{<\gamma >}
$$
with $Q_{~<\alpha ><\beta >}^{<\gamma >}$ from (6.73).

The action of operator $\nabla _{<\alpha >}^{(1)}$ on d-spinor
tensors of
type $\chi _{\underline{\alpha }_1\underline{\alpha }_2\underline{\alpha }%
_3...}^{\qquad \quad \underline{\beta }_1\underline{\beta
}_2...}$ must be
constructed by using formula (6.72) for every upper index $\underline{\beta }%
_1\underline{\beta }_2...$ and formula (6.73) for every lower index $%
\underline{\alpha }_1\underline{\alpha }_2\underline{\alpha
}_3...$ .

\subsection{Infeld--van der Wa\-er\-den d--co\-ef\-fi\-ci\-ents}

Let \index{Infeld--van der Waerden co\-ef\-fi\-ci\-ents}
$$
\delta _{\underline{{\bf \alpha }}}^{\quad \underline{\alpha
}}=\left(
\delta _{\underline{{\bf 1}}}^{\quad \underline{i}},\delta _{\underline{{\bf %
2}}}^{\quad \underline{i}},...,\delta _{\underline{{\bf
N(n)}}}^{\quad
\underline{i}},\delta _{\underline{{\bf 1}}}^{\quad \underline{a}},\delta _{%
\underline{{\bf 2}}}^{\quad \underline{a}},...,\delta _{\underline{{\bf N(m)}%
}}^{\quad \underline{i}}\right)
$$
be a d--spinor basis. The dual to it basis is denoted as
\index{D--spinor basis}
$$
\delta _{\underline{\alpha }}^{\quad \underline{{\bf \alpha
}}}=\left(
\delta _{\underline{i}}^{\quad \underline{{\bf 1}}},\delta _{\underline{i}%
}^{\quad \underline{{\bf 2}}},...,\delta _{\underline{i}}^{\quad \underline{%
{\bf N(n)}}},\delta _{\underline{i}}^{\quad \underline{{\bf 1}}},\delta _{%
\underline{i}}^{\quad \underline{{\bf 2}}},...,\delta _{\underline{i}%
}^{\quad \underline{{\bf N(m)}}}\right) .
$$
A d-spinor $\kappa ^{\underline{\alpha }}\in {\cal \sigma }$ $^{\underline{%
\alpha }}$ has components $\kappa ^{\underline{{\bf \alpha }}}=\kappa ^{%
\underline{\alpha }}\delta _{\underline{\alpha }}^{\quad \underline{{\bf %
\alpha }}}.$ Taking into account that
$$
\delta _{\underline{{\bf \alpha }}}^{\quad \underline{\alpha }}\delta _{%
\underline{{\bf \beta }}}^{\quad \underline{\beta }}\nabla _{\underline{%
\alpha }\underline{\beta }}=\nabla _{\underline{{\bf \alpha }}\underline{%
{\bf \beta }}},
$$
we write out the components $\nabla _{\underline{\alpha
}\underline{\beta }}$
$\kappa ^{\underline{\gamma }}$ as%
$$
\delta _{\underline{{\bf \alpha }}}^{\quad \underline{\alpha }}~\delta _{%
\underline{{\bf \beta }}}^{\quad \underline{\beta }}~\delta _{\underline{%
\gamma }}^{\quad \underline{{\bf \gamma }}}~\nabla _{\underline{\alpha }%
\underline{\beta }}\kappa ^{\underline{\gamma }}=\delta _{\underline{{\bf %
\epsilon }}}^{\quad \underline{\tau }}~\delta _{\underline{\tau
}}^{\quad
\underline{{\bf \gamma }}}~\nabla _{\underline{{\bf \alpha }}\underline{{\bf %
\beta }}}\kappa ^{\underline{{\bf \epsilon }}}+\kappa ^{\underline{{\bf %
\epsilon }}}~\delta _{\underline{\epsilon }}^{\quad \underline{{\bf \gamma }}%
}~\nabla _{\underline{{\bf \alpha }}\underline{{\bf \beta }}}\delta _{%
\underline{{\bf \epsilon }}}^{\quad \underline{\epsilon }}=
$$
$$
\nabla _{\underline{{\bf \alpha }}\underline{{\bf \beta }}}\kappa ^{%
\underline{{\bf \gamma }}}+\kappa ^{\underline{{\bf \epsilon }}}\gamma _{~%
\underline{{\bf \alpha }}\underline{{\bf \beta }}\underline{{\bf \epsilon }}%
}^{\underline{{\bf \gamma }}},\eqno(6.74)
$$
where the coordinate components of the d--spinor connection $\gamma _{~%
\underline{{\bf \alpha }}\underline{{\bf \beta }}\underline{{\bf \epsilon }}%
}^{\underline{{\bf \gamma }}}$ are defined as \index{D--spinor
connection} \index{Connection!d--spinor}
$$
\gamma _{~\underline{{\bf \alpha }}\underline{{\bf \beta }}\underline{{\bf %
\epsilon }}}^{\underline{{\bf \gamma }}}\doteq \delta _{\underline{\tau }%
}^{\quad \underline{{\bf \gamma }}}~\nabla _{\underline{{\bf \alpha }}%
\underline{{\bf \beta }}}\delta _{\underline{{\bf \epsilon
}}}^{\quad \underline{\tau }}.\eqno(6.75)
$$
We call the Infeld - van der Waerden d-symbols a set of $\sigma $-objects ($%
\sigma _{{\bf \alpha }})^{\underline{{\bf \alpha
}}\underline{{\bf \beta }}}$ paramet\-ri\-zed with respect to a
coordinate d-spinor basis. Defining
$$
\nabla _{<{\bf \alpha >}}=(\sigma _{<{\bf \alpha >}})^{\underline{{\bf %
\alpha }}\underline{{\bf \beta }}}~\nabla _{\underline{{\bf \alpha }}%
\underline{{\bf \beta }}},
$$
introducing denotations
$$
\gamma ^{\underline{{\bf \gamma }}}{}_{<{\bf \alpha >\underline{\tau }}%
}\doteq \gamma ^{\underline{{\bf \gamma }}}{}_{{\bf \underline{\alpha }%
\underline{\beta }\underline{\tau }}}~(\sigma _{<{\bf \alpha >}})^{%
\underline{{\bf \alpha }}\underline{{\bf \beta }}}
$$
and using properties (6.74) we can write relations%
$$
l_{<{\bf \alpha >}}^{<\alpha >}~\delta _{\underline{\beta
}}^{\quad
\underline{{\bf \beta }}}~\nabla _{<\alpha >}\kappa ^{\underline{\beta }%
}=\nabla _{<{\bf \alpha >}}\kappa ^{\underline{{\bf \beta }}}+\kappa ^{%
\underline{{\bf \delta }}}~\gamma _{~<{\bf \alpha >}\underline{{\bf \delta }}%
}^{\underline{{\bf \beta }}}\eqno(6.76)
$$
and%
$$
l_{<{\bf \alpha >}}^{<\alpha >}~\delta _{\underline{{\bf \beta
}}}^{\quad
\underline{\beta }}~\nabla _{<\alpha >}~\mu _{\underline{\beta }}=\nabla _{<%
{\bf \alpha >}}~\mu _{\underline{{\bf \beta }}}-\mu
_{\underline{{\bf \delta
}}}\gamma _{~<{\bf \alpha >}\underline{{\bf \beta }}}^{\underline{{\bf %
\delta }}}\eqno(6.77)
$$
for d-covariant derivations $~\nabla _{\underline{\alpha }}\kappa ^{%
\underline{\beta }}$ and $\nabla _{\underline{\alpha }}~\mu _{\underline{%
\beta }}.$

We can consider expressions similar to (6.76) and (6.77) for
values having
both types of d-spinor and d-tensor indices, for instance,%
$$
l_{<{\bf \alpha >}}^{<\alpha >}~l_{<\gamma >}^{<{\bf \gamma >}}~\delta _{%
\underline{{\bf \delta }}}^{\quad \underline{\delta }}~\nabla
_{<\alpha
>}\theta _{\underline{\delta }}^{~<\gamma >}=\nabla _{<{\bf \alpha >}}\theta
_{\underline{{\bf \delta }}}^{~<{\bf \gamma >}}-\theta _{\underline{{\bf %
\epsilon }}}^{~<{\bf \gamma >}}\gamma _{~<{\bf \alpha >}\underline{{\bf %
\delta }}}^{\underline{{\bf \epsilon }}}+\theta _{\underline{{\bf \delta }}%
}^{~<{\bf \tau >}}~\Gamma _{\quad <{\bf \alpha ><\tau >}}^{~<{\bf
\gamma >}}
$$
(we can prove this by a straightforward calculation).

Now we shall consider some possible relations between components
of
d-connec\-ti\-ons $\gamma _{~<{\bf \alpha >}\underline{{\bf \delta }}}^{%
\underline{{\bf \epsilon }}}$ and $\Gamma _{\quad <{\bf \alpha ><\tau >}}^{~<%
{\bf \gamma >}}$ and derivations of $(\sigma _{<{\bf \alpha >}})^{\underline{%
{\bf \alpha }}\underline{{\bf \beta }}}$ . According to
definitions (6.12)
we can write%
$$
\Gamma _{~<{\bf \beta ><\gamma >}}^{<{\bf \alpha >}}=l_{<\alpha >}^{<{\bf %
\alpha >}}\nabla _{<{\bf \gamma >}}l_{<{\bf \beta >}}^{<\alpha >}=
$$
$$
l_{<\alpha >}^{<{\bf \alpha >}}\nabla _{<{\bf \gamma >}}(\sigma _{<{\bf %
\beta >}})^{\underline{\epsilon }\underline{\tau }}=l_{<\alpha >}^{<{\bf %
\alpha >}}\nabla _{<{\bf \gamma >}}((\sigma _{<{\bf \beta >}})^{\underline{%
{\bf \epsilon }}\underline{{\bf \tau }}}\delta _{\underline{{\bf \epsilon }}%
}^{~\underline{\epsilon }}\delta _{\underline{{\bf \tau }}}^{~\underline{%
\tau }})=
$$
$$
l_{<\alpha >}^{<{\bf \alpha >}}\delta _{\underline{{\bf \alpha }}}^{~%
\underline{\alpha }}\delta _{\underline{{\bf \epsilon }}}^{~\underline{%
\epsilon }}\nabla _{<{\bf \gamma >}}(\sigma _{<{\bf \beta >}})^{\underline{%
{\bf \alpha }}\underline{{\bf \epsilon }}}+l_{<\alpha >}^{<{\bf \alpha >}%
}(\sigma _{<{\bf \beta >}})^{\underline{{\bf \epsilon
}}\underline{{\bf \tau
}}}(\delta _{\underline{{\bf \tau }}}^{~\underline{\tau }}\nabla _{<{\bf %
\gamma >}}\delta _{\underline{{\bf \epsilon }}}^{~\underline{\epsilon }%
}+\delta _{\underline{{\bf \epsilon }}}^{~\underline{\epsilon }}\nabla _{<%
{\bf \gamma >}}\delta _{\underline{{\bf \tau
}}}^{~\underline{\tau }})=
$$
$$
l_{\underline{{\bf \epsilon }}\underline{{\bf \tau }}}^{<{\bf \alpha >}%
}~\nabla _{<{\bf \gamma >}}(\sigma _{<{\bf \beta >}})^{\underline{{\bf %
\epsilon }}\underline{{\bf \tau }}}+l_{\underline{{\bf \mu }}\underline{{\bf %
\nu }}}^{<{\bf \alpha >}}\delta _{\underline{\epsilon }}^{~\underline{{\bf %
\mu }}}\delta _{\underline{\tau }}^{~\underline{{\bf \nu }}}(\sigma _{<{\bf %
\beta >}})^{\underline{\epsilon }\underline{\tau }}(\delta _{\underline{{\bf %
\tau }}}^{~\underline{\tau }}\nabla _{<{\bf \gamma >}}\delta _{\underline{%
{\bf \epsilon }}}^{~\underline{\epsilon }}+\delta
_{\underline{{\bf \epsilon
}}}^{~\underline{\epsilon }}\nabla _{<{\bf \gamma >}}\delta _{\underline{%
{\bf \tau }}}^{~\underline{\tau }}),
$$
where $l_{<\alpha >}^{<{\bf \alpha >}}=(\sigma _{\underline{{\bf \epsilon }}%
\underline{{\bf \tau }}})^{<{\bf \alpha >}}$ , from which it follows%
$$
(\sigma _{<{\bf \alpha >}})^{\underline{{\bf \mu }}\underline{{\bf \nu }}%
}(\sigma _{\underline{{\bf \alpha }}\underline{{\bf \beta }}})^{<{\bf \beta >%
}}\Gamma _{~<{\bf \gamma ><\beta >}}^{<{\bf \alpha >}}=$$
 $$(\sigma _{\underline{%
{\bf \alpha }}\underline{{\bf \beta }}})^{<{\bf \beta >}}\nabla _{<{\bf %
\gamma >}}(\sigma _{<{\bf \alpha >}})^{\underline{{\bf \mu }}\underline{{\bf %
\nu }}}+\delta _{\underline{{\bf \beta }}}^{~\underline{{\bf \nu
}}}\gamma
_{~<{\bf \gamma >\underline{\alpha }}}^{\underline{{\bf \mu }}}+\delta _{%
\underline{{\bf \alpha }}}^{~\underline{{\bf \mu }}}\gamma _{~<{\bf \gamma >%
\underline{\beta }}}^{\underline{{\bf \nu }}}.
$$
Connecting the last expression on \underline{${\bf \beta }$} and \underline{$%
{\bf \nu }$} and using an orthonormalized d-spinor basis when $$\gamma _{~<%
{\bf \gamma >\underline{\beta }}}^{\underline{{\bf \beta }}}=0$$
(a consequence from (6.75)) we have
$$\gamma _{~<{\bf \gamma >\underline{\alpha }}}^{\underline{{\bf \mu }}}=
$$ $$\frac
1{N(n)+N(m_1)+...+N(m_z)}(\Gamma _{\quad <{\bf \gamma >~\underline{\alpha }%
\underline{\beta }}}^{\underline{{\bf \mu }}\underline{{\bf \beta }}%
}-(\sigma _{\underline{{\bf \alpha }}\underline{{\bf \beta
}}})^{<{\bf \beta
>}}\nabla _{<{\bf \gamma >}}(\sigma _{<{\bf \beta >}})^{\underline{{\bf \mu }%
}\underline{{\bf \beta }}}),\eqno(6.78)
$$
where
$$
\Gamma _{\quad <{\bf \gamma >~\underline{\alpha }\underline{\beta }}}^{%
\underline{{\bf \mu }}\underline{{\bf \beta }}}=(\sigma _{<{\bf \alpha >}})^{%
\underline{{\bf \mu }}\underline{{\bf \beta }}}(\sigma _{\underline{{\bf %
\alpha }}\underline{{\bf \beta }}})^{{\bf \beta }}\Gamma _{~<{\bf
\gamma
><\beta >}}^{<{\bf \alpha >}}.\eqno(6.79)
$$
We also note here that, for instance, for the canonical (see
(6.18) and (6.19)) and Berwald (see (6.20)) connections,
Christoffel d-symbols (see (6.21)) we can express d-spinor
connection (6.79) through corresponding locally adapted
derivations of components of metric and N-connection by
introducing corresponding coefficients instead of $\Gamma
_{~<{\bf \gamma
><\beta >}}^{<{\bf \alpha >}}$ in (6.79) and than in (6.78).

\subsection{ D--spinors of ha--space curvature and torsion}

\index{D--spinor!ha--space curvature} \index{D--spinor!ha--space
torsion} The d-tensor indices of the commutator (6.29), $\Delta
_{<\alpha ><\beta >},$
can be transformed into d-spinor ones:%
$$
\Box _{\underline{\alpha }\underline{\beta }}=(\sigma ^{<\alpha ><\beta >})_{%
\underline{\alpha }\underline{\beta }}\Delta _{\alpha \beta }=(\Box _{%
\underline{i}\underline{j}},\Box
_{\underline{a}\underline{b}})=\eqno(6.80)
$$
$$
(\Box _{\underline{i}\underline{j}},\Box _{\underline{a}_1\underline{b}%
_1},...,\Box _{\underline{a}_p\underline{b}_p},...,\Box _{\underline{a}_z%
\underline{b}_z}),
$$
with h- and v$_p$-components,
$$
\Box _{\underline{i}\underline{j}}=(\sigma ^{<\alpha ><\beta >})_{\underline{%
i}\underline{j}}\Delta _{<\alpha ><\beta >}\mbox{ and }\Box _{\underline{a}%
\underline{b}}=(\sigma ^{<\alpha ><\beta >})_{\underline{a}\underline{b}%
}\Delta _{<\alpha ><\beta >},
$$
being symmetric or antisymmetric in dependence of corresponding
values of dimensions $n\,$ and $m_p$ (see eight-fold
parametizations (6.50) and
(6.51)). Considering the actions of operator (6.80) on d-spinors $\pi ^{%
\underline{\gamma }}$ and $\mu _{\underline{\gamma }}$ we
introduce the
d-spinor curvature $X_{\underline{\delta }\quad \underline{\alpha }%
\underline{\beta }}^{\quad \underline{\gamma }}\,$ as to satisfy equations%
$$
\Box _{\underline{\alpha }\underline{\beta }}\ \pi ^{\underline{\gamma }}=X_{%
\underline{\delta }\quad \underline{\alpha }\underline{\beta
}}^{\quad \underline{\gamma }}\pi ^{\underline{\delta
}}\eqno(6.81)
$$
and%
$$
\Box _{\underline{\alpha }\underline{\beta }}\ \mu _{\underline{\gamma }}=X_{%
\underline{\gamma }\quad \underline{\alpha }\underline{\beta
}}^{\quad \underline{\delta }}\mu _{\underline{\delta }}.
$$
The gravitational d-spinor $\Psi _{\underline{\alpha }\underline{\beta }%
\underline{\gamma }\underline{\delta }}$ is defined by a
corresponding
symmetrization of d-spinor indices:%
$$
\Psi _{\underline{\alpha }\underline{\beta }\underline{\gamma }\underline{%
\delta }}=X_{(\underline{\alpha }|\underline{\beta }|\underline{\gamma }%
\underline{\delta })}.\eqno(6.82)
$$
We note that d-spinor tensors $X_{\underline{\delta }\quad
\underline{\alpha
}\underline{\beta }}^{\quad \underline{\gamma }}$ and $\Psi _{\underline{%
\alpha }\underline{\beta }\underline{\gamma }\underline{\delta
}}\,$ are transformed into similar 2-spinor objects on locally
isotropic spaces
 [181,182] if we consider vanishing of the N-connection structure and a
limit to a locally isotropic space.

Putting $\delta _{\underline{\gamma }}^{\quad {\bf
\underline{\gamma }}}$ instead of $\mu _{\underline{\gamma }}$ in
(6.81) and using (6.82) we can express respectively the curvature
and gravitational d-spinors as
$$
X_{\underline{\gamma }\underline{\delta }\underline{\alpha
}\underline{\beta
}}=\delta _{\underline{\delta }\underline{{\bf \tau }}}\Box _{\underline{%
\alpha }\underline{\beta }}\delta _{\underline{\gamma }}^{\quad
{\bf \underline{\tau }}}
$$
and%
$$
\Psi _{\underline{\gamma }\underline{\delta }\underline{\alpha }\underline{%
\beta }}=\delta _{\underline{\delta }\underline{{\bf \tau }}}\Box _{(%
\underline{\alpha }\underline{\beta }}\delta _{\underline{\gamma
})}^{\quad {\bf \underline{\tau }}}.
$$

The d-spinor torsion $T_{\qquad \underline{\alpha }\underline{\beta }}^{%
\underline{\gamma }_1\underline{\gamma }_2}$ is defined similarly
as for d-tensors (see (6.30)) by using the d-spinor commutator
(6.80) and equations
$$
\Box _{\underline{\alpha }\underline{\beta }}f=T_{\qquad \underline{\alpha }%
\underline{\beta }}^{\underline{\gamma }_1\underline{\gamma }_2}\nabla _{%
\underline{\gamma }_1\underline{\gamma }_2}f.\eqno(6.82)
$$

The d-spinor components $R_{\underline{\gamma
}_1\underline{\gamma }_2\qquad
\underline{\alpha }\underline{\beta }}^{\qquad \underline{\delta }_1%
\underline{\delta }_2}$ of the curvature d-tensor $R_{\gamma
\quad \alpha \beta }^{\quad \delta }$ can be computed by using
relations (6.79), and (6.80) and (6.82) as to satisfy the
equations (the d-spinor analogous of
equations (6.31) )%
$$
(\Box _{\underline{\alpha }\underline{\beta }}-T_{\qquad \underline{\alpha }%
\underline{\beta }}^{\underline{\gamma }_1\underline{\gamma }_2}\nabla _{%
\underline{\gamma }_1\underline{\gamma }_2})V^{\underline{\delta }_1%
\underline{\delta }_2}=R_{\underline{\gamma }_1\underline{\gamma
}_2\qquad
\underline{\alpha }\underline{\beta }}^{\qquad \underline{\delta }_1%
\underline{\delta }_2}V^{\underline{\gamma }_1\underline{\gamma }_2},%
$$
here d-vector $V^{\underline{\gamma }_1\underline{\gamma }_2}$ is
considered
as a product of d-spinors, i.e. $V^{\underline{\gamma }_1\underline{\gamma }%
_2}=\nu ^{\underline{\gamma }_1}\mu ^{\underline{\gamma }_2}$. We
find

$$
R_{\underline{\gamma }_1\underline{\gamma }_2\qquad \underline{\alpha }%
\underline{\beta }}^{\qquad \underline{\delta }_1\underline{\delta }%
_2}=\left( X_{\underline{\gamma }_1~\underline{\alpha }\underline{\beta }%
}^{\quad \underline{\delta }_1}+T_{\qquad \underline{\alpha }\underline{%
\beta }}^{\underline{\tau }_1\underline{\tau }_2}\quad \gamma
_{\quad
\underline{\tau }_1\underline{\tau }_2\underline{\gamma }_1}^{\underline{%
\delta }_1}\right) \delta _{\underline{\gamma }_2}^{\quad \underline{\delta }%
_2}+
$$
$$
\left( X_{\underline{\gamma }_2~\underline{\alpha }\underline{\beta }%
}^{\quad \underline{\delta }_2}+T_{\qquad \underline{\alpha }\underline{%
\beta }}^{\underline{\tau }_1\underline{\tau }_2}\quad \gamma
_{\quad
\underline{\tau }_1\underline{\tau }_2\underline{\gamma }_2}^{\underline{%
\delta }_2}\right) \delta _{\underline{\gamma }_1}^{\quad \underline{\delta }%
_1}.\eqno(6.83)
$$

It is convenient to use this d-spinor expression for the
curvature d-tensor
$$
R_{\underline{\gamma }_1\underline{\gamma }_2\qquad \underline{\alpha }_1%
\underline{\alpha }_2\underline{\beta }_1\underline{\beta
}_2}^{\qquad
\underline{\delta }_1\underline{\delta }_2}=\left( X_{\underline{\gamma }_1~%
\underline{\alpha }_1\underline{\alpha }_2\underline{\beta }_1\underline{%
\beta }_2}^{\quad \underline{\delta }_1}+T_{\qquad \underline{\alpha }_1%
\underline{\alpha }_2\underline{\beta }_1\underline{\beta }_2}^{\underline{%
\tau }_1\underline{\tau }_2}~\gamma _{\quad \underline{\tau }_1\underline{%
\tau }_2\underline{\gamma }_1}^{\underline{\delta }_1}\right) \delta _{%
\underline{\gamma }_2}^{\quad \underline{\delta }_2}+
$$
$$
\left( X_{\underline{\gamma }_2~\underline{\alpha }_1\underline{\alpha }_2%
\underline{\beta }_1\underline{\beta }_2}^{\quad \underline{\delta }%
_2}+T_{\qquad \underline{\alpha }_1\underline{\alpha }_2\underline{\beta }_1%
\underline{\beta }_2~}^{\underline{\tau }_1\underline{\tau
}_2}\gamma
_{\quad \underline{\tau }_1\underline{\tau }_2\underline{\gamma }_2}^{%
\underline{\delta }_2}\right) \delta _{\underline{\gamma
}_1}^{\quad \underline{\delta }_1}
$$
in order to get the d--spinor components of the Ricci d-tensor%
\index{D--spinor!Ricci} \index{Ricci d--spinor}
$$
R_{\underline{\gamma }_1\underline{\gamma }_2\underline{\alpha }_1\underline{%
\alpha }_2}=R_{\underline{\gamma }_1\underline{\gamma }_2\qquad \underline{%
\alpha }_1\underline{\alpha }_2\underline{\delta }_1\underline{\delta }%
_2}^{\qquad \underline{\delta }_1\underline{\delta }_2}=
$$
$$
X_{\underline{\gamma }_1~\underline{\alpha }_1\underline{\alpha }_2%
\underline{\delta }_1\underline{\gamma }_2}^{\quad \underline{\delta }%
_1}+T_{\qquad \underline{\alpha }_1\underline{\alpha }_2\underline{\delta }_1%
\underline{\gamma }_2}^{\underline{\tau }_1\underline{\tau
}_2}~\gamma
_{\quad \underline{\tau }_1\underline{\tau }_2\underline{\gamma }_1}^{%
\underline{\delta }_1}+X_{\underline{\gamma }_2~\underline{\alpha }_1%
\underline{\alpha }_2\underline{\delta }_1\underline{\gamma
}_2}^{\quad
\underline{\delta }_2}+T_{\qquad \underline{\alpha }_1\underline{\alpha }_2%
\underline{\gamma }_1\underline{\delta }_2~}^{\underline{\tau }_1\underline{%
\tau }_2}\gamma _{\quad \underline{\tau }_1\underline{\tau }_2\underline{%
\gamma }_2}^{\underline{\delta }_2}\eqno(6.84)
$$
and this d-spinor decomposition of the scalar curvature:%
$$
q\overleftarrow{R}=R_{\qquad \underline{\alpha }_1\underline{\alpha }_2}^{%
\underline{\alpha }_1\underline{\alpha }_2}=X_{\quad ~\underline{~\alpha }%
_1\quad \underline{\delta }_1\underline{\alpha }_2}^{\underline{\alpha }_1%
\underline{\delta }_1~~\underline{\alpha }_2}+T_{\qquad ~~\underline{\alpha }%
_2\underline{\delta }_1}^{\underline{\tau }_1\underline{\tau }_2\underline{%
\alpha }_1\quad ~\underline{\alpha }_2}~\gamma _{\quad \underline{\tau }_1%
\underline{\tau }_2\underline{\alpha }_1}^{\underline{\delta }_1}+
$$
$$
X_{\qquad \quad \underline{\alpha }_2\underline{\delta }_2\underline{\alpha }%
_1}^{\underline{\alpha }_2\underline{\delta }_2\underline{\alpha }%
_1}+T_{\qquad \underline{\alpha }_1\quad ~\underline{\delta }_2~}^{%
\underline{\tau }_1\underline{\tau }_2~~\underline{\alpha }_2\underline{%
\alpha }_1}\gamma _{\quad \underline{\tau }_1\underline{\tau }_2\underline{%
\alpha }_2}^{\underline{\delta }_2}.\eqno(6.85)
$$

Putting (6.84) and (6.85) into (6.34) and, correspondingly,
(6.35) we find the d--spinor components of the Einstein and $\Phi
_{<\alpha ><\beta >}$ \index{Einstein!d--spinor}
\index{D--spinor!Einstein} \index{D--spinor!$\Phi _{<\alpha
><\beta >}$}
d-tensors:%
$$
\overleftarrow{G}_{<\gamma ><\alpha >}=\overleftarrow{G}_{\underline{\gamma }%
_1\underline{\gamma }_2\underline{\alpha }_1\underline{\alpha }_2}=X_{%
\underline{\gamma }_1~\underline{\alpha }_1\underline{\alpha }_2\underline{%
\delta }_1\underline{\gamma }_2}^{\quad \underline{\delta
}_1}+T_{\qquad
\underline{\alpha }_1\underline{\alpha }_2\underline{\delta }_1\underline{%
\gamma }_2}^{\underline{\tau }_1\underline{\tau }_2}~\gamma
_{\quad
\underline{\tau }_1\underline{\tau }_2\underline{\gamma }_1}^{\underline{%
\delta }_1}+
$$
$$
X_{\underline{\gamma }_2~\underline{\alpha }_1\underline{\alpha }_2%
\underline{\delta }_1\underline{\gamma }_2}^{\quad \underline{\delta }%
_2}+T_{\qquad \underline{\alpha }_1\underline{\alpha }_2\underline{\gamma }_1%
\underline{\delta }_2~}^{\underline{\tau }_1\underline{\tau
}_2}\gamma
_{\quad \underline{\tau }_1\underline{\tau }_2\underline{\gamma }_2}^{%
\underline{\delta }_2}-
$$
$$
\frac 12\varepsilon _{\underline{\gamma }_1\underline{\alpha
}_1}\varepsilon
_{\underline{\gamma }_2\underline{\alpha }_2}[X_{\quad ~\underline{~\beta }%
_1\quad \underline{\mu }_1\underline{\beta }_2}^{\underline{\beta }_1%
\underline{\mu }_1~~\underline{\beta }_2}+T_{\qquad ~~\underline{\beta }_2%
\underline{\mu }_1}^{\underline{\tau }_1\underline{\tau }_2\underline{\beta }%
_1\quad ~\underline{\beta }_2}~\gamma _{\quad \underline{\tau }_1\underline{%
\tau }_2\underline{\beta }_1}^{\underline{\mu }_1}+
$$
$$
X_{\qquad \quad \underline{\beta }_2\underline{\mu }_2\underline{\nu }_1}^{%
\underline{\beta }_2\underline{\mu }_2\underline{\nu
}_1}+T_{\qquad
\underline{\beta }_1\quad ~\underline{\delta }_2~}^{\underline{\tau }_1%
\underline{\tau }_2~~\underline{\beta }_2\underline{\beta
}_1}\gamma _{\quad
\underline{\tau }_1\underline{\tau }_2\underline{\beta }_2}^{\underline{%
\delta }_2}]\eqno(6.86)
$$
and%
$$
\Phi _{<\gamma ><\alpha >}=\Phi _{\underline{\gamma }_1\underline{\gamma }_2%
\underline{\alpha }_1\underline{\alpha }_2}=\frac
1{2(n+m_1+...+m_z)}\varepsilon _{\underline{\gamma }_1\underline{\alpha }%
_1}\varepsilon _{\underline{\gamma }_2\underline{\alpha }_2}[X_{\quad ~%
\underline{~\beta }_1\quad \underline{\mu }_1\underline{\beta }_2}^{%
\underline{\beta }_1\underline{\mu }_1~~\underline{\beta }_2}+
$$
$$
T_{\qquad ~~\underline{\beta }_2\underline{\mu }_1}^{\underline{\tau }_1%
\underline{\tau }_2\underline{\beta }_1\quad ~\underline{\beta
}_2}~\gamma
_{\quad \underline{\tau }_1\underline{\tau }_2\underline{\beta }_1}^{%
\underline{\mu }_1}+X_{\qquad \quad \underline{\beta }_2\underline{\mu }_2%
\underline{\nu }_1}^{\underline{\beta }_2\underline{\mu }_2\underline{\nu }%
_1}+T_{\qquad \underline{\beta }_1\quad ~\underline{\delta }_2~}^{\underline{%
\tau }_1\underline{\tau }_2~~\underline{\beta }_2\underline{\beta
}_1}\gamma
_{\quad \underline{\tau }_1\underline{\tau }_2\underline{\beta }_2}^{%
\underline{\delta }_2}]-
$$
$$
\frac 12[X_{\underline{\gamma }_1~\underline{\alpha }_1\underline{\alpha }_2%
\underline{\delta }_1\underline{\gamma }_2}^{\quad \underline{\delta }%
_1}+T_{\qquad \underline{\alpha }_1\underline{\alpha }_2\underline{\delta }_1%
\underline{\gamma }_2}^{\underline{\tau }_1\underline{\tau
}_2}~\gamma
_{\quad \underline{\tau }_1\underline{\tau }_2\underline{\gamma }_1}^{%
\underline{\delta }_1}+
$$
$$
X_{\underline{\gamma }_2~\underline{\alpha }_1\underline{\alpha }_2%
\underline{\delta }_1\underline{\gamma }_2}^{\quad \underline{\delta }%
_2}+T_{\qquad \underline{\alpha }_1\underline{\alpha }_2\underline{\gamma }_1%
\underline{\delta }_2~}^{\underline{\tau }_1\underline{\tau
}_2}\gamma
_{\quad \underline{\tau }_1\underline{\tau }_2\underline{\gamma }_2}^{%
\underline{\delta }_2}].\eqno(6.87)
$$

The components of the conformal Weyl d-spinor can be computed by
putting d-spinor values of the curvature (6.83) and Ricci (6.84)
d-tensors into corresponding expression for the d-tensor (6.33).
We omit this calculus in this work.

\section{ Field Equations on Ha-Spaces}

The problem of formulation gravitational and gauge field
equations on different types of la-spaces is considered, for
instance, in
 [161,29,18]
 and
 [272]. In this section we shall introduce the basic
field equations for gravitational and matter field
la-interactions in a generalized form for generic higher order
anisotropic spaces.

\subsection{ Locally anisotropic scalar field interactions}

Let $\varphi \left( u\right) =(\varphi _1\left( u\right) ,\varphi
_2\left( u\right) \dot ,...,\varphi _k\left( u\right) )$ be a
complex k-component scalar field of mass $\mu $ on ha-space
${\cal E}^{<z>}{\cal .}$ The \index{Scalar field
interactions!locally anisotropic} d-covariant generalization of
the conformally invariant (in the massless
case) scalar field equation 
 [181,182] can be defined by using the
d'Alambert locally anisotropic operator
 [11,262] $\Box =D^{<\alpha
>}D_{<\alpha >}$, where $D_{<\alpha >}$ is a d-covariant derivation on $%
{\cal E}^{<z>}$ satisfying conditions (6.14) and (6.15) and
constructed, for simplicity, by using Christoffel d--symbols
(6.21) (all formulas for field equations and conservation values
can be deformed by using corresponding deformations d--tensors
$P_{<\beta ><\gamma >}^{<\alpha >}$ from the Cristoffel
d--symbols, or the canonical d--connection to a general
d-connection into consideration):

$$
(\Box +\frac{n_E-2}{4(n_E-1)}\overleftarrow{R}+\mu ^2)\varphi
\left( u\right) =0,\eqno(6.88)
$$
where $n_E=n+m_1+...+m_z.$We must change d-covariant derivation
$D_{<\alpha
>}$ into $^{\diamond }D_{<\alpha >}=D_{<\alpha >}+ieA_{<\alpha >}$ and take
into account the d-vector current
$$
J_{<\alpha >}^{(0)}\left( u\right) =i(\left( \overline{\varphi
}\left(
u\right) D_{<\alpha >}\varphi \left( u\right) -D_{<\alpha >}\overline{%
\varphi }\left( u\right) )\varphi \left( u\right) \right)
$$
if interactions between locally anisotropic electromagnetic field
( d-vector potential $A_{<\alpha >}$ ), where $e$ is the
electromagnetic constant, and charged scalar field $\varphi $ are
considered. The equations (6.88) are (locally adapted to the
N-connection structure) Euler equations for the
Lagrangian%
$${\cal L}^{(0)}\left( u\right) =$$ $$
\sqrt{|g|}\left[ g^{<\alpha ><\beta >}\delta _{<\alpha
>}\overline{\varphi }\left( u\right) \delta _{<\beta >}\varphi
\left( u\right) -\left( \mu ^2+\frac{n_E-2}{4(n_E-1)}\right) \overline{%
\varphi }\left( u\right) \varphi \left( u\right) \right]
,\eqno(6.89)
$$
where $|g|=detg_{<\alpha ><\beta >}.$

The locally adapted variations of the action with Lagrangian
(6.89) on variables $\varphi \left( u\right) $ and
$\overline{\varphi }\left( u\right) $ leads to the locally
anisotropic generalization of the energy-momentum
tensor,%
$$
E_{<\alpha ><\beta >}^{(0,can)}\left( u\right) =\delta _{<\alpha >}\overline{%
\varphi }\left( u\right) \delta _{<\beta >}\varphi \left(
u\right) +$$
$$ \delta
_{<\beta >}\overline{\varphi }\left( u\right) \delta _{<\alpha
>}\varphi
\left( u\right) -\frac 1{\sqrt{|g|}}g_{<\alpha ><\beta >}{\cal L}%
^{(0)}\left( u\right) ,\eqno(6.90)
$$
and a similar variation on the components of a d-metric (6.12)
leads to a
symmetric metric energy-momentum d-tensor,%
$$
E_{<\alpha ><\beta >}^{(0)}\left( u\right) =E_{(<\alpha ><\beta
>)}^{(0,can)}\left( u\right) -\eqno(6.91)
$$
$$
\frac{n_E-2}{2(n_E-1)}\left[ R_{(<\alpha ><\beta >)}+D_{(<\alpha
>}D_{<\beta
>)}-g_{<\alpha ><\beta >}\Box \right] \overline{\varphi }\left( u\right)
\varphi \left( u\right) .
$$
Here we note that we can obtain a nonsymmetric energy-momentum
d-tensor if we firstly vary on $G_{<\alpha ><\beta >}$ and than
impose constraints of type (6.10) in order to have a
compatibility with the N-connection structure. We also conclude
that the existence of a N-connection in dv-bundle ${\cal
E}^{<z>}$ results in a nonequivalence of energy-momentum
d-tensors (6.90) and (6.91), nonsymmetry of the Ricci tensor (see
(6.29)),
nonvanishing of the d-covariant derivation of the Einstein d-tensor (6.34), $%
D_{<\alpha >}\overleftarrow{G}^{<\alpha ><\beta >}\neq 0$ and, in
consequence, a corresponding breaking of conservation laws on
ha-spaces when $D_{<\alpha >}E^{<\alpha ><\beta >}\neq 0\,$ . The
problem of formulation of conservation laws on la-spaces is
discussed in details and two variants of its solution (by using
nearly autoparallel maps and tensor integral formalism on locally
anisotropic and higher order multispaces) are proposed
in 
 [262] (see Chapter 8). In this section we shall present only
straightforward generalizations of field equations and necessary
formulas for energy-momentum d-tensors of matter fields on ${\cal
E}^{<z>}$ considering that it is naturally that the conservation
laws (usually being consequences of global, local and/or
intrinsic symmetries of the fundamental space-time and of the
type of field interactions) have to be broken on locally
anisotropic spaces.

\subsection{ Proca equations on ha--spaces}

Let consider a d-vector field $\varphi _{<\alpha >}\left(
u\right) $ with mass $\mu ^2$ (locally anisotropic Proca field )
interacting with exterior \index{Proca equations!on ha--spaces}
la-gravitational field. From the Lagrangian
$$
{\cal L}^{(1)}\left( u\right) =\sqrt{\left| g\right| }\left[ -\frac 12{%
\overline{f}}_{<\alpha ><\beta >}\left( u\right) f^{<\alpha
><\beta >}\left( u\right) +\mu ^2{\overline{\varphi }}_{<\alpha
>}\left( u\right) \varphi ^{<\alpha >}\left( u\right) \right]
,\eqno(6.92)
$$
where $f_{<\alpha ><\beta >}=D_{<\alpha >}\varphi _{<\beta
>}-D_{<\beta
>}\varphi _{<\alpha >},$ one follows the Proca equations on higher order
anisotropic spaces
$$
D_{<\alpha >}f^{<\alpha ><\beta >}\left( u\right) +\mu ^2\varphi
^{<\beta
>}\left( u\right) =0.\eqno(6.93)
$$
Equations (6.93) are a first type constraints for $\beta =0.$ Acting with $%
D_{<\alpha >}$ on (6.93), for $\mu \neq 0$ we obtain second type constraints%
$$
D_{<\alpha >}\varphi ^{<\alpha >}\left( u\right) =0.\eqno(6.94)
$$

Putting (6.94) into (6.93) we obtain second order field equations
with
respect to $\varphi _{<\alpha >}$ :%
$$
\Box \varphi _{<\alpha >}\left( u\right) +R_{<\alpha ><\beta
>}\varphi
^{<\beta >}\left( u\right) +\mu ^2\varphi _{<\alpha >}\left( u\right) =0.%
\eqno(6.95)
$$
The energy-momentum d-tensor and d-vector current following from
the (6.95) can be written as
$$
E_{<\alpha ><\beta >}^{(1)}\left( u\right) =-g^{<\varepsilon
><\tau >}\left(
{\overline{f}}_{<\beta ><\tau >}f_{<\alpha ><\varepsilon >}+{\overline{f}}%
_{<\alpha ><\varepsilon >}f_{<\beta ><\tau >}\right) +
$$
$$
\mu ^2\left( {\overline{\varphi }}_{<\alpha >}\varphi _{<\beta >}+{\overline{%
\varphi }}_{<\beta >}\varphi _{<\alpha >}\right) -\frac{g_{<\alpha ><\beta >}%
}{\sqrt{\left| g\right| }}{\cal L}^{(1)}\left( u\right) .
$$
and%
$$
J_{<\alpha >}^{\left( 1\right) }\left( u\right) =i\left( {\overline{f}}%
_{<\alpha ><\beta >}\left( u\right) \varphi ^{<\beta >}\left( u\right) -{%
\overline{\varphi }}^{<\beta >}\left( u\right) f_{<\alpha ><\beta
>}\left( u\right) \right) .
$$

For $\mu =0$ the d-tensor $f_{<\alpha ><\beta >}$ and the
Lagrangian (6.92) are invariant with respect to locally
anisotropic gauge transforms of type
$$
\varphi _{<\alpha >}\left( u\right) \rightarrow \varphi _{<\alpha
>}\left( u\right) +\delta _{<\alpha >}\Lambda \left( u\right) ,
$$
where $\Lambda \left( u\right) $ is a d-differentiable scalar
function, and we obtain a locally anisot\-rop\-ic variant of
Maxwell theory.

\subsection{ Higher order an\-i\-sot\-rop\-ic gravitons}

Let a massless d-tensor field $h_{<\alpha ><\beta >}\left(
u\right) $ is \index{Gravitons!higher order an\-i\-sot\-rop\-ic}
interpreted as a small perturbation of the locally anisotropic
background metric d-field $g_{<\alpha ><\beta >}\left( u\right)
.$ Considering, for simplicity, a torsionless background we have
locally anisotropic Fierz--Pauli
equations%
\index{Fierz--Pauli equations}
$$
\Box h_{<\alpha ><\beta >}\left( u\right) +2R_{<\tau ><\alpha
><\beta ><\nu
>}\left( u\right) ~h^{<\tau ><\nu >}\left( u\right) =0
$$
and d--gauge conditions%
\index{D--gauge conditions}
$$
D_{<\alpha >}h_{<\beta >}^{<\alpha >}\left( u\right) =0,\quad
h\left( u\right) \equiv h_{<\beta >}^{<\alpha >}(u)=0,
$$
where $R_{<\tau ><\alpha ><\beta ><\nu >}\left( u\right) $ is
curvature d-tensor of the la-background space (these formulae can
be obtained by using a perturbation formalism with respect to
$h_{<\alpha ><\beta >}\left( u\right) $ developed in
 [94]; in our case we must take into account
the distinguishing of geometrical objects and operators on
ha--spaces).

\subsection{ Higher order anisotropic Dirac equations}

Let denote the Dirac d--spinor field on ${\cal E}^{<z>}$ as $\psi
\left( u\right) =\left( \psi ^{\underline{\alpha }}\left(
u\right) \right) $ and consider as the generalized Lorentz
transforms the group of automorphysm of \index{Dirac d--spinor
field} the metric $G_{<\widehat{\alpha }><\widehat{\beta }>}$
(see the ha-frame decomposition of d-metric (6.12), (6.68) and
(6.69) ).The d-covariant derivation of field $\psi \left(
u\right) $ is written as
$$
\overrightarrow{\nabla _{<\alpha >}}\psi =\left[ \delta _{<\alpha
>}+\frac 14C_{\widehat{\alpha }\widehat{\beta }\widehat{\gamma
}}\left( u\right)
~l_{<\alpha >}^{\widehat{\alpha }}\left( u\right) \sigma ^{\widehat{\beta }%
}\sigma ^{\widehat{\gamma }}\right] \psi ,\eqno(6.96)
$$
where coefficients $C_{\widehat{\alpha }\widehat{\beta }\widehat{\gamma }%
}=\left( D_{<\gamma >}l_{\widehat{\alpha }}^{<\alpha >}\right) l_{\widehat{%
\beta }<\alpha >}l_{\widehat{\gamma }}^{<\gamma >}$ generalize
for ha-spaces the corresponding Ricci coefficients on Riemannian
spaces
[79]. Using $%
\sigma $-objects $\sigma ^{<\alpha >}\left( u\right) $ (see
(6.44) and (6.60)--(6.62)) we define the Dirac equations on
ha--spaces: \index{Dirac equations!on ha--spaces}
$$
(i\sigma ^{<\alpha >}\left( u\right) \overrightarrow{\nabla _{<\alpha >}}%
-\mu )\psi =0,
$$
which are the Euler equations for the Lagrangian%
$$
{\cal L}^{(1/2)}\left( u\right) =\sqrt{\left| g\right| }\{[\psi
^{+}\left( u\right) \sigma ^{<\alpha >}\left( u\right)
\overrightarrow{\nabla _{<\alpha
>}}\psi \left( u\right) -
$$
$$
(\overrightarrow{\nabla _{<\alpha >}}\psi ^{+}\left( u\right)
)\sigma ^{<\alpha >}\left( u\right) \psi \left( u\right) ]-\mu
\psi ^{+}\left( u\right) \psi \left( u\right) \},\eqno(6.97)
$$
where $\psi ^{+}\left( u\right) $ is the complex conjugation and
transposition of the column $~\psi \left( u\right) .$

From (6.97) we obtain the d-metric energy-momentum d-tensor%
$$
E_{<\alpha ><\beta >}^{(1/2)}\left( u\right) =\frac i4[\psi
^{+}\left( u\right) \sigma _{<\alpha >}\left( u\right)
\overrightarrow{\nabla _{<\beta
>}}\psi \left( u\right) +\psi ^{+}\left( u\right) \sigma _{<\beta >}\left(
u\right) \overrightarrow{\nabla _{<\alpha >}}\psi \left( u\right)
-
$$
$$
(\overrightarrow{\nabla _{<\alpha >}}\psi ^{+}\left( u\right)
)\sigma _{<\beta >}\left( u\right) \psi \left( u\right)
-(\overrightarrow{\nabla _{<\beta >}}\psi ^{+}\left( u\right)
)\sigma _{<\alpha >}\left( u\right) \psi \left( u\right) ]
$$
and the d-vector source%
$$
J_{<\alpha >}^{(1/2)}\left( u\right) =\psi ^{+}\left( u\right)
\sigma _{<\alpha >}\left( u\right) \psi \left( u\right) .
$$
We emphasize that la-interactions with exterior gauge fields can
be introduced by changing the higher order anisotropic partial
derivation from
(6.96) in this manner:%
$$
\delta _\alpha \rightarrow \delta _\alpha +ie^{\star }B_\alpha ,
$$
where $e^{\star }$ and $B_\alpha $ are respectively the constant
d-vector potential of locally anisotropic gauge interactions on
higher order
anisotropic spaces (see 
 [272] and the next subsection).

\subsection{ D--spinor locally anisotropic Yang--Mills fields}

\index{D--spinor!Yang--Mills fields}
We consider a dv-bundle ${\cal B}_E,~\pi _B:{\cal B\rightarrow E}^{<z>}{\cal %
,}$ on ha-space ${\cal E}^{<z>}{\cal .\,}$ Additionally to
d-tensor and d-spinor indices we shall use capital Greek letters,
$\Phi ,\Upsilon ,\Xi ,\Psi ,...$ for fibre (of this bundle)
indices (see details in
 [181,182] for the case when the base space of the v-bundle $\pi _B$ is a
locally isotropic space-time). Let $\underline{\nabla }_{<\alpha
>}$ be, for simplicity, a torsionless, linear connection in
${\cal B}_E$ satisfying conditions:
$$
\underline{\nabla }_{<\alpha >}:{ \Upsilon }^\Theta \rightarrow { %
\Upsilon }_{<\alpha >}^\Theta \quad \left[ \mbox{or }{ \Xi
}^\Theta \rightarrow { \Xi }_{<\alpha >}^\Theta \right] ,
$$
$$
\underline{\nabla }_{<\alpha >}\left( \lambda ^\Theta +\nu ^\Theta \right) =%
\underline{\nabla }_{<\alpha >}\lambda ^\Theta +\underline{\nabla
}_{<\alpha
>}\nu ^\Theta ,
$$
$$
\underline{\nabla }_{<\alpha >}~(f\lambda ^\Theta )=\lambda
^\Theta \underline{\nabla }_{<\alpha >}f+f\underline{\nabla
}_{<\alpha >}\lambda ^\Theta ,\quad f\in { \Upsilon }^\Theta
~[\mbox{or }{ \Xi }^\Theta ],
$$
where by ${ \Upsilon }^\Theta ~\left( { \Xi }^\Theta \right) $ we
denote the module of sections of the real (complex) v-bundle
${\cal B}_E$
provided with the abstract index $\Theta .$ The curvature of connection $%
\underline{\nabla }_{<\alpha >}$ is defined as
$$
K_{<\alpha ><\beta >\Omega }^{\qquad \Theta }\lambda ^\Omega
=\left(
\underline{\nabla }_{<\alpha >}\underline{\nabla }_{<\beta >}-\underline{%
\nabla }_{<\beta >}\underline{\nabla }_{<\alpha >}\right) \lambda
^\Theta .
$$

For Yang-Mills fields as a rule one considers that ${\cal B}_E$
is enabled with a unitary (complex) structure (complex
conjugation changes mutually the
upper and lower Greek indices). It is useful to introduce instead of $%
K_{<\alpha ><\beta >\Omega }^{\qquad \Theta }$ a Hermitian matrix $%
F_{<\alpha ><\beta >\Omega }^{\qquad \Theta }=i$ $K_{<\alpha
><\beta >\Omega }^{\qquad \Theta }$ connected with components of
the Yang-Mills d-vector potential $B_{<\alpha >\Xi }^{\quad \Phi
}$ according the formula:

$$
\frac 12F_{<\alpha ><\beta >\Xi }^{\qquad \Phi }=\underline{\nabla }%
_{[<\alpha >}B_{<\beta >]\Xi }^{\quad \Phi }-iB_{[<\alpha
>|\Lambda |}^{\quad \Phi }B_{<\beta >]\Xi }^{\quad \Lambda
},\eqno(6.98)
$$
where the la-space indices commute with capital Greek indices.
The gauge transforms are written in the form:

$$
B_{<\alpha >\Theta }^{\quad \Phi }\mapsto B_{<\alpha >\widehat{\Theta }%
}^{\quad \widehat{\Phi }}=B_{<\alpha >\Theta }^{\quad \Phi
}~s_\Phi ^{\quad \widehat{\Phi }}~q_{\widehat{\Theta }}^{\quad
\Theta }+is_\Theta ^{\quad \widehat{\Phi }}\underline{\nabla
}_{<\alpha >}~q_{\widehat{\Theta }}^{\quad \Theta },
$$
$$
F_{<\alpha ><\beta >\Xi }^{\qquad \Phi }\mapsto F_{<\alpha ><\beta >\widehat{%
\Xi }}^{\qquad \widehat{\Phi }}=F_{<\alpha ><\beta >\Xi }^{\qquad
\Phi }s_\Phi ^{\quad \widehat{\Phi }}q_{\widehat{\Xi }}^{\quad
\Xi },
$$
where matrices $s_\Phi ^{\quad \widehat{\Phi }}$ and $q_{\widehat{\Xi }%
}^{\quad \Xi }$ are mutually inverse (Hermitian conjugated in the
unitary
case). The Yang-Mills equations on torsionless la-spaces 
 [272] (see
details in the next Chapter) are written in this form:%
$$
\underline{\nabla }^{<\alpha >}F_{<\alpha ><\beta >\Theta
}^{\qquad \Psi }=J_{<\beta >\ \Theta }^{\qquad \Psi },\eqno(6.99)
$$
$$
\underline{\nabla }_{[<\alpha >}F_{<\beta ><\gamma >]\Theta
}^{\qquad \Xi }=0.\eqno(6.100)
$$
We must introduce deformations of connection of type  $\underline{\nabla }%
_\alpha ^{\star }~\longrightarrow \underline{\nabla }_\alpha
+P_\alpha ,$ (the deformation d-tensor $P_\alpha $ is induced by
the torsion in dv-bundle ${\cal B}_E)$ into the definition of the
curvature of ha-gauge fields (6.98) and motion equations (6.99)
and (6.100) if interactions are modeled on a generic ha-space.

\subsection{D--spinor Einstein--Cartan equ\-a\-ti\-ons }

Now we can write out the field equations of the Einstein-Cartan
theory in \index{D--spinor!Einstein--Cartan equ\-a\-ti\-ons} the
d-spinor form. So, for the Einstein equations (6.34) we have

$$
\overleftarrow{G}_{\underline{\gamma }_1\underline{\gamma }_2\underline{%
\alpha }_1\underline{\alpha }_2}+\lambda \varepsilon _{\underline{\gamma }_1%
\underline{\alpha }_1}\varepsilon _{\underline{\gamma }_2\underline{\alpha }%
_2}=\kappa E_{\underline{\gamma }_1\underline{\gamma }_2\underline{\alpha }_1%
\underline{\alpha }_2},
$$
with $\overleftarrow{G}_{\underline{\gamma }_1\underline{\gamma }_2%
\underline{\alpha }_1\underline{\alpha }_2}$ from (6.86), or, by
using the d-tensor (6.87),

$$
\Phi _{\underline{\gamma }_1\underline{\gamma }_2\underline{\alpha }_1%
\underline{\alpha }_2}+(\frac{\overleftarrow{R}}4-\frac \lambda
2)\varepsilon _{\underline{\gamma }_1\underline{\alpha }_1}\varepsilon _{%
\underline{\gamma }_2\underline{\alpha }_2}=-\frac \kappa 2E_{\underline{%
\gamma }_1\underline{\gamma }_2\underline{\alpha
}_1\underline{\alpha }_2},
$$
which are the d-spinor equivalent of the equations (6.35). These
equations must be completed by the algebraic equations (6.36) for
the d-torsion and d-spin density with d-tensor indices changed
into corresponding d-spinor ones.

\section{Outlook on Ha--Spinors}

We have developed the spinor differential geometry of
distinguished vector bundles provided with nonlinear and
distinguished connections and metric structures and shown in
detail the way of formulation the theory of fundamental field
(gravitational, gauge and spinor) interactions on generic locally
an\-isot\-rop\-ic spaces.

We investigated the problem of definition of spinors on spaces
with higher order anisotropy. Our approach is based on the
formalism of Clifford d-algebras. We introduced spinor structures
on ha-spaces as Clifford d-module structures on dv-bundles. We
also proposed the second definition, as distinguished spinor
structures, by using Clifford fibrations. It was shown that
$H^{2n}$-spaces admit as a proper characteristic the almost
complex spinor structures. We argued that one of the most
important properties of spinors in both dv-bundles with
compatible N-connection, d-connection and metric and in
$H^{2n}$-spaces is the periodicity 8 on the dimension of the base
and on the dimension of the typical fiber spaces.

It should be noted that we introduced 
 [256,255,275] d-spinor
structures in an algebraic topological manner, and that in our
considerations the compatibility of d-connection and metric,
adapted to a given N-connection, plays a crucial role. The Yano
and Ishihara method of lifting of geometrical objects in the
total spaces of tangent bundles
 [295] and the general formalism for vector bundles of Miron and Anastasiei
 [160,161] clearing up the possibility and way of definition of
spinors on higher order anisotropic spaces. Even a straightforward
definition of spinors on Finsler and Lagrange spaces, and, of
course, on various theirs extensions, with general noncompatible
connection and metric structures, is practically impossible (if
spinors are introduced locally with respect to a given metric
quadratic form, the spinor constructions will not be invariant on
parallel transports), we can avoid this difficulty by lifting in
a convenient manner the geometric objects and physical values
from the base of a la-space on the tangent bundles of v- and
t-bundles under consideration. We shall introduce corresponding
discordance laws and values and define nonstandard spinor
structures by using nonmetrical d-tensors (see such constructions
for locally isotropic curved spaces with torsion and
nonmetricity in 
 [154]).

The distinguishing by a N-connection structure of the
multidimensional space into horizontal and vertical subbundles
points out to the necessity to start up the spinor constructions
for la-spaces with a study of distinguished Clifford algebras for
vector spaces split into h- and v-subspaces. The d-spinor objects
exhibit a eight-fold periodicity on dimensions of the
mentioned subspaces. As it was shown in 
 [256,255], see also the
sections 6.2 -6.4 of this work, a corresponding d-spinor
technique can be developed, which is a generalization for higher
dimensional with N-connection structure of that proposed by
Penrose and Rindler
 [180,181,182] for locally isotropic curved spaces, if the locally
adapted to the N-connection structures d-spinor and d-vector
frames are used. It is clear the d-spinor calculus is more
tedious than the 2-spinor one for Einstein spaces because of
multidimensional and multiconnection character of generic
ha-spaces. The d-spinor differential geometry formulated in
section 6.6 can be considered as a branch of the geometry of
Clifford fibrations for v-bundles provided with N-connection,
d-connection and metric structures. We have emphasized only the
features containing d-spinor torsions and curvatures which are
necessary for a d-spinor formulation of la-gravity. To develop a
conformally invariant d-spinor calculus is possible only for a
particular class of ha-spaces when the Weyl d-tensor (6.33) is
defined by the N-connection and d-metric structures. In general,
we have to extend the class of conformal transforms to that of
nearly autoparallel maps of
ha-spaces (see 
 [275,276,279,263] and Chapter 8 in this monograph).

Having fixed compatible N-connection, d-connection and metric
structures on a dv--bundle  ${\cal E}^{<z>}$ we can develop
physical models on this space by using a covariant variational
d-tensor calculus as on Riemann-Cartan spaces (really there are
specific complexities because, in general, the Ricci d-tensor is
not symmetric and the locally anisotropic frames are
nonholonomic). The systems of basic field equations for
fundamental matter (scalar, Proca and Dirac) fields and gauge and
gravitational fields have been introduced in a geometric manner
by using d-covariant operators and la-frame decompositions of
d-metric. These equations and expressions for energy-momentum
d-tensors and d-vector currents can be established by using the
standard variational procedure, but correspondingly adapted to the
N-connection structure if we work by using la-bases.



\chapter{Gauge and Gravitational Ha--Fields}

Despite the charm and success of general relativity there are some
fundamental problems still unsolved in the framework of this
theory. Here we point out the undetermined status of
singularities, the problem of formulation of conservation laws in
curved spaces, and the unrenormalizability of quantum field
interactions. To overcome these defects a number of authors (see,
for example, refs.
 [240,285,194,3])
tended to reconsider and reformulate gravitational theories as a
gauge model similar to the theories of weak, electromagnetic, and
strong forces. But, in spite of theoretical arguments and the
attractive appearance of different proposed models of gauge
gravity, the possibility and manner of interpretation of gravity
as a kind of gauge interaction remain unclear.

The work of Popov and Daikhin 
 [195,196] is distinguished among other gauge
approaches to gravity. Popov and Dikhin did not advance a gauge
extension, or modification, of general relativity; they obtained
an equivalent reformulation (such as well-known tetrad or spinor
variants) of the Einstein equations as Yang-Mills equations for
correspondingly induced Cartan connections
 [40] in the affine frame bundle on the pseudo-Riemannian
space time. This result was used in solving some specific
problems in mathematical physics, for example, for formulation of
a twistor-gauge interpretation of gravity and of nearly
autoparallel conservation laws on
curved spaces 
 [246,250,252,249,233]. It has also an important
conceptual role. On one hand, it points to a possible unified
treatment of gauge and gravitational fields in the language of
linear connections in corresponding vector bundles. On the other,
it emphasizes that the types of fundamental interactions
mentioned essentially differ one from another, even if we admit
for both of them a common gauge like formalism, because if to
Yang-Mills fields one associates semisimple gauge groups, to gauge
treatments of Einstein gravitational fields one has to introduce
into consideration nonsemisimple gauge groups.

Recent developments in theoretical physics suggest the idea that
a more adequate description of radiational, statistical, and
relativistic optic effects in classical and quantum gravity
requires extensions of the geometric background of theories
 [282,163,13,14,15,28,29,118,119,120,122,212,235,236,238,\\ 280,41] by
introducing into consideration spaces with local anisotropy and
formulating corresponding variants of Lagrange and Finsler
gravity and theirs extensions
 to higher order anisotropic spaces 
 [295,162,266,267,268].

The aim of this Chapter is twofold. The first objective is to
present our
results 
 [272,258,259] on formulation of a geometrical approach to
interactions of Yang-Mills fields on spaces with higher order
anisotropy in the framework of the theory of linear connections
in vector bundles (with semisimple structural groups) on
ha-spaces. The second objective is to extend the  formalism in a
manner including theories with nonsemisimple groups which permit
a unique fiber bundle treatment for both locally anisotropic
Yang-Mills field and gravitational interactions. In
general lines, we shall follow the ideas and geometric methods proposed in \\
refs.
 [240,195,196,194,40] but we shall apply them in a form convenient
for introducing into consideration geometrical constructions
 [160,161] and physical theories on ha-spaces.

There is a number of works on gauge models of interactions on
Finsler spaces and theirs extensions(see, for instance,
 [17,18,19,28,164,177]). One has introduced different variants of
generalized gauge transforms, postulated corresponding
Lagrangians for gravitational, gauge and matter field
interactions and formulated variational calculus (here we note
the approach developed by A. Bejancu
 [30,32,29]). The main problem of such models is the dependence of
the basic equations on chosen definition of gauge "compensation"
symmetries and on type of space and field interactions
anisotropy. In order to avoid the ambiguities connected with
particular characteristics of possible la-gauge theories we
consider a "pure" geometric approach to gauge theories (on both
locally isotropic and anisotropic spaces) in the framework of the
theory of fiber bundles provided in general with different types
of nonlinear and linear multiconnection and metric structures.
This way, based on global geometric methods, holds also good for
nonvariational, in the total spaces of bundles, gauge theories
(in the case of gauge gravity based on Poincare or affine gauge
groups); physical values and motion (field) equations have
adequate geometric interpretation and do not depend on the type
of local anisotropy of space-time background. It should be
emphasized here that extensions for higher order anisotropic
spaces which will be presented in this Chapter can be realized in
a straightforward manner.

The presentation in the Chapter is organized as follows:

In section 7.1 we give a geometrical interpretation of gauge
(Yang-Mills) fields on general ha-spaces. Section 7.2 contains a
geometrical definition of anisotropic Yang-Mills equations; the
variational proof of gauge field equations is considered in
connection with the "pure" geometrical method of introducing
field equations. In section 7.3 the ha--gravity is reformulated
as a gauge theory for nonsemisimple groups. A model of nonlinear
de Sitter gauge gravity with local anisotropy is formulated in
section 7.4. We study gravitational gauge instantons with trivial
local anisotropy in section 7.5. Some remarks are given in
section 7.6.

\section{Gauge Fields on Ha-Spaces}

This section is devoted to formulation of the geometrical
background for gauge field theories on spaces with higher order
anisotropy.

Let $\left( P,\pi ,Gr,{\cal E}^{<z>}\right) $ be a principal
bundle
\index{Bundle!principal} on base $%
{\cal E}^{<z>}$ (being a ha-space) with structural group $Gr$ and
surjective map $\pi :P\rightarrow {\cal E}^{<z>}{\cal .\,}$At
every point $u=\left(
x,y_{(1)},...,y_{(z)}\right) \in {\cal E}^{<z>}$ there is a vicinity ${\cal %
U\subset E}^{<z>}{\cal ,}u\in {\cal U,}$ with trivializing $P$
diffeomorphisms $f$ and $\varphi :$%
$$
f_{{\cal U}}:\pi ^{-1}\left( {\cal U}\right) \rightarrow {\cal U\times }%
Gr,\qquad f\left( p\right) =\left( \pi \left( p\right) ,\varphi
\left( p\right) \right) ,
$$
$$
\varphi _{{\cal U}}:\pi ^{-1}\left( {\cal U}\right) \rightarrow
Gr,\varphi (pq)=\varphi \left( p\right) q,\quad \forall q\in
Gr,~p\in P.
$$
We remark that in the general case for two open regions%
$$
{\cal U,V}\subset {\cal E}^{<z>}{\cal ,U\cap V}\neq \emptyset ,f_{{\cal U|}%
_p}\neq f_{{\cal V|}_p},\mbox{ even }p\in {\cal U\cap V.}
$$

Transition functions $g_{{\cal UV}}$ are defined as
$$
g_{{\cal UV}}:{\cal U\cap V\rightarrow }Gr,g_{{\cal UV}}\left(
u\right) =\varphi _{{\cal U}}\left( p\right) \left( \varphi
_{{\cal V}}\left( p\right) ^{-1}\right) ,\pi \left( p\right) =u.
$$

Hereafter we shall omit, for simplicity, the specification of
trivializing regions of maps and denote, for example, $f\equiv
f_{{\cal U}},\varphi \equiv \varphi _{{\cal U}},$ $s\equiv
s_{{\cal U}},$ if this will not give rise to ambiguities.

Let $\theta \,$ be the canonical left invariant 1-form on $Gr$
with values
in algebra Lie ${\cal G}$ of group $Gr$ uniquely defined from the relation $%
\theta \left( q\right) =q,\forall q\in {\cal G,}$ and consider a 1-form $%
\omega $ on ${\cal U\subset E}^{<z>}$ with values in ${\cal G.}$ Using $%
\theta $ and $\omega ,$ we can locally define the connection form
$\Omega $
in $P$ as a 1-form:%
$$
\Omega =\varphi ^{*}\theta +Ad~\varphi ^{-1}\left( \pi ^{*}\omega
\right) \eqno(7.1)
$$
where $\varphi ^{*}\theta $ and $\pi ^{*}\omega $ are,
respectively, forms induced on $\pi ^{-1}\left( {\cal U}\right) $
and $P$ by maps $\varphi $ and $\pi $ and $\omega =s^{*}\Omega .$
The adjoint action on a form $\lambda $ with values in ${\cal G}$
is defined as
$$
\left( Ad~\varphi ^{-1}\lambda \right) _p=\left( Ad~\varphi
^{-1}\left( p\right) \right) \lambda _p
$$
where $\lambda _p$ is the value of form $\lambda $ at point $p\in
P.$

Introducing a basis $\{\Delta _{\widehat{a}}\}$ in ${\cal G}$ (index $%
\widehat{a}$ enumerates the generators making up this basis), we
write the 1-form $\omega $ on ${\cal E}^{<z>}$ as
$$
\omega =\Delta _{\widehat{a}}\omega ^{\widehat{a}}\left( u\right) ,~\omega ^{%
\widehat{a}}\left( u\right) =\omega _{<\mu >}^{\widehat{a}}\left(
u\right) \delta u^{<\mu >}\eqno(7.2)
$$
where $\delta u^{<\mu >}=\left( dx^i,\delta y^{<a>}\right) $ and
the Einstein summation rule on indices $\widehat{a}$ and $<\mu >$
is used. Functions $\omega _{<\mu >}^{\widehat{a}}\left( u\right)
$ from (7.2) will
be called the components of Yang-Mills fields on ha-space ${\cal E}^{<z>}%
{\cal .}$ Gauge transforms of $\omega $ can be geometrically
interpreted as transition relations for $\omega _{{\cal U}}$ and
$\omega _{{\cal V}},$ when
$u\in {\cal U\cap V,}$%
$$
\left( \omega _{{\cal U}}\right) _u=\left( g_{{\cal
UV}}^{*}\theta \right)
_u+Ad~g_{{\cal UV}}\left( u\right) ^{-1}\left( \omega _{{\cal V}}\right) _u.%
\eqno(7.3)
$$

To relate $\omega _{<\mu >}^{\widehat{a}}$ with a covariant
derivation we shall consider a vector bundle $\Upsilon $
associated to $P.$ Let $\rho
:Gr\rightarrow GL\left( {\cal R}^s\right) $ and $\rho ^{\prime }:{\cal G}%
\rightarrow End\left( E^s\right) $ be, respectively, linear
representations of group $Gr$ and Lie algebra ${\cal G}$ (in a
more general case we can consider ${\cal C}^s$ instead of ${\cal
R}^s).$ Map $\rho $ defines a left
action on $Gr$ and associated vector bundle $\Upsilon =P\times {\cal R}%
^s/Gr,~\pi _E:E\rightarrow {\cal E}^{<z>}{\cal .}$ Introducing
the standard
basis $\xi _{\underline{i}}=\{\xi _{\underline{1}},\xi _{\underline{2}%
},...,\xi _{\underline{s}}\}$ in ${\cal R}^s,$ we can define the
right action on $P\times $ ${\cal R}^s,\left( \left( p,\xi
\right) q=\left( pq,\rho \left( q^{-1}\right) \xi \right) ,q\in
Gr\right) ,$ the map induced
from $P$%
$$
p:{\cal R}^s\rightarrow \pi _E^{-1}\left( u\right) ,\quad \left(
p\left( \xi \right) =\left( p\xi \right) Gr,\xi \in {\cal
R}^s,\pi \left( p\right) =u\right)
$$
and a basis of local sections $e_{\underline{i}}:U\rightarrow \pi
_E^{-1}\left( U\right) ,~e_{\underline{i}}\left( u\right)
=s\left( u\right)
\xi _{\underline{i}}.$ Every section $\varsigma :{\cal E}^{<z>}{\cal %
\rightarrow }\Upsilon $ can be written locally as $\varsigma
=\varsigma ^ie_i,\varsigma ^i\in C^\infty \left( {\cal U}\right)
.$ To every vector field $X$ on ${\cal E}^{<z>}$ and Yang-Mills
field $\omega ^{\widehat{a}}$
on $P$ we associate operators of covariant derivations:%
$$
\nabla _X\zeta =e_{\underline{i}}\left[ X\zeta
^{\underline{i}}+B\left( X\right)
_{\underline{j}}^{\underline{i}}\zeta ^{\underline{j}}\right]
$$
where
$$
B\left( X\right) =\left( \rho ^{\prime }X\right) _{\widehat{a}}\omega ^{%
\widehat{a}}\left( X\right) .\eqno(7.4)
$$
Transformation laws (7.3) and operators (7.4) are interrelated by
these transition transforms for values $e_{\underline{i}},\zeta
^{\underline{i}},$
and $B_{<\mu >}:$%
$$
e_{\underline{i}}^{{\cal V}}\left( u\right) =\left[ \rho g_{{\cal
UV}}\left(
u\right) \right] _{\underline{i}}^{\underline{j}}e_{\underline{i}}^{{\cal U}%
},~\zeta _{{\cal U}}^{\underline{i}}\left( u\right) =\left[ \rho g_{{\cal UV}%
}\left( u\right) \right] _{\underline{i}}^{\underline{j}}\zeta _{{\cal V}}^{%
\underline{i}},
$$
$$
B_{<\mu >}^{{\cal V}}\left( u\right) =\left[ \rho g_{{\cal
UV}}\left( u\right) \right] ^{-1}\delta _{<\mu >}\left[ \rho
g_{{\cal UV}}\left( u\right) \right] +\left[ \rho g_{{\cal
UV}}\left( u\right) \right] ^{-1}B_{<\mu >}^{{\cal U}}\left(
u\right) \left[ \rho g_{{\cal UV}}\left( u\right) \right]
,\eqno(7.5)
$$
where $B_{<\mu >}^{{\cal U}}\left( u\right) =B^{<\mu >}\left(
\delta /du^{<\mu >}\right) \left( u\right) .$

Using (7.5), we can verify that the operator $\nabla _X^{{\cal
U}},$ acting on sections of $\pi _\Upsilon :\Upsilon \rightarrow
{\cal E}^{<z>}$
according to definition (7.4), satisfies the properties%
$$
\begin{array}{c}
\nabla _{f_1X+f_2Y}^{
{\cal U}}=f_1\nabla _X^{{\cal U}}+f_2\nabla _X^{{\cal U}},~\nabla _X^{{\cal U%
}}\left( f\zeta \right) =f\nabla _X^{{\cal U}}\zeta +\left(
Xf\right) \zeta
, \\ \nabla _X^{{\cal U}}\zeta =\nabla _X^{{\cal V}}\zeta ,\quad u\in {\cal %
U\cap V,}f_1,f_2\in C^\infty \left( {\cal U}\right) .
\end{array}
$$

So, we can conclude that the Yang--Mills connection
\index{Yang--Mills connection} in the vector bundle $%
\pi _\Upsilon :\Upsilon \rightarrow {\cal E}^{<z>}$ is not a
general one, but is induced from the principal bundle $\pi
:P\rightarrow {\cal E}^{<z>}$ with structural group $Gr.$

The curvature ${\cal K}$ of connection $\Omega $ from (7.1) is defined as%
$$
{\cal K}=D\Omega ,~D=\widehat{H}\circ d\eqno(7.6)
$$
where $d$ is the operator of exterior derivation acting on ${\cal
G}$-valued forms as\\ $d\left( \Delta _{\widehat{a}}\otimes \chi
^{\widehat{a}}\right) =\Delta _{\widehat{a}}\otimes d\chi
^{\widehat{a}}$ and $\widehat{H}\,$ is
the horizontal projecting operator act\-ing, for example, on the 1-form $%
\lambda $ as $\left( \widehat{H}\lambda \right) _P\left(
X_p\right) =\lambda
_p\left( H_pX_p\right) ,$ where $H_p$ projects on the horizontal subspace $%
{\cal H}_p\in P_p [ X_p\in {\cal H}_p$ is equivalent to $\Omega
_p\left( X_p\right) =0] .$ We can express (7.6) locally as
$$
{\cal K}=Ad~\varphi _{{\cal U}}^{-1}\left( \pi ^{*}{\cal K}_{{\cal U}%
}\right) \eqno(7.7)
$$
where
$$
{\cal K}_{{\cal U}}=d\omega _{{\cal U}}+\frac 12\left[ \omega _{{\cal U}%
},\omega _{{\cal U}}\right] .\eqno(7.8)
$$
The exterior product of ${\cal G}$-valued form (7.8) is defined as
$$
\left[ \Delta _{\widehat{a}}\otimes \lambda ^{\widehat{a}},\Delta _{\widehat{%
b}}\otimes \xi ^{\widehat{b}}\right] =\left[ \Delta _{\widehat{a}},\Delta _{%
\widehat{b}}\right] \otimes \lambda ^{\widehat{a}}\bigwedge \xi ^{\widehat{b}%
},
$$
where the antisymmetric tensorial product is%
$$
\lambda ^{\widehat{a}}\bigwedge \xi ^{\widehat{b}}=\lambda
^{\widehat{a}}\xi ^{\widehat{b}}-\xi ^{\widehat{b}}\lambda
^{\widehat{a}}.
$$

Introducing structural coefficients
$f_{\widehat{b}\widehat{c}}^{\quad \widehat{a}}$ of ${\cal G}$
satisfying
$$
\left[ \Delta _{\widehat{b}},\Delta _{\widehat{c}}\right] =f_{\widehat{b}%
\widehat{c}}^{\quad \widehat{a}}\Delta _{\widehat{a}}
$$
we can rewrite (7.8) in a form more convenient for local considerations:%
$$
{\cal K}_{{\cal U}}=\Delta _{\widehat{a}}\otimes {\cal K}_{<\mu ><\nu >}^{%
\widehat{a}}\delta u^{<\mu >}\bigwedge \delta u^{<\nu >}\eqno(7.9)
$$
where%
$$
{\cal K}_{<\mu ><\nu >}^{\widehat{a}}=\frac{\delta \omega _{<\nu >}^{%
\widehat{a}}}{\partial u^{<\mu >}}-\frac{\delta \omega _{<\mu >}^{\widehat{a}%
}}{\partial u^{<\nu >}}+\frac 12f_{\widehat{b}\widehat{c}}^{\quad \widehat{a}%
}\left( \omega _{<\mu >}^{\widehat{b}}\omega _{<\nu
>}^{\widehat{c}}-\omega _{<\nu >}^{\widehat{b}}\omega _{<\mu
>}^{\widehat{c}}\right) .
$$

This section ends by considering the problem of reduction of the
local an\-i\-sot\-rop\-ic gauge symmetries and gauge fields to
isotropic ones. For local trivial considerations we suppose that
the vanishing of dependencies on $y$ variables leads to isotropic
Yang-Mills fields with the same gauge group as in the anisotropic
case. Global geometric constructions require a more rigorous
topological study of possible obstacles for reduction of total
spaces and structural groups on anisotropic bases to their
analogous on isotropic (for example, pseudo-Riemannian) base
spaces.

\section{Yang-Mills Equations on Ha-spaces}

Interior gauge (nongravitational) symmetries are associated to
semisimple
structural groups. On the principal bundle $\left( P,\pi ,Gr,{\cal E}%
^{<z>}\right) $ with nondegenerate Killing form for semisimple
group $Gr$ we \index{Killing form}
can define the generalized Lagrange metric%
$$
h_p\left( X_p,Y_p\right) =G_{\pi \left( p\right) }\left( d\pi
_PX_P,d\pi _PY_P\right) +K\left( \Omega _P\left( X_P\right)
,\Omega _P\left( X_P\right) \right) ,\eqno(7.10)
$$
where $d\pi _P$ is the differential of map $\pi :P\rightarrow {\cal E}^{<z>}%
{\cal ,}$ $G_{\pi \left( p\right) }$ is locally generated as the
ha-metric
(6.12), and $K$ is the Killing form on ${\cal G:}$%
$$
K\left( \Delta _{\widehat{a}},\Delta _{\widehat{b}}\right) =f_{\widehat{b}%
\widehat{d}}^{\quad \widehat{c}}f_{\widehat{a}\widehat{c}}^{\quad \widehat{d}%
}=K_{\widehat{a}\widehat{b}}.
$$

Using the metric $G_{<\alpha ><\beta >}$ on ${\cal E}^{<z>}$
$\left[
h_P\left( X_P,Y_P\right) \mbox{ on }P\right] ,$ we can introduce operators $%
*_G$ and $\widehat{\delta }_G$ acting in the space of forms on ${\cal E}%
^{<z>}$ ($*_H$ and $\widehat{\delta }_H$ acting on forms on ${\cal E}^{<z>}%
{\cal ).}$ Let $e_{<\mu >}$ be orthonormalized frames on ${\cal U\subset E}%
^{<z>}$ and $e^{<\mu >}$ the adjoint coframes. Locally%
$$
G=\sum\limits_{<\mu >}\eta \left( <\mu >\right) e^{<\mu >}\otimes
e^{<\mu
>},
$$
where $\eta _{<\mu ><\mu >}=\eta \left( <\mu >\right) =\pm 1,$
$<\mu
>=1,2,...,n_E,n_E=1,...,n+m_1+...+m_z,$ and the Hodge operator $*_G$ can be
\index{Hodge operator} defined as $*_G:\Lambda ^{\prime }\left(
{\cal E}^{<z>}\right) \rightarrow \Lambda ^{n+m_1+...+m_z}\left(
{\cal E}^{<z>}\right) ,$ or, in explicit
form, as%
$$
*_G\left( e^{<\mu _1>}\bigwedge ...\bigwedge e^{<\mu _r>}\right)
=\eta \left( \nu _1\right) ...\eta \left( \nu _{n_E-r}\right)
\times \eqno(7.11)
$$
$$
sign\left(
\begin{array}{ccccc}
1 & 2 & ...r & r+1 & ...n_E \\
<\mu _{1>} & <\mu _2> & ...<\mu _r> & <\nu _1> & ...\nu _{n_E-r}
\end{array}
\right) \times
$$
$$
e^{<\nu _1>}\bigwedge ...\bigwedge e^{<\nu _{n_E-r}>}.
$$
Next, define the operator%
$$
*_{G^{}}^{-1}=\eta \left( 1\right) ...\eta \left( n_E\right)
\left( -1\right) ^{r\left( n_E-r\right) }*_G
$$
and introduce the scalar product on forms $\beta _1,\beta
_2,...\subset
\Lambda ^r\left( {\cal E}^{<z>}\right) $ with compact carrier:%
$$
\left( \beta _1,\beta _2\right) =\eta \left( 1\right) ...\eta
\left( n_E\right) \int \beta _1\bigwedge *_G\beta _2.
$$
The operator $\widehat{\delta }_G$ is defined as the adjoint to
$d$ associated to the scalar product for forms, specified for
$r$-forms as
$$
\widehat{\delta }_G=\left( -1\right) ^r*_{G^{}}^{-1}\circ d\circ *_G.%
\eqno(7.12)
$$

We remark that operators $*_H$ and $\delta _H$ acting in the total space of $%
P$ can be defined similarly to (7.11) and (7.12), but by using
metric (7.10). Both these operators also act in the space of
${\cal G}$-valued
forms:%
$$
*\left( \Delta _{\widehat{a}}\otimes \varphi ^{\widehat{a}}\right) =\Delta _{%
\widehat{a}}\otimes (*\varphi ^{\widehat{a}}),
$$
$$
\widehat{\delta }\left( \Delta _{\widehat{a}}\otimes \varphi ^{\widehat{a}%
}\right) =\Delta _{\widehat{a}}\otimes (\widehat{\delta }\varphi ^{\widehat{a%
}}).
$$

The form $\lambda $ on $P$ with values in ${\cal G}$ is called
horizontal if $\widehat{H}\lambda =\lambda $ and equivariant if
$R^{*}\left( q\right) \lambda =Ad~q^{-1}\varphi ,~\forall g\in
Gr,R\left( q\right) $ being the right shift on $P.$ We can verify
that equivariant and horizontal forms also
satisfy the conditions%
$$
\lambda =Ad~\varphi _{{\cal U}}^{-1}\left( \pi ^{*}\lambda
\right) ,\qquad \lambda _{{\cal U}}=S_{{\cal U}}^{*}\lambda ,
$$
$$
\left( \lambda _{{\cal V}}\right) _{{\cal U}}=Ad~\left( g_{{\cal
UV}}\left( u\right) \right) ^{-1}\left( \lambda _{{\cal
U}}\right) _u.
$$

Now, we can define the field equations for curvature (7.7) and
connection
(7.1) :%
$$
\Delta {\cal K}=0,\eqno(7.13)
$$
$$
\nabla {\cal K}=0,\eqno(7.14)
$$
where $\Delta =\widehat{H}\circ \widehat{\delta }_H.$ Equations
(7.13) are similar to the well-known Maxwell equations and for
non-Abelian gauge fields are called Yang-Mills equations. The
structural equations (7.14) are called \index{Yang--Mills
equations} Bianchi identities. \index{Bianchi identities}

The field equations (7.13) do not have a physical meaning because
they are written in the total space of bundle $\Upsilon $ and not
on the base anisotropic space-time ${\cal E}^{<z>}{\cal .}$ But
this dificulty may be
obviated by projecting the mentioned equations on the base. The 1-form $%
\Delta {\cal K}$ is horizontal by definition and its equivariance
follows
from the right invariance of metric (7.10). So, there is a unique form $%
(\Delta {\cal K})_{{\cal U}}$ satisfying
$$
\Delta {\cal K=}Ad~\varphi _{{\cal U}}^{-1}\pi ^{*}(\Delta {\cal K})_{{\cal U%
}}.
$$
Projection of (7.13) on the base can be written as $(\Delta {\cal K})_{{\cal %
U}}=0.$ To calculate $(\Delta {\cal K})_{{\cal U}},$ we use the
equality
 [40,196]
$$
d\left( Ad~\varphi _{{\cal U}}^{-1}\lambda \right) =Ad~~\varphi _{{\cal U}%
}^{-1}~d\lambda -\left[ \varphi _{{\cal U}}^{*}\theta ,Ad~\varphi _{{\cal U}%
}^{-1}\lambda \right]
$$
where $\lambda $ is a form on $P$ with values in ${\cal G.}$ For
r-forms we have
$$
\widehat{\delta }\left( Ad~\varphi _{{\cal U}}^{-1}\lambda \right)
=Ad~\varphi _{{\cal U}}^{-1}\widehat{\delta }\lambda -\left(
-1\right)
^r*_H\{\left[ \varphi _{{\cal U}}^{*}\theta ,*_HAd~\varphi _{{\cal U}%
}^{-1}\lambda \right]
$$
and, as a consequence,%
$$
\widehat{\delta }{\cal K}=Ad~\varphi _{{\cal U}}^{-1}\{\widehat{\delta }%
_H\pi ^{*}{\cal K}_{{\cal U}}+*_H^{-1}[\pi ^{*}\omega _{{\cal U}},*_H\pi ^{*}%
{\cal K}_{{\cal U}}]\}-
$$
$$
-*_H^{-1}\left[ \Omega ,Ad~\varphi _{{\cal U}}^{-1}*_H\left( \pi ^{*}{\cal K}%
\right) \right] .\eqno(7.15)
$$
By using straightforward calculations in the adapted dual basis
on $\pi
^{-1}\left( {\cal U}\right) $ we can verify the equalities%
$$
\left[ \Omega ,Ad~\varphi _{{\cal U}}^{-1}~*_H\left( \pi ^{*}{\cal K}_{{\cal %
U}}\right) \right] =0,\widehat{H}\delta _H\left( \pi ^{*}{\cal K}_{{\cal U}%
}\right) =\pi ^{*}\left( \widehat{\delta }_G{\cal K}\right) ,
$$
$$
*_H^{-1}\left[ \pi ^{*}\omega _{{\cal U}},*_H\left( \pi ^{*}{\cal K}_{{\cal U%
}}\right) \right] =\pi ^{*}\{*_G^{-1}\left[ \omega _{{\cal U}},*_G{\cal K}_{%
{\cal U}}\right] \}.\eqno(7.16)
$$
From (7.15) and (7.16) it follows that
$$
\left( \Delta {\cal K}\right) _{{\cal U}}=\widehat{\delta }_G{\cal K}_{{\cal %
U}}+*_G^{-1}\left[ \omega _{{\cal U}},*_G{\cal K}_{{\cal U}}\right] .%
\eqno(7.17)
$$

Taking into account (7.17) and (7.12), we prove that projection
on ${\cal E}$ of equations (7.13) and (7.14) can be expressed
respectively as
$$
*_G^{-1}\circ d\circ *_G{\cal K}_{{\cal U}}+*_G^{-1}\left[ \omega _{{\cal U}%
},*_G{\cal K}_{{\cal U}}\right] =0.\eqno(7.18)
$$
$$
d{\cal K}_{{\cal U}}+\left[ \omega _{{\cal U}},{\cal K}_{{\cal U}}\right] =0.%
$$

Equations (7.18) (see (7.17)) are gauge-invariant because%
$$
\left( \Delta {\cal K}\right) _{{\cal U}}=Ad~g_{{\cal
UV}}^{-1}\left( \Delta {\cal K}\right) _{{\cal V}}.
$$

By using formulas (7.9)-(7.12) we can rewrite (7.18) in coordinate form%
$$
D_{<\nu >}\left( G^{<\nu ><\lambda >}{\cal K}_{~<\lambda ><\mu >}^{\widehat{a%
}}\right) +f_{\widehat{b}\widehat{c}}^{\quad
\widehat{a}}G^{<v><\lambda
>}\omega _{<\lambda >}^{~\widehat{b}}{\cal K}_{~<\nu ><\mu >}^{\widehat{c}%
}=0,\eqno(7.19)
$$
where $D_{<\nu >}$ is, for simplicity, a compatible with metric
covariant derivation on ha-space ${\cal E}^{<z>}.$

We point out that for our bundles with semisimple structural
groups the Yang-Mills equations (7.13) (and, as a consequence,
their horizontal
projections (7.18) or (7.19)) can be obtained by variation of the action%
$$
I=\int {\cal K}_{~<\mu ><\nu >}^{\widehat{a}}{\cal K}_{~<\alpha ><\beta >}^{%
\widehat{b}}G^{<\mu ><\alpha >}G^{<\nu ><\beta >}K_{\widehat{a}\widehat{b}%
}\times \eqno(7.20)
$$
$$
\left| G_{<\alpha ><\beta >}\right| ^{1/2}dx^1...dx^n\delta
y_{(1)}^1...\delta y_{(1)}^{m_1}...\delta y_{(z)}^1...\delta
y_{(z)}^{m_z}.
$$
Equations for extremals of (7.20) have the form
$$
K_{\widehat{r}\widehat{b}}G^{<\lambda ><\alpha >}G^{<\kappa
><\beta
>}D_{<\alpha >}{\cal K}_{~<\lambda ><\beta >}^{\widehat{b}}-
$$
$$
K_{\widehat{a}\widehat{b}}G^{<\kappa ><\alpha >}G^{<\nu ><\beta >}f_{%
\widehat{r}\widehat{l}}^{\quad \widehat{a}}\omega _{<\nu >}^{\widehat{l}}%
{\cal K}_{~<\alpha ><\beta >}^{\widehat{b}}=0,
$$
which are equivalent to ''pure'' geometric equations (7.19) (or
(7.18) due to nondegeneration of the Killing form
$K_{\widehat{r}\widehat{b}}$ for semisimple groups.

To take into account gauge interactions with matter fields
(section of vector bundle $\Upsilon $ on ${\cal E}$ ) we have to
introduce a source \index{Source 1--form}
1--form ${\cal J}$ in equations (7.13) and to write them as%
$$
\Delta {\cal K}={\cal J}\eqno(7.21)
$$

Explicit constructions of ${\cal J}$ require concrete definitions
of the bundle $\Upsilon ;$ for example, for spinor fields an
invariant formulation of the Dirac equations on la-spaces is
necessary. We omit spinor considerations in this Chapter (see
section 6.7), but we shall present the definition of the source
${\cal J}$ for gravitational interactions (by using the
energy-momentum tensor of matter on la--space) in the next
section.

\section{Gauge HA--Gravity}

A considerable body of work on the formulation of gauge
gravitational models on isotropic spaces is based on using
nonsemisimple groups, for example, Poincare and affine groups, as
structural gauge groups (see critical
analysis and original results in 
 [285,240,155,194]. The main
impediment to developing such models is caused by the
degeneration of Killing forms for nonsemisimple groups, which
make it impossible to construct consistent variational gauge
field theories (functional (7.20) and extremal equations are
degenarate in these cases). There are at least two possibilities
to get around the mentioned difficulty.\ The first is to realize
a minimal extension of the nonsemisimple group to a semisimple
one, similar to the extension of the Poincare group to the de
Sitter group
considered in 
 [195,196,240] (in the next section we shall use this
operation for the definition of locally anisotropic gravitational
instantons). The second possibility is to introduce into
consideration the bundle of adapted affine frames on la-space
${\cal E}^{<z>},$ to use an auxiliary nondegenerate bilinear form
$a_{\widehat{a}\widehat{b}}$ instead of the degenerate Killing
form $K_{\widehat{a}\widehat{b}}$ and to consider a ''pure''
geometric method, illustrated in the previous section, of
defining gauge field equations. Projecting on the base ${\cal
E}^{<z>},$ we shall obtain gauge gravitational field equations on
ha-space having a form similar to Yang-Mills equations.

The goal of this section is to prove that a specific
parametrization of components of the Cartan connection in the
bundle of adapted affine frames \index{Cartan!connection} on
${\cal E}^{<z>}$ establishes an equivalence between Yang-Mills
equations (7.21) and Einstein equations on ha-spaces.

\subsection{Bundles of linear ha--frames}

Let $\left( X_{<\alpha >}\right) _u=\left( X_i,X_{<a>}\right)
_u=\left(
X_i,X_{a_1},...,X_{a_z}\right) _u$ be an adapted frame (see (7.4) at point $%
u\in {\cal E}^{<z>}.$ We consider a local right distinguished
action of
matrices%
\index{Bundles!of linear ha--frames}
$$
A_{<\alpha ^{\prime }>}^{\quad <\alpha >}=\left(
\begin{array}{cccc}
A_{i^{\prime }}^{\quad i} & 0 & ... & 0 \\
0 & B_{a_1^{\prime }}^{\quad a_1} & ... & 0 \\
... & ... & ... & ... \\
0 & 0 & ... & B_{a_z^{\prime }}^{\quad a_z}
\end{array}
\right) \subset GL_{n_E}=
$$
$$
GL\left( n,{\cal R}\right) \oplus GL\left( m_1,{\cal R}\right)
\oplus ...\oplus GL\left( m_z,{\cal R}\right) .
$$
Nondegenerate matrices $A_{i^{\prime }}^{\quad i}$ and
$B_{j^{\prime }}^{\quad j}$ respectively transforms linearly
$X_{i|u}$ into $X_{i^{\prime
}|u}=A_{i^{\prime }}^{\quad i}X_{i|u}$ and $X_{a_p^{\prime }|u}$ into $%
X_{a_p^{\prime }|u}=B_{a_p^{\prime }}^{\quad a_p}X_{a_p|u},$ where $%
X_{<\alpha ^{\prime }>|u}=A_{<\alpha ^{\prime }>}^{\quad <\alpha
>}X_{<\alpha >}$ is also an adapted frame at the same point $u\in {\cal E}%
^{<z>}.$ We denote by $La\left( {\cal E}^{<z>}\right) $ the set
of all adapted frames $X_{<\alpha >}$ at all points of ${\cal
E}^{<z>}$ and consider the surjective map $\pi $ from $La\left(
{\cal E}^{<z>}\right) $ to
${\cal E}^{<z>}$ transforming every adapted frame $X_{\alpha |u}$ and point $%
u$ into point $u.$ Every $X_{<\alpha ^{\prime }>|u}$ has a unique
representation as $X_{<\alpha ^{\prime }>}=A_{<\alpha ^{\prime
}>}^{\quad <\alpha >}X_{<\alpha >}^{\left( 0\right) },$ where
$X_{<\alpha >}^{\left(
0\right) }$ is a fixed distinguished basis in tangent space $T\left( {\cal E}%
^{<z>}\right) .$ It is obvious that $\pi ^{-1}\left( {\cal U}\right) ,{\cal U%
}\subset {\cal E}^{<z>},$ is bijective to ${\cal U}\times
GL_{n_E}\left( {\cal R}\right) .$ We can transform $La\left(
{\cal E}^{<z>}\right) $ in a differentiable manifold taking
$\left( u^{<\beta >},A_{<\alpha ^{\prime }>}^{\quad <\alpha
>}\right) $ as a local coordinate system on $\pi
^{-1}\left( {\cal U}\right) .$ Now, it is easy to verify that $${\cal {L}}a(%
{\cal E}^{<z>})=(La({\cal E}^{<z>},{\cal E}^{<z>},GL_{n_E}({\cal
R})))$$ is a principal bundle. We call ${\cal {L}}a({\cal
E}^{<z>})$ the bundle of linear adapted frames on ${\cal
E}^{<z>}.$

The next step is to identify the components of, for simplicity,
compatible
d-connection $\Gamma _{<\beta ><\gamma >}^{<\alpha >}$ on ${\cal E}^{<z>}:$%
$$
\Omega _{{\cal U}}^{\widehat{a}}=\omega ^{\widehat{a}}=\{\omega
_{\quad <\lambda >}^{\widehat{\alpha }\widehat{\beta }}\doteq
\Gamma _{<\beta
><\gamma >}^{<\alpha >}\}.\eqno(7.22)
$$
Introducing (7.22) in (7.17), we calculate the local 1-form%
$$
\left( \Delta {\cal R}^{(\Gamma )}\right) _{{\cal U}}=\Delta _{\widehat{%
\alpha }\widehat{\alpha }_1}\otimes (G^{<\nu ><\lambda >}D_{<\lambda >}{\cal %
R}_{\qquad <\nu ><\mu >}^{<\widehat{\alpha }><\widehat{\gamma }>}+
$$
$$
f_{\qquad <\widehat{\beta }><\widehat{\delta }><\widehat{\gamma }><\widehat{%
\varepsilon }>}^{<\widehat{\alpha }><\widehat{\gamma }>}G^{<\nu
><\lambda
>}\omega _{\qquad <\lambda >}^{<\widehat{\beta }><\widehat{\delta }>}{\cal R}%
_{\qquad <\nu ><\mu >}^{<\widehat{\gamma }><\widehat{\varepsilon
}>})\delta u^{<\mu >},\eqno(7.23)
$$
where
$$
\Delta _{\widehat{\alpha }\widehat{\beta }}=\left(
\begin{array}{cccc}
\Delta _{\widehat{i}\widehat{j}} & 0 & ... & 0 \\
0 & \Delta _{\widehat{a}_1\widehat{b}_1} & ... & 0 \\
... & ... & ... & ... \\
0 & 0 & ... & \Delta _{\widehat{a}_z\widehat{b}_z}
\end{array}
\right)
$$
is the standard distinguished basis in Lie algebra of matrices ${{\cal {G}}l}%
_{n_E}\left( {\cal R}\right) $ with $\left( \Delta _{\widehat{i}\widehat{k}%
}\right) _{jl}=\delta _{ij}\delta _{kl}$ and $\left( \Delta _{\widehat{a}_p%
\widehat{c}_p}\right) _{b_pd_p}=\delta _{a_pb_p}\delta _{c_pd_p}$
be\-ing
res\-pec\-ti\-ve\-ly the stand\-ard bas\-es in ${\cal {G}}l\left( {\cal R}%
^{n_E}\right) .$ We have denoted the curvature of connection
(7.22), \index{Lie algebra!of matrices} considered in (7.23), as
$$
{\cal R}_{{\cal U}}^{(\Gamma )}=\Delta _{\widehat{\alpha }\widehat{\alpha }%
_1}\otimes {\cal R}_{\qquad <\nu ><\mu >}^{\widehat{\alpha }\widehat{\alpha }%
_1}X^{<\nu >}\bigwedge X^{<\mu >},
$$
where ${\cal R}_{\qquad <\nu ><\mu >}^{\widehat{\alpha }\widehat{\alpha }%
_1}=R_{<\alpha _1>\quad <\nu ><\mu >}^{\quad <\alpha >}$ (see
curvatures (6.28)).

\subsection{Bundles of affine ha--frames and Einstein equations}

Besides ${\cal {L}}a\left( {\cal E}^{<z>}\right) $ with ha-space ${\cal E}%
^{<z>},$ another bundle is naturally related, the bundle of
adapted affine \index{Bundles!of affine ha--frames} frames with
structural group $Af_{n_E}\left( {\cal R}\right) =GL_{n_E}\left(
{\cal E}^{<z>}\right)$ $\otimes {\cal R}^{n_E}.$ Because as
linear space the
Lie Algebra $af_{n_E}\left( {\cal R}\right) $ is a direct sum of ${{\cal {G}}%
l}_{n_E}\left( {\cal R}\right) $ and ${\cal R}^{n_E},$ we can
write forms on ${\cal {A}}a\left( {\cal E}^{<z>}\right) $ as
$\Theta =\left( \Theta _1,\Theta _2\right) ,$ where $\Theta _1$
is the ${{\cal {G}}l}_{n_E}\left( {\cal R}\right) $ component and
$\Theta _2$ is the ${\cal R}^{n_E}$
component of the form $\Theta .$ Connection (7.22), $\Omega $ in ${{\cal {L}}%
a}\left( {\cal E}^{<z>}\right) ,$ induces the Cartan connection $\overline{%
\Omega }$ in ${{\cal {A}}a}\left( {\cal E}^{<z>}\right) ;$ see
the isotropic
case in 
 [195,196,40]. This is the unique connection on ${{\cal {A}}a}%
\left( {\cal E}^{<z>}\right) $ represented as $i^{*}\overline{\Omega }%
=\left( \Omega ,\chi \right) ,$ where $\chi $ is the shifting form and $i:{%
{\cal {A}}a}\rightarrow {{\cal {L}}a}$ is the trivial reduction
of bundles. If $s_{{\cal U}}^{(a)}$ is a local adapted frame in
${{\cal {L}}a}\left( {\cal E}^{<z>}\right) ,$ then
$\overline{s}_{{\cal U}}^{\left( 0\right)
}=i\circ s_{{\cal U}}$ is a local section in${{\cal {A}}a}\left( {\cal E}%
^{<z>}\right) $ and
$$
\left( \overline{\Omega }_{{\cal U}}\right) =s_{{\cal U}}\Omega
=\left( \Omega _{{\cal U}},\chi _{{\cal U}}\right) ,\eqno(7.24)
$$
$$
\left( \overline{{\cal R}}_{{\cal U}}\right) =s_{{\cal U}}\overline{{\cal R}}%
=\left( {\cal R}_{{\cal U}}^{(\Gamma )},T_{{\cal U}}\right) ,
$$
where $\chi =e_{\widehat{\alpha }}\otimes \chi _{\quad <\mu >}^{\widehat{%
\alpha }}X^{<\mu >},G_{<\alpha ><\beta >}=\chi _{\quad <\alpha >}^{\widehat{%
\alpha }}\chi _{\quad <\beta >}^{\widehat{\beta }}\eta _{\widehat{\alpha }%
\widehat{\beta }}\quad (\eta _{\widehat{\alpha }\widehat{\beta
}}$ is diagonal with $\eta _{\widehat{\alpha }\widehat{\alpha
}}=\pm 1)$ is a frame decomposition of metric (6.12) on ${\cal
E}^{<z>},e_{\widehat{\alpha }}$ is the standard distinguished
basis on ${\cal R}^{n_E},$ and the projection of torsion ,
$T_{{\cal U}},$ on base ${\cal E}^{<z>}$ is defined as
$$
T_{{\cal U}}=d\chi _{{\cal U}}+\Omega _{{\cal U}}\bigwedge \chi _{{\cal U}%
}+\chi _{{\cal U}}\bigwedge \Omega _{{\cal U}}=\eqno(7.25)
$$
$$
e_{\widehat{\alpha }}\otimes \sum\limits_{<\mu ><<\nu >}T_{\quad
<\mu ><\nu
>}^{\widehat{\alpha }}X^{<\mu >}\bigwedge X^{<\nu >}.
$$
For a fixed local adapted basis on ${\cal U}\subset {\cal
E}^{<z>}$ we can identify components $T_{\quad <\mu ><\nu
>}^{\widehat{a}}$ of torsion (7.25) with components of torsion
(6.25) on ${\cal E}^{<z>},$ i.e. $T_{\quad <\mu
><\nu >}^{\widehat{\alpha }}=T_{\quad <\mu ><\nu >}^{<\alpha >}.$ By
straightforward calculation we obtain
$$
{(\Delta \overline{{\cal R}})}_{{\cal U}}=[{(\Delta {\cal R}^{(\Gamma )})}_{%
{\cal U}},\ {(R\tau )}_{{\cal U}}+{(Ri)}_{{\cal U}}],\eqno(7.26)
$$
where%
$$
\left( R\tau \right) _{{\cal U}}=\widehat{\delta }_GT_{{\cal U}%
}+*_G^{-1}\left[ \Omega _{{\cal U}},*_GT_{{\cal U}}\right] ,\quad
\left(
Ri\right) _{{\cal U}}=*_G^{-1}\left[ \chi _{{\cal U}},*_G{\cal R}_{{\cal U}%
}^{(\Gamma )}\right] .
$$
Form $\left( Ri\right) _{{\cal U}}$ from (7.26) is locally
constructed by using components of the Ricci tensor (see (6.29))
as follows from decomposition on the local adapted basis $X^{<\mu
>}=\delta u^{<\mu >}:$
$$
\left( Ri\right) _{{\cal U}}=e_{\widehat{\alpha }}\otimes \left(
-1\right)
^{n_E+1}R_{<\lambda ><\nu >}G^{\widehat{\alpha }<\lambda >}\delta u^{<\mu >}%
$$

We remark that for isotropic torsionless pseudo-Riemannian spaces
the requirement that $\left( \Delta \overline{{\cal R}}\right)
_{{\cal U}}=0,$ i.e., imposing the connection (7.22) to satisfy
Yang-Mills equations (7.13) (equivalently (7.18) or (7.19) we
obtain
 [195,196,3] the equivalence of
the mentioned gauge gravitational equations with the vacuum
Einstein equations $R_{ij}=0.\,$ In the case of ha--spaces with
arbitrary given torsion, even considering vacuum gravitational
fields, we have to introduce a source for gauge gravitational
equations in order to compensate for the contribution of torsion
and to obtain equivalence with the Einstein equations.

Considerations presented in previous two subsections constitute
the\\ proof of the following result

\begin{theorem}
The Einstein equations (6.34) for la-gravity are equivalent to
Yang-Mills
equations%
$$
\left( \Delta \overline{{\cal R}}\right) =\overline{{\cal
J}}\eqno(7.27)
$$
for the induced Cartan connection $\overline{\Omega }$ (see
(7.22), (7.24)) \index{Cartan connection}
in the bundle of local adapted affine frames ${\cal A}a\left( {\cal E}%
\right) $ with source $\overline{{\cal J}}_{{\cal U}}$
constructed locally
by using the same formulas (7.26) (for $\left( \Delta \overline{{\cal R}}%
\right) $), where $R_{<\alpha ><\beta >}$ is changed by the matter source ${%
\tilde E}_{<\alpha ><\beta >}-\frac 12G_{<\alpha ><\beta
>}{\tilde E},$ where ${\tilde E}_{<\alpha ><\beta >}=kE_{<\alpha
><\beta >}-\lambda G_{<\alpha ><\beta >}.$
\end{theorem}

We note that this theorem is an extension for higher order
anisotropic
spaces of the Popov and Dikhin results 
 [196] with respect to a possible
gauge like treatment of the Einstein gravity. Similar theorems
have been \index{Gauge like!Einstein gravity}
proved for locally anisotropic gauge gravity 
 [258,259,272] and in
the framework of some variants of locally (and higher order)
anisotropic \index{Gauge gravity!locally anisotropic}
supergravity 
 [260,267].

\section{Nonlinear De Sitter Gauge Ha--Gravity}

The equivalent reexpression of the Einstein theory as a gauge
like theory implies, for both locally isotropic and anisotropic
space-times, the \index{Gauge Ha--gravity!De Sitter}
nonsemisimplicity of the gauge group, which leads to a
nonvariational theory in the total space of the bundle of locally
adapted affine frames. A variational gauge gravitational theory
can be formulated by using a minimal
extension of the affine structural group ${{\cal A}f}_{n_E}\left( {\cal R}%
\right) $ to the de Sitter gauge group $S_{n_E}=SO\left(
n_E\right) $ acting on distinguished ${\cal R}^{n_E+1}$ space.

\subsection{Nonlinear gauge theories of de Sitter group}

Let us consider the de Sitter space $\Sigma ^{n_E}$ as a
hypersurface given by the equations $\eta _{AB}u^Au^B=-l^2$ in
the (n+m)-dimensional spaces
enabled with diagonal metric $\eta _{AB},\eta _{AA}=\pm 1$ (in this section $%
A,B,C,...=1,2,...,n_E+1),(n_E=n+m_1+...+m_z),$ where $\{u^A\}$
are global Cartesian coordinates in ${\cal R}^{n_E+1};l>0$ is the
curvature of de Sitter space. The de Sitter group $S_{\left( \eta
\right) }=SO_{\left( \eta \right) }\left( n_E+1\right) $ is
defined as the isometry group of $\Sigma ^{n_E}$-space with
$\frac{n_E}2\left( n_E+1\right) $ generators of Lie
\index{Group!de Sitter} algebra ${{\it s}o}_{\left( \eta \right)
}\left( n_E+1\right) $ satisfying
the commutation relations%
$$
\left[ M_{<AB},M_{CD}\right] =\eta _{AC}M_{BD}-\eta
_{BC}M_{AD}-\eta _{AD}M_{BC}+\eta _{BD}M_{AC}.\eqno(7.28)
$$

Decomposing indices $A,B,...$ as $A=\left( \widehat{\alpha
},n_E+1\right) ,B=\left( \widehat{\beta },n_E+1\right) ,$ \\
$...,$ the metric $\eta _{AB}$ as $\eta _{AB}=\left( \eta
_{\widehat{\alpha }\widehat{\beta }},\eta _{\left( n_E+1\right)
\left( n_E+1\right) }\right) ,$ and operators $M_{AB}$
as $M_{\widehat{\alpha }\widehat{\beta }}={\cal F}_{\widehat{\alpha }%
\widehat{\beta }}$ and $P_{\widehat{\alpha
}}=l^{-1}M_{n_E+1,\widehat{\alpha }},$ we can write (7.28) as
$$
\left[ {\cal F}_{\widehat{\alpha }\widehat{\beta }},{\cal F}_{\widehat{%
\gamma }\widehat{\delta }}\right] =\eta _{\widehat{\alpha }\widehat{\gamma }}%
{\cal F}_{\widehat{\beta }\widehat{\delta }}-\eta _{\widehat{\beta }\widehat{%
\gamma }}{\cal F}_{\widehat{\alpha }\widehat{\delta }}+\eta
_{\widehat{\beta
}\widehat{\delta }}{\cal F}_{\widehat{\alpha }\widehat{\gamma }}-\eta _{%
\widehat{\alpha }\widehat{\delta }}{\cal F}_{\widehat{\beta
}\widehat{\gamma }},
$$
$$
\left[ P_{\widehat{\alpha }},P_{\widehat{\beta }}\right] =-l^{-2}{\cal F}_{%
\widehat{\alpha }\widehat{\beta }},\quad \left[ P_{\widehat{\alpha }},{\cal F%
}_{\widehat{\beta }\widehat{\gamma }}\right] =\eta _{\widehat{\alpha }%
\widehat{\beta }}P_{\widehat{\gamma }}-\eta _{\widehat{\alpha }\widehat{%
\gamma }}P_{\widehat{\beta }},
$$
where we have indicated the possibility to decompose ${{\it
s}o}_{\left( \eta \right) }\left( n_E+1\right) $ into a direct
sum, ${{\it s}o}_{\left( \eta \right) }\left( n_E+1\right) ={{\it
s}o}_{\left( \eta \right) }(n_E)\oplus V_{n_E},$ where $V_{n_E}$
is the vector space stretched on vectors $P_{\widehat{\alpha }}.$
We remark that $\Sigma ^{n_E}=S_{\left( \eta \right) }/L_{\left(
\eta \right) },$ where $L_{\left( \eta \right) }=SO_{\left( \eta
\right) }\left( n_E\right) .$ For $\eta _{AB}=diag\left(
1,-1,-1,-1\right) $ and $S_{10}=SO\left( 1,4\right) ,L_6=SO\left(
1,3\right) $ is the group of Lorentz rotations. \index{Group!of
Lorentz rotations}

Let $W\left( {\cal E},{\cal R}^{n_E+1},S_{\left( \eta \right)
},P\right) $ be the vector bundle associated with principal
bundle $P\left( S_{\left( \eta \right) },{\cal E}\right) $ on
la-spaces. The action of the structural group $S_{\left( \eta
\right) }$ on $E\,$ can be realized by using $\left( n_E\right)
\times \left( n_E\right) $ matrices with a parametrization
distinguishing subgroup $L_{\left( \eta \right) }:$%
$$
B=bB_L,\eqno(7.29)
$$
where%
$$
B_L=\left(
\begin{array}{cc}
L & 0 \\
0 & 1
\end{array}
\right) ,
$$
$L\in L_{\left( \eta \right) }$ is the de Sitter bust matrix
transforming the vector $\left( 0,0,...,\rho \right) \in {\cal
R}^{n_E+1}$ into the arbitrary point $\left(
V^1,V^2,...,V^{n_E+1}\right) \in \Sigma _\rho ^{n_E}\subset {\cal
R}^{n_E+1}$ with curvature $\rho \quad \left( V_AV^A=-\rho
^2,V^A=t^A\rho \right) .$ Matrix $b$ can be expressed as
$$
b=\left(
\begin{array}{cc}
\delta _{\quad \widehat{\beta }}^{\widehat{\alpha }}+\frac{t^{\widehat{%
\alpha }}t_{\widehat{\beta }}}{\left( 1+t^{n_E+1}\right) } & t^{
\widehat{\alpha }} \\ t_{\widehat{\beta }} & t^{n_E+1}
\end{array}
\right) .
$$

The de Sitter gauge field is associated with a linear connection
in $W$, i.e., with a ${{\it s}o}_{\left( \eta \right) }\left(
n_E+1\right) $-valued connection 1-form on ${\cal E}^{<z>}:$

$$
\widetilde{\Omega }=\left(
\begin{array}{cc}
\omega _{\quad \widehat{\beta }}^{\widehat{\alpha }} & \widetilde{\theta }^{%
\widehat{\alpha }} \\ \widetilde{\theta }_{\widehat{\beta }} & 0
\end{array}
\right) ,\eqno(7.30)
$$
where $\omega _{\quad \widehat{\beta }}^{\widehat{\alpha }}\in
so(n_E)_{\left( \eta \right) },$ $\widetilde{\theta
}^{\widehat{\alpha }}\in
{\cal R}^{n_E},\widetilde{\theta }_{\widehat{\beta }}\in \eta _{\widehat{%
\beta }\widehat{\alpha }}\widetilde{\theta }^{\widehat{\alpha }}.$

Because $S_{\left( \eta \right) }$-transforms mix $\omega _{\quad \widehat{%
\beta }}^{\widehat{\alpha }}$ and $\widetilde{\theta
}^{\widehat{\alpha }}$ fields in (7.30) (the introduced
para\-met\-ri\-za\-ti\-on is invariant on action on $SO_{\left(
\eta \right) }\left( n_E\right) $ group we cannot
identify $\omega _{\quad \widehat{\beta }}^{\widehat{\alpha }}$ and $%
\widetilde{\theta }^{\widehat{\alpha }},$ respectively, with the connection $%
\Gamma _{~<\beta ><\gamma >}^{<\alpha >}$ and the fundamental
form $\chi ^{<\alpha >}$ in ${\cal E}^{<z>}$ (as we have for
(7.22) and (7.24)). To
avoid this difficulty we consider 
 [240,194] a nonlinear gauge
realization of the de Sitter group $S_{\left( \eta \right) },$
namely, we
introduce into consideration the nonlinear gauge field%
\index{Nonlinear gauge field}
$$
\Omega =b^{-1}\Omega b+b^{-1}db=\left(
\begin{array}{cc}
\Gamma _{~\widehat{\beta }}^{\widehat{\alpha }} & \theta ^{
\widehat{\alpha }} \\ \theta _{\widehat{\beta }} & 0
\end{array}
\right) ,\eqno(7.31)
$$
where
$$
\Gamma _{\quad \widehat{\beta }}^{\widehat{\alpha }}=\omega _{\quad \widehat{%
\beta }}^{\widehat{\alpha }}-\left( t^{\widehat{\alpha }}Dt_{\widehat{\beta }%
}-t_{\widehat{\beta }}Dt^{\widehat{\alpha }}\right) /\left(
1+t^{n_E+1}\right) ,
$$
$$
\theta ^{\widehat{\alpha }}=t^{n_E+1}\widetilde{\theta }^{\widehat{\alpha }%
}+Dt^{\widehat{\alpha }}-t^{\widehat{\alpha }}\left( dt^{n_e+1}+\widetilde{%
\theta }_{\widehat{\gamma }}t^{\widehat{\gamma }}\right) /\left(
1+t^{n_E+1}\right) ,
$$
$$
Dt^{\widehat{\alpha }}=dt^{\widehat{\alpha }}+\omega _{\quad \widehat{\beta }%
}^{\widehat{\alpha }}t^{\widehat{\beta }}.
$$

The action of the group $S\left( \eta \right) $ is nonlinear,
yielding transforms\\ $\Gamma ^{\prime }=L^{\prime }\Gamma \left(
L^{\prime }\right) ^{-1}+L^{\prime }d\left( L^{\prime }\right)
^{-1},\theta ^{\prime }=L\theta , $ where the nonlinear
matrix-valued function\\ $L^{\prime }=L^{\prime }\left(
t^{<\alpha >},b,B_T\right) $ is defined from $B_b=b^{\prime
}B_{L^{\prime }}$ (see parametrization (7.29)).

Now, we can identify components of (7.31) with components of
$\Gamma
_{~<\beta ><\gamma >}^{<\alpha >}$ and $\chi _{\quad <\alpha >}^{\widehat{%
\alpha }}$ on ${\cal E}^{<z>}$ and induce in a consistent manner
on the base of bundle\\ $W\left( {\cal E},{\cal
R}^{n_E+1},S_{\left( \eta \right) },P\right) $ the la-geometry.

\subsection{Dynamics of the nonlinear $S\left( \eta \right) $ ha--gravity}

Instead of the gravitational potential (7.22), we introduce the
gravitational connection (similar to (7.31))
$$
\Gamma =\left(
\begin{array}{cc}
\Gamma _{\quad \widehat{\beta }}^{\widehat{\alpha }} &
l_0^{-1}\chi ^{ \widehat{\alpha }} \\ l_0^{-1}\chi
_{\widehat{\beta }} & 0
\end{array}
\right) \eqno(7.32)
$$
where
$$
\Gamma _{\quad \widehat{\beta }}^{\widehat{\alpha }}=\Gamma _{\quad \widehat{%
\beta }<\mu >}^{\widehat{\alpha }}\delta u^{<\mu >},
$$
$$
\Gamma _{\quad \widehat{\beta }<\mu >}^{\widehat{\alpha }}=\chi
_{\quad
<\alpha >}^{\widehat{\alpha }}\chi _{\quad <\beta >}^{\widehat{\beta }%
}\Gamma _{\quad <\beta ><\gamma >}^{<\alpha >}+\chi _{\quad <\alpha >}^{%
\widehat{\alpha }}\delta _{<\mu >}\chi _{\quad \widehat{\beta
}}^{<\alpha
>},
$$
$\chi ^{\widehat{\alpha }}=\chi _{\quad \mu }^{\widehat{\alpha
}}\delta
u^\mu ,$ and $G_{\alpha \beta }=\chi _{\quad \alpha }^{\widehat{\alpha }%
}\chi _{\quad \beta }^{\widehat{\beta }}\eta _{\widehat{\alpha }\widehat{%
\beta }},$ and $\eta _{\widehat{\alpha }\widehat{\beta }}$ is
parametrized as
$$
\eta _{\widehat{\alpha }\widehat{\beta }}=\left(
\begin{array}{cccc}
\eta _{ij} & 0 & ... & 0 \\
0 & \eta _{a_1b_1} & ... & 0 \\
... & ... & ... & ... \\
0 & 0 & ... & \eta _{a_zb_z}
\end{array}
\right) ,
$$
$\eta _{ij}=\left( 1,-1,...,-1\right) ,...\eta _{ij}=\left( \pm
1,\pm 1,...,\pm 1\right) ,...,l_0$ is a dimensional constant.

The curvature of (7.32), ${\cal R}^{(\Gamma )}=d\Gamma +\Gamma
\bigwedge
\Gamma ,$ can be written as%
$$
{\cal R}^{(\Gamma )}=\left(
\begin{array}{cc}
{\cal R}_{\quad \widehat{\beta }}^{\widehat{\alpha }}+l_0^{-1}\pi _{\widehat{%
\beta }}^{\widehat{\alpha }} & l_0^{-1}T^{ \widehat{\alpha }} \\
l_0^{-1}T^{\widehat{\beta }} & 0
\end{array}
\right) ,\eqno(7.33)
$$
where
$$
\pi _{\widehat{\beta }}^{\widehat{\alpha }}=\chi ^{\widehat{\alpha }%
}\bigwedge \chi _{\widehat{\beta }},{\cal R}_{\quad \widehat{\beta }}^{%
\widehat{\alpha }}=\frac 12{\cal R}_{\quad \widehat{\beta }<\mu ><\nu >}^{%
\widehat{\alpha }}\delta u^{<\mu >}\bigwedge \delta u^{<\nu >},
$$
and
$$
{\cal R}_{\quad \widehat{\beta }<\mu ><\nu >}^{\widehat{\alpha }}=\chi _{%
\widehat{\beta }}^{\quad <\beta >}\chi _{<\alpha >}^{\quad \widehat{\alpha }%
}R_{\quad <\beta ><\mu ><\nu >}^{<\alpha >}
$$
(see (6.27) and (6.28), the components of d-curvatures). The de
Sitter gauge group is semisimple and we are able to construct a
variational gauge gravitational locally anisotropic theory
(bundle metric (7.10) is nondegenerate). The Lagrangian of the
theory is postulated as
$$
L=L_{\left( G\right) }+L_{\left( m\right) }
$$
where the gauge gravitational Lagrangian is defined as
$$
L_{\left( G\right) }=\frac 1{4\pi }Tr\left( {\cal R}^{(\Gamma )}\bigwedge *_G%
{\cal R}^{(\Gamma )}\right) ={\cal L}_{\left( G\right) }\left|
G\right| ^{1/2}\delta ^{n_E}u,
$$
$$
{\cal L}_{\left( G\right) }=\frac 1{2l^2}T_{\quad <\mu ><\nu >}^{\widehat{%
\alpha }}T_{\widehat{\alpha }}^{\quad <\mu ><\nu >}+\eqno(7.34)
$$
$$
\frac 1{8\lambda }{\cal R}_{\quad \widehat{\beta }<\mu ><\nu >}^{\widehat{%
\alpha }}{\cal R}_{\quad \widehat{\alpha }}^{\widehat{\beta
}\quad <\mu
><\nu >}-\frac 1{l^2}\left( {\overleftarrow{R}}\left( \Gamma \right)
-2\lambda _1\right) ,
$$
$T_{\quad <\mu ><\nu >}^{\widehat{\alpha }}=\chi _{\quad <\alpha >}^{%
\widehat{\alpha }}T_{\quad <\mu ><\nu >}^{<\alpha >}$ (the
gravitational constant $l^2$ in (7.34) satisfies the relations
$l^2=2l_0^2\lambda ,\lambda _1=-3/l_0],\quad Tr$ denotes the
trace on $\widehat{\alpha },\widehat{\beta }
$ indices, and the matter field Lagrangian is defined as%
$$
L_{\left( m\right) }=-1\frac 12Tr\left( \Gamma \bigwedge *_G{\cal I}\right) =%
{\cal L}_{\left( m\right) }\left| G\right| ^{1/2}\delta ^{n_E}u,
$$
$$
{\cal L}_{\left( m\right) }=\frac 12\Gamma _{\quad \widehat{\beta }<\mu >}^{%
\widehat{\alpha }}S_{\quad <\alpha >}^{\widehat{\beta }\quad <\mu
>}-t_{\quad \widehat{\alpha }}^{<\mu >}l_{\quad <\mu >}^{\widehat{\alpha }}.%
\eqno(7.35)
$$
The matter field source ${\cal I}$ is obtained as a variational
derivation of ${\cal L}_{\left( m\right) }$ on $\Gamma $ and is
parametrized as
$$
{\cal I}=\left(
\begin{array}{cc}
S_{\quad \widehat{\beta }}^{\widehat{\alpha }} & -l_0t^{
\widehat{\alpha }} \\ -l_0t_{\widehat{\beta }} & 0
\end{array}
\right) \eqno(7.36)
$$
with $t^{\widehat{\alpha }}=t_{\quad <\mu >}^{\widehat{\alpha
}}\delta u^{<\mu >}$ and $S_{\quad \widehat{\beta
}}^{\widehat{\alpha }}=S_{\quad \widehat{\beta }<\mu
>}^{\widehat{\alpha }}\delta u^{<\mu >}$ being respectively the
canonical tensors of energy-momentum and spin density.
Because of the contraction of the ''interior'' indices $\widehat{\alpha },%
\widehat{\beta }$ in (7.34) and (7.35) we used the Hodge operator
$*_G$ instead of $*_H$ (hereafter we consider $*_G=*).$

Varying the action
$$
S=\int \left| G\right| ^{1/2}\delta ^{n_E}u\left( {\cal
L}_{\left( G\right) }+{\cal L}_{\left( m\right) }\right)
$$
on the $\Gamma $-variables (7.26), we obtain the
gauge-gravitational field
equations:%
$$
d\left( *{\cal R}^{(\Gamma )}\right) +\Gamma \bigwedge \left( *{\cal R}%
^{(\Gamma )}\right) -\left( *{\cal R}^{(\Gamma )}\right)
\bigwedge \Gamma =-\lambda \left( *{\cal I}\right) .\eqno(7.37)
$$

Specifying the variations on $\Gamma _{\quad \widehat{\beta }}^{\widehat{%
\alpha }}$ and $l^{\widehat{\alpha }}$-variables, we rewrite
(7.37) as
$$
\widehat{{\cal D}}\left( *{\cal R}^{(\Gamma )}\right) +\frac{2\lambda }{l^2}%
\left( \widehat{{\cal D}}\left( *\pi \right) +\chi \bigwedge
\left( *T^T\right) -\left( *T\right) \bigwedge \chi ^T\right)
=-\lambda \left( *S\right) ,\eqno(7.38)
$$
$$
\widehat{{\cal D}}\left( *T\right) -\left( *{\cal R}^{(\Gamma
)}\right)
\bigwedge \chi -\frac{2\lambda }{l^2}\left( *\pi \right) \bigwedge \chi =%
\frac{l^2}2\left( *t+\frac 1\lambda *\tau \right) ,\eqno(7.39)
$$
where
$$
T^t=\{T_{\widehat{\alpha }}=\eta _{\widehat{\alpha }\widehat{\beta }}T^{%
\widehat{\beta }},~T^{\widehat{\beta }}=\frac 12T_{\quad <\mu ><\nu >}^{%
\widehat{\beta }}\delta u^{<\mu >}\bigwedge \delta u^{<\nu >}\},
$$
$$
\chi ^T=\{\chi _{\widehat{\alpha }}=\eta _{\widehat{\alpha }\widehat{\beta }%
}\chi ^{\widehat{\beta }},~\chi ^{\widehat{\beta }}=\chi _{\quad <\mu >}^{%
\widehat{\beta }}\delta u^{<\mu >}\},\qquad \widehat{{\cal D}}=d+\widehat{%
\Gamma }
$$
($\widehat{\Gamma }$ acts as $\Gamma _{\quad \widehat{\beta }<\mu >}^{%
\widehat{\alpha }}$ on indices $\widehat{\gamma },\widehat{\delta
},...$ and as $\Gamma _{\quad <\beta ><\mu >}^{<\alpha >}$ on
indices $<\gamma
>,<\delta >,...).$ In (7.39), $\tau $ defines the energy-momentum tensor of
the $S_{\left( \eta \right) }$-gauge gravitational field $\widehat{\Gamma }:$%
$$
\tau _{<\mu ><\nu >}\left( \widehat{\Gamma }\right) =\frac 12Tr\left( {\cal R%
}_{<\mu ><\alpha >}{\cal R}_{\quad <\nu >}^{<\alpha >}-\frac 14{\cal R}%
_{<\alpha ><\beta >}{\cal R}^{<\alpha ><\beta >}G_{<\mu ><\nu >}\right) .%
\eqno(7.40)
$$

Equations (7.37) (or equivalently (7.38),(7.39)) make up the
complete system of variational field equations for nonlinear de
Sitter gauge gravity with higher order  anisotropy. They can be
interpreted as a generalization of
gauge like  equations for la-gravity 
 [272] (equivalently, of gauge
gravitational equations (7.27)] to a system of gauge field
equations with dynamical torsion and corresponding spin-density
source.

A. Tseytlin 
 [240] presented a quantum analysis of the isotropic version
of equations (7.38) and (7.39). Of course, the problem of
quantizing gravitational interactions is unsolved for both
variants of locally anisotropic and isotropic gauge de Sitter
gravitational theories, but we
think that the generalized Lagrange version of $S_{\left( \eta \right) }$%
-gravity is more adequate for studying quantum radiational and
statistical gravitational processes. This is a matter for further
investigations.

Finally, we remark that we can obtain a nonvariational Poincare
gauge gravitational theory on la-spaces if we consider the
contraction of the gauge potential (7.32) to a potential with
values in the Poincare Lie algebra
$$
\Gamma =\left(
\begin{array}{cc}
\Gamma _{\quad \widehat{\beta }}^{\widehat{\alpha }} &
l_0^{-1}\chi ^{ \widehat{\alpha }} \\ l_0^{-1}\chi
_{\widehat{\beta }} & 0
\end{array}
\right) \rightarrow \Gamma =\left(
\begin{array}{cc}
\Gamma _{\quad \widehat{\beta }}^{\widehat{\alpha }} &
l_0^{-1}\chi ^{ \widehat{\alpha }} \\ 0 & 0
\end{array}
\right) .
$$
Isotropic Poincare gauge gravitational theories are studied in a
number of
papers (see, for example, 
 285,240,155,194]). In a manner similar to
considerations presented in this work, we can generalize Poincare
gauge models for spaces with local anisotropy.

\section{La--Gravitational Gauge Instantons}

The existence of self-dual, or instanton, topologically
nontrivial solutions of Yang-Mills equations is a very important
physical consequence of gauge theories. All known instanton-type
Yang-Mills and gauge gravitational \index{Gauge
instantons!la--gravitational}
solutions (see, for example, 
 [240,194]) are locally isotropic. A
variational gauge-gravitational extension of la-gravity makes
possible a straightforward application of techniques of
constructing solutions for first order gauge equations for the
definition of locally anisotropic gravitational instantons. This
section is devoted to the study of some particular instanton
solutions of the gauge gravitational theory on la-space.

Let us consider the Euclidean formulation of the $S_{\left( \eta \right) }$%
-gauge gravitational theory by changing gauge structural groups
and flat
metric:%
$$
SO_{(\eta )}(n_E+1)\rightarrow SO(n_E+1),SO_{(\eta
)}(n_E)\rightarrow SO(n_E),\eta _{AB}\rightarrow -\delta _{AB}.
$$
Self-dual (anti-self-dual) conditions for the curvature (7.33)
$$
{\cal R}^{(\Gamma )}=*{\cal R}^{(\Gamma )}\quad (-*{\cal R}^{(\Gamma )})%
$$
can be written as a system of equations
$$
\left( {\cal R}_{\quad \widehat{\beta }}^{\widehat{\alpha
}}-l_0^{-2}\pi
_{\quad \widehat{\beta }}^{\widehat{\alpha }}\right) =\pm *\left( {\cal R}%
_{\quad \widehat{\beta }}^{\widehat{\alpha }}-l_0^{-2}\pi _{\quad \widehat{%
\beta }}^{\widehat{\alpha }}\right) \eqno(7.41)
$$
$$
T^{\widehat{\alpha }}=\pm *T^{\widehat{\alpha }}\eqno(7.42)
$$
(the ''-'' refers to the anti-self-dual case), where the ''-'' before $%
l_0^{-2}$ appears because of the transition of the Euclidean
negatively defined metric $-\delta _{<\alpha ><\beta >},$ which
leads to $\chi _{\quad
<\alpha >}^{\widehat{\alpha }}\rightarrow i\chi _{\quad <\alpha >|(E)}^{%
\widehat{\alpha }},\pi \rightarrow -\pi _{(E)}$ (we shall omit
the index $(E) $ for Euclidean values).

For solutions of (7.41) and (7.42) the energy-momentum tensor
(7.40) is identically equal to zero. Vacuum equations (7.37) and
(7.38), when source is ${\cal I\equiv 0}$ (see (7.36)), are
satisfied as a consequence of generalized Bianchi identities for
the curvature (7.33). The mentioned solutions of (7.41) and
(7.42) realize a local minimum of the Euclidean
action%
$$
S=\frac 1{8\lambda }\int \left| G^{1/2}\right| \delta
^{n_E}\{(R\left( \Gamma \right) -l_0^{-2}\pi )^2+2T^2\},
$$
where $T^2=T_{\quad <\mu ><\nu >}^{\widehat{\alpha }}T_{\widehat{\alpha }%
}^{\quad <\mu ><\nu >}$ is extremal on the topological invariant
(Pontryagin
index)%
$$
p_2=-\frac 1{8\pi ^2}\int Tr\left( {\cal R}^{(\Gamma )}\bigwedge {\cal R}%
^{(\Gamma )}\right) =--\frac 1{8\pi ^2}\int Tr\left( \widehat{{\cal R}}%
\bigwedge \widehat{{\cal R}}\right) .
$$

For the Euclidean de Sitter spaces, when%
$$
{\cal R}=0\quad \{T=0,R_{\quad <\mu ><\nu >}^{\widehat{\alpha }\widehat{%
\beta }}=-\frac 2{l_0^2}\chi _{\quad <\mu >}^{[\alpha }\chi
_{\quad <\nu
>}^{\beta ]}\}\eqno(7.43)
$$
we obtain the absolute minimum, $S=0.$

We emphasize that for $$R_{<\beta >\quad <\mu ><\beta >}^{\quad
<\alpha
>}=\left( 2/l_0^2\right) \delta _{[<\mu >}^{<\alpha >}G_{<\nu >]<\beta >}$$
torsion vanishes. Torsionless instantons also have another
interpretation.\ For $$T_{\quad <\beta ><\gamma >}^{<\alpha
>}=0$$ contraction of equations (7.41) leads to Einstein
equations with cosmological $\lambda $-term (as a
consequence of generalized Ricci identities):%
$$
R_{<\alpha ><\beta ><\mu ><\nu >}-R_{<\mu ><\nu ><\alpha ><\beta
>}=\frac 32\{R_{[<\alpha ><\beta ><\mu >]<\nu >}-
$$
$$
R_{[<\alpha ><\beta ><\nu >]<\mu >}+R_{[<\mu ><\nu ><\alpha
>]<\beta
>}-R_{[<\mu ><\nu ><\beta >]<\alpha >}\}.
$$
So, in the Euclidean case the locally anisotropic vacuum Einstein
equations are a subset of instanton solutions.

Now, let us study the $SO\left( n_E\right) $ solution of
equations (7.41) and\ (7.42). We consider the spherically
symmetric ansatz (in order to point out the connection between
high-dimensional gravity and la-gravity the N-connection
structure is chosen to be trivial, i.e. ${N}_j^a\left( u\right)
\equiv 0):$%
$$
\Gamma _{\quad \widehat{\beta }<\mu >}^{\widehat{\alpha
}}=a\left( u\right)
\left( u^{\widehat{\alpha }}\delta _{\widehat{\beta }<\mu >}-u_{\widehat{%
\beta }}\delta _{<\mu >}^{\widehat{\alpha }}\right) +q\left(
u\right) \epsilon _{\quad \widehat{\beta }<\mu ><\nu
>}^{\widehat{\alpha }}u^{<\mu
>},
$$
$$
\chi _{<\alpha >}^{\widehat{\alpha }}=f\left( u\right) \delta
_{\quad
<\alpha >}^{\widehat{\alpha }}+n\left( u\right) u^{\widehat{\alpha }%
}u_\alpha ,\eqno(7.44)
$$
where $u=u^{<\alpha >}u^{<\beta >}G_{<\alpha ><\beta >}=x^{\widehat{i}}x_{%
\widehat{i}}+y^{\widehat{a}}y_{\widehat{a}},$ and $a\left(
u\right) ,q\left( u\right) ,f\left( u\right) $ and $n\left(
u\right) $ are some scalar functions. Introducing (7.44) into
(7.41) and (7.42), we obtain,
respectively,%
$$
u\left( \pm \frac{dq}{du}-a^2-q^2\right) +2\left( a\pm q\right) +l_0^{-1}f^2,%
\eqno(7.45)
$$
$$
2d\left( a\mp q\right) /du+\left( a\mp q\right)
^2-l_0^{-1}fn=0,\eqno(7.46)
$$
$$
2\frac{df}{du}+f\left( a\mp 2q\right) +n\left( au-1\right)
=0.\eqno(7.47)
$$

The traceless part of the torsion vanishes because of the
parametrization (7.44), but in the general case the trace and
pseudo-trace of the torsion
are not identical to zero:%
$$
T^\mu =q^{\left( 0\right) }u^\mu \left( -2df/du+n-a\left(
f+un\right) \right) ,
$$
$$
\overbrace{T}^\mu =q^{\left( 1\right) }u^\mu \left( 2qf\right) ,
$$
$q^{\left( 0\right) }$ and $q^{\left( 0\right) }$ are constant.
Equation
(7.42) or (7.47) establishes the proportionality of $T^\mu $ and $\overbrace{%
T}^\mu .$ As a consequence we obtain that the $SO\left(
n+m\right) $ solution of (3.52) is torsionless if $q\left(
u\right) =0$ of $f\left( u\right) =0.$

Let first analyze the torsionless instantons, $T_{\quad <\alpha
><\beta
>}^{<\mu >}=0.$ If $f=0,$ then from (7.46) one has two possibilities: (a) $%
n=0$ leads to nonsense because $\chi _\alpha ^{\widehat{\alpha }}=0$ or $%
G_{\alpha \beta }=0.$ b) $a=u^{-1}$ and $n\left( u\right) $ is an
arbitrary
scalar function; we have from (7.46) $a\mp q=2/\left( a+C^2\right) $ or $%
q=\pm 2/u\left( u+C^2\right) ,$ where $C=const.$ If $q\left(
u\right) =0,$ we obtain the de Sitter space (7.43) because
equations (7.45) and (7.46)
impose vanishing of both self-dual and anti-self-dual parts of $\left( {\cal %
R}_{\quad \widehat{\beta }}^{\widehat{\alpha }}-l_0^2\pi _{\quad \widehat{%
\beta }}^{\widehat{\alpha }}\right) ,$ so, as a consequence, ${\cal R}%
_{\quad \widehat{\beta }}^{\widehat{\alpha }}-l_0^2\pi _{\quad \widehat{%
\beta }}^{\widehat{\alpha }}\equiv 0.$ There is an infinite number of $%
SO\left( n_E\right) $-symmetric solutions of (7.43):%
$$
f=l_0\left[ a\left( 2-au\right) \right] ^{1/2},\quad n=l_0\{2\frac{da}{du}+%
\frac{a^2}{\left[ a\left( 2-au\right) \right] ^{1/2}}\},
$$
$a(u)$ is a scalar function.

To find instantons with torsion, $T_{\quad \beta \gamma }^\alpha
\neq 0,$ is also possible. We present the $SO\left( 4\right) $
one-instanton solution,
obtained in 
 [194] (which in the case of $H^4$-space parametrized by
local coordinates $\left( x^{\widehat{1}},x^{\widehat{2}},y^{\widehat{1}},y^{%
\widehat{2}}\right) ,$ with $u=x^{\widehat{1}}x_{\widehat{1}}+x^{\widehat{2}%
}x_{\widehat{2}}+y^{\widehat{1}}y_{\widehat{1}}+y^{\widehat{2}}y_{\widehat{2}%
}):$%
$$
a=a_0\left( u+c^2\right) ^{-1},q=\mp q_0\left( u+c^2\right) ^{-1}
$$
$$
f=l_0\left( \alpha u+\beta \right) ^{1/2}/\left( u+c^2\right)
,n=c_0/\left( u+c^2\right) \left( \gamma u+\delta \right) ^{1/2}
$$
where%
$$
a_0=-1/18,q_0=5/6,\alpha =266/81,\beta =8/9,$$ $$\gamma
=10773/11858,\delta =1458/5929.
$$
We suggest that local regions with $$T_{\ \beta \gamma }^\alpha
\neq 0$$ are similarly to Abrikosv vortexes in superconductivity
and the appearance of torsion is a possible realization of the
Meisner effect in gravity (for details and discussions on the
superconducting or Higgs-like interpretation
of gravity see 
 [240,194]). In our case we obtain a locally anisotropic
superconductivity and we think that the formalism of gauge locally
anisotropic theories may even work for some models of anisotropic
low- and
high-temperature superconductivity 
 [248].

\section{Remarks}

In this Chapter we have reformulated the fiber bundle formalism
for both Yang-Mills and gravitational fields in order to include
into consideration space-times with higher order anisotropy. We
have argued that our approach has the advantage of making
manifest the relevant structures of the theories with local
anisotropy and putting greater emphasis on the analogy with
anisotropic models than the standard coordinate formulation in
Finsler geometry.

Our geometrical approach to higher order anisotropic gauge and
gravitational interactions are refined in such a way as to shed
light on some of the more specific properties and common and
distinguishing features of the Yang-Mills and Einstein fields. As
we have shown, it is possible to make a gauge like treatment for
both models with local anisotropy (by using correspondingly
defined linear connections in bundle spaces with semisimple
structural groups, with variants of nonlinear realization and
extension to semisimple structural groups, for gravitational
fields). Here, we note that same geometric approach holds good
for superbundles; some models of gauge higher order anisotropic
supergravity proposed in Chapter 2 are constructed in a similar
manner as for gauge la-fields.

We have extended our models proposed in 
 [272,258,259] for
generalized Lagrange spaces to the case on ha-spaces modeled as
 higher order vector
bundles provided with N-connection, d-connection and d-metric
structures. The same geometric machinery can be used for
developing of gauge field theories on spaces with higher order
anisotropy
 [162,295] or on Jet
bundles 
 [19].



\chapter{Na-Maps and Conservation Laws}

Theories of field interactions on locally anisotropic curved
spaces form a new branch of modern theoretical and mathematical
physics. They are used for modelling in a self--consistent manner
physical processes in locally anisotropic, stochastic and
turbulent media with beak radiational reaction and diffusion
 [161,13,14,282]. The first model of locally
anisotropic space was proposed by P.Finsler 
 [78] as a generalization
of  Riemannian geometry; here we also cite the fundamental
contribution made by E. Cartan
 [55] and mention that in monographs
 [213,159,17,19,29]
detailed  bibliographies are contained. In this Chapter we follow
R. Miron and M. Anastasiei
 [160,161] conventions and
base our investigations on their general model of locally
anisotropic (la) gravity (in brief we shall write la-gravity) on
vector bundles, v--bundles, provided with nonlinear and
distinguished connection and metric structures (we call a such
type of v--bundle as a la-space if connections and metric are
compatible).

The study of models of classical and quantum field interactions
on la-spaces is in order of the day. For instance, the problem of
definition of spinors
on la-spaces is already solved (see 
 [256,275,264] and Chapter
6 and some models of locally anisotropic Yang--Mills and gauge
like
gravitational interactions are analyzed (see 
 [272,263] and Chapter 7
and alternative approaches in
 [17,29,122,119]). The development of
this direction entails great difficulties because of
problematical character of the possibility and manner of
definition of conservation laws on la-spaces. It will be recalled
that, for instance, the conservation laws of energy--momentum
type are a consequence of existence of a global group of
automorphisms of the fundamental Mikowski spaces (for
(pseudo)Riemannian spaces the tangent space' automorphisms and
particular cases when there are symmetries generated by existence
of Killing vectors are considered). No global or local
automorphisms exist on generic la-spaces and in result of this
fact the formulation of la-conservation laws is sophisticate and
full of ambiguities. R. Miron and M. Anastasiei firstly pointed
out the nonzero divergence of the matter energy-momentum
d--tensor, the source in Einstein equations on la-spaces, and
considered an original approach to the geometry of
time--dependent Lagrangians
 [12,160,161]. Nevertheless, the
rigorous definition of energy-momentum values for
la-gravitational and matter fields and the form of conservation
laws for such values have not been considered in present--day
studies of the mentioned problem.

The aim of this Chapter is to develop a necessary geometric
background (the theory of nearly autoparallel maps, in brief
na-maps,  and tensor integral formalism on la-multispaces) for
formulation and a detailed investigation of conservation laws on
locally isotropic and anisotropic curved spaces. We shall
summarize our results on formulation of na-maps for generalized
affine
spaces (GAM-spaces) 
 [249,251,273], Einstein-Cartan and Einstein spaces
  [250,247,278] bundle spaces 
  [250,247,278] and different
classes of la-spaces [279,276,102,263]
 and present an extension of
the na-map theory for superspaces. For simplicity we shall
restrict our considerations only with the "first" order
anisotropy (the basic  results on higher order anisotropies are
presented in a supersymmetric manner in Chapter 3. Comparing the
geometric constructions from both Chapters on  na--map theory we
assure ourselves that the developed methods hold good  for all
type of curved spaces (with or not torsion, locally isotropic or
even with local anisotropy and being, or not, supesymmetric). In
order  to make the reader more familiar with na--maps and theirs
applications  and to point to some common features, as well to
proper particularities of  the supersymmetric and higher order
constructions, we shall recirculate some basic definitions,
theorems and proofs from Chapter 3.

The question of definition of tensor integration as the inverse
operation of
covariant derivation was posed and studied by A.Mo\'or 
 [167].
Tensor--integral and bitensor formalisms turned out to be very
useful in solving certain problems connected with conservation
laws in general relativity
 [100,247].  In order to extend tensor--integral
constructions we have proposed 
 [273,278] to  take into consideration
nearly autoparallel
 [249,247,250] and nearly geodesic 
 [230] maps, ng--maps,
 which forms a subclass of local 1--1 maps of  curved spaces
with deformation of the connection and metric structures. A
generalization of the Sinyukov's ng--theory for spaces with local
anisotropy was proposed by considering maps with deformation of
connection for Lagrange spaces (on Lagrange spaces see
 [136,160,161]) and generalized Lagrange spaces
 [263,279,276,275,101]. Tensor integration formalism for  generalized
Lagrange spaces was developed in
[255,102,263]. One of the main purposes of this Chapter is to
synthesize the results obtained in the mentioned  works and to
formulate them for a very general class of la--spaces. As the
next step the la--gravity and analysis of la--conservation laws
are considered.

We note that proofs of our theorems are mechanical, but, in most
cases, they are rather tedious calculations similar to those
presented in
 [230,252,263]. Some of them, on la-spaces, will be given in detail the
rest, being similar, or consequences, will be only sketched or
omitted.

Section 8.1 is  devoted to the formulation of the theory of nearly
autoparallel maps of  la--spaces. The classification of na--maps
and formulation of their invariant conditions are given in
section 8.2. In section 8.3 we define the nearly  autoparallel
tensor--integral on locally anisotropic multispaces. The problem
of formulation of conservation laws on spaces with local
anisotropy is studied in section 8.4 . We present a definition of
conservation laws for la--gravitational fields on na--images of
la--spaces in section 8.5. Finally,  in this Chapter, section
8.6, we analyze the locally isotropic limit, to  the Einstein
gravity and it generalizations, of the na-conservation laws.

\section{Nearly Autoparallel Maps of La--Spaces}

In this section we shall extend the ng-- [230] 
 [249,250,252,273,274,278] theory by introducing into consideration maps
of vector bundles provided with compatible N--connection,
d--connection and metric structures.

Our geometric arena consists from pairs of open regions $( U,
{\underline U})
$ of la--spaces, $U{\subset}{\xi},\, {\underline U}{\subset}{\underline {\xi}%
}$, and 1--1 local maps $f : U{\to}{\underline U}$ given by functions $%
f^{a}(u)$ of smoothly class $C^r( U) \, (r>2, $ or $r={\omega}$~
for
analytic functions) and their inverse functions $f^{\underline a}({%
\underline u})$ with corresponding non--zero Jacobians in every
point $u{\in} U$ and ${\underline u}{\in}{\underline U}.$

We consider that two open regions $U$~ and ${\underline U}$~ are
attributed to a common for f--map coordinate system if this map
is realized on the principle of coordinate equality
$q(u^{\alpha}) {\to} {\underline q} (u^{\alpha})$~ for every
point $q {\in} U$~ and its f--image ${\underline q} {\in}
{\underline U}.$ We note that all calculations included in this
work will be local in nature and taken to refer to open subsets
of mappings of
type
 ${\xi} {\supset} U {\longrightarrow } {\underline U}
 {\subset} {\underline {\xi}}.$
 For simplicity, we suppose that in
a fixed
common coordinate system for $U$ and ${\underline U}$ spaces $\xi$ and ${%
\underline {\xi}}$ are characterized by a common N--connection
structure (in consequence of (6.10) by a corresponding
concordance of d--metric structure), i.e.
$$
N^{a}_{j}(u)={\underline N}^{a}_{j}(u)={\underline N}^{a}_{j} ({\underline u}%
),%
$$
which leads to the possibility to establish common local bases,
adapted to a given N--connection, on both regions $U$ and
${\underline U.}$ We consider that on $\xi$ it is defined the
linear d--connection structure with
components ${\Gamma}^{\alpha}_{{.}{\beta}{\gamma}}.$ On the space $%
\underline {\xi}$ the linear d--connection is considered to be a
general one with torsion
$$
{\underline T}^{\alpha}_{{.}{\beta}{\gamma}}={\underline
{\Gamma}}^{\alpha}_ {{.}{\beta}{\gamma}}-{\underline
{\Gamma}}^{\alpha}_{{.}{\gamma}{\beta}}+
w^{\alpha}_{{.}{\beta}{\gamma}}
$$
and nonmetricity
$$
{\underline K}_{{\alpha}{\beta}{\gamma}}={{\underline D}_{\alpha}} {%
\underline G}_{{\beta}{\gamma}}. \eqno(8.1)
$$

Geometrical objects on ${\underline {\xi}}$ are specified by
underlined
symbols (for example, ${\underline A}^{\alpha}, {\underline B}^{{\alpha}{%
\beta}})$~ or underlined indices (for example, $A^{\underline a}, B^{{%
\underline a}{\underline b}}).$

For our purposes it is convenient to introduce auxiliary
sym\-met\-ric
d--con\-nec\-ti\-ons, ${\gamma}^{\alpha}_{{.}{\beta}{\gamma}}={\gamma}%
^{\alpha}_{{.}{\gamma}{\beta}} $~ on $\xi$ and ${\underline {\gamma}}%
^{\alpha}_{.{\beta}{\gamma}}= {\underline {\gamma}}^{\alpha}_{{.}{\gamma}{%
\beta}}$ on ${\underline {\xi}}$ defined, correspondingly, as
$$
{\Gamma}^{\alpha}_{{.}{\beta}{\gamma}}= {\gamma}^{\alpha}_{{.}{\beta}{\gamma}%
}+  T^{\alpha}_{{.}{\beta}{\gamma}}\quad {\rm \; and}\quad
{\underline
{\Gamma}}^{\alpha}_{{.}{\beta}{\gamma}}= {\underline {\gamma}}^{\alpha}_{{.}{%
\beta}{\gamma}}+ {\underline T}^{\alpha}_{{.}{\beta}{\gamma}}.%
$$

We are interested in definition of local 1--1 maps from $U$ to ${\underline U%
}$ characterized by symmetric, $P^{\alpha}_{{.}{\beta}{\gamma}},$
and antisymmetric, $Q^{\alpha}_{{.}{\beta}{\gamma}}$,~
deformations:
$$
{\underline {\gamma}}^{\alpha}_{{.}{\beta}{\gamma}} ={\gamma}^{\alpha}_{{.}{%
\beta}{\gamma}}+ P^{\alpha}_{{.}{\beta}{\gamma}} \eqno(8.2)%
$$
and
$$
{\underline T}^{\alpha}_{{.}{\beta}{\gamma}}= T^{\alpha}_{{.}{\beta}{\gamma}%
}+ Q^{\alpha}_{{.}{\beta}{\gamma}}. \eqno(8.3)
$$
The auxiliary linear covariant derivations induced by ${\gamma}^{\alpha}_{{.}%
{\beta}{\gamma}}$ and ${\underline
{\gamma}}^{\alpha}_{{.}{\beta}{\gamma}}$~
are denoted respectively as $^{({\gamma})}D$~ and $^{({\gamma})}{\underline D%
}.$~

Let introduce this local coordinate parametrization of curves on
$U$~:
$$
u^{\alpha}=u^{\alpha}({\eta})=(x^{i}({\eta}), y^{i}({\eta}),~{\eta}_{1}<{\eta%
}<{\eta}_{2},%
$$
where corresponding tangent vector field is defined as
$$
v^{\alpha}={\frac{{du^{\alpha}} }{d{\eta}}}= ({\frac{{dx^{i}({\eta})} }{{d{%
\eta}}}}, {\frac{{dy^{j}({\eta})} }{d{\eta}}}).%
$$

\begin{definition}
Curve $l$~ is called auto parallel, a--parallel, on $\xi $ if its
tangent vector field $v^\alpha $~ satisfies a--parallel equations:
$$
vDv^\alpha =v^\beta {^{({\gamma })}D}_\beta v^\alpha ={\rho }({\eta }%
)v^\alpha ,\eqno(8.4)
$$
where ${\rho }({\eta })$~ is a scalar function on $\xi $.
\end{definition}
\index{Auto parallel!a--parallel}

Let curve ${\underline l} {\subset} {\underline {\xi}}$ is given
in
parametric form as $u^{\alpha}=u^{\alpha}({\eta}),~{\eta}_1 < {\eta} <{\eta}%
_2$ with tangent vector field $v^{\alpha} = {\frac{{du^{\alpha}} }{{d{\eta}}}%
} {\ne} 0.$ We suppose that a 2--dimensional distribution $E_2({\underline l}%
)$ is defined along ${\underline l} ,$ i.e. in every point $u {\in} {%
\underline l}$ is fixed a 2-dimensional vector space $E_{2}({\underline l}) {%
\subset} {\underline {\xi}}.$ The introduced distribution
$E_{2}({\underline
l})$~ is coplanar along ${\underline l}$~ if every vector ${\underline p}%
^{\alpha}(u^{b}_{(0)}) {\subset} E_{2}({\underline l}), u^{\beta}_{(0)} {%
\subset} {\underline l}$~ rests contained in the same
distribution after
parallel transports along ${\underline l},$~ i.e. ${\underline p}%
^{\alpha}(u^{\beta}({\eta})) {\subset} E_{2} ({\underline l}).$

\begin{definition}
A curve ${\underline{l}}$~ is called nearly autoparallel, or in
brief an
na--parallel, on space ${\underline{\xi }}$~ if a coplanar along ${%
\underline{l}}$~ distribution $E_2({\underline{l}})$ containing tangent to ${%
\underline{l}}$~ vector field $v^\alpha ({\eta })$,~ i.e. $v^\alpha ({\eta })%
{\subset }E_2({\underline{l}}),$~ is defined.
\end{definition}
\index{Nearly autoparallel!na--parallel} We can define nearly
autoparallel maps of la--spaces as an anisotropic
generalization (see also 
ng--
 [230]  and na--maps 
 [249,273,278,274,247]:

\begin{definition}
Nearly autoparallel maps, na--maps, of la--spaces are defined as
local 1--1 mappings of v--bundles, $\xi {\to }{\underline{\xi
}},$ changing every a--parallel on $\xi $ into a na--parallel on
${\underline{\xi }}.$
\end{definition}
\index{nearly autoparallel maps!na--maps}

Now we formulate the general conditions when deformations (8.2)
and (8.3) charac\-ter\-ize na-maps : Let a-parallel $l{\subset
}U$~ is given by
func\-ti\-ons\\ $u^\alpha =u^{({\alpha })}({\eta }),v^\alpha ={\frac{{%
du^\alpha }}{d{\eta }}}$, ${\eta }_1<{\eta }<{\eta }_2$,
satisfying equations (8.4). We suppose that to this a--parallel
corresponds a na--parallel ${\underline{l}}\subset
{\underline{U}}$ given by the same parameterization in a common
for a chosen na--map coordinate system on $U$~
and ${\underline{U}}.$ This condition holds for vectors ${\underline{v}}%
_{(1)}^\alpha =v{\underline{D}}v^\alpha $~ and $v_{(2)}^\alpha =v{\underline{%
D}}v_{(1)}^\alpha $ satisfying equality
$$
{\underline{v}}_{(2)}^\alpha ={\underline{a}}({\eta })v^\alpha +{\underline{b%
}}({\eta }){\underline{v}}_{(1)}^\alpha \eqno(8.5)
$$
for some scalar functions ${\underline{a}}({\eta })$~ and ${\underline{b}}({%
\eta })$~ (see Definitions 8.4 and 8.5). Putting splittings (8.2)
and (8.3)
into expressions for $${\underline{v}}_{(1)}^\alpha \mbox{ and }{%
\underline{v}}_{(2)}^\alpha $$ in (8.5) we obtain:
$$
v^\beta v^\gamma v^\delta (D_\beta P_{{.}{\gamma }{\delta }}^\alpha +P_{{.}{%
\beta }{\tau }}^\alpha P_{{.}{\gamma }{\delta }}^\tau +Q_{{.}{\beta }{\tau }%
}^\alpha P_{{.}{\gamma }{\delta }}^\tau )=bv^\gamma v^\delta P_{{.}{\gamma }{%
\delta }}^\alpha +av^\alpha ,\eqno(8.6)
$$
where
$$
b({\eta },v)={\underline{b}}-3{\rho },\qquad \mbox{and}\qquad a({\eta },v)={%
\underline{a}}+{\underline{b}}{\rho }-v^b{\partial }_b{\rho }-{\rho }^2%
\eqno(8.7)
$$
are called the deformation parameters of na--maps.
\index{Na--maps!deformation parameters}

The algebraic equations for the deformation of torsion $Q_{{.}{\beta }{\tau }%
}^\alpha $ should be written as the compatibility conditions for
a given
nonmetricity tensor ${\underline{K}}_{{\alpha }{\beta }{\gamma }}$~ on ${%
\underline{\xi }}$ ( or as the metricity conditions if d--connection ${%
\underline{D}}_\alpha $~ on ${\underline{\xi }}$~ is required to be metric) :%
$$
D_\alpha G_{{\beta }{\gamma }}-P_{{.}{\alpha }({\beta }}^\delta G_{{{\gamma }%
)}{\delta }}-{\underline{K}}_{{\alpha }{\beta }{\gamma }}=Q_{{.}{\alpha }({%
\beta }}^\delta G_{{\gamma }){\delta }},\eqno(8.8)
$$
where $({\quad })$ denotes the symmetric alternation.

So, we have proved this

\begin{theorem}
The na--maps from la--space $\xi $ to la--space ${\underline{\xi
}}$~ with a fixed common nonlinear connection
$N_j^a(u)={\underline{N}}_j^a(u)$ and given d--connections,
${\Gamma }_{{.}{\beta }{\gamma }}^\alpha $~ on $\xi $~
and ${\underline{\Gamma }}_{{.}{\beta }{\gamma }}^\alpha $~ on ${\underline{%
\xi }}$ are locally parametrized by the solutions of equations
(8.6) and
(8.8) for every point $u^\alpha $~ and direction $v^\alpha $~ on $U{\subset }%
{\xi }.$
\end{theorem}

We call (8.6) and (8.8) the basic equations for na--maps of
la--spaces. They \index{Na--maps!basic equations}
generalize the corresponding Sinyukov's equations 
 [230] for isotropic
spaces provided with symmetric affine connection structure.

\section{ Classification of Na--Maps }

Na--maps are classed on possible polynomial parametrizations on variables $%
v^{\alpha}$~ of deformations parameters $a$ and $b$ (see (8.6)
and (8.7) ).

\begin{theorem}
There are four classes of na--maps characterized by corresponding
deformation parameters and tensors and basic equations:
\index{Na--maps!classification}

\begin{enumerate}
\item  for $na_{(0)}$--maps, ${\pi }_{(0)}$--maps,
$$
P_{{\beta }{\gamma }}^\alpha (u)={\psi }_{{(}{\beta }}{\delta }_{{\gamma }%
)}^\alpha
$$
(${\delta }_\beta ^\alpha $~ is Kronecker symbol and ${\psi }_\beta ={\psi }%
_\beta (u)$~ is a covariant vector d--field);

\item  for $na_{(1)}$--maps
$$
a(u,v)=a_{{\alpha }{\beta }}(u)v^\alpha v^\beta ,\quad
b(u,v)=b_\alpha (u)v^\alpha
$$
and $P_{{.}{\beta }{\gamma }}^\alpha (u)$~ is the solution of
equations
$$
D_{({\alpha }}P_{{.}{\beta }{\gamma })}^\delta +P_{({\alpha
}{\beta }}^\tau
P_{{.}{\gamma }){\tau }}^\delta -P_{({\alpha }{\beta }}^\tau Q_{{.}{\gamma })%
{\tau }}^\delta =b_{({\alpha }}P_{{.}{\beta }{\gamma })}^{{\delta }}+a_{({%
\alpha }{\beta }}{\delta }_{{\gamma })}^\delta ;\eqno(8.9)
$$

\item  for $na_{(2)}$--maps
$$
a(u,v)=a_\beta (u)v^\beta ,\quad b(u,v)={\frac{{b_{{\alpha }{\beta }%
}v^\alpha v^\beta }}{{{\sigma }_\alpha (u)v^\alpha }}},\quad {\sigma }%
_\alpha v^\alpha {\neq }0,
$$
$$
P_{{.}{\alpha }{\beta }}^\tau (u)={{\psi }_{({\alpha }}}{\delta }_{{\beta }%
)}^\tau +{\sigma }_{({\alpha }}F_{{\beta })}^\tau
$$
and $F_\beta ^\alpha (u)$~ is the solution of equations
$$
{D}_{({\gamma }}F_{{\beta })}^\alpha +F_\delta ^\alpha
F_{({\gamma }}^\delta
{\sigma }_{{\beta })}-Q_{{.}{\tau }({\beta }}^\alpha F_{{\gamma })}^\tau ={%
\mu }_{({\beta }}F_{{\gamma })}^\alpha +{\nu }_{({\beta }}{\delta
}_{{\gamma })}^\alpha \eqno(8.10)
$$
$({\mu }_\beta (u),{\nu }_\beta (u),{\psi }_\alpha (u),{\sigma }_\alpha (u)$%
~ are covariant d--vectors);

\item  for $na_{(3)}$--maps
$$
b(u,v)={\frac{{{\alpha }_{{\beta }{\gamma }{\delta }}v^\beta
v^\gamma v^\delta }}{{{\sigma }_{{\alpha }{\beta }}v^\alpha
v^\gamma }}},
$$
$$
P_{{.}{\beta }{\gamma }}^\alpha (u)={\psi }_{({\beta }}{\delta }_{{\gamma }%
)}^\alpha +{\sigma }_{{\beta }{\gamma }}{\varphi }^\alpha ,
$$
where ${\varphi }^\alpha $~ is the solution of equations

$$
D_\beta {\varphi }^\alpha ={\nu }{\delta }_\beta ^\alpha +{\mu }_\beta {%
\varphi }^\alpha +{\varphi }^\gamma Q_{{.}{\gamma }{\delta }}^\alpha ,%
\eqno(8.11)
$$
${\alpha }_{{\beta }{\gamma }{\delta }}(u),{\sigma }_{{\alpha }{\beta }}(u),{%
\psi }_\beta (u),{\nu }(u)$~ and ${\mu }_\beta (u)$~ are
d--tensors.
\end{enumerate}
\end{theorem}

{\it Proof.} We sketch the proof respectively for every point in
the theorem
 (see similar considerations for Theorem 3.2 for supersymmetric formulas):

\begin{enumerate}
\item  It is easy to verify that a--parallel equations (8.4) on $\xi $
transform into similar ones on $\underline{\xi }$ if and only if
deformations (8.2) with deformation d--tensors of type ${P^\alpha
}_{\beta \gamma }(u)={\psi }_{(\beta }{\delta }_{\gamma )}^\alpha
$ are considered.

\item  Using corresponding to $na_{(1)}$--maps parametrizations of $a(u,v)$
and $b(u,v)$ (see conditions of the theorem) for arbitrary
$v^\alpha \neq 0$ on $U\in \xi $ and after a redefinition of
deformation parameters we obtain that equations (8.6) hold if and
only if ${P^\alpha }_{\beta \gamma }$ satisfies (8.3).

\item  In a similar manner we obtain basic $na_{(2)}$--map equations (8.10)
from (8.6) by considering $na_{(2)}$--parametrizations of
deformation parameters and d--tensor.

\item  For $na_{(3)}$--maps we mast take into consideration deformations of
torsion (8.3) and introduce $na_{(3)}$--parametrizations for $b(u,v)$ and ${%
P^\alpha }_{\beta \gamma }$ into the basic na--equations (8.6).
The last ones for $na_{(3)}$--maps are equivalent to equations
(8.11) (with a corresponding redefinition of deformation
parameters). \qquad $\Box $
\end{enumerate}

We point out that for ${\pi}_{(0)}$-maps we have not differential
equations on $P^{\alpha}_{{.}{\beta}{\gamma}}$ (in the  isotropic
case one considers a
first order system of differential equations on metric [230]; 
 we omit
constructions  with deformation of metric in this section).\

To formulate invariant conditions for reciprocal na--maps (when
every \index{Na--maps!invariant conditions} a-parallel on
${\underline {\xi}}$~ is also transformed into na--parallel on
$\xi$ ) it is convenient to introduce into consideration the
curvature and
Ricci tensors defined for auxiliary connection ${\gamma}^{\alpha}_{{.}{\beta}%
{\gamma}}$:
$$
r^{{.}{\delta}} _{{\alpha}{.}{\beta}{\tau}}={\partial}_{[{\beta}}{\gamma}%
^{\delta}_ {{.}{\tau}]{\alpha}}+{\gamma}^{\delta}_{{.}{\rho}[{\beta}}{\gamma}%
^{\rho}_ {{.}{\tau}]{\alpha}} + {{\gamma}^{\delta}}_{\alpha \phi} {w^{\phi}}%
_{\beta \tau}%
$$
and, respectively, $r_{{\alpha}{\tau}}=r^{{.}{\gamma}} _{{\alpha}{.}{\gamma}{%
\tau}} $, where $[\quad ]$ denotes antisymmetric alternation of
indices, and to define values:
$$
^{(0)}T^{\mu}_{{.}{\alpha}{\beta}}=
{\Gamma}^{\mu}_{{.}{\alpha}{\beta}} -
T^{\mu}_{{.}{\alpha}{\beta}}- {\frac{1 }{(n+m + 1)}}({\delta}^{\mu}_{({\alpha%
}}{\Gamma}^{\delta}_ {{.}{\beta}){\delta}}-{\delta}^{\mu}_{({\alpha}%
}T^{\delta}_ {{.}{\beta}){\gamma}}),
$$
$$
{}^{(0)}{W}^{\cdot \tau}_{\alpha \cdot \beta \gamma} = {r}^{\cdot
\tau}_{\alpha \cdot \beta \gamma} + {\frac{1}{n+m+1}} [ {\gamma}%
^{\tau}_{\cdot \varphi \tau} {\delta}^{\tau}_{( \alpha}
{w^{\varphi}}_{\beta
) \gamma} - ( {\delta}^{\tau}_{\alpha}{r}_{[ \gamma \beta ]} + {\delta}%
^{\tau}_{\gamma} {r}_{[ \alpha \beta ]} - {\delta}^{\tau}_{\beta}
{r}_{[
\alpha \gamma ]} )] -%
$$
$$
{\frac{1}{{(n+m+1)}^2}} [ {\delta}^{\tau}_{\alpha}  ( 2 {\gamma}%
^{\tau}_{\cdot \varphi \tau} {w^{\varphi}}_{[ \gamma \beta ] } - {\gamma}%
^{\tau}_{\cdot \tau [ \gamma } {w^{\varphi}}_{\beta ] \varphi} ) + {\delta}%
^{\tau}_{\gamma}  ( 2 {\gamma}^{\tau}_{\cdot \varphi \tau} {w^{\varphi}}%
_{\alpha \beta} -{\gamma}^{\tau}_{\cdot \alpha \tau}
{w^{\varphi}}_{\beta \varphi}) -
$$
$$
{\delta}^{\tau}_{\beta}  ( 2 {\gamma}^{\tau}_{\cdot \varphi \tau} {%
w^{\varphi}}_{\alpha \gamma} - {\gamma}^{\tau}_{\cdot \alpha \tau} {%
w^{\varphi}}_{\gamma \varphi} ) ],%
$$

$$
{^{(3)}T}^{\delta}_{{.}{\alpha}{\beta}}= {\gamma}^{\delta}_{{.}{\alpha}{\beta%
}}+ {\epsilon}{\varphi}^{\tau}{^{({\gamma})}D}_{\beta}q_{\tau}+ {\frac{1 }{%
n+m}}({\delta}^{\gamma}_{\alpha}- {\epsilon}{\varphi}^{\delta}q_{\alpha})[{%
\gamma}^{\tau}_{{.}{\beta}{\tau}}+ {\epsilon}{\varphi}^{\tau}{^{({\gamma})}D}%
_{\beta}q_{\tau}+%
$$
$$
{\frac{1 }{{n+m
-1}}}q_{\beta}({\epsilon}{\varphi}^{\tau}{\gamma}^{\lambda}_
{{.}{\tau}{\lambda}}+ {\varphi}^{\lambda}{\varphi}^{\tau}{^{({\gamma})}D}%
_{\tau}q_{\lambda})]- $$ $${\frac{1 }{n+m}}({\delta}^{\delta}_{\beta}-{\epsilon}{%
\varphi}^{\delta} q_{\beta})[{\gamma}^{\tau}_{{.}{\alpha}{\tau}}+ {\epsilon}{%
\varphi}^{\tau} {^{({\gamma})}D}_{\alpha}q_{\tau}+%
$$
$$
{\frac{1 }{{n+m -1}}}q_{\alpha}({\epsilon}{\varphi}^{\tau}{\gamma}%
^{\lambda}_ {{.}{\tau}{\lambda}}+ {\varphi}^{\lambda}{\varphi}^{\tau} {%
^{(\gamma)}D} _{\tau}q_{\lambda})],%
$$

$$
{^{(3)}W}^\alpha {{.}{\beta }{\gamma }{\delta }}={\rho }_{{\beta
}{.}{\gamma
}{\delta }}^{{.}{\alpha }}+{\epsilon }{\varphi }^\alpha q_\tau {\rho }_{{%
\beta }{.}{\gamma }{\delta }}^{{.}{\tau }}+({\delta }_\delta
^\alpha -
$$
$$
{\epsilon }{\varphi }^\alpha q_\delta )p_{{\beta }{\gamma }}-({\delta }%
_\gamma ^\alpha -{\epsilon }{\varphi }^\alpha q_\gamma )p_{{\beta }{\delta }%
}-({\delta }_\beta ^\alpha -{\epsilon }{\varphi }^\alpha q_\beta )p_{[{%
\gamma }{\delta }]},
$$
$$
(n+m-2)p_{{\alpha }{\beta }}=-{\rho }_{{\alpha }{\beta }}-{\epsilon }q_\tau {%
\varphi }^\gamma {\rho }_{{\alpha }{.}{\beta }{\gamma
}}^{{.}{\tau }}+{\frac 1{n+m}}[{\rho }_{{\tau }{.}{\beta }{\alpha
}}^{{.}{\tau }}-{\epsilon }q_\tau
{\varphi }^\gamma {\rho }_{{\gamma }{.}{\beta }{\alpha }}^{{.}{\tau }}+{%
\epsilon }q_\beta {\varphi }^\tau {\rho }_{{\alpha }{\tau }}+
$$
$$
{\epsilon }q_\alpha (-{\varphi }^\gamma {\rho }_{{\tau }{.}{\beta }{\gamma }%
}^{{.}{\tau }}+{\epsilon }q_\tau {\varphi }^\gamma {\varphi }^\delta {\rho }%
_{{\gamma }{.}{\beta }{\delta }}^{{.}{\tau }}]),
$$
where $q_\alpha {\varphi }^\alpha ={\epsilon }=\pm 1,$
$$
{{\rho }^\alpha }_{\beta \gamma \delta }=r_{\beta \cdot \gamma
\delta
}^{\cdot \alpha }+{\frac 12}({\psi }_{(\beta }{\delta }_{\varphi )}^\alpha +{%
\sigma }_{\beta \varphi }{\varphi }^\tau ){w^\varphi }_{\gamma
\delta }
$$
( for a similar value on $\underline{\xi }$ we write ${\quad }{\underline{%
\rho }}_{\cdot \beta \gamma \delta }^\alpha
={\underline{r}}_{\beta \cdot
\gamma \delta }^{\cdot \alpha }-{\frac 12}({\psi }_{(\beta }{\delta }_{{%
\varphi })}^\alpha -{\sigma }_{\beta \varphi }{\varphi }^\tau ){w^\varphi }%
_{\gamma \delta }{\quad })$ and ${\rho }_{\alpha \beta }={\rho
}_{\cdot \alpha \beta \tau }^\tau .$

Similar values,
$$
^{(0)}{\underline T}^{\alpha}_{{.}{\beta}{\gamma}}, ^{(0)}{\underline W}%
^{\nu}_{{.}{\alpha}{\beta}{\gamma}}, {\hat T}^{\alpha} _{{.}{\beta}{\gamma}%
}, {\check T}^{\alpha}_ {{.}{\beta}{\tau}}, {\hat W}^{\delta}_{{.}{\alpha}{%
\beta}{\gamma}}, {\check W}^{\delta}_ {{.}{\alpha}{\beta}{\gamma}}, ^{(3)}{%
\underline T}^{\delta} _{{.}{\alpha}{\beta}},%
$$
and $^{(3)}{\underline W}^ {\alpha}_{{.}{\beta}{\gamma}{\delta}}
$ are
given, correspondingly, by auxiliary connections ${\quad}{\underline {\Gamma}%
}^{\mu}_{{.}{\alpha}{\beta}},$~
$$
{\star {\gamma}}^{\alpha}_{{.}{\beta}{\lambda}}={\gamma}^{\alpha} _{{.}{\beta%
}{\lambda}} +
{\epsilon}F^{\alpha}_{\tau}{^{({\gamma})}D}_{({\beta}}
F^{\tau}_{{\lambda})}, \quad {\check {\gamma}}^{\alpha}_{{.}{\beta}{\lambda}%
}= {\widetilde {\gamma}}^{\alpha}_{{.}{\beta}{\lambda}} + {\epsilon}%
F^{\lambda} _{\tau} {\widetilde D}_{({\beta}}F^{\tau}_{{\lambda})},%
$$
$$
{\widetilde {\gamma}}^{\alpha}_{{.}{\beta}{\tau}}={\gamma}^{\alpha} _{{.}{%
\beta}{\tau}}+ {\sigma}_{({\beta}}F^{\alpha}_{{\tau})}, \quad {\hat {\gamma}}%
^{\alpha}_{{.}{\beta}{\lambda}}={\star {\gamma}}^{\alpha}_ {{.}{\beta}{%
\lambda}} + {\widetilde {\sigma}}_{({\beta}}{\delta}^{\alpha}_ {{\lambda})},%
$$
where ${\widetilde
{\sigma}}_{\beta}={\sigma}_{\alpha}F^{\alpha}_{\beta}.$

\begin{theorem}
Four classes of reciprocal na--maps of la--spaces are
characterized by corresponding invariant criterions:
\index{Na--maps!invariant criterions}

\begin{enumerate}
\item  for a--maps $^{(0)}T_{{.}{\alpha }{\beta }}^\mu =^{(0)}{\underline{T}}%
_{{.}{\alpha }{\beta }}^\mu ,$
$$
{}^{(0)}W_{{.}{\alpha }{\beta }{\gamma }}^\delta =^{(0)}{\underline{W}}_{{.}{%
\alpha }{\beta }{\gamma }}^\delta ;\eqno(8.12)
$$

\item  for $na_{(1)}$--maps
$$
3({^{({\gamma })}D}_\lambda P_{{.}{\alpha }{\beta }}^\delta +P_{{.}{\tau }{%
\lambda }}^\delta P_{{.}{\alpha }{\beta }}^\tau )=r_{({\alpha }{.}{\beta }){%
\lambda }}^{{.}{\delta }}-{\underline{r}}_{({\alpha }{.}{\beta }){\lambda }%
}^{{.}{\delta }}+\eqno(8.13)
$$
$$
[T_{{.}{\tau }({\alpha }}^\delta P_{{.}{\beta }{\lambda })}^\tau +Q_{{.}{%
\tau }({\alpha }}^\delta P_{{.}{\beta }{\lambda })}^\tau +b_{({\alpha }}P_{{.%
}{\beta }{\lambda })}^\delta +{\delta }_{({\alpha }}^\delta a_{{\beta }{%
\lambda })}];
$$

\item  for $na_{(2)}$--maps ${\hat T}_{{.}{\beta }{\tau }}^\alpha ={\star T}%
_{{.}{\beta }{\tau }}^\alpha ,$
$$
{\hat W}_{{.}{\alpha }{\beta }{\gamma }}^\delta ={\star W}_{{.}{\alpha }{%
\beta }{\gamma }}^\delta ;\eqno(8.14)
$$

\item  for $na_{(3)}$--maps $^{(3)}T_{{.}{\beta }{\gamma }}^\alpha =^{(3)}{%
\underline{T}}_{{.}{\beta }{\gamma }}^\alpha ,$
$$
{}^{(3)}W_{{.}{\beta }{\gamma }{\delta }}^\alpha =^{(3)}{\underline{W}}_{{.}{%
\beta }{\gamma }{\delta }}^\alpha .\eqno(8.15)
$$
\end{enumerate}
\end{theorem}

{\it Proof. }

We note that Theorem 3.3 is a supersymmetric higher order
generalization
 of this one. Nevetheless we consider a detailed proof of this particular
 (non supersymmetric) case because a number of formulas are important
 in formulation of conservation laws in general relativity and
 different variants of gravitation with torsion and nonmetricity in
   locally isotropic spaces.

\begin{enumerate}
\item  Let us prove that a--invariant conditions (8.12) hold. Deformations
of d--connections of type
$$
{}^{(0)}{\underline{\gamma }}_{\cdot \alpha \beta }^\mu ={{\gamma }^\mu }%
_{\alpha \beta }+{\psi }_{(\alpha }{\delta }_{\beta )}^\mu
\eqno(8.16)
$$
define a--applications. Contracting indices $\mu $ and $\beta $
we can write
$$
{\psi }_\alpha ={\frac 1{m+n+1}}({{\underline{\gamma }}^\beta
}_{\alpha \beta }-{{\gamma }^\beta }_{\alpha \beta }).\eqno(8.17)
$$
Introducing d--vector ${\psi }_\alpha $ into previous relation and
expressing
$$
{{\gamma }^\alpha }_{\beta \tau }=-{T^\alpha }_{\beta \tau }+{{\Gamma }%
^\alpha }_{\beta \tau }
$$
and similarly for underlined values we obtain the first invariant
conditions from (8.12).

Putting deformation (8.16) into the formula for
$$
{\underline{r}}_{\alpha \cdot \beta \gamma }^{\cdot \tau }\quad \mbox{and}%
\quad {\underline{r}}_{\alpha \beta }={\underline{r}}_{\alpha
\tau \beta \tau }^{\cdot \tau }
$$
we obtain respectively relations
$$
{\underline{r}}_{\alpha \cdot \beta \gamma }^{\cdot \tau
}-r_{\alpha \cdot \beta \gamma }^{\cdot \tau }={\delta }_\alpha
^\tau {\psi }_{[\gamma \beta ]}+{\psi }_{\alpha [\beta }{\delta
}_{\gamma ]}^\tau +{\delta }_{(\alpha }^\tau {\psi }_{\varphi
)}{w^\varphi }_{\beta \gamma }\eqno(8.18)
$$
and
$$
{\underline{r}}_{\alpha \beta }-r_{\alpha \beta }={\psi
}_{[\alpha \beta ]}+(n+m-1){\psi }_{\alpha \beta }+{\psi
}_\varphi {w^\varphi }_{\beta \alpha }+{\psi }_\alpha {w^\varphi
}_{\beta \varphi },\eqno(8.19)
$$
where
$$
{\psi }_{\alpha \beta }={}^{({\gamma })}D_\beta {\psi }_\alpha -{\psi }%
_\alpha {\psi }_\beta .
$$
Putting (8.16) into (8.19) we can express ${\psi }_{[\alpha \beta
]}$ as

$${\psi }_{[\alpha \beta ]}=
{\frac 1{n+m+1}}[{\underline{r}}_{[\alpha \beta ]}+$$ $$
{\frac 2{n+m+1}}{\underline{\gamma }}_{\cdot \varphi \tau }^\tau {w^\varphi }%
_{[\alpha \beta ]}-{\frac 1{n+m+1}}{\underline{\gamma }}_{\cdot
\tau [\alpha }^\tau {w^\varphi }_{\beta ]\varphi }]- {\frac
1{n+m+1}}[r_{[\alpha \beta ]}+$$
$${\frac 2{n+m+1}}{{\gamma }^\tau }%
_{\varphi \tau }{w^\varphi }_{[\alpha \beta ]}-{\frac 1{n+m+1}}{{\gamma }%
^\tau }_{\tau [\alpha }{w^\varphi }_{\beta ]\varphi }].\eqno(8.20)
$$
To simplify our consideration we can choose an a--transform,
 pa\-ra\-met\-riz\-ed by
corresponding $\psi $--vector from (8.16), (or fix a local
coordinate cart) the antisymmetrized relations (8.20) to be
satisfied by d--tensor
$$
{\psi }_{\alpha \beta }={\frac 1{n+m+1}}[{\underline{r}}_{\alpha \beta }+{%
\frac 2{n+m+1}}{\underline{\gamma }}_{\cdot \varphi \tau }^\tau {w^\varphi }%
_{\alpha \beta }-$$ $${\frac 1{n+m+1}}{\underline{\gamma
}}_{\cdot \alpha \tau }^\tau {w^\varphi }_{\beta \varphi
}-r_{\alpha \beta }-
$$
$$
{\frac 2{n+m+1}}{{\gamma }^\tau }_{\varphi \tau }{w^\varphi
}_{\alpha \beta }+{\frac 1{n+m+1}}{{\gamma }^\tau }_{\alpha \tau
}{w^\varphi }_{\beta \varphi }]\eqno(8.21)
$$
Introducing expressions (8.16),(8.20) and (8.21) into deformation
of curvature (8.17) we obtain the second conditions (8.12) of
a-map invariance:
$$^{(0)}W_{\alpha \cdot \beta \gamma }^{\cdot \delta }=
{}^{(0)}{\underline{W}}%
_{\alpha \cdot \beta \gamma }^{\cdot \delta },
$$
where the Weyl d--tensor on $\underline{\xi }$ (the extension of
the usual \index{Weyl d--tensor} one for geodesic maps on
(pseudo)--Riemannian spaces to the ca\-se of v--bund\-les
pro\-vid\-ed with N--connection structure) is defined as
$$
{}^{(0)}{\underline{W}}_{\alpha \cdot \beta \gamma }^{\cdot \tau }={%
\underline{r}}_{\alpha \cdot \beta \gamma }^{\cdot \tau }+{\frac 1{n+m+1}}[{%
\underline{\gamma }}_{\cdot \varphi \tau }^\tau {\delta }_{(\alpha }^\tau {%
w^\varphi }_{\beta )\gamma }-$$ $$({\delta }_\alpha ^\tau {\underline{r}}%
_{[\gamma \beta ]}+{\delta }_\gamma ^\tau {\underline{r}}_{[\alpha \beta ]}-{%
\delta }_\beta ^\tau {\underline{r}}_{[\alpha \gamma ]})]-
$$
$$
{\frac 1{{(n+m+1)}^2}}[{\delta }_\alpha ^\tau
(2{\underline{\gamma }}_{\cdot
\varphi \tau }^\tau {w^\varphi }_{[\gamma \beta ]}-{\underline{\gamma }}%
_{\cdot \tau [\gamma }^\tau {w^\varphi }_{\beta ]\varphi
})+{\delta }_\gamma
^\tau (2{\underline{\gamma }}_{\cdot \varphi \tau }^\tau {w^\varphi }%
_{\alpha \beta }-{\underline{\gamma }}_{\cdot \alpha \tau }^\tau {w^\varphi }%
_{\beta \varphi })-
$$
$$
{\delta }_\beta ^\tau (2{\underline{\gamma }}_{\cdot \varphi \tau }^\tau {%
w^\varphi }_{\alpha \gamma }-{\underline{\gamma }}_{\cdot \alpha
\tau }^\tau {w^\varphi }_{\gamma \varphi })]
$$
The formula for $^{(0)}W_{\alpha \cdot \beta \gamma }^{\cdot \tau
}$ written similarly with respect to non--un\-der\-li\-ned values
is presented in subsection 2.1.2 .

\item  To obtain $na_{(1)}$--invariant conditions we rewrite $na_{(1)}$%
--equations (8.9) as to consider in explicit form covariant derivation $^{({%
\gamma })}D$ and deformations (8.2) and (8.3):
$$
2({}^{({\gamma })}D_\alpha {P^\delta }_{\beta \gamma }+{}^{({\gamma }%
)}D_\beta {P^\delta }_{\alpha \gamma }+{}^{({\gamma })}D_\gamma {P^\delta }%
_{\alpha \beta }+{P^\delta }_{\tau \alpha }{P^\tau }_{\beta
\gamma }+
$$
$$
{P^\delta }_{\tau \beta }{P^\tau }_{\alpha \gamma }+{P^\delta
}_{\tau \gamma }{P^\tau }_{\alpha \beta })={T^\delta }_{\tau
(\alpha }{P^\tau }_{\beta \gamma )}+
$$
$$
{H^\delta }_{\tau (\alpha }{P^\tau }_{\beta \gamma )}+b_{(\alpha }{P^\delta }%
_{\beta \gamma )}+a_{(\alpha \beta }{\delta }_{\gamma )}^\delta
.\eqno(8.22)
$$
Alternating the first two indices in (8.22) we have
$$
2({\underline{r}}_{(\alpha \cdot \beta )\gamma }^{\cdot \delta
}-r_{(\alpha
\cdot \beta )\gamma }^{\cdot \delta })=2({}^{(\gamma )}D_\alpha {P^\delta }%
_{\beta \gamma }+
$$
$$
{}^{(\gamma )}D_\beta {P^\delta }_{\alpha \gamma }-2{}^{(\gamma )}D_\gamma {%
P^\delta }_{\alpha \beta }+{P^\delta }_{\tau \alpha }{P^\tau
}_{\beta \gamma }+{P^\delta }_{\tau \beta }{P^\tau }_{\alpha
\gamma }-2{P^\delta }_{\tau \gamma }{P^\tau }_{\alpha \beta }).
$$
Substituting the last expression from (8.22) and rescalling the
deformation parameters and d--tensors we obtain the conditions
(8.9).

\item  Now we prove the invariant conditions for $na_{(0)}$--maps satisfying
conditions
$$
\epsilon \neq 0\quad \mbox{and}\quad \epsilon -F_\beta ^\alpha
F_\alpha ^\beta \neq 0
$$
Let define the auxiliary d--connection
$$
{\tilde \gamma }_{\cdot \beta \tau }^\alpha ={\underline{\gamma
}}_{\cdot
\beta \tau }^\alpha -{\psi }_{(\beta }{\delta }_{\tau )}^\alpha ={{\gamma }%
^\alpha }_{\beta \tau }+{\sigma }_{(\beta }F_{\tau )}^\alpha
\eqno(8.23)
$$
and write
$$
{\tilde D}_\gamma ={}^{({\gamma })}D_\gamma F_\beta ^\alpha +{\tilde \sigma }%
_\gamma F_\beta ^\alpha -{\epsilon }{\sigma }_\beta {\delta
}_\gamma ^\alpha ,
$$
where ${\tilde \sigma }_\beta ={\sigma }_\alpha F_\beta ^\alpha
,$ or, as a consequence from the last equality,
$$
{\sigma }_{(\alpha }F_{\beta )}^\tau ={\epsilon }F_\lambda ^\tau ({}^{({%
\gamma })}D_{(\alpha }F_{\beta )}^\alpha -{\tilde D}_{(\alpha
}F_{\beta )}^\lambda )+{\tilde \sigma }_{(}{\alpha }{\delta
}_{\beta )}^\tau .
$$
Introducing auxiliary connections
$$
{\star {\gamma }}_{\cdot \beta \lambda }^\alpha ={\gamma }_{\cdot
\beta \lambda }^\alpha +{\epsilon }F_\tau ^\alpha {}^{({\gamma
})}D_{(\beta }F_{\lambda )}^\tau
$$
and
$$
{\check \gamma }_{\cdot \beta \lambda }^\alpha ={\tilde \gamma
}_{\cdot \beta \lambda }^\alpha +{\epsilon }F_\tau ^\alpha
{\tilde D}_{(\beta }F_{\lambda )}^\tau
$$
we can express deformation (8.23) in a form characteristic for
a--maps:
$$
{\hat \gamma }_{\cdot \beta \gamma }^\alpha ={\star {\gamma
}}_{\cdot \beta
\gamma }^\alpha +{\tilde \sigma }_{(\beta }{\delta }_{\lambda )}^\alpha .%
\eqno(8.24)
$$
Now it's obvious that $na_{(2)}$--invariant conditions (8.24) are
equivalent with a--invariant conditions (8.12) written for
d--connection (8.24). As a matter of principle we can write
formulas for such $na_{(2)}$--invariants in terms of
''underlined'' and ''non--underlined'' values by expressing
consequently all used auxiliary connections as deformations of
''prime'' connections on $\xi $ and ''final'' connections on
$\underline{\xi }.$ We omit such tedious calculations in this
work.

\item  Let us  prove the last statement, for $na_{(3)}$--maps, of the
theorem 8.3. We consider
$$
q_\alpha {\varphi }^\alpha =e=\pm 1,\eqno(8.25)
$$
where ${\varphi }^\alpha $ is contained in
$$
{\underline{\gamma }}_{\cdot \beta \gamma }^\alpha ={{\gamma }^\alpha }%
_{\beta \gamma }+{\psi }_{(\beta }{\delta }_{\gamma )}^\alpha +{\sigma }%
_{\beta \gamma }{\varphi }^\alpha .\eqno(8.26)
$$
Acting with operator $^{({\gamma })}{\underline{D}}_\beta $ on
(8.25) we write
$$
{}^{({\gamma })}{\underline{D}}_\beta q_\alpha ={}^{({\gamma
})}D_\beta
q_\alpha -{\psi }_{(\alpha }q_{\beta )}-e{\sigma }_{\alpha \beta }.%
\eqno(8.27)
$$
Contracting (8.27) with ${\varphi }^\alpha $ we  express
$$
e{\varphi }^\alpha {\sigma }_{\alpha \beta }={\varphi }^\alpha
({}^{({\gamma
})}D_\beta q_\alpha -{}^{({\gamma })}{\underline{D}}_\beta q_\alpha )-{%
\varphi }_\alpha q^\alpha q_\beta -e{\psi }_\beta .
$$
Putting the last formula in (8.26) and contracting on indices $\alpha $ and $%
\gamma $ we obtain
$$
(n+m){\psi }_\beta ={\underline{\gamma }}_{\cdot \alpha \beta }^\alpha -{{%
\gamma }^\alpha }_{\alpha \beta }+e{\psi }_\alpha {\varphi
}^\alpha q_\beta +e{\varphi }^\alpha {\varphi }^\beta
({}^{({\gamma })}{\underline{D}}_\beta -{}^{({\gamma })}D_\beta
).\eqno(8.28)
$$
From these relations, taking into consideration (8.25), we have%
$$
(n+m-1){\psi }_\alpha {\varphi }^\alpha =
$$
$$
{\varphi }^\alpha ({\underline{\gamma }}_{\cdot \alpha \beta }^\alpha -{{%
\gamma }^\alpha }_{\alpha \beta })+e{\varphi }^\alpha {\varphi }^\beta ({}^{(%
{\gamma })}{\underline{D}}_\beta q_\alpha -{}^{({\gamma
})}D_\beta q_\alpha )
$$

Using the equalities and identities (8.27) and (8.28) we can
express deformations (8.26) as the first $na_{(3)}$--invariant
conditions from (8.15).

To prove the second class of $na_{(3)}$--invariant conditions we
introduce two additional d--tensors:
$$
{{\rho }^\alpha }_{\beta \gamma \delta }=r_{\beta \cdot \gamma
\delta
}^{\cdot \alpha }+{\frac 12}({\psi }_{(\beta }{\delta }_{\varphi )}^\alpha +{%
\sigma }_{\beta \varphi }{\varphi }^\tau ){w^\varphi }_{\gamma \delta }{%
\quad }
$$
and
$$
{\underline{\rho }}_{\cdot \beta \gamma \delta }^\alpha ={\underline{r}}%
_{\beta \cdot \gamma \delta }^{\cdot \alpha }-{\frac 12}({\psi }_{(\beta }{%
\delta }_{{\varphi })}^\alpha -{\sigma }_{\beta \varphi }{\varphi }^\tau ){%
w^\varphi }_{\gamma \delta }.\eqno(8.29)
$$
Using deformation (8.26) and (8.29) we write relation
$$
{\tilde \sigma }_{\cdot \beta \gamma \delta }^\alpha ={\underline{\rho }}%
_{\cdot \beta \gamma \delta }^\alpha -{\rho }_{\cdot \beta \gamma
\delta
}^\alpha ={\psi }_{\beta [\delta }{\delta }_{\gamma ]}^\alpha -{\psi }_{[{%
\gamma }{\delta }]}{\delta }_\beta ^\alpha -{\sigma }_{\beta \gamma \delta }{%
\varphi }^\alpha ,\eqno(8.30)
$$
where
$$
{\psi }_{\alpha \beta }={}^{({\gamma })}D_\beta {\psi }_\alpha +{\psi }%
_\alpha {\psi }_\beta -({\nu }+{\varphi }^\tau {\psi }_\tau ){\sigma }%
_{\alpha \beta },
$$
and
$$
{\sigma }_{\alpha \beta \gamma }={}^{({\gamma })}D_{[\gamma }{\sigma }_{{%
\beta }]{\alpha }}+{\mu }_{[\gamma }{\sigma }_{{\beta }]{\alpha }}-{\sigma }%
_{{\alpha }[{\gamma }}{\sigma }_{{\beta }]{\tau }}{\varphi }^\tau
.
$$
Let multiply (8.30) on $q_\alpha $ and write (taking into account
relations (8.25)) the relation
$$
e{\sigma }_{\alpha \beta \gamma }=-q_\tau {\tilde \sigma }_{\cdot
\alpha \beta \delta }^\tau +{\psi }_{\alpha [\beta }q_{\gamma
]}-{\psi }_{[\beta \gamma ]}q_\alpha .\eqno(8.31)
$$
The next step is to express ${\psi }_{\alpha \beta }$ trough d--objects on ${%
\xi }.$ To do this we contract indices $\alpha $ and $\beta $ in
(8.30) and obtain
$$
(n+m){\psi }_{[\alpha \beta ]}=-{\sigma }_{\cdot \tau \alpha
\beta }^\tau
+eq_\tau {\varphi }^\lambda {\sigma }_{\cdot \lambda \alpha \beta }^\tau -e{%
\tilde \psi }_{[\alpha }{\tilde \psi }_{\beta ]}.
$$
Then contracting indices $\alpha $ and $\delta $ in (8.30) and
using (8.31) we write
$$
(n+m-2){\psi }_{\alpha \beta }={\tilde \sigma }_{\cdot \alpha
\beta \tau }^\tau -eq_\tau {\varphi }^\lambda {\tilde \sigma
}_{\cdot \alpha \beta
\lambda }^\tau +{\psi }_{[\beta \alpha ]}+e({\tilde \psi }_\beta q_\alpha -{%
\hat \psi }_{(\alpha }q_{\beta )},\eqno(8.32)
$$
where ${\hat \psi }_\alpha ={\varphi }^\tau {\psi }_{\alpha \tau
}.$ If the both parts of (8.32) are contracted with ${\varphi
}^\alpha ,$ it results that
$$
(n+m-2){\tilde \psi }_\alpha ={\varphi }^\tau {\sigma }_{\cdot
\tau \alpha
\lambda }^\lambda -eq_\tau {\varphi }^\lambda {\varphi }^\delta {\sigma }%
_{\lambda \alpha \delta }^\tau -eq_\alpha ,
$$
and, in consequence of ${\sigma }_{\beta (\gamma \delta )}^\alpha
=0,$ we have
$$
(n+m-1){\varphi }={\varphi }^\beta {\varphi }^\gamma {\sigma
}_{\cdot \beta \gamma \alpha }^\alpha .
$$
By using the last expressions we can write
$$
(n+m-2){\underline{\psi }}_\alpha ={\varphi }^\tau {\sigma
}_{\cdot \tau
\alpha \lambda }^\lambda -eq_\tau {\varphi }^\lambda {\varphi }^\delta {%
\sigma }_{\cdot \lambda \alpha \delta }^\tau -e{(n+m-1)}^{-1}q_\alpha {%
\varphi }^\tau {\varphi }^\lambda {\sigma }_{\cdot \tau \lambda
\delta }^\delta .\eqno(8.33)
$$
Contracting (8.32) with ${\varphi }^\beta $ we have
$$
(n+m){\hat \psi }_\alpha ={\varphi }^\tau {\sigma }_{\cdot \alpha
\tau \lambda }^\lambda +{\tilde \psi }_\alpha
$$
and taking into consideration (8.33) we can express ${\hat \psi
}_\alpha $ through ${\sigma }_{\cdot \beta \gamma \delta }^\alpha
.$

As a consequence of (8.31)--(8.33) we obtain this formulas for d--tensor ${%
\psi }_{\alpha \beta }:$
$$
(n+m-2){\psi }_{\alpha \beta }={\sigma }_{\cdot \alpha \beta \tau
}^\tau -eq_\tau {\varphi }^\lambda {\sigma }_{\cdot \alpha \beta
\lambda }^\tau +
$$
$$
{\frac 1{n+m}}\{-{\sigma }_{\cdot \tau \beta \alpha }^\tau
+eq_\tau {\varphi
}^\lambda {\sigma }_{\cdot \lambda \beta \alpha }^\tau -q_\beta (e{\varphi }%
^\tau {\sigma }_{\cdot \alpha \tau \lambda }^\lambda -q_\tau {\varphi }%
^\lambda {\varphi }^\delta {\sigma }_{\cdot \alpha \lambda \delta
}^\tau )+
$$
$$ eq_\alpha [{\varphi }^\lambda {\sigma }_{\cdot \tau \beta \lambda }^\tau
-eq_\tau {\varphi }^\lambda {\varphi }^\delta {\sigma }_{\cdot
\lambda \beta \delta }^\tau -$$
$${\frac e{n+m-1}}q_\beta ({\varphi }^\tau {\varphi }^\lambda {%
\sigma }_{\cdot \tau \gamma \delta }^\delta -eq_\tau {\varphi }^\lambda {%
\varphi }^\delta {\varphi }^\varepsilon {\sigma }_{\cdot \lambda
\delta \varepsilon }^\tau )]\}.
$$

Finally, putting the last formula and (8.31) into (8.30) and
after a rearrangement of terms we obtain the second group of
$na_{(3)}$-invariant conditions (8.15). If necessary we can
rewrite these conditions in terms of geometrical objects on $\xi
$ and $\underline{\xi }.$ To do this we mast introduce splittings
(8.29) into (8.15). \qquad $\Box $
\end{enumerate}

For the particular case of $na_{(3)}$--maps when
$$
{\psi}_{\alpha}=0 , {\varphi}_{\alpha} = g_{\alpha \beta}
{\varphi}^{\beta} = {\frac{\delta }{\delta u^{\alpha}}} ( \ln
{\Omega} ) , {\Omega}(u) > 0
$$
and
$$
{\sigma}_{\alpha \beta} = g_{\alpha \beta}%
$$
we define a subclass of conformal transforms ${\underline
g}_{\alpha \beta} (u) = {\Omega}^2 (u)  g_{\alpha \beta}$ which,
in consequence of the fact that d--vector ${\varphi}_{\alpha}$
must satisfy equations (8.11),
generalizes the class of concircular transforms (see 
 [230] for
references and details on concircular mappings of Riemannaian
spaces) .

We emphasize that basic na--equations (8.9)--(8.11) are systems
of first order partial differential equations. The study of their
geometrical properties and definition of integral varieties,
general and particular
solutions are possible by using the formalism of Pffaf systems 
 [278].
\index{Pfaff systems} Here we point out that by using algebraic
methods we can always verify if systems of na--equations of type
(8.9)--(8.11) are, or not, involute, even to find their explicit
solutions it is a difficult task (see more detailed
considerations for isotropic ng--maps in 
 [230] and, on language of
Pffaf systems for na--maps, in 
 247]). We can also formulate the
Cauchy problem for na--equations on $\xi $~ and choose deformation
parameters (8.7) as to make involute mentioned equations for the
case of maps to a given background space ${\underline{\xi }}$. If
a solution, for
example, of $na_{(1)}$--map equations exists, we say that space $\xi $ is $%
na_{(1)}$--projective to space ${\underline{\xi }}.$ In general,
we have to introduce chains of na--maps in order to obtain
involute systems of
equations for maps (superpositions of na-maps) from $\xi $ to ${\underline{%
\xi }}:$
$$U \buildrel {ng<i_{1}>} \over \longrightarrow  {U_{\underline 1}}
\buildrel ng<i_2> \over \longrightarrow \cdots \buildrel
ng<i_{k-1}> \over \longrightarrow U_{\underline {k-1}} \buildrel
ng<i_k> \over \longrightarrow {\underline U} $$
where $U\subset {\xi },U_{\underline{1}}\subset {\xi }_{\underline{1}%
},\ldots ,U_{k-1}\subset {\xi }_{k-1},{\underline{U}}\subset {\xi
}_k$ with corresponding splittings of auxiliary symmetric
connections
$$
{\underline{\gamma }}_{.{\beta }{\gamma }}^\alpha =_{<i_1>}P_{.{\beta }{%
\gamma }}^\alpha +_{<i_2>}P_{.{\beta }{\gamma }}^\alpha +\cdots +_{<i_k>}P_{.%
{\beta }{\gamma }}^\alpha
$$
and torsion
$$
{\underline{T}}_{.{\beta }{\gamma }}^\alpha =T_{.{\beta }{\gamma
}}^\alpha
+_{<i_1>}Q_{.{\beta }{\gamma }}^\alpha +_{<i_2>}Q_{.{\beta }{\gamma }%
}^\alpha +\cdots +_{<i_k>}Q_{.{\beta }{\gamma }}^\alpha
$$
where cumulative indices $<i_1>=0,1,2,3,$ denote possible types
of na--maps.

\begin{definition}
Space $\xi $~ is nearly conformally projective to space ${\underline{\xi }},{%
\quad }nc:{\xi }{\to }{\underline{\xi }},$~ if there is a finite
chain of na--maps from $\xi $~ to ${\underline{\xi }}.$
\end{definition}

For nearly conformal maps we formulate : \index{Nearly conformal
maps}

\begin{theorem}
For every fixed triples $(N_j^a,{\Gamma }_{{.}{\beta }{\gamma
}}^\alpha
,U\subset {\xi }$ and $(N_j^a,{\underline{\Gamma }}_{{.}{\beta }{\gamma }%
}^\alpha $, ${\underline{U}}\subset {\underline{\xi }})$,
components of
nonlinear connection, d--connection and d--metric being of class $C^r(U),C^r(%
{\underline{U}})$, $r>3,$ there is a finite chain of na--maps $nc:U\to {%
\underline{U}}.$
\end{theorem}

Proof is similar to that for isotropic maps 
 [249,273,252] (we have to
introduce a finite number of na-maps with corresponding
components of deformation parameters and deformation tensors in
order to transform step by step coefficients of d-connection
${\Gamma}^{\alpha}_{\gamma \delta}$ into ${\underline
{\Gamma}}^{\alpha}_{\beta \gamma} ).$

Now we introduce the concept of the Category of la--spaces,
${\cal C}({\xi}). $ The elements of ${\cal C}({\xi})$ consist
from $Ob{\cal C}({\xi})=\{{\xi}, {\xi}_{<i_{1}>},
{\xi}_{<i_{2}>},{\ldots}, \}$ being la--spaces, for simplicity in
this work, having common N--connection structures, and\\ $Mor
{\cal C}({\xi})=\{ nc ({\xi}_{<i_{1}>}, {\xi}_{<i_{2}>})\}$ being
chains of na--maps interrelating la--spaces. We point out that we
can consider equivalent models of physical theories on every
object of ${\cal C}({\xi})$ (see details for isotropic
gravitational models in
 [249,252,278,250,273,274] and anisotropic gravity in
 [263,276,279]). One of the main purposes of this chapter is to develop a
d--tensor and variational formalism on ${\cal C}({\xi}),$ i.e. on
la--multispaces, interrelated with nc--maps. Taking into account
the distinguished character of geometrical objects on la--spaces
we call tensors on ${\cal C}({\xi})$ as distinguished tensors on
la--space Category, or dc--tensors.

Finally, we emphasize that presented in that section definitions
and theorems can be generalized for v--bundles with arbitrary
given structures of nonlinear connection, linear d--connection
and metric structures. Proofs
are similar to those from 
 [251,230].

\section{Na-Tensor-Integral on La-Spaces}

The aim of this section is to define tensor integration not only
for bitensors, objects defined on the same curved space, but for
dc--tensors, defined on two spaces, $\xi$ and ${\underline
{\xi}}$, even it is necessary on la--multispaces. A. Mo\'or
tensor--integral formalism having a lot of
\index{Tensor--integral} applications in classical and quantum
gravity
 [234,289,100] was
extended for locally isotropic multispaces in 
 [278,273]. The
unispacial locally anisotropic version is given in 
 [258,102].

Let $T_{u}{\xi}$~ and $T_{\underline u}{\underline {\xi}}$ be
tangent spaces in corresponding points $u {\in} U {\subset}
{\xi}$ and ${\underline u} {\in}
{\underline U} {\subset} {\underline {\xi}}$ and, respectively, $T^{\ast}_{u}%
{\xi}$ and $T^{\ast}_{\underline u}{\underline {\xi}} $ be their
duals (in general, in this section we shall not consider that a
common coordinatization is introduced for open regions $U$ and
${\underline U}$ ). We call as the dc--tensors on the pair of
spaces (${\xi}, {\underline {\xi}}$ ) the elements of
distinguished tensor algebra
$$
( {\otimes}_{\alpha} T_{u}{\xi}) {\otimes} ({\otimes}_{\beta} T^{\ast}_{u} {%
\xi}) {\otimes}({\otimes}_{\gamma} T_{\underline u}{\underline {\xi}}) {%
\otimes} ({\otimes}_{\delta} T^{\ast}_{\underline u} {\underline {\xi}})%
$$
defined over the space ${\xi}{\otimes} {\underline {\xi}}, $ for
a given $nc : {\xi} {\to} {\underline {\xi}} $.

We admit the convention that underlined and non--underlined
indices refer,
respectively, to the points ${\underline u}$ and $u$. Thus $Q_{{.}{%
\underline {\alpha}}}^{\beta}, $ for instance, are the components
of dc--tensor $Q{\in} T_{u}{\xi} {\otimes} T_{\underline
u}{\underline {\xi}}.$

Now, we define the transport dc--tensors. Let open regions $U$ and ${%
\underline U}$ be homeomorphic to sphere ${\cal R}^{2n}$ and
introduce
isomorphism ${\mu}_{{u},{\underline u}}$ between $T_{u}{\xi}$ and $%
T_{\underline u}{\underline {\xi}}$ (given by map $nc : U {\to} {\underline U%
}).$ We consider that for every d--vector $v^{\alpha} {\in}
T_{u}{\xi}$ corresponds the vector ${\mu}_{{u},{\underline u}}
(v^{\alpha})=v^{\underline {\alpha}} {\in} T_{\underline u}{\underline {\xi}}%
,$ with components $v^{\underline {\alpha}}$ being linear functions of $%
v^{\alpha}$:
$$
v^{\underline {\alpha}}=h^{\underline {\alpha}}_{\alpha}(u,
{\underline u})
v^{\alpha}, \quad v_{\underline {\alpha}}= h^{\alpha}_{\underline {\alpha}}({%
\underline u}, u)v_{\alpha},%
$$
where $h^{\alpha}_{\underline {\alpha}}({\underline u}, u)$ are
the components of dc--tensor associated with
${\mu}^{-1}_{u,{\underline u}}$. In a similar manner we have
$$
v^{\alpha}=h^{\alpha}_{\underline {\alpha}}({\underline u}, u)
v^{\underline {\alpha}}, \quad v_{\alpha}=h^{\underline
{\alpha}}_{\alpha} (u, {\underline
u})v_{\underline {\alpha}}.%
$$

In order to reconcile just presented definitions and to assure
the identity for trivial maps ${\xi }{\to }{\xi
},u={\underline{u}},$ the transport dc-tensors must satisfy
conditions :
$$
h_\alpha ^{\underline{\alpha }}(u,{\underline{u}})h_{\underline{\alpha }%
}^\beta ({\underline{u}},u)={\delta }_\alpha ^\beta ,h_\alpha ^{\underline{%
\alpha }}(u,{\underline{u}})h_{\underline{\beta }}^\alpha ({\underline{u}}%
,u)={\delta }_{\underline{\beta }}^{\underline{\alpha }}
$$
and ${\lim }_{{({\underline{u}}{\to }u})}h_\alpha ^{\underline{\alpha }}(u,{%
\underline{u}})={\delta }_\alpha ^{\underline{\alpha }},\quad {\lim }_{{({%
\underline{u}}{\to }u})}h_{\underline{\alpha }}^\alpha ({\underline{u}},u)={%
\delta }_{\underline{\alpha }}^\alpha .$

Let ${\overline S}_{p} {\subset} {\overline U} {\subset}
{\overline {\xi}}$ is a homeomorphic to $p$-dimensional sphere
and suggest that chains of na--maps are used to connect regions :
$$ U \buildrel nc_{(1)} \over \longrightarrow {\overline S}_p
     \buildrel nc_{(2)} \over \longrightarrow {\underline U}.$$

\begin{definition}
The tensor integral in ${\overline{u}}{\in }{\overline{S}}_p$ of a
dc--tensor $$N_{{\varphi }{.}{\underline{\tau }}{.}{\overline{\alpha }}_1{%
\cdots }{\overline{\alpha }}_p}^{{.}{\gamma }{.}{\underline{\kappa }}}$$ $({%
\overline{u}},u),$ completely antisymmetric on the indices ${{\overline{%
\alpha }}_1},{\ldots },{\overline{\alpha }}_p,$ over domain ${\overline{S}}%
_p,$ is defined as
$$
N_{{\varphi }{.}{\underline{\tau }}}^{{.}{\gamma }{.}{\underline{\kappa }}}({%
\underline{u}},u)=I_{({\overline{S}}_p)}^{\underline{U}}N_{{\varphi }{.}{%
\overline{\tau }}{.}{\overline{\alpha }}_1{\ldots }{\overline{\alpha }}_p}^{{%
.}{\gamma }{.}{\overline{\kappa }}}({\overline{u}},{\underline{u}})dS^{{%
\overline{\alpha }}_1{\ldots }{\overline{\alpha }}_p}=
$$
$$
{\int }_{({\overline{S}}_p)}h_{\underline{\tau }}^{\overline{\tau }}({%
\underline{u}},{\overline{u}})h_{\overline{\kappa }}^{\underline{\kappa }}({%
\overline{u}},{\underline{u}})N_{{\varphi }{.}{\overline{\tau }}{.}{%
\overline{\alpha }}_1{\cdots }{\overline{\alpha }}_p}^{{.}{\gamma }{.}{%
\overline{\kappa }}}({\overline{u}},u)d{\overline{S}}^{{\overline{\alpha }}_1%
{\cdots }{\overline{\alpha }}_p},\eqno(8.34)
$$
where $dS^{{\overline{\alpha }}_1{\cdots }{\overline{\alpha }}_p}={\delta }%
u^{{\overline{\alpha }}_1}{\land }{\cdots }{\land }{\delta }u_p^{\overline{%
\alpha }}$.
\end{definition}

Let suppose that transport dc--tensors $h_\alpha
^{\underline{\alpha }}$~ \index{Transport dc--tensors} and
$h_{\underline{\alpha }}^\alpha $~ admit covariant derivations of
or\-der two and pos\-tu\-la\-te ex\-is\-ten\-ce of
de\-for\-ma\-ti\-on dc--ten\-sor\\ $B_{{\alpha }{\beta
}}^{{..}{\gamma }}(u,{\underline{u}})$~ satisfying relations
$$
D_\alpha h_\beta ^{\underline{\beta
}}(u,{\underline{u}})=B_{{\alpha }{\beta
}}^{{..}{\gamma }}(u,{\underline{u}})h_\gamma ^{\underline{\beta }}(u,{%
\underline{u}})\eqno(8.35)
$$
and, taking into account that $D_\alpha {\delta }_\gamma ^\beta
=0,$

$$
D_\alpha h_{\underline{\beta }}^\beta ({\underline{u}},u)=-B_{{\alpha }{%
\gamma }}^{{..}{\beta }}(u,{\underline{u}})h_{\underline{\beta }}^\gamma ({%
\underline{u}},u).
$$
By using formulas  for torsion and, respectively, curvature of
connection ${\Gamma }_{{\beta }{\gamma }}^\alpha $~ we can
calculate next commutators:
$$
D_{[{\alpha }}D_{{\beta }]}h_\gamma ^{\underline{\gamma }}=-(R_{{\gamma }{.}{%
\alpha }{\beta }}^{{.}{\lambda }}+T_{{.}{\alpha }{\beta }}^\tau B_{{\tau }{%
\gamma }}^{{..}{\lambda }})h_\lambda ^{\underline{\gamma
}}.\eqno(8.36)
$$
On the other hand from (8.35) one follows that
$$
D_{[{\alpha }}D_{{\beta }]}h_\gamma ^{\underline{\gamma }}=(D_{[{\alpha }}B_{%
{\beta }]{\gamma }}^{{..}{\lambda }}+B_{[{\alpha }{|}{\tau }{|}{.}}^{{..}{%
\lambda }}B_{{\beta }]{\gamma }{.}}^{{..}{\tau }})h_\lambda ^{\underline{%
\gamma }},\eqno(8.37)
$$
where ${|}{\tau }{|}$~ denotes that index ${\tau }$~ is excluded
from the action of antisymmetrization $[{\quad }]$. From (8.36)
and (8.37) we obtain
$$
D_{[{\alpha }}B_{{\beta }]{\gamma }{.}}^{{..}{\lambda }}+B_{[{\beta }{|}{%
\gamma }{|}}B_{{\alpha }]{\tau }}^{{..}{\lambda }}=(R_{{\gamma }{.}{\alpha }{%
\beta }}^{{.}{\lambda }}+T_{{.}{\alpha }{\beta }}^\tau B_{{\tau }{\gamma }}^{%
{..}{\lambda }}).\eqno(8.38)
$$

Let ${\overline{S}}_p$~ be the boundary of
${\overline{S}}_{p-1}$. The Stoke's type formula for
tensor--integral (8.34) is defined as

$$
I_{{\overline{S}}_p}N_{{\varphi }{.}{\overline{\tau }}{.}{\overline{\alpha }}%
_1{\ldots }{\overline{\alpha }}_p}^{{.}{\gamma }{.}{\overline{\kappa }}}dS^{{%
\overline{\alpha }}_1{\ldots }{\overline{\alpha }}_p}=
I_{{\overline{S}}_{p+1}}{^{{\star
}{(p)}}{\overline{D}}}_{[{\overline{\gamma
}}{|}}N_{{\varphi }{.}{\overline{\tau }}{.}{|}{\overline{\alpha }}_1{\ldots }%
{{\overline{\alpha }}_p]}}^{{.}{\gamma }{.}{\overline{\kappa }}}dS^{{%
\overline{\gamma }}{\overline{\alpha }}_1{\ldots }{\overline{\alpha }}_p},%
\eqno(8.39)
$$
where
$$
{^{{\star }{(p)}}D}_{[{\overline{\gamma }}{|}}N_{{\varphi }{.}{\overline{%
\tau }}{.}{|}{\overline{\alpha }}_1{\ldots }{\overline{\alpha }}_p]}^{{.}{%
\gamma }{.}{\overline{\kappa }}}=
D_{[{\overline{\gamma }}{|}}N_{{\varphi }{.}{\overline{\tau }}{.}{|}{%
\overline{\alpha }}_1{\ldots }{\overline{\alpha }}_p]}^{{.}{\gamma }{.}{%
\overline{\kappa }}}+ $$ $$
pT_{{.}[{\overline{\gamma }}{\overline{\alpha }}_1{|}}^{%
\underline{\epsilon }}N_{{\varphi }{.}{\overline{\tau }}{.}{\overline{%
\epsilon }}{|}{\overline{\alpha }}_2{\ldots }{\overline{\alpha }}_p]}^{{.}{%
\gamma }{.}{\overline{\kappa }}}-B_{[{\overline{\gamma }}{|}{\overline{\tau }%
}}^{{..}{\overline{\epsilon }}}N_{{\varphi }{.}{\overline{\epsilon }}{.}{|}{%
\overline{\alpha }}_1{\ldots }{\overline{\alpha }}_p]}^{{.}{\gamma }{.}{%
\overline{\kappa }}}+B_{[{\overline{\gamma }}{|}{\overline{\epsilon }}}^{..{%
\overline{\kappa }}}N_{{\varphi }{.}{\overline{\tau }}{.}{|}{\overline{%
\alpha }}_1{\ldots }{\overline{\alpha }}_p]}^{{.}{\gamma }{.}{\overline{%
\epsilon }}}.\eqno(8.40)
$$
We define the dual element of the hypersurfaces element $dS^{{j}_1{\ldots }{j%
}_p}$ as
$$
d{\cal S}_{{\beta }_1{\ldots }{\beta }_{q-p}}={\frac 1{{p!}}}{\epsilon }_{{%
\beta }_1{\ldots }{\beta }_{k-p}{\alpha }_1{\ldots }{\alpha }_p}dS^{{\alpha }%
_1{\ldots }{\alpha }_p},\eqno(8.41)
$$
where ${\epsilon }_{{\gamma }_1{\ldots }{\gamma }_q}$ is
completely antisymmetric on its indices and
$$
{\epsilon }_{12{\ldots }(n+m)}=\sqrt{{|}G{|}},G=det{|}G_{{\alpha }{\beta }{|}%
},
$$
$G_{{\alpha }{\beta }}$ is taken from (6.12). The dual of dc--tensor $N_{{%
\varphi }{.}{\overline{\tau }}{.}{\overline{\alpha }}_1{\ldots }{\overline{%
\alpha }}_p}^{{.}{\gamma }{\overline{\kappa }}}$ is defined as the
dc--tensor  ${\cal N}_{{\varphi }{.}{\overline{\tau }}}^{{.}{\gamma }{.}{%
\overline{\kappa }}{\overline{\beta }}_1{\ldots }{\overline{\beta
}}_{n+m-p}} $ satisfying
$$
N_{{\varphi }{.}{\overline{\tau }}{.}{\overline{\alpha }}_1{\ldots }{%
\overline{\alpha }}_p}^{{.}{\gamma }{.}{\overline{\kappa }}}={\frac 1{{p!}}}%
{\cal N}_{{\varphi }{.}{\overline{\tau }}}^{{.}{\gamma
}{.}{\overline{\kappa
}}{\overline{\beta }}_1{\ldots }{\overline{\beta }}_{n+m-p}}{\epsilon }_{{%
\overline{\beta }}_1{\ldots }{\overline{\beta }}_{n+m-p}{\overline{\alpha }}%
_1{\ldots }{\overline{\alpha }}_p}.\eqno(8.42)
$$
Using (8.16), (8.41) and (8.42) we can write
$$
I_{{\overline{S}}_p}N_{{\varphi }{.}{\overline{\tau }}{.}{\overline{\alpha }}%
_1{\ldots }{\overline{\alpha }}_p}^{{.}{\gamma }{.}{\overline{\kappa }}}dS^{{%
\overline{\alpha }}_1{\ldots }{\overline{\alpha }}_p}={\int }_{{\overline{S}}%
_{p+1}}{^{\overline{p}}D}_{\overline{\gamma }}{\cal N}_{{\varphi }{.}{%
\overline{\tau }}}^{{.}{\gamma }{.}{\overline{\kappa }}{\overline{\beta }}_1{%
\ldots }{\overline{\beta }}_{n+m-p-1}{\overline{\gamma }}}d{\cal S}_{{%
\overline{\beta }}_1{\ldots }{\overline{\beta
}}_{n+m-p-1}},\eqno(8.43)
$$
where
$$
{^{\overline{p}}D}_{\overline{\gamma }}{\cal N}_{{\varphi }{.}{\overline{%
\tau }}}^{{.}{\gamma }{.}{\overline{\kappa }}{\overline{\beta }}_1{\ldots }{%
\overline{\beta }}_{n+m-p-1}{\overline{\gamma }}}=
{\overline{D}}_{\overline{\gamma }}{\cal N}_{{\varphi }{.}{\overline{\tau }}%
}^{{.}{\gamma }{.}{\overline{\kappa }}{\overline{\beta }}_1{\ldots }{%
\overline{\beta }}_{n+m-p-1}{\overline{\gamma }}}+$$
$$(-1)^{(n+m-p)}(n+m-p+1)T_{{%
.}{\overline{\gamma }}{\overline{\epsilon }}}^{[{\overline{\epsilon }}}{\cal %
N}_{{\varphi }{.}{{\overline{\tau }}}}^{{.}{|}{\gamma }{.}{\overline{\kappa }%
}{|}{\overline{\beta }}_1{\ldots }{\overline{\beta }}_{n+m-p-1}]{\overline{%
\gamma }}}-
$$
$$
B_{{\overline{\gamma }}{\overline{\tau }}}^{{..}{\overline{\epsilon }}}{\cal %
N}_{{\varphi }{.}{\overline{\epsilon }}}^{{.}{\gamma }{.}{\overline{\kappa }}%
{\overline{\beta }}_1{\ldots }{\overline{\beta
}}_{n+m-p-1}{\overline{\gamma
}}}+B_{{\overline{\gamma }}{\overline{\epsilon }}}^{{..}{\overline{\kappa }}}%
{\cal N}_{{\varphi }{.}{\overline{\tau }}}^{{.}{\gamma }{.}{\overline{%
\epsilon }}{\overline{\beta }}_1{\ldots }{\overline{\beta }}_{n+m-p-1}{%
\overline{\gamma }}}.
$$
To verify the equivalence of (8.42) and (8.43) we must take in
consideration that $ D_\gamma {\epsilon }_{{\alpha }_1{\ldots
}{\alpha }_k}=0$ and
$$ {\epsilon }_{{\beta }_1{\ldots }{\beta }_{n+m-p}{\alpha }_1{\ldots }{\alpha }%
_p}{\epsilon }^{{\beta }_1{\ldots }{\beta }_{n+m-p}{\gamma }_1{\ldots }{%
\gamma }_p}=p!(n+m-p)!{\delta }_{{\alpha }_1}^{[{\gamma
}_1}{\cdots }{\delta }_{{\alpha }_p}^{{\gamma }_p]}.
$$
The developed in this section tensor integration formalism will
be used in the next section for definition of conservation laws
on spaces with local anisotropy.

\section{On Conservation Laws on La--Spaces}

To define conservation laws on locally anisotropic spaces is a
challenging \index{Conservation laws!on la--spaces} task because
of absence of global and local groups of automorphisms of such
spaces. Our main idea is to use chains of na--maps from a given,
called hereafter as the fundamental la--space to an auxiliary one
with trivial curvatures and torsions admitting a global group of
automorphisms. The aim of this section is to  formulate
conservation laws for la-gravitational fields by using
dc--objects and tensor--integral values, na--maps and variational
calculus on the Category of la--spaces. \index{Category!of
la--spaces}

\subsection{Nonzero divergence of energy--mo\-men\-tum d-\-ten\-sor}

R. Miron and M. Anastasiei 
 [160,161] pointed to this specific form of
conservation laws of matter on la--spaces: They calculated the
divergence of
the energy--momentum d--tensor on la--space $\xi ,$%
$$
D_\alpha {E}_\beta ^\alpha ={\frac 1{\ {\kappa }_1}}U_\alpha
,\eqno(8.44)
$$
and concluded that d--vector
$$
U_\alpha ={\frac 12}(G^{\beta \delta }{{R_\delta }^\gamma }_{\phi \beta }{T}%
_{\cdot \alpha \gamma }^\phi -G^{\beta \delta }{{R_\delta
}^\gamma }_{\phi \alpha }{T}_{\cdot \beta \gamma }^\phi +{R_\phi
^\beta }{T}_{\cdot \beta \alpha }^\phi )
$$
vanishes if and only if d--connection $D$ is without torsion.

No wonder that conservation laws, in usual physical theories
being a consequence of global (for usual gravity of local)
automorphisms of the fundamental space--time, are more
sophisticate on the spaces with local anisotropy. Here it is
important to emphasize the multiconnection character of
la--spaces. For example, for a d--metric (6.12) on $\xi $ we can
equivalently introduce another (see (6.21)) metric linear connection $%
\tilde D$ The Einstein equations
$$
{\tilde R}_{\alpha \beta }-{\frac 12}G_{\alpha \beta }{\tilde R}={\kappa }_1{%
\tilde E}_{\alpha \beta }\eqno(8.45)
$$
constructed by using connection (8.23) have vanishing divergences
$$
{\tilde D}^\alpha ({{\tilde R}_{\alpha \beta }}-{\frac 12}G_{\alpha \beta }{%
\tilde R})=0\mbox{ and }{\tilde D}^\alpha {\tilde E}_{\alpha
\beta }=0,
$$
similarly as those on (pseudo)Riemannian spaces. We conclude that
by using the connection (6.21) we construct a model of
la--gravity which looks like locally isotropic on the total space
$E.$ More general gravitational models
with local anisotropy can be obtained by using deformations of connection ${%
\tilde \Gamma }_{\cdot \beta \gamma }^\alpha ,$
$$
{{\Gamma }^\alpha }_{\beta \gamma }={\tilde \Gamma }_{\cdot \beta
\gamma }^\alpha +{P^\alpha }_{\beta \gamma }+{Q^\alpha }_{\beta
\gamma },
$$
were, for simplicity, ${{\Gamma }^\alpha }_{\beta \gamma }$ is
chosen to be also metric and satisfy Einstein equations (8.45).
We can consider deformation d--tensors ${P^\alpha }_{\beta \gamma
}$ generated (or not) by deformations of type (8.9)--(8.11) for
na--maps. In this case d--vector $U_\alpha $ can be interpreted
as a generic source of local anisotropy on $\xi $ satisfying
generalized conservation laws (8.44).

\subsection{ D--tensor conservation laws}

From (8.34) we obtain a tensor integral on ${\cal C}({\xi})$ of a
d--tensor:
$$
N_{{\underline {\tau}}}^{{.}{\underline {\kappa}}}(\underline u)= I_{{%
\overline S}_{p}}N^{{..}{\overline {\kappa}}}_ {{\overline {\tau}}{..}{%
\overline {\alpha}}_{1}{\ldots}{\overline {\alpha}}_{p}} ({\overline u})h^{{%
\overline {\tau}}}_{{\underline {\tau}}}({\underline u}, {\overline u})h^{{%
\underline {\kappa}}}_{{\overline {\kappa}}} ({\overline u}, {\underline u}%
)dS^{{\overline {\alpha}}_{1}{\ldots} {\overline {\alpha}}_{p}}.%
$$

We point out that tensor--integral can be defined not only for
dc--tensors but and for d--tensors on $\xi $. Really, suppressing
indices ${\varphi }$~ and ${\gamma }$~ in (8.42) and (8.43),
considering instead of a deformation dc--tensor a deformation
tensor
$$
B_{{\alpha }{\beta }}^{{..}{\gamma }}(u,{\underline{u}})=B_{{\alpha }{\beta }%
}^{{..}{\gamma }}(u)=P_{{.}{\alpha }{\beta }}^\gamma
(u)\eqno(8.46)
$$
(we consider deformations induced by a nc--transform) and integration\\ $%
I_{S_p}{\ldots }dS^{{\alpha }_1{\ldots }{\alpha }_p}$ in
la--space $\xi $ we obtain from (8.34) a tensor--integral on
${\cal C}({\xi })$~ of a d--tensor:
$$
N_{{\underline{\tau }}}^{{.}{\underline{\kappa }}}({\underline{u}}%
)=I_{S_p}N_{{\tau }{.}{\alpha }_1{\ldots }{\alpha }_p}^{.{\kappa }}(u)h_{{%
\underline{\tau }}}^\tau ({\underline{u}},u)h_\kappa ^{\underline{\kappa }%
}(u,{\underline{u}})dS^{{\alpha }_1{\ldots }{\alpha }_p}.
$$
Taking into account (8.38) we can calculate that curvature
$$
{\underline{R}}_{{\gamma }{.}{\alpha }{\beta }}^{.{\lambda }}=D_{[{\beta }%
}B_{{\alpha }]{\gamma }}^{{..}{\lambda }}+B_{[{\alpha }{|}{\gamma }{|}}^{{..}%
{\tau }}B_{{\beta }]{\tau }}^{{..}{\lambda }}+T_{{.}{\alpha }{\beta }}^{{%
\tau }{..}}B_{{\tau }{\gamma }}^{{..}{\lambda }}
$$
of connection ${\underline{\Gamma }}_{{.}{\alpha }{\beta }}^\gamma (u)={%
\Gamma }_{{.}{\alpha }{\beta }}^\gamma (u)+B_{{\alpha }{\beta }{.}}^{{..}{%
\gamma }}(u),$ with $B_{{\alpha }{\beta }}^{{..}{\gamma }}(u)$~
taken from
(8.46), vanishes, ${\underline{R}}_{{\gamma }{.}{\alpha }{\beta }}^{{.}{%
\lambda }}=0.$ So, we can conclude that la--space $\xi $ admits a
tensor integral structure on ${\cal {C}}({\xi })$ for d--tensors
associated to deformation tensor $B_{{\alpha }{\beta
}}^{{..}{\gamma }}(u)$ if the nc--image ${\underline{\xi }}$~ is
locally parallelizable. That way we generalize the one space
tensor integral constructions in
 [100,102,258], were the possibility to introduce
 tensor integral structure on a curved
space was restricted by the condition that this space is locally
parallelizable. For $q=n+m$~ relations (8.43), written for d--tensor ${\cal N%
}_{\underline{\alpha }}^{{.}{\underline{\beta
}}{\underline{\gamma }}}$ (we
change indices ${\overline{\alpha }},{\overline{\beta }},{\ldots }$ into ${%
\underline{\alpha }},{\underline{\beta }},{\ldots })$ extend the
Gauss formula on ${\cal {C}}({\xi })$:
$$
I_{S_{q-1}}{\cal N}_{\underline{\alpha }}^{{.}{\underline{\beta }}{%
\underline{\gamma }}}d{\cal S}_{\underline{\gamma }}=I_{{\underline{S}}_q}{^{%
\underline{q-1}}D}_{{\underline{\tau }}}{\cal N}_{{\underline{\alpha }}}^{{.}%
{\underline{\beta }}{\underline{\tau
}}}d{\underline{V}},\eqno(8.47)
$$
where $d{\underline{V}}={\sqrt{{|}{\underline{G}}_{{\alpha }{\beta }}{|}}}d{%
\underline{u}}^1{\ldots }d{\underline{u}}^q$ and
$$
{^{\underline{q-1}}D}_{{\underline{\tau }}}{\cal N}_{\underline{\alpha }}^{{.%
}{\underline{\beta }}{\underline{\tau }}}=D_{{\underline{\tau }}}{\cal N}_{%
\underline{\alpha }}^{{.}{\underline{\beta }}{\underline{\tau }}}-T_{{.}{%
\underline{\tau }}{\underline{\epsilon }}}^{{\underline{\epsilon }}}{\cal N}%
_{{\underline{\alpha }}}^{{\underline{\beta }}{\underline{\tau }}}-B_{{%
\underline{\tau }}{\underline{\alpha }}}^{{..}{\underline{\epsilon }}}{\cal N%
}_{{\underline{\epsilon }}}^{{.}{\underline{\beta }}{\underline{\tau }}}+B_{{%
\underline{\tau }}{\underline{\epsilon }}}^{{..}{\underline{\beta }}}{\cal N}%
_{{\underline{\alpha }}}^{{.}{\underline{\epsilon }}{\underline{\tau }}}.%
\eqno(8.48)
$$

Let consider physical values $N_{{\underline{\alpha }}}^{{.}{\underline{%
\beta }}}$ on ${\underline{\xi }}$~ defined on its density ${\cal N}_{{%
\underline{\alpha }}}^{{.}{\underline{\beta }}{\underline{\gamma
}}},$ i. e.
$$
N_{{\underline{\alpha }}}^{{.}{\underline{\beta }}}=I_{{\underline{S}}_{q-1}}%
{\cal N}_{{\underline{\alpha }}}^{{.}{\underline{\beta }}{\underline{\gamma }%
}}d{\cal S}_{{\underline{\gamma }}}\eqno(8.49)
$$
with this conservation law (due to (8.47)):%
$$
{^{\underline{q-1}}D}_{{\underline{\gamma }}}{\cal N}_{{\underline{\alpha }}%
}^{{.}{\underline{\beta }}{\underline{\gamma }}}=0.\eqno(8.50)
$$
We note that these conservation laws differ from covariant
conservation laws for well known physical values such as density
of electric current or of
energy-- momentum tensor. For example, taking density ${E}_\beta ^{{.}{%
\gamma }},$ with corresponding to (8.48) and (8.50) conservation
law,
$$
{^{\underline{q-1}}D}_{{\underline{\gamma }}}{E}_{{\underline{\beta }}}^{{%
\underline{\gamma }}}=D_{{\underline{\gamma }}}{E}_{{\underline{\beta }}}^{{%
\underline{\gamma }}}-T_{{.}{\underline{\epsilon }}{\underline{\tau }}}^{{%
\underline{\tau }}}{E}_{{\underline{\beta }}}^{{.}{\underline{\epsilon }}%
}-B_{{\underline{\tau }}{\underline{\beta }}}^{{..}{\underline{\epsilon }}}{E%
}_{\underline{\epsilon }}^{{\underline{\tau }}}=0,\eqno(8.51)
$$
we can define values (see (8.47) and (8.49))
$$
{\cal P}_\alpha =I_{{\underline{S}}_{q-1}}{E}_{{\underline{\alpha }}}^{{.}{%
\underline{\gamma }}}d{\cal S}_{{\underline{\gamma }}}.
$$
Defined conservation laws (8.51) for ${E}_{{\underline{\beta }}}^{{.}{%
\underline{\epsilon }}}$ have nothing to do with those for
energy--momentum tensor $E_\alpha ^{{.}{\gamma }}$ from Einstein
equations for the almost
Hermitian gravity 
 [160,161] or with ${\tilde E}_{\alpha \beta }$ from
(8.45) with vanishing divergence $D_\gamma {\tilde E}_\alpha ^{{.}{\gamma }%
}=0.$ So ${\tilde E}_\alpha ^{{.}{\gamma }}{\neq }{E}_\alpha
^{{.}{\gamma }}. $ A similar conclusion was made in
 [100] for unispacial locally
isotropic tensor integral. In the case of multispatial tensor
integration we
have another possibility (firstly pointed in 
 [278,258] for
Einstein-Cartan spaces), namely, to identify ${E}_{{\underline{\beta }}}^{{.}%
{\underline{\gamma }}}$ from (8.51) with the na-image of ${E}_\beta ^{{.}{%
\gamma }}$ on la--space $\xi .$ We shall consider this
construction in the next section.

\section{Na--Conservation Laws in La--Gravity}

It is well known that the standard pseudo--tensor description of
the \index{Na--conservation laws} energy--momen\-tum values for
the Einstein gravitational fields is full of ambiguities. Some
light was shed by introducing additional geometrical
structures on curved space--time (bimetrics 
 [211,143], biconnections
 [59], by taking into account background spaces
 [94,289], or
formulating variants of general relativity theory on flat space
  [152,99]).  We emphasize here that rigorous mathematical investigations
based on two (fundamental and background) locally anisotropic, or
isotropic, spaces should use well--defined, motivated from
physical point of view, mappings of these spaces. Our na--model
largely contains both attractive features of the mentioned
approaches; na--maps establish a local 1--1 correspondence
between the fundamental la--space and auxiliary la--spaces on
which biconnection (or even multiconnection) structures are
induced. But these structures are not a priory postulated as in a
lot of gravitational theories, we tend to specify them to be
locally reductible to the locally isotropic Einstein theory
 [94,152].

Let us consider a fixed background la--space $\underline {\xi}$
with given
metric ${\underline G}_{\alpha \beta} = ({\underline g}_{ij} , {\underline h}%
_{ab} )$ and d--connection ${\underline {\tilde
{\Gamma}}}^{\alpha}_{\cdot \beta\gamma},$ for simplicity in this
subsection we consider compatible metric and connections  being
torsionless and with vanishing curvatures. Supposing that there
is an nc--transform from the fundamental la--space $\xi$ to the
auxiliary $\underline {\xi} .$ we are interested in the
equivalents of the Einstein equations (8.45) on $\underline {\xi}
.$

We consider that a part of gravitational degrees of freedom is
"pumped out"
into the dynamics of deformation d--tensors for d--connection, ${P^{\alpha}}%
_{\beta \gamma},$ and metric, $B^{\alpha \beta} =  ( b^{ij} ,
b^{ab} ) .$
The remained part of degrees of freedom is coded into the metric ${%
\underline G}_{\alpha \beta}$ and d--connection ${\underline
{\tilde {\Gamma}}}^{\alpha}_{\cdot \beta \gamma} .$

Following 
 [94,252] we apply the first order formalism and consider $%
B^{\alpha \beta }$ and ${P^\alpha }_{\beta \gamma }$ as
independent variables on $\underline{\xi }.$ Using notations
$$
P_\alpha ={P^\beta }_{\beta \alpha },\quad {\Gamma }_\alpha ={{\Gamma }%
^\beta }_{\beta \alpha },
$$
$$
{\hat B}^{\alpha \beta }=\sqrt{|G|}B^{\alpha \beta },{\hat
G}^{\alpha \beta
}=\sqrt{|G|}G^{\alpha \beta },{\underline{\hat G}}^{\alpha \beta }=\sqrt{|%
\underline{G}|}{\underline{G}}^{\alpha \beta }
$$
and making identifications
$$
{\hat B}^{\alpha \beta }+{\underline{\hat G}}^{\alpha \beta }={\hat G}%
^{\alpha \beta },{\quad }{\underline{\Gamma }}_{\cdot \beta \gamma }^\alpha -%
{P^\alpha }_{\beta \gamma }={{\Gamma }^\alpha }_{\beta \gamma },
$$
we take the action of la--gravitational field on $\underline{\xi
}$ in this form:
$$
{\underline{{\cal S}}}^{(g)}=-{(2c{\kappa }_1)}^{-1}\int {\delta }^qu{}{%
\underline{{\cal L}}}^{(g)},\eqno(8.52)
$$
where
$$
{\underline{{\cal L}}}^{(g)}={\hat B}^{\alpha \beta }(D_\beta
P_\alpha
-D_\tau {P^\tau }_{\alpha \beta })+({\underline{\hat G}}^{\alpha \beta }+{%
\hat B}^{\alpha \beta })(P_\tau {P^\tau }_{\alpha \beta }-{P^\alpha }%
_{\alpha \kappa }{P^\kappa }_{\beta \tau })
$$
and the interaction constant is taken ${\kappa }_1={\frac{4{\pi }}{{c^4}}}k,{%
\quad }(c$ is the light constant and $k$ is Newton constant) in
order to obtain concordance with the Einstein theory in the
locally isotropic limit.

We construct on $\underline{\xi }$ a la--gravitational theory
with matter fields (denoted as ${\varphi }_A$ with $A$ being a
general index) interactions by postulating this Lagrangian
density for matter fields
$$
{\underline{{\cal L}}}^{(m)}={\underline{{\cal L}}}^{(m)}[{\underline{\hat G}%
}^{\alpha \beta }+{\hat B}^{\alpha \beta };{\frac \delta {\delta u^\gamma }}(%
{\underline{\hat G}}^{\alpha \beta }+{\hat B}^{\alpha \beta });{\varphi }_A;{%
\frac{\delta {\varphi }_A}{\delta u^\tau }}].\eqno(8.53)
$$

Starting from (8.52) and (8.53) the total action of la--gravity on $%
\underline{\xi }$ is written as
$$
{\underline{{\cal S}}}={(2c{\kappa }_1)}^{-1}\int {\delta }^qu{\underline{%
{\cal L}}}^{(g)}+c^{-1}\int {\delta }^{(m)}{\underline{{\cal L}}}^{(m)}.%
\eqno(8.54)
$$
Applying variational procedure on $\underline{\xi },$ similar to
that
presented in 
 [94] but in our case adapted to N--connection by using
derivations (2.4) instead of partial derivations, we derive from
(8.54) the la--gravitational field equations
$$
{\bf {\Theta }}_{\alpha \beta }={{\kappa }_1}({\underline{{\bf
t}}}_{\alpha \beta }+{\underline{{\bf T}}}_{\alpha \beta
})\eqno(8.55)
$$
and matter field equations
$$
{\frac{{\triangle }{\underline{{\cal L}}}^{(m)}}{\triangle {\varphi }_A}}=0,%
\eqno(8.56)
$$
where $\frac{\triangle }{\triangle {\varphi }_A}$ denotes the
variational derivation.

In (8.55) we have introduced these values: the energy--momentum
d--tensor for la--gravi\-ta\-ti\-on\-al field
$$
{\kappa }_1{\underline{{\bf t}}}_{\alpha \beta }=({\sqrt{|G|}})^{-1}{\frac{%
\triangle {\underline{{\cal L}}}^{(g)}}{\triangle G^{\alpha \beta }}}%
=K_{\alpha \beta }+{P^\gamma }_{\alpha \beta }P_\gamma -{P^\gamma
}_{\alpha \tau }{P^\tau }_{\beta \gamma }+
$$
$$
{\frac 12}{\underline{G}}_{\alpha \beta }{\underline{G}}^{\gamma \tau }({%
P^\phi }_{\gamma \tau }P_\phi -{P^\phi }_{\gamma \epsilon }{P^\epsilon }%
_{\phi \tau }),\eqno(8.57)
$$
(where
$$
K_{\alpha \beta }={\underline{D}}_\gamma K_{\alpha \beta }^\gamma
,
$$
$$
2K_{\alpha \beta }^\gamma =-B^{\tau \gamma }{P^\epsilon }_{\tau (\alpha }{%
\underline{G}}_{\beta )\epsilon }-B^{\tau \epsilon }{P^\gamma
}_{\epsilon (\alpha }{\underline{G}}_{\beta )\tau }+
$$
$$
{\underline{G}}^{\gamma \epsilon }h_{\epsilon (\alpha }P_{\beta )}+{%
\underline{G}}^{\gamma \tau }{\underline{G}}^{\epsilon \phi }{P^\varphi }%
_{\phi \tau }{\underline{G}}_{\varphi (\alpha }B_{\beta )\epsilon }+{%
\underline{G}}_{\alpha \beta }B^{\tau \epsilon }{P^\gamma }_{\tau
\epsilon }-B_{\alpha \beta }P^\gamma {\quad }),
$$
$$
2{\bf \Theta }={\underline{D}}^\tau {\underline{D}}_{tau}B_{\alpha \beta }+{%
\underline{G}}_{\alpha \beta }{\underline{D}}^\tau
{\underline{D}}^\epsilon
B_{\tau \epsilon }-{\underline{G}}^{\tau \epsilon }{\underline{D}}_\epsilon {%
\underline{D}}_{(\alpha }B_{\beta )\tau }
$$
and the energy--momentum d--tensor of matter
\index{Energy--momentum d--tensor}
$$
{\underline{{\bf T}}}_{\alpha \beta }=2{\frac{\triangle {\cal L}^{(m)}}{%
\triangle {\underline{\hat G}}^{\alpha \beta
}}}-{\underline{G}}_{\alpha
\beta }{\underline{G}}^{\gamma \delta }{\frac{\triangle {\cal L}^{(m)}}{%
\triangle {\underline{\hat G}}^{\gamma \delta }}}.\eqno(8.58)
$$
As a consequence of (8.56)--(8.58) we obtain the d--covariant on $\underline{%
\xi }$ conservation laws
$$
{\underline{D}}_\alpha ({\underline{{\bf t}}}^{\alpha \beta }+{\underline{%
{\bf T}}}^{\alpha \beta })=0.\eqno(8.59)
$$
We have postulated the Lagrangian density of matter fields (8.53)
in a form
as to treat ${\underline{{\bf t}}}^{\alpha \beta }+{\underline{{\bf T}}}%
^{\alpha \beta }$ as the source in (8.55).

Now we formulate the main results of this section:

\begin{proposition}
The dynamics of the Einstein la--gravitational fi\-elds,
mo\-del\-ed as solutions of equations (8.45) and matter fields on
la--space $\xi ,$ can be equivalently locally modeled on a
background la--space $\underline{\xi }$ provided with a trivial
d-connection and metric structures having zero d--tensors of
torsion and curvature by field equations (8.55) and (8.56) on
condition that deformation tensor ${P^\alpha }_{\beta \gamma }$
is a solution of the Cauchy problem posed for basic equations for
a chain of na--maps from $\xi $ to $\underline{\xi }.$
\end{proposition}

\begin{proposition}
Local, d--tensor, conservation laws for Einstein
la--gra\-vi\-ta\-ti\-on\-al fi\-elds can be written in form
(8.59) for la--gravita\-ti\-on\-al (8.57) and matter (8.58)
energy--momentum d--tensors. These laws are d--covariant on the
background space $\underline{\xi }$ and must be completed with
invariant conditions of type (8.12)-(8.15) for every deformation
parameters of a chain of na--maps from $\xi $ to $\underline{\xi
}.$
\end{proposition}

The above presented considerations consist proofs of both
propositions.

We emphasize that nonlocalization of both locally anisotrop\-ic
and isot\-rop\-ic gravitational energy--momentum values on the
fundamental (locally an\-isot\-rop\-ic or isotropic) space $\xi $
is a consequence of the absence of global group automorphisms for
generic curved spaces. Considering gravitational theories from
view of multispaces and their mutual maps
 (directed by the basic geometric structures on $\xi  $ such as
 N--con\-nec\-ti\-on, d--con\-nec\-ti\-on, d--tor\-si\-on and
 d--cur\-va\-tu\-re components, see
coefficients for basic na--equations (8.9)-(8.11)), we can
formulate local d--tensor conservation laws on auxiliary globally
automorphic spaces being related with space $\xi $ by means of
chains of na--maps. Finally, we remark that as a matter of
principle we can use d--connection deformations in order to
modelate the la--gravitational interactions with nonvanishing
torsion and nonmetricity. In this case we must introduce a
corresponding source in (8.59) and define generalized
conservation laws as in (8.44) \quad (see similar details for
locally isotropic generalizations of the Einstein
gravity in Refs 
  [274,278,250]).

\section{Na--maps in Einstein Gravity}

The nearly autoparallel map methods plays also an important role
in \index{Na--maps!in Einstein gravity} formulation of
conversation laws for locally isotropic gravity
  [249,233,274]
(Einstein  gravity), 
 [252] (Einstein-Cartan theory),
 [273,278] (gauge gravity  and gauge fields), and 
 [273,251]
(gravitation with torsion and nonmetricity). We shall present
some examples of na-maps for solutions of Einstein equations (for
trivial N-connection structures the equations (8.45) are usual
gravitational field equations in general relativity) and analyze
the  problem of formulation laws on curved locally anisotropic
spaces).

\subsection{Nc--flat solutions of Einstein equations}

In order to illustrate some applications of the na-map theory we
consider three particular solutions of Einstein equations.
\index{Einstein equations!nc--flat solutions}

\subsubsection{Example 1.}

In works 
 [42,43] one introduced te metric
$$
d s^2 = g_{ij} d x^i d x^j =exp (2 \Omega (q)) (d t^2 -d x^2) -%
$$
$$
q^2 (exp(2\beta (q)) dy^2 + exp(-2\beta (q)) d z^2), \eqno(8.60)
$$
where $q=t-x .$ If functions $\Omega (q)$ and $\beta (q) ,$
depending only the variable $q,$ solve equations $2 d\Omega /dq
=q (d\beta (q)/dq),$ the metric (8.60) satisfies the vacuum
Einstein equations. In general this metric describes curved
spaces, but if function $\beta (q)$ is a solution of equations
$$
q(d^2 \beta /dq^2 ) + 2 d\beta /dq =q^2 (d\beta /dq)^3%
$$
we have a flat space, i. e. $R^{\ i}_{j\ kl} =0.$

1a)\ Let us show that metric (8.60) admits also $na_{(3)} +
na_{(1)}$-maps to Min\-kow\-ski space. If $q=0$, using conformal
rescaling
$$
{\tilde g}_{ij} =q^{-2} exp(-2\beta ) {\tilde g}_{ij},%
$$
and introducing new variables $x^a = (p,v,y,z) (a=0,1,2,3),$ where
$$
v=t+p,\ p= \frac{1}{2} \int dq q^{-2} exp[2(\Omega - \beta )]%
$$
and $S(p) = -4 {\beta} (q)_{|q=q(p)},$ we obtain
$$
na_{(3)} : ds^2 \to d{\tilde s}^2 =2 dp dv -dy^2 - exp{S(p)} dz^2
\eqno(8.61)
$$
with one non-zero component of connection ${\tilde {\Gamma}}^3_{03} = \frac{1%
}{2}{(dS/dp)} = \frac{1}{2} {\dot S}.$ A next possible step will
be a na-map to the flat space with metric $ds^2 = dp^2 - dv^2 -
dy^2 - dz^2. $ Really, the set of values ${\tilde P}^3_{03} = -
\frac{1}{2} {\dot R},$ the rest of components of the deformation
tensor being equal to zero, and $b_0 = [2 \ddot S - (\dot S)^2 ]/
S, b_1 = b_2 = 0, b_3 =-\dot S , a_{ij}=0,
Q^i_{jk}=0$ solves equations (8.45) for a map $na_{(1)}: d {\tilde s}^2 \to d%
{\underline s}^2.$

1b)The metric $d {\tilde s}$ from (8.61) also admits a
$na_{(2)}$-map to the
flat space (as N. Sinyukov pointed out 
 [230] intersections of type of
na-maps are possible). For example, the set of values $r=(r_0 (p),0,0,0), {%
\nu}_b =0, \sigma = (0,0,0, {\sigma}_3 ),$
$$
{\sigma}_3 = -(2 F^3_{0(0)})^{-1} {\dot S}exp [-\int r_0 (p) dp +
S(p)],
F^3_0 =F^3_{0(0)} exp [\int r_0 (p)dp -  S(p)],%
$$
constant $F^3_{0(0)}\neq 0,$ the rest of the affinor components
being  equal to zero, and $r(p)$ is an arbitrary function on $p$
satisfies basic equations with $e=0$ for a map $na_{(2)}: ds^2
\to d{\tilde s}^2.$

1c) Metric $d{\tilde s}^2$ from (8.61) does not admit
$na_{(3)}$-maps to the flat space because, in general, conditions
$^{(3)}{\tilde W}^a_{bcd}=0,$ are not satisfied, for example,
$^{(3)}{\tilde W}^1_{001} = \frac{1}{4} ( \ddot S +\frac{1}{2}
\dot S ).$

\subsubsection{Example 2.}

The Peres's solution of Einstein equations 
 [183] is given by metric
$$
d s^2 = -dx^2 -dy^2 + 2 d\eta d\lambda - Q(x,y) N(\zeta ) d
{\zeta}^2 , \eqno(8.62)
$$
where $x=x^1 , y = x^2 , \lambda = x^3 , \zeta = x^4 .$ Parametrizing $%
Q(x,y)=\frac{1}{4} (x^2 - y^2 ), N(\zeta ) =sin \zeta $ we obtain
plane wave solution of Einstein equations. Choosing
$$
Q(x,y) = - \frac{xy}{2(x^2 +y^2)^2} \mbox{ and } N\left( \zeta
\right) =\{
\begin{array}{c}
\exp \left[ \left( b^2-\zeta ^2\right) \right] , \mbox{ if
}\left| \zeta \right| <b; \\ 0\qquad ,\qquad \mbox{if }\left|
\zeta \right| \geq 0 ,
\end{array}
$$
where $b$ is a real constant, we obtain a wave packed solutions
of Einstein equations. By straightforward calculations we can
convince ourselves that values
$$
{\sigma}_1 = \frac{1}{2} Q_1 N, {\sigma}_2 = \frac{1}{2} Q_2 N,
{\sigma}_3 =0, {\sigma}_4 = \frac{1}{2} Q {\dot N}, r_b =0,
{\nu}_b =0,
$$
$$
Q_1 = {\frac{\partial Q}{\partial x}},\qquad {\dot N} = dN/d\zeta
$$
solve basic equations for a map $na_{(2)}: d{\underline s}^2 \to
d s^2, $ where\\ ${\underline g}_{ij} =diag(-1,-1,-1,1).$ So,
metrics of type (8.62),
being of type II, according to Petrov's classification 
 [185], admits $%
na_{(2)}$-maps to the flat space.

\subsubsection{Example 3.}

Now, we shall show that by using superpositions of na-maps and
imbeddings into pseudo--Riemannian spaces with dimension $d>2$ we
can construct metrics satisfying Einstein equations. Let consider
a 2-dimensional metric
$$
d s^2_{(2)} = -d^2 (ln t) + (m-n)^{-2} d t^2 ,%
$$
where $m$ and $n$ (only in this subsection) are constants and
$t>0.$ Firstly, we use the conformal rescaling:
$$
na_{(3)} : d s^2_{(2)} \to d {\tilde s}^2_{(2)}=e^{2x}
t^{2(m+n-1)} ds^2_{(2)}
$$
and the canonical imbedding into a 3-dimensional space,
$$
d {\tilde s}^2_{(2)} \subset d s^2_{(3)} = e^{2x} t^{2(m+n+1)}
(-d^2(lnt) +
(m-n)^{-2} d x^2 ) + d y^2 .%
$$
Then we consider a conformal rescaling of this 3-dimensional
space,
$$
na_{(3)}: d s^2_{(3)} \to d {\tilde s}^2_{(3)} = e^{-4x} d s^2_{(3)}%
$$
and, the next, the canonical imbedding into a 4 dimensional
pseudo--Ri\-e\-man\-ni\-an space
$$
d {\tilde s}^2_{(3)} \subset d {\tilde s}^2_{(4)} =%
$$
$$
e^{-4x} [e^{2x} t^{2(m+n+1)}(-d^2(lnt)+(m-n)^{-2}d x^2 ) + d y^2] + d z^2 .%
$$
Finally, after a conformal mapping from $d {\tilde s}^2_{(4)},$
$$
na_{(3)}: d{\tilde s}^2_{(4)} \to d s^2 =%
$$
$$
e^{2x} t^{-2(n+m)} \{ e^{-4x}[t^{2(m+n+1)}(-d^2 lnt) + (m-n)^{-2}
d x^2 ) +
dy^2 ] + d z^2 \},%
$$
we obtain a metric $d s^2 $ describing a class of space-times
with ideal liquid matter (as a rule, some relations between
constants $n$ and $m$ and the physical parameters of the liquid
medium are introduced, see details in
 [145]).

Similar considerations, but in general containing investigations
of more complicated systems of first order Pfaff equations, show
that a very large class of the Einstein equations solutions can
be locally generated by using imbeddings and chains of na-maps
from flat auxiliary background spaces of lower dimensions than
that of the fundamental space-time.

\subsection{Na--conservation laws in general relativity}

We have considered two central topics in the Chapter 8: the first
was the \index{Na--conservation laws!in general relativity}
generalization of the nearly autoparallel map theory to the case
of spaces with local anisotropy and the second was the
formulation of the gravitational models on locally anisotropic
(or in particular cases, isotropic) na-backgrounds. For trivial
N-connection and d-torsion structures and pseudo--Riemannian
metric the la-gravitational field equations (8.45) became usual
Einstein equations for general relativity. The results of section
8.5 hold good in simplified locally isotropic version. In this
subsection we pay  a special attention to the very important
problem of definition of conservation  laws for gravitational
fields.

A brief historical note is in order. A. Z. Petrov at the end of
60th
initiated 
 [184,186,187,188] a programme of modeling field
interactions and modeling general relativity on arbitrary
pseudo--Riemannian spaces. He advanced an original approach to
the gravitational energy-momentum problem.  After Petrov's death
(1972) N. S. Sinykov finally
elaborated 
 [230] the  geometrical theory of nearly geodesics maps
(ng-maps; on (pseudo-) Riemannian spaces geodesics coincide with
autoparallels) of affine connected spaces as an  extension for the
(n-2)-projective space theory 
 [284,166], concircular  geometry 
 [294], holomorphic projective mappings of Kahlerian spaces 
 [179] and
holomorphic projective correspondences for almost complex spaces
 [121,237]. A part of Petrov's geometric purposes have been achieved, but
Sinyukov's works practically do not contain investigations and
applications in gravitational theories. In addition to the
canonical formulation of the general relativity theory in the
framework of the pseudo-Riemannian geometry, we tried to
elaborate well-defined criterions for  conditions when for
gravitational (locally isotropic or anisotropic) field theories
equivalent reformulations in arbitrary (curved or flat) spaces
are possible
 [249,233,274,278,273,251].

The concept of spaces na-maps implies a kind of space-field
Poincare
conventionality 
 [191,192] which, in our case, states the possibility
to  formulate physical theories equivalently on arbitrary curved,
or flat, spaces (na-backgrounds, being the images of na-map
chains) with a further choice of one of them to be the space-time
from a view of convenience or simplicity. If the existence of
ideal probing bodies (not destroying the space-time structure by
measurements) is postulated, the fundamental pseudo-Riemannian,
or a corresponding generalization, structure can be established
experimentally. Mappings to other curved or flat spaces would be
considered as mathematical "tricks" for illustrating some
properties and possible transformations of field equations. We
remark a different situation in quantum gravity where ideal
probing bodies are not introduced and one has to consider all
na-backgrounds as equal in rights. Perhaps, it is preferred
to concern quantum gravitational problems on the space-time category ${\cal C%
}( \eta )$ (see section 8.2, Theorem 8.4 ; to analyze the
renormalization of field interactions we shall extend our
considerations on  the tangent bundle, or bundles of higher
tangence) in order to develop quantum field theories with
gravitational interactions on multispaces interrelated with
"quantized" na-maps, on the category theory see 
 [51].

In spite of some common features between our approach and those
based on background methods
 [289,94,99] the na-method gives rise to a
different class of gravitational theories. As a rule, background
investigations  are carried out in the linear approximation and
do not take into consideration the "back reaction" of
perturbations. Sometimes one uses successive approximations. In
the framework of the na-map theory we do not postulate the
existence of any {\it a priori} given background space-time. The
auxiliary spaces (curved, or flat) being considered by us, should
belong to the set of spaces which are images of na-map chains
from the fundamental space-time. As a matter of principle such
nc-transforms can be constructed as to be directed by solutions
of the Einstein equations; if the Cauchy problem is posed for
both field and basic na-maps equations the set of possible
na-backgrounds can be defined exactly and without any
approximations and additional suppositions. Physical processes
can be modeled locally equivalently on every admissible
na-background space.
According to A. Z. Petrov 
 [184,186,187,188], if we tend to model
the gravitational  interactions, for example, in the Minkowski
space, the space-time curvature  should be "pumped out" into a
"gravitational force". Free gravitating ideal  probing bodies do
not "fall" along geodesics in a such auxiliary flat space.  We
have to generalize the concept of geodesics by introducing nearly
geodesics or, more generally, nearly autoparallels in order to
describe equivalently trajectories of point probing mass as well
field interactions on  the fundamental space-time and auxiliary
na-backgrounds.

So, we can conclude that to obtain an equivalent formulation of
the general relativity theory on na-backgrounds one has to
complete the gravitational field equations (8.55) with equations
for na-map chains (minimal chains are pre\-fer\-red). Of
cour\-se, one also has to consider two types of conservation
laws: for values of energy-momentum type, for gravitational
fields, see (8.59), and conditions of type (8.12),(8.13),(8.14)
and (8.15) for na--maps invariants.

We point out the solitarity of mappings to the Minkowski
space-time. This space  is not only a usual and convenient arena
for developing new variants of theoretical models and
investigating conservation laws in already known manners.
Perhaps, the "simplest" pseudo-Euclidean background should be
considered as a primary essence from which, by using different
types of maps, more generalized curved spaces (auxiliary, or
being fundamental Einstein space-times) can be constructed. Here
the question arises:\ May be the  supplementary relations to the
background field equations and conservation  laws are motivated
only by our requirement that mappings from the curved space-time
to a flat space to be obtained only by the mean of na-maps and do
not reflect any, additional to the background formulations,
properties of  the Einstein gravitational fields? The answer is
related with the problem of  concordance of geometrical
structures on all spaces taken into consideration.  Really, on
backgrounds we have biconnection (or even multi--connection
structures). So, various types of geodesics, nearly geodesics,
motion and field equations, and a lot of other properties, being
characteristic for such spaces, can be defined. That is why, we
need a principle to base and describe  the splitting of
geometrical structures by 1-1 local mappings from one space  to
another (which we have advocated to be the transformation of
geodesics into nearly autoparallels). Of course, other types of
space mappings can be advanced. But always equations defined for
maps splittings of geometrical structures and corresponding
invariants must be introduced. This is the price  we shall pay if
the general relativity is considered out of its natural
pseudo-Riemannian geometry. Sometimes such digressions from the
fundamental space-time geometry are very useful and, as it was
illustrated, gives us the possibility to formulate, in a
well-known for flat spaces manner, the conservation laws for
gravitational fields. However, all introduced ma-map  equations
and invariant conditions are not only motivated by a chosen type
of  mappings' and auxiliary spaces' artifacts. Being determined
by the fundamental objects on curved space-time, such as metric,
connection and curvature, these na-relations pick up a part of
formerly unknown properties and symmetries of gravitational
fields an can be used (see section 8.4) for a new type  (with
respect to na-maps flexibility) classification of curved spaces.

Finally, we remark that because na-maps consists a class of
nearest extensions of conformal rescalings we suppose them to
play a preferred role among others being more sophisticate and,
at present time, less justified from physical point of view.


\chapter{HA--Strings}

The relationship between two dimensional $\sigma $-models and
strings has been considered
 [153,80,53,229,7]
in order to discuss the effective low energy field equations for
the massless models of strings. In this Chapter we shall study
some of the problems associated with the theory of higher order
anisotropic strings being a natural generalization to higher
order anisotropic  backgrounds (we shall write in brief
ha--backgrounds, ha--spaces and ha--geometry) of the Polyakov's
covariant functional--integral
approach to string theory 
 [193]. Our aim is to show that a
corresponding low--energy string dynamics contains the motion
equations for field equations on ha-spaces; models of ha--gravity
could be more adequate for investigation of quantum gravitational
and Early Universe cosmology.

The plan of the Chapter is as follows. We begin, section 9.1,
with a study of the nonlinear $\sigma$--model and ha--string
propagation by developing the d--covariant method of
ha--background field. Section 9.2 is devoted to
problems of regularization and renormalization of the locally anisotropic $%
\sigma$--model and a corresponding analysis of one- and two--loop
diagrams of this model. Scattering of ha-gravitons and duality
are considered in section 9.3, and a summary and conclusions are
drawn in section 9.4.\

\section{HA--$\sigma$--Models}

In this section we present a generalization of some necessary
results on nonlinear $\sigma$--model and string propagation to
the case of \index{S!$\sigma$--model} ha--backgrounds.
Calculations on both type of locally anisotropic and isotropic
spaces are rather similar if we accept the Miron and Anastasiei
 [160,161] geometric formalism. We emphasize that on ha--backgrounds we
have to take into account the distinguished character, by
N--connection, of geometric objects.

\subsection{Action of nonlinear $\sigma$--model and torsion of ha--space}

Let a map of a two--dimensional (2d), for simplicity, flat space
$M^2$ into ha-space $\xi $ defines a $\sigma $--model field
$$u^{<\mu >}\left( z\right) =\left( x^i\left( z\right)
,y^{<a>}\left( z\right) \right) =\left( x^i\left( z\right)
,y^{a_1}\left( z\right) ,...,y^{a_z}\left( z\right) \right) ,$$
where $z=\left\{ z^A,A=0,1\right\} $ are two--dimensional complex
coordinates on $M^2.$ The moving of a bosonic string in ha--space
is governed by the \index{Bosonic string!in ha--space} nonlinear
$\sigma $-model action (see, for instance,
 [153,80,53,229]
for details on locally isotropic spaces):
$$
I=\frac 1{\lambda ^2}\int d^2z[\frac 12\sqrt{\gamma }\gamma
^{AB}\partial _Au^{<\mu >}\left( z\right) \partial _Bu^{<\nu
>}\left( z\right) G_{<\mu
><\nu >}\left( u\right) +
$$
$$
\frac{\widetilde{n}}3\epsilon ^{AB}\partial _Au^{<\mu >}\partial
_Bu^{<\nu
>}b_{<\mu ><\nu >}\left( u\right) +\frac{\lambda ^2}{4\pi }\sqrt{\gamma }%
R^{(2)}\Phi \left( u\right) ],\eqno(9.1)
$$
where $\lambda ^2$ and $\widetilde{n}\,$ are interaction
constants, $\Phi \left( u\right) $ is the dilaton field,
$R^{(2)}$ is the curvature of the 2d world sheet provided with
metric $\gamma _{AB},\gamma =\det \left( \gamma
_{AB}\right) $ and $\partial _A=\frac \partial {\partial z^A},$ tensor $%
\epsilon ^{AB}$ and d-tensor $b_{<\mu ><\nu >}$ are antisymmetric.

From the viewpoint of string theory we can interpret $b_{<\alpha
><\beta >}$ as the vacuum expectation of the antisymmetric, in
our case locally anisotropic, d-tensor gauge field $B_{<\alpha
><\beta ><\gamma >}$ (see considerations for locally isotropic
models in
 [244,115] and the
Wess-Zumino-Witten model 
 [292,287], which lead to the conclusion 
 [44] that $\widetilde{n}$ takes only integer values and that in the
perturbative quantum field theory the effective quantum action
depends only on B$_{...}$ and does not depend on b$_{...}$ ).

In order to obtain compatible with N--connection motions of
ha--strings we
consider these relations between d--tensor $b_{<\alpha ><\beta >},$ strength $$%
B_{<\alpha ><\beta ><\gamma >}=\delta _{[<\alpha >}b_{<\beta
><\gamma >]}$$
and torsion $T_{.<\beta ><\gamma >}^{<\alpha >}:$%
$$
\delta _{<\alpha >}b_{<\beta ><\gamma >}=T_{<\alpha ><\beta ><\gamma >},%
\eqno(9.2)
$$
with the integrability conditions \index{Integrability conditions}
$$
\Omega _{a_fb_f}^{<a_p>}\delta _{<a_p>}b_{<\beta ><\gamma
>}=\delta _{[a_f}T_{a_f]<\beta ><\gamma >},(0\leq f<p\leq
z),\eqno(9.3)
$$
where $\Omega _{a_fb_f}^{<a_p>}$ are the coefficients of the
N-connection curvature. In this case we can express $B_{<\alpha
><\beta ><\gamma
>}=T_{[<\alpha ><\beta ><\gamma >]}.$ Conditions (9.2) and (9.3) define a
simplified model of ha--strings when the $\sigma $-model
antisymmetric strength is induced from the ha--background
torsion. More general constructions are possible by using normal
coordinates adapted to both N-connection and torsion structures
on ha--background space. For simplicity, we omit such
considerations in this work.

Choosing the complex (conformal) coordinates $z=\iota ^0+i\iota ^1,\overline{%
z}=\iota ^0-i\iota ^1,$ where $\iota ^A,A=0,1$ are real
coordinates, on the world sheet we can represent the
two-dimensional metric in the conformally flat form:
$$
ds^2=e^{2\varphi }dzd\overline{z},\eqno(9.4)
$$
where $\gamma _{z\overline{z}}=\frac 12e^{2\varphi }$ and $\gamma
_{zz}=\gamma _{\overline{z}\overline{z}}=0.$

Let us consider an ha-field $U\left( u\right) ,u\in {\cal
E}^{<z>}$ taking values in ${\cal G}$ being the Lie algebra of a
compact and semi simple Lie
group,%
$$
U\left( u\right) =\exp [i\varphi \left( u\right) ],\varphi \left(
u\right) =\varphi ^{\underline{\alpha }}\left( u\right)
q^{\underline{\alpha }},
$$
where $q^{\underline{\alpha }}$ are generators of the Lie algebra
with
antisymmetric structural constants $f^{\underline{\alpha }\underline{\beta }%
\underline{\gamma }}$ satisfying conditions%
$$
[q^{\underline{\alpha }},q^{\underline{\beta }}]=2if^{\underline{\alpha }%
\underline{\beta }\underline{\gamma }}q^{\underline{\gamma }},\quad tr(q^{%
\underline{\alpha }}q^{\underline{\beta }})=2\delta ^{\underline{\alpha }%
\underline{\beta }}.
$$

The action of the Wess-Zumino-Witten type ha--model should be
written as
$$
I\left( U\right) =\frac 1{4\lambda ^2}\int d^2z\ tr\left( \partial
_AU\partial ^AU^{-1}\right) +\widetilde{n}\Gamma \left[ U\right]
,\eqno(9.5)
$$
where $\Gamma \left[ U\right] $ is the standard topologically
invariant
functional 
 [44]. For perturbative calculations in the framework of the
model (9.1) it is enough to know that as a matter of principle we
can
represent the action of our theory as (9.5) and to use d-curvature $%
r_{<\beta >.<\gamma ><\delta >}^{.<\alpha >}$ for a torsionless
d-connection $\tau _{.<\beta ><\gamma >}^{<\alpha >},$ and
strength $B_{<\alpha ><\beta
><\gamma >}$ respectively expressed as
$$
R_{<\alpha ><\beta ><\gamma ><\delta >}=f_{\underline{\alpha }\underline{%
\beta }\underline{\tau }}f_{\underline{\gamma }\underline{\delta }\underline{%
\tau }}V_{<\alpha >}^{\underline{\alpha }}V_{<\beta >}^{\underline{\beta }%
}V_{<\gamma >}^{\underline{\gamma }}V_{<\delta
>}^{\underline{\delta }}
$$
and%
$$
B_{<\alpha ><\beta ><\gamma >}=\eta f_{\underline{\alpha }\underline{\beta }%
\underline{\tau }}V_{<\alpha >}^{\underline{\alpha }}V_{<\beta >}^{%
\underline{\beta }}V_\tau ^{\underline{\tau }},
$$
where a new interaction constant $\eta \equiv \frac{\widetilde{n}\lambda ^2}{%
2\pi }$ is used and $V_{<\alpha >}^{\underline{\alpha }}$ is a
locally
adapted vielbein associated to the metric (6.12):%
$$
G_{<\alpha ><\beta >}=V_{<\alpha >}^{\underline{\alpha }}V_{<\beta >}^{%
\underline{\beta }}\delta ^{\underline{\alpha }\underline{\beta }}
$$
and
$$
G^{<\alpha ><\beta >}V_{<\alpha >}^{\underline{\alpha }}V_{<\beta >}^{%
\underline{\beta }}=\delta ^{\underline{\alpha }\underline{\beta }}.%
\eqno(9.6)
$$
For simplicity, we shall omit underlining of indices if this will
not give rise to ambiguities.

Finally, in this subsection, we remark that for $\eta =1$ we
obtain a conformally invariant two-dimensional quantum field
theory (being similar to those developed in
 [34]).

\subsection{The d--covariant method for ha--bac\-kgro\-und fi\-elds }

Suggesting the compensation of all anomalies we can fix the gauge
for the two--dimensional metric when action (9.1) is written as
$$
I\left[ u\right] =\frac 1{2\lambda ^2}\int d^2z\{G_{<\alpha
><\beta >}\eta ^{AB}+\frac 23b_{<\alpha ><\beta >}\epsilon
^{AB}\}\partial _Au^{<\alpha
>}\partial _Bu^{<\beta >},\eqno(9.1a)
$$
where $\eta ^{AB}$ and $\epsilon ^{AB}$ are, respectively,
constant two-dimensional metric and antisymmetric tensor. The
covariant method of background field, as general references see
 [139,140,290,5] can be
extended for ha--spaces. Let consider a curve in ${\cal E}^{<z>}$
parameterized as $\rho ^{<\alpha >}\left( z,s\right) ,s\in [0,1],$
satisfying autoparallel equations%
$$
\frac{d^2\rho ^{<\alpha >}\left( z,s\right) }{ds^2}+\Gamma
_{.<\beta
><\gamma >}^{<\alpha >}
\left[ \rho \right] \frac{d\rho ^{<\beta >}}{ds}\frac{%
d\rho ^{<\gamma >}}{ds}=
$$
$$
\frac{d^2\rho ^{<\alpha >}\left( z,s\right) }{ds^2}+\tau
_{.<\beta ><\gamma
>}^{<\alpha >}\left[ \rho \right] \frac{d\rho ^{<\beta >}}{ds}\frac{d\rho
^{<\gamma >}}{ds}=0,
$$
with boundary con\-di\-tions
$$
\rho \left( z,s=0\right) =u\left( z\right) \ \mbox{and}\ \rho
\left( z,s=1\right) u\left( z\right) +v\left( z\right) .
$$
For simplicity, hereafter we shall consider that d-con\-nec\-ti\-on ${{%
\Gamma }^{<\alpha >}}_{<\beta ><\gamma >}$ is defined by
d--metric (6.12) and N-con\-nec\-ti\-on structures, i.e.
$$
{{\Gamma }^{<\alpha >}}_{<\beta ><\gamma >}={}^{\circ }{{\Gamma }^{<\alpha >}%
}_{<\beta ><\gamma >}.
$$
The tangent d-vector $\zeta ^{<\alpha >}=\frac d{ds}\rho
^{<\alpha >},$ where $\zeta ^{<\alpha >}\mid _{s=0}=\zeta
_{\left( 0\right) }^{<\alpha >}$ is chosen as the quantum
d-field. Then the expansion of action $I[u+v\left( \zeta \right)
],$ see (9.1a), as power series on $\zeta ,$
$$
I[u+v\left( \zeta \right) ]=\sum_{k=0}^\infty I_k,
$$
where
$$
I_k=\frac 1{k!}\frac{d^k}{ds^k}I\left[ \rho \left( s\right)
\right] \mid _{s=0},
$$
defines d-covariant, de\-pend\-ing on the background d-field,
in\-ter\-ac\-tion vortexes of locally an\-isot\-rop\-ic $\sigma
$-model.

In order to compute $I_k\,$ it is useful to consider relations%
$$
\frac d{ds}\partial _A\rho ^{<\alpha >}=\partial _A\zeta
^{<\alpha >},\frac d{ds}G_{<\alpha ><\beta >}=\zeta ^{<\tau
>}\delta _{<\tau >}G_{<\alpha
><\beta >},
$$
$$
\partial _AG_{<\alpha ><\beta >}=\partial _A\rho ^{<\tau >}\delta _{<\tau
>}G_{<\alpha ><\beta >},
$$
to introduce auxiliary operators%
$$
\left( \widehat{\nabla }_A\zeta \right) ^{<\alpha >}=\left(
\nabla _A\zeta \right) ^{<\alpha >}-G^{<\alpha ><\tau >}T_{\left[
<\alpha ><\beta ><\gamma
>\right] }\left[ \rho \right] \epsilon _{AB}\partial ^B\rho ^{<\gamma
>}\zeta ^{<\beta >},
$$
$$
\left( \nabla _A\zeta \right) ^{<\alpha >}=\left[ \delta _{<\beta
>}^{<\alpha >}\partial _A+\tau _{.<\beta ><\gamma >}^{<\alpha >}\partial
_A\rho ^{<\gamma >}\right] \zeta ^{<\beta >},
$$
$$
\nabla \left( s\right) \xi ^{<\lambda >}=\zeta ^{<\alpha >}\nabla
_{<\alpha
>}\xi ^{<\lambda >}=\frac d{ds}\xi ^{<\lambda >}+\tau _{.<\beta ><\gamma
>}^{<\lambda >}\left[ \rho \left( s\right) \right] \zeta ^{<\beta >}\xi
^{<\gamma >},\eqno(9.7)
$$
having properties%
$$
\nabla \left( s\right) \zeta ^{<\alpha >}=0,\nabla \left(
s\right) \partial _A\rho ^{<\alpha >}=\left( \nabla _A\zeta
\right) ^{<\alpha >},
$$
$$
\nabla ^2\left( s\right) \partial _A\rho ^{<\alpha >}=r_{<\beta
>.<\gamma
><\delta >}^{.<\alpha >}\zeta ^{<\beta >}\zeta ^{<\gamma >}\partial _A\rho
^{<\delta >},
$$
and to use the curvature d-tensor of d-connection (9.7),
$$
\widehat{r}_{<\beta ><\alpha ><\gamma ><\delta >}=r_{<\beta
><\alpha
><\gamma ><\delta >}-$$ $$\nabla _{<\gamma >}T_{\left[ <\alpha ><\beta ><\delta
>\right] }+\nabla _{<\delta >}T_{\left[ <\alpha ><\beta ><\gamma >\right] }-
$$
$$
T_{\left[ <\tau ><\alpha ><\gamma >\right] }G^{<\tau ><\lambda
>}T_{\left[ <\lambda ><\delta ><\beta >\right] }+T_{\left[ <\tau
><\alpha ><\delta
>\right] }G^{<\tau ><\lambda >}T_{\left[ <\lambda ><\gamma ><\beta >\right]
}.
$$

Values $I_k$ can be computed in a similar manner as in
 [115,44,34]
but in our case by using corresponding d-connections and
d-objects. Here we
present the first four terms in explicit form:%
$$
I_1=\frac 1{2\lambda ^2}\int d^2z2G_{<\alpha ><\beta >}\left( \widehat{%
\nabla }_A\zeta \right) ^{<\alpha >}\partial ^Au^{<\beta
>},\eqno(9.8)
$$
$$
I_2=\frac 1{2\lambda ^2}\int d^2z\{\left( \widehat{\nabla
}_A\zeta \right) ^2+
$$
$$
\widehat{r}_{<\beta ><\alpha ><\gamma ><\delta >}\zeta ^{<\beta
>}\zeta ^{<\gamma >}\left( \eta ^{AB}-\epsilon ^{AB}\right)
\partial _Au^{<\alpha
>}\partial _Bu^{<\beta >}\},
$$
$$
I_3=\frac 1{2\lambda ^2}\int d^2z\{\frac 43(r_{<\beta ><\alpha
><\gamma
><\delta >}-
$$
$$
G^{<\tau ><\epsilon >}T_{\left[ <\epsilon ><\alpha ><\beta
>\right]
}T_{\left[ <\tau ><\gamma ><\delta >\right] })\partial _Au^{<\alpha >}(%
\widehat{\nabla }^A\zeta ^{<\delta >})\zeta ^{<\beta >}\zeta
^{<\gamma >}+
$$
$$
\frac 43\nabla _{<\alpha >}T_{\left[ <\delta ><\beta ><\gamma
>\right] }\partial _Au^{<\beta >}\epsilon ^{AB}(\widehat{\nabla
}_B\zeta ^{<\gamma
>})\zeta ^{<\alpha >}\zeta ^{<\delta >}+
$$
$$
\frac 23T_{\left[ <\alpha ><\beta ><\gamma >\right] }(\widehat{\nabla }%
_A\zeta ^{<\alpha >})\epsilon ^{AB}(\widehat{\nabla }\zeta
^{<\beta >})\zeta ^{<\gamma >}+
$$
$$
\frac 13\left( \nabla _{<\lambda >}r_{<\beta ><\alpha ><\gamma
><\delta
>}+4G^{<\tau ><\epsilon >}T_{\left[ <\epsilon ><\lambda ><\alpha >\right]
}\nabla _{<\beta >}T_{\left[ <\gamma ><\delta ><\tau >\right]
}\right) \times
$$
$$
\partial _Au^{<\alpha >}\partial ^Au^{<\delta >}\zeta ^{<\gamma >}\zeta
^{<\lambda >}+
$$
$$
\frac 13(\nabla _{<\alpha >}\nabla _{<\beta >}T_{\left[ <\tau
><\gamma
><\delta >\right] }+
$$
$$
2G^{<\lambda ><\epsilon >}G^{<\varphi ><\phi >}T_{\left[ <\alpha
><\lambda
><\varphi >\right] }T_{[<\epsilon ><\beta ><\delta >]}T_{\left[ <\phi ><\tau
><\gamma >\right] }+
$$
$$
2r_{<\alpha >.<\beta ><\gamma >}^{.<\lambda >}T_{\left[ <\alpha
><\tau
><\delta >\right] })\partial _Au^{<\gamma >}\epsilon ^{AB}\left( \partial
_B\zeta ^{<\tau >}\right) \zeta ^{<\alpha >}\zeta ^{<\beta >}\},
$$
$$
I_4=\frac 1{4\lambda ^2}\int d^2z\{(\frac 12\nabla _{<\alpha
>}r_{<\gamma
><\beta ><\delta ><\tau >}-
$$
$$
G^{<\lambda ><\epsilon >}T_{\left[ <\epsilon ><\beta ><\gamma
>\right] }\nabla _{<\alpha >}T_{\left[ <\lambda ><\delta ><\tau
>\right] })\times
$$
$$
\partial _Au^{<\beta >}(\widehat{\nabla }^A\zeta ^{<\tau >})\zeta ^{<\alpha
>}\zeta ^{<\gamma >}\zeta ^{<\delta >}+
$$
$$
\frac 13r_{<\beta ><\alpha ><\gamma ><\delta >}(\widehat{\nabla
}_A\zeta ^{<\alpha >})(\widehat{\nabla }^A\zeta ^{<\delta
>})\zeta ^{<\beta >}\zeta ^{<\gamma >}+
$$
$$
(\frac 1{12}\nabla _{<\alpha >}\nabla _{<\beta >}r_{<\delta
><\gamma ><\tau
><\lambda >}+\frac 13r_{<\delta >.<\tau ><\gamma >}^{.<\kappa >}r_{<\beta
><\kappa ><\alpha ><\lambda >}-
$$
$$
\frac 12(\nabla _{<\alpha >}\nabla _{<\beta >}T_{\left[ <\gamma
><\tau
><\epsilon >\right] })G^{<\epsilon ><\pi >}T_{\left[ <\pi ><\delta ><\lambda
>\right] }-
$$
$$
\frac 12r_{<\alpha >.<\beta ><\gamma >}^{.<\kappa >}T_{\left[
<\kappa ><\tau
><\epsilon >\right] }G^{<\epsilon ><\pi >}T_{\left[ <\pi ><\delta ><\lambda
>\right] }+
$$
$$
\frac 16r_{<\alpha >.<\beta ><\epsilon >}^{.<\kappa >}T_{\left[
<\kappa
><\gamma ><\tau >\right] }G^{<\epsilon ><\pi >}T_{\left[ <\epsilon ><\delta
><\lambda >\right] })\times
$$
$$
\partial _Au^{<\gamma >}\partial ^Au^{<\lambda >}\zeta ^{<\alpha >}\zeta
^{<\beta >}\zeta ^{<\delta >}\zeta ^{<\tau >}+
$$
$$
[\frac 1{12}\nabla _{<\alpha >}\nabla _{<\beta >}\nabla _{<\gamma
>}T_{\left[ <\lambda ><\delta ><\tau >\right] }+
$$
$$
\frac 12\nabla _{<\alpha >}\left( G^{<\kappa ><\epsilon
>}T_{\left[ <\epsilon ><\lambda ><\delta >\right] }\right)
r_{<\beta ><\kappa ><\gamma
><\tau >}+
$$
$$
\frac 12\left( \nabla _{<\alpha >}T_{\left[ <\pi ><\beta ><\kappa
>\right] }\right) G^{<\pi ><\epsilon >}G^{<\kappa ><\nu
>}T_{\left[ <\epsilon
><\gamma ><\delta >\right] }T_{\left[ <\nu ><\lambda ><\tau >\right] }-
$$
$$
\frac 13G^{<\kappa ><\epsilon >}T_{\left[ <\epsilon ><\lambda
><\delta
>\right] }\nabla _{<\alpha >}r_{<\beta ><\kappa ><\delta ><\tau >}]\times
$$
$$
\partial _Au^{<\delta >}\epsilon ^{AB}\partial _Bu^{<\tau >}\zeta ^{<\lambda
>}\zeta ^{<\alpha >}\zeta ^{<\beta >}\zeta ^{<\gamma >}+
$$
$$
\frac 12[\nabla _{<\alpha >}\nabla _{<\beta >}T_{[<\tau ><\gamma
><\delta
>]}+T_{[<\kappa ><\delta ><\tau >]}r_{<\alpha >.<\beta ><\gamma >}^{.<\kappa
>}+
$$
$$
r_{<\alpha >.<\beta ><\delta >}^{.<\kappa >}T_{\left[ <\kappa
><\gamma
><\tau >\right] }]
$$
$$
\times \partial _Au^{<\gamma >}\epsilon ^{AB}(\widehat{\nabla
}_B\zeta ^{<\delta >})\zeta ^{<\alpha >}\zeta ^{<\beta >}\zeta
^{<\tau >}+
$$
$$
\frac 12\nabla _{<\alpha >}T_{\left[ <\delta ><\beta ><\gamma >\right] }(%
\widehat{\nabla }_A\zeta ^{<\beta >})\epsilon
^{AB}(\widehat{\nabla }_B\zeta ^{<\gamma >})\zeta ^{<\alpha
>}\zeta ^{<\delta >}\}.
$$

Now we con\-struct the d-covariant ha-background fun\-cti\-o\-nal
(we use meth\-ods, in our case cor\-res\-pond\-ing\-ly adapted to
the N-con\-nect\-i\-on structure, developed in
 [115,140,112]). The standard
quantization technique is based on the functional integral
\index{Functional integral}
$$
Z\left[ J\right] =\exp \left( iW\left[ J\right] \right) =\int
d[u]\exp \{i\left( I+uJ\right) \},\eqno(9.9)
$$
with source $J^{<\alpha >}$ (we use condensed denotations and
consider that
computations are made in the Euclidean space). The generation functional $%
\Gamma $ of one-particle irreducible (1PI) Green functions is
defined as \index{Green functions}
$$
\Gamma \left[ \overline{u}\right] =W\left[ J\left(
\overline{u}\right) \right] -\overline{u}\cdot J\left[
\overline{u}\right] ,
$$
where $\overline{u}=\frac{\Delta W}{\delta J}$ is the mean field.
For explicit perturbative calculations it is useful to connect
the source only with the covariant quantum d-field $\zeta $ and
to use instead of (9.9) the new functional
$$
\exp \left( iW\left[ \overline{u},J\right] \right) =\int \left[
d\zeta \right] \exp \{i\left( I\left[ \overline{u}+v\left( \zeta
\right) \right] +J\cdot \zeta \right) \}.\eqno(9.10)
$$

It is clear that Feynman diagrams obtained from this functional
are \index{Feynman diagrams} d-covariant.

Defining the mean d-field $\overline{\zeta }\left( u\right) =\frac{\Delta W}{%
\delta J\left( u\right) }$ and introducing the auxiliary d-field
$\zeta ^{\prime }=\zeta -\overline{\zeta }$ we obtain from (6.9)
a double expansion
on both classical and quantum ha--backgrounds:%
\index{Ha--background!classical and quantum}
$$
\exp \left( i\overline{\Gamma [}\overline{u},\overline{\zeta
]}\right) =
\int \left[ d\zeta ^{\prime }\right] \exp \{i\left( I\left[ \overline{u}%
+v\left( \zeta ^{\prime }+\overline{\zeta }\right) \right] -\zeta ^{\prime }%
\frac{\Delta \Gamma }{\delta \overline{\zeta }}\right)
\}.\eqno(9.11)
$$

The manner of fixing the measure in the functional (9.10) (and as
a consequence in (9.11) ) is obvious :
$$
[d\zeta ]=\prod_u\sqrt{\mid G\left( u\right) \mid }\prod_{<\alpha
>=0}^{n_E-1}d\zeta ^{<\alpha >}\left( u\right) .\eqno(9.12)
$$

Using vielbein fields (9.6) we can rewrite the measure (9.12) in
the form
$$
\left[ d\zeta \right] =\prod_u\prod_{\alpha =0}^{n_E-1}d\zeta ^{\underline{%
\alpha }}\left( u\right) .
$$
The structure of renormalization of $\sigma $-models of type
(9.10) (or
(9.11)) is analyzed, for instance, in 
 [115,140,112]. For ha--spaces we
must take into account the N--connection structure.

\section{Regularization and $\beta $--Funct\-i\-ons}

The aim of this section is to study the problem of regularization
and quantum ambiguities in $\beta$--functions of the
renormalization group and \index{B!$\beta$--functions}
\index{Renormalization group} to present the results on one- and
two--loop calculus for the ha--$\sigma$--model (HAS--model).
\index{Ha--$\sigma$--model} \index{HAS--model}

\subsection{Renormalization group beta func\-ti\-ons}

Because our $\sigma $-model is a two-dimensional and massless
locally anisotropic theory we have to con\-sider both types of
in\-frared and ul\-tra\-vio\-let reg\-u\-lar\-iza\-tions (in
brief, IR- and UV-reg\-u\-lar\-iza\-tion). In order to regularize
IR-divergences and distinguish them from UV-divergences we can
use a standard mass term in the
action (9.1) of the HAS--model%
\index{Action!of the HAS--model}
$$
I_{\left( m\right) }=-\frac{\tilde m^2}{2\lambda ^2}\int
d^2zG_{<\alpha
><\beta >}u^{<\alpha >}u^{<\beta >}.
$$
For regularization of UF-divergences it is convenient to use the
dimensional regularization. For instance, the regularized
propagator of quantum d--fields
$\zeta $ looks like%
$$
<\zeta ^{<\alpha >}\left( u_1\right) \zeta ^{<\beta >}\left(
u_2\right)
>=\delta ^{<\alpha ><\beta >}G\left( u_1-u_2\right) =
$$
$$
i\lambda ^2\delta ^{<\alpha ><\beta >}\int \frac{d^qp}{\left( 2\pi \right) ^q%
}\frac{\exp \left[ -ip\left( u_1-u_2\right) \right] }{\left(
p^2-\tilde m^2+i0\right) },
$$
where $q=2-2\epsilon .$

The d-covariant dimensional regularization of UF-divergences is
complicated because of existence of the antisymmetric symbol
$\epsilon ^{AB}.$ One
introduces 
 [137,140] this general prescription:%
$$
\epsilon ^{LN}\eta _{NM}\epsilon ^{MR}=\psi \left( \epsilon
\right) \eta ^{LR}
$$
and
$$
\epsilon ^{MN}\epsilon ^{RS}=\omega \left( \epsilon \right)
\left[ \eta ^{MS}\eta ^{NR}-\eta ^{MR}\eta ^{NS}\right] ,
$$
where $\eta ^{MN}$ is the q-dimensional Minkowski metric, and
$\psi \left( \epsilon \right) $ and $\omega \left( \epsilon
\right) $ are arbitrary d-functions satisfying conditions $\psi
\left( 0\right) =\omega \left( 0\right) =1$ and depending on the
type of renormalization.

We use the standard dimensional regularization , with
dimensionless scalar d-field $u^{<\alpha >}\left( z\right) ,$
when expressions for unrenormalized $G_{<\alpha ><\beta
>}^{\left( ur\right) }$ and $B_{<\alpha ><\beta
>}^{(k,l)}$ have a d-tensor character, i.e. they are polynoms on d-tensors
of curvature and torsion and theirs d-covariant derivations (for
simplicity in this subsection we consider $\lambda ^2=1;$ in
general one-loop 1PI-diagrams must be proportional to $\left(
\lambda ^2\right) ^{l-1}$).

RG $\beta $-functions are defined by relations (for simplicity we
shall omit index R for renormalized values)
$$
\mu \frac d{d\mu }G_{<\alpha ><\beta >}=\beta _{\left( <\alpha
><\beta
>\right) }^G\left( G,B\right) ,\mu \frac d{d\mu }B_{\left[ <\alpha ><\beta
>\right] }=\beta _{\left[ <\alpha ><\beta >\right] }^B\left( G,B\right) ,
$$
$$
\beta _{<\alpha ><\beta >}=\beta _{\left( <\alpha ><\beta
>\right) }^G+\beta _{\left[ <\alpha ><\beta >\right] }^B.
$$

By using the scaling property of the one--loop counter--term
under global \index{Counter--term!one--loop}
conformal transforms%
$$
G_{<\alpha ><\beta >}^G\rightarrow \Lambda ^{\left( l-1\right)
}G_{<\alpha
><\beta >}^{\left( k,l\right) },B_{<\alpha ><\beta >}^{\left( k,l\right)
}\rightarrow \Lambda ^{\left( l-1\right) }B_{<\alpha ><\beta
>}^{\left( k,l\right) }
$$
we obtain%
$$
\beta _{\left( <\alpha ><\beta >\right)
}^G=-\sum_{l=1}^{(1,l)}lG_{(<\alpha
><\beta >)}^{(1,l)},\beta _{[<\alpha ><\beta >]}^B=-\sum_{l=1}^\infty
lB_{[<\alpha ><\beta >]}^{(1,l)}
$$
in the leading order on $\epsilon $ (compare with the usual
perturbative
calculus from 
 [107]).

The d-covariant one-loop counter-term is taken as%
$$
\Delta I^{\left( l\right) }=\frac 12\int d^2zT_{<\alpha ><\beta
>}^{\left( l\right) }\left( \eta ^{AB}-\epsilon ^{AB}\right)
\partial _Au^{<\alpha
>}\partial _Bu^{<\beta >},
$$
where
$$
T_{<\alpha ><\beta >}^{\left( l\right) }=\sum_{k=1}^l\frac
1{\left( 2\epsilon ^k\right) }T_{<\alpha ><\beta >}^{\left(
k,l\right) }\left( G,B\right) .\eqno(9.13)
$$
For instance, in the three-loop approximation we have
$$
\beta _{<\alpha ><\beta >}=T_{<\alpha ><\beta
>}^{(1,1)}+2T_{<\alpha ><\beta
>}^{(1,2)}+3T_{<\alpha ><\beta >}^{(1,3)}.\eqno(9.14)
$$

In the next subsection we shall also consider constraints on the
structure of $\beta $-functions connected with conditions of
integrability (caused by conformal invariance of the
two-dimensional world-sheet).

\subsection{1--loop divergences and ha--renorm gro\-up equ\-ati\-ons}

We generalize the one-loop results
 to the case of ha-backgrounds.
If in locally isotropic models one considers an one-loop diagram,
for the \index{One--loop divergences} HAS-model the distinguished
by N-connection character of ha-interactions leads to the
necessity to consider one-loop diagrams (see Fig. 1).To these
diagrams one corresponds counter-terms:%
$$
I_1^{\left( c\right)
}=I_1^{(c,x^2)}+I_1^{(c,y^2)}+I_1^{(c,xy)}+I_1^{(c,yx)}=
$$
$$
-\frac 12I_1\int d^2z\widehat{r}_{ij}\left( \eta ^{AB}-\epsilon
^{AB}\right)
\partial _Ax^i\partial _Bx^j-
$$
$$
\frac 12I_1\int d^2z\widehat{r}_{<a><b>}\left( \eta ^{AB}-\epsilon
^{AB}\right) \partial _Ay^{<a>}\partial _By^{<b>}-
$$
$$
\frac 12I_1\int d^2z\widehat{r}_{i<a>}\left( \eta ^{AB}-\epsilon
^{AB}\right) \partial _Ax^i\partial _By^{<a>}-
$$
$$
\frac 12I_1\int d^2z\widehat{r}_{<a>i}\left( \eta ^{AB}-\epsilon
^{AB}\right) \partial _Ay^{<a>}\partial _Bx^i,
$$
where $I_1$ is the standard integral%
$$
I_1=\frac{G\left( 0\right) }{\lambda ^2}=i\int \frac{d^qp}{(2\pi
)^2}\frac 1{p^2-\tilde m^2}=\frac{\Gamma \left( \epsilon \right)
}{4\pi ^{\frac q2}(\tilde m^2)^\epsilon }=
$$
$$
\frac 1{4\pi \epsilon }-\frac 1\pi \ln \tilde m+\mbox{finite
counter terms}.
$$

There are one--loops on the base and fiber spaces or describing
quantum interactions between fiber and base components of
d-fields. If the ha-background d-connection is of distinguished
Levi-Civita type we obtain only two one-loop diagrams (on the
base and in the fiber) because in this case the Ricci d-tensor is
symmetric. It is clear that this four-multiplying (doubling for
the Levi-Civita d-connection) of the number of one-loop diagrams
is caused by the ''indirect'' interactions with the N-connection
field. Hereafter, for simplicity, we shall use a compactified
(non-distinguished on x- and y-components) form of writing out
diagrams and corresponding formulas and emphasize that really all
expressions containing components of d-torsion generate
irreducible types of diagrams (with respective interaction
constants) and that all expressions containing components of
d-curvature give rise in a similar manner to irreducible types of
diagrams. We shall take into consideration these details in the
subsection where we shall write the two-loop effective action.
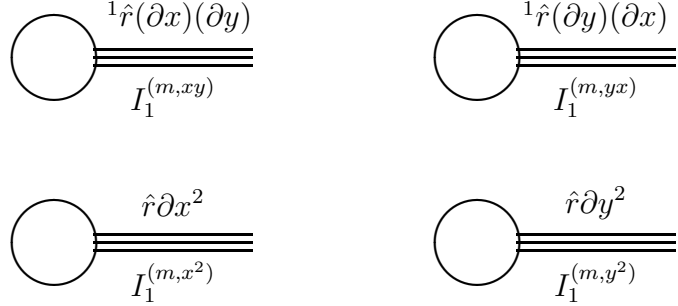
\begin{figure}[htbp]
\begin{picture}(380,150) \setlength{\unitlength}{1pt}
\thicklines \put(105,35){\circle{30}} \put(120,35){\line(1,0){60}}
\put(120,32){\line(1,0){60}} \put(120,38){\line(1,0){60}}
\put(130,10){\makebox(40,20){$I_1^{(m,x^2)}$}}
\put(125,40){\makebox(50,20){$\hat r {\partial x}^2$}}

\put(265,35){\circle{30}} \put(280,35){\line(1,0){60}}
\put(280,32){\line(1,0){60}} \put(280,38){\line(1,0){60}}
\put(290,10){\makebox(40,20){$I_1^{(m,y^2)}$}}
\put(285,40){\makebox(50,20){$\hat r {\partial y}^2$}}

\put(105,105){\circle{30}} \put(120,105){\line(1,0){60}}
\put(120,102){\line(1,0){60}} \put(120,108){\line(1,0){60}}
\put(130,80){\makebox(40,20){$I_1^{(m,xy)}$}}
\put(125,110){\makebox(55,20){$^1{\hat r}(\partial x)(\partial
y)$}}

\put(265,105){\circle{30}} \put(280,105){\line(1,0){60}}
\put(280,102){\line(1,0){60}} \put(280,108){\line(1,0){60}}
\put(290,80){\makebox(40,20){$I_1^{(m,yx)}$}}
\put(285,110){\makebox(50,20){$^1{\hat r}(\partial y)(\partial
x)$}}

\end{picture}
\caption{Feyn\-man di\-a\-grams for  1--loop $\beta$--functions}
\end{figure}

Subtracting in a trivial manner $I_1,$
$$
I_1+subtractions=\frac 1{4\pi \epsilon },
$$
we can write the one-loop $\beta $-function in the form:%
$$
\beta _{<\alpha ><\beta >}^{\left( 1\right) }=\frac 1{2\pi }\widehat{r}%
_{<\alpha ><\beta >}=\frac 1{2\pi }(r_{<\alpha ><\beta >}-
$$
$$
G_{<\alpha ><\tau >}T^{\left[ <\tau ><\gamma ><\phi >\right]
}T_{\left[ <\beta ><\gamma ><\phi >\right] }+G^{<\tau ><\mu
>}\nabla _{<\mu >}T_{\left[ <\alpha ><\beta ><\tau >\right] }).
$$

We also note that the mass term in the action generates the mass
one-loop counter-term
$$
\Delta I_1^{(m)}=\frac{\tilde m^2}2I_1\int d^2z\{\frac
13r_{<\alpha ><\beta
>}u^{<\alpha >}u^{<\beta >}-u_{<\alpha >}\tau _{.<\beta ><\gamma >}^{<\alpha
>}G^{<\beta ><\gamma >}\}.
$$

The last two formulas can be used for a study of effective
charges as in
 [44] where some solutions of RG-equations are analyzed. We shall not
consider in this work such methods connected with the theory of
differential equations.

\subsection{Two-loop $\beta $-functions for the HAS-model}

In order to obtain two--loops of the HAS--model we add to the
list (9.9) the \index{Two--loops!the HAS--model} expansion
$$
\Delta I_{1\mid 2}^{\left( c\right) }=-\frac 12I_1\int d^2z\{\widehat{r}%
_{<\alpha ><\beta >}\left( \eta ^{AB}-\epsilon ^{AB}\right) \left( \widehat{%
\nabla }_A\zeta ^{<\alpha >}\right) \left( \widehat{\nabla
}_B\zeta ^{<\beta
>}\right) +
$$
$$
(\nabla _\tau \widehat{r}_{<\alpha ><\beta
>}+\widehat{r}_{<\alpha ><\gamma
>}G^{<\gamma ><\delta >}T_{\left[ <\delta ><\beta ><\tau >\right] })\times
$$
$$
\left( \eta ^{AB}-\epsilon ^{AB}\right) \zeta ^{<\tau >}(\widehat{\nabla }%
_A\zeta ^{<\alpha >})\partial _Bu^{<\beta >}+
$$
$$
\left( \nabla _{<\tau >}\widehat{r}_{<\alpha ><\beta >}-T_{\left[
<\alpha
><\tau ><\gamma >\right] }\widehat{r}_{.<\beta >}^{<\gamma >}\right) (\eta
^{AB}-\epsilon ^{AB})\partial _Au^{<\alpha >}\left(
\widehat{\nabla }_B\zeta ^{<\beta >}\right) \zeta ^{<\tau >}+
$$
$$
\left( \eta ^{AB}-\epsilon ^{AB}\right) (\frac 12\nabla _{<\gamma
>}\nabla _{<\tau >}\widehat{r}_{<\alpha ><\beta >}+\frac
12\widehat{r}_{<\epsilon
><\beta >}r_{<\gamma >.<\tau ><\alpha >}^{.<\epsilon >}+
$$
$$
\frac 12\widehat{r}_{<\alpha ><\epsilon >}r_{<\gamma >.<\tau
><\beta
>}^{.<\epsilon >}+
$$
$$
T_{\left[ <\alpha ><\tau ><\epsilon >\right]
}\widehat{r}^{<\epsilon
><\delta >}T_{\left[ <\delta ><\gamma ><\beta >\right] }+G^{<\mu ><\nu
>}T_{\left[ <\nu ><\beta ><\gamma >\right] }\nabla _{<\tau >}\widehat{r}%
_{<\alpha ><\mu >}-
$$
$$
G^{<\mu ><\nu >}T_{\left[ <\alpha ><\gamma ><\nu >\right] }\nabla _{<\tau >}%
\widehat{r}_{<\mu ><\beta >})\partial _Au^{<\alpha >}\partial
_Bu^{<\beta
>}\zeta ^{<\tau >}\zeta ^{<\gamma >}\}
$$
and the d--covariant part of the expansion for the one--loop mass
counter-term%
$$
\Delta I_{1\mid 2}^{\left( m\right) }=\frac{\left( \tilde m\right) ^2}%
2I_1\int d^2z\frac 13r_{<\alpha ><\beta >}\zeta ^{<\alpha >}\zeta
^{<\beta
>}.
$$

The non-distinguished diagrams defining two--loop divergences are
illustrated \index{Two--loop divergences} in Fig.2. We present
the explicit form of corresponding counter-terms computed by
using, in our case adapted to ha-backgrounds, methods developed
in 
 [140,137]:

For counter-term of the diagram $(\alpha )$ we obtain
$$
(\alpha )=-\frac 12\lambda ^2(I_i)^2\int d^2z\{(\frac 14\Delta
r_{<\delta
><\varphi >}-\frac 1{12}\nabla _{<\delta >}\nabla _{<\varphi >}%
\overleftarrow{r}+
$$
$$
\frac 12r_{<\delta ><\alpha >}r_{<\varphi >}^{<\alpha >}-\frac
16r_{.<\delta
>.<\psi >}^{<\alpha >.<\beta >}r_{<\alpha ><\beta >+}\frac 12r_{.<\delta
>}^{<\alpha >.<\beta ><\gamma >}r_{<\alpha ><\varphi ><\beta ><\gamma >}+
$$
$$
\frac 12G_{<\delta ><\tau >}T^{\left[ <\tau ><\alpha ><\beta
>\right] }\Delta T_{\left[ <\varphi ><\alpha ><\beta >\right] }+
$$
$$
\frac 12G_{<\varphi ><\tau >}r_{<\delta >}^{<\alpha >}T_{\left[
<\alpha
><\beta ><\gamma >\right] }T^{\left[ <\tau ><\beta ><\gamma >\right] }-
$$
$$
\frac 16G^{<\beta ><\tau >}T_{\left[ <\delta ><\alpha ><\beta
>\right] }T_{\left[ <\varphi ><\gamma ><\tau >\right] }r^{<\alpha
><\gamma >}+
$$
$$
G^{<\gamma ><\tau >}T_{[<\delta ><\alpha ><\tau >]}\nabla
^{(<\alpha
>}\nabla ^{<\beta >)}T_{\left[ <\varphi ><\beta ><\gamma >\right] }+
$$
$$
\frac 34G^{<\kappa ><\tau >}r_{.<\delta >}^{<\alpha >.<\beta
><\gamma
>}T_{\left[ <\beta ><\gamma ><\kappa >\right] }T_{\left[ <\alpha ><\varphi
><\tau >\right] }-
$$
$$
\frac 14r^{<\kappa ><\alpha ><\beta ><\gamma >}T_{\left[ <\delta
><\beta
><\gamma >\right] }T_{\left[ <\kappa ><\alpha ><\varphi >\right] })\partial
_Au^{<\delta >}\partial ^Au^{<\varphi >}+
$$
$$
\frac 14[\nabla ^{<\beta >}\Delta T_{\left[ <\delta ><\varphi
><\beta
>\right] }-3r_{...<\delta >}^{<\gamma ><\beta ><\alpha >}\nabla _{<\alpha
>}T_{\left[ <\beta ><\gamma ><\varphi >\right] }-
$$
$$
3T_{\left[ <\alpha ><\beta ><\delta >\right] }\nabla ^{<\gamma
>}r_{..<\gamma ><\psi >}^{<\beta ><\alpha >}+
$$
$$
\frac 14r^{<\alpha ><\gamma >}\nabla _{<\alpha >}T_{\left[
<\gamma ><\delta
><\varphi >\right] }+\frac 16T_{\left[ <\delta ><\varphi ><\alpha >\right]
}\nabla ^{<\alpha >}r-
$$
$$
4G^{<\gamma ><\tau >}T_{\left[ <\tau ><\beta ><\delta >\right]
}T_{\left[ <\alpha ><\kappa ><\varphi >\right] }\nabla ^{<\beta
>}(G_{<\gamma
><\epsilon >}T^{\left[ <\alpha ><\kappa ><\epsilon >\right] })+
$$
$$
2G_{<\delta ><\tau >}G^{<\beta ><\epsilon >}\nabla _{<\alpha
>}(G^{<\alpha
><\nu >}T_{\left[ <\nu ><\beta ><\gamma >\right] })\times
$$
$$
T^{\left[ <\gamma ><\kappa ><\tau >\right] }T_{\left[ <\epsilon
><\kappa
><\varphi >\right] }]\epsilon ^{AB}\partial _Au^{<\delta >}\partial
_Bu^{<\varphi >}\}.
$$

In order to computer the counter-term for diagram $(\beta )$ we
use
integrals:%
$$
\lim \limits_{u\rightarrow v}i<\partial _A\zeta \left( u\right)
\partial ^A\zeta \left( v\right) >=i\int \frac{d^qp}{(2\pi
)^q}\frac{p^2}{p^2-\tilde m^2}=\tilde m^2I_1
$$
(containing only a IR-divergence) and
$$
J\equiv i\int \frac{d^2p}{(2\pi )^2}\frac 1{(p^2-\tilde
m^2)^2}=-\frac 1{(2\pi )^2}\int d^2k_E\frac 1{(k_E^2+\tilde
m^2)^2}.
$$
(being convergent). In result we can express%
$$
(\beta )=\frac 16\lambda ^2\left( I_1^2+2\tilde mI_1J\right)
\times
$$
$$
\int d^2z\widehat{r}_{<\beta ><\alpha ><\gamma ><\delta
>}r^{<\beta ><\gamma
>}\left( \eta ^{AB}-\epsilon ^{AB}\right) \partial _Au^{<\alpha >}\partial
_Bu^{<\beta >}.
$$
In our further considerations we shall use identities (we can
verify them by
straightforward calculations):%
$$
\widehat{r}_{(<\alpha >.<\beta >)}^{.[<\gamma >.<\delta
>]}=-\nabla _{(<\alpha >}(G_{<\beta >)<\tau >}T^{[<\tau ><\gamma
><\delta >]}),
$$
$$
\widehat{r}_{\left[ <\beta ><\alpha ><\gamma ><\delta >\right]
}=2G^{<\kappa
><\tau >}T_{\left[ <\tau >[<\alpha ><\beta >\right] }T_{\left[ <\gamma
><\delta >]<\kappa >\right] ,}
$$
in the last expression we have three type of antisymmetrizations
on indices, $\left[ <\tau ><\alpha ><\beta >\right] ,$ $\left[
<\gamma ><\delta ><\kappa
>\right] $ and $\left[ <\alpha ><\beta ><\gamma ><\delta >\right] ,$
$$
\widehat{\nabla }_{<\delta >}T_{\left[ <\alpha ><\beta ><\gamma >\right] }%
\widehat{\nabla }_{<\varphi >}T^{\left[ <\alpha ><\beta ><\gamma
>\right]
}=\frac 9{16}\left( \widehat{r}_{\left[ \beta \alpha \gamma \right] \delta }-%
\widehat{r}_{\delta \left[ \alpha \beta \gamma \right] }\right)
\times \eqno(9.15)
$$
$$
\left( \widehat{r}_{...........<\varphi >}^{\left[ <\beta
><\alpha ><\gamma
>\right] }-\widehat{r}_{...........<\varphi >}^{.\left[ <\alpha ><\beta
><\gamma >\right] }\right) -\frac 94\widehat{r}_{\left[ <\alpha ><\beta
><\gamma ><\delta >\right] }\widehat{r}_{.........<\varphi >]}^{[<\alpha
><\beta ><\gamma >}+
$$
$$
\frac 94\widehat{r}_{......[<\delta >}^{<\alpha ><\beta ><\gamma >}\widehat{r%
}_{\left[ <\varphi >]<\alpha ><\beta ><\gamma >\right] }+\frac 94\widehat{r}%
_{.[<\delta >}^{<\alpha >.<\beta ><\gamma >}\widehat{r}_{\left[
<\varphi
>]<\alpha ><\beta ><\gamma >\right] }.
$$

The momentum integral for the first of diagrams $(\gamma )$
$$
\int \frac{d^qpd^qp^{\prime }}{(2\pi
)^{2q}}\frac{p_Ap_B}{(p^2-\tilde m^2)([k+q]^2-\tilde
m^2)([p+q]^2-\tilde m^2)}
$$
diverges for a vanishing exterior momenta $k_{<\mu >}.$The
explicit calculus of the corresponding counter-term results in
$$
\gamma _1=-\frac{2\lambda ^2}{3q}I_1^2\int d^2z\{(r_{<\alpha
>(<\beta
><\gamma >)<\delta >}+\eqno(9.16)
$$
$$
G^{<\varphi ><\tau >}T_{[<\tau ><\alpha >(<\beta >]}T_{[<\gamma
>)<\delta
><\varphi >]})\times
( r_{.<\mu >}^{<\beta >.<\gamma ><\delta >}-$$
$$ G_{<\mu ><\tau >}G_{<\kappa
><\epsilon >}T^{[<\tau ><\beta ><\kappa >]}T^{\left[ <\gamma ><\delta
><\epsilon >\right] }) \partial ^Au^{<\alpha >}\partial _Au^{<\mu >}+
$$
$$
(\nabla _{(<\beta >}T_{[<\delta >)<\alpha ><\gamma >]})\nabla
^{<\beta
>}(G_{<\mu ><\tau >}T^{[<\tau ><\gamma ><\delta >]})$$
$$\epsilon ^{LN}\eta
_{NM}\epsilon ^{MR}\partial _Lu^{<\alpha >}\partial _Ru^{<\mu >}-
$$
$$
2(r_{<\alpha >(<\beta ><\gamma >)<\delta >}+G^{<\varphi ><\tau
>}T_{[<\alpha
><\tau >(<\beta >]}T_{[<\gamma >)<\delta ><\varphi >]})\times
$$
$$
\nabla ^{<\beta >}(G_{<\mu ><\epsilon >}T^{[<\epsilon ><\delta
><\gamma
>]})\epsilon ^{MR}\partial _Mu^{<\alpha >}\partial _Ru^{<\mu >}\}.
$$
The counter-term of the sum of next two $(\gamma )$-diagrams is
chosen to be
the ha-extension of that introduced in 
 [140,137]
$$
\gamma _2+\gamma _3=-\lambda ^2\omega \left( \epsilon \right) \frac{10-7q}{%
18q}I_1^2\times \eqno(9.17)
$$
$$
\int d^2z\{\widehat{\nabla }_AT_{[<\alpha ><\beta ><\gamma >]}\widehat{%
\nabla }^AT^{[<\alpha ><\beta ><\gamma >]}+
$$
$$
6G^{<\tau ><\epsilon >}T_{[<\delta ><\alpha ><\tau
>]}T_{[<\epsilon ><\beta
><\gamma >]}\widehat{\nabla }_{<\varphi >}T^{[<\alpha ><\beta ><\gamma
>]}\times
$$
$$
\left( \eta ^{AB}-\epsilon ^{AB}\right) \partial _Au^{<\delta
>}\partial _Bu^{<\varphi >}\}.
$$
In a similar manner we can computer the rest part of counter-terms:%
$$
\delta =\frac 12\lambda ^2(I_1^2+m^2I_1J)\times
$$
$$
\int d^2z\widehat{r}_{<\alpha >(<\beta ><\gamma >)<\delta >}\widehat{r}%
^{<\beta ><\gamma >}\left( \eta ^{AB}-\epsilon ^{AB}\right)
\partial _Au^{<\alpha >}\partial _Bu^{<\delta >},
$$
$$
\epsilon =\frac 14\lambda ^2I_1^2\int d^2z\left( \eta
^{AB}-\epsilon ^{AB}\right) \times
$$
$$
[\Delta \widehat{r}_{<\delta ><\varphi >}+r_{<\delta >}^{<\alpha >}\widehat{r%
}_{<\alpha ><\varphi >}+r_{<\varphi >}^{<\alpha
>}\widehat{r}_{<\delta
><\alpha >}
$$
$$
-2(G^{<\alpha ><\tau >}T_{\left[ <\delta ><\beta ><\alpha >\right]
}T_{\left[ <\varphi ><\gamma ><\tau >\right] }\widehat{r}^{<\beta
><\gamma
>}-
$$
$$
G_{<\varphi ><\tau >}T^{\left[ <\tau ><\alpha ><\beta >\right]
}\nabla _{<\alpha >}\widehat{r}_{<\delta ><\beta >}+
$$
$$
G_{<\delta ><\tau >}T^{[<\tau ><\alpha ><\beta >]}\nabla _{<\alpha >}%
\widehat{r}_{<\beta ><\varphi >}]\partial _Au^{<\delta >}\partial
_Bu^{<\varphi >},
$$
$$
\iota =\frac 16\lambda ^2\tilde m^2I_1J\times
$$
$$
\int d^2z\widehat{r}_{<\beta ><\alpha ><\gamma ><\delta
>}r^{<\beta ><\gamma
>}\left( \eta ^{AB}-\epsilon ^{AB}\right) \partial _Au^{<\alpha >}\partial
_Bu^{<\delta >},
$$
$$
\eta =\frac 14\lambda ^2\omega \left( \epsilon \right) \left(
I_1^2+2\tilde m^2I_1J\right) \times
$$
$$
\int d^2z\widehat{r}_{.<\alpha ><\gamma ><\delta >}^{<\beta
>}T_{\left[ <\beta ><\varphi ><\tau >\right] }T^{\left[ <\gamma
><\varphi ><\tau
>\right] }\left( \eta ^{AB}-\epsilon ^{AB}\right) \partial _Au^{<\alpha
>}\partial _Bu^{<\delta >}.
$$

By using relations (9.5) we can represent terms (9.6) and (9.7)
in the canonical form (9.1) from which we find the contributions
in the $\beta
_{<\delta ><\varphi >}$-function (9.2):%
$$
\gamma _1:-\frac 2{3(2\pi )^2}\widehat{r}_{<\delta >(<\alpha
><\beta
>)<\gamma >}\widehat{r}_{........<\varphi >}^{<\gamma >(<\alpha ><\beta >)}-%
\eqno(9.18)
$$
$$
\frac{(\omega _1-1)}{(2\pi )^2}\{\frac 43\widehat{r}_{[<\gamma
>(<\alpha
><\beta >)<\delta >]}\widehat{r}_{........<\varphi >]}^{[<\gamma >(<\alpha
><\beta >)}+
$$
$$
\widehat{r}_{[<\alpha ><\beta ><\gamma ><\delta >]}\widehat{r}%
_{.......<\varphi >]}^{[<\alpha ><\beta ><\gamma >}\},
$$
$$
\gamma _2+\gamma _3:\frac{(4\omega _1-5)}{9(2\pi )^2}\{\widehat{\nabla }%
_{<\delta >}T_{\left[ <\alpha ><\beta ><\gamma >\right] }\widehat{\nabla }%
_{<\varphi >}T^{\left[ <\alpha ><\beta ><\gamma >\right] }+
$$
$$
6G^{<\tau ><\epsilon >}T_{\left[ <\delta ><\alpha ><\tau >\right]
}T_{\left[ <\epsilon ><\beta ><\gamma >\right] }\widehat{\nabla
}_{<\varphi >}T^{\left[ <\alpha ><\beta ><\gamma >\right] }-
$$
$$
\frac{(\omega _1-1)}{(2\pi )^2}\widehat{r}_{\left[ <\alpha
><\beta ><\gamma
><\delta >\right] }\widehat{r}_{.......<\varphi >]}^{[<\alpha ><\beta
><\gamma >}\},
$$
$$
\eta :\frac{\omega _1}{(2\pi )^2}\widehat{r}^\alpha ._{\delta
\beta \varphi }T_{\left[ \alpha \tau \epsilon \right] }T^{\left[
\beta \tau \epsilon \right] }.
$$
{\normalsize
\begin{figure}[htbp]
\begin{picture}(350,470) \setlength{\unitlength}{1pt}
\thicklines

\put(91,55){\circle{30}} \put(67,55){\circle{16}}
\put(67,47){\line(0,1){16}} \put(19,55){\line(1,0){40}}
\put(19,52){\line(1,0){40}} \put(19,58){\line(1,0){40}}
\put(107,52){\line(1,0){60}} \put(107,55){\line(1,0){60}}
\put(107,58){\line(1,0){60}} \put(20,60){\makebox(40,20){$m^2
\hat r$}} \put(111,60){\makebox(60,20){$\hat r (\eta -
\varepsilon ) {\partial u}^2$}}
\put(50,00){\makebox(70,20){Diagram $(\iota )$}}

\put(130,150){\line(1,0){40}} \put(130,147){\line(1,0){40}}
\put(130,153){\line(1,0){40}} \put(115,150){\circle{30}}
\put(92,150){\circle{16}} \put(14,150){\line(1,0){70}}
\put(14,147){\line(1,0){70}} \put(14,153){\line(1,0){70}}
\put(86,144){\line(1,1){12}} \put(86,155){\line(1,-1){12}}
\put(50,110){\makebox(70,20){Diagram $(\varepsilon )$}}
\put(14,154){\makebox(70,20){$\hat r (\eta -\varepsilon
)^2(\partial u)^2$}} \put(131,154){\makebox(40,20){$\hat r(\eta
-\varepsilon )$}}

\put(335,140){\circle{30}} \put(311,140){\circle{16}}
\put(172,140){\line(1,0){130}} \put(172,137){\line(1,0){130}}
\put(172,143){\line(1,0){130}} \put(305,134){\line(1,1){12}}
\put(305,145){\line(1,-1){12}}
\put(260,95){\makebox(70,15){Diagram $(\delta )$}}
\put(172,145){\makebox(150,15){$(\nabla ^2\hat r +r\hat r +
T^2\hat r + T\nabla \hat r)$}}
\put(175,120){\makebox(100,15){$(\eta - \varepsilon )(\partial
u)^2$}}

\put(285,55){\circle{30}} \put(230,55){\line(1,0){110}}
\put(300,52){\line(1,0){40}} \put(300,58){\line(1,0){40}}
\put(230,58){\line(1,0){40}} \put(230,52){\line(1,0){40}}
\put(285,25){\line(0,1){15}} \put(288,25){\line(0,1){15}}
\put(282,25){\line(0,1){15}}
\put(230,60){\makebox(40,20){$\varepsilon T$}}
\put(300,60){\makebox(50,20){$\varepsilon T$}}
\put(290,26){\makebox(70,15){$(\eta - \varepsilon ) (\partial
u)^2$}} \put(250,00){\makebox(70,15){Diagram $(\eta )$}}

\put(320,250){\circle{30}} \put(335,250){\line(1,0){15}}
\put(335,247){\line(1,0){15}} \put(335,253){\line(1,0){15}}
\put(290,250){\line(1,0){15}} \put(290,247){\line(1,0){15}}
\put(290,253){\line(1,0){15}} \put(320,220){\line(0,1){15}}
\put(317,220){\line(0,1){15}} \put(323,220){\line(0,1){15}}
\put(300,200){\makebox(40,15){$\varepsilon T(\partial u)$}}
\put(335,255){\makebox(15,15){$\varepsilon T$}}
\put(290,255){\makebox(15,15){$\varepsilon T$}}
\put(260,297){\makebox(5,10){+}}

\put(300,300){\circle{30}} \put(270,300){\line(1,0){60}}
\put(270,303){\line(1,0){15}} \put(270,297){\line(1,0){15}}
\put(315,303){\line(1,0){15}} \put(315,297){\line(1,0){15}}
\put(316,305){\makebox(15,15){$\varepsilon T$}}
\put(270,305){\makebox(15,15){$\varepsilon T$}}
\put(340,297){\makebox(5,10){+}}

\put(100,225){\makebox(70,15){Diagrams $(\gamma )$}}

\put(00,300){\line(1,0){240}} \put(120,300){\circle{30}}
\put(00,303){\line(1,0){105}} \put(00,297){\line(1,0){105}}
\put(135,303){\line(1,0){105}} \put(135,297){\line(1,0){105}}
\put(00,305){\makebox(105,20){$(r+T^2+\varepsilon \nabla
T)\partial u$}}
\put(136,305){\makebox(105,20){$(r+T^2+\varepsilon \nabla
T)\partial u$}}

\put(00,470){\line(1,0){230}} \put(230,485){\circle{28}}
\put(230,455){\circle{28}} \put(00,467){\line(1,0){230}}
\put(00,473){\line(1,0){230}}
\put(05,450){\makebox(200,15){$(\nabla ^2r+r^2+T\nabla ^2T=rT^2)
(\partial u)^2+$}} \put(05,430){\makebox(200,15){$\varepsilon
(\nabla ^3T+r\nabla T+T^2\nabla T+ T\nabla r)(\partial u)^2$}}
\put(80,390){\makebox(70,15){Diagram $(\alpha )$}}

\put(270,445){\circle{28}} \put(270,415){\circle{28}}
\put(245,430){\line(1,0){30}} \put(245,433){\line(1,0){30}}
\put(245,427){\line(1,0){30}}
\put(220,350){\makebox(70,15){Diagram $(\beta )$}}
\put(285,445){\line(1,0){65}} \put(285,448){\line(1,0){65}}
\put(285,442){\line(1,0){65}} \put(250,433){\makebox(5,10){$r$}}
\put(285,425){\makebox(64,15){$\hat r(\eta -\varepsilon
)(\partial u)^2$}}

\end{picture}
\caption{Two-loops diagrams for  HAS-models}
\end{figure}}

Finally, in this subsection we remark that two-loop $\beta
$-function can not be written only in terms of curvature
$\widehat{r}_{<\alpha ><\beta
><\gamma ><\delta >}$ and d-derivation $\widehat{\nabla }_{<\alpha >}$
(similarly as in the locally isotropic case 
 [140,137] ).

\subsection{Low--energy effective action for la--strings}

The conditions of vanishing of $\beta $-functions describe the
propagation \index{Action effective!low--energy for la--strings}
of string in the background of ha-fields $G_{<\alpha ><\beta >}$ and $%
b_{<\alpha ><\beta >}.$ (in this section we chose the canonic d-connection $%
^{\circ }{\Gamma }_{\cdot <\beta ><\gamma >}^{<\alpha >}$ on ${\cal E}^{<z>}%
{\cal ).}$ The $\beta $-functions are proportional to d-field
equations obtained from the on--shell string effective action
$$
I_{eff}=\int du\sqrt{|\gamma |}L_{eff}\left( \gamma ,b\right)
.\eqno(9.19)
$$
The adapted to N-connection variations of (9.17) with respect to
$G^{<\mu
><\nu >}$ and $b^{<\mu ><\nu >}$ can be written as
$$
\frac{\Delta I_{eff}}{\delta G^{<\alpha ><\beta >}}=W_{<\alpha
><\beta
>}+\frac 12G_{<\alpha ><\beta >}(L_{eff}+{\mbox{complete derivation)})},
$$
$$
\frac{\Delta I_{eff}}{\delta b^{<\alpha ><\beta >}}=0.
$$

The invariance of action (9.17) with respect to N-adapted
diffeomorfisms
gives rise to the identity%
$$
\nabla _{<\beta >}W^{<\alpha ><\beta >}-T^{\left[ <\alpha ><\beta
><\gamma
>\right] }\frac{\Delta I_{eff}}{\delta b^{<\beta ><\gamma >}}=
$$
$$
-\frac 12\nabla ^{<\alpha >}(L_{eff}+\mbox{complete derivation)}
$$
(in the locally isotropic limit we obtain the well-known results
from
 [297,\\ 54]). This points to the possibility to write out the integrability
conditions as
$$
\nabla ^{<\beta >}\beta _{(<\alpha ><\beta >)}-G_{<\alpha ><\tau
>}T^{\left[ <\tau ><\beta ><\gamma >\right] }\beta _{[<\beta
><\gamma >]}=-\frac 12\nabla _{<\alpha >}L_{eff}.\eqno(9.20)
$$
For one-loop $\beta $-function, $\beta _{<\alpha ><\beta
>}^{\left( 1\right) }=\frac 1{2\pi }\widehat{r}_{<\alpha ><\beta
>},$ we find from the last
equations%
$$
\nabla ^{<\beta >}\beta _{(<\delta ><\beta >)}^{(1)}-G_{<\delta
><\tau
>}T^{[<\tau ><\beta ><\gamma >]}\beta _{[<\beta ><\gamma >]}^{(1)}=
$$
$$
\frac 1{4\pi }\nabla _{<\delta >}\left( \overleftarrow{R}+\frac
13T_{\left[ <\alpha ><\beta ><\gamma >\right] }T^{\left[ <\alpha
><\beta ><\gamma
>\right] }\right)
$$
We can take into account two-loop $\beta$--functions by fixing an
explicit form of
$$
\omega (\epsilon )=1+2\omega _1\epsilon +4\omega _2\epsilon ^2+...
$$
when
$$
\omega _{HVB}\left( \epsilon \right) =\frac 1{(1-\epsilon
)^2},\omega _1^{HVB}=1,\omega _2^{HVB}=\frac 34
$$
(the t'Hooft-Veltman-Bos prescription
 [108]). Putting values (9.18)
into (9.20) we obtain the two--loop approximation for ha--field
equations
$$
\nabla ^{<\beta >}\beta _{(<\delta ><\beta >)}^{(2)}-G_{<\delta
><\tau
>}T^{[<\tau ><\beta ><\gamma >]}\beta _{[<\beta ><\gamma >]}^{(2)}=
$$
$$
\frac 1{2(2\pi )^2}\nabla _\delta [-\frac 18r_{<\alpha ><\beta
><\gamma
><\delta >}r^{<\alpha ><\beta ><\gamma ><\delta >}+
$$
$$
\frac 14r_{<\alpha ><\beta ><\gamma ><\delta >}G_{<\tau ><\epsilon
>}T^{[<\alpha ><\beta ><\tau >]}T^{[<\epsilon ><\gamma ><\delta >]}+
$$
$$
\frac 14G^{<\beta ><\epsilon >}G_{<\alpha ><\kappa >}T^{[<\alpha
><\tau
><\sigma >]}T_{[<\epsilon ><\tau ><\sigma >]}T^{[<\kappa ><\mu ><\nu
>]}T_{[<\beta ><\mu ><\nu >]}-
$$
$$
\frac 1{12}G^{<\beta ><\epsilon >}G_{<\gamma ><\lambda
>}T_{[<\alpha ><\beta
><\tau >]}T^{[<\alpha ><\gamma ><\varphi >]}T_{[<\epsilon ><\varphi ><\kappa
>]}T^{[<\lambda ><\tau ><\kappa >]}],
$$
which can be obtained from effective action%
$$
I^{eff}\sim \int \delta ^{n_E}u\sqrt{|\gamma
|}[-\overleftarrow{r}+\frac 13T_{[<\alpha ><\beta ><\gamma
>]}T^{[<\alpha ><\beta ><\gamma >]}-
$$
$$
\frac{\alpha ^{\prime }}4(\frac 12r_{<\alpha ><\beta ><\gamma
>\delta }r^{<\alpha ><\beta ><\gamma ><\delta >}-
$$
$$
G_{<\tau ><\epsilon >}r_{<\alpha ><\beta ><\gamma ><\delta
>}T^{[<\alpha
><\beta ><\tau >]}T^{[<\epsilon ><\gamma ><\delta >]}-
$$
$$
G^{<\beta ><\epsilon >}G_{<\alpha ><\upsilon >}T^{[<\alpha
><\gamma ><\kappa
>]}T_{[<\epsilon ><\gamma ><\kappa >]}T^{[<\upsilon ><\sigma ><\varsigma
>]}T_{[<\beta ><\sigma ><\varsigma >]}+
$$
$$
\frac 13G^{<\beta ><\epsilon >}G_{<\delta ><\kappa >}T_{[<\alpha
><\beta
><\gamma >]}T^{[<\alpha ><\delta ><\varphi >]}T_{[<\varphi ><\upsilon
><\epsilon >]}T^{[<\gamma ><\upsilon ><\kappa >]})]\eqno(9.21)
$$

The action (9.21) (for 2$\pi \alpha ^{\prime }=1\,$ and in
locally isotropic limit) is in good concordance with the similar
ones on usual closed strings
 [91,140].

We note that the existence of an effective action is assured by
the
Zamolodchikov c-theorem 
 [296] which was generalized 
 [242] for the
case of bosonic nonlinear $\sigma $-model with dilaton
connection. In a similar manner we can prove that such results
hold good for ha-backgrounds.

\section{Scattering of Ha--Gravitons}

The quantum theory of ha-strings can be naturally considered by
using the formalism of functional integrals on ''hypersurfaces''
(see Polyakov's works
 [193]). In this section we study the structure of scattering
amplitudes of ha-gravitons. Questions on duality of ha-string
theories will \index{Scattering of Ha--gravitons} be also
analyzed.

\subsection{Ha--string amplitudes for ha-gravitons scattering}

We introduce the Green function of ha--tachyons, the fundamental
state of \index{Ha--string amplitudes} \index{Green function!of
ha--tachyons}
ha--string, as an integral ( after Weeck rotation in the Euclidean space )%
$$
G_t(p_{1,...,}p_N)=$$
$$\int [D\gamma _{AB}(\zeta )][Du^{<\alpha >}(z)]\exp
(-\frac 1{4\pi \alpha ^{\prime }}\int d^2z\sqrt{|\gamma |}\gamma
^{AB}\partial _Au^{<\alpha >}\partial _Bu^{<\alpha >})
$$
$$
\int \left[ \prod_AD^2z_A\right] \sqrt{|\gamma (z_B)|}\exp
(ip_B^{<\alpha
>}u(z_B)),\eqno(9.22)
$$
where the integration measure on $\gamma _{AB}$ includes the
standard ghost Fadeev-Popov determinant corresponding to the
fixation of the
reparametrization invariance and%
$$
\left[ \prod_AD^2z_A\right] =\prod_{A\neq
M,N,K}d^2z_A|z_M-z_{N\,}|^2\left| z_K-z_M\right| ^2\left|
z_K-z_N\right| ^2.
$$
Because the string quantum field theory can be uncontradictory
formulated for spaces of dimension d=26 we consider that in
formula (9.22) $\alpha $ takes values from 0 to 25. Formula
(9.22) leads to dual amplitudes for
ha-tachyon scatterings for $p^2=\frac 4{\alpha ^{\prime }}$ (see 
 [228]
for details and references on usual locally isotropic tachyon
scattering).

The generating functional of Green functions (9.22) in the
coordinate u-repre\-sen\-ta\-ti\-on can be formally written as a
hyper surface mean value
$$
\Gamma ^0\left[ \Phi \right] =<\exp \left( -\frac 1{2\pi \alpha
^{\prime }}\int d^2z\sqrt{|\gamma |}\Phi \left[ u\left( z\right)
\right] \right) >.
$$
In order to conserve the reparametrization invariance we define
the ha-graviton source as
$$
\Gamma ^0\left[ G\right] =<\exp \left( -\frac 1{4\pi \alpha
^{\prime }}\int d^2z\sqrt{|\gamma |}\gamma ^{AB}\partial
_Au^{<\alpha >}\partial _Bu^{<\beta
>}G_{<\alpha ><\beta >}[u\left( z\right) ]\right) >
$$
from which we obtain the Green function of a number of $K$
elementary perturbations of the closed ha-string ($K$
ha--gravitons) \index{Ha--gravitons}
$$
G_g(u_1,...,u_K)=
<\frac 12\int [\prod_{[j]}D^2z_{[j]}]\sqrt{|\gamma (z)|}G^{AB\;}(z_{[j]})%
$$
$$\partial _Au^{<\alpha >}(z_{[j]})\partial _Bu^{<\beta >}(z_{[j]\;})\chi
_{<\alpha ><\beta >}^{\left( j\right) }\delta
^{(d)}(u_{[j]}-u(z_{[j]}))>,
$$
where $\chi _{<\alpha ><\beta >}^{\left( j\right) }$ are
polarization d-tensors of exterior ha-gravitons and
$[j]=1,2,...K.$ Applying the Fourier transform we obtain
$$
G_g(p_1,...,p_K)=
$$
$$
\int [D\gamma _{AB}\left( z\right) ][Du^{<\alpha >}\left(
z\right) ]\exp \left( -\frac 1{2\pi \alpha ^{\prime }}\int
d^2z\sqrt{|\gamma |}\frac 12G^{AB}\partial _Au^{<\alpha
>}\partial _Bu^{<\alpha >}\right)
$$
$$
\int \left[ \prod_{[j]}D^2z_{[j]}\right] \frac 12\sqrt{|\gamma (z_{[j])}|}%
\gamma ^{AB}(z_{[j]})\partial _Au^{<\beta >}(z_{[j]})\partial
_Bu^{<\gamma
>}(z_{[j]})\chi _{<\alpha ><\beta >}^{(j)}$$
$$\exp [ip_{[j]}^\delta u^\delta
(z)].
$$
Integrating the last expression on $G_{<\mu ><\nu >}\,$ and
$u^{<\alpha >}$
for d=26, when there are not conformal anomalies, we have%
$$
G_g(p_1,...,p_K)= \int \left[ \prod_{[j]}D^2z_{[j]}\right]
(\partial _A\frac \partial {\partial p_{[j]}}\cdot \chi
^{(j)}.\partial ^A\frac \partial {\partial p_{[j]}})\times
$$
$$
\int [D\sigma (z)]\exp [-\pi \alpha ^{\prime
}\sum_{[i,j]}p_{[i]}p_{[j]}V(z_{[i]},z_{[j]},\sigma )],\eqno(9.23)
$$
where V is the Green function of the Laplacian for the
conformally-flat
metric $G_{AB}=e^\sigma \delta _{AB}:$%
$$
\partial _A(\sqrt{|\gamma |}\gamma ^{AB}\partial _A)V=-\delta
^2(z_{[i]}-z_{[j]})
$$
which can be represented as
$$
V(z_{[i]},z_{[j]},\sigma )=-\frac 1{4\pi }\ln
|z_{[i]}-z_{[j]}|^2,z_{[i]}\neq z_{[j]},
$$
$$
V\left( z_{[k]},z_{[k]}^p,\sigma \right) =\frac 1{4\pi }(\sigma
(z_{[k]})-\ln (\frac 1\epsilon )),
$$
for $\epsilon \,$ being the cutting parameter. \ Putting the last
expression
into (9.23) we compute the Green function of ha-gravitons:%
$$
G_g(p_1,...p_K)=
$$
$$
\int \prod_{[j]\neq
[p],[q],[s]}d^2z_{[j]}|z_{[q]}-z_{[p]}|^2|z_{[s]}-z_{[q]}|^2|z_{[s]}-z_{[p]}|^2(\partial
_A\frac \partial {\partial p_{[j]}}\cdot \chi ^{(j)}\cdot
\partial ^A\frac \partial {\partial p_{[j]}})
$$
$$
\int d\sigma (z)\prod_{[i]<[m]}|z_{[i]}- z_{[m]}|^{\alpha
^{\prime }p_{[i]}p_{[j]}} $$
$$\exp \left[ -\frac 14\alpha
^{\prime }\sum_{[k]}p_{[k]}^2\sigma (z_{[k]})\right] \exp \left[ \frac{%
\alpha ^{\prime }}4\ln (\frac 1\epsilon
)\sum_{[k]}p_{[k]}^2\right] ,
$$
where the definition of integration on $\sigma (z)$ is extended as%
$$
\int d\sigma (z)\equiv \lim \limits_{\sigma
_d(z_{[j]})\rightarrow -\infty }^{\sigma _s(z_{[j]})\rightarrow
+\infty }\prod_{[j]}\int_{\sigma _d(z_{[j]})}^{\sigma
_s(z_{[j]})}d\sigma (z_{[j]}).
$$
So the scattering amplitude
$$
A_g(p_1,...,p_K)= \lim \limits_{p_{[j]}^2\rightarrow
0}\prod_{[j]}p_{[j]}^2G_g(p_{1,}...,p_K)
$$
is finite if%
$$
\lim \limits_{\sigma _d(z_{[j]})\to -\infty ,p_{[j]}^2\to
0}|p_{[j]}^2\sigma (z_{[j]})|=const<\infty .
$$
$G_g(p_1,...,p_{K)}\,$ has poles on exterior momenta
corresponding to massless higher order  anisotropic modes of spin
2 (ha-gravitons). The final result for the scattering amplitute
of ha-gravitons is of the form
$$
A_g(p_1,...,p_K)\sim \int \prod_{[j]\neq
[p],[q],[s]}d^2z_{[j]}|z_{[p]}-z_{[q]}|^2|z_{[s]}-z_{[q]}|^2|z_{[s]}-z_{[p]}|^2
$$
$$
\partial _A\frac \partial {\partial p_{[j]}^{<\alpha >}}\chi _{<\alpha
><\beta >}^{(j)}\partial ^A\frac \partial {\partial p_{[j]}^{<\beta
>}}\prod_{[m]<[n]}|z_{[m]}-z_{[n]}|^{\alpha ^{\prime }p_{[m]}\cdot p_{[n]}}.
$$
If instead of polarization d-tensor $\chi _{<\alpha ><\beta
}^{(j)}$ the graviton polarization tensor $\chi _{ik}^{(j)}$ is
taken we obtain the well known results on scattering of gravitons
in the framework of the first
quantization of the string theory 
[80,138]
\subsection{Duality of Ha--$\sigma$--mo\-dels}

Two theories are dual if theirs non-equivalent second order
actions can be \index{Duality!of Ha--$\sigma$--mo\-dels}
generated by the same first order action. The action principle
assures the equivalence of the classical dual theories. But, in
general, the duality transforms affects the quantum conformal
properties
 [52]. In this
subsection we shall prove this for the ha-$\sigma $-model (9.1) when metric $%
\gamma $ and the torsion potential b on ha-background ${\cal
E}^{<z>}$ do not depend on coordinate u$^0.$ If such conditions
are satisfied we can write for (9.1) the first order action
$$
I=\frac 1{4\pi \alpha ^{\prime }}\int d^2z\{\sum_{<\alpha >,<\beta
>=1}^{n_E-1}[\sqrt{|\gamma |}\gamma ^{AB}(G_{00}V_AV_B+2G_{0\alpha
}V_A\left( \partial _Bu^{<\alpha >}\right) +\eqno(9.24)
$$
$$
G_{<\alpha ><\beta >}(\partial _Au^{<\alpha >})\left( \partial
_Bu^{<\beta
>}\right) )+\epsilon ^{AB}(b_{0<\alpha >}V_B(\partial _Au^{<\alpha >})+
$$
$$
b_{<\alpha ><\beta >}(\partial _Au^{<\alpha >})(\partial
_Bu^{<\beta >}))]+
\epsilon ^{AB}\widehat{u}^0(\partial _AV_B)+\alpha ^{\prime }\sqrt{|\gamma |}%
R^{(2)}\Phi (u)\},
$$
where string interaction constants from (9.1) and (9.24) are related as $%
\lambda ^2=2\pi \alpha ^{\prime }.$

This action will generate an action of type (9.1) if we shall
exclude the Lagrange multiplier $\widehat{u}^0.$ The dual to
(9.24) action can be constructed by substituting \ V$_A$
expressed from the motion equations for
fields V$_A$ (also obtained from action (9.23)):%
$$
\widehat{I}=\frac 1{4\pi \alpha ^{\prime }}\int
d^2z\{\sqrt{|\gamma |}\gamma ^{AB}\widehat{G}_{<\alpha ><\beta
>}(\partial _A\widehat{u}^{<\alpha
>})(\partial _B\widehat{u}^{<\beta >})+
$$
$$
\epsilon ^{AB}\widehat{b}_{<\alpha ><\beta >}(\partial _A\widehat{u}%
^{<\alpha >})(\partial _B\widehat{u}^{<\beta >})+\alpha ^{\prime }\sqrt{%
|\gamma |}R^{(2)}\Phi (u)\},
$$
where the knew metric and torsion potential are introduced
respectively as
$$
\widehat{G}_{00}=\frac 1{G_{00}},\widehat{G}_{0<\alpha
>}=\frac{b_{0<\alpha
>}}{G_{00}},
$$
$$
\widehat{G}_{<\alpha ><\beta >}=G_{<\alpha ><\beta
>}-\frac{G_{0<\alpha
>}G_{0<\beta >}-b_{0<\alpha >}b_{0<\beta >}}{G_{00}}
$$
and
$$
\widehat{b}_{0<\alpha >}=-\widehat{b}_{<\alpha >0}=\frac{G_{0<\alpha >}}{%
G_{00}},
$$
$$
\widehat{b}_{<\alpha ><\beta >}=b_{<\alpha ><\beta
>}+\frac{G_{0<\alpha
>}b_{0<\beta >}-b_{0<\alpha >}G_{0<\beta >}}{G_{00}}
$$
(in the formulas for the new metric and torsion potential indices
$\alpha $ and $\beta $ take values $1,2,...n_E-1$).

If the model (9.1) satisfies the conditions of one--loop
conformal invariance (see details for locally isotropic
backgrounds in
 [53], one holds
these ha-field equations%
$$
\frac 1{\alpha ^{\prime }}\frac{n_E-25}3+ [4(\nabla \Phi
)^2-4\nabla ^2\Phi -r-\frac 13T_{[<\alpha ><\beta ><\gamma
>]}T^{[<\alpha ><\beta ><\gamma >]}]=0,
$$
$$
\widehat{r}_{(<\alpha ><\beta >)}+2\nabla _{(<\alpha >}\nabla
_{<\beta
>)}\Phi =0,
$$
$$
\widehat{r}_{[<\alpha ><\beta >]}+2T_{[<\alpha ><\beta ><\gamma
>]}\nabla ^{<\gamma >}\Phi =0.\eqno(9.25)
$$
By straightforward calculations we can show that the dual theory
has the same conformal properties and satisfies the conditions
(9.25) if the dual transform is completed by the shift of dilaton
field $ \widehat{\Phi }=\Phi -\frac 12\log G_{00}. $

The system of ha--field equations (9.25), obtained as a
low-energy limit of \index{Equations!ha--field} the ha--string
theory, is similar to Einstein--Cartan equations (6.34) and
(6.36). We note that the explicit form of locally anisotropic
ener\-gy--mo\-ment\-um source in (9.25) is defined from well
defined principles and symmetries of string interactions and this
form is not postulated, as in usual locally isotropic field
models, from some general considerations in order to satisfy the
necessary conservation laws on ha-space whose formulation is very
sophisticated because of nonexistence of global and even local
group of symmetries of such type of spaces. Here we also remark
that the higher order anisotropic model with dilaton field
interactions does not generate in the low-energy limit the
Einstein-Cartan ha-theory because the first system of equations
from (9.25) represents some constraints (being a consequence of
the two-dimensional symmetry of the model) on torsion and scalar
curvature which can not be interpreted as some algebraic
relations of type (6.36) between locally anisotropic spin-matter
source and torsion. As a matter of principle we can generalize
our constructions by introducing interactions with gauge
ha-fields and considering a variant of locally
anisotropic chiral $\sigma $-model 
 [126] in order to get a system of
equations quite similar to (6.36). However, there are not
exhaustive arguments for favoring the Einstein-Cartan theory and
we shall not try in this work to generate it necessarily from
ha-strings.

\section{Summary and Conclusions}

Let us try to summarize our results, discuss their possible
implications and make the basic conclusions. Firstly, we have
shown that the Einstein-Cartan theory has a natural extension for
a various class of ha-spaces. Following
the R. Miron and M. Anastesiei approach 
 [160,161] to the geometry of
la-spaces it becomes evident the possibility and manner of
formulation of classical and quantum field theories on such
spaces. Here we note that in la-theories we have an additional
geometric structure, the N-connection. From physical point of
view it can be interpreted, for instance, as a fundamental field
managing the dynamics of splitting of high-dimensional space-time
into the four-dimensional and compactified ones. We can also
consider the N-connection as a generalized type of gauge field
which reflects some specifics of ha-field interactions and
possible intrinsic structure of ha-spaces. It was convenient to
analyze the geometric structure of different variants of
ha-spaces (for instance, Finsler, Lagrange and generalized
Lagrange spaces) in order to make obvious physical properties and
compare theirs perspectives in developing of new physical models.

According to modern-day views the theories of fundamental field
interactions should be a low energy limit of the string theory.
One of the main results of this work is the proof of the fact
that in the framework of ha-string theory is contained a more
general, locally anisotropic, gravitational physics. To do this
we have developed the locally anisotropic nonlinear sigma model
and studied it general properties. We shown that the condition of
self consistent propagation of string on ha-background impose
corresponding constraints on the N-connection curvature, ha-space
torsion and antisymmetric d-tensor. Our extension of background
field method for ha-spaces has a distinguished by N-connection
character and the main advantage of this formalism is doubtlessly
its universality for all types of locally isotropic or
anisotropic spaces.

The presented one- and two-loop calculus for the higher order
anisotropic string model and used in this work d-covariant
dimensional regularization are developed for ha-background spaces
modelled as vector bundles provided with compatible N-connection,
d-connection and metric structures. In the locally isotropic
limit we obtain the corresponding formulas for the usual
nonlinear sigma model.

Finally, it should be stressed that we firstly calculated the
amplitudes for scattering of ha-gravitons and that the duality
properties of the formulated in this work higher order
anisotropic models are similar to those of models considered for
locally isotropic strings


\chapter{Stochastic Processes on HA--Spaces}

The purpose of investigations in this Chapter 
 [253,262] is to
extend the formalism of stochastic calculus to the case of spaces
with \index{Stochastic Processes!on ha--spaces} higher order
anisotropy (distinguished vector bundles with compatible
nonlinear and distinguished connections and metric structures and
generalized Lagrange and Finsler spaces). We shall examine
nondegenerate diffusions on the mentioned spaces and theirs
horizontal lifts.

Probability theorists, physicists, biologists and financiers are
already familiar with classical and quantum statistical and
geometric methods applied in va\-ri\-ous bran\-ches of science
and economy
 [13,14,171,189,131,84,141]. We note that modeling of diffusion processes
in nonhomogerneous media and formulation of nonlinear
thermodynamics in physics, or of dynamics of evolution of species
in biology, requires a more extended geometrical background than
that used in the theory of stochastic differential equations and
diffusion processes on Riemann, Lorentz
manifolds 
 [117,74,75,76] and in Rieman--Cartan--Weyl spaces 
 [197,199].

Our aim is to formulate the theory of diffusion processes on
spaces with local anisotropy. As a model of such spaces we choose
vector bundles on space-times provided with nonlinear and
distinguished connections and metric structures
 [160,161]. Transferring our considerations on tangent
bundles we shall formulate the theory of stochastic differential
equations on generalized Lagrange spaces which contain as
particular cases Lagrange and Finsler spaces
 [213,17,18,29,159].

The plan of the presentation in the Chapter is as follow: We
present a brief introduction into the theory of stochastic
differential equations and diffusion processes on Euclidean
spaces in section 10.1. In section 10.2 we give a brief summary
of the geometry of higher order anisotropic spaces. Section 10.3
is dedicated to the formulation of the theory of stochastic
differential equations in distinguished vector bundle spaces.
This section also concerns the basic aspects of stochastic
calculus in curved spaces. In section 10.4 the heat equations in
bundle spaces are analyzed. The nondegenerate diffusion on spaces
with higher order anisotropy is defined in section 10.5. We shall
generalize in section 10.6 the results of section 10.4 to the
case of heat equations for distinguished tensor fields in vector
bundles with (or not) boundary conditions. Section 10.7 contains
concluding remarks and a discussion of the obtained results.\

\section{ Diffusion Processes on Euclidean Spa\-ces}

We summarize the results necessary for considerations in the next
sections. Details on stochastic calculus and diffusion can be
found, for example, in works
 [117,62,90,132,123,124,71].

\subsection{ Basic concepts and notations}

Let consider the probability space $\left( \Omega ,{\cal F},
P\right) ,$ where $\left( \Omega ,{\cal F}\right) $ is a
measurable space, called the sample space, on which a probability
measure $P$ can be placed. A stochastic
process is a collection of random variables $X=\{X_L;0\leq t<\infty \}$ on $%
\left( \Omega ,{\cal F}\right) ,$ which take values in a second
measurable space $\left( S,{\cal B\ }\right) ,\,$ called the
state space. For our purposes we suggest that $\left( S,{\cal B\
}\right) $ is locally a r-dimensional Euclidean space equipped
with a $\sigma $-field of Borel sets, when we have the
isomorphism $S\cong {\cal R}^r$ and ${\cal B}\cong B\left(
{\cal R}^r\right) $ , where ${\cal B\left( U\right) }$ denotes the smallest $%
\sigma $-field containing all open sets of a topological space
${\cal U.}$ The index $t\in [0,\infty )$ of the random variables
$X_t$ will admit a convenient interpretation as time.

We equip the sample space$\left( \Omega ,{\cal F}\right) $ with a
filtration, i.e. we consider a non decreasing family ${\cal
\{F,}$ $t\geq 0\} $ of sub $\sigma $-fields of ${\cal F}:{\cal
F}_s \subseteq {\cal F}_t
\subseteq {\cal F} $ for 0$\leq s<t<\infty .$ We set ${\cal F}_\infty =$ $%
\sigma \left( \bigcup\limits_{t\geq 0}{\cal F}\right) .$

One says that a sequence $X_n$ converges almost surely (one
writes in brief a.s.) to $X$ if for all $\omega \in \Omega ,$
excepting subsets of zero
probability, one holds the convergence%
$$
\lim \limits_{n\rightarrow \infty }X_n\left( \omega \right)
=X\left( \omega \right) .
$$

A random variable $X_t$ is called p-integrable if%
$$
\int\limits_\Omega \left| X\left( \omega \right) \right|
^pP\left( d\omega \right) <\infty ,p>0,\omega \in \Omega
,a.s.\eqno(10.1)
$$
($X_t$ is integrable if (10.1) holds for $p=1).$ For an integrable variable $%
X$ the number
$$
E\left( X\right) =\int\limits_\Omega X\left( w\right) P\left(
d\omega \right)
$$
is the mathematical expectation of $X$ with respect to the
probability measure $P$ on $\left( \Omega ,{\cal F}\right) .$

Using a sub- $\sigma $-field ${\cal G}$ of $\sigma $-field ${\cal
F}$ we can
define the value%
$$
E\left( X,{\cal G}\right) =\int\limits_{{\cal G\ }}X\left( \omega
\right) d\omega
$$
called as the conditional mathematical expectation of $X$ with respect to $%
{\cal G.}$

Smooth random processes are modeled by the set of all smooth functions $%
w:[0,\infty )\ni t\rightarrow w\left( t\right) \in {\cal R}^r ,$ denoted as $%
W^r=C\left( [0,\infty )\rightarrow {\cal R}^r\ \right) .\,$ Set
$W^r$ is complete and separable with respect to the metric
$$
\rho \left( w_1,w_2\right) =\sum\limits_{n=1}^\infty 2^{-n}\left[
\left( \max \limits_{0\leq t\leq n}\left| w_1\left( t\right)
-w_2\left( 2\right) \right| \right) \bigwedge 1\right] ,
$$
w$_1,w_2\in W^r,$ where $a\bigwedge 1=\min \{a,1\}.$

Let ${\cal B\ }$ ($W^r)$ be a topological $\sigma $-field. As a
Borel cylindrical set we call a set $B\subset W^r,$ defined as
$B=\{w:\left( w\left( t_1\right) ,w\left( t_2\right) ,...,w\left(
t_n\right) \right) $ and
$E\subset {\cal B}\left( {\cal R}^{nr}\right) .$ We define as ${\cal %
C\subset B\ }$ ($W^r)$ the set of all Borel cylindrical sets.

\begin{definition}
A process $\{X_t,{\cal F}_t,$ $0\leq t<\infty \}$ is said to be a
submartingale (or supermartingale ) if for every $0\leq
s<t<\infty ,$ we
have $\dot P-$a.s. $E\left( X_t|{\cal F}\right) \geq X_s$ ( or $E\left( X_t|%
{\cal F}\right) \leq X_s).$ We shall say that $\{X_t,{\cal F}_t,$
$0\leq t<\infty \}$ is a martingale if it is both a submartingale
and a supermartingale.
\end{definition}
\index{Submartingale} \index{Supermartingale} \index{Martingale}

Let the function $p\left( t,x\right) ,t>0,x\in {\cal R}^r$ is
defined as
$$
p\left( t,x\right) =\left( 2\pi t\right) ^{-\frac r2}\exp \left[ -\frac{%
\left| x\right| ^2}{2t}\right]
$$
and $X=\left( X_t\right) _{t\in [0,\infty )}$ is a r-dimensional
process that for all $0<t_1<...<t_m$ and $E_i\in {\cal B}\left(
{\cal R}^r \right) ,$ $i=1,2,...,m,$

$$
P\{X_{t_1}\in E_1,X_{t_2}\in E_2,...,X_{t_m}\in E_m\}=\int\limits_{{\cal %
R^r\ }}\mu \left( dx\right)
$$
$$
\int\limits_{E_1}p\left( t_1,x_1-x\right)
dx_1\int\limits_{E_2}p\left( t_2-t_1,x_2-x_1\right) dx_2...
$$
$$
\int\limits_{E_m}p\left( t_m-t_{m-1},x_m-x_{m-1}\right)
dx_m,\eqno(10.2)
$$
where $\mu $ is the probability measure on$\left( {\cal R}^r,B\left( {\cal R}%
^r\right) \right) .$

\begin{definition}
A process $X=\left( X_t\right) $ with the above stated properties
is called a r--dimensional Brownian motion (or a Wiener process
with initial distribution $\mu .$ A probability $P^X$ on  $\left(
W^r,{\cal B}\left( W^r\right) \right) ,$ \\ where $P\{w:w\left(
t_1\right) \in E_1,w\left( t_2\right) \in E_2,...,w\left(
t_m\right) \in E_m\}$ is given by the right part of (10.2) is
called a r-dimensional Wiener distribution with initial
distribution $\mu .$
\end{definition}
\index{Brownian motion!r--dimensional} \index{Wiener process}
Now, let us suggest that on the probability space $\left( \Omega
,{\cal F,} \mbox{ \thinspace }P\right) $ a filtration $\left(
{\cal F}_t \right) ,t\in
[0,\infty $ ) is given. We can introduce a r--dimensional $\left( {\cal F}%
_t\right) $--Brownian motion as a d--dimensional smooth process
$X=\left(
X\left( t\right) \right) _{t\in [0,\infty )}, \left( {\cal F}_t \right) $%
--co\-or\-di\-nat\-ed and satisfying condition
$$
E\left[ \exp \left[ i<\xi ,X_t-X_s>\right] |{\cal F}_s\ \right]
=\exp \left[ -\left( t-s\right) \left| \xi \right| ^2/2\right] {
a.s.\ }
$$
for every $\xi \in {\cal R}^r$ and $0\leq s<t.$

The next step is the definition of the Ito stochastic integral
 [123,117,62,\\ 90,132]. Let denote as ${\cal L}_2$ the space of all real
measurable processes $\Phi =\{\Phi \left( t,u\right) \}_{t\geq 0}$ on $%
\Omega ,$ $\left( {\cal F}_t\right) $ -adapted for every $T>0,$

$$
\parallel \Phi \parallel _{2,T}^2\doteq E\left[ \int\limits_0^T\Phi ^2\left(
s,\omega \right) ds<\infty \right] ,
$$
where ''$\doteq "$ means ''is defined to be''. For $\Phi \in
{\cal L}_2$ we
write%
$$
\parallel \Phi \parallel _2\doteq \sum\limits_{n=1}^\infty 2^{-2}\left(
\parallel \Phi \parallel _{2,n}\bigwedge 1\right) .
$$
We restrict our considerations to processes of type%
$$
\Phi \left( t,\omega \right) =f_0\left( \omega \right)
I_{\{t=0\}}+\sum\limits_{i=0}^\infty f_i\left( \omega \right)
I_{(t_it_{i+1}]}\left( t\right) ,\eqno(10.3)
$$
where $I_A\left( B\right) =1,$ if $A\subset B$ and $I_{A\,}\left(
B\right) =0,$ if $A\subseteq B.$

Let denote ${\cal M}_2=$\ \{$X=\left( X_t\right) _{t\geq 0};X$ is
a quadratic integrable martingale on $\left( \Omega ,{\cal F,}
P\right) $
referring to $\left( {\cal F}_t\right) _{t\geq 0}$ and $X_0=0$ a.s. \} and $%
{\cal M}_2^c=\{ $ $X\in {\cal M}_2;\ $ \thinspace $t\rightarrow
X$ is smooth a.s.\}. For $X\in {\cal M}_2\ $ we use denotations
$$
\left| X\right| _T\doteq E\left[ X_T^2\right] ^{\frac 12},T>0,
$$
and $\left| X\right| =\sum\limits_{n=1}^\infty 2^{-n}\left(
\left| X\right| _n\bigwedge 1\right) .$

Now we can define stochastic integral \index{Stochastic integral}
 on $\left( {\cal F}_t \right) $%
-Brownian motion $B\left( t\right) $ on\\ $\left( \Omega ,{\cal
F}, P\right) $ as a map
$$
{\cal L}_2 {\ni }\Phi \rightarrow I\left( \Phi \right) \in {\cal M}_2^c.%
$$

For a process of type (10.3) we postulate
$$
I\left( \Phi \right) \left( t,\omega \right)
=\sum\limits_{i=0}^{n-1}[f_i\left( \omega \right) (B\left(
t_{i+1},\omega \right) -B\left( t_i,\omega \right) )+f_n\left(
\omega \right) (B\left( t,w\right) -$$
$$B\left( t_n,\omega \right) )]=
\sum\limits_{i=0}^\infty f_i\left( B\left( t\bigwedge
t_{i+1}\right) -B\left( t\bigwedge t_i\right) \right) \eqno(10.4)
$$
for $t_n\leq t\leq t_{n+1},n=0,1,2,....$

Process $I\left( \Phi \right) \in {\cal M}_2^c$ defined by (10.4)
is called the stochastic integral of $\Phi \in {\cal L}_2$ on
Brownian motion $B\left( t\right) $ and is denoted as
$$
I\left( \Phi \right) \left( t\right) =\int\limits_0^t\Phi \left(
s,\omega \right) dB\left( s,w\right) =\int\limits_0^t\Phi \left(
s\right) dB\left( s\right) .\eqno(10.5)
$$
It is easy to verify that the integral (10.5) satisfies properties:%
$$
\left| I\left( \Phi \right) \right| _T=\parallel \Phi \parallel
_{2,T}=\parallel \Phi \parallel _2,
$$
$$
E\left( I\left( \Phi \right) \left( t\right) ^2\right)
=\sum\limits_{i=0}^\infty E\left[ f_i^2\left( t\bigwedge
t_{i+1}-t\bigwedge t_i\right) \right] =E\left[
\int\limits_0^t\Phi ^2\left( s,w\right) ds\right]
$$
and
$$
I\left( \alpha \Phi +\beta \Psi \right) \left( t\right) =\alpha
I\left( \Phi \right) \left( t\right) +\beta I\left( \Psi \right)
\left( t\right) ,
$$
for every $\Phi ,\Psi \in {\cal L}_2$ $\left( \alpha ,\beta \in {\cal R}%
\right) $ and $t\geq 0.$

Consider a measurable space $\left( \Omega ,{\cal F}\right) $
equipped with a filtration $\left( {\cal F}_t\ \right) .$ A
random time $T$ is a stopping
time of this filtration if the event $\{T\leq t\}$ belongs to the $\sigma $%
-field$\left( {\cal F}_t\ \right) $ for every $t\geq 0.$ A random
time is an
optional time of the given filtration if $\{T\leq t\}$ $\in \left( {\cal F}%
_t\ \right) $ for every $t\geq 0.$ A real random process $X=\left(
X_t\right) _{t>0}$ on $\left( \Omega ,{\cal F,} P\right) $ is
called a local (${\cal F}_t)$ -martingale if it is adapted to
$\left( {\cal F}_t\ \right) $
and there is a sequence of stopping $\left( {\cal F}_t\ \right) $-moments $%
\sigma _n$ with $\sigma _n<\infty ,\sigma _n\uparrow \infty $ and $%
X_n=\left( X_n(t)\right) $ is a $\left( {\cal F}_t\ \right)
$-martingale for every $n=1,2,...,$ where $X_n=X\left( t\bigwedge
\sigma _n\right) .$ If $X_n$ is a quadratic integrable martingale
for every $n,$ than $X$ is called a local quadratic integrable
$\left( {\cal F}_t\ \right) $-martingale .

Let denote ${\cal M}_2^{loc}=\{$ $X=\left( X_t\right) _{t\geq
0},X$ is a
locally quadratic integrable $\left( {\cal F}_t\ \right) $-martingale and $%
X_0=0$ a.s.\}\thinspace and ${\cal M}_2^{c,loc}$=\{$X\in {\cal M}_2^{loc}:$ $%
t\rightarrow X_t$ is smooth a.s.\}. In a similar manner with the
Brownian motion we can define stochastic integrals for processes
$\Phi \in {\cal L}_2$ and $\Phi \in {\cal L}_2^{loc}$ on
$M\subset {\cal M}_2^{loc}\ $(we have to introduce $M\left(
t_j,\omega \right) $ instead of $B\left( t_j,\omega \right) $
respectively for $t_j=t_{i+1},t_j=t_i,t_i=t_n,t_j=1$ in formulas
(10.4)). In this case one denotes the stochastic integral as%
$$
I^M\left( \Phi \right) \left( t\right) =\int\limits_0^t\Phi
\left( s\right) dM\left( s\right) .
$$

A great part of random processes can be expressed as a sum of a
mean motion and fluctuations ( It\^o processes ) \index{It\^o
processes}
$$
X\left( t\right) =X\left( 0\right) +\int\limits_0^tf\left(
s\right) ds+\int\limits_0^tg\left( s\right) dB\left( s\right) ,
$$
where $\int\limits_0^tf\left( s\right) ds$ defines a mean motion , $%
\int\limits_0^tg\left( s\right) dB\left( s\right) $ defines
fluctuations and $\int dB$ is a stochastic integral on Brownian
motion $B\left( t\right) .$ In general such processes are sums of
processes with limited variations and martingales. Here we
consider the so-called smooth semimartingales
introduced on a probability space with a given filtration $\left( {\cal F}%
_t\right) _{t\geq 0},$ $M^i\left( t\right) \in {\cal M}_2^{c,loc}$ and $%
A^i\left( t\right) $ being smooth (${\cal F}_t)$-adapted
processes with trajectories having a limited variation and
$A^i(0)=0$ (i=1,2,...,r). So a smooth r-dimensional
semimartingale can be written as
$$
X^i\left( t\right) =X^i\left( 0\right) +M^i\left( t\right)
+A^i\left( t\right) .\eqno(10.6)
$$
For processes of type (10.6) one holds the It\^o formula
 [123,117,62,90,132]
 which gives us a differential-integral calculus for
paths of random processes:%
$$
F\left( X\left( t\right) \right) -F\left( X\left( 0\right)
\right) =
$$
$$
\int\limits_0^tF_i^{\prime }\left( X\left( s\right) \right)
dM^i\left( s\right) +\frac 12\int\limits_0^tF_{ij}^{\prime \prime
}\left( X\left( s\right) \right) d<M^iM^j>\left( s,\right)
,\eqno(10.7)
$$
where $F_i^{\prime }=\frac{\partial F}{\partial x^i},F_{ij}^{\prime \prime }=%
\frac{\partial ^2F}{\partial x^i\partial x^j},<M^iM^j>$ is the
quadratic covariation of processes $M^i$ ,$M^j\in {\cal M}_2,\ $
which really represents a random process $A=A_t,$ parametrized as
a difference of two
natural integrable processes with the property that $M_tN_t-A_t$ is a $%
\left( {\cal F}_t\right) $-martingale. Here we remark that a
process $Q=Q_t$
is an increasing integrable process if it is $\left( {\cal F}_t\right) $%
-adapted, $Q_0=0,$ the map $t\rightarrow Q_1$ is left-continuous
, $Q_t\geq 0 $ and $E\left( Q_t\right) <\infty $ for every $t\in
[0,\infty ).$ A process $Q$ is called natural if for every
bounded martingale $M=\left( M_t\right) $ and every $t\in
[0,\infty )$
$$
E\left[ \int\limits_0^\tau M_sdA_s\right] =E\left[
\int\limits_0^tM_sdA_s\right] .
$$

There is another way of definition of stochastic integration
instead of It\^o integral, the so-called Fisk--Stratonovich
integral, which obeys the \index{Fisk--Stratonovich integral}
usual rules of mathematical analysis. Let introduce denotations :
${\cal A}$
is the set of all such smooth $\left( {\cal F}_t\right) $-adapted processes $%
A=\left( A_t\right) $ that $A_0=0$ and $t\rightarrow A_t$ is a
function with limited variation on every finite integral a.s.;
${\cal B}$ is the set of all such $\left( {\cal F}_t\right)
$-predictable processes $\Phi =\left( \Phi _t\right) $ that with
the probability one the function $t\rightarrow \Phi _t$ is a
bounded function on every finite interval and $\left( t,\omega
\right) \rightarrow X_t\left( \omega \right) $ is ${\cal C}/B\left({\cal R}%
^r \right) $-measurable; ${\cal O\ }$ is the set of
quasimartingales (for every $X\subset {\cal O\ }$ we have the
martingale part $M_X$ and the part with limited variation ). For
every $\Phi \in {\cal B}$ and $X\in {\cal O}$ one defines the
scalar product
$$
\left( \Phi \circ X\right) =X\left( 0\right) +\int\limits_0^t\Phi
\left( s,w\right) dM_X\left( s\right) +\int\limits_0^t\Phi \left(
s,w\right) dA_X\left( s\right) ,t\geq 0,
$$
as an element of ${\cal O}$ . One introduces an element $\Phi \circ dX\in d%
{\cal O}$ as
$$
\Phi dX=\Phi \circ dX=d\left( \Phi \circ X\right)
$$
in order to define the symmetric Q-product :%
$$
Y\circ dX=YdX+\frac 12dXdY
$$
for $dX\in d{\cal O}$ and $Y\in {\cal O.\ }$ The stochastic integral $%
\int\limits_0^tY\circ dX$ is called the symmetric
Fisk-Stratonovich integral.

\subsection{ Stochastic dif\-fe\-ren\-ti\-al equ\-a\-ti\-ons}

\index{Stochastic!dif\-fe\-ren\-ti\-al equ\-a\-ti\-ons} Let
denote as ${\cal A}^{c,r}$ the set of functions satisfying the
conditions : $\alpha \left( t,w\right) :[0,\infty )\times
W^r\rightarrow {\cal R}^r\times {\cal R}^r$ are ${\cal B}$
$\left( [0,\infty )\right)
\times {\cal B}$ $\left( W^r\right) /{\cal B}\left({\cal R}^r\otimes {\cal R}%
^r\right)$-measurable, for every $t\in [0,\infty )$ a function\\
$W^r\ni
w\rightarrow \alpha \left( t,w\right) \in {\cal R}^r \times {\cal R}^r$ is $%
{\cal B}_t \left( W^r\right) /{\cal B} \left( {\cal R}^d \otimes
{\cal R}^c \right) $-measurable, where \\ ${\cal R}^r \times
{\cal R}^c$ denotes the set of $r\times c$ matrices, ${\cal B}
\left( {\cal R}^r\otimes {\cal R}^r \right) $ is the topological
$\sigma $-field on ${\cal R}^r\times {\cal R}^c$ , obtained in
the result of identification of ${\cal R}^r\times {\cal R}^c$
with $rc $-dimensional Euclidean space.

Suppose that values $\alpha \in {\cal A}^{r,c}$ and $\beta \in {\cal A}%
^{r,1} $ are given and consider the next stochastic differential
equations for a r-dimensional smooth process\\ $X=\left( X\left(
t\right) \right) _{t\geq 0}:$

$$
dX_t^\epsilon =\sum\limits_{j=1}^r\alpha _\gamma ^\epsilon \left(
t,X\right) dB^\gamma \left( t\right) +\beta ^\epsilon \left(
t,X\right) dt,\eqno(10.8)
$$
$$
(\epsilon =1,2,...,r),
$$
for simplicity also written as%
$$
dX_t=\alpha \left( t,X\right) dB\left( t\right) +\beta \left(
t,X\right) dt.
$$

As a weak solution (with respect to a c--dimensional Brownian motion $%
B\left( t\right) ,B\left( 0\right) =0$ a.s.) of the equations
(10.8) we mean a r--dimensional smooth process $X=\left( X\left(
t\right) \right) _{t\geq 0}, $ defined on the probability space
$\left( \Omega ,{\cal F},P\right) $
with such a filtration of $\sigma $--algebras ${\left( {\cal F}_t\right) }%
_{t\geq 0}$ that $X=X\left( t\right) $ is adapted to $\left( {\cal F}%
_t\right) _{t\geq 0},$ i.e. a map $\omega \in \Omega \rightarrow
X(\omega
)\in W^r$ is defined and for every $t\in [0,\infty )$ this map is ${\cal F}%
_t/{\cal B}_t$ $(W^{r\ })$--measurable; we can define processes
$\Phi _\beta ^\delta \left( t,\omega \right) = \alpha _\beta
^\delta \left( t,X\left( \omega \right) \right) \subset {\cal
L}_2^{loc}$ and $\Psi ^\delta \left( t,\omega \right) =\beta
^\delta \left( t,X\left( \omega \right) \right) \subset {\cal
L}_1^{loc}$; values $X\left( t\right) =\left( X^1\left( t\right)
,X^2\left( t\right) ,...,X^r\left( t\right) \right) $ and $B\left(
t\right) =\left( B^1\left( t\right) ,B^2\left( t\right)
,...,B^c\left( t\right) \right) $ satisfy equations
$$
X^i\left( t\right) -X^i\left( 0\right) =\sum\limits_{\beta
=1}^c\int\limits_0^t\alpha _\beta ^\delta \left( s,X\right)
dB^\beta \left( s\right) +\int\limits_0^t\beta ^\delta \left(
s,X\right) ds,\eqno(10.9)
$$
with the unit probability, where the integral on $dB^\beta \left(
s\right) $ is considered as the It\^o integral (10.7). The first
and the second terms in (10.9) are called correspondingly as the
martingale and drift terms.

Let $\sigma \left( t,x\right) =\sigma _j^i\left( t,x\right) $ be
a Borel function $\left( t,x\right) \in [0,\infty )\times {\cal
R}^r\rightarrow {\cal R}^r\otimes {\cal R}^r$ and $b\left(
t,x\right) =\{b^i\left( t,x\right) \}$ be a Borel function
$\left( t,x\right) \in [0,\infty )\times {\cal R}^r\rightarrow
{\cal R}^r.$ Then $\alpha \left( t,w\right) =\sigma \left(
t,w\left( t\right) \right) \subset {\cal A}^{r,c}$ and $\beta
\left( t,w\right) =b\left( t,w\left( t\right) \right) \in {\cal
A}^{r,1}.$ In this case the stochastic differential equations
(10.8) are of the Markov type and
can be written in the form%
$$
dX^{i\;}\left( t\right) =\sum\limits_{k=1}^r\sigma _k^i\left(
t,X\left( t\right) \right) dB^k\left( t\right) +b^i\left(
t,X\left( t\right) \right) dt.
$$
If $\sigma $ and $b$ depend only on $x\in {\cal R^c}$ we obtain a
equation with homogeneous in time $t$ coefficients.

Function
 $$\Phi \left( x,w\right) : {\cal R}^r \times W_0^c\rightarrow
W^r,W_0^c=\{w \in {\cal C} \left( [0,\infty )\rightarrow {\cal
R}^r \right) ; w\left( 0\right) =0\}$$ is called $\widehat{{\cal
E}}\left( {\cal R}^r \times W_0^c\right) $--measurable if for
every Borel probability measure $\mu $ on ${\cal R}^r\ $ there is
a function $\widetilde{\Phi }_\mu \left(
x,w\right) :{\cal R}^r \times W_0^c\rightarrow W^c,$ which is\\ $\overline{%
{\cal B}({\cal R}^r\times W_0^c)}^{\mu \times P^W}/{\cal B} \left(
W^r\right) $--measurable for all $x\left( \mu \right), \Phi
\left( x,w\right) =\widetilde{\Phi }_\mu \left( x,w\right) $ and
$P^W$--almost all $w $ ($P^W$ is the c--dimensional Wiener
measure on $W_0^c,$ i.e. a distribution $B$ ).

A solution $X=\left( X\left( t\right) \right) $ of the equations
(10.8) with a Brownian motion $B=B\left( t\right) $ is called a
strong solution if there is a function $F\left( x,w\right) :{\cal
R}^r$ $\times W_0^c\rightarrow W^r,$
which is  $\widehat{{\cal E}}\left( {\cal R}^r\times W_0^c\right) $%
--measurable for every $x\in {\cal R}^r$ , $w\rightarrow F\left(
x,w\right) $ is $\overline{{\cal B}_t\left( W_0^c\right)
}^{P^W}/{\cal B}_t\left( W^c\right) $--measurable for every
$t\geq 0$ and $X=F\left( X\left( 0\right) ,B\right) $ a.s.

We obtain a unique strong solution if for every r--dimensional
 $\left( {\cal F}%
_t\right) $--Brownian motion $B=B\left( t\right)$ $ \left(
B\left( 0\right)
=0\right) $ on the probability space with the filtration ${\left( {\cal F}%
_t\right) }$ and arbitrary $\left( {\cal F}_0\right) $--measurable ${\cal R}%
^r $--valued random vector $X=F\left( \xi ,B\right) $ is a
solution of (10.8) on this space with $X\left( 0\right) =\xi $
a.s. So, a strong solution can be considered as a function
$F\left( x,w\right) $ which generates a solution
$X$ of equation (10.8) if and only if we shall fix the initial value $%
X\left( 0\right) $ and Brownian motion $B.$

\subsection{ Diffusion Processes}

As the diffusion processes one names the class of processes which
are \index{Diffusion processes} \index{Markov property}
characterized by the Markov property and smooth paths (see
details in
 [124,71,\\ 117]. Here we restrict ourselves with the definition of
diffusion processes generated by second order differential operators on $%
{\cal R}^r:$

$$
Af\left( x\right) =\frac 12\sum\limits_{i,j=1}^ra^{ij}\left( x\right) \frac{%
\partial ^2f}{\partial x^i\partial x^j}\left( x\right)
+\sum\limits_{i=1}^rb^i\left( x\right) \frac{\partial f}{\partial
x}\left( x\right) ,\eqno(10.10)
$$
where $a^{ij}\left( x\right) $ and $b^i\left( x\right) $ are real
smooth functions on ${\cal R}^r$, matrix $a^{ij}\left( x\right) $
is symmetric and positively defined . Let denote by
$\widehat{{\cal R}}^r={\cal R}^r$ $\cup \{\Delta \}$ the point
compactification of ${\cal R}^r.$ Every function $f$
on ${\cal R}^r$ is considered as a function on $\widehat{{\cal R}}^r$ with $%
f\left( \Delta \right) =0.\,$ The region of definition of the
operator (10.10) is taken the set of doubly differentiable
functions with compact carrier, denoted as $C_K^2\left( {\cal
R}^r\right) .$ Let ${\cal B}$ $\left( \widehat{W}^r\right) $ be
the $\sigma $-field generated by the Borel cylindrical sets,
where $\widehat{W}^r=\{w:[0,\infty )\ni t\rightarrow w\left(
t\right) \in $ $\widehat{{\cal R}}^r$ is smooth and if $w\left(
t\right) =\Delta $ , then $w\left( t^{\prime }\right) =\Delta $ for all $%
t^{\prime }\geq t.$ The value $e\left( w\right) =\inf \{t;w\left(
t\right) =\Delta ,w\in \widehat{W}^r\}$ is called the explosion
time of the path $w.$

\begin{definition}
A system of Markov probability distributions $\{P_x,x\in $ ${\cal
R}^r\}$ {\it on $\left( \widehat{W}^r,{\cal B}\left(
\widehat{W}^r\right) \right) ,$}
which satisfy conditions : $P_X\{w:w(0)=x\}=1$ {\it for every }$x\in {\cal R}%
^r;$ $f\left( w\left( t\right) \right) -f\left( w\left( 0\right)
\right) -\int\limits_0^t\left( Af\right) \left( w\left( s\right)
\right) ds$ {\it is
a $\left( P_x,{\cal B}_t\ (\widehat{W}^r)\right) $}--martingale for every $%
f\in C_K^2\left( {\cal R}^r\ \right) $ and $x\in {\cal R}^r$
defines a diffusion measure generated by an operator $A$ ( or
$A$- diffusion).
\end{definition}

\begin{definition}
A random process $X=\left( X\left( t\right) \right) $ on ${\cal
R}^r$ is
said to be a diffusion process, generated by the operator $A$ ( or simply a $%
A$--diffusion process) if almost all paths $\left[ t\rightarrow
X\left( t\right) \right] \in \widehat{W}^r$ and probability law
of the process $X$ coincides with $P_\mu \left( \cdot \right)
=\int\limits_{{\it \ }{\cal R}^r\ }P_x\left( \cdot \right) \mu
\left( dx\right) ,$ where $\mu $ is the diffusion measure
generated by the operator $A$ and $\{P_x\}$ is the probability
law of $X\left( 0\right) .$
\end{definition}

To a given A-diffusion we can associate a corresponding stochastic
differential equation. Let the matrix function $\sigma \left(
x\right) =\left( \sigma _j^i\left( x\right) \right) \in {\cal
R}^r\times {\cal R}^r$ defines $a^{ij}\left( x\right)
=\sum\limits_{k=1}^r\sigma _k^i\left(
x\right) $ $\sigma _k^j\left( x\right) $ and consider the equations%
$$
dX^i\left( t\right) =\sum\limits_{k=1}^r\sigma _k^i\left( X\left(
t\right)
\right) dB^k\left( t\right) +b^i\left( X\left( t\right) \right) dt.%
\eqno(10.11)
$$
There is an ex\-ten\-sion of $\left( \Omega ,{\cal F,}P\right) $
with a fil\-tra\-tion $\left( {\cal F}_t\right) $ of the
prob\-a\-bil\-ity space
$$
\left( \widehat{W}^r,{\cal B}\left( \widehat{W}^r\right)
,P_X\right)
$$
and with a filtration ${\cal B}_t\left( \widehat{W}^r\right) $
and a $\left( {\cal F}_t\right) $-Brownian motion $B\left(
t\right) $ (see
 [117] and
the previous subsections) that putting $X\left( t\right) =w\left(
t\right) $ and $e=e\left( w\right) \,$ one obtains for $t\in
[0,e)$
$$
X^i\left( t\right) =x^i+\sum\limits_{k=1}^r\int\limits_0^t\sigma
_k^i\left( X\left( s\right) \right) dB^k\left( s\right)
+\int\limits_0^tb^i\left( X\left( s\right) \right) ds.
$$
So $\left( X\left( t\right) ,B\left( t\right) \right) $ is the
solution of the equations (10.11) with $X\left( 0\right) =x.$

If bounded regions are considered, diffusion is described by
second order
partial differential operators with boundary conditions. Let denote $D={\cal %
R}_t^r =$ $\{ x=(x^1,x^2,....x^r);x^r\geq 0 \},$ $\partial
D=\{x\in D,x^r=0\},D^0=\{x\in D;x^r>0\}.$ The Wentzell bound
operator is defined as a
map from $C_K^2\left( L\right) $ to the space of smooth functions on $%
\partial D$ of this type:

$$
Lf\left( x\right) =\frac 12\sum\limits_{i,j=1}^{r-1}\alpha
^{ij}\left( x\right) \frac{\partial ^2f}{\partial x^i\partial
x^j}\left( x\right)
+\sum\limits_{i=1}^{r-1}\beta ^i\left( x\right) \frac{\partial f}{\partial x}%
\left( x\right) +
$$
$$
\mu \left( x\right) \frac{\partial f}{\partial x^r}\left(
x\right) -\rho \left( x\right) Af\left( x\right) ,\eqno(10.12)
$$
where $x\in \partial D,\alpha ^{ij}\left( x\right) ,\beta
^i\left( x\right) ,\mu \left( x\right) $ and $\rho \left(
x\right) $ are bounded smooth functions on $\partial D,\alpha
^{ij}\left( x\right) $ is a symmetric and nondegenerate matrix,
$\mu \left( x\right) \geq 0$ and $\rho \left( x\right) \geq 0.$

A diffusion process defined by the operators (10.10) and (10.12)
is called a $\left( A,L\right) $-diffusion.

\section{Ha--Frames and Horizontal Lifts}

We emphasize that in this Chapter all geometric constructions and
results on stochastic calculus will be formulated for the general
case of ha-spaces.

On a ha--space ${\cal E}^{<z>}$ we can consider arbitrary
compatible with metric $G$ d-con\-ne-ti\-ons $\Gamma _{<\beta
><\gamma >}^{<\alpha >},$ which are analogous of the affine
connections on locally isotropic spaces (with or not torsion). On
${\cal E}^{<z>}$ it is defined the canonical d-struc\-tu\-re
$\overrightarrow{\Gamma }_{<\beta ><\gamma >}^{<\alpha >}$ with
coefficients generated by components of metric and
N-con\-nec\-ti\-on (we shall consider the metric and
d--connection to be induced in a
Riemannian manner but on a ha--space)%
$$
\overrightarrow{\Gamma }_{jk}^i=L_{jk}^i,\overrightarrow{\Gamma }%
_{j<a>}^i=C_{j<a>}^i,\overrightarrow{\Gamma }_{<a>j}^i=0,\overrightarrow{%
\Gamma }_{<a><b>}^i=0,\eqno(10.13)
$$
$$
\overrightarrow{\Gamma }_{jk}^{<a>}=0,\overrightarrow{\Gamma }%
_{j<b>}^{<a>}=0,\overrightarrow{\Gamma }_{<b>k}^{<a>}=L_{<b>k}^{<a>},%
\overrightarrow{\Gamma }_{<b><c>}^{<a>}=C_{<b><c>,}^{<a>}
$$
where%
$$
L_{jk}^i=\frac 12g^{ip}(\delta _kg_{pi}+\delta _jg_{pk}-\delta
_pg_{jk}),
$$
$$
L_{b_Pd_f}^{a_p}=\delta _{b_p}N_{d_f}^{a_p}+ \frac
12h^{a_p<c>}(\delta _{d_f}h_{b_p<c>}-\delta
_{b_p}N_{d_f}^{d_p}h_{d_p<c>}-\delta
_{<c>}N_{d_f}^{d_p}h_{d_pb_f}),
$$
$$
C_{b_pc_f}^{a_p}=\frac 12g^{a_p<k>}\delta _{c_f}g_{b_p<k>},
$$
$$
C_{b_pc_p}^{a_p}=\frac 12h^{a_pd_p}(\delta _{c_p}h_{d_pb_p}+\delta
_{b_p}h_{d_pc_p}-\delta _{d_p}h_{b_pc_p}),
$$
where $0\leq f<p<z.$ In formulas (10.13) we have used matrices $g^{ij}$ and $%
h^{<a><b>}\,$ which are respectively inverse to matrices $g_{ij}$ and $%
h_{<a><b>}.$

We also present the explicit formulas for unholonomy
coefficients\\ $w_{<\beta
><\gamma >}^{<\alpha >},$ of the adapted frame basis (6.4):
$$
w_{ij}^k=0,w_{a_pc_p}^{b_f}=0, w_{c_fa_p}^{b_f}=0, $$ $$
w_{a_pb_p}^{c_f}=0,w_{a_pb_p}^{c_p}=0,%
\eqno(10.14)
$$
$$
w_{b_fc_f}^{a_p}=R_{b_fc_f}^{a_p},w_{a_pc_f}^{b_p}=-\delta
_{a_p}N_{c_f}^{b_p},w_{c_fa_p}^{b_p}=\delta _{a_p}N_{c_f}^{b_p}.
$$
Putting (10.13) and (10.14) into, correspondingly, (6.25) and
(6.28) we can computer the components of canonical torsion
$\overrightarrow{T}_{<\beta
><\gamma >}^{<\alpha >}$ and curvature\\ $\overrightarrow{R}_{<\beta
><\gamma ><\delta >}^{<\alpha >}$ with respect to the locally adapted bases
(6.5) and (6.6) 
 [160,161].

Really, on every la-space ${\cal E}^{<z>}$ a linear
multiconnection d-structure is defined. We can consider at the
same time some ways of local
transports of d-tensors by using, for instance, an arbitrary d-connection $%
\Gamma _{<\beta ><\gamma >}^{<\alpha >},$ the canonical one $\overrightarrow{%
\Gamma }_{<\beta ><\gamma >}^{<\alpha >},$ or the so-called
Christoffel
d--symbols (see (1.49) for arbitrary signatures) defined as%
$$
\{\frac{<\alpha >}{<\beta ><\gamma >}\}=\eqno(10.15)
$$
$$
\frac 12G^{<\alpha ><\tau >}(\delta _{<\gamma >}G_{<\beta ><\tau
>}+\delta _{<\beta >}G_{<\gamma ><\tau >}-\delta _{<\tau
>}G_{<\beta ><\gamma >}).
$$
Every compatible with metric d-connection $\Gamma _{<\beta
><\gamma
>}^{<\alpha >}$ can be characterized by a corresponding deformation d-tensor
with respect, for simplicity, to $\{\frac{<\alpha >}{<\beta ><\gamma >}\}:$%
$$
P_{<\beta ><\gamma >}^{<\alpha >}=\Gamma _{<\beta ><\gamma >}^{<\alpha >}-\{%
\frac{<\alpha >}{<\beta ><\gamma >}\}\eqno(10.16)
$$
(the deformation of the canonical d-connection is written as
$$
\overrightarrow{P}_{<\beta ><\gamma >}^{<\alpha >}=\overrightarrow{\Gamma }%
_{<\beta ><\gamma >}^{<\alpha >}-\{\frac{<\alpha >}{<\beta ><\gamma >}\}).%
\eqno(10.17)
$$
Perhaps, it is more convenient to consider physical models and
geometric
constructions with respect to the torsionless d-connection $\{\frac{<\alpha >%
}{<\beta ><\gamma >}\}.$ The more general ones will be obtained
by using deformations of connections of type (10.16). But
sometimes it is possible to write out d-covariant equations on
${\cal E}^{<z>}$ by changing respectively components
$\{\frac{<\alpha >}{<\beta ><\gamma >}\}$ on $\Gamma _{<\beta
><\gamma >}^{<\alpha >}$ . This holds for definition of stochastic
differential equations on la-spaces (see, in particular,
 [13,14] on
diffusion on Finsler spaces) and in our works we use the last way.

Let suppose that ${\cal E}^{<z>}$ is locally trivial and $\sigma
$-compact. In this case ${\cal E}^{<z>}$ is a paracompact
manifold and has a countable
open base. We denote as $F({\cal E}^{<z>}{\cal \ )}$ the set of all real $%
C^\infty $-functions on ${\cal E}^{<z>}$ and as $F_0\left( {\cal E}%
^{<z>}\right) $ the subclass of $F\left( {\cal E}^{<z>}\right) $
consisting from functions with compact carriers. $F_0\left( {\cal
E}^{<z>}\right) $ and
$F\left( {\cal E}^{<z>}\right) $ are algebras on the field of real numbers $%
{\cal R\ }$ with usual operations $f+q,fq$ and $\lambda f(f,q\in F({\cal E}%
^{<z>}{\cal \ )}$ or $F_0\left( {\cal E}^{<z>}\right) ,\lambda
\in {\cal R).} $

Vector fields on ${\cal E}^{<z>}$ are defined as maps%
$$
V:u\in {\cal E}^{<z>}{\cal \ }\rightarrow V\left( u\right) \in
T_u\left( {\cal E}^{<z>}\right) .
$$
Vectors $\left( \partial _{<\alpha >}\right) _u,(<\alpha
>=0,1,2,...,n_E-1),$ form a local linear basis in\\ $T_u\left(
{\cal E}^{<z>}\right) .$ We shall also use decompositions on
locally adapted basis,\\ $\left( \delta _{<\alpha >}\right) ,$
and denote by {\bf X$\left( {\cal E}^{<z>}\right) $} the set of
$C^\infty $-vector fields on ${\cal E}^{<z>}{\cal .}$

Now, let introduce the bundle of linear adapted frames $GL\left( {\cal E}%
^{<z>}\right) $ on ${\cal E}^{<z>}.$ As a linear adapted frame $%
e=[e_{<\underline{\alpha }>}],(<\underline{\alpha
}>=0,1,...,n_E-1),$ in
point $u\in {\cal E_N}$ we mean a linear independent system of vectors $e_{<%
\underline{\alpha }>}\in T_u\left( {\cal E}^{<z>}\right) $
obtained as a
linear distinguished transform of local adapted basis, i.e. $$e_{<%
\underline{\alpha }>}=e_{<\underline{\alpha }>}^{<\alpha >}\delta
_{<\alpha
>},$$ where
$$
e_{<\underline{\alpha }>}^{<\alpha >}\in GL^d\left( {\cal
R}\right) =GL\left( n,{\cal R\ }\right) \oplus GL\left( m_1,{\cal
R}\right) \oplus
...\oplus GL\left( m_z,{\cal R}\right) .%
$$
Then $GL\left( {\cal E}^{<z>}\right) $ is defined as the set of
all linear adapted frames in all points $u\in {\cal E}^{<z>}:$

$GL\left( {\cal E}^{<z>}\right) =\{r=(u,e),u\in {\cal E}^{<z>}$
and $e$ is a linear adapted frame in the point $u\}.$

Local coordinate carts on $GL\left( {\cal E}^{<z>}\right) $ are defined as $%
\left( \widetilde{{\cal U\ }}_{<\alpha >},\widetilde{\varphi
}_{<\alpha
>}\right) ,$ where $\widetilde{{\cal U\ }}_{<\alpha >}=\{r=(u,e)\in GL\left(
{\cal E}^{<z>}\right) ,u\in {\cal U_\alpha \ \subset E}^{<z>}$
and $e$ is a
linear adapted frame in the point $u\}$ , $\widetilde{\varphi _{<\alpha >}}%
\left( r\right) =\left( \varphi _{<\alpha >}\left( u\right)
=(u^{<\alpha
>}),e_{<\underline{\alpha }>}^{<\alpha >}\right) $ and $e_{<\underline{%
\alpha }>}=e_{<\underline{\alpha }>}^{<\alpha >}\delta _{<\alpha
>}\mid _u.$
So $GL\left( {\cal E}^{<z>}\right) $ has the structure of $C^\infty $%
-manifold of dimension $n+m_1+...+m_z+n^2+m_1^2+...+m_z^2.$
Elements $a\in GL^d\left( {\cal R}\right) $ act on $GL\left(
{\cal E}^{<z>}\right) $
according the formula $T_a(u,e)=(u,ea),$ where $(ea)_{<\underline{\alpha }%
>}=a_{<\underline{\alpha }>}^{<\underline{\beta }>}e_{<\underline{\beta >}}.$
The surjective projection $\pi :GL\left( {\cal E}^{<z>}\right)
\rightarrow {\cal E}^{<z>}{\cal \ }$is defined in a usual manner
by the equality $\pi (u,e)=u.$

Every vector field $L\in {\bf X}({\bf {\cal E}}^{<z>})$ induces a
vector
field $\widetilde{L}$ on $GL\left( {\cal E}^{<z>}\right) .$ Really, for $%
f\in {\bf X}({\bf {\cal E}}^{<z>})$ we can consider
$$
\left( \widetilde{L}f\right) (r)=\frac d{dt}f\left( (\exp
tL)u,(\exp tL)_{*}e\right) \mid _{t=0},\eqno(10.18)
$$
where $r=(u,e)$ and
$$
(\exp tL)_{*}e=[(\exp tL)_{*}e_1,(\exp tL)_{*}e_2,...,(\exp
tL)_{*}e_{n_E}],
$$
is the differential (an isomorphism $T_u\left( {\cal E}^{<z>}{\cal \ }%
\right) \rightarrow T_{(\exp tL)_u}{\cal E}^{<z>}$ for every $u\in {\cal E}%
^{<z>}{\cal )}$ of $\exp tL$ and the local diffeomorphism
$u\rightarrow
v(t,u)$ is defined by differential equations%
$$
\frac{dv^{<\alpha >}}{dt}(t,u)=a^{<\alpha >}(v(t,u)),\eqno(10.19)
$$
$$
\left( L=a^{<\alpha >}(u)\delta _{<\alpha >}\right) ,v(0,u)=u.
$$

Let $L\in {\bf X}({\bf {\cal E}}^{<z>})$ and introduce functions $%
f_L^{<\alpha >}(r)\in F\left( GL({\cal E}^{<z>}{\cal )}\right) $ for every $%
\alpha =0,1,2,...,n_E-1$ by the equalities%
$$
f_L^{<\alpha >}(r)=\left( e^{-1}\right) _{<\underline{\alpha
}>}^{<\alpha
>}a^{<\underline{\alpha }>}(u)\eqno(10.20)
$$
written in locally adapted coordinates $r=\left( u^{<\alpha >},e=(e_{<%
\underline{\alpha }>}^{<\alpha >})\right) $ on the manifold $GL\left( {\cal E%
}^{<z>}\right) ,$ where $L=a^{<\alpha >}(u)\delta _{<\alpha >}$
and $e^{-1}$ is the matrix inverse to $e.$ Because the equality
(10.20) does not depend on
local coordinates, we have defined a global function on $GL\left( {\cal E}%
^{<z>}\right) .$ It's obvious that for $L_{(1)},L_{(2)}\in {\bf
X}({\bf {\cal E}}^{<z>})$ we have
$$
\left( \widetilde{L}_{(1)}f_{L_{(2)}}^{<\alpha >}\right) \left(
r\right) =f_{[L_{(1)}L_{(2)}]}^{<\alpha >}\left( r\right) ,
$$
where $\widetilde{L}_{(1)}$ and $\widetilde{L}_{(2)}$ are
constructed
similarly to operator (10.18), and%
$$
[L_{(1)},L_{(2)}]=L_{(1)}L_{(2)}-L_{(2)}L_{(1)}.
$$
A distinguished connection $\Gamma _{<\beta ><\gamma >}^{<\alpha
>}$ defines the covariant derivation of d-tensors in ${\cal
E}^{<z>}$ in a usual manner. For example, we can introduce a
d-covariant derivation $DB$ of a d-tensor field $B\left( u\right)
=B_{<\beta _1><\beta _2>...<\beta _q>}^{<\alpha
_1><\alpha _2>...<\alpha _p>}\left( u\right) $ in the form%
$$
D_{<\gamma >}B_{<\beta _1><\beta _2>...<\beta _q>}^{<\alpha
_1><\alpha _2>...<\alpha _p>}(u)=B_{<\beta _1><\beta _2>...<\beta
_q>;<\gamma
>}^{<\alpha _1><\alpha _2>...<\alpha _p>}(u)=
$$
$$
\delta _{<\gamma >}B_{<\beta _1><\beta _2>...<\beta _q>}^{<\alpha
_1><\alpha _2>...<\alpha _p>}(u)+\sum_{<\epsilon >=1}^p\Gamma
_{<\gamma ><\delta
>}^{<\alpha _\epsilon >}\left( u\right) B_{<\beta _1><\beta _2>...<\beta
_q>}^{<\alpha _1><\alpha _2>...<\delta >...<\alpha _p>}(u)-
$$
$$
\sum_{\tau =1}^q\Gamma _{<\gamma ><\beta _\tau >}^{<\delta
>}\left( u\right) B_{<\beta _1><\beta _2>...<\delta >...<\beta
_q>}^{<\alpha _1><\alpha _2>...<\alpha _p>}(u),\eqno(10.21)
$$
or the covariant derivative $D_YB$ in the direction $Y=Y^{<\alpha
>}\delta
_{<\alpha >}\in {\bf X}({\bf {\cal E}}^{<z>}){\bf {\cal \ }}$ ,%
$$
(D_YB)_{<\beta _1><\beta _2>...<\beta _q>}^{<\alpha _1><\alpha
_2>...<\alpha _p>}(u)=Y^{<\delta >}B_{<\beta _1><\beta
_2>...<\beta _q>;<\delta
>}^{<\alpha _1><\alpha _2>...<\alpha _p>}(u)
$$
and the parallel transport along a (piecewise) smooth curve
$c:{\cal R\
\supset \,}I=(t_{1,}t_2)\ni t\rightarrow c\left( t\right) $ (considering $%
B\left( t\right) =B\left( c(t)\right) $ )%
$$
\frac d{dt}B_{<\beta _1><\beta _2>...<\beta _q>}^{<\alpha
_1><\alpha _2>...<\alpha _p>}(t)+\eqno(10.22)
$$
$$
\sum_{\epsilon =1}^p\Gamma _{<\gamma ><\delta >}^{<\alpha
_\epsilon >}\left( c(t)\right) B_{<\beta _1><\beta _2>...<\beta
_q>}^{<\alpha _1><\alpha _2>...<\delta >...<\alpha
_p>}(c(t))\frac{dc^{<\gamma >}}{dt}-
$$
$$
\sum_{\tau =1}^q\Gamma _{<\gamma ><\beta _\tau >}^{<\delta
>}\left( c(t)\right) B_{<\beta _1><\beta _2>...<\delta >...<\beta
_q>}^{<\alpha _1><\alpha _2>...<\alpha
_p>}(c(t))\frac{dc^{<\gamma >}}{dt}=0.
$$
For every $r\in GL\left( {\cal E}^{<z>}\right) $ we can define the
horizontal subspace
$$
H_r=\{U=a^{<\alpha >}\delta _{<\alpha >}\mid _u-\Gamma _{<\beta
><\gamma
>}^{<\alpha >}\left( u\right) e_{<\underline{\gamma }>}^{<\gamma >}a^{<\beta
>}\frac{<\partial >}{<\partial e_{\underline{\gamma }}^\alpha >},a^{<\alpha
>}\in {\cal R}^{n_E}\}
$$
of $T_u\left( GL({\cal E}^{<z>}{\cal )}\right) .$ Vector $U\in
H_r\,$ is called horizontal. Let $\xi \in T_u\left( {\cal
E}^{<z>,}{\cal \ }\right) ,$ then $\widetilde{\xi }\in T_r\left(
GL({\cal E}^{<z>}{\cal )}\right) \,$is a horizontal lift of $\xi
$ if the vector $\widetilde{\xi }$ is horizontal, i.e. $\pi
\left( r\right) =u$ and $(d\pi )_r\widetilde{\xi }=\xi .$ If $r$
is given as to satisfy $\pi \left( r\right) =u$ the
$\widetilde{\xi }$ is
uniquely defined. So, for given $U\in {\bf X}({\bf {\cal E}}^{<z>})$%
there is a unique $\widetilde{U}\in {\bf X\ }\left( GL\left( {\cal E}%
^{<z>}\right) \right) ,$ where $\widetilde{U}_r$ is the horizontal lift $%
U_{\pi \left( r\right) }$ for every $r\in GL\left( {\cal E}^{<z>}\right) .%
\widetilde{U}$ is called the horizontal lift of vector field $U.$
In local coordinates $\widetilde{U}=U^{<\alpha >}\left( u\right)
\delta _{<\alpha
>}-\Gamma _{<\alpha ><\beta >}^{<\delta >}\left( u\right) U^{<\alpha
>}\left( u\right) e_{<\underline{\epsilon }>}^{<\beta >}\frac \partial
{\partial e_{<\underline{\epsilon }>}^{<\delta >}}$ if
$U=U^{<\alpha
>}\left( u\right) \delta _{<\alpha >}.$

In a sim\-i\-lar man\-ner we can de\-fine the hor\-i\-zon\-tal
lift
$$
\widetilde{c}\left( t\right) =\left( c(t),e(t)\right)
=[e_0(t),e_1(t),...,e_{q-1}(t)]\in GL\left( {\cal E}^{<z>}\right)
$$
of a curve $c\left( t\right) \in {\cal E}^{<z>}$ with the
property that $\pi
\left( \widetilde{c}(t)\right) =c(t)\,$ for $t\in I\,$ and $\frac{d%
\widetilde{c}}{dt}(t)$ is horizontal. For every $<\underline{\alpha }%
>=0,1,...q-1$ there is a unique vector field $\widetilde{L}\in {\bf X}\left(
GL({\cal E}^{<z>}{\cal )}\right) ,$ for which $\left( \widetilde{L}_{%
\underline{\alpha }}\right) _r$ is the horizontal lift of vector $e_{<%
\underline{\alpha }>}\in T_u\left( {\cal E}^{<z>}\right) $ for every $%
r=\left( u,e=\left[ e_0,e_1,...,e_{q-1}\right] \right) .$ In coordinates $%
\left( u^{<\alpha >},e_{<\underline{\beta }>}^{<\beta >}\right) $
we can
express%
$$
\widetilde{L}_{<\underline{\alpha }>}=e_{<\underline{\alpha
}>}^{<\alpha
>}\delta _{<\alpha >}-\Gamma _{<\beta ><\gamma >}^{<\alpha >}e_{<\underline{%
\alpha }>}^{<\beta >}e_{<\underline{\beta }>}^{<\gamma >}\frac
\partial {\partial e_{<\underline{\beta >}}^{<\alpha >}}.
$$
Vector fields $\widetilde{L}_{<\underline{\alpha }>}$ form the
system of canonical horizontal vector fields.

Let $B\left( u\right) =B_{<\beta _1><\beta _2>...<\beta
_s>}^{<\alpha _1><\alpha _2>...<\alpha _p>}\left( u\right) $ be a
(p,s)--tensor field and define a system of smooth functions
$$
F_B\left( r\right) =\{F_{B<\underline{\beta _1}><\underline{\beta _2}>...<%
\underline{\beta _s}>}^{<\underline{\alpha _1}><\underline{\alpha _2}>...<%
\underline{\alpha _p}>}\left( r\right) =
$$
$$
B_{<\delta _1><\delta _2>...<\delta _s>}^{<\gamma _1><\gamma
_2>...<\gamma _p>}\left( u\right) e_{<\gamma
_1>}^{<\underline{\alpha _1}>}e_{<\gamma
_2>}^{<\underline{\alpha _2}>}...e_{<\gamma _p>}^{<\underline{\alpha _p}%
>}e_{<\underline{\beta _1}>}^{<\delta _1>}e_{<\underline{\beta _2}%
>}^{<\delta _2>}...e_{<\underline{\beta _s}>}^{<\delta _s>}\}
$$
(the scalarization of the d--tensor field $B\left( r\right) $
with the respect to the locally adapted basis $e)$ on $GL\left(
{\cal E}^{<z>}\right)
,$ where we consider that%
$$B\left( u\right) = $$
$$F_{B<\underline{\beta _1}><\underline{\beta _2}>...<%
\underline{\beta _s}>}^{<\underline{\alpha _1}><\underline{\alpha _2>}...<%
\underline{\alpha _p}>}\left( u\right) e_{<\underline{\alpha
_1}>}\otimes e_{<\underline{\alpha _2}>}\otimes
...e_{<\underline{\alpha _p}>}\otimes
e_{*}^{<\underline{\beta _1}>}\otimes e_{*}^{<\underline{\beta _2}%
>}...\otimes e_{*}^{<\underline{\beta _s}>},
$$
the matrix $e_{<\underline{\beta }>}^{<\delta >}$ is inverse to the matrix $%
e_{<\gamma >}^{<\underline{\alpha }>},$ basis $e_{*}$ is dual to $e$ and $%
r=(u,e).$ It is easy to verify that
$$
\widetilde{L}_{<\underline{\alpha }>}(F_{B<\underline{\beta _1}><\underline{%
\beta _2}>...<\underline{\beta _s}>}^{<\underline{\alpha _1}><\underline{%
\alpha _2}>...<\underline{\alpha _p}>})\left( r\right) =(F_{\nabla B})_{<%
\underline{\beta _1}><\underline{\beta _2}>...<\underline{\beta _s>};<%
\underline{\alpha }>}^{<\underline{\alpha _1}><\underline{\alpha _2}>...<%
\underline{\alpha _p}>}\left( r\right) \eqno(10.23)
$$
where the covariant derivation $\nabla _{<\underline{\alpha }>}A^{<%
\underline{\beta }>}=A_{;<\underline{\alpha
}>}^{<\underline{\beta }>}$ is
taken by using connection%
$$
\Gamma _{<\underline{\beta }><\underline{\gamma }>}^{<\underline{\alpha }%
>}=e_{<\alpha >}^{<\underline{\alpha }>}e_{<\underline{\beta }>}^{<\beta
>}e_{<\underline{\gamma }>}^{<\gamma >}\Gamma _{<\beta ><\gamma >}^{<\alpha
>}+e_{<\underline{\gamma }>}^{<\gamma >}e_{<\sigma >}^{<\underline{\alpha }%
>}\delta _{<\gamma >}e_{<\underline{\beta }>}^{<\sigma >}
$$
in $GL\left( {\cal E}^{<z>}\right) $ (induced from ${\cal
E}^{<z>}{\cal ).}$

In our further considerations we shall also use the bundle of
orthonormalized adapted frames on ${\cal E}^{<z>},$ defined as a
subbundle of $GL\left( {\cal E}^{<z>}\right) $ satisfying
conditions:

$O\left( {\cal E}^{<z>}\right) =\{r=(u,e)\in GL\left( {\cal
E}^{<z>}\right) ,e$ is a orthonormalized basis in\\ $T_u\left(
{\cal E}^{<z>}\right) \}.$

So $r=(u^{<\alpha >},e_{<\underline{\alpha }>}^{<\alpha >}\in O\left( {\cal E%
}^{<z>}\right) )$ if and only if%
$$
G_{<\alpha ><\beta >}e_{<\underline{\alpha }>}^{<\alpha >}e_{<\underline{%
\beta }>}^{<\beta >}=\delta _{<\underline{\alpha }><\underline{\beta }>}%
\eqno(10.24)
$$
or, equivalently,%
$$
\sum_{\underline{\alpha }=0}^{q-1}e_{<\underline{\alpha }>}^{<\alpha >}e_{<%
\underline{\alpha }>}^{<\beta >}=G^{<\alpha ><\beta >},
$$
where the matrix $G^{<\alpha ><\beta >}$ is inverse to the matrix $%
G_{<\alpha ><\beta >}$ from (6.12).

\section{ Stochastic Differenti\-al D--Equa\-ti\-ons}

\index{Stochastic differenti\-ak d--equa\-ti\-ons} In this
section we assume that the reader is familiar with the concepts
and basic results on stochastic calculus, Brownian motion and
diffusion processes (an excellent presentation can be found in
 [117,74,75,62,90,132], see also a brief introduction into the
mentioned subjects in the previous section of this Chapter and
considerations for the higher order anisotropic diffusion in
Chapter 5. The purpose of the section is to extend the theory of
stochastic differential equations on Riemannian spaces
 [117,74,75] to the case of spaces
with general anisotropy, defined in the previous section as
v-bundles.

Let $A_{\widehat{0}},A_{\widehat{1}},...,A_{\widehat{r}}\in {\bf
X}({\bf
{\cal E}}^{<z>})$ and consider stochastic differential equations%
$$
dU\left( t\right) =A_{<\widehat{\alpha }>}\circ dB^{<\widehat{\alpha }%
>}\left( t\right) +A_{\widehat{0}}\left( U(t)\right) dt,\eqno(10.25)
$$
where $<\widehat{\alpha }>=1,2,...,r$ and $\circ \,$ is the
symmetric Q-product (see subsection 10.1.1). We shall use the
point compactification
of space ${\cal E}^{<z>}$ and write $\widehat{{\cal E\ }}^{<z>}={\cal E}%
^{<z>}{\cal \ }$ or $\widehat{{\cal E\ }}^{<z>}={\cal E}^{<z>}{\cal \ }$ $%
\cup \{\Delta \}$ in dependence of that if ${\cal E}^{<z>}{\cal \
}$ is compact or noncompact. By $\widehat{W}\left( {\cal
E}^{<z>}\right) $ we denote the space of paths in ${\cal
E}^{<z>}{\cal \ ,\ }$ defined as

$\widehat{W}\left( {\cal E}^{<z>}\right) =\{w:w$ is a smooth map
[0,$\infty )\rightarrow \widehat{{\cal E\ }}^{<z>}$ with the
property that $w\left( 0\right) \in {\cal E}^{<z>}$ and
$w(t)=\Delta ,w(t^{\prime })=\Delta $ for all $t^{\prime }\ge t\}$

and by ${\cal B}$ $\left( \widehat{W}\left( {\cal E}^{<z>}\right)
\right) $ the $\sigma $-field generated by Borel cylindrical sets.
\index{Borel cylindrical sets}

The explosion moment $e(w)$ is defined as
$$
e\left( w\right) =\inf \{t,w(t)=\Delta \}
$$

\begin{definition}
The solution $U=U\left( t\right) $ of equation (10.25) in v-bundle space $%
{\cal E}^{<z>}{\cal \ }$ is defined as such a ${({\cal F}_t)}%
\mbox{-compatible }\widehat{W}\left( {\cal E}^{<z>}\right)
$-valued random element (i.e. as a smooth process in ${\cal
E}^{<z>}{\cal \ }$ with the trap $\Delta ),$ given on the
probability space with filtration $({\cal F}_t)$ and
$r$-dimensional ${\cal F}_t)$- Brownian motion $B=B(t),$ with
$B(0)=0,$ for which

$$
f\left( U\left( t\right) \right) -f\left( U\left( 0\right)
\right) =
$$
$$
\int_0^tA_{<\widehat{\alpha }>}\left( t\right) \left( U\left(
s\right) \right) \delta B^{<\widehat{\alpha }>}\left( s\right)
+\int_0^t\left( A\circ f\right) \left( U\left( s\right) \right)
ds\eqno(10.26)
$$
for every $f\in F_0\left( {\cal E}^{<z>}\right) $ (we consider
$f\left( \Delta \right) =0),$ \ where the first term is
understood as a Fisk-Stratonovich integral.
\end{definition}

In (10.26) we use $\delta B^{<\widehat{\alpha }>}\left( s\right)
$ instead of $dB^{<\widehat{\alpha }>}\left( s\right) $ because
on ${\cal E}^{<z>}$ the Brownian motion must be adapted to the
N-connection structure.

In a manner similar to that for stochastic equations on
Riemannian spaces
 [117] we can construct the unique strong solution to the equations
(10.25). To do this we have to use the space of paths in ${\cal
R}^r$
starting in point 0, denoted as $W_0^r,$ the Wiener measure $P^W$ on $%
W_0^r,\sigma $-field ${\cal B}_t\ (W_0^r)$-generated by Borel
cylindrical sets up to moment $t$ and the similarly defined
$\sigma $-field.

\begin{theorem}
There is a function $F:{\cal E}^{<z>}\times W_0^r\rightarrow \widehat{W}%
\left( {\cal E}^{<z>}\right) $ being\\ $\bigcap_\mu {\cal B}\left( {\cal E}%
^{<z>}\right) \times {\cal B}_t\ $ $\left( W_0^r\right) ^{\mu \times P^W}/ $%
{\it ${\cal B}_t\ $} $\left( \widehat{W}\left( {\cal F}_t\right) \right) $%
-measurable (index $\mu $ runs all probabilities in $\left( {\cal E}^{<z>}%
{\cal B}({\cal E}^{<z>}) \right) $ ) for every $t\geq 0$ and
having properties:

1) For every $U(t)$ and Brownian motion $B=B\left( t\right) $ the equality\\
$U=F(U\left( 0\right) ,B)$ a.s. is satisfied.

2) For every r-dimensional $({\cal F}_t)$--Brownian motion
$B=B\left( t\right) $ with $B=B\left( 0\right) ,$ defined on the
probability space with
filtration ${\cal F}_t,\ $\mbox{ and } ${\cal E}^{<z>}$--valued ${\cal F_0}$%
--measurable random element $\xi ,$ the function $U=F(\xi ,B)$ is
the solution of the differential equation (10.26) with $U\left(
0\right) =\xi ,$ a.s.
\end{theorem}

{\it Sketch of the proof.} Let take a compact coordinate vicinity
${\cal V}$
with respect to a locally adapted basis $\delta _{<\alpha >}$ and express $%
A_{<\underline{\alpha }>}=\sigma _{<\underline{\alpha
}>}^{<\alpha >}\left(
u\right) \delta _{<\alpha >},$ where functions $\sigma _{<\underline{\alpha }%
>}^{<\alpha >}\left( u\right) $ are considered as bounded smooth functions
in ${\cal R}^{n_E}$ , and consider on ${\cal V}$ the stochastic
differential
equation%
$$
dU_t^{<\alpha >}=\sigma _{<\widehat{\alpha }>}^{<\alpha >}\left(
U_t^{<\alpha >}\right) \circ \delta B^{<\widehat{\alpha }>}\left(
t\right) +\sigma _0^{<\alpha >}\left( U_t\right) dt,\eqno(10.27)
$$
$$
U_0^{<\alpha >}=u^{<\alpha >},(<\alpha >=0,1,...q-1).
$$
Equations (10.27) are equivalent to%
$$
dU_t^{<\alpha >}=\sigma _{<\widehat{\alpha }>}^{<\alpha >}\left(
U_t^{<\alpha >}\right) dB^{<\widehat{\alpha }>}\left( t\right) +\overline{%
\sigma }_0^{<\alpha >}\left( U_t\right) dt,
$$
$$
U_0^{<\alpha >}=u^{<\alpha >},
$$
where $\overline{\sigma }_0^{<\alpha >}\left( u\right) =\sigma
_0^{<\alpha
>}\left( u\right) +\frac 12\sum_{<\widehat{\alpha }>=1}^r\left( \frac{\delta
\sigma _{<\widehat{\alpha }>}^{<\alpha >}\left( u\right) }{\delta
u^{<\beta
>}}\right) \sigma _{<\beta >}^{<\widehat{\alpha }>}\left( u\right) .$ It's
known
 [117] that (10.27) has a unique strong solution $F:{\cal R}%
^{n_E}\ \times W_0\rightarrow \widehat{W}^{n_E}$ or $F(u,w)=\left(
U(t,u,w)\right) .$ Taking $\tau _{{\cal V}}(w)=\inf
\{t:U(t,u,w)\in {\cal V\ \}}$ we define
$$
U_{{\cal V\ }}(t,u,w)=U(t\bigwedge \tau _{{\cal V\ }}\left( w\right) ,u,w)%
\eqno(10.28)
$$
In a point $u\in {\cal V\cap \widetilde{V},}$ where ${\cal
\widetilde{V}\ }$ is covered by local coordinates
$u^{<\widetilde{\alpha }>},$ we have to consider transformations
$$
\sigma _{<\underline{\alpha }>}^{<\widetilde{\alpha }>}\left( \widetilde{u}%
\left( u\right) \right) =\sigma _{<\underline{\alpha }>}^{<\alpha >}(u)\frac{%
\partial u^{<\widetilde{\alpha }>}}{\partial u^{<\alpha >}},
$$
where coordinate transforms $u^{\widetilde{\alpha }}\left(
u^\alpha \right) $ satisfy the properties (6.1). The global
solution of (10.27) can be constructed by gluing together
functions (10.28) defined on corresponding coordinate regions
covering ${\cal E}^{<z>}$. $\Box $

Let $P_u$ be a probability law on $\widehat{W}\left( {\cal
E}^{<z>}\right) $ of a solution $U=U\left( t\right) $ of equation
(10.25) with initial conditions $U_{(0)}=u.$ Taking into account
the uniqueness of the mentioned solution we can prove that
$U=U\left( t\right) $ is a A--diffusion and
satisfy the Markov property 
 [117] (see also subsection 10.1.3).
Really, because for every $f\in F_0\left( {\cal E}^{<z>}\right) $

$$
df\left( U\left( t\right) \right) =\left( A_{<\widehat{\alpha
}>}f\right) \left( U\left( t\right) \right) \circ
dw^{<\widehat{\alpha }>}+\left( A_0f\right) \left( U\left(
t\right) \right) dt=
$$

$$
\left( A_{<\widehat{\alpha }>}f\right) \left( U\left( t\right) \right) dw^{<%
\widehat{\alpha }>}+\left( A_0f\right) \left( U\left( t\right)
\right) d+\frac 12d\left( A_{<\widehat{\alpha }>}f\right) \left(
U\left( t\right) \right) \cdot dw^{<\widehat{\alpha }>}\left(
t\right)
$$
and%
$$d\left( A_{<\widehat{\beta }>}f\right) \left( U\left( t\right) \right) =
$$ $$A_{<%
\widehat{\alpha }>}\left( A_{<\widehat{\beta }>}f\right) \left(
U\left(
t\right) \right) \circ dw^{<\widehat{\alpha }>}\left( t\right) +\left( A_{%
\widehat{0}}A_{<\widehat{\beta }>}f\right) \left( U\left(
t\right) \right) dt,
$$
we have%
$$
d\left( A_{<\widehat{\alpha }>}f\right) \left( U\left( t\right)
\right) \cdot dw^{<\widehat{\alpha }>}\left( t\right)
=\sum\limits_{<\widehat{\alpha }>=1}^rA_{<\widehat{\alpha
}>}\left( A_{<\widehat{\alpha }>}f\right) \left( U\left( t\right)
\right) dt.
$$
Consequently, it follows that%
$$
df\left( U\left( t\right) \right) =\left( A_{<\widehat{\alpha
}>}f\right) \left( U\left( t\right) \right) dw^{<\widehat{\alpha
}>}\left( t\right) +\left( Af\right) \left( U\left( t\right)
\right) dt,
$$
i.e. the operator $\left( Af\right) ,$ defined by the equality
$$
Af=\frac 12\sum\limits_{<\widehat{\alpha }>=1}^rA_{<\widehat{\alpha }%
>}\left( A_{<\widehat{\alpha }>}f\right) +A_0f\eqno(10.29)
$$
generates a diffusion process $\{ P_u\},u\in {\cal E}^{<z>}.$

The above presented results are summarized in this form:

\begin{theorem}
A second order differential operator $Af\,$ generates a A--dif\-fu\-si\-on on $%
\widehat{W}$ $({\cal E}^{<z>}{\cal )}$ of a solution $U=U\left(
t\right) \,\, $ it of the equation (10.29) with initial condition
$U\left( 0\right) =u.\,$
\end{theorem}

Using similar considerations as in flat spaces 
 [117] on carts covering $%
{\cal E}^{<z>},$ we can prove the uniqueness of A--diffusion
$$\{ P_u\},u\in {\cal E}^{<z>} \mbox{ on } \widehat{W}  ({\cal E}^{<z>}
 ).$$

\section{ Heat Equations and Flows of Diffeomorfisms}

Let v-bundle ${\cal E}^{<z>}{\cal \ }$ be a com\-pact man\-i\-fold of class $%
C^\infty .$ \index{Heat equations} \index{Flows of diffeomorfisms}
We con\-sider op\-er\-a\-tors
$$
A_0,A_1,...,A_r\in {\bf X}({\bf {\cal E}}^{<z>})
$$
and suppose that the property
$$
E[Sup_{t\in [0,1]}Sup_{u\in {\cal U\ }}|D^{<\underline{\alpha
}>}\{f\left( U\left( t,u,w\right) \right) \}|]<\infty
$$
is satisfied for all $f\in F_0\left( {\cal E}^{<z>}\right) $ and
every multiindex $<\underline{\alpha }>$ in the coordinate
vicinity ${\cal U\ }$ with $\overline{{\cal U}}$ being compact
for every $T>0.$ The heat equation in $F_0\left( {\cal
E}^{<z>}\right) $ is written as
$$
\frac{\partial \nu }{\partial t}(t,u)=A\nu (t,u),\eqno(10.30)
$$
$$
\lim \limits_{t\downarrow 0,\overline{u}\rightarrow u}\nu (t,\overline{u}%
)=f\left( u\right) ,
$$
where operator $A$ acting on $F_{}\left( F_0\left( {\cal
E}^{<z>}\right) \right) $ $\,$ is defined in (10.29).

We denote by $C^{1,2}\left( [0,\infty )\times F_0\left( {\cal E}%
^{<z>}\right) \right) $ the set of all functions $f(t,u)$ on
$[0,\infty )\times {\cal E}^{<z>}$ being smoothly differentiable
on $t$ and twice differentiable on $u.$

The existence and properties of solutions of equations (10.30)
are stated according the theorem:

\begin{theorem}
The function
$$
\zeta (t,u)=E[f(U(t,u,w)]\in C^\infty [0.\infty )\times {\cal
E}^{<z>} ],
$$
$f\in F_0\left( {\cal E}^{<z>}\right) $ satisfies heat equation
(10.30). Inversely, if a bounded function $\nu (t,u)\in
C^{1,2}\left( [0,\infty )\times {\cal E}^{<z>}\right) $ solves
equation (10.30) and satisfies the
condition%
$$
\lim \limits_{k\uparrow \infty }E[\nu (t-\sigma _k,U\left( \sigma
_k,u,w\right) ):\sigma _k\leq t]=0\eqno(10.31)
$$
for every $t>0$ and $u\in {\cal E}^{<z>},$ where $\sigma _k=\inf
\{t,U(t,u,w)\in D_k\}$ and $D_k$ is an increasing sequence with
respect to
closed sets in ${\cal E}^{<z>},$ $\bigcup\limits_kD_k=$ ${\cal E}%
^{<z>}.$
\end{theorem}

{\it Sketch of the proof.}{\rm \ }The function $\zeta (t,u)$ is a
function on ${\cal E}^{<z>}{\cal ,\ }$ because $u\rightarrow
f\left( U\left( t,u,w\right) \right) \in C^\infty $ and in this
case the derivation under
mathematical expectation symbol is possible. According to (10.26) we have%
$$
f\left( U\left( t,u,w\right) \right) -f\left( u\right)
=martingale+\int_0^t\left( At\right) \left( U\left( s,u,w\right)
\right) ds
$$
for every $u\in {\cal E}^{<z>},$ i.e.%
$$
\zeta (t,u)=f\left( u\right) +\int_0^tE[\left( Af\right) \left(
U\left( s,u,w\right) \right) ds.\eqno(10.32)
$$
Because $A^nf\in F_0\left( {\cal E}^{<z>}\right) ,\left(
n=1,2,...\right) ,$
we can write%
$$
\zeta (t,u)=f\left( u\right) +t\left( Af\right) \left( u\right)
+\int_0^tdt_1\int_0^{t_1}E\left[ \left( A^2f\right) \left( U\left(
t_2,u,w\right) \right) \right] dt_2=
$$
$$
f\left( u\right) +t\left( Af\right) \left( u\right)
+\frac{t^2}2\left( A^2f\right) \left( u\right)+ $$ $$
\int_0^tdt_1\int_0^{t_1}dt_2\int_0^{t_2}E[\left( A^3f\right)
\left( U\left( t_3,u,w\right) \right) ]dt_3=
$$
$$
f\left( u\right) +t\left( Af\right) \left( u\right)
+\frac{t^2}2\left( A^2f\right) \left( u\right) +...
$$
$$
+\int_0^tdt_1\int_0^{t_1}dt_2...\int_0^{t_{n-1}}E[\left(
A^nf\right) \left( U\left( t_n,u,w\right) \right) dt_n
$$

from which it is clear that
$$
\zeta \left( t,u\right) \in C^\infty \left( [0,\infty )\times {\cal E}%
^{<z>}\right) .
$$
In Chapter V, section 3 of the monograph 
 [117] it is proved the
equality
$$
\left( A\zeta _t\right) \left( u\right) =E[\left( Af)U\left(
t,u,w\right) \right) ]\eqno(10.33)
$$
for every $t\geq 0,$ where $\zeta _t\left( u\right) =\zeta \left(
t,u\right) .$

From (10.32) and (10.33) one follows that
$$
\zeta \left( t,u\right) =f\left( u\right) +\int_0^t\left( A\zeta
\right) \left( s,u\right) ds
$$
and%
$$
\frac{\partial \zeta }{\partial t}\left( t,u\right) =A\zeta \left(
t,u\right) ,
$$
i.e. $\zeta =\zeta \left( t,u\right) $ satisfies the heat
equation (10.30).

Inversely, let $\nu \left( t,u\right) $ $\in C^{1,2}\left(
[0,\infty )\times {\cal E}^{<z>}\right) $ be a bounded solution
of the equation (10.30). Taking into account that $P\left(
e[U\left( \cdot ,u,w\right) \right) ]=\infty )=1$ for every $u\in
{\cal E}^{<z>}$ and using the It\^o formula
(see (10.6) ) we obtain that for every $t_0>0$ and $0\leq t\leq t_0\,$ .%
$$
E[\nu \left( t_0-t\bigwedge \sigma _n,U\left( t\bigwedge \sigma
_n,u,w\right) \right) ]-\nu \left( t_0,u\right) =
$$
$$
E[\int_0^{t\bigwedge \sigma _n}\{\left( A\nu \right) \left(
t_0-s,U\left( s,u,w\right) \right) -\frac{\partial \nu }{\partial
t}\left( t_0-s,U\left( s,u,w\right) \right) \}ds].
$$
Supposing that conditions (10.31) are satisfied and considering
$n\uparrow \infty $ we obtain
$$
E=\{\nu \left( t_0-t,U\left( t,u,w\right) \right) ;e[U\left( \cdot
,u,w\right) ]>t\}=\nu \left( t_0,u\right) .
$$
For $t\uparrow t_0$ we have $E[f\left( U\left( t_0,u,w\right)
\right) ]=\zeta \left( t_0,u\right) ,$ i.e. $\nu \left(
t,u\right) =\zeta \left( t,u\right) .\Box $

{\bf Remarks; 1.}{\it \ The conditions (10.31) are necessary in
order to select a unique solution of (10.30).}

2.{\it \ Defining
$$
\zeta \left( t,u\right) =E[\exp \{\int_0^tC\left( U\left(
s,u,w\right) \right) ds\}f\left( U\left( t,u,w\right) \right) ]
$$
instead of (10.30) we generate the solution of the generalized
heat equation in }${\cal E}^{<z>}:$
$$
\frac{\partial \nu }{\partial t}\left( t,u\right) =\left( A\nu
\right) \left( t,u\right) +C\left( u\right) \nu \left( t,u\right)
,
$$
$$
\lim \limits_{t\downarrow 0,\overline{u}\rightarrow u}\nu \left( t,\overline{%
u}\right) =f\left( u\right) .
$$

For given vector fields $A_{\left( \alpha \right) }\in {\bf X(}{\cal E}%
^{<z>},\left( \alpha \right) =0,1,...,r$ in section 10.4 we have
constructed
the map%
$$
U=\left( U\left( t,u,w\right) \right) :{\cal E}^{<z>}{\cal \ \times }%
W_0^r\ni (u,w)\rightarrow U\left( \cdot ,u,w\right) \in \widehat{W}({\cal E}%
^{<z>},
$$
which can be constructed as a map of type
$$
[0,\infty )\times {\cal E}^{<z>}{ \times }W_0^r\ni
(u,w)\rightarrow U\left( t,u,w\right) \in \widehat{{\cal
E}^{<z>}}.
$$
Let us show that map $u\in {\cal E}^{<z>}\rightarrow U\left(
t,u,w\right) \in \widehat{{\cal E}^{<z>}}$ is a local
diffeomorphism of the manifold ${\cal E}^{<z>}$ for every fixed
$t\geq 0$ and almost every $w$ that $\ \in {\cal E}^{<z>}$ .

We first consider the case when ${\cal E}^{<z>}$ $\cong {\cal R}%
^{n_E},\sigma \left( u\right) =\left( \sigma _{<\beta >}^{<\alpha
>}\left( u\right) \right) \in {\cal R}^{n_E}\otimes {\cal
R}^{n_E}$ and $b\left( u\right) =\left( b^{<\alpha >}\left(
u\right) \right) \in {\cal R}^{n_E}$ are given smooth functions
(i.e. $C^\infty $-functions) on ${\cal R}^{n_E}$ , $\parallel
\sigma \left( u\right) \parallel +\parallel b\left( u\right)
\parallel \leq K\left( 1+|u|\right) $ for a constant $K>0$ and all
derivations of $\sigma ^{<\alpha >}$ and $b^{<\alpha >}\,$ are
bounded. It
is known 
 [117] that there is a unique solution $U=U\left( t,u,w\right) $
, with the property that $E[\left( U\left( t\right) \right)
^p]<\infty $ for
all $p>1,$ of the equation%
$$
dU_t^{<\alpha >}=\sigma _{<\widehat{\alpha }>}^{<\alpha >}\left(
U_t\right)
dw^{<\widehat{\alpha }>}\left( t\right) +b^{<\alpha >}\left( U_t\right) dt,%
\eqno(10.34)
$$
$$
U_0=u,(\alpha =1,2,...,n_E-1),
$$
defined on the space $\left( W_0^r,P^W\right) $ with the flow $\left( {\cal F%
}_t^0\right) .$

In order to show that the map $u\rightarrow U\left( t,u,w\right)
$ is a diffeomorphism of ${\cal R}^{n_E}$ it is more convenient
to use the Fisk-Stratonovich differential and to write the
equation (10.34) equivalently as
$$
dU_t^{<\alpha >}=\sigma _{<\widehat{\alpha }>}^{<\alpha >}\left(
U_t\right) \circ \delta w^{<\widehat{\alpha }>}\left( t\right)
+\overline{b}^{<\alpha
>}\left( U_t\right) dt,\eqno(10.35)
$$
$$
U_0=u,
$$
by considering that
$$
\overline{b}^{<\alpha >}\left( u\right) =b^{<\alpha >}\left(
u\right) +\frac
12\sum_{<\widehat{\alpha }>=1}^r\left( \delta _{<\beta >}\sigma _{<\widehat{%
\alpha }>}^{<\alpha >}\right) \sigma _{<\widehat{\alpha
}>}^{<\beta >}\left( u\right) .
$$
We emphasize that for solutions of equations of type (10.35) one
holds the usual derivation rules as in mathematical analysis.

Let introduce matrices
$$
\sigma _{<\widehat{\alpha }>}^{\prime }=\left( \sigma ^{\prime
}\left( u\right) _{<\widehat{\alpha }><\beta >}^{<\alpha >}=\frac
\delta {\delta u^{<\beta >}}\sigma _{<\widehat{\alpha
}>}^{<\alpha >}\left( u\right) \right) ,b^{\prime }\left(
u\right) =\left( b^{\prime }\left( u\right)
_{<\beta >}^{<\alpha >}=\frac{\delta b^{<\alpha >}}{\partial u^{<\beta >}}%
\right) ,
$$
$$
I=\delta _{<\beta >}^{<\alpha >}
$$
and the Jacobi matrix
$$
Y\left( t\right) =\left( Y_{<\beta >}^{<\alpha >}\left( t\right) =\frac{%
\delta U^{<\alpha >}}{\delta u^{<\beta >}}\left( t,u,w\right)
\right) ,
$$
which satisfy the matrix equation%
$$
Y\left( t\right) =I+\int_0^t\sigma _{<\widehat{\alpha }>}^{\prime
}\left(
U\left( s\right) \right) Y\left( s\right) \circ dw^{<\widehat{\alpha }%
>}\left( s\right) +
$$
$$
\int_0^tb^{\prime }\left( U\left( s\right) \right) Y\left( s\right) ds.%
\eqno(10.36)
$$
As a modification of a process $U\left( t,u,w\right) $ one means
a such process $\widehat{U}\left( t,u,w\right) $ that
$P^W\{\widehat{U}\left( t,u,w\right) =U\left( t,u,w\right) $ for
all $t\geq 0\}=1$ a.s.

It is known this result for flows of diffeomorphisms of flat
spaces
 [83,156,\\ 74]:

\begin{theorem}
Let $U\left( t,u,w\right) $ be the solution of the equation
(10.35)\\ (or (10.34)) on Wiener space $\left( W_0^r,P^W\right)
.$ Then we can choose a modification $\widehat{U}\left(
t,u,w\right) $ of this solution when the map $u\rightarrow
U\left( t,u,w\right) $ is a diffeomorphism ${\cal R}^{n_E}$ a.s.
for every $t\in [0,\infty ).$
\end{theorem}

Process $u=\widehat{U}\left( t,u,w\right) $ is constructed by
using equations
$$
dU_t^{<\alpha >}=\sigma _{<\widehat{\alpha }>}^{<\alpha >}\left(
U_t\right) \circ \delta w^{<\widehat{\alpha }>}\left( t\right)
-b^{<\alpha >}\left( U_t\right) dt,
$$
$$
U_0=u.
$$
Then for every fixed $T>0$ we have
$$
U\left( T-t,u,w\right) =\widehat{U}\left( t,U\left( T,u,w\right) ,\widehat{w}%
\right)
$$
for every 0$\leq t\leq T$ and $u$ $P^W$-a.s., where the Wiener process $%
\widehat{w}$ is defined as $\widehat{w}\left( t\right) =w\left(
T-t\right) -w\left( T\right) ,0\leq t\leq T.$

Now we can extend the results on flows of diffeomorphisms of
stochastic processes to v-bundles. The solution $U\left(
t,u,w\right) $ of the equation (10.25) can be considered as the
set of maps $U_t:u\rightarrow U\left(
t,u,w\right) $ from ${\cal E}^{<z>}$ to $\hat E^{<z>}={\cal E}^{<z>}{\cal %
\cup \{\bigtriangleup \}.}$

\begin{theorem}
A process $|U|\left( t,u,w\right) $\ has such a modification, for
simplicity
let denote it also as $U\left( t,u,w\right) ,$ that the map $%
U_t(w):u\rightarrow U\left( t,u,w\right) $ belongs to the class
$C^\infty $ for every $f\in F_0\left( {\cal E}^{<z>}\right) $ and
all fixed $t\in [0,\infty )$ a.s. In addition, for every $u\in U$
and $t\in [0,\infty )$ the differential of map $u\rightarrow
U\left( t,u,w\right) ,$

$$
U\left( t,u,w\right) _{*}:T_{u\,}\left( U\left( t,u,w\right)
\right) \rightarrow T_{U\left( t,u,w\right) }\left( {\cal
E}^{<z>}\right) ,
$$
is an isomorphism, a.s., in the set $\{w:U\left( t,u,w\right) \in {\cal E}%
^{<z>}{\cal \ \}.}$
\end{theorem}

{\it Proof.}{\rm \ }Let $u_0\in ${\it \ }${\cal E}^{<z>}$ and fix
$t\in
[0,\infty )$ . We can find a sequence of coordinate carts $%
U_1,U_2,...,U_p\subset {\cal E}^{<z>}$ that for almost all $w$
that $U\left( t,u_0\subset w\right) \in {\cal E}^{<z>}$ there is
an integer $p>0$ that $\{U\left( s,u_0,w\right) :s\in [\left(
k-1\right) t/p,kt/p]\}\subset {\cal U_k\ ,}\left(
k=1,2,...p\right) .$ According to the theorem 10.2 we can
conclude that for every coordinate cart ${\cal U\ }$ and
$\{U\left( s,u_0,w\right) ;s\in [0,1]\}\subset {\cal U\ }$ a map
$v\rightarrow U\left( t_0,v,w\right) $ is a diffeomorphism in the
neighborhood of $v_{0.}$ The proof of the theorem follows from
the relation $U\left( t,w_0,w\right) =[U_{t/p}\left( \theta
_{\left( p-1\right) t/p}w\right) \circ ...\circ U_{t/p}\left(
\theta _{t/p}w\right) \circ U_{t/p}]\left( u_0\right) ,$ where
$\theta _t:W_0^r\rightarrow W_0^r$ is defined as $\left( \theta
_tw\right) \left( s\right) =w\left( t+s\right) -w\left( t\right)
.\Box $

Let $A_0,A_1,...,A_r\in {\bf X}({\bf {\cal E}}^{<z>})$ and
$U_t=\left(
U\left( t,u,w\right) \right) $ is a flow of diffeomorphisms on ${\bf {\cal E}%
}^{<z>}$ . Then $\widetilde{A}_0,\widetilde{A}_1,...,\widetilde{A}_r\in {\bf %
X}\left( GL\left( {\bf {\cal E}}^{<z>}\right) \right) $ define a
flow of diffeomorphisms $r_t=\left( r\left( t,r,w\right) \right)
$ on $GL\left( {\bf
{\cal E}}^{<z>}\right) $ with $\left( r\left( t,r,w\right) \right) =$\\ $%
\left( U\left( t,u,w\right) ,e\left( t,u,w\right) \right) ,$
where $r=\left( u,e\right) $ and $e\left( t,r,u\right) =U\left(
t,u,w\right) _{*}e$ is the differen\-tial of the map
$u\rightarrow U\left( t,u,w\right) $ sat\-is\-fying the prop\-erty
$$
U\left( t,u,w\right) _{*}e=[U\left( t,u,w\right) _{*}e_0,U\left(
t,u,w\right) _{*}e_1,...,U\left( t,u,w\right) _{*}e_{q-1}].
$$
In lo\-cal co\-or\-di\-nates
$$
A_{<\widehat{\alpha }>}\left( u\right) =\sigma _{<\widehat{\alpha }%
>}^{<\alpha >}\delta _{<\alpha >},\left( <\alpha >=1,2,...,r\right) ,
$$
$$
A_0\left( u\right) =b^{<\alpha >}\left( u\right) \delta _{<\alpha >},e_{<%
\underline{\beta }>}^{<\alpha >}\left( t,u,w\right) =Y_{<\underline{\gamma }%
>}^{<\alpha >}\left( t,u,w\right) e_{<\underline{\beta }>}^{<\underline{%
\gamma }>},
$$
where $Y_{<\underline{\gamma }>}^{<\alpha >}\left( t,u,w\right) $
is defined
from (10.36). So we can construct flows of diffeomorphisms of the bundle $%
{\bf {\cal E}}^{<z>}$ .

\section{ Nondegenerate Dif\-fu\-si\-on in La--Spa\-ces}

Let a dv-bundle ${\bf {\cal E}}^{<z>}{\bf {\cal \ }}$ be provided
with a \index{Nondegenerate diffusion!in la--spaces} positively
defined metric of type (6.12) being compatible with a d-
connection $D=\{\Gamma _{<\beta ><\gamma >}^{<\alpha >}\}.$ The connection $%
D $ allows us to roll ${\bf {\cal E}}^{<z>}{\bf {\cal \ }}$ along a curve $%
\gamma \left( t\right) $ $\subset {\cal R}^{n_E}$ in order to
draw the curve $c\left( t\right) $ on ${\bf {\cal E}}^{<z>}$ as
the trace of $\gamma \left( t\right) .\,$ More exactly, let
$\gamma :[0,\infty )\ni t\rightarrow \gamma
\left( t\right) \subset {\cal R}^{n_E}\ $ be a smooth curve in ${\cal R}%
^{n_E},\ r=\left( u,e\right) \in O\left( {\bf {\cal
E}}^{<z>}\right) .$ We define a curve $\widetilde{c}\left(
t\right) =\left( c\left( t\right) ,e\left( t\right) \right) $ in
$O\left( {\bf {\cal E}}^{<z>}\right) $ by
using the equalities%
$$
\frac{dc^{<\alpha >}\left( t\right) }{dt}=e_{<\underline{\alpha
}>}^{<\alpha
>}\left( t\right) \frac{d\gamma ^{<\underline{\alpha }>}}{dt},\eqno(10.37)
$$
$$
\frac{de_{<\underline{\alpha }>}^{<\alpha >}\left( t\right)
}{dt}=-\Gamma
_{<\beta ><\gamma >}^{<\alpha >}\left( c\left( t\right) \right) e_{<%
\underline{\alpha }>}^{<\gamma >}\left( t\right) \frac{dc^{<\beta
>}}{dt},
$$
$$
c^{<\alpha >}\left( 0\right) =u^{<\alpha >},e_{<\underline{\alpha }%
>}^{<\alpha >}\left( 0\right) =e_{<\underline{\alpha }>}^{<\alpha >}.
$$
Equations (10.37) can be written as
$$
\frac{d\widetilde{c}\left( t\right) }{dt}=\widetilde{L}_{<\alpha
>}\left( \widetilde{c}\left( t\right) \right) d\gamma ^{<\alpha
>},
$$
$$
\widetilde{c}\left( 0\right) =r,
$$
where$\{\widetilde{L}_{<\alpha >}\}$ is the system of canonical
horizontal
vector fields (see (10.29)). Curve $c\left( t\right) =\pi \left( \widetilde{c%
}\left( t\right) \right) $ on ${\cal E}^{<z>}{\cal \ }$ depends
on fixing of the initial frame $p$ in a point $u;$ this curve is
parametrized as $c\left( t\right) =c\left( t,r,\gamma \right)
,r=r\left( u,e\right) .$

Let $w\left( t\right) =\left( w^{\underline{\alpha }}\left(
t\right) \right) $ is the canonical realization of a
n+m-dimensional Wiener process.\ We can define the random curve
$U\left( t\right) \subset $ ${\cal E}^{<z>}$ in a similar manner.
Consider $r\left( t\right) =\left( r\left( t,r,w\right) \right) $
as the solution of stochastic differential equations
$$
dr\left( t\right) =\widetilde{L}_{<\underline{\alpha }>}\left(
r\left(
t\right) \right) \circ \delta w^{<\underline{\alpha }>}\left( t\right) ,%
\eqno(10.38)
$$
$$
r\left( 0\right) =r,
$$
where $r\left( t,r,w\right) $ is the flow of diffeomorphisms on
$O\left( {\cal E}^{<z>}{\cal \ }\right) $ corresponding to the
canonical horizontal vector fields
$\widetilde{L}_1,\widetilde{L}_2,...,\widetilde{L}_{q-1}$ and
vanishing drift field $\widetilde{L}_0=0.$ In local coordinates
the equations (10.38) are written as
$$
dU^{<\alpha >}\left( t\right) =e_{<\underline{\alpha }>}^{<\alpha
>}\left( t\right) \circ \delta w^{<\underline{\alpha }>}\left(
t\right) ,
$$
$$
de_{<\underline{\alpha }>}^{<\alpha >}\left( t\right) =-\Gamma
_{<\beta
><\gamma >}^{<\alpha >}\left( U\left( t\right) \right) e_{<\underline{\alpha
}>}^{<\gamma >}\circ \delta u^{<\beta >},
$$
where $r\left( t\right) =\left( U^{<\alpha >}\left( t\right) ,e_{<\underline{%
\alpha }>}^{<\alpha >}\left( t\right) \right) .$ It is obvious
that $$r\left(
t\right) =\left( U^{<\alpha >}\left( t\right) ,e_{<\underline{\alpha }%
>}^{<\alpha >}\left( t\right) \right) \in O\left( {\cal E}^{<z>}
\right) $$ if $r\left( 0\right) \in O\left( {\cal E}^{<z>}\right)
$
because $\widetilde{L}_{<\underline{\alpha }>}$ are vector fields on $%
O\left( {\cal E}^{<z>}{\cal \ }\right) .$ The random curve
$\{U^{<\alpha
>}\left( t\right) \}$ on ${\cal E}^{<z>}$ is defined as $U\left(
t\right) =\pi \left[ r\left( t\right) \right] .$ We point out that $%
aw=\left( aw\left( t\right) \right) $ is another
$n_E$-dimensional Wiener process and as a consequence the
probability law $U\left( \cdot ,r,w\right) $ does not depend on
$a\in O\left( n_E\right) .$ It depends only on $u=\pi \left(
r\right) .$ This law is denoted as $P_w$ and should be mentioned
that it is a Markov process because a similar property has
$r\left( \cdot ,r,w\right) .$

{\bf Remark 10.1.\thinspace }{\it \ We can define }$r\left(
t,r,w\right) $ {\it as a flow of diffeomorphisms on} $GL\left(
{\cal E}^{<z>}\right) $
{\it for every d-connection on } ${\cal E}^{<z>}.$ {\it In this case $%
\pi \left[ r\left( \cdot ,r,w\right) \right] $} {\it does not
depend only on }$u=\pi \left( t\right) $ {\it and in consequence
we do not obtain a Markov process by projecting on } ${\cal
E}^{<z>}.$ {\it The Markov property of diffusion processes on
}${\cal E}^{<z>}$ {\it is assumed by the conditions of
compatibility of metric and d--connection (10.13)
 (or linear connection (10.15)) and of vanishing
of torsion.}

Now let us show that a diffusion $\{P_u\}$ on ${\cal E}^{<z>}$
can be considered as an A-diffusion process with the differential
operator
$$
A=\frac 12\Delta _{{\cal E\ }}+b,\eqno(10.39)
$$
where $\Delta _{{\cal E\ }}$ is the Laplace--Beltrami operator on ${\cal E}%
^{<z>},$ \index{Laplace--Beltrami!operator on ${\cal E}^{<z>}$}
$$
\Delta _{{\cal E\ }}f=G^{<\alpha ><\beta >}\overrightarrow{D}_{<\alpha >}%
\overrightarrow{D}_{<\beta >}f=\eqno(10.40)
$$
$$
G^{<\alpha ><\beta >}\frac{\delta ^2f}{\delta u^{<\alpha >}\delta
u^{<\beta
>}}-\{\frac{<\alpha >}{<\gamma ><\beta >}\}\frac{\delta f}{\delta u^{<\alpha
>}},
$$
where operator $\overrightarrow{D}_\alpha $ is constructed by
using Christoffel d--symbols (10.15) and $b$ is the vector
d-field with components
$$
b^{<\alpha >}=\frac 12G^{<\beta ><\gamma >}\left( \{\frac{<\alpha
>}{<\beta
><\gamma >}\}-\Gamma _{<\beta ><\gamma >}^{<\alpha >}\right) \eqno(10.41)
$$

\begin{theorem}
The solution of stochastic differential equation (10.38) on\\ $O\left( {\cal %
E}^{<z>}\right) $ defines a flow of diffeomorphisms $r\left(
t\right) =\left( r\left( t,r,w\right) \right) $ on $O\left( {\cal
E}^{<z>}\right) $ and its projection $U\left( t\right) =\pi
\left( r\left( t\right) \right) $ defines a diffusion process on
${\cal E}^{<z>}$ corresponding to the differential operator
(10.39).
\end{theorem}

{\it Proof.}{\rm \ } Considering $f\left( r\right) \equiv f\left(
u\right) $
for $r=\left( u,e\right) $ we obtain%
$$
f\left( U(t)\right) -f\left( U\left( 0\right) \right) =f\left(
r\left( t\right) \right) -f\left( r\left( 0\right) \right) =
$$
$$
\int_0^r\left( \widetilde{L}_{<\underline{\alpha }>}f\right)
\left( r\left( s\right) \right) \circ \delta
w^{<\underline{\alpha }>}=
$$
$$
\int_0^t\widetilde{L}_{<\underline{\alpha }>}f\left( r\left(
s\right)
\right) \delta w^{<\underline{\alpha }>}+\frac 12\int_0^t\sum\limits_{%
\underline{\alpha }=0}^{q-1}\widetilde{L}_{<\underline{\alpha
}>}\left( \widetilde{L}_{<\underline{\alpha }>}f\right) \left(
r\left( s\right) \right) ds.
$$
Let us show that $\frac 12\sum\limits_{<\underline{\alpha }>=0}^{q-1}%
\widetilde{L}_{<\underline{\alpha }>}\left( \widetilde{L}_{<\underline{%
\alpha }>}f\right) =Af.$ Really, because the operator (10.39) can
be written as
$$
A=\frac 12G^{<\alpha ><\beta >}\overrightarrow{D}_{<\alpha >}\overrightarrow{%
D}_{<\beta >}=
$$
$$
\frac 12(G^{<\alpha ><\beta >}\frac{\delta ^2}{\delta u^{<\alpha
>}\delta u^{<\beta >}}-\{\frac{<\alpha >}{<\gamma ><\beta
>}\}\frac \delta {\delta u^{<\alpha >}})
$$
and taking into account (10.32) we have
$$
\widetilde{L}_{<\underline{\alpha }>}\left( \widetilde{L}_{<\underline{%
\alpha }>}f\right) =\widetilde{L}_{<\underline{\alpha }>}\left(
F_{\nabla f}\right) _{<\underline{\alpha }>}=
$$
$$
\left( F_{\nabla \nabla f}\right) _{<\underline{\alpha }><\underline{\alpha }%
>}=\left( \nabla _{<\gamma >}\nabla _{<\delta >}f\right) e_{<\underline{%
\alpha }>}^{<\gamma >}e_{<\underline{\alpha }>}^{<\delta >}.
$$
Now we can write
$$
\sum\limits_{<\underline{\alpha
}>=0}^{q-1}\widetilde{L}_{<\underline{\alpha
}>}\left( \widetilde{L}_{<\underline{\alpha >}}f\right) =%
$$
$$
\sum\limits_{<\underline{\alpha }>=0}^{q-1}(\overrightarrow{D}_{<\alpha >}%
\overrightarrow{D}_{<\beta >}f\ )e_{<\underline{\alpha }>}^{<\alpha >}e_{<%
\underline{\beta }>}^{<\beta >}=G^{<\alpha ><\beta >}\overrightarrow{D}%
_{<\alpha >}\overrightarrow{D}_{<\beta >}f\
$$
(see (10.33)), which complete our proof. $\Box $

\begin{definition}
The process $r\left( t\right) =\left( r\left( t,r,w\right)
\right) $ from the theorem 10.5 is called the horizontal lift of
the A--diffusion $U\left( t\right) $ {\it on }${\cal E}^{<z>}.$
\end{definition}

\begin{proposition}
For every d-vector field $b=b^{<\alpha >}\left( u\right) \delta
_{<\alpha >}$ on ${\cal E}^{<z>}{\cal \ }$ provided with the
canonical d--connection structure there is a d-connection
$D=\{\Gamma _{<\beta ><\gamma >}^{<\alpha
>}\}$ on ${\cal E}^{<z>}{\cal \ ,}$ compatible with d--metric $G_{<\alpha
><\beta >},$ which satisfies the equality (10.41).
\end{proposition}

{\it Proof. }Let define
$$
\Gamma _{<\beta ><\gamma >}^{<\alpha >}=\{\frac{<\alpha >}{<\beta ><\gamma >}%
\}+\frac 2{q-1}\left( \delta _{<\beta >}^{<\alpha >}b_{<\gamma
>}-G_{<\beta
><\gamma >}b^{<\alpha >}\right) ,\eqno(10.42)
$$
where $b_{<\alpha >}=G_{<\alpha ><\beta >}b^{<\beta >}.$ By
straightforward calculations we can verify that d-connec\-ti\-on
(10.42) satisfies the
metricity conditions%
$$
\delta _{<\gamma >}G_{<\alpha ><\beta >}-G_{<\tau ><\beta
>}\Gamma _{<\gamma
><\alpha >}^{<\tau >}-G_{<\alpha ><\tau >}\Gamma _{<\gamma ><\beta >}^{<\tau
>}=0
$$
and that
$$
\frac 12G^{<\alpha ><\beta >}\left( \{\frac{<\gamma >}{<\alpha ><\beta >}%
\}-\Gamma _{<\alpha ><\beta >}^{<\gamma >}\right) =b^{<\gamma >}.
$$
$\Box $

We note that a similar proposition is proved in 
 [117] for,
respectively, metric and affine connections on Riemannian and
affine connected manifolds: M. Anastasiei proposed
 [11] to define
Laplace-Beltrami operator (10.40) by using the canonical
d--connection (6.21) in generalized Lagrange spaces. Taking into
account (10.16) and (10.17) and a corresponding redefinition of
components of d-vector fields
(10.41), because of the existence of multiconnection structure on the space $%
{\cal E}^{<z>}$ , we conclude that we can equivalently formulate
the theory of d--diffusion on ${\cal E}^{<z>}$--space by using
both variants of Christoffel d--symbols and canonical
d--connection.

\begin{definition}
For $A=\frac 12\Delta _{{\cal E\ }}$ an A-diffusion $U\left(
t\right) $ is called a Riemannian motion on ${\cal E}^{<z>}{\cal
.}$
\end{definition}

Let an A-differential operator on ${\cal E}^{<z>}$ is expressed
locally as
$$
Af\left( u\right) =\frac 12a^{<\alpha ><\beta >}\left( u\right)
\frac{\delta ^2f}{\delta u^{<\alpha >}\delta u^{<\beta >}}\left(
u\right) +b^{<\alpha
>}\left( u\right) \frac{\delta f}{\delta u^{<\alpha >}}\left( u\right) ,
$$
where $f\in F\left( {\cal E}^{<z>}\right) ,$ matrix $a^{<\alpha
><\beta >}$ is symmetric and nonegatively defined . If
$a^{<\alpha ><\beta >}\left( u\right) $ $\xi _{<\alpha >}\xi
_{<\beta >}>0$ for all $u$ and $\xi =\left(
\xi _{<\alpha >}\right) \in {\cal R}^q\backslash \{0\},$ than the operator $%
A $ is nondegenerate and the corresponding diffusion is called
nondegenerate.

By using a vector d-field $b_{<\alpha >}$ we can define the 1-form%
$$
\omega _{\left( b\right) }=b_{<\alpha >}\left( u\right) \delta
u^{<\alpha
>},
$$
where $b=b^{<\alpha >}\delta _{<\alpha >}$ and $b_{<\alpha
>}=G_{<\alpha
><\beta >}b^{<\beta >}$ in local coordinates. According the de Rham-Codaira
theorem 
 [201] we can write%
$$
\omega _{\left( b\right) }=dF+\widehat{\delta }\beta +\alpha
\eqno(10.43)
$$
where $F\in F\left( {\cal E}^{<z>}\right) ,\beta $ is a 2-form
and $\alpha $
is a harmonic 1-form. The scalar product of p-forms $\Lambda _p\left( {\cal E%
}^{<z>}{\cal \ }\right) $ on ${\cal E}^{<z>}$ is introduced as%
$$
(\alpha ,\beta )_B=\int_{{\cal E}^{<z>}}<\alpha ,\beta >\delta u,
$$
where%
$$
\alpha =\sum\limits_{<\gamma _1><<\gamma _2><...<<\gamma
_p>}\alpha _{<\gamma _1><\gamma _2>...<\gamma _p>}
$$
$$
\delta u^{<\gamma _1>}\bigwedge \delta u^{<\gamma _2>}\bigwedge
...\bigwedge \delta u^{<\gamma _p>},
$$
$$
\beta =\sum\limits_{<\gamma _1><<\gamma _2><...<<\gamma _p>}\beta
_{<\gamma _1><\gamma _2>...<\gamma _p>}
$$
$$
\delta u^{<\gamma _1>}\bigwedge \delta u^{<\gamma _2>}\bigwedge
...\bigwedge \delta u^{<\gamma _p>},
$$
$$
\beta ^{<\gamma _1><\gamma _2>...<\gamma _p>}=G^{<\gamma _1><\tau
_1>}G^{<\gamma _2><\tau _2>}...G^{<\gamma _p><\tau _p>}\beta
_{<\tau _1><\tau _2>...<\tau _p>},
$$
$$
<\alpha ,\beta >=\sum\limits_{<\gamma _1><<\gamma _2><...<<\gamma
_p>}\alpha _{<\gamma _1><\gamma _2>...<\gamma _p>}\left( u\right)
\beta ^{<\gamma _1><\gamma _2>...<\gamma _p>}\left( u\right) ,
$$
$$
\delta u=\sqrt{|\det G_{<\alpha ><\beta >}|}\delta u^0\delta
u^1...\delta u^{q-1}.
$$
The operator $\widehat{\delta }:\Lambda _p\left( {\cal
E}^{<z>}\right) \rightarrow \Lambda _{p-1}\left( {\cal
E}^{<z>}\right) $ from (10.43) is
defined by the equality%
$$
\left( d\alpha ,\beta \right) _p=\left( \alpha ,\widehat{\delta
}\beta \right) _{p-1},\alpha \in \Lambda _{p-1}\left( {\cal
E}^{<z>}\right) ,\beta \in \Lambda _p\left( {\cal E}^{<z>}\right)
.
$$
De Rham--Codaira Laplacian $\Box :\Lambda _p\left( {\cal
E_N}\right)
\rightarrow \Lambda _p\left( {\cal E_N}\right) $ is defined by the equality%
\index{De Rham--Codaira Laplacian}
$$
\Box =-\left( d\widehat{\delta }+\widehat{\delta }d\right)
.\eqno(10.44)
$$

A form $\alpha \in \Lambda _p\left( {\cal E}^{<z>}\right) $ is
called as harmonic if $\Box \alpha =0$ . It is known that $\Box
\alpha =0$ if and only
if $d\alpha =0$ and $\widehat{\delta }\alpha =0.$For $f\in F\left( {\cal E}%
^{<z>}\right) $ and $U\in {\bf X}\left( {\cal E}^{<z>}\right) $
we can
define the operators $gradf\in {\bf X}$ $\left( {\cal E}^{<z>}\right) $ and $%
divU\in F\left( {\cal E}^{<z>}\right) $ by using correspondingly
the
equalities%
$$
gradf=G^{<\alpha ><\beta >}\delta _{<\alpha >}\delta _{<\beta >}f
$$
and
$$
divU=-\widehat{\delta }\omega _U=\frac 1{\sqrt{|\det G|}}\delta
_{<\alpha
>}\left( U^{<\alpha >}\sqrt{|\det G|}\right) .
$$

The Laplace-Beltrami operator (10.39) can be also written as
$$
\bigtriangleup _{{\cal E\ }}f=div(gradf)=-\widehat{\delta }\widehat{\delta }f%
\eqno(10.45)
$$
for $F\left( M\right) .$

Let suggest that ${\cal E}^{<z>}$ is compact and oriented and
$\{P_u\}$ be
the system of diffusion measures defined by a A--operator (10.39). Because $%
{\cal E}^{<z>}$ is compact $P_u$ is the probability measure on the set $%
\widehat{W}\left( {\cal E}^{<z>}\right) =W\left( {\cal
E}^{<z>}\right) $ of all continuous paths in ${\cal E}^{<z>}$ .

\begin{definition}
The transition semigroup $T_t$ of A--diffusion is defined by the equality%
$$
\left( T_tf\right) \left( u\right) =\int\limits_{W({\cal E}^{<z>}{\cal )\ }%
}f\left( w\left( t\right) \right) P_u\left( dw\right) ,f\in C\left( {\cal E}%
^{<z>}\right) .
$$
\end{definition}

For a connected open region $\Omega \subset {\cal E}^{<z>}{\cal \
}$ we
define $\rho ^\Omega w\in \widehat{W}\left( \Omega \right) ,w\in \widehat{W}%
\left( {\cal E}^{<z>}\right) $ by the equality%
$$
(\rho ^\Omega w)\left( t\right) =\langle _{\Delta ,{ if\ }t\geq
\tau _\Omega \left( w\right) ,}^{w(t),{ if\ }t<\tau _\Omega
\left( w\right) ,}
$$
where $\tau _\Omega \left( w\right) =\inf \{t:w\left( t\right)
\notin \Omega
\}.$ We denote the image-measure $P_u\left( u\in \Omega \right) $ on map $%
\rho ^\Omega $ as $P_u^\Omega $ ; this way we define a
probability measure on $\widehat{W}\left( \Omega \right) $ which
will be called as the minimal A-diffusion on $\Omega .$ The
transition group of this diffusion is introduced as
$$
\left( T_t^\Omega f\right) \left( u\right) =\int\limits_{W\left(
\Omega \right) }f\left( w\left( t\right) \right) P_u^\Omega
\left( dw\right) =
$$
$$
\int\limits_{W\left( {\cal E}^{<z>}{\cal \ }\right) }f\left(
w\left( t\right) \right) I_{\{\tau _\Omega \left( w\right)
>t\}}P_u\left( dw\right) ,f\in C_p\left( \Omega \right) .
$$

\begin{definition}
The Borel measure $\mu \left( du\right) $ on ${\cal E}^{<z>}$ is
called an \index{Borel measure} invariant measure on A-diffusion
$\{P_u\}$ if
$$
\int\limits_{{\cal E}^{<z>}}T_tf\left( u\right) \mu \left(
du\right) =\int\limits_{{\cal E}^{<z>}}f\left( u\right) \mu
\left( du\right)
$$
for all $f\in C\left( {\cal E}^{<z>}\right) .$
\end{definition}

\begin{definition}
An A--diffusion $\{P_u\}$ is called symmetrizable (locally \\
symmetrizable) if there is a Borel measure
$$\nu \left( du\right) \mbox{ on } {\cal E}^{<z>} (\nu
^\Omega \left( du\right) ) \mbox{ on } (\Omega )$$ that
$$
\int\limits_{{\cal E}^{<z>}}T_tf\left( u\right) g\left( u\right)
\nu \left( du\right) =\int\limits_{{\cal E}^{<z>}}f\left(
u\right) T_tg\left( u\right) \nu \left( du\right)
$$
for all $f,g\in C\left( {\cal E}^{<z>}\right) $ and
$$
(\int\limits_\Omega T_t^\Omega f\left( u\right) g\left( u\right)
\nu ^\Omega \left( du\right) =\int\limits_\Omega f\left( u\right)
T_t^\Omega g\left( u\right) \nu ^\Omega \left( du\right)
$$
for all $f,g\in C\left( \Omega \right) ).$
\end{definition}

The fundamental properties of A-diffusion measures are satisfied
by the following theorem and corollary:

\begin{theorem}
a) An A-diffusion is symmetrizable if and only if
$\widehat{\delta }\beta =\alpha =0$ (see (10.44)); this condition
is equivalent to the condition that $b=gradF,F\in F\left( {\cal
E}^{<z>}\right) $ and in this case the
invariant measures are of type $C\exp [2F\left( u\right) ]du,$ where $%
C=const.$

b) An A-diffusion is locally symmetrizable if and only if $\widehat{\delta }%
\beta =0$ (see (10.19)) or, equivalently, $dw_{\left( b\right)
}=0.$

c) A measure $cdu$ (constant $c>0)$ is an invariant measure of an
A-diffusion if and only if $dF=0$ (see (10.44)) or, equivalently, $\widehat{%
\delta }w_{\left( b\right) }=-divb=0.$
\end{theorem}

\begin{corolarry}
An A--diffusion is symmetric with respect to a Riemannian volume
$du$ (i.e. is symmetrizable and the measure $\nu $ in (10.45)
coincides with $du)$ if and only if it is a Brownian motion on
${\cal E}^{<z>}.$
\end{corolarry}
\index{A--diffusion}

We omit the proofs of the theorem 10.6 and corollary 10.1 because
they are
similar to those presented in 
 [117] for Riemannian manifolds. In our
case we have to change differential forms and measures on
Riemannian spaces into similar objects on ${\cal E}^{<z>}$

\section{ Heat Equations for D--Tensor Fields}

To generalize the results presented in section 10.5 to the case
of d-tensor \index{Heat equations!for d--tensor fields} fields in
${\cal E}^{<z>}$ we use the It\^o idea of stochastic
parallel transport 
 [201,125] \\ (correspond\-ing\-ly adapted to transports in
vector bundles provided with N-connection structure).

\subsection{ Scalarized tensor d-fields and heat equations}

Consider a compact bundle ${\cal E}^{<z>}$ and the bundle of
orthonormalized adapted frames on ${\cal E}^{<z>}$ denoted as
$O\left( {\cal E}^{<z>}\right) .$ Let
$\{\widetilde{L}_0,\widetilde{L}_1,...,\widetilde{L}_{q-1}\}$ be
the
system of canonical horizontal vector fields on $O\left( {\cal E}%
^{<z>}\right) $ (with respect to canonical d--connection $\overrightarrow{%
\Gamma }_{<\beta ><\gamma >}^{<\alpha >}.$ The flow of diffeomorphisms $%
r\left( t\right) =r\left( t,r,w\right) $ on $O\left( {\cal
E}^{<z>}\right) $ is defined through the solution of equations
$$
dr\left( t\right) =\widetilde{L}_{<\underline{\alpha }>}\left(
r\left( t\right) \right) \circ \delta w^{<\underline{\alpha
}>}\left( t\right) ,
$$
$$
r\left( 0\right) =r,
$$
and this flow defines a diffusion process, the horizontal
Brownian motion on $O\left( {\cal E}^{<z>}\right) ,$ which
corresponds to the differential operator
$$
\frac 12\Delta _{O\left( {\cal E}^{<z>}\right) }=\frac 12\sum\limits_{<%
\underline{\alpha }>}\widetilde{L}_{<\underline{\alpha }>}\left( \widetilde{L%
}_{<\underline{\alpha }>}\right) .\eqno(10.46)
$$
For a tensor d-field $S\left( u\right) =S_{<\beta _1><\beta
_2>...<\beta _q>}^{<\alpha _1><\alpha _2>...<\alpha _p>}\left(
u\right) $ we can define
its scalarization $F_S\left( r\right) =F_{S<\underline{\beta }_1><\underline{%
\beta }_2>...<\underline{\beta }_q>}^{<\underline{\alpha }_1><\underline{%
\alpha }_2>...<\underline{\alpha }_p>}$ (a system of smooth functions on $%
O\left( {\cal E}^{<z>}\right) $) similarly as we have done in
section 10.3, but in our case by using frames satisfying
conditions (10.22) in order to deal with bundle $O\left( {\cal
E}^{<z>}\right) .$

The action of Laplace-Beltrami operator on d-tensor fields is defined as%
$$
\left( \Delta T\right) _{<\beta _1><\beta _2>...<\beta
_q>}^{<\alpha _1><\alpha _2>...<\alpha _p>}=
$$
$$
G^{<\alpha ><\beta >}\left( \overrightarrow{D}_{<\alpha >}\left(
\overrightarrow{D}_{<\beta >}T\right) \right) _{<\beta _1><\beta
_2>...<\beta _q>}^{<\alpha _1><\alpha _2>...<\alpha _p>}=
$$
$$
G^{<\alpha ><\beta >}T_{<\beta _1><\beta _2>...<\beta _q>;<\alpha
><\beta
>}^{<\alpha _1><\alpha _2>...<\alpha _p>},
$$
where $\overrightarrow{D}T$ is the covariant derivation with respect to $%
\overrightarrow{\Gamma }_{<\beta ><\gamma >}^{<\alpha >}.$ We can
calculate (by putting formula (10.21) into (10.46) that
$$
\Delta _{O\left( {\cal E}^{<z>}\right) }(F_{S<\beta _1><\beta
_2>...<\beta _q>}^{<\alpha _1><\alpha _2>...<\alpha _p>})=\left(
F_{\Delta S}\right) _{<\beta _1><\beta _2>...<\beta _q>}^{<\alpha
_1><\alpha _2>...<\alpha _p>}.
$$

For a given d-tensor field $S=S\left( u\right) $ let be defined
this system of functions on $[0,\infty )\times O\left( {\cal
E}^{<z>}\right) $:

$$
V_{<\beta _1><\beta _2>...<\beta _q>}^{<\alpha _1><\alpha
_2>...<\alpha _p>}\left( t,r\right) =E\left[ F_{S<\beta _1><\beta
_2>...<\beta _q>}^{<\alpha _1><\alpha _2>...<\alpha _p>}\left(
r\left( t,r,w\right) \right) \right] .
$$

According to the theorem 10.5 $V_{<\beta _1><\beta _2>...<\beta
_q>}^{<\alpha _1><\alpha _2>...<\alpha _p>}$ is a unique solution
of heat equation
$$
\frac{\partial V}{\partial t}=\frac 12\Delta _{O\left( {\cal
E}^{<z>}\right) }V,\eqno(10.47)
$$
$$
V_{|t=0}=F_{S<\beta _1><\beta _2>...<\beta _q>}^{<\alpha
_1><\alpha _2>...<\alpha _p>}.
$$

In a similar manner we can construct unique solutions of heat
equations
(10.47) for the case when instead of differential forms one considers ${\cal %
R}^{n_E}$-tensors (see 
 [117] for details concerning Riemannian
manifolds).\ We have to take into account the torsion components
of the canonical d-connection on ${\cal E}^{<z>}$ .

\subsection{ Boundary conditions}

We analyze the heat equations for differential forms on a bounded space $%
{\cal E}^{<z>}{\cal :}$

$$
\frac{\partial \alpha }{\partial t}=\frac 12\Box \alpha
,\eqno(10.48)
$$
$$
\alpha _{|t=0}=f,
$$
$$
\alpha _{norm}=0,(d\alpha )_{norm}=0,\eqno(10.49)
$$
where $\Box $ is the de Rham-Codaira Laplacian (10.44),%
$$
\alpha _{norm}=\theta _{q-1}\left( u\right) du^{q-1},
$$
$$
\left( d\alpha \right) _{norm}=\sum\limits_{\gamma =1}^q\left(
\frac{\delta
\alpha }{\delta u^{q-1}}-\frac{\delta \alpha _{q-1}}{\delta u^{<\gamma >}}%
\right) \delta u^{q-1}\bigwedge \delta u^{<\gamma >}.
$$
We consider the boundary of ${\cal E}^{<z>}$to be a manifold of
dimension $q=n_E$ and denote by $\overbrace{{\cal E_N}}$ the
interior part
of ${\cal E}^{<z>}$ and as $\partial {\cal E_N}$ the boundary of ${\cal E}%
^{<z>}.$ In the vicinity $\overbrace{{\cal U}}$ of the boundary we
introduce the system of local coordinates $u=\{\left( u^\alpha
\right) ,u^{q-1}\geq 0\}$ for every $u\in {\cal U\ }$and $u\in
{\cal U\ }\cap
\partial {\cal E_N}$ if and only if $u^{q-1}=0.$

The scalarization of 1-form $\alpha $ is defined as%
$$
\left[ F_{<\alpha >}\right] _{<\underline{\beta }>}\left(
r\right) =\theta _{<\beta >}\left( u\right) e_{<\underline{\beta
}>}^{<\beta >},r=\left(
u^{<\beta >},e_{<\underline{\beta }>}^{<\beta >}\right) \in {\cal O(E}^{<z>})%
{\cal .}
$$
Conditions (10.49) are satisfied if and only if
$$
e_{q-1}^{<\alpha >}\left[ F_{<\alpha >}\right] _{<\underline{\alpha }%
>}\left( r\right) =0
$$
and
$$
e_{<\beta >}^{<\underline{\beta }>}\frac \delta {\delta
u^{q-1}}\left[ F_{<\alpha >}\right] _{<\underline{\beta }>}\left(
r\right) =0,
$$
\underline{$\alpha $}=0,1,2,...,q-1, where $e_{<\beta
>}^{<\underline{\alpha }>}$ is inverse to $e_{<\underline{\beta
}>.}^{<\alpha >}.$

Now we can formulate the Cauchy problem for differential 1-forms
\\ (10.48) and (10.49) as a corresponding problem for ${\cal
R}^{n_E}$--valued equivariant functions $V_{<\underline{\alpha
}>}\left( t,r\right) $ on ${\cal O(E}^{<z>}:$
$$
\frac{\partial V_{<\underline{\alpha }>}}{\partial r}\left(
t,r\right)
=\frac 12\{\Delta _{{\cal O(E}^{<z>})}V_{<\underline{\alpha }%
>}\left( t,r\right) +R_{<\underline{\alpha }>}^{<\underline{\beta }>}\left(
r\right) V_{<\underline{\beta }>}\left( t,r\right) ,\eqno(10.50)
$$
$$
V_{<\underline{\alpha }>}\left( 0,r\right) =\left( F_f\right) _{<\underline{%
\alpha }>}\left( r\right) ,
$$
$$
(<\beta >=0,1,...,q-2),(<\underline{\alpha }>,<\underline{\beta >}%
=0,1,...,q-1),
$$
$$
e_{<\beta >}^{<\underline{\beta }>}\frac \delta {\delta u^{q-1}}V_{<%
\underline{\beta >}}\left( t,r\right) _{|\partial {\cal O(E}_N)}=0,f_{q-1}^{%
\underline{\beta }}V_{\underline{\beta }}\left( t,r\right) _{|\partial {\cal %
O\left( E_N\right) \ }}=0,
$$
where $R_{<\underline{\alpha }>}^{<\underline{\beta }>}\left(
r\right) $ is the scalarization of the Ricci d-tensor and
$\partial {\cal O\left(
E_N\right) =}$ $\{r=(u,e)\in {\cal O(E}^{<z>}{\cal ,}$ $u\in \partial {\cal %
E_N}$ \}.

The Cauchy problem (10.50) can be solved by using the stochastic
differential equations for the process $\left( U\left( t\right)
,c\left( t\right) \right) $ on ${\cal R}_{+}^{n_E}\times {\cal
R}^{{(n}_E{)}^2}:$

$$
dU_t^{<\alpha >}=e_{<\widehat{\beta }>}^{<\alpha >}\left(
t\right) \circ \delta B^{<\widehat{\beta }>}\left( t\right)
+\delta _{q-1}^{<\alpha
>}\delta \varphi \left( t\right) ,
$$
$$
de_{<\underline{\beta }>}^{<\alpha >}\left( t\right) =-\overrightarrow{%
\Gamma }_{<\beta ><\gamma >}^{<\alpha >}\left( U\left( t\right) \right) e_{<%
\underline{\beta }>}^{<\gamma >}\left( t\right) \circ \delta U^{<\underline{%
\beta }>}\left( t\right) =
$$
$$
-\overrightarrow{\Gamma }_{<\beta ><\gamma >}^{<\alpha >}\left(
U\left(
t\right) \right) e_{<\underline{\beta }>}^{<\gamma >}\left( t\right) e_{<%
\widehat{\tau }>}^{<\beta >}\left( t\right) \circ \delta B^{<\widehat{\tau }%
>}\left( t\right) -
$$
$$
\overrightarrow{\Gamma }_{q-1<\tau >}^{<\alpha >}\left( U\left(
t\right) \right) e_{<\underline{\beta }>}^{<\tau >}\left(
t\right) \delta \varphi \left( t\right) ,
$$
$$
\left( <\widehat{\beta }>,<\widehat{\tau }>=1,2,...,q-1\right)
$$
where $B^{<\alpha >}\left( t\right) $ is a $\left( n_E\right)
$-dimensional Brownian motion, $U\left( t\right) $ is a
nondecreasing process which
increase only if $U\left( t\right) \in \partial {\cal E_N}$ . In 
 [117]
(Chapter IV,7) it is proved that for every Borel probability
measure $\mu $ on ${\cal R}_{+}^{n_E}\times {\cal R}^{\left(
n_E\right) ^2}$ there is a unique solution $\left( U\left(
t\right) ,c\left( t\right) \right) $ of
equations (10.6) with initial distribution $\mu .$ Because if%
$$
G_{<\alpha ><\beta >}\left( U\left( 0\right) \right) e_{<\underline{\alpha }%
>}^{<\alpha >}\left( 0\right) e_{<\underline{\beta }>}^{<\beta >}\left(
0\right) =\delta _{<\underline{\alpha }><\underline{\beta }>}
$$
then for every $t\geq 0$
$$
G_{<\alpha ><\beta >}\left( U\left( t\right) \right) e_{<\underline{\alpha }%
>}^{<\alpha >}\left( t\right) e_{<\underline{\beta }>}^{<\beta >}\left(
t\right) =\delta _{<\underline{\alpha }><\underline{\beta }>}a.s.
$$
(this is a consequence of the metric compatibility criterions of
type (1.34) or (1.36)) we obtain a diffusion process $r\left(
t\right) =\left( U\left( t\right) ,c\left( t\right) \right) $ on
${\cal O(E}^{<z>}).$ This
process is called the horizontal Brownian motion on the bundle ${\cal O(E}%
^{<z>})$ with a reflecting bound. Let introduce the canonical
horizontal
fields (as in (10.21))%
$$
\left( \widetilde{L}_{<\underline{\alpha }>}F\right) \left( r\right) =e_{<%
\underline{\alpha }>}^{<\alpha >}\frac{\delta F\left( r\right)
}{\partial u^{<\alpha >}}-\overrightarrow{\Gamma }_{<\beta
><\gamma >}^{<\alpha
>}\left( u\right) e_{<\underline{\alpha }>}^{<\gamma >}e_{<\underline{\tau }%
>}^{<\beta >}\frac{\partial F\left( r\right) }{\partial e_{<\underline{\tau }%
>}^{<\alpha >}},r=\left( u,e\right) ,
$$
define the Bochner Laplacian as%
\index{Bochner Laplacian}
$$
\Delta _{{\cal O(E}^{<z>})}=\sum\limits_{\underline{\alpha }=1}^q%
\widetilde{L}_{<\underline{\alpha }>}\left( \widetilde{L}_{<\underline{%
\alpha }>}\right)
$$
and put%
$$
\alpha ^{q-1q-1}\left( r\right) =G^{q-1q-1}(u),\alpha _{<\underline{\beta }%
>}^{q-1<\beta >}=-e_{<\underline{\beta }>}^{<\tau >}\overrightarrow{\Gamma }%
_{q-1<\tau >}^{<\beta >}\left( u\right) G^{q-1q-1}.
$$

\begin{theorem}
Let $r\left( t\right) =\left( U\left( t\right) ,c\left( t\right)
\right) $ be a horizontal Brownian motion with reflecting bound
giving as a solution of equations (7.60). Then for every smooth
function $\;S\left( t,r\right) $ on $[0,\infty )\times $ ${\cal
O(E}^{<z>})){\cal \ }$ we have
$$
dS\left( t,r(t)\right) =\widetilde{L}_{<\widehat{\alpha
}>}S\left( t,r\left( t\right) \right) \delta B^{<\widehat{\alpha
}>}+
$$
$$
\{\frac 12\left( \Delta _{{\cal O(E}^{<z>})}S\right) \left(
t,r\left( t\right) \right) +\frac{\partial S}{\partial t}\left(
t,r\left( t\right) \right) \}dt+\left(
\widetilde{U}_{q-1}S\right) \left( t,r\left( t\right) \right)
\delta \varphi \left( t\right) ,
$$
where $\widetilde{U}_{q-1}$ is the horizontal lift of the vector field $%
U_{q-1}=\frac \delta {\delta u^{q-1}}$ defined as
$$
\left( \widetilde{U}_{q-1}S\right) \left( t,r\right)
=\frac{\delta S}{\delta
u^{q-1}}\left( t,r\right) +\frac{\alpha _{<\underline{\beta }>}^{q-1<\beta >}%
}{\alpha ^{q-1q-1}\left( r\right) }\frac{\partial S}{\partial e_{<\underline{%
\beta }>}^{<\beta >}}\left( t,r\right)
$$
and
$$
\delta U^{q-1}\left( t\right) \delta U^{q-1}\left( t\right)
=\alpha ^{q-1q-1}\left( r\left( t\right) \right) dt,
$$
$$
\delta U^{q-1}\left( t\right) \delta e_{<\underline{\beta
}>}^{<\beta
>}\left( t\right) =\alpha _{<\underline{\beta }>}^{q-1<\beta >}\left(
r\left( t\right) \right) dt.
$$
\end{theorem}

The proof of this theorem is a straightforward consequence of the
It\^o
formula (see (10.6) ) and of the property that $\sum_{<\underline{\alpha }%
>}e_{<\underline{\alpha }>}^{<\alpha >}\left( t\right) e_{<\underline{\alpha
}>}^{<\beta >}\left( t\right) =G^{<\alpha ><\beta >}\left(
U\left( t\right) \right) \,$ (see (10.24)).

Finally, in this subsection, we point out that for diffusion
processes we are also dealing with the so-called (A,L)-diffusion
for bounded manifolds
(see, for example, 
 [117] and formula (10.11)) which is defined by
second order operators $A$ and $L$ given correspondingly on
${\cal E_N}$ and $\partial {\cal E_N}$ .

\section{ Discussion}

In the present Chapter we have given a geometric evidence for a
generalization of stochastic calculus on spaces with higher order
anisotropy . It was possible a consideration rather similar to
that for Riemannian manifolds by using adapted to nonlinear
connection lifts to
tangent bundles 
 [295] and restricting our analysis to the case of
v-bundles provided with compatible N-connection, d-connection and
metric structures.  As a matter of principle we can construct
diffusion processes on every space ${\cal E}^{<z>}$ provided with
arbitrary d-connection structure. In this case we can formulate
all results with respect to an auxiliary convenient d-connection,
for instance, induced by the Christoffel d-symbols (10.15), and
then by using deformations of type (10.16) (or (10.17)) we shall
find the deformed analogous of stochastic differential equations
and theirs solutions.

We cite here  the pioneer works on the theory of diffusion on
Finsler manifolds with applications in biology by P. L. Antonelli
and T. J.
 Zastavniak
 [13,14] and  remark that because on Finsler spaces the metric in
general is not compatible with connection the definition of
stochastic processes is very sophisticate. Perhaps, the
uncompatible metric and connection structures are more convenient
for modeling of stochastic processes in biology and this is
successfully exploited by the mentioned authors in spite of the
fact that in general it is still unclear the possibility and
manner of definition of metric relations in biology. As for
formulation of physical models of diffusion in anisotropic media
and on locally anisotropic spaces we have to pay a due attention
to the mutual concordance of the laws of transport (i.e. of
connections) and of metric properties of the space, which in
physics plays a crucial role. This allows us to define the
Laplace-Beltrami, gradient and divergence operators and in
consequence to give the mathematical definition of diffusion
process on la-spaces in a standard manner.


\part{ Generalized Isofinsler Gravity }

\chapter{Basic Notions on Isotopies}

\section{Introduction}
 This Part is devoted to a generalization 
 [271] of the geometry of
 Santilli's locally anisotropic and inhomogeneous
 isospaces 
  [217,219,220,218,221,222,\\ 223,224]
 to the geometry of vector isobundles  provided with nonlinear and
 distinguished isoconnections and isometric  structures. We present,
 apparently for the first time, the isotopies of Lagrange,
Finsler and Kaluza--Klein
 spaces. We also continue the study of the
 interior, locally anisotropic and inhomogeneous gravitation by extending
 the isoriemannian space's constructions and presenting a geometric
 background  for the theory of isofield interactions  in
  generalized isolagrange and isofinsler spaces.

 The main purpose of this Part is to formulate a synthesis of
the Santilli isotheory and the approach on modeling locally
anisotropic geometries and physical models on bundle spaces
provided with nonlinear connection and distinguished connection
and metric structures
 [160,161,295].
 The isotopic
variants of generalized Lagrange and Finsler geometry will be
analyzed. Basic geometric constructions such as nonlinear
isoconnections in vector isobundles, the isotopic curvatures and
torsions of distinguished isoconnections and theirs structure
equations and invariant values will be defined. A model of
locally anisotropic and inhomogeneous gravitational isotheory
will be constructed.

 Our study of Santilli's isospaces and isogeometries over isofields  will
 be treated via the isodifferential calculus according to their
latest formulation 
 [224] (we extend this calculus
 for isospaces provided with nonlinear isoconnection structure). We shall
also use Kadeisvili's notion of isocontinuity
 [129,130] and the
novel Santilli--Tsagas--Sourlas isodifferential topology
 [223,239,232].

After reviewing the basic elements for completeness as well as
for notational convenience, we shall extend  Santilli's
foundations of the isosympletic
 geometry 
 [223] to isobundles and related aspects
 (by applying, in an isotopic manner, the methods summarized in
 Miron and Anastasiei  
 [160,161] and
 Yano and Ishihara 
 [295] monographs).
 We shall  apply our results
on isotopies of Lagrange, Finsler and Kaluza--Klein geometries
 to further
 studies of the isogravitational theories (for isoriemannian spaces
 firstly considered by Santilli 
 [223]) on vector isobundle
 provided with
 compatible nonlinear and distinguished isoconnections and isometric
 structures. Such  isogeometrical models of isofield interaction isotheories
 are in general nonlinear, nonlocal and
 nonhamiltonian and contain a very large class of local anisotropies
 and inhomogeneities induced  by four fundamental  isostructures:
 the partition of unity,  nonlinear isoconnection, distinguished
 isoconnections and isometric.

The novel geometric profile emerging from all the above studies
is rather
 remarkable inasmuch as the first class of all isotopies herein
considered (called Kadeisvili's Class I
 [129,130]) preserves
the abstract axioms of conventional formulations, yet permits a
clear broadening of their applicability, and actually result to be
''directly universal'' 
 [223] for a number of possible well
 behaved nonlinear, nonlocal and nonhamiltonian systems. In turn, this
 permits a number of geometric unification such as that of all possible
 metrics (on isospaces with trivial nonlinear isoconnection structure)
 of a given dimension into Santilli's isoeuclidean metric, the
 unification of  exterior and interior gravitational problems despite
 their sizable structural differences and other unification.

\section{Isotopies of the unit and isospaces}

A number of physical problems connected with the general interior
dynamics of deformable particles while moving within
inhomogeneous and anisotropic physical media result in a study of
the most general known systems which are nonlinear in coordinates
$x$ and their derivatives $\dot{x}, \ddot{x},...,$ on wave
functions and $\psi$ and their derivatives $\partial \psi ,
\partial\partial \psi ,... .$ Such systems are also nonlocal because of
possible integral dependencies on all of the proceeding
quantities and noncanonical with violation of integrability
conditions for the existence of a Lagrangian or a Hamiltonian
 [217,219,220,218].

The mathematical methods for a quantitative treatment of the
latter nonlinear, nonlocal and nonhamiltonian systems have been
identified by Santilli in a series of contributions beginning
the late 1970's
 [217,219,220,\\ 218,221,222,223,224].
 under the name of isotopies, and
 include axiom preserving liftings of fields of numbers, vector and metric
 spaces, differential and integral calculus, algebras and geometries. These
 studies were then continued by a number of authors (see ref.
 [25]  for a comprehensive literature up to 1985, and monographs
 [129,130,151,217,219,220,218,222,224,226]
 for subsequent literature).

In this section we shall mainly recall some necessary fundamental
notions
and refer to works 
 [223,129,130] for details and
references on Lie--Santilli isotheory.

 For simplicity, we consider that maps $I\rightarrow
\widehat{I}$ are of necessary Kadeisvili Class I (II), the Class
III being considered as the union of the first two, i. e. they
are sufficiently smooth, bounded, nowhere degenerate, Hermitian
and positive (negative) definite, characterizing isotopies
(isodualities).

One demands a compatible lifting of all associative products $AB$
of some generic quantities $A$ and $B$ into the isoproduct $A*B$
satisfying the
properties:%
$$
AB\Rightarrow A*B=A\widehat{T}B,~I A=A I \equiv A\rightarrow
\widehat{I}*A=A*\widehat{I}\equiv A,
$$
$$
A\left( BC\right) =\left( AB\right) C\rightarrow A*\left(
B*C\right) =\left( A*B\right) *C,
$$
where the fixed and invertible matrix $\widehat{T}$ is called the
isotopic element.

To follow our outline, a conventional field $F\left( a,+,\times
\right) ,$ for instance of real, complex or quaternion numbers,
with elements $a$, conventional sum + and product $a\times
b\doteq ab,\,$ must be lifted into the so--called isofield
$\widehat{F}\left( \widehat{a},+,*\right) ,$
satisfying properties%
$$
F\left( a,+,*\right) \rightarrow \widehat{F}\left( \widehat{a},+,*\right) ,~%
\widehat{a}=a\widehat{I}
$$
$$
\widehat{a}*\widehat{b}=\widehat{a}\widehat{T}\widehat{b}=\left(
ab\right) \widehat{I},~\widehat{I}=\widehat{T}^{-1}
$$
with elements $\widehat{a}$ called isonumbers, $+$ and $*$ are
conventional sum and isoproduct preserving the axioms of the
former field $F\left(
a,+,\times \right) .$ All operations in $F$ are generalized for $\widehat{F}%
, $ for instance we have isosquares $\widehat{a}^{\widehat{2}}=\widehat{a}*%
\widehat{a}=\widehat{A}\widehat{T}\widehat{a}=a^2\widehat{I},$ isoquotient $%
\widehat{a}\widehat{/}\widehat{b}=\left( a/b\right) \widehat{I},$
isosquare roots $\widehat{a}^{1/2}=a^{1/2}\widehat{I},...;$\
$\widehat{a}*A\equiv aA.\, $ We note that in the literature one
uses two types of denotation for
isotopic product $*$ or $\widehat{\times }$ (in our work we shall consider $%
*\equiv \widehat{\times }).$

Let us consider, for example, the main lines of the isotopies of a $n$%
--di\-men\-si\-o\-nal Euclidean space $E^n\left( x,g,{\cal
R}\right) ,$ where ${\cal R}\left( n,+,\times \right) $ is the
real number field,
provided with a local coordinate chart $x=\{x^k\},k=1,2,...,n,$ and $n$%
--dimensional metric $\rho =\left( \rho _{ij}\right) =diag\left(
1,1,...,1\right) .$ The scalar product of two vectors $x,y\in
E^n$ is defined as
$$
\left( x-y\right) ^2=\left( x^i-y^i\right) \rho _{ij}\left(
x^j-y^j\right) \in {\cal R}\left( n,+,\times \right)
$$
were the Einstein summation rule on repeated indices is assumed
hereon.

The {\bf Santilli's isoeuclidean} spaces
 $\widehat{E}\left( \widehat{x},\widehat{\rho },%
\widehat{R}\right) $ of Class III are introduced as
$n$--dimensional metric
spaces defined over an isoreal isofield $\widehat{R}\left( \widehat{n},+,%
\widehat{\times }\right) $ with an $n\times n$--dimensional
real--valued and symmetrical isounit $\widehat{I}=\widehat{I}^t$
of the same class, equipped
with the ''isometric''%
$$
\widehat{\rho }\left( t,x,v,a,\mu ,\tau ,...\right) =\left( \widehat{\rho }%
_{ij}\right) =\widehat{T}\left( t,x,v,a,\mu ,\tau ,...\right) \times \rho =%
\widehat{\rho }^t,
$$
where $\widehat{I}=\widehat{T}^{-1}=\widehat{I}^t.$

A local coordinate cart on $\widehat{E}\left( \widehat{x},\widehat{\rho },%
\widehat{R}\right) $ can be defined in contravariant
$$
\widehat{x}=\{\widehat{x}^k=x^{\widehat{k}}\}=\{x^k\times \widehat{I}_{~k}^{%
\widehat{k}}\}
$$
or covariant form%
$$
\widehat{x}_k=\widehat{\rho
}_{kl}\widehat{x}^l=\widehat{T}_k^r\rho _{ri}x^i\times
\widehat{I},
$$
where $x^k,x_k\in \widehat{E}$. The square of ''isoeuclidean
distance'' between two points $\widehat{x},\widehat{y}\in
\widehat{E}$ is defined as
$$
\left( \widehat{x}-\widehat{y}\right) ^{\widehat{2}}=\left[ \left( \widehat{x%
}^i-\widehat{y}^i\right) \times \widehat{\rho }_{ij}\times \left( \widehat{x}%
^j-\widehat{y}^j\right) \right] \times \widehat{I}\in \widehat{R}
$$
and the isomultiplication is given by
$$
\widehat{x}^{\widehat{2}}=\widehat{x}^k\widehat{\times
}\widehat{x}_k=\left( x^k\times \widehat{I}\right) \times
\widehat{T}\times \left( x_k\times \widehat{I}\right) =\left(
x^k\times x_k\right) \times \widehat{I}=n\times \widehat{I} .
$$

Whenever confusion does not arise isospaces can be practically
treated via
the conventional coordinates $x^k$ rather than the isotopic ones $\widehat{x}%
^k=x^k\times \widehat{I}.$ The symbols $x,v,a,...$ will be used
for conventional spaces while symbols
$\widehat{x},\widehat{v},\widehat{a},...$ will be used for
isospaces; the letter $\widehat{\rho }\left( x,v,a,...\right) $
refers to the projection of the isometric $\widehat{\rho }$ in
the original space.

We note that an isofield of Class III, explicitly denoted as $\widehat{F}%
_{III}\left( \widehat{a},+,\widehat{\times }\right) $ is a union
of two
disjoint isofields, one of Class I, $\widehat{F}_I\left( \widehat{a},+,%
\widehat{\times }\right) ,$ in which the isounit is positive
definite, and
one of Class II, $\widehat{F}_{II}\left( \widehat{a},+,\widehat{\times }%
\right) ,$ in which the isounit is negative--definite. The Class
II\ of
isofields is usually written as $\widehat{F}^d\left( \widehat{a}^d,+,%
\widehat{\times }^d\right) $ and called isodual fields with isodual unit $%
\widehat{I}^d=-\widehat{I}<0,$ isodual isonumbers
$\widehat{a}^d=a\times \widehat{I}^d=-\widehat{a},$ isodual
isoproduct $\widehat{\times }^d=\times \widehat{T}^d\times
=-\widehat{\times },$ etc. For simplicity, in our further
considerations we shall use the general
 terms isofields, isonumbers even for isodual fields, isodual numbers and
so on if this will not give rise to ambiguities.

\section{Isocontinuity and isotopology}

The isonorm of an isofield of Class III is defined as%
$$
\left\uparrow \widehat{a}\right\uparrow =\left| a\right| \times
\widehat{I}
$$
where $\left| a\right| $ is the conventional norm. Having defined
a function $\widehat{f}\left( \widehat{x}\right) $ on isospace
$\widehat{E}\left(
\widehat{x},\widehat{\delta },\widehat{R}\right) $ over isofield $\widehat{R}%
\left( \widehat{n},+,\widehat{\times }\right) $ one introduces
(see details
and references in 
 [129,130]) the isomodulus
$$
\left\uparrow \widehat{f}\left( \widehat{x}\right) \right\uparrow
=\left| \widehat{f}\left( \widehat{x}\right) \right| \times
\widehat{I}
$$
where $\left| \widehat{f}\left( \widehat{x}\right) \right| $ is
the conventional modulus.

One says that an infinite sequence of isofunctions of Class I\ $\widehat{f}%
_1,\widehat{f}_2,...$ is ''strongly isoconvergent'' to the isofunction $%
\widehat{f}$ of the same class if
$$
\lim _{k\rightarrow \infty }\left\uparrow \widehat{f}_k-\widehat{f}%
\right\uparrow =\widehat{0}.
$$
The Cauchy isocondition is expressed as
$$
\left\uparrow \widehat{f}_m-\widehat{f}_n\right\uparrow <\widehat{\rho }%
=\rho \times \widehat{I}
$$
where $\delta $ is real and $m$ and $n$ are greater than a suitably chosen $%
N\left( \rho \right) .$ Now the isotopic variants of continuity,
limits, series, etc, can be easily constructed in a traditional
manner.

The notion of $n$--dimensional isomanifold was studied by Tsagas
and Sourlas
(we refer the reader for details in 
 [239,232]). Their
constructions are based on idea that every isounit of Class III
can always be diagonalized into the form
$$
\widehat{I}=diag\left( B_1,B_2,...,B_n\right) ,B_k\left(
x,...\right) \neq 0,k=1,2,...,n .
$$
In result of this one defines an isotopology $\widehat{\tau }$ on $\widehat{R%
}^n$ which coincides everywhere with the conventional topology $\tau $ on $%
R^n$ except at the isounit $\widehat{I}.$ In particular,
$\widehat{\tau }$ is everywhere local--differential, except at
$\widehat{I}$ which can incorporate integral terms. The above
structure is called the Tsagas--Sourlas isotopology or an
integro--differential topology. Finally, in this subsection, we
note that Prof. Tsagas and Sourlas used a
 conventional topology on isomanifolds. The isotopology was first introduced
 by Prof. Santilli in ref. 
  [223].

\section{Isodifferential and isointegral calculus}

Now we are able to introduce isotopies of the ordinary
differential calculus, i.e. the isodifferential calculus (for
short).

The {\bf isodifferentials} of Class I of the contravariant and
covariant
coordinates $\widehat{x}^k=x^{\widehat{k}}$ and $\widehat{x}_k=x_{\widehat{k}%
}$ on an isoeuclidean space $\widehat{E}$ of the same class is
given by
$$
\widehat{d}\widehat{x}^k=\widehat{I}_{~i}^k\left( x,...\right) dx^i,~%
\widehat{d}\widehat{x}_k=\widehat{T}_k^{~i}\left( x,...\right)
dx_i\eqno(11.1)
$$
where $\widehat{d}\widehat{x}^k$ and $\widehat{d}\widehat{x}_k$
are defined
on $\widehat{E}$ while the $\widehat{I}_{~i}^kdx^i$ and $\widehat{T}%
_k^{~~i}dx_i$ are the projections on the conventional Euclidean
space.

For a sufficiently smooth isofunction $\widehat{f}\left(
\widehat{x}\right) $ on a closed domain $\widehat{U}\left(
\widehat{x}^k\right) $ covered by contravariant
iso\-co\-or\-di\-na\-tes $\widehat{x}^k$ we can define the
partial iso\-de\-ri\-va\-ti\-ves $\widehat{\partial }_{\widehat{k}}=\frac{%
\widehat{\partial }}{\widehat{\partial }\widehat{x}^k}$ at a point $\widehat{%
x}_{(0)}^k\in \widehat{U}\left( \widehat{x}^k\right) $ by
considering the
limit%
$$
\widehat{f^{\prime }}\left( \widehat{x}_{(0)}^k\right) =\widehat{\partial }_{%
\widehat{k}}\widehat{f}\left( \widehat{x}\right) \mid _{\widehat{x}_{(0)}^k}=%
\frac{\widehat{\partial }\widehat{f}\left( \widehat{x}\right) }{\widehat{%
\partial }\widehat{x}^k}\mid _{\widehat{x}_{(0)}^k}=\widehat{T}_k^{~i}\frac{%
\partial f\left( x\right) }{\partial x^i}\mid _{\widehat{x}_{(0)}^k}=%
\eqno(11.2)
$$
$$
\lim _{\widehat{d}\widehat{x}^k\rightarrow \widehat{0}^k}\frac{\widehat{f}%
\left( \widehat{x}_{(0)}^k+\widehat{d}\widehat{x}^k\right) -\widehat{f}%
\left( \widehat{x}_{(0)}^k\right) }{\widehat{d}x^k}
$$
where $\widehat{\partial }\widehat{f}\left( \widehat{x}\right) /\widehat{%
\partial }\widehat{x}^k$ is computed on $\widehat{E}$ and $\widehat{T}%
_k^{~i}\partial f\left( x\right) /\partial x^i$ is the projection
in $E.$

In a similar manner we can define the{\bf \ partial isoderivatives} $%
\widehat{\partial }^{\widehat{k}}=\frac{\widehat{\partial }}{\widehat{%
\partial }\widehat{x}_k}$ with respect to a covariant variable $\widehat{x}%
_k:$%
$$
\widehat{f^{\prime }}\left( \widehat{x}_{k(0)}\right) =\widehat{\partial }^{%
\widehat{k}}\widehat{f}\left( \widehat{x}\right) \mid _{\widehat{x}_{k(0)}}=%
\frac{\widehat{\partial }\widehat{f}\left( \widehat{x}\right) }{\widehat{%
\partial }\widehat{x}_k}\mid _{\widehat{x}_{k(0)}}=\widehat{T}_k^{~i}\frac{%
\partial f\left( x\right) }{\partial x^i}\mid _{\widehat{x}_{k(0)}}=%
\eqno(11.3)
$$
$$
\lim _{\widehat{d}\widehat{x}_{\widehat{k}}\rightarrow \widehat{0}_k}\frac{%
\widehat{f}\left( \widehat{x}_{k(0)}+\widehat{d}\widehat{x}_k\right) -%
\widehat{f}\left( \widehat{x}_{k(0)}\right) }{\widehat{d}x_k}.
$$

The isodifferentials of an isofunction of contravariant or
covariant coordinates, $\widehat{x}^k$ or $\widehat{x}_k,$ are
defined according the
formulas%
$$
\widehat{d}\widehat{f}\left( \widehat{x}\right) \mid _{contrav}=\widehat{%
\partial }_{\widehat{k}}\widehat{f}\widehat{d}\widehat{x}^k=\widehat{T}%
_k^{~i}\frac{\partial f\left( x\right) }{\partial x^i}\widehat{I}_{~j}^kdx^j=%
\frac{\partial f\left( x\right) }{\partial
x^k}dx^k=\frac{\partial f\left( x\right) }{\partial
x^i}\widehat{T}_j^{~i}dx^j
$$
and
$$
\widehat{d}\widehat{f}\left( \widehat{x}\right) \mid _{covar}=\widehat{%
\partial }^{\widehat{k}}\widehat{f}\widehat{d}\widehat{x}_k=\widehat{I}%
_{~i}^k\frac{\partial f\left( x\right) }{\partial x_i}\widehat{T}_k^{~j}dx_j=%
\frac{\partial f\left( x\right) }{\partial x_k}dx_k=\frac{\partial f}{%
\partial x_j}\widehat{I}_{~j}^idx_i.
$$

The second order isoderivatives there are introduced by iteration
of the
notion of isoderivative:%
$$
\widehat{\partial }_{\widehat{i}\widehat{j}}^2\widehat{f}(\widehat{x})=\frac{%
\widehat{\partial }^2\widehat{f}(\widehat{x})}{\widehat{\partial }\widehat{x}%
^i\widehat{\partial }\widehat{x}^j}=\widehat{T}_{\widehat{i}}^{~i}\widehat{T}%
_{\widehat{j}}^{~j}\frac{\partial ^2f\left( x\right) }{\partial
x^i\partial x^j},
$$
$$
\widehat{\partial }^{2\widehat{i}\widehat{j}}\widehat{f}(\widehat{x})=\frac{%
\widehat{\partial }^2\widehat{f}(\widehat{x})}{\widehat{\partial }\widehat{x}%
_i\widehat{\partial }\widehat{x}_j}=\widehat{I}_{~i}^{\widehat{i}}\widehat{I}%
_{~j}^{\widehat{j}}\frac{\partial ^2f\left( x\right) }{\partial
x_i\partial x_j},
$$
$$
\widehat{\partial }_{\widehat{i}}^{2~\widehat{j}}\widehat{f}(\widehat{x})=%
\frac{\widehat{\partial }^2\widehat{f}(\widehat{x})}{\widehat{\partial }%
\widehat{x}^i\widehat{\partial }\widehat{x}_{\widehat{j}}}=\widehat{T}_{%
\widehat{i}}^{~i}\widehat{I}_j^{~\widehat{j}}\frac{\partial
^2f\left( x\right) }{\partial x^i\partial x_j}.
$$

The Laplace isooperator on Euclidean space $\widehat{E}\left( \widehat{x},%
\widehat{\delta },\widehat{R}\right) $ is given by
$$
\widehat{\Delta }=\widehat{\partial }_k\widehat{\partial }^k=\widehat{%
\partial }^i\rho _{ij}\widehat{\partial }^j=\widehat{I}_{~k}^{\widehat{i}%
}\partial ^k\rho _{ij}\partial ^j\eqno(11.4)
$$
where there are also used usual partial derivatives $\partial
^j=\partial /\partial x_j$ and $\partial _k=\partial /\partial
x^k.$

The isodual isodifferential calculus is characterized by the
following
isodual differentials and isodual isoderivatives%
$$
\widehat{d}^{(d)}\widehat{x}^{(d)k}=\widehat{I}_{\quad i}^{(d)k}d\widehat{x}%
^{(d)i}\equiv \widehat{d}x^k,\quad \widehat{\partial }^{(d)}/\widehat{%
\partial }\widehat{x}^{(d)i}\widehat{T}_k^{~i(d)}\partial /\partial \widehat{%
x}^{i(d)}\equiv \widehat{T}_k^{~i}\partial /\partial
\widehat{x}^i.
$$

The formula (11.4) is different from the expression for the
Laplace operator
$$
\Delta =\widehat{\rho }^{-1/2}\partial _i\widehat{\rho }^{1/2}\widehat{\rho }%
^{ij}\partial _j
$$
even though the Euclidean isometric $\widehat{\rho }\left(
x,v,a,...\right) $ is more general than the Riemannian metric
$g\left( x\right) .$ For partial
isoderivations one follows the next properties:%
$$
\frac{\widehat{\partial }\widehat{x}^i}{\widehat{\partial }\widehat{x}^j}%
=\delta _{~j}^i,~\frac{\widehat{\partial }\widehat{x}_i}{\widehat{\partial }%
\widehat{x}_j}=\delta _i^{~j},~\frac{\widehat{\partial }\widehat{x}_i}{%
\widehat{\partial }\widehat{x}^j}=\widehat{T}_i^{~j},~\frac{\widehat{%
\partial }\widehat{x}^i}{\widehat{\partial }\widehat{x}_j}=\widehat{I}%
_{~j}^i.
$$

Here we remark that isointegration (the inverse to
isodifferential) is
defined 
 [223] as to satisfy conditions
$$
\int^{\symbol{94}}\widehat{d}\widehat{x}=\int
\widehat{T}\widehat{I}dx=\int dx=x,
$$
where $\int^\symbol{94}=\int \widehat{T}.$

\section{Santilli's isoriemannian isospaces}

Let consider ${\cal R=R}\left( x,g,{\sl R}\right) $ a (pseudo)\
Riemannian space over the reals ${\sl R}\left( n,+,\times \right)
$ with local coordinates $x=\{x^\mu \}$ and nonwere singular,
symmetrical and real--valued metric $g\left( x\right) =\left(
g_{\mu \nu }\right) =g^t$ and the tangent
flat space $M\left( x,\eta ,{\sl R}\right) $ provided with flat real metric $%
\eta $ (for a corresponding signature and dimension we can
consider $M$ as the well known Minkowski space). The metric
properties of the Riemannian
spaces are defined by scalar square of a tangent vector $x,$%
$$
x^2=x^\mu g_{\mu \nu }\left( x\right) x^v\in {\sl R}
$$
or, in infinitesimal form by the line element%
$$
ds^2=dx^\mu g_{\mu \nu }\left( x\right) dx^\nu
$$
and related formalism of covariant derivation (see for instance
 [165]).

The isotopies of the Riemannian spaces and geometry,
 were first studied and applied by 
 [223] and
 are called Santilli's isoriemannian spaces and geometry.
 In this section we consider isoriemannian spaces equipped with
  the Santilli--Tsagas--Sourlas isotopology [223,239,232]
 in a similar manner as we have done
in the previous subsection for isoeuclidean spaces but with
respect to a
general, non flat, isometric. A isoriemannian space $\widehat{{\cal R}}{\cal %
=}\widehat{{\cal R}}\left( \widehat{x},\widehat{g},\widehat{{\sl
R}}\right) ,
$ over the isoreals $\widehat{R}=\widehat{R}\left( \widehat{n},+,\widehat{x}%
\right) $ with common isounits $\widehat{I}=\left(
\widehat{I}_{~\nu }^\mu
\right) =\widehat{T}^{-1},$ is provided with local isocoordinates $\widehat{x%
}=\{\widehat{x}^\mu \}=\{x^\mu \}$ and isometric
$\widehat{g}\left( x,v,a,\mu ,\tau ,...\right) =\widehat{T}\left(
x,v,\mu ,\tau ,...\right) g\left( x\right) ,$ where
$\widehat{T}=\left( \widehat{T}_\mu ^{~\nu }\right) $ is nowhere
singular, real valued and symmetrical matrix of Class I with
$C^\infty $ elements. The corresponding isoline and infinitesimal
elements are written as
$$
\widehat{x}^{\widehat{2}}=[\widehat{x}^\mu \widehat{g}_{\mu \nu
}\left(
x,v,a,\mu ,\tau ,...\right) \widehat{x}^\nu ]\times \widehat{I}\in \widehat{R%
}
$$
with infinitesimal version
$$
d\widehat{s}^2=(\widehat{d}\widehat{x}^\mu \widehat{g}_{\mu \nu
}\left( x\right) \widehat{d}\widehat{x}^\nu )\times
\widehat{I}\in \widehat{R}.
$$

The {\bf covariant isodifferential calculus} has been introduced
in ref.
 [223] via the expression
$$
\widehat{D}\widehat{X}^\beta =\widehat{d}\widehat{X}^\beta +\widehat{\Gamma }%
_{\alpha \gamma }^\beta \widehat{X}^\alpha
\widehat{d}\widehat{x}^\gamma
$$
with corresponding covariant isoderivative%
$$
\widehat{X}_{\uparrow \mu }^\beta =\widehat{\partial }_\mu
\widehat{X}^\beta +\widehat{\Gamma }_{\alpha \mu }^\beta
\widehat{X}^\alpha
$$
with the {\bf isocristoffel symbols} written as
$$
\widehat{\{\alpha \beta \gamma \}}=\frac 12\left(
\widehat{\partial }_\alpha \widehat{g}_{\beta \gamma
}+\widehat{\partial }_\gamma \widehat{g}_{\alpha
\beta }-\widehat{\partial }_\beta \widehat{g}_{\alpha \gamma }\right) =%
\widehat{\{\gamma \beta \alpha \}},\eqno(11.5)
$$
$$
\widehat{\Gamma }_{\alpha \gamma }^\beta =\widehat{g}^{\beta \rho }\widehat{%
\{\alpha \rho \gamma \}}=\widehat{\Gamma }_{\alpha \gamma }^\beta
,
$$
where $\widehat{g}^{\beta \rho }$ is inverse to
$\widehat{g}_{\alpha \beta }. $

The crucial difference between Riemannian spaces and
iso\-spa\-ces is obvious if the corresponding auto--parallel
equations
$$
\frac{Dx_\beta }{Ds}=\frac{dv_\beta }{ds}+\{\alpha \beta \gamma
\}\left( x\right) \frac{dx^\alpha }{ds}\frac{dx^\gamma
}{ds}=0\eqno(11.6)
$$
and auto--isoparallel equations
$$
\frac{\widehat{D}\widehat{x}_\beta }{\widehat{D}\widehat{s}}=\frac{\widehat{d%
}v_\beta }{\widehat{d}\widehat{s}}+\widehat{\{\alpha \beta \gamma
\}}\left(
\widehat{x},\widehat{v},\widehat{a},...\right) \frac{\widehat{d}\widehat{x}%
^\alpha }{\widehat{d}\widehat{s}}\frac{\widehat{d}\widehat{x}^\gamma }{%
\widehat{d}\widehat{s}}=0\eqno(11.7)
$$
where $\widehat{v}=\widehat{d}\widehat{x}/\widehat{d}\widehat{s}=\widehat{I}%
_S\times dx/ds,\widehat{s}$ is the proper isotime and
$\widehat{I}_S$ is the related one--dimensional isounit, can be
identified by observing that equations (11.6) are at most
quadratic in the velocities while the isotopic equations (11.7)
are arbitrary nonlinear in the velocities and another possible
variables and parameters $\left( \widehat{a},...\right) .$

By using coefficients $\widehat{\Gamma }_{\alpha \gamma }^\beta $
we
introduce the next isotopic values 
 [223]:

the {\bf isocurvature tensor}
$$
\widehat{R}_{\alpha ~\gamma \delta }^{\quad \beta }=\widehat{\partial }%
_\delta \widehat{\Gamma }_{\alpha \gamma }^\beta
-\widehat{\partial }_\gamma \widehat{\Gamma }_{\alpha \delta
}^\beta +\widehat{\Gamma }_{\varepsilon
\delta }^\beta \widehat{\Gamma }_{\alpha \gamma }^\varepsilon -\widehat{%
\Gamma }_{\varepsilon \gamma }^\beta \widehat{\Gamma }_{\alpha
\delta }^\varepsilon ;\eqno(11.8)
$$

the {\bf isoricci tensor} $\widehat{R}_{\alpha \gamma
}=\widehat{R}_{\alpha ~\gamma \beta }^{\quad \beta };$

the {\bf isocurvature scalar} $\widehat{R}=%
\widehat{g}^{\alpha \gamma }\widehat{R}_{\alpha \gamma };$

the {\bf isoeinstein tensor}
$$
\widehat{G}_{\mu \nu }=\widehat{R}_{\mu \nu }-\frac 12\widehat{g}_{\mu \nu }%
\widehat{R}\eqno(11.9)
$$

and the {\bf istopic isoscalar}%
$$
\widehat{\Theta }=\widehat{g}^{\alpha \beta }\widehat{g}^{\gamma
\delta }\left( \widehat{\{\rho \alpha \delta \}}\widehat{\Gamma
}_{\gamma \beta }^\rho -\widehat{\{\rho \alpha \beta
\}}\widehat{\Gamma }_{\gamma \delta }^\rho \right) \eqno(11.10)
$$
(the later is a new object for the\ Riemannian isometry).

The isotopic lifting of the Einstein equations (see the history,
details and
references in 
 [223]) is written as
$$
\widehat{R}^{\alpha \beta }-\frac 12\widehat{g}^{\alpha \beta }(\widehat{R}+%
\widehat{\Theta })=\widehat{t}^{\alpha \beta }-\widehat{\tau
}^{\alpha \beta },\eqno(11.11)
$$
where $\widehat{t}^{\alpha \beta }$ is a {\bf source isotensor} and $%
\widehat{\tau }^{\alpha \beta }$ is the {\bf stress--energy
isotensor} and
there is satisfied the Freud isoidentity 
 [223]
$$
\widehat{G}_{~\beta }^\alpha -\frac 12\delta _{~\beta }^\alpha \widehat{%
\Theta }=\widehat{U}_{~\beta }^\alpha +\widehat{\partial }_\rho \widehat{V}%
_{\quad \beta }^{\alpha \rho },\eqno(11.12)
$$
$$
\widehat{U}_{~\beta }^\alpha =-\frac 12\frac{\widehat{\partial }\widehat{%
\Theta }}{\widehat{\partial }\widehat{g}_{\quad \uparrow \alpha
}^{\gamma \delta }}\widehat{g}_{\quad \uparrow \beta }^{\gamma
\delta },
$$
$$
\widehat{V}_{\quad \beta }^{\alpha \rho }=\frac
12[\widehat{g}^{\gamma \delta }\left( \delta _{~\beta }^\alpha
\widehat{\Gamma }_{\alpha \delta }^\rho -\delta _{~\delta
}^\alpha \widehat{\Gamma }_{\alpha \beta }^\rho \right) +
\widehat{g}^{\rho \gamma }\widehat{\Gamma }_{\beta \gamma }^\alpha -\widehat{%
g}^{\alpha \gamma }\widehat{\Gamma }_{\beta \gamma }^\rho +\left(
\delta _{~\beta }^\rho \widehat{g}^{\alpha \gamma }-\delta
_{~\beta }^\alpha \widehat{g}^{\rho \gamma }\right)
\widehat{\Gamma }_{\gamma \rho }^\rho ].
$$
Finally, we remark that for antiautomorphic maps of isoduality we
have to modify correspondingly the above presented formulas
holding true for
Riemannian isodual spaces $\widehat{{\cal R}}^{(d)}=\widehat{{\cal R}}%
^{(d)}\left( \widehat{x}^{(d)},\widehat{g}^{(d)},\widehat{{\sl R}}%
^{(d)}\right) ,$ over the isodual reals $\widehat{R}^{(d)}=\widehat{R}%
^{(d)}\left( \widehat{n}^{(d)},+,\widehat{\times }^{(d)}\right) $
with curvature, Ricci, Einstein and so on isodual tensors. For
simplicity we omit such details in this work.

\chapter{Isobundle Spaces}

This chapter serves the twofold purpose of establishing of
abstract index denotations and starting the geometric backgrounds
of isotopic locally an\-i\-sot\-rop\-ic extensions of the
isoriemannian spaces which are used in the next chapters of the
work.

\section{Lie--Santilli isoalgebras and isogroups}

The Lie--Santilli isotheory is based on a generalization of the
very notion of numbers and fields. If the Lie's theory is
centrally dependent on the basic $n$--dimensional unit
$I=diag\left( 1,1,...,1\right) $ in, for instance, enveloping
algebras, Lie algebras, Lie groups, representation theory, and so
on, the Santilli's main idea  is the reformulation of the entire
conventional theory with respect to the most general possible,
integro--differential isounit. In this section we introduce some
necessary definitions and formulas on Lie--Santilli
isoalgebra and isogroups following 
 [129,130] where details,
developments and basic references on Santilli original result are
contained.
A Lie--Santilli algebra is defined as a finite--dimensional isospaces $%
\widehat{L}$ over the isofield $\widehat{F}$ of isoreal or
isocomplex numbers with isotopic element $T$ and isounit
$\widehat{I}=T^{-1}.$ In brief one uses the term isoalgebra (when
there is not confusion with isotopies of
non--Lie algebras) which is defined by isolinear isocommutators of type $[A,%
\symbol{94}B]\in \widehat{L}$ satisfying the conditions:%
$$
[A,\symbol{94}B]=-[B,\symbol{94}A],
$$
$$
[A,\symbol{94}[B,\symbol{94}C]]+[B,\symbol{94}[C,
\symbol{94}A]]+[C,\symbol{94%
}[A,\symbol{94}B]]=0,
$$
$$
[A*B,\symbol{94}C]=A*[B,\symbol{94}C]+[A,\symbol{94}C]*B
$$
for all $A,B,C\in \widehat{L}.$ The structure functions
$\widehat{C}$ of the
Lie--Santilli algebras are introduced according the relations%
$$
[X_i,\symbol{94}X_j]=X_i*X_j-X_j*X_i=
$$
$$
X_iT\left( x,...\right) X_j-X_jT\left( x,...\right) X_i=\widehat{C}%
_{ij}^{\quad k}\left( x,\dot x,\ddot x,...\right) *X_k.
$$
It should be noted that, in fact, the basis $e_k,(k=1,2,...,N)$
of a Lie
algebra $L$ is not changed under isotopy except the renormalization factors $%
\widehat{e}_k:$ the isocommutation rules of the isotopies
$\widehat{L}$ are
$$
\left[ \widehat{e}_i,\widehat{e}_j\right] =\widehat{e}_iT\widehat{e}_j-%
\widehat{e}_jT\widehat{e}_i=\widehat{C}_{ij}^{~k}\left( x,\dot
x,\ddot x,...\right) \widehat{e}_k
$$
where $\widehat{C}=CT.$

An isomatrix $\hat M$ is and ordinary matrix whose elements are
isoscalars.
 All operations among isomatrices are therefore isotopic.

 The isotrace of a isomatrix $A$ is introduced by using
 the unity $\widehat{I}:$
$$
\widehat{Tr}A=\left( TrA\right) \widehat{I}\in \widehat{F}
$$
where $TrA$ is the usual trace. One holds properties%
$$
\widehat{Tr}(A*B)=\left( \widehat{Tr}A\right) *\left(
\widehat{Tr}B\right)
$$
and
$$
\widehat{Tr}A=\widehat{Tr}\left( BAB^{-1}\right) .
$$
The Killing isoform is determined by the isoscalar product%
$$
\left( A,\symbol{94}B\right) =\widehat{Tr}\left[ \left(
\widehat{Ad}X\right) *\left( \widehat{Ad}B\right) \right]
$$
where the isolinear maps are introduced as $\widehat{ad}A\left( B\right) =[A,%
\symbol{94}B],\forall A,B\in \widehat{L}.$ Let $e_k,k=1,2,...,N$
be the basis of a Lie algebra with an isomorphic map
$e_k\rightarrow \widehat{e}_k$ to the basis $\widehat{e}_k$ of a
Lie--Santilli isoalgebra $\widehat{L}.$ We can write the elements
in $\widehat{L}$ in local coordinate form. For
instance, considering $A=x^i\widehat{e}_i,B=y^j\widehat{e}_j$ and $C=z^k%
\widehat{e}_k=[A,\symbol{94}B]$ we have%
$$
C=z^k\widehat{e}_k=\left[ A,\symbol{94}B\right] =x^iy^j[\widehat{e}_i,%
\symbol{94}\widehat{e}_j]=x^ix^j\widehat{C}_{ij}^{~k}\widehat{e}_k
$$
and
$$
\left[ \widehat{Ad}A\left( B\right) \right] ^k=\left[
A,\symbol{94}B\right] ^k=x^ix^j\widehat{C}_{ij}^{~k}.
$$
In standard manner there is introduced the {\bf isocartan tensor}
$$
\widehat{q}_{ij}\left( x,\dot x,\ddot x,...\right) =\widehat{C}_{ip}^{~k}%
\widehat{C}_{ik}^{~p}\in \widehat{L}
$$
via the definition
$$
\left( A,\symbol{94}B\right) =\widehat{q}_{ij}x^iy^j.
$$
Considering that $\widehat{L}$ is an isoalgebra with generators
$X_k$ and
isounit $\widehat{I}=T^{-1}>0$ the isodual Lie--Santilli algebras $\widehat{L%
}^d$ of $\widehat{L}$ (we note that $\widehat{L}$ and
$\widehat{L}^d$ are (anti) isomorphic).

The conventional structure of the Lie theory admits a
conventional isotopic lifting. Let give some examples. The
general isolinear and isocomplex Lie--Santilli algebras
$\widehat{gl}\left( n,\widehat{C}\right) $ are
introduced as the vector isospaces of all $n\times n$ isocomplex matrices over $%
\widehat{C}.$ For the isoreal numbers $\widehat{R}$ we shall write $\widehat{%
gl}\left( n,\widehat{R}\right) .$ By using ''hats'' we denote
respectively
the special, isocomplex, isolinear isoalgebra $\widehat{sl}\left( n,\widehat{C}%
\right) $ and the isoorthogonal algebra $\widehat{o}\left(
n\right) .$

A right Lie--Santilli isogroup $\widehat{G}r$ on an isospace $\widehat{S}%
\left( x,\widehat{F}\right) $ over an isofield $\widehat{F},\widehat{I}%
=T^{-1}$ (in brief isotransformation group or isogroup) is
introduced in standard form but with respect to isonumbers and
isofields as a group which maps each element $x\in
\widehat{S}\left( x,\widehat{F}\right) $ into a new element
$x^{\prime }\in \widehat{S}\left( x,\widehat{F}\right) $ via the
isotransformations $x^{\prime }=\widehat{U}*x=\widehat{U}Tx,$
where $T$ is fixed such that

1. The map $\left( U,x\right) \rightarrow \widehat{U}*x$ of $\widehat{G}%
r\times \widehat{S}\left( x,\widehat{F}\right) $ onto $\widehat{S}\left( x,%
\widehat{F}\right) $ is isodifferentiable;

2.
$\widehat{I}*\widehat{U}=\widehat{U}*\widehat{I}=\widehat{U},~\forall
\widehat{U}\in \widehat{G}r;$

3. $\widehat{U}_1*(\widehat{U}_2*x)=\left( \widehat{U}_1*\widehat{U}%
_2\right) *x,~\forall x\in \widehat{S}\left( x,\widehat{F}\right) $ and $%
\widehat{U}_1,\widehat{U}_2\in \widehat{G}r.$

We can define accordingly a left isotransformation group.

\section{Fiber isobundles}

Prof. Santilli identified the foundations of the isosympletic
geometry in
 the work 
 [223]. In this section we present, apparently for
 the first time, the isotopies of fibre bundles and related topics.

The notion of locally trivial fiber isobundle naturally
generalizes that of the isomanifold. The fiber isobundles will be
used to get some results in isogeometry as well as to build
geometrical models for physical isotheories. In general the
proofs, being corresponding reformulation in isotopic manner of
standard results, will be omitted. The reader is referred to some
well--known books containing the theory of fibre bundles and the
mathematical foundations of the Lie--Santilli isotheory.

Let $\widehat{G}r$ be a Lie--Santilli isogroup which acts
isodifferentiably
and effectively on a isomanifold $\widehat{V},$ i.e. every element $%
\widehat{q}\in \widehat{G}r$ defines an isotopic diffeomorphism $L_{\widehat{%
G}r}:\widehat{V}\rightarrow \widehat{V}.$

As a rule, all isomanifolds are assumed to be isocontinuous,
finite dimensional and having the isotopic variants of the
conditions to be Hausdorff, paracompact and isoconnected; all
isomaps are isocontinous.

A locally trivial {\bf fibre isobundle} is defined by the data\\
$\left(
\widehat{E},\widehat{p},\widehat{M},\widehat{V},\widehat{G}r\right)
,$ where $\widehat{M}$ (the base isospace) and $\widehat{E}$ (the
total isospace) are isomanifolds,
$\widehat{E},\widehat{p}:\widehat{E}\rightarrow \widehat{M}$ is a
surjective isomap and the following conditions are satisfied:

1/ the isomanifold $\widehat{M}$ can be covered by a set ${\cal
E}$ of open
isotopic sets $\widehat{U},\widehat{W},...$ such that for every open set $%
\widehat{U}$ there exist a bijective isomap $\widehat{\varphi }_{\widehat{U}%
}:\widehat{p}^{-1}\left( \widehat{U}\right) \rightarrow
\widehat{U}\times \widehat{V}$ so that $\widehat{p}\left( \varphi
_{\widehat{U}}^{-1}\left( \widehat{x},\widehat{y}\right) \right)
=\widehat{x},~\forall \widehat{x}\in \widehat{U},\forall
\widehat{y}\in \widehat{V};$

2/ if $\widehat{x}\in \widehat{U}\cap \widehat{W}\neq \oslash ,$ than $%
\widehat{\varphi }_{\widehat{W},x}\circ \widehat{\varphi }_{\widehat{U}%
,x}^{-1}:\widehat{V}$ $\rightarrow \widehat{V}$ is an isotopic
diffeomorphism $L_{\widehat{g}r}$ with $\widehat{g}r\in \widehat{G}r$ where $%
\widehat{\varphi }_{\widehat{U},x}$ denotes the restriction of $\widehat{%
\varphi }_{\widehat{U}}$ to $p^{-1}\left( \widehat{x}\right) $ and $\widehat{%
U},\widehat{W}\in {\cal E;}$

3/ the isomap $q_{\widehat{U}\widehat{V}}:\widehat{U}\cap \widehat{V}%
\rightarrow \widehat{G}r$ defined by structural isofunctions $q_{\widehat{U}%
\widehat{V}}\left( \widehat{x}\right) =\widehat{\varphi }_{\widehat{W}%
,x}\circ \widehat{\varphi }_{\widehat{U},x}^{-1}$ is isocontinous.

Let $In_U$ and $In_V$ be sets of indices and denote by $\left( \widehat{U}%
_\alpha ,\widehat{\varphi }_\alpha \right) _{\alpha \in In_U}$
and $\left( \widehat{V}_\beta ,\widehat{\psi }_\beta \right)
_{\beta \in In_V}$ be
correspondingly isocontinous atlases on $\widehat{U}_\alpha $ and $\widehat{V%
}_\beta .\,$ One obtains an isotopic topology on $\widehat{E}$
for which the
bijections $\widehat{\varphi }_{\widehat{U}},\widehat{\varphi }_{\widehat{W}%
,x}...$ become isotopic homeomorphisms. Denoting respectively by
$n$ and $m$ the dimensions of the isomanifolds $\widehat{M}$ and
$\widehat{V}$ we can define the isotopic maps
$$
\phi _{\alpha \beta }:\varpi _{\alpha \beta }\rightarrow \widehat{R}%
^{n+m},\phi _{\alpha \beta }=\left( \widehat{\varphi }_\alpha
\times \widehat{\psi }_\beta \right) \circ \varphi
_{\widehat{U}}^{\alpha \beta }
$$
where $\varphi _{\widehat{U}}^{\alpha \beta }$ is the restriction
to the isomap $\widehat{\varphi }_{\widehat{U}}$ to $\varpi
_{\alpha \beta }.$ Than the set\\ $\left( \varpi _{\alpha \beta
},\phi _{\alpha \beta }\right) _{\left( \alpha ,\beta \right) \in
In_U\times In_V}$ is a isocontinous atlas on $\widehat{E}.$

A locally trivial {\bf principal isobundle} $\left(
\widehat{P},\widehat{\pi
},\widehat{M},\widehat{G}r\right) $ is a fibre isobundle $\left( \widehat{E},%
\widehat{p},\widehat{M},\widehat{V},\widehat{G}r\right) $ for
which the type fibre coincides with the structural group,
$\widehat{V}=\widehat{G}r$ and the action of $\widehat{G}r$ on
$\widehat{G}r$ is given by the left isotransform
$\widehat{L}_q\left( a\right) =qa,\forall q,a\in \widehat{G}r.$

The structural functions of the principal isobundle $\left( \widehat{P},%
\widehat{\pi },\widehat{M},\widehat{G}r\right) $ are
$$
q_{\widehat{U}\widehat{W}}:\widehat{U},\widehat{W}\rightarrow \widehat{G}%
r,~q_{\widehat{U}\widehat{W}}\left( \widehat{\pi }\left( u\right) \right) =%
\widehat{\varphi }_{\widehat{W}}\left( u\right) \circ \widehat{\varphi }_{%
\widehat{U}}^{-1}\left( u\right) ,u\in \widehat{\pi }^{-1}\left( \widehat{U}%
\cap \widehat{W}\right) .
$$

A {\bf morphism of principal isobundles} $\left( \widehat{P},\widehat{\pi },%
\widehat{M},\widehat{G}r\right) $ and \\ $\left( \widehat{P}^{\prime },%
\widehat{\pi }^{\prime },\widehat{M}^{\prime
},\widehat{G}r^{\prime }\right) $ is a pair $\left(
\widehat{f},\widehat{f}^{\prime }\right) $ of isomaps for which
the following conditions hold:

1/ $\widehat{f}:\widehat{P}\rightarrow \widehat{P}^{\prime }$ is a
isocontinous isomap,

2/ $\widehat{f}:\widehat{G}r\rightarrow \widehat{G}r^{\prime }$
is an isotopic morphism of Lie--Santilli isogroups.

3/ $\widehat{f}\left( \widehat{u}\widehat{q}\right) =f\left( \widehat{u}%
\right) f^{\prime }\left( \widehat{q}\right) ,\widehat{u}\in \widehat{P},%
\widehat{q}\in \widehat{G}r.$

We can define isotopic isomorphisms, automorphisms and subbundles
in a usual manner but with respect to isonumbers, isofields,
isogroups and isomanifold when corresponding isotopic transforms
and maps provide the isotopic properties.

A {\bf isotopic subbundle} $\left( \widehat{P},\widehat{\pi },\widehat{M},%
\widehat{G}r\right) $ of the principal isobundle\\ $\left( \widehat{P}%
^{\prime },\widehat{\pi }^{\prime },\widehat{M}^{\prime },\widehat{G}%
r^{\prime }\right) $ is called a reduction of the structural isogroup $%
\widehat{G}r^{\prime }$ to $\widehat{G}r.$

An isotopic frame (isoframe) in a point $\widehat{x}\in
\widehat{M}$ is a
set of $n$ linearly independent isovectors tangent to $\widehat{M}$ in $%
\widehat{x}.$ The set $\widehat{L}\left( \widehat{M}\right) $ of
all isoframes in all points of $\widehat{M}$ can be naturally
provided (as in
the non isotopic case, see, for instance, 
 [142]) with an
isomanifold structure. The principal isobundle $\left(
\widehat{L}\left(
\widehat{M}\right) ,\widehat{\pi },\widehat{M},\widehat{Gl}\left( n,\widehat{%
R}\right) \right) $of isoframes on $\widehat{M},$ denoted in brief also by $%
\widehat{L}\left( \widehat{M}\right) ,$ has $\widehat{L}\left( \widehat{M}%
\right) $ as the total space and the general linear isogroup\\ $\widehat{Gl}%
\left( n,\widehat{R}\right) $ as the structural isogroup.

Having introduced the isobundle $\widehat{L}\left(
\widehat{M}\right) $ we
can give define an {\bf isotopic }$G${\bf --structure} on a isomanifold $%
\widehat{M}$ is a subbundle $\left( \widehat{P},\widehat{\pi },\widehat{M},%
\widehat{G}r\right) $ of the principal isobundle $\widehat{L}\left( \widehat{%
M}\right) $ being an isotopic reduction of the structural isogroup $\widehat{%
Gl}\left( n,\widehat{R}\right) $ to a isotopic subgroup
$\widehat{G}r$ of it.

A very important class of bundle spaces used for modeling of
locally anisotropic interactions is that of vector bundles. We
present here the necessary isotopic generalizations.

An locally trivial {\bf iso\-vec\-tor bund\-le} (equi\-valently,
{\bf vector
iso\-bundle, v--isobundle}) $\left( \widehat{E},\widehat{p},\widehat{M},%
\widehat{V},\widehat{G}r\right) $ is defined as a corresponding
fibre isobundle if $\widehat{V}$ is a linear iso\-spa\-ce and
$\widehat{G}r$ is the Lie--Santilli isogroup of iso\-top\-ic
authomorphisms of $\widehat{V}.$

For $\widehat{V}=\widehat{R}^m$ and $\widehat{G}r=\widehat{Gl}\left( m,%
\widehat{R}\right) $ the v--isobundle\\ $\left( \widehat{E},\widehat{p},%
\widehat{M},\widehat{R}^m,\widehat{Gl}\left( m,\widehat{R}\right)
\right) $
is denoted shortly as $\widehat{\xi }=\left( \widehat{E},\widehat{p},%
\widehat{M}\right) .$ Here we also note that the transformations
of the isocoordinates $\left( \widehat{x}^k,\widehat{y}^a\right)
\rightarrow \left(
\widehat{x}^{k^{\prime }},\widehat{y}^{a^{\prime }}\right) $ on $\widehat{%
\xi }$ are of the form
$$
\widehat{x}^{k^{\prime }}=\widehat{x}^{k^{\prime }}\left( \widehat{x}^1,...,%
\widehat{x}^n\right) ,\quad rank\left( \frac{\widehat{\partial }\widehat{x}%
^{k^{\prime }}}{\widehat{\partial }\widehat{x}^k}\right) =n
$$
$$
\widehat{y}^{a^{\prime }}=\widehat{Y}_a^{a^{\prime }}\left(
x\right) y^a,\quad \widehat{Y}_a^{a^{\prime }}\left( x\right) \in
\widehat{Gl}\left( m,\widehat{R}\right) .
$$

A local isocoordinate parametrization of $\widehat{\xi }$
naturally defines
an isocoordinate basis%
$$
\frac \partial {\partial \widehat{u}^\alpha }=\left( \frac
\partial {\partial x^i},\frac \partial {\partial y^a}\right)
,\eqno(12.1)
$$
in brief we shall write $\widehat{\partial }_\alpha =(\widehat{\partial }_i,%
\widehat{\partial }_a),$ and the reciprocal to (12.1) coordinate
basis
$$
d\widehat{u}^\alpha =(d\widehat{x}^i,d\widehat{y}^a),
$$
or, in brief, $\widehat{d}^\alpha
=(\widehat{d}^i,\widehat{d}^a),$ which is
uniquely defined from the equations%
$$
\widehat{d}^\alpha \circ \widehat{\partial }_\beta =\delta _\beta
^\alpha ,
$$
where $\delta _\beta ^\alpha $ is the Kronecher symbol and by
$"\circ $$"$
we denote the inner (scalar) product in the isotangent isobundle $\widehat{%
T\xi }$ (see the definition of isodifferentials and partial
isoderivations in (11.1)--(11.3)). Here we note that the tangent
isobundle (in brief t--isobundle) of a isomanifold $\widehat{M},$
denoted as $\widehat{TM}=\cup _{x\in \widehat{M}}\widehat{T_xM},$
where $\widehat{T_xM}$ is tangent isospaces of tangent isovectors
in the point $\widehat{x}\in $ $\widehat{M},$
is defined as a v--isobundle $\widehat{E}=\widehat{TM}.$ By $\widehat{TM}%
^{*} $ we define the dual (not confusing with isotopic dual) of
the
t--isobundle $\widehat{TM}.$ We note that for $\widehat{TM}$ and $\widehat{TM%
}^{*}$ isobundles the fibre and the base have both the same
dimension and it is not necessary to distinguish always the fiber
and base indices by different letters.

\section{ Nonlinear and Distinguished Isocon\-nec\-ti\-ons}

The concept of {\bf nonlinear connection,} in brief,
N--connection, is fundamental in the geometry of locally
anisotropic spaces (in brief,
 la--spaces,
see a detailed study and basic references in 
 [160,161]). Here
 it should be noted that we consider the term la--space
in a more general context than
 G. Yu. Bogoslovsky 
 [41] which uses it for a class of Finsler
spaces and Finsler gravitational theories. In our works
 [272,258,259,256,255,264,269,261,254,260,\\ 265,266,267,262,268,249,250,
251,252,279,263,278,277] the la--spaces and la--superspaces are
 respectively modelled on vector bundles and vector superbundles
 enabled with compatible nonlinear and distinguished connections
 and metric structures (as particular cases, on tangent bundles and
 superbundles, for corresponding classes of metrics and nonlinear
 connections, one constructs  generalized Lagrange and Finsler spaces and
 superspaces). The geometrical objects on a la--space are called
{\bf distinguished} (see detailes in 
 [160,161]) if they are
 compatible with the N--connection structure (one considers,
 for instance, distinguished connections and distinguished tensors,
 in brief, d--connections and d--tensors).

 In this section we study, apparently for the first time, the isogeometry
 of N--connection in vector isobundle. We note that a type of
 generic nonlinearity is contained by definition in the  structure
 of isospace (it can be associated to a corresponding class of nonlinear
 isoconnections which can be turned into linear ones under corresponding
 isotopic transforms). As to a general N--connection introduced as
 a  global decompositions of a vector isobundle into  horizontal and vertical
  isotopic subbundles (see below) it can not be isolinearized if its
 isocurvature is nonzero.

Let consider a v--isobundle $\widehat{\xi }=\left( \widehat{E},\widehat{p},%
\widehat{M}\right) $ whose type fibre is $\widehat{R}^m$ and $\widehat{p}^T:%
\widehat{TE}\rightarrow \widehat{TM}$ is the isodifferential of the isomap $%
\widehat{p}.$ The kernel of the isomap $\widehat{p}^T$ (which is a
fibre--preserving isotopic morphism of the t--isobundle $\left( \widehat{TE},%
\widehat{\tau }_E,\widehat{E}\right) $ to $\widehat{E}$ and of t--isobundle $%
\left( \widehat{TM},\widehat{\tau }_M,\widehat{M}\right) $ to
$\widehat{M})$
defines the vertical isotopic subbundle $\left( \widehat{VE},\widehat{\tau }%
_V,\widehat{E}\right) $ over $\widehat{E}$ being an isovector
subbundle of the v--isobundle $\left( \widehat{TE},\widehat{\tau
}_E,\widehat{E}\right) .$

An isovector $\widehat{X}_u$ tangent to $\widehat{E}$ in a point $\widehat{u}%
\in \widehat{E}$ locally defined by the decomposition $\widehat{X}^i\widehat{%
\partial }_i+\widehat{Y}^a\widehat{\partial }_a$ is locally represented by
the isocoordinates%
$$
\widehat{X}_u=\left(
\widehat{x},\widehat{y},\widehat{X},\widehat{Y}\right) =\left(
\widehat{x}^i,\widehat{y}^a,\widehat{X}^i,\widehat{Y}^a\right) .
$$
Since $\widehat{p}^T\left( \widehat{\partial }_a\right) =0$ it results that $%
\widehat{p}^T\left(
\widehat{x},\widehat{y},\widehat{X},\widehat{Y}\right) =\left(
\widehat{x},\widehat{X}\right) .$ We also consider the isotopic
imbedding map $\widehat{i}:\widehat{VE}\rightarrow \widehat{TE}$
and the
isobundle of inverse image $\widehat{p}^{*}\widehat{TM}$ of the $\widehat{p}:%
\widehat{E}\rightarrow \widehat{M}$ and define in result the isomap $%
\widehat{p}!:\widehat{TE}\rightarrow $ $\widehat{p}^{*}\widehat{TM},\widehat{%
p}!\left( \widehat{X}_u\right) =\left(
\widehat{u},\widehat{p}^T\left(
\widehat{X}_u\right) \right) $ for which one holds $Ker~\widehat{p}!=Ker~%
\widehat{p}^T=\widehat{VE}.$

A {\bf nonlinear isoconnection,} (in brief, {\bf
N--isoconnection)} in the iso\-vec\-t\-or bundle $\widehat{\xi
}=\left( \widehat{E},\widehat{p},\widehat{M}\right) $ is defined
as a splitting on
the left of the exact sequence of isotopic maps%
$$
0\longrightarrow \widehat{VE}\stackrel{\widehat{i}}{\longrightarrow }%
\widehat{TE}\stackrel{\widehat{p}!}{\longrightarrow }\widehat{TE}/\widehat{VE%
}\longrightarrow 0
$$
that is an isotopic morphism of vector isobundles $\widehat{C}:\widehat{TE}%
\rightarrow \widehat{VE}$ such that $\widehat{C}\circ
\widehat{i}$ is the identity on $\widehat{VE}.$

The kernel of the isotopic morphism $\widehat{C}$ is a isovector
subbundle of $\left( \widehat{TE},\widehat{\tau
}_E,\widehat{E}\right) $ and will be
called the horizontal isotopic subbundle\\ $\left( \widehat{HE},\widehat{%
\tau }_H,\widehat{E}\right) .$

As a consequence of the above presented definition
 we can consider that a N--isoconnection in
v--isobundle $\widehat{E}$ is a isotopic distribution $\{\widehat{N}:%
\widehat{E}_u\rightarrow
H_u\widehat{E},T_u\widehat{E}=H_u\widehat{E}\oplus
V_u\widehat{E}\}$ on $\widehat{E}$ such that it is defined a
global decomposition, as a Whitney sum, into
horizontal,$\widehat{HE},$ and vertical, $\widehat{VE},$
subbundles of the tangent isobundle $\widehat{TE}:$
$$
\widehat{TE}=\widehat{HE}\oplus \widehat{VE}.\eqno(12.2)
$$

Locally a N--isoconnection in $\widehat{\xi }$ is given by its components $%
\widehat{N}_i^a(\widehat{u}) =
\widehat{N}_i^a(\widehat{x},\widehat{y})$ $=
N_{\widehat{i}}^{\widehat{a}}(\widehat{x},\widehat{y})$ with
respect to local isocoordinate bases (11.14) and (11.15)):
$$
\widehat{{\bf N}}=\widehat{N}_i^a(\widehat{u})\widehat{d}^i\otimes \widehat{%
\partial }_a.
$$

We note that a linear isoconnection in a v--isobundle
$\widehat{\xi }$ can be
considered as a particular case of a N--isoconnection when $\widehat{N}_i^a(%
\widehat{x},\widehat{y})=\widehat{K}_{bi}^a\left(
\widehat{x}\right)
\widehat{y}^b,$ where isofunctions $\widehat{K}_{ai}^b\left( \widehat{x}%
\right) $ on the base $\widehat{M}$ are called the isochristoffel
coefficients.

To coordinate locally geometric constructions with the global
splitting of isobundle defined by a N--isoconnection structure,
we have to introduce a locally adapted isobasis (la---isobasis,
la---isoframe):
$$
\frac{\widehat{\delta }}{\delta \widehat{u}^\alpha }=\left( \frac{\widehat{%
\delta }}{\delta \widehat{x}^i}=\widehat{\partial
}_i-\widehat{N}_i^a\left( \widehat{u}\right) \widehat{\partial
}_a,\frac{\widehat{\partial }}{\partial \widehat{y}^a}\right)
,\eqno(12.3)
$$
or, in brief, $\widehat{\delta }_\alpha =\delta _{\widehat{\alpha
}}=\left(
\widehat{\delta }_i,\widehat{\partial }_a\right) ,$ and its dual la-isobasis%
$$
\widehat{\delta }\widehat{u}^\alpha =\left( \widehat{\delta }\widehat{x}^i=%
\widehat{d}\widehat{x}^i,\widehat{\delta }\widehat{y}^a+\widehat{N}%
_i^a\left( \widehat{u}\right) \widehat{d}\widehat{x}^i\right)
,\eqno(12.4)
$$
or, in brief,{\bf \ }$\ \widehat{\delta }\ ^\alpha =$ $\left( \widehat{d}^i,%
\widehat{\delta }^a\right) .$ We note that isooperators (12.3)
and (12.4) generalize correspondingly the partial isoderivations
and isodifferentials (11.1)--(11.3) for the case when a
N--isoconnection is defined.

The{\bf \ nonholonomic isocoefficients}$\widehat{{\bf \ w}}=\{\widehat{w}%
_{\beta \gamma }^\alpha \left( \widehat{u}\right) \}$ of
la--isoframes are
defined as%
$$
\left[ \widehat{\delta }_\alpha ,\widehat{\delta }_\beta \right] =\widehat{%
\delta }_\alpha \widehat{\delta }_\beta -\widehat{\delta }_\beta \widehat{%
\delta }_\alpha =\widehat{w}_{\beta \gamma }^\alpha \left( \widehat{u}%
\right) \widehat{\delta }_\alpha .
$$

The {\bf algebra of tensorial distinguished isofields}
$D\widehat{T}\left( \widehat{\xi }\right) $ (d--iso\-fi\-elds,
d--isotensors, d--tensor isofield,
d--isobjects) on $\widehat{\xi }$ is introduced as the tensor algebra ${\cal %
T}=\{\widehat{{\cal T}}_{qs}^{pr}\}$ of the v--isobundle
$\widehat{\xi }_{\left( d\right) },$
$\widehat{p}_d:\widehat{HE}\oplus \widehat{VE}\rightarrow
\widehat{E}.$ An element $\widehat{{\bf t}}\in \widehat{{\cal
T}}_{qs}^{pr},$ d--tensor isofield of type $\left(
\begin{array}{cc}
p & r \\
q & s
\end{array}
\right) ,$ can be written in local form as%
$$
\widehat{{\bf t}}=\widehat{t}_{j_1...j_qb_1...b_r}^{i_1...i_pa_1...a_r}%
\left( u\right) \widehat{\delta }_{i_1}\otimes ...\otimes \widehat{\delta }%
_{i_p}\otimes \widehat{\partial }_{a_1}\otimes ...\otimes \widehat{\partial }%
_{a_r}\otimes
$$
$$
\widehat{d}^{j_1}\otimes ...\otimes \widehat{d}^{j_q}\otimes
\widehat{\delta }^{b_1}...\otimes \widehat{\delta }^{b_r}.
$$

We shall respectively use denotations ${\cal X}\left(
\widehat{\xi }\right) $
(or ${\cal X}{\left( \widehat{M}\right) ),\ }\Lambda ^p\left( \widehat{\xi }%
\right) $\\ (or $\Lambda ^p\left( \widehat{M}\right) )$ and
${\cal F}\left( \widehat{\xi }\right) $ (or ${\cal F}\left(
\widehat{M}\right) $) for the isotopic module of d-vector
isofields on $\widehat{\xi }$ (or $\widehat{M}$ ), the exterior
algebra of p-forms on $\widehat{\xi }$ (or $\widehat{M}$ ) and
the set of real functions on $\widehat{\xi }$ (or $\widehat{M}$ ).

In general, d--objects on $\widehat{\xi }$ are introduced as
geometric objects with various isogroup and isocoordinate
transforms coordinated with the N--connection isostructure on
$\widehat{\xi }.$ For example, a d--connection $\widehat{D}$ on
$\widehat{\xi } $ is defined as a isolinear connection
$\widehat{D}$ on $\widehat{E}$ conserving under a parallelism the
global decomposition (12.2) into horizontal and vertical subbundles of $%
\widehat{TE}.$

A N--connection in $\widehat{\xi }$ induces a corresponding
decomposition of d--iso\-ten\-sors into sums of horizontal and
vertical parts, for example, for every d--isovector
$\widehat{X}\in {\cal X}\left( \widehat{\xi }\right) $ and
1--form $\widetilde{X}\in \Lambda ^1\left( \widehat{\xi }\right)
$ we have respectively
$$
\widehat{X}=\widehat{hX}+\widehat{vX}{\bf \ \quad }\mbox{and \quad }%
\widetilde{X}=h\widetilde{X}+v\widetilde{X}~.
$$
In consequence, we can associate to every d--covariant isoderivation $%
\widehat{D_X}$ $=\widehat{X}\circ \widehat{D}$ two new operators
of h- and v-covariant isoderivations defined respectively as
$$
\widehat{D}_X^{(h)}\widehat{Y}=\widehat{D}_{hX}\widehat{Y}\quad
\mbox{ and \quad }\widehat{D}_X^{\left( v\right) }\widehat{Y}=\widehat{D}%
_{vX}\widehat{Y}{\bf ,\quad }\forall \widehat{Y}{\bf \in }{\cal
X}\left( \widehat{\xi }\right)
$$
for which the following conditions hold:%
$$
\widehat{D}_X\widehat{Y}{\bf =}\widehat{D}_X^{(h)}\widehat{Y}{\bf \ +}%
\widehat{D}_X^{(v)}\widehat{Y}{\bf ,}\eqno(12.5)
$$
$$
\widehat{D}_X^{(h)}f=(h\widehat{X}{\bf )}f\mbox{ \quad and\quad }\widehat{D}%
_X^{(v)}f=(v\widehat{X}{\bf )}f,\quad \widehat{X},\widehat{Y}{\bf \in }{\cal %
X}\left( \widehat{\xi }\right) ,f\in {\cal F}\left(
\widehat{M}\right) .
$$

An {\bf isometric structure } $\widehat{{\bf G}}$ in the total
space $\widehat{E} $
of v--isobundle\\ $\widehat{\xi } = $ $\left( \widehat{E},\widehat{p},\widehat{%
M}\right) $ over a connected and paracompact base $\widehat{M}$ is
introduced as a symmetrical covariant tensor isofield of type
$\left( 0,2\right) $, $\widehat{G}_{\alpha \beta ,}$ being
nondegenerate and of constant signature on $\widehat{E}.$

Nonlinear isoconnection $\widehat{{\bf N}}$ and isometric
$\widehat{{\bf G}}$ structures on $\widehat{\xi }$ are mutually
compatible it there are satisfied the conditions:
$$
\widehat{{\bf G}}\left( \widehat{\delta }_i,\widehat{\partial
}_a\right) =0
$$
which in component form are written as
$$
\widehat{G}_{ia}\left( \widehat{u}\right) -\widehat{N}_i^b\left( \widehat{u}%
\right) \widehat{h}_{ab}\left( \widehat{u}\right) =0,\eqno(12.6)
$$
where $\widehat{h}_{ab}=\widehat{{\bf G}}\left( \widehat{\partial }_a,%
\widehat{\partial }_b\right) $ and
$\widehat{G}_{ia}=\widehat{{\bf G}}\left(
\widehat{\partial }_i,\widehat{\partial }_a\right) $ (the matrix $\widehat{h}%
^{ab}$ is inverse to $\widehat{h}_{ab}).$

In consequence one obtains the following decomposition of isotopic metric :%
$$
\widehat{{\bf G}}(\widehat{X},\widehat{Y}){\bf =}\widehat{{\bf hG}}(\widehat{%
X},\widehat{Y})+\widehat{{\bf vG}}(\widehat{X},\widehat{Y})
$$
where the d-tensor $\widehat{{\bf hG}}(\widehat{X},\widehat{Y}){\bf =}%
\widehat{{\bf G}}(\widehat{hX},\widehat{hY})$ is of type $\left(
\begin{array}{cc}
0 & 0 \\
2 & 0
\end{array}
\right) $ and the d--isotensor
 $\widehat{{\bf vG}}(\widehat{X},\widehat{Y}){\bf =%
}\widehat{{\bf G}}(v\widehat{X},v\widehat{Y})$ is of type $\left(
\begin{array}{cc}
0 & 0 \\
0 & 2
\end{array}
\right) .$ With respect to la--isobasis (12.4) the d--isometric
is written
as%
$$
\widehat{{\bf G}}=\widehat{g}_{\alpha \beta }\left(
\widehat{u}\right)
\widehat{\delta }^\alpha \otimes \widehat{\delta }^\beta =\widehat{g}%
_{ij}\left( \widehat{u}\right) \widehat{d}^i\otimes \widehat{d}^j+\widehat{h}%
_{ab}\left( \widehat{u}\right) \widehat{\delta }^a\otimes \widehat{\delta }%
^b,\eqno(12.7)
$$
where $\widehat{g}_{ij}=\widehat{{\bf G}}\left( \widehat{\delta }_i,\widehat{%
\delta }_j\right) .$

A metric isostructure of type (12.7) on $\widehat{E}$ with
components satisfying constraints (12.4)) defines an adapted to
the given N--isoconnection inner (d--scalar) isoproduct on the
tangent isobundle $\widehat{TE}.$

A d--isoconnection $\widehat{D}_X$ is {\bf compatible } with an isometric $%
\widehat{{\bf G}}$ on $\widehat{\xi }$ if%
$$
\widehat{D}_X\widehat{{\bf G}}=0,\forall \widehat{X}{\bf \in }\widehat{{\cal %
X}}\left( \widehat{\xi }\right) .
$$
Locally adapted components $\widehat{\Gamma }_{\beta \gamma
}^\alpha $ of a
d--isoconnection $\widehat{D}_\alpha =(\widehat{\delta }_\alpha \circ \widehat{D}%
)$ are defined by the equations%
$$
\widehat{D}_\alpha \widehat{\delta }_\beta =\widehat{\Gamma
}_{\alpha \beta }^\gamma \widehat{\delta }_\gamma ,
$$
from which one immediately follows%
$$
\widehat{\Gamma }_{\alpha \beta }^\gamma \left(
\widehat{u}\right) =\left(
\widehat{D}_\alpha \widehat{\delta }_\beta \right) \circ \widehat{\delta }%
^\gamma .\eqno(12.8)
$$
The operations of h- and v--covariant
 isoderivations, $\widehat{D}_k^{(h)}=\{%
\widehat{L}_{jk}^i,\widehat{L}_{bk\;}^a\}$ and $\widehat{D}_c^{(v)}=\{%
\widehat{C}_{jk}^i,\widehat{C}_{bc}^a\}$ (see (4.4)), are
introduced as
corresponding h-- and v--parametrizations of (4.7):%
$$
\widehat{L}_{jk}^i=\left( \widehat{D}_k\widehat{\delta }_j\right)
\circ \widehat{d}^i,\quad \widehat{L}_{bk}^a=\left(
\widehat{D}_k\widehat{\partial }_b\right) \circ \widehat{\delta
}^a\eqno(12.9)
$$
and%
$$
\widehat{C}_{jc}^i=\left( \widehat{D}_c\widehat{\delta }_j\right)
\circ \widehat{d}^i,\quad \widehat{C}_{bc}^a=\left(
\widehat{D}_c\widehat{\partial }_b\right) \circ \widehat{\delta
}^a.\eqno(12.10)
$$
Components (12.9) and (12.10), $\widehat{D}\widehat{\Gamma }=\left( \widehat{L}%
_{jk}^i,\widehat{L}_{bk}^a,\widehat{C}_{jc}^i,\widehat{C}_{bc}^a\right)
,$
completely defines the local action of a d--isoconnection $\widehat{D}$ in $%
\widehat{\xi }.$ For instance, taken a d--tensor isofield of type
$\left(
\begin{array}{cc}
1 & 1 \\
1 & 1
\end{array}
\right) ,$
$$
\widehat{{\bf t}}=\widehat{t}_{jb}^{ia}\widehat{\delta }_i\otimes \widehat{%
\partial }_a\otimes \widehat{\partial }^j\otimes \widehat{\delta }^b,
$$
and a d-vector $\widehat{{\bf X}}=\widehat{X}^i\widehat{\delta }_i+\widehat{X%
}^a\widehat{\partial }_a$ we have%
$$
\widehat{D}_X\widehat{{\bf t}}{\bf =}\widehat{D}_X^{(h)}\widehat{{\bf t}}%
{\bf +}\widehat{D}_X^{(v)}\widehat{{\bf t}}{\bf =}\left( \widehat{X}^k%
\widehat{t}_{jb|k}^{ia}+\widehat{X}^c\widehat{t}_{jb\perp
c}^{ia}\right) \widehat{\delta }_i\otimes \widehat{\partial
}_a\otimes \widehat{d}^j\otimes \widehat{\delta }^b,
$$
where the{\bf \ h--covariant and v--covariant isoderivatives} are
written
respectively as%
$$
\widehat{t}_{jb|k}^{ia}=\frac{\widehat{\delta
}\widehat{t}_{jb}^{ia}}{\delta
\widehat{x}^k}+\widehat{L}_{hk}^i\widehat{t}_{jb}^{ha}+\widehat{L}_{ck}^a%
\widehat{t}_{jb}^{ic}-\widehat{L}_{jk}^h\widehat{t}_{hb}^{ia}-\widehat{L}%
_{bk}^c\widehat{t}_{jc}^{ia}
$$
and
$$
\widehat{t}_{jb\perp c}^{ia}=\frac{\widehat{\partial }\widehat{t}_{jb}^{ia}}{%
\partial \widehat{y}^c}+\widehat{C}_{hc}^i\widehat{t}_{jb}^{ha}+\widehat{C}%
_{dc}^a\widehat{t}_{jb}^{id}-\widehat{C}_{jc}^h\widehat{t}_{hb}^{ia}-%
\widehat{C}_{bc}^d\widehat{t}_{jd}^{ia}.
$$
For a scalar isofunction $f\in {\cal F(}\widehat{\xi })$ we have
$$
\widehat{D}_k^{(h)}=\frac{\widehat{\delta }f}{\delta \widehat{x}^k}=\frac{%
\widehat{\partial }f}{\partial \widehat{x}^k}-\widehat{N}_k^a\frac{\widehat{%
\partial }f}{\partial \widehat{y}^a}\mbox{ and }\widehat{D}_c^{(v)}f=\frac{%
\widehat{\partial }f}{\partial \widehat{y}^c}.
$$
We emphasize that the geometry of connections in a v--isobundle $\widehat{%
\xi }$ is very reach. For instance, if a triple of fundamental
isogeometric
objects $(\widehat{N}_i^a\left( \widehat{u}\right) ,$ $\widehat{\Gamma }%
_{\beta \gamma }^\alpha \left( \widehat{u}\right) ,$
$\widehat{G}_{\alpha \beta }\left( \widehat{u}\right) )$ is fixed
on $\widehat{\xi },$ a multi--isoconnection structure (with
corresponding rules of covariant isoderivation, which are, or
not, mutually compatible and with the same, or not, induced
d--scalar products in $\widehat{TE})$ is defined.

Let enumerate some of isoconnections and covariant isoderivations
which can present interest in investigation of locally
anisotropic and
 homogeneous gravitational and
matter field isotopic interactions:

\begin{enumerate}
\item  Every N--isoconnection in $\widehat{\xi }$ with coefficients $\widehat{N}%
_i^a\left( \widehat{x},\widehat{y}\right) $ being
isodifferentiable on
y-variables induces a structure of isolinear isoconnection $\widetilde{N}%
_{\beta \gamma }^\alpha ,$ where $\widetilde{N}_{bi}^a=\frac{\widehat{%
\partial }\widehat{N}_i^a}{\partial \widehat{y}^b}$ and $\widetilde{N}%
_{bc}^a\left( \widehat{x},\widehat{y}\right) =0.$ For some $\widehat{Y}%
\left( \widehat{u}\right) =\widehat{Y}^i\left( \widehat{u}\right) \widehat{%
\partial }_i+\widehat{Y}^a\left( \widehat{u}\right) \widehat{\partial }_a$
and $\widehat{B}\left( \widehat{u}\right) =\widehat{B}^a\left( \widehat{u}%
\right) \widehat{\partial }_a$ one writes%
$$
\widehat{D}_Y^{(\widetilde{N})}\widehat{B}=\left[ \widehat{Y}^i\left( \frac{%
\widehat{\partial }\widehat{B}^a}{\partial \widehat{x}^i}+\widetilde{N}%
_{bi}^a\widehat{B}^b\right) +\widehat{Y}^b\frac{\widehat{\partial }\widehat{B%
}^a}{\partial \widehat{y}^b}\right] \frac{\widehat{\partial
}}{\partial \widehat{y}^a}.
$$

\item  The d--isoconnection of Berwald type [39]
$$
\widehat{\Gamma }_{\beta \gamma }^{(B)\alpha }=\left( \widehat{L}_{jk}^i,%
\frac{\widehat{\partial }\widehat{N}_k^a}{\partial \widehat{y}^b},0,\widehat{%
C}_{bc}^a\right) ,
$$
where
$$
\widehat{L}_{.jk}^i\left( \widehat{x},\widehat{y}\right) =\frac 12\widehat{g}%
^{ir}\left( \frac{\widehat{\delta }\widehat{g}_{jk}}{\delta \widehat{x}^k}+%
\frac{\widehat{\delta }\widehat{g}_{kr}}{\delta \widehat{x}^j}-\frac{%
\widehat{\delta }\widehat{g}_{jk}}{\delta \widehat{x}^r}\right)
,\eqno(12.11)
$$
$$
\widehat{C}_{.bc}^a\left( \widehat{x},\widehat{y}\right) =\frac 12\widehat{h}%
^{ad}\left( \frac{\widehat{\partial }\widehat{h}_{bd}}{\partial \widehat{y}^c%
}+\frac{\widehat{\partial }\widehat{h}_{cd}}{\partial \widehat{y}^b}-\frac{%
\partial \widehat{h}_{bc}}{\partial \widehat{y}^d}\right) ,
$$
 is {\bf hv-isometric,} i.e. $\widehat{D}_k^{(B)}\widehat{g}_{ij}=0$
and $\widehat{D}_c^{(B)}\widehat{h}_{ab}=0.$

\item  The isocanonical d--isoconnection $\widehat{{\bf \Gamma }}{\bf ^{(c)}}$
is associated to a isometric $\widehat{{\bf G}}$ of type (12.6) $\widehat{%
\Gamma }_{\beta \gamma }^{(c)\alpha }=(\widehat{L}_{jk}^{(c)i},\widehat{L}%
_{bk}^{(c)a},\widehat{C}_{jc}^{(c)i},\widehat{C}_{bc}^{(c)a}),$
with
coefficients (see (12.11))%
$$
\widehat{L}_{jk}^{(c)i}=\widehat{L}_{.jk}^i,\widehat{C}_{bc}^{(c)a}=\widehat{%
C}_{.bc}^a\eqno(12.12))
$$
$$
\widehat{L}_{bi}^{(c)a}=\widetilde{N}_{bi}^a+\frac
12\widehat{h}^{ac}\left(
\frac{\widehat{\delta }\widehat{h}_{bc}}{\delta \widehat{x}^i}-\widetilde{N}%
_{bi}^d\widehat{h}_{dc}-\widetilde{N}_{ci}^d\widehat{h}_{db}\right)
,
$$
$$
\widehat{C}_{jc}^{(c)i}=\frac 12\widehat{g}^{ik}\frac{\widehat{\partial }%
\widehat{g}_{jk}}{\partial \widehat{y}^c}.
$$
This is a isometric d--isoconnection which satisfies
compatibility conditions
$$
\widehat{D}_k^{(c)}\widehat{g}_{ij}=0,\widehat{D}_c^{(c)}\widehat{g}_{ij}=0,%
\widehat{D}_k^{(c)}\widehat{h}_{ab}=0,\widehat{D}_c^{(c)}\widehat{h}_{ab}=0.
$$

\item  We can consider N--adapted {\bf isochristoffel distinguished symbols}
(as in (11.5))%
$$
\widetilde{\Gamma }_{\beta \gamma }^\alpha =\frac
12\widehat{G}^{\alpha \tau
}\left( \widehat{\delta }_\gamma \widehat{G}_{\tau \beta }+\widehat{\delta }%
_\beta \widehat{G}_{\tau \gamma }-\widehat{\delta }_\tau
\widehat{G}_{\beta \gamma }\right) ,\eqno(12.13)
$$
which have the components of d--connection $\widetilde{\Gamma
}_{\beta \gamma }^\alpha =\left(
\widehat{L}_{jk}^i,0,0,\widehat{C}_{bc}^a\right) ,$
with $\widehat{L}_{jk}^i$ and $\widehat{C}_{bc}^a$ as in (12.11) if $\widehat{%
G}_{\alpha \beta }$ is taken in the form (12.7).
\end{enumerate}

Arbitrary isolinear isoconnections on a v--isobundle
$\widehat{\xi }$ can be also characterized by theirs deformation
isotensors with respect, for
instance, to d--isoconnection (12.13):%
$$
\widehat{\Gamma }_{\beta \gamma }^{(B)\alpha }=\widetilde{\Gamma
}_{\beta
\gamma }^\alpha +\widehat{P}_{\beta \gamma }^{(B)\alpha },\widehat{\Gamma }%
_{\beta \gamma }^{(c)\alpha }=\widetilde{\Gamma }_{\beta \gamma }^\alpha +%
\widehat{P}_{\beta \gamma }^{(c)\alpha }
$$
or, in general,%
$$
\widehat{\Gamma }_{\beta \gamma }^\alpha =\widetilde{\Gamma
}_{\beta \gamma }^\alpha +\widehat{P}_{\beta \gamma }^\alpha
,\eqno(12.14)
$$
where $\widehat{P}_{\beta \gamma }^{(B)\alpha
},\widehat{P}_{\beta \gamma }^{(c)\alpha }$ and
$\widehat{P}_{\beta \gamma }^\alpha $ are corresponding
deformation d--isotensors of d--iso\-con\-nec\-ti\-ons.

\section{ Isotorsions and Isocurvatures}

The notions of isotorsion and isocurvature were introduced in the
ref.
 [223] for an isoriemannian spaces.  In this section we
 reformulate these notions on isobundles provided with N--isoconnection and
 d--isoconnection structures.

The {\bf isocurvature} $\widehat{{\bf \Omega }}\,$ of a {\bf
nonlinear
 isoconnection} $\widehat{{\bf N}}$ in a v--iso\-bund\-le
 $\widehat{\xi }$ can be
defined as the Nijenhuis tensor isofield $\widehat{N}_v\left( \widehat{X},%
\widehat{Y}\right) $ associated to $\widehat{{\bf N}}$ (this is
an isotopic transform for N--curvature considered, for instance,
in
 [160,161]):
$$
\widehat{{\bf \Omega }}=\widehat{N}_v={\bf \left[ v\widehat{X},v\widehat{Y}%
\right] +v\left[ \widehat{X},\widehat{Y}\right] -v\left[ v\widehat{X},%
\widehat{Y}\right] -v\left[ \widehat{X},v\widehat{Y}\right] ,}\widehat{{\bf X%
}}{\bf ,}\widehat{{\bf Y}}\in {\cal X}\left( \widehat{\xi }\right)
$$
having this local representation%
$$
\widehat{{\bf \Omega }}=\frac 12\widehat{\Omega }_{ij}^a\widehat{d}%
^i\bigwedge \widehat{d}^j\otimes \widehat{\partial }_a,
$$
where%
$$
\widehat{\Omega }_{ij}^a=\frac{\widehat{\partial
}\widehat{N}_i^a}{\partial
\widehat{x}^j}-\frac{\widehat{\partial }\widehat{N}_j^a}{\partial \widehat{x}%
^i}+\widehat{N}_i^b\widetilde{N}_{bj}^a-\widehat{N}_j^b\widetilde{N}_{bi}^a.%
\eqno(12.15)
$$

The {\bf isotorsion} $\widehat{{\bf T}}$ of a {\bf
d--isoconnection}
 $\widehat{%
{\bf D}}$ in $\widehat{\xi }$ is defined by the equation%
$$
\widehat{{\bf T}}{\bf \left( \widehat{X},\widehat{Y}\right) =}\widehat{D}_X%
\widehat{{\bf Y}}{\bf -}\widehat{D}_Y\widehat{{\bf X}}{\bf \
-\left[ \widehat{X},\widehat{Y}\right] }. \eqno(12.16)
$$
One holds the following h- and v--decompositions%
$$
\widehat{{\bf T}}{\bf \left( \widehat{X},\widehat{Y}\right) =}\widehat{{\bf T%
}}{\bf \left( h\widehat{X},h\widehat{Y}\right) +}\widehat{{\bf T}}{\bf %
\left( h\widehat{X},v\widehat{Y}\right) +}\widehat{{\bf T}}{\bf \left( v%
\widehat{X},h\widehat{Y}\right) +}\widehat{{\bf T}}{\bf \left( v\widehat{X},v%
\widehat{Y}\right) .}\eqno(12.17)
$$
We consider the projections:
$$
{\bf h}\widehat{{\bf T}}{\bf \left( \widehat{X},\widehat{Y}\right) ,v}%
\widehat{{\bf T}}{\bf \left( h\widehat{X},h\widehat{Y}\right) ,h}\widehat{%
{\bf T}}{\bf \left( h\widehat{X},h\widehat{Y}\right) ,...}
$$
and say that, for instance, ${\bf h}\widehat{{\bf T}}{\bf \left( h\widehat{X}%
,h\widehat{Y}\right) }$ is the h(hh)--isotorsion of $\widehat{{\bf D}},$\\ $%
{\bf v}\widehat{{\bf T}}{\bf \left(
h\widehat{X},h\widehat{Y}\right) }$ is the v(hh)--isotorsion of
$\widehat{{\bf D}}$ and so on.

The isotorsion (12.16) is locally determined by five d--tensor
isofields, isotorsions, defined as
$$
\widehat{T}_{jk}^i={\bf h}\widehat{{\bf T}}\left( \widehat{\delta }_k,%
\widehat{\delta }_j\right) \cdot \widehat{d}^i,\quad \widehat{T}_{jk}^a={\bf %
v}\widehat{{\bf T}}\left( \widehat{\delta }_k,\widehat{\delta
}_j\right) \cdot \widehat{\delta }^a,
$$
$$
\widehat{P}_{jb}^i={\bf h}\widehat{{\bf T}}\left( \widehat{\partial }_b,%
\widehat{\delta }_j\right) \cdot \widehat{d}^i,\quad \widehat{P}_{jb}^a={\bf %
v}\widehat{{\bf T}}\left( \widehat{\partial }_b,\widehat{\delta
}_j\right) \cdot \widehat{\delta }^a,
$$
$$
\widehat{S}_{bc}^a={\bf v}\widehat{{\bf T}}\left( \widehat{\partial }_c,%
\widehat{\partial }_b\right) \cdot \widehat{\delta ^a}.
$$
Using formulas (12.3), (12.4), (12.14) and (12.16) we compute in
explicit form the components of isotorsions (12.17)
 for a d--isoconnection of type (12.9) and (12.10):
$$
\widehat{T}_{.jk}^i=\widehat{T}_{jk}^i=\widehat{L}_{jk}^i-\widehat{L}%
_{kj}^i,\quad \widehat{T}_{ja}^i=\widehat{C}_{.ja}^i,\widehat{T}_{aj}^i=-%
\widehat{C}_{ja}^i,\eqno(12.18)
$$
$$
\widehat{T}_{.ja}^i=0,\widehat{T}_{.bc}^a=\widehat{S}_{.bc}^a=\widehat{C}%
_{bc}^a-\widehat{C}_{cb}^a,
$$
$$
\widehat{T}_{.ij}^a=\frac{\widehat{\delta }N_i^a}{\delta \widehat{x}^j}-%
\frac{\widehat{\delta }\widehat{N}_j^a}{\delta \widehat{x}^i},\quad \widehat{%
T}_{.bi}^a=\widehat{P}_{.bi}^a=\frac{\widehat{\partial }\widehat{N}_i^a}{%
\partial \widehat{y}^b}-\widehat{L}_{.bj}^a,\quad \widehat{T}_{.ib}^a=-%
\widehat{P}_{.bi}^a.
$$

The {\bf isocurvature} $\widehat{{\bf R}}$ of a {\bf d--isoconnection} in $%
\widehat{\xi }$ is defined by the equation
$$
\widehat{{\bf R}}{\bf \left( \widehat{X},\widehat{Y}\right) }\widehat{{\bf Z}%
}{\bf =}\widehat{D}_X\widehat{D}_Y\widehat{{\bf Z}}-\widehat{D}_Y\widehat{D}%
_X\widehat{{\bf Z}}{\bf -}\widehat{D}_{[X,Y]}\widehat{{\bf Z}}{\bf .}%
\eqno(12.19)
$$
One holds the next properties for the h- and v--decompositions of
isocurvature:%
$$
{\bf v}\widehat{{\bf R}}{\bf \left( \widehat{X},\widehat{Y}\right) h}%
\widehat{{\bf Z}}{\bf =0,\ h}\widehat{{\bf R}}{\bf \left( \widehat{X},%
\widehat{Y}\right) v}\widehat{{\bf Z}}{\bf =0,}
$$
$$
\widehat{{\bf R}}{\bf \left( \widehat{X},\widehat{Y}\right) }\widehat{{\bf Z}%
}{\bf =h}\widehat{{\bf R}}{\bf \left( \widehat{X},\widehat{Y}\right) h}%
\widehat{{\bf Z}}{\bf +v}\widehat{{\bf R}}{\bf \left( \widehat{X},\widehat{Y}%
\right) v}\widehat{{\bf Z}}{\bf .}
$$
From (12.19) and the equation $\widehat{{\bf R}}{\bf \left( \widehat{X},%
\widehat{Y}\right) =-}\widehat{{\bf R}}{\bf \left( \widehat{Y},\widehat{X}%
\right) }$ we conclude that the curvature of a d-con\-nec\-ti\-on $\widehat{%
{\bf D}}$ in $\widehat{\xi }$ is completely determined by the
following six
d--tensor isofields:%
$$
\widehat{R}_{h.jk}^{.i}=\widehat{d}^i\cdot \widehat{{\bf R}}\left( \widehat{%
\delta }_k,\widehat{\delta }_j\right) \widehat{\delta }_h,~\widehat{R}%
_{b.jk}^{.a}=\widehat{\delta }^a\cdot \widehat{{\bf R}}\left( \widehat{%
\delta }_k,\widehat{\delta }_j\right) \widehat{\partial
}_b,\eqno(12.20)
$$
$$
\widehat{P}_{j.kc}^{.i}=\widehat{d}^i\cdot \widehat{{\bf R}}\left( \widehat{%
\partial }_c,\widehat{\partial }_k\right) \widehat{\delta }_j,~\widehat{P}%
_{b.kc}^{.a}=\widehat{\delta }^a\cdot \widehat{{\bf R}}\left( \widehat{%
\partial }_c,\widehat{\partial }_k\right) \widehat{\partial }_b,
$$
$$
\widehat{S}_{j.bc}^{.i}=\widehat{d}^i\cdot \widehat{{\bf R}}\left( \widehat{%
\partial }_c,\widehat{\partial }_b\right) \widehat{\delta }_j,~\widehat{S}%
_{b.cd}^{.a}=\widehat{\delta }^a\cdot \widehat{{\bf R}}\left( \widehat{%
\partial }_d,\widehat{\partial }_c\right) \widehat{\partial }_b.
$$
By a direct computation, using (12.3),(12.4),(12.9),(12.10)
 and (12.20) we get:
$$
\widehat{R}_{h.jk}^{.i}=\frac{\widehat{\delta
}\widehat{L}_{.hj}^i}{\delta
\widehat{x}^h}-\frac{\widehat{\delta }\widehat{L}_{.hk}^i}{\delta \widehat{x}%
^j}+\widehat{L}_{.hj}^m\widehat{L}_{mk}^i-\widehat{L}_{.hk}^m\widehat{L}%
_{mj}^i+\widehat{C}_{.ha}^i\widehat{R}_{.jk}^a,\eqno(12.21)
$$
$$
\widehat{R}_{b.jk}^{.a}=\frac{\widehat{\delta
}\widehat{L}_{.bj}^a}{\delta
\widehat{x}^k}-\frac{\widehat{\delta }\widehat{L}_{.bk}^a}{\delta \widehat{x}%
^j}+\widehat{L}_{.bj}^c\widehat{L}_{.ck}^a-\widehat{L}_{.bk}^c\widehat{L}%
_{.cj}^a+\widehat{C}_{.bc}^a\widehat{R}_{.jk}^c,
$$
$$
\widehat{P}_{j.ka}^{.i}=\frac{\widehat{\partial }\widehat{L}_{.jk}^i}{%
\partial \widehat{y}^k}-\left( \frac{\widehat{\partial }\widehat{C}_{.ja}^i}{%
\partial \widehat{x}^k}+\widehat{L}_{.lk}^i\widehat{C}_{.ja}^l-\widehat{L}%
_{.jk}^l\widehat{C}_{.la}^i-\widehat{L}_{.ak}^c\widehat{C}_{.jc}^i\right) +%
\widehat{C}_{.jb}^i\widehat{P}_{.ka}^b,
$$
$$
\widehat{P}_{b.ka}^{.c}=\frac{\widehat{\partial }\widehat{L}_{.bk}^c}{%
\partial \widehat{y}^a}-\left( \frac{\widehat{\partial }\widehat{C}_{.ba}^c}{%
\partial \widehat{x}^k}+\widehat{L}_{.dk}^{c\,}\widehat{C}_{.ba}^d-\widehat{L%
}_{.bk}^d\widehat{C}_{.da}^c-\widehat{L}_{.ak}^d\widehat{C}_{.bd}^c\right) +%
\widehat{C}_{.bd}^c\widehat{P}_{.ka}^d,
$$
$$
\widehat{S}_{j.bc}^{.i}=\frac{\widehat{\partial }\widehat{C}_{.jb}^i}{%
\partial \widehat{y}^c}-\frac{\widehat{\partial }\widehat{C}_{.jc}^i}{%
\partial \widehat{y}^b}+\widehat{C}_{.jb}^h\widehat{C}_{.hc}^i-\widehat{C}%
_{.jc}^h\widehat{C}_{hb}^i,
$$
$$
\widehat{S}_{b.cd}^{.a}=\frac{\widehat{\partial }\widehat{C}_{.bc}^a}{%
\partial \widehat{y}^d}-\frac{\widehat{\partial }\widehat{C}_{.bd}^a}{%
\partial \widehat{y}^c}+\widehat{C}_{.bc}^e\widehat{C}_{.ed}^a-\widehat{C}%
_{.bd}^e\widehat{C}_{.ec}^a.
$$

We note that isotorsions (12.18) and isocurvatures (12.21) can be
computed by
particular cases of d--isoconnections when d--isoconnections\\
(12.12), or (12.13) are used instead of (12.9) and (12.10). The
above presented
 formulas are similar
to (11.8),(11.9) and (11.10) being distinguished (in the case of
 locally anisotropic and inhomogeneous  isospaces) by N--isoconnection
structure.

For our further considerations it is useful to compute
deformations of isotorsion (12.16) and isocurvature (12.19) under
 deformations of d--con\-nec\-ti\-ons
(12.14). Putting the splitting (12.14),
 $\widehat{{\Gamma }}{{^\alpha }_{\beta
\gamma }}={{\tilde \Gamma }_{\cdot \beta \gamma }^\alpha }+\widehat{{P}}{{%
^\alpha }_{\beta \gamma }},$into (12.16) and (12.19)
 we can express isotorsion $%
\widehat{{T}}{^\alpha }_{\beta \gamma }$ and isocurvature $\widehat{{R}}{{%
_\beta }^\alpha }_{\gamma \delta }$ of a d--isoconnection
 $\widehat{{\Gamma }}{%
^\alpha }_{\beta \gamma }$ as respective deformations of
isotorsion
 ${{\tilde T}%
^\alpha }_{\beta \gamma }$ and isotorsion ${\tilde R}_{\beta
\cdot \gamma \delta }^{\cdot \alpha }$ for connection ${{\tilde
\Gamma }^\alpha }_{\beta \gamma }{\quad }:$
$$
{{T^\alpha }_{\beta \gamma }}={{\tilde T}_{\cdot \beta \gamma }^\alpha }+{{%
\ddot T}_{\cdot \beta \gamma }^\alpha }
$$
and
$$
{{{R_\beta }^\alpha }_{\gamma \delta }}={{\tilde R}_{\beta \cdot
\gamma \delta }^{\cdot \alpha }}+{{\ddot R}_{\beta \cdot \gamma
\delta }^{\cdot \alpha }},
$$
where
$$
{{\tilde T}^\alpha }_{\beta \gamma }={{\tilde \Gamma }^\alpha
}_{\beta \gamma }-{{\tilde \Gamma }^\alpha }_{\gamma \beta
}+{w^\alpha }_{\gamma
\delta },\qquad {{\ddot T}^\alpha }_{\beta \gamma }={{\ddot \Gamma }^\alpha }%
_{\beta \gamma }-{{\ddot \Gamma }^\alpha }_{\gamma \beta },
$$
and
$$
{{\tilde R}_{\beta \cdot \gamma \delta }^{\cdot \alpha }}={{\delta }_\delta }%
{{\tilde \Gamma }^\alpha }_{\beta \gamma }-{{\delta }_\gamma
}{{\tilde \Gamma }^\alpha }_{\beta \delta }+{{{\tilde \Gamma
}^\varphi }_{\beta \gamma
}}{{{\tilde \Gamma }^\alpha }_{\varphi \delta }}-{{{\tilde \Gamma }^\varphi }%
_{\beta \delta }}{{{\tilde \Gamma }^\alpha }_{\varphi \gamma
}}+{{\tilde \Gamma }^\alpha }_{\beta \varphi }{w^\varphi
}_{\gamma \delta },
$$
$$
{{\ddot R}_{\beta \cdot \gamma \delta }^{\cdot \alpha }}={{\tilde D}_\delta }%
{{P^\alpha }_{\beta \gamma }}-{{\tilde D}_\gamma }{{P^\alpha
}_{\beta \delta }}+{{P^\varphi }_{\beta \gamma }}{{P^\alpha
}_{\varphi \delta }}-{{P^\varphi }_{\beta \delta }}{{P^\alpha
}_{\varphi \gamma }}+{{P^\alpha }_{\beta \varphi }}{{w^\varphi
}_{\gamma \delta }}.
$$

\section{Isobianchi and Isoricci Identities}

The isobianchi and isoricci identities were first studied by
Santilli
 [223] on an isoriemannian space. On spaces with  N--connection
 structures the general formulas for Bianchi and Ricci identities
(for osculator and vector bundles,
 generalized Lagrange and Finsler geometry) have been considered by
 Miron and Anastasiei 
 [160,161] and Miron and Atanasiu
[162]. We have extended the Miron--Anastasiei--Atanasiu
   constructions for superspaces with local and higher order anisotropy
in refs. 
 [260,265,266,267,270].
The purpose of
 this section is to consider distinguished isobianchi and isorichi
 for vector isobundles.

The isotorsion and isocurvature of every linear isoconnection
$\widehat{D}$ on a v--isobundle satisfy the following {\bf
generalized isobianchi identities:}
$$
\sum {[(}\widehat{{D}}{_{\widehat{X}}}\widehat{{T}}{)(}\widehat{{Y}}{,}%
\widehat{{Z}}{)-}\widehat{{R}}{(}\widehat{{X}}{,}\widehat{{Y}}{)}\widehat{{Z}%
}{+}\widehat{{T}}{(}\widehat{{T}}{(}\widehat{{X}}{,}\widehat{{Y}}{),}%
\widehat{{Z}}{)]}=0,\eqno(12.22)
$$
$$
\sum {[(}\widehat{{D}}{_{\widehat{X}}}\widehat{{R}}{)(}\widehat{{U}}{,}%
\widehat{{Y}}{,}\widehat{{Z}}{)+}\widehat{{R}}{(}\widehat{{T}}{(}\widehat{{X}%
}{,}\widehat{{Y}}{)}\widehat{{Z}}{)}\widehat{{U}}{]}=0,
$$
where $\sum $ means the respective cyclic sum over $\widehat{X},\widehat{Y},%
\widehat{Z}$ and $\widehat{U}.$ Using the property that
$$
v(\widehat{D}_X\widehat{{R}})(\widehat{U},\widehat{Y},h\widehat{Z})=0,{\quad
}h(\widehat{{D}}{_X}\widehat{R}(\widehat{U},\widehat{Y},v\widehat{Z})=0,
$$
the identities (12.22) become
$$
\sum [{h(}\widehat{{D}}{_X}\widehat{{T}}{)(}\widehat{{Y}}{,}\widehat{{Z}}{)-h%
}\widehat{{R}}{(}\widehat{{X}}{,}\widehat{{Y}}{)}\widehat{{Z}}+\eqno(12.23)
$$
$$
{h}\widehat{{T}}{(h}\widehat{{T}}{(}\widehat{{X}}{,}\widehat{{Y}}{),}%
\widehat{{Z}}{)+h}\widehat{{T}}{(v}\widehat{{T}}{(}\widehat{{X}}{,}\widehat{{%
Y}}{),}\widehat{{Z}}{)]}=0,
$$
$$
\sum {[v(}\widehat{{D}}{_X}\widehat{{T}}{)(}\widehat{{Y}}{,}\widehat{{Z}}{)-v%
}\widehat{{R}}{(}\widehat{{X}}{,}\widehat{{Y}}{)}\widehat{{Z}}{+}
$$
$$
{v}\widehat{{T}}{(h}\widehat{{T}}{(}\widehat{{X}}{,}\widehat{{Y}}{),}%
\widehat{{Z}}{)+v}\widehat{{T}}{(v}\widehat{{T}}{(}\widehat{{X}}{,}\widehat{{%
Y}}{),}\widehat{{Z}}{)]}=0,
$$
$$
\sum {[h(}\widehat{{D}}{_X}\widehat{{R}}{)(}\widehat{{U}}{,}\widehat{{Y}}{,}%
\widehat{{Z}}{)+h}\widehat{{R}}{(h}\widehat{{T}}{(}\widehat{{X}}{,}\widehat{{%
Y}}{),}\widehat{{Z}}{)}\widehat{{U}}{+h}\widehat{{R}}{(v}\widehat{{T}}{(}%
\widehat{{X}}{,}\widehat{{Y}}{),}\widehat{{Z}}{)}\widehat{{U}}{]}=0,
$$
$$
\sum {[v(}\widehat{{D}}{_X}\widehat{{R}}{)(}\widehat{{U}}{,}\widehat{{Y}}{,}%
\widehat{{Z}}{)+v}\widehat{{R}}{(h}\widehat{{T}}{(}\widehat{{X}}{,}\widehat{{%
Y}}{),}\widehat{{Z}}{)}\widehat{{U}}{+v}\widehat{{R}}{(v}\widehat{{T}}{(}%
\widehat{{X}}{,}\widehat{{Y}}{),}\widehat{{Z}}{)}\widehat{{U}}{]}=0.
$$
The local adapted form of these identities is obtained by
inserting in (12.23)
the necessary values of triples $(\widehat{X},\widehat{Y},\widehat{Z})$,($=(%
\widehat{{\delta }}{_i},\widehat{{\delta }}{_k},\widehat{{\delta
}}{_l}),$
or $(\widehat{{\partial }}{_d},\widehat{{\partial }}{_c},\widehat{{\partial }%
}{_b}),$) and putting successively $\widehat{U}=\widehat{{\delta }}_h$ and $%
\widehat{U}=\widehat{{\partial }}_a.$ Taking into account
(12.3),(12.4) and (12.23) we obtain:
$$
\sum [\widehat{{T}}{_{jk|h}^i+}\widehat{{T}}{^m}_{jk}\widehat{{T}}{^j}_{hm}+%
\widehat{{R}}{^a}_{jk}\widehat{{C}}{^i}_{ha}-\widehat{{R}}{{_j}^i}_{kh}]=0,%
\eqno(12.24)
$$
$$
\sum [\widehat{{R}}{{^a}_{jk{\mid h}}}+\widehat{{T}}{^m}_{jk}\widehat{{R}}{^a%
}_{hm}+\widehat{{R}}{^b}_{jk}\widehat{{P}}{^a}_{hb}]=0,
$$
$$
\widehat{{C}}{^i}_{jb{\mid }k}-\widehat{{C}}{^i}_{kb{\mid }j}-\widehat{{T}}{%
^i}_{jk{\mid }b}+\widehat{{C}}{^m}_{jb}\widehat{{T}}{^i}_{km}-\widehat{C}{^m}%
_{kb}\widehat{{T}}{^i}_{jm}+\widehat{{T}}{^m}_{jk}\widehat{{C}}{^i}_{mb}+
$$
$$
\widehat{{P}}{^d}_{jb}\widehat{{C}}{^i}_{kd}-\widehat{{P}}{^d}_{kb}\widehat{{%
C}}{^i}_{jd}+\widehat{{P}}{{_j}^i}_{kb}-\widehat{{P}}{{_k}^i}_{jb}=0,
$$
$$
\widehat{{P}}{^a}_{jb{\mid }k}-\widehat{{P}}{^a}_{kb{\mid }j}-\widehat{{R}}{%
^a}_{jk\perp b}+\widehat{{C}}{^m}_{jb}\widehat{{R}}{^a}_{km}-\widehat{{C}}{^m%
}_{kb}\widehat{{R}}{^a}_{jm}+
$$
$$
\widehat{{T}}{^m}_{jk}\widehat{{P}}{^a}_{mb}+\widehat{{P}}{^d}_{db}\widehat{{%
P}}{^a}_{kd}-\widehat{{P}}{^d}_{kb}\widehat{{P}}{^a}_{jd}-\widehat{{R}}{{^d}%
_{jk}}\widehat{{S}}{{^a}_{bd}}+\widehat{{R}}{_{b\cdot jk}^{\cdot
a}}=0,
$$
$$
\widehat{{C}}{^i}_{jb\perp c}-\widehat{{C}}{^i}_{jc\perp b}+\widehat{{C}}{^m}%
_{jc}\widehat{{C}}{^i}_{mb}-
$$
$$
\widehat{{C}}{^m}_{jb}\widehat{{C}}{^i}_{mc}+\widehat{{S}}{^d}_{bc}\widehat{{%
C}}{^i}_{jd}-\widehat{{S}}{_{j\cdot bc}^{\cdot i}}=0,
$$
$$
\widehat{{P}}{^a}_{jb\perp c}-\widehat{{P}}{^a}_{jc\perp b}+\widehat{{S}}{^a}%
_{bc\mid j}+\widehat{{C}}{^m}_{jc}\widehat{{P}}{^a}_{mb}-\widehat{{C}}{^m}%
_{jb}\widehat{{P}}{^a}_{mc}+
$$
$$
\widehat{{P}}{^d}_{jb}\widehat{{S}}{^a}_{cd}-\widehat{{P}}{^d}_{jc}\widehat{{%
S}}{^a}_{bd}+\widehat{{S}}{^d}_{bc}\widehat{{P}}{^a}_{jd}+\widehat{{P}}{{_b}%
^a}_{jc}-\widehat{{P}}{{_c}^a}_{jb}=0,
$$
$$
\sum [\widehat{{S}}{^a}_{bc\perp d}+\widehat{{S}}{^f}_{bc}\widehat{{S}}{^a}%
_{df}-\widehat{{S}}{{_f}^a}_{cd}]=0,
$$
$$
\sum [\widehat{{R}}{{_k}^i}_{hj\mid l}-\widehat{{T}}{^m}_{hj}\widehat{{R}}{{%
_k}^i}_{lm}-\widehat{{R}}{{^a}_{hj}}\widehat{{P}}{_{k\cdot la}^{\cdot i}}%
]=0,
$$
$$
\sum [\widehat{{R}}{_{d\cdot hj\mid l}^{\cdot a}}-\widehat{{T}}{{^m}_{hj}}%
\widehat{{R}}{_{d\cdot lm}^{\cdot a}}-\widehat{{R}}{{^c}_{hj}}\widehat{{P}}{{%
{_d}^a}_{lc}}]=0,
$$
$$
\widehat{{P}}{_{k\cdot jd\mid l}^{\cdot
i}}-\widehat{{P}}{_{k\cdot ld\mid
j}^{\cdot i}}+{{R_k}^i}_{lj\perp d}+{C^m}_{ld}{{R_k}^i}_{jm}-{C^m}_{jd}{{R_k}%
^i}_{lm}-
$$
$$
\widehat{{T}}{^m}_{jl}\widehat{{P}}{_{k\cdot md}^{\cdot i}}+\widehat{{P}}{^a}%
_{ld}\widehat{{P}}{_{k\cdot jl}^{\cdot i}}-\widehat{{P}}{{^a}_{jd}}\widehat{{%
P}}{_{k\cdot la}^{\cdot
i}}-\widehat{{R}}{{^a}_{jl}}\widehat{{S}}{_{k\cdot ad}^{\cdot
i}}=0,
$$
$$
\widehat{{P}}{{_c}^a}_{jd\mid l}-\widehat{{P}}{{_c}^a}_{ld\mid j}+\widehat{{R%
}}{_{c\cdot lj\mid d}^{\cdot a}}+\widehat{{C}}{^m}_{ld}\widehat{{R}}{{_c}^a}%
_{jm}-\widehat{{C}}{^m}_{jd}\widehat{{R}}{{_c}^a}_{lm}-
$$
$$
\widehat{{T}}{^m}_{jl}\widehat{{P}}{{_c}^a}_{md}+\widehat{{P}}{^f}_{ld}%
\widehat{{P}}{{_c}^a}_{jf}-\widehat{{P}}{^f}_{jd}\widehat{{P}}{{_c}^a}_{lf}-%
\widehat{{R}}{^f}_{jl}\widehat{{S}}{{_c}^a}_{fd}=0,
$$
$$
\widehat{{P}}{_{k\cdot jd\perp c}^{\cdot
i}}-\widehat{{P}}{_{k\cdot jc\perp
d}^{\cdot i}}+\widehat{{S}}{{_k}^i}_{dc\mid j}+\widehat{{C}}{^m}_{jd}%
\widehat{{P}}{_{k\cdot mc}^{\cdot i}}-\widehat{{C}}{^m}_{jc}\widehat{{P}}{%
_{k\cdot md}^{\cdot i}}+
$$
$$
\widehat{{P}}{^a}_{jc}\widehat{{S}}{_{k\cdot da}^{\cdot i}}-\widehat{{P}}{^a}%
_{jd}\widehat{{S}}{_{k\cdot ca}^{\cdot i}}+\widehat{{S}}{^a}_{cd}\widehat{{P}%
}{_{k\cdot ja}^{\cdot i}}=0,
$$
$$
\widehat{{P}}{{_b}^a}_{jd\perp c}-\widehat{{P}}{{_b}^a}_{jc\perp d}+\widehat{%
{S}}{{_b}^a}_{cd\mid
j}+\widehat{{C}}{^m}_{jd}\widehat{{P}}{{_b}^a}_{mc}-
$$
$$
\widehat{{C}}{^m}_{jc}\widehat{{P}}{{_b}^a}_{md}+\widehat{{P}}{^f}_{jc}%
\widehat{{S}}{{_b}^a}_{df}-\widehat{{P}}{^f}_{jd}\widehat{{S}}{{_b}^a}_{cf}+%
\widehat{{S}}{^f}_{cd}\widehat{{P}}{{_b}^a}_{jf}=0,
$$
$$
\sum_{[b,c,d]}\widehat{{S}}{{{_k}^i}_{bc\perp d}-}\widehat{{S}}{{^a}_{bc}}%
\widehat{{S}}{{{{_k}^i}_{da}}}
$$
$$
\sum_{[b,c,d]}{[}\widehat{{S}}{{{_f}^a}_{bc\perp d}-}\widehat{{S}}{{^e}_{bc}}%
\widehat{{S}}{{{{_f}^a}_{de}}]}=0,
$$
where, for instance, ${\sum_{[b,c,d]}}$ means the cyclic sum over indices $%
b,c,d.$

Identities (12.24) are isotopic generalizations of the
corresponding formulas
presented in 
 [160,161], or equivalently, an extension of
 Santilli's  
 [223] formulas to the case of d--isoconnections.

As a consequence of a corresponding rearrangement of (12.23) we
obtain the {\bf isoricci identities} (for simplicity we establish
them only for distinguished vector isofields, although they may
be written for every distinguished tensor isofield):
$$
\widehat{{D}}{_{[X}^{(h)}}\widehat{{D}}{_{Y\}}^{(h)}}h\widehat{Z}=\widehat{R}%
(h\widehat{X},h\widehat{Y})h\widehat{Z}+\widehat{{D}}{_{[hX,hY\}}^{(h)}}h%
\widehat{Z}+\widehat{{D}}{_{[hX,hY\}}^{(v)}}h\widehat{Z},\eqno(12.25)
$$
$$
\widehat{{D}}{_{[X}^{(v)}}{D_{Y\}}^{(h)}}h\widehat{Z}=\widehat{R}(v\widehat{X%
},h\widehat{Y})h\widehat{Z}+\widehat{{D}}{_{[vX,hY\}}^{(h)}}h\widehat{Z}+%
\widehat{{D}}{_{[vX,hY\}}^{(v)}}h\widehat{Z},
$$
$$
\widehat{{D}}{_{[X}^{(v)}}\widehat{{D}}{_{Y\}}^{(v)}}h\widehat{Z}=\widehat{R}%
(v\widehat{X},v\widehat{Y})h\widehat{Z}+\widehat{{D}}{_{[vX,vY\}}^{(v)}}h%
\widehat{Z}
$$
and
$$
\widehat{{D}}{_{[X}^{(h)}}\widehat{{D}}{_{Y\}}^{(h)}}v\widehat{Z}=\widehat{R}%
(h\widehat{X},h\widehat{Y})v\widehat{Z}+{D_{[hX,hY\}}^{(h)}}v\widehat{Z}+{%
D_{[hX,hY\}}^{(v)}}v\widehat{Z},\eqno(12.26)
$$
$$
\widehat{{D}}{_{[X}^{(v)}}\widehat{{D}}{_{Y\}}^{(h)}}v\widehat{Z}=\widehat{R}%
(v\widehat{X},h\widehat{Y})v\widehat{Z}+\widehat{{D}}{_{[vX,hY\}}^{(v)}}v%
\widehat{Z}+\widehat{{D}}{_{[vX,hY\}}^{(v)}}v\widehat{Z},
$$
$$
\widehat{{D}}{_{[X}^{(v)}}\widehat{{D}}{_{Y\}}^{(v)}}v\widehat{Z}=\widehat{R}%
(v\widehat{X},v\widehat{Y})v\widehat{Z}+\widehat{{D}}{_{[vX,vY\}}^{(v)}}v%
\widehat{Z}.
$$
For $\widehat{X}=\widehat{{X}}{^i}(\widehat{u}){\frac{\widehat{\delta }}{%
\delta
\widehat{x}^i}}+\widehat{{X}}{^a}(\widehat{u})\frac{\widehat{\partial
}}{\partial \widehat{x}^a}$ and (12.3),(12.4),(12.18) and (12.21)
we can express respectively identities (12.25) and (12.26) in
this form:
$$
\widehat{{X}}{^a}_{\mid k\mid l}-\widehat{{X}}{^a}_{\mid l\mid k}=\widehat{{R%
}}{{{_B}^a}_{kl}}\widehat{{X}}{^b}-\widehat{{T}}{^h}_{kl}\widehat{{X}}{^a}%
_{\mid h}-\widehat{{R}}{^b}_{kl}\widehat{{X}}{^a}_{\perp b},
$$
$$
\widehat{{X}}{^i}_{\mid k\perp d}-\widehat{{X}}{^i}_{\perp d\mid k}=\widehat{%
{P}}{_{h\cdot kd}^{\cdot i}}\widehat{{X}}{^h}-\widehat{{C}}{^h}_{kd}\widehat{%
{X}}{^i}_{\mid h}-\widehat{{P}}{^a}_{kd}\widehat{{X}}{^i}_{\perp
a},
$$
$$
\widehat{{X}}{^i}_{\perp b\perp c}-\widehat{{X}}{^i}_{\perp c\perp b}=%
\widehat{{S}}{_{h\cdot bc}^{\cdot i}}\widehat{{X}}{^h}-\widehat{{S}}{^a}_{bc}%
\widehat{{X}}{^i}_{\perp a}
$$
and
$$
\widehat{{X}}{^a}_{\mid k\mid l}-\widehat{{X}}{^a}_{\mid l\mid k}=\widehat{{R%
}}{{_b}^a}_{kl}\widehat{{X}}{^b}-\widehat{{T}}{^h}_{kl}\widehat{{X}}{^a}%
_{\mid h}-\widehat{{R}}{^b}_{kl}\widehat{{X}}{^a}_{\perp b},
$$
$$
\widehat{{X}}{^a}_{\mid k\perp b}-\widehat{{X}}{^a}_{\perp b\mid k}=\widehat{%
{P}}{{_b}^a}_{kc}\widehat{{X}}{^c}-\widehat{{C}}{^h}_{kb}\widehat{{X}}{^a}%
_{\mid h}-\widehat{{P}}{^d}_{kb}\widehat{{X}}{^a}_{\perp d},
$$
$$
\widehat{{X}}{^a}_{\perp b\perp c}-\widehat{{X}}{^a}_{\perp c\perp b}=%
\widehat{{S}}{{_d}^a}_{bc}\widehat{{X}}{^d}-\widehat{{S}}{^d}_{bc}\widehat{{X%
}}{^a}_{\perp d}.
$$

For some considerations it is useful to use an alternative way of
definition isotorsion (12.16) and isocurvature (12.19) by using
the commutator
$$
\widehat{\Delta }_{\alpha \beta }\doteq \widehat{\nabla }_\alpha \widehat{%
\nabla }_\beta -\widehat{\nabla }_\beta \widehat{\nabla }_\alpha =2\widehat{%
\nabla }_{[\alpha }\widehat{\nabla }_{\beta ]}.\eqno(12.27)
$$
For components (12.27) of d--isotorsion we have
$$
\widehat{\Delta }_{\alpha \beta }\widehat{f}=\widehat{T}_{.\alpha
\beta }^\gamma \widehat{\nabla }_\gamma \widehat{f}
$$
for every scalar function $\widehat{f}\,$ on $\widehat{\xi }.$
Curvature can
be introduced as an operator acting on arbitrary d--isovector $\widehat{V}%
^\delta :$
$$
(\widehat{\Delta }_{\alpha \beta }-\widehat{T}_{.\alpha \beta
}^\gamma \widehat{\nabla }_\gamma )\widehat{V}^\delta
=\widehat{R}_{~\gamma .\alpha \beta }^{.\delta }\widehat{V}^\gamma
$$
(in this work we are following conventions similar to Miron and
Anastasiei
 [160,161] on d--isotensors; we can obtain corresponding Penrose
and Rindler abstract index formulas 
 [180,181,182] just for a trivial
N--connection structure and by changing denotations for
components of isotorsion and isocurvature in this manner:\
$T_{.\alpha \beta }^\gamma \rightarrow T_{\alpha \beta }^{\quad
\gamma }$ and $R_{~\gamma .\alpha \beta }^{.\delta }\rightarrow
R_{\alpha \beta \gamma }^{\qquad \delta }).$

\section{Structure Equations of a  d--Isocon\-nec\-ti\-on}

Let us, for instance, consider d--tensor isofield:
$$
\widehat{t}=\widehat{{t}}{_a^i}\widehat{{\delta }}_I{\otimes }\widehat{{%
\delta }}{^a}.
$$
We introduce the so--called d--connection 1--forms ${\omega }_j^i$ and ${{%
\tilde \omega }_b^a}$ as%
$$
\widehat{D}\widehat{t}=(\widehat{D}\widehat{{t}}{_a^i})\widehat{{\delta }}_I{%
\otimes }\widehat{{\delta }}^a
$$
with
$$
\widehat{D}\widehat{t}_a^i=\widehat{d}\widehat{t}_a^i+{\omega }_j^i\widehat{{%
t}}{_a^j}-{{\tilde \omega }_a^b}\widehat{{t}}{_b^i}=\widehat{t}_{a\mid j}^i%
\widehat{{d}}\widehat{{x}}{^j}+\widehat{t}_{a\perp b}^I\widehat{{\delta }}%
\widehat{y}^b.
$$
For the d--isoconnection 1--forms of a
 d--isoconnection $\widehat{D}$ on $\widehat{%
\xi }$ defined by ${{\omega }_j^i}$ and ${{\tilde \omega }_b^a}$
one holds
the following {\bf structure isoequations: }%
$$
d(\widehat{{d}}^i)-\widehat{{d}}{^h}\wedge {\omega }_h^i=-\widehat{{\Omega }}%
,
$$
$$
d{(}\widehat{{\delta }}{{^a})}-\widehat{{\delta }}{^a}\wedge {\omega _b^a}=-%
\widehat{{\Omega }}{^a},
$$
$$
d{{\omega }_j^i}-{{\omega }_j^h}\wedge {{\omega }_h^i}=-\widehat{{\Omega }}{%
_j^i},
$$
$$
d{\omega _b^a}-{\omega _b^c}\wedge {\omega
_c^a}=-\widehat{{\Omega }}{_b^a},
$$
in which the isotorsion 2--forms $\widehat{{\Omega }}^i$ and $\widehat{{%
\Omega }}{^i}$ are given respectively by formulas:
$$
\widehat{{\Omega }}{^i}={\frac 12}\widehat{{T}}{^i}_{jk}\widehat{{d}}{^j}%
\wedge \widehat{{d}}{^k}+{\frac 12}\widehat{{C}}{^i}_{jk}\widehat{{d}}{^j}%
\wedge \widehat{{\delta }}{^c},
$$
$$
\widehat{{\Omega }}{^a}={\frac 12}\widehat{{R}}{^a}_{jk}\widehat{{d^j}}%
\wedge \widehat{{d}}{^k}+{\frac 12}\widehat{{P}}{^a}_{jc}\widehat{{d}}{^j}%
\wedge \widehat{{\delta }}{^c}+{\frac 12}\widehat{{S}}{^s}_{bc}\widehat{{%
\delta }}{^b}\wedge \widehat{{\delta }}{^c},
$$
and
$$
\widehat{{\Omega }}{_j^i}={\frac 12}\widehat{{R}}{{_j}^i}_{kh}\widehat{{d}}{%
^k}\wedge \widehat{{d}}{^h}+{\frac 12}\widehat{P}{_{j\cdot kc}^{\cdot i}}%
\widehat{{d}}{^k}\wedge \widehat{{\delta }}{^c}+{\frac 12}\widehat{S}{%
_{j\cdot kc}^{\cdot i}}\widehat{{\delta }}{^b}\wedge \widehat{{\delta }}{^c}%
,
$$
$$
\widehat{\Omega }{_b^a}={\frac 12}\widehat{{R}}{_{b\cdot kh}^{\cdot a}}%
\widehat{{d}}{^k}\wedge \widehat{{d}}{^h}+\frac 12\widehat{P}_{b.kc}^{.a}%
\widehat{d}^k\wedge \widehat{\delta }^c+\frac 12\widehat{S}_{b.cd}^{.a}%
\widehat{{\delta }}{^c}\wedge \widehat{{\delta }}{^d}
$$
The just presented formulas are very similar to those for usual
locally
anisotropic spaces 
 [160,161] but in our case they are written
for isotopic values and generalize the isoriemannian
Santilli's formulas 
 [223].

\chapter{The Isogeometry of Tangent Iso\-bund\-les}

The aim of this Chapter is to formulate some results in the
isogeometry of tangent isobundle, t--isobundle, $\widehat{TM}$
and to use them in order to develop the geometry of Finsler and
Lagrange isospaces.

\section{\qquad Notions of Isotopies of Generalized Lagrange and
Finsler Spaces} All results presented in the preceding Chapter
on v--isobundles provided with N--isoconnection, d--isoconnection
and isometric
 structures hold
good for $\widehat{TM}.$ In this case the dimension of the base
isospace and of typical isofibre coincides and we can write
locally, for instance, isovectors as
$$
\widehat{X}=\widehat{{X}}{^i}\widehat{{\delta }}_i+\widehat{{Y}}{^i}\widehat{%
{\partial }}_i=\widehat{X}^i\widehat{{\delta }}_i+\widehat{Y}^{(i)}\widehat{{%
\partial }}_{(i)},
$$
where $\widehat{u}^\alpha =(\widehat{x}^i,\widehat{y}^j)=(\widehat{x}^i,%
\widehat{y}^{(j)}).$

On t-isobundles we can define a global map
$$
\widehat{J}:{\cal X}\left( \widehat{TM}\right) \to {\cal X}\left( \widehat{TM%
}\right) \eqno(13.1)
$$
which does not depend on N--isoconnection structure:
$$
\widehat{J}({\frac{\widehat{\delta }}{\delta \widehat{x}^i}})={\frac{%
\widehat{\partial }}{\partial \widehat{y}^i}}
$$
and%
$$
\widehat{J}({\frac{\widehat{\partial }}{\partial
\widehat{y}^i}})=0.
$$
This endomorphism is called the {\bf natural (or canonical)
almost tangent isostructure} on $\widehat{TM}$; it has the
properties:
$$
1)\widehat{J}^2=0,{\quad
}2)Im\widehat{J}=Ker\widehat{J}=V\widehat{TM}
$$
and 3) the {\bf Nigenhuis isotensor,}
$$
{N_J}(\widehat{X},\widehat{Y})=[J\widehat{X},J\widehat{Y}\}-J[J\widehat{X},%
\widehat{Y}\}-J[\widehat{X},J\widehat{Y}] \
(\widehat{X},\widehat{Y}\in {\cal X}\left( \widehat{TM}\right) )
$$
identically vanishes, i.e. the natural almost tangent isostructure $J$ on $%
\widehat{TM}$ is isointegrable.

Let $\widehat{M}$ be a isosmooth $(2n)$--dimensional isomanifold and $(%
\widehat{TM},{\tau },\widehat{M})$ its t--isobundle. For
isospaces we
 define a {\bf generalized isolagrange space,} GIL--space,  as a pair ${G}%
\widehat{{L}}^{n,m}=(\widehat{M},\widehat{g}_{ij}(\widehat{x},\widehat{y}))$%
, where $\widehat{g}_{ij}(\widehat{x},\widehat{y})$ is a
d--tensor isofield on ${\tilde TM}=\widehat{TM}-\{0\},$ of
isorank $(2n),$ and is called as the fundamental d--isotensor, or
metric d--isotensor, of GIL--space.

Let denote as a normal d--isoconnection that defined by using $N$
and being
adapted to the almost tangent isostructure (6.1) as $\widehat{D}{\Gamma }=(%
\widehat{{L}}{^a}_{jk},\widehat{{C}}{^a}_{jk}).$ This
d--isoconnection is
compatible with isometric $\widehat{g}_{ij}(\widehat{x},\widehat{y})$ if $%
\widehat{g}_{ij\mid k}=0$ and $\widehat{g}_{ij\perp k}=0.$

There exists an unique d--isoconnection $C\widehat{\Gamma }(N)$
which is compatible with $\widehat{g}_{ij}{(}\widehat{{u}}{)}$
and has vanishing isotorsions $\widehat{{T}}{^i}_{jk}$ and
$\widehat{{S}}{^i}_{jk}$ (see formulas (5.4) rewritten for
t--isobundles). This isoconnection, depending only
on $\widehat{g}_{ij}{(}\widehat{{u}}{)}$ and $\widehat{{N}}_j^i{(}\widehat{{u%
}}{)}$ is called the canonical metric d--isoconnection of
GIL--space. It has coefficients
$$
\widehat{{L}}{^i}_{jk}={\frac 12}\widehat{{g}}{^{ih}}(\widehat{{\delta }}_j%
\widehat{{g}}{_{hk}}+\widehat{{\delta }}_h\widehat{{g}}{_{jk}}-\widehat{{%
\delta }}_h\widehat{{g}}{_{jk}}),\eqno(13.2)
$$
$$
\widehat{{C}}{^i}_{jk}={\frac 12}\widehat{{g}}{^{ih}}(\widehat{{\partial }}_j%
\widehat{{g}}{_{hk}}+\widehat{{\partial }}_h\widehat{{g}}{_{jk}}-\widehat{{%
\partial }}_h\widehat{{g}}{_{jk}}).
$$
Of course, metric d--isoconnections different from
$C\widehat{\Gamma }(N)$ may
be found. For instance, there is a unique normal d--isoconnection $\widehat{D}%
\Gamma (N)=({\bar L}_{\cdot jk}^i,{\bar C}_{\cdot jk}^i)$ which
is metric
and has a priori given isotorsions $\widehat{{T}}{^i}_{jk}$ and $\widehat{{S}%
}{^i}_{jk}.$ The coefficients of $\widehat{D}\Gamma (N)$ are the
following ones:
$$
{\bar L}_{\cdot jk}^i=\widehat{{L}}{^i}_{jk}-\frac 12\widehat{g}^{ih}(%
\widehat{g}_{jr}\widehat{{T}}{^r}_{hk}+\widehat{g}_{kr}\widehat{{T}}{^r}%
_{hj}-\widehat{g}_{hr}\widehat{{T}}{^r}_{kj}),
$$
$$
{\bar C}_{\cdot jk}^i=\widehat{{C}}{^i}_{jk}-\frac 12\widehat{g}^{ih}(%
\widehat{g}_{jr}\widehat{{S}}{^r}_{hk}+\widehat{g}_{kr}\widehat{{S}}{^r}%
_{hj}-\widehat{g}_{hr}\widehat{{S}}{^r}_{kj}),
$$
where $\widehat{{L}}{^i}_{jk}$ and $\widehat{{C}}{^i}_{jk}$ are
the same as for the $C\widehat{\Gamma }(N)$--isoconnection (13.2).

The Lagrange spaces were introduced in order to geometrize the
concept of Lagrangian in mechanics (the Lagrange geometry is
studied in details, see also basic references, in Miron and
Anastasiei
 [160,161]). For isospaces we present
this generalization:

A {\bf isolagrange space}, IL--space, $\widehat{L}^n=(\widehat{M},\widehat{%
g}_{ij}),$ is defined as a particular case of GIL--space when the
d--isometric on $\widehat{M}$ can be expressed as
$$
\widehat{g}_{ij}{(}\widehat{{u}}{)}={\frac 12}{\frac{\widehat{{\partial }}^2%
{\cal L}}{{\partial }\widehat{{y}}{{^i}{\partial }}\widehat{{y}}{^j}}},%
\eqno(13.3)
$$
where ${\cal L}:\widehat{TM}\to \widehat{\Lambda },$ is a
isodifferentiable function called a iso--Lagrangian on
$\widehat{M}$.

Now we consider the isotopic extension of the Finsler space:

A {\bf isofinsler isometric} on $\widehat{M}$ is a function $F_S:\widehat{TM}%
\to \widehat{\Lambda }$ having the properties:

1. The restriction of $F_S$ to ${\tilde
{TM}}=\widehat{TM}\setminus \{0\}$ is of the class $G^\infty $
and F is only isosmooth on the image of the null cross--section
in the t--isobundle to $\widehat{M}$.

2. The restriction of $\widehat{F}$ to ${\tilde {TM}}$ is
positively
homogeneous of degree 1 with respect to ${(}\widehat{{y}}{^i)}$, i.e. $%
\widehat{F}(\widehat{x},\widehat{{\lambda }}\widehat{y})=\widehat{{\lambda }}%
\widehat{F}(\widehat{x},\widehat{y}),$ where $\widehat{{\lambda
}}$ is a real positive number.

3. The restriction of $\widehat{F}$ to the even subspace of
$\tilde {TM}$ is a positive function.

4. The quadratic form on ${\Lambda }^n$ with the coefficients
$$
\widehat{g}_{ij}{(}\widehat{{u}}{)}={\frac 12}{\frac{\widehat{{\partial }}^2%
\widehat{F}^2}{{\partial }\widehat{{y}}{{^i}{\partial }}\widehat{{y}}{^j}}}%
\eqno(13.4)
$$
defined on $\tilde {TM}$ is nondegenerate.

A pair $\widehat{F}^n=(\widehat{M},\widehat{F})$ which consists
from a continuous isomanifold $\widehat{M}$ and a isofinsler
isometric is called a {\bf isofinsler space}, IF--space.

It's obvious that IF--spaces form a particular class of
IL--spaces with iso-Lagrangian ${\cal L}=\widehat{{F}}{^2}$ and a
particular class of GIL--spaces with metrics of type (13.4).

For a IF--space we can introduce the isotopic variant of
nonlinear Cartan
connection 
 [160,161]:
$$
\widehat{N}_j^i{(}\widehat{{x}}{,}\widehat{{y}}{)}={\frac{\widehat{\partial }%
}{\partial \widehat{y}^j}}\widehat{G}^{*I},
$$
where
$$
\widehat{G}^{*i}={\frac 14}\widehat{g}^{*ij}({\frac{\widehat{{\partial }}^2{%
\varepsilon }}{{\partial }\widehat{{y}}{^i}{\partial }\widehat{{x}}{^k}}}%
\widehat{{y}}{^k}-{\frac{\widehat{\partial }{\varepsilon
}}{\partial
\widehat{x}^j}}),{\quad }{\varepsilon }{(}\widehat{{u}}{)}=\widehat{g}_{ij}{(%
}\widehat{{u}}{)}\widehat{y}^i\widehat{y}^j,
$$
and $\widehat{g}^{*ij}$ is inverse to $\widehat{g}_{ij}^{*}{(}\widehat{{u}}{)%
}={\frac 12}{\frac{\widehat{{\partial }}^2\varepsilon }{{\partial }\widehat{{%
y}}{{^i}{\partial }}\widehat{{y}}{^j}}}.$ In this case the
coefficients of canonical metric d--isoconnection (13.2) gives
the isotopic variants of coefficients of the Cartan connection of
Finsler spaces. A similar remark applies to the isolagrange
spaces.

\section{The Isotopic Almost Hermitian Model of  GIL--Spaces}

Consider a GIL--space endowed with the canonical metric d--isoconnection $C%
\widehat{\Gamma }(N).$ Let $\widehat{{\delta }}_\alpha =(\widehat{{\delta }}%
_\alpha ,\widehat{{\dot \partial }}_I)$ be a usual adapted frame
(13.2) on TM and $\widehat{{\delta }}^\alpha =(\widehat{{\partial
}}^I,\widehat{{\dot \delta }}^I)$ its dual, see (13.3). The
linear operator
$$
\widehat{F}:\Xi ({\tilde {TM}})\to \Xi ({\tilde {TM}}),
$$
acting on $\widehat{{\delta }}_\alpha $ by $\widehat{F}(\widehat{{\delta }}%
_i)=-\widehat{{\partial }}_i,\widehat{F}(\widehat{{\dot \partial }}_i)=%
\widehat{{\delta }}_i,$ defines an almost complex isostructure on ${T}\widehat{{%
M}}.$ We shall obtain a complex isostructure if and only if the
even component of the horizontal distribution $\widehat{N}$ is
integrable. For isospaces, in general with even and odd
components, we write the isotopic almost Hermitian property
(almost Hermitian isostructure) as
$$
\widehat{{F}}{_\beta ^\alpha }\widehat{{F}}{_\delta ^\beta }=-{{\delta }%
_\beta ^\alpha }.
$$

The isometric
$\widehat{g}_{ij}{(}\widehat{{x}}{,}\widehat{{y}}{)}$ on
GIL--spaces induces on $\dot {T\widehat{M}}$ the following
isometric:
$$
\widehat{G}=\widehat{g}_{ij}{(}\widehat{{u}}{)}\widehat{d}\widehat{x}%
^i\otimes \widehat{d}\widehat{x}^j+\widehat{g}_{ij}{(}\widehat{{u}}{)}%
\widehat{{\delta }}\widehat{{y}}{^i}\otimes \widehat{{\delta }}\widehat{{y}}{%
^j}.\eqno(13.5)
$$
We can verify that pair $(\widehat{G},\widehat{F})$ is an almost
Hermitian isostructure on ${\dot {T\widehat{M}}}$ with the
associated supersymmetric 2--form
$$
\widehat{\theta }=\widehat{g}_{ij}{(}\widehat{{x}}{,}\widehat{{y}}{)}%
\widehat{{\delta }}\widehat{{y}}{^i}\wedge
\widehat{d}\widehat{x}^j.
$$

The almost Hermitian isospace $\widehat{H}^{2n}=(T\widehat{M},\widehat{G},%
\widehat{F}),$ provided with a isometric of type (13.5)
 is called the lift on $T%
\widehat{M}$, or the almost Hermitian isomodel, of GIL--space $G\widehat{L}%
^n.$ We say that a linear isoconnection $\widehat{D}$ on ${\dot
{T\widehat{M}}}$ is almost Hermitian isotopic of Lagrange type if
it preserves by parallelism the vertical distribution $V$ and is
compatible with the almost Hermitian isostructure
$(\widehat{G},\widehat{F})$, i.e.
$$
\widehat{D}_X\widehat{G}=0,\quad
\widehat{D}_X\widehat{F}=0,\eqno(13.6)
$$
for every $X\in \widehat{{\cal X}}\left( T\widehat{M}\right) .$

There exists an unique almost Hermitian isoconnection of Lagrange type $%
\widehat{D}^{(c)}$ having h(hh)- and v(vv)--isotorsions equal to
zero. We
can prove (similarly as in 
 [160,161]) that coefficients ${(}%
\widehat{{L}}{{^i}_{jk},}\widehat{{C}}{{^i}_{jk})}$ of
$\widehat{D}^{(c)}$ in the adapted basis $(\widehat{{\delta
}}_i,\widehat{{\dot \delta }}_j)$
are just the coefficients (13.2) of the canonical metric d--isoconnection $C%
\widehat{\Gamma }(N)$ of the GIL--space $G\widehat{L}^n.$
Inversely, we can say that $C\widehat{\Gamma }(N)$--connection
determines on ${\tilde {TM}}$ and isotopic almost Hermitian
connection of Lagrange type with vanishing
h(hh)- and v(vv)-isotorsions. If instead of GIL--space isometric $\widehat{g}%
_{ij}$ in (13.4) the isolagrange (or isofinsler) isometric (13.2)
(or (13.3)) is taken, we obtain the almost Hermitian isomodel of
isolagrange (or isofinsler) isospaces $\widehat{L}^n$ (or
$\widehat{{F}}{^n}).$

We note that the natural compatibility conditions (13.6) for the
isometric
(13.5) and $C\widehat{\Gamma }(N)$--con\-nec\-ti\-ons on $\widehat{H}^{2n}$%
--spaces plays an important role for developing physical models on
la--isospaces. In the case of usual locally anisotropic spaces
geometric constructions and d--covariant calculus are very
similar to those for the Riemann and Einstein--Cartan spaces.
This is exploited for formulation in a selfconsistent manner the
theory of spinors on la--spaces
 [256,255,264]
for introducing a geometric background for locally anisotropic
Yang--Mills
and gauge like gravitational interactions 
 [272,258,259] and for
extending the theory of stochastic processes and diffusion to the
case of locally anisotropic spaces and interactions on such spaces
 [257,262,268].
 In a similar manner we can introduce N--lifts to v- and
t--isobundles in order to investigate isotopic gravitational
la--models.

\chapter{Locally Anisotropic and Inhomogeneous Isogravity}

The conventional Riemannian geometry can be generally assumed to
be exactly valid for the exterior gravitational problem in vacuum
where  bodies can
 be well approximated as being massive points, thus implying the validity
 of conventional and calculus.

On the contrary, there have been serious doubt dating back to E.
Cartan
 on the same exact validity of the Riemannian geometry for interior
gravitational problem because the latter imply internal effects
which are
 arbitrary nonlinear in the velocities and other variables, nonlocal
 integral and of general non--(first)--order Lagrangian type.

Santilli 
 [221,222,224,223]
 constructed his isoriemannian geometry and proposed the related
 isogravitation theory precisely to resolve the latter shortcoming. In
 fact, the  isometric acquires an arbitrary functional; dependence thus
 being   able to   represent directly the locally anisotropic and
 inhomogeneous character of interior gravitational problems.

A remarkable aspect of the latter advances is that they were
achieved by
 preserving the  abstract geometric axioms of the exterior gravitation.
 In fact, exterior and interior gravitation are unified in the above
geometric approach and are merely differentiated by the selected
unit,
 the trivial value $I=diag(1,1,1,1)$ yielding the conventional gravitation
 in vacuum while more general realization of the unit yield interior
 conditions under the same abstract axioms (see ref.
 [129,130] for an independent study).

A number of applications of the isogeometries for interior
problems have
already been  identified, such as (see ref. 
 [224] for an
 outline): the representation of the local variation of the speed of light
 within physical media such as atmospheres or chromospheres; the
 representation of the large difference between cosmological redshift
 between certain quasars and their associated galaxies when physically
 connected according to spectroscopic evidence; the initiation of the
 study of the origin  of the   gravitation via its identification with
 the field originating the mass of elementary  constituents.

As we have shown 
 [269,261,254,270]
the low energy limits of string
 and superstring theories  give  also rise
 to models of (super)field interactions
 with locally anisotropic and even higher order  anisotropic interactions.
The N--connection field can be treated as a corresponding
nonlinear
 gauge field managing the dynamics of ''step by step'' splitting (reduction)
 of higher dimensional spaces to lower dimensional ones. Such (super)string
 induced (super)gravitational models have a generic local anisotropy
 and, in consequence, a more sophisticate  form of field equations and
 conservation laws and of corresponding theirs stochastic and quantum
 modifications. Perhaps similar considerations are in right for isotopic
 versions of sting theories. That it is why we are interested in a study
 of models of isogravity with nonvanishing nonlinear isoconnection,
 distinguished isotorsion and, in general, non--isometric fields.

\section{Generalized Isofinsler gravity}
To begin our presentation let us consider
 a v--isobundle $\widehat{\xi }=(\widehat{E},{\pi },\widehat{M})$
provided with some compatible nonlinear isoconnection
$\widehat{N}$, d--isoconnection $\widehat{D}$ and isometric
$\widehat{G}$ structures.For a locally N--adapted isoframe we
write
$$
\widehat{D}_{({\frac \delta {\delta u^\gamma }})}{\frac{\widehat{\delta }}{%
\delta \widehat{u}^\beta }}=\widehat{{\Gamma }}_{\beta \gamma }^\alpha {%
\frac{\widehat{\delta }}{\delta \widehat{u}^\alpha }},
$$
where the d--isoconnection $\widehat{D}$ has the following
coefficients:
$$
\widehat{{\Gamma }}{{^i}_{jk}}=\widehat{{L}}{{^i}_{jk}},\widehat{{\Gamma }}{{%
^i}_{ja}}=\widehat{{C}}{{^i}_{ja}},\widehat{{\Gamma }}{^i}_{aj}=0,\widehat{{%
\Gamma }}{^i}_{ab}=0,\eqno(14.1)
$$
$$
\widehat{{\Gamma }}{^a}_{jk}=0,\widehat{{\Gamma }}{^a}_{jb}=0,\widehat{{%
\Gamma }}{^a}_{bk}=\widehat{{L}}{^a}_{bk},\widehat{{\Gamma }}{{^a}_{bc}}=%
\widehat{{C}}{{^a}_{bc}}.
$$
The nonholonomy isocoefficients $\widehat{{w}}{{^\gamma }_{\alpha
\beta }}$ are as follows:
$$
\widehat{{w}}{{^k}_{ij}}=0,\widehat{{w}}{{^k}_{aj}}=0,\widehat{{w}}{{^k}_{ia}%
}=0,\widehat{{w}}{{^k}_{ab}}=0,\widehat{{w}}{{^a}_{ij}}=\widehat{{R}}{^a}%
_{ij},
$$
$$
\widehat{{w}}{{^b}_{ai}}=-{\frac{\widehat{\partial }\widehat{{N}}{_a^b}}{%
\partial \widehat{y}^a}},\widehat{{w}}{{^b}_{ia}}={\frac{\widehat{\partial }%
\widehat{{N}}{_a^b}}{\partial
\widehat{y}^a}},\widehat{{w}}{{^c}_{ab}}=0.
$$
By straightforward calculations we can obtain respectively these
components
of isotorsion, ${\cal T}(\widehat{{\delta }}_\gamma ,\widehat{{\delta }}%
_\beta )={{\cal T}_{\cdot \beta \gamma }^\alpha }\widehat{{\delta
}}_\alpha
, $ and isocurvature, ${\cal R}(\widehat{{\delta }}_\beta ,\widehat{{\delta }%
}_\gamma )\widehat{{\delta }}_\tau ={{\cal R}_{\beta \cdot \gamma
\tau }^{\cdot \alpha }}\widehat{{\delta }}{_\alpha },$
d--isotensors:
$$
{\cal T}_{\cdot jk}^i=\widehat{{T}}{^i}_{jk},{\cal T}_{\cdot ja}^I=\widehat{{%
C}}{^i}_{ja},{\cal T}_{\cdot ja}^I=-\widehat{{C}}{^i}_{ja},{\cal
T}_{\cdot ab}^i=0,\eqno(14.2)
$$
$$
{\cal T}_{\cdot ij}^a=\widehat{{R}}{^a}_{ij},{\cal T}_{\cdot ib}^a=-\widehat{%
{P}}{^a}_{bi},{\cal T}_{\cdot bi}^a=\widehat{{P}}{^a}_{bi},{\cal
T}_{\cdot bc}^a=\widehat{{S}}{^a}_{bc}
$$
and
$$
{\cal R}_{i\cdot kl}^{\cdot j}=\widehat{{R}}{{_j}^i}_{kl},{\cal
R}_{b\cdot kl}^{\cdot j}=0,{\cal R}_{j\cdot kl}^{\cdot a}=0,{\cal
R}_{b\cdot kl}^{\cdot a}=\widehat{{R}}_{b\cdot kl}^{\cdot
a},\eqno(14.3)
$$
$$
{\cal R}_{j\cdot kd}^{\cdot i}=\widehat{{P}}{{_j}^i}_{kd},{\cal
R}_{b\cdot kd}^{\cdot a}=0,{\cal R}_{j\cdot kd}^{\cdot a}=0,{\cal
R}_{b\cdot kd}^{\cdot a}=\widehat{{P}}{{_b}^a}_{kd},
$$
$$
{\cal R}_{j\cdot dk}^{\cdot i}=-\widehat{{P}}{{_j}^i}_{kd},{\cal
R}_{b\cdot dk}^{\cdot i}=0,{\cal R}_{j\cdot dk}^{\cdot a}=0,{\cal
R}_{b\cdot dk}^{\cdot h}=-\widehat{{P}}{{_b}^a}_{kd},
$$
$$
{\cal R}_{j\cdot cd}^{\cdot i}=\widehat{{S}}{{_j}^i}_{cd},{\cal
R}_{b\cdot cd}^{\cdot i}=0,{\cal R}_{j\cdot cd}^{\cdot a}=0,
{\cal R}_{b\cdot cd}^{\cdot a}=\widehat{{S}}{{_b}^a}_{cd}
$$
(for explicit dependencies of components of isotorsions and
isocurvatures on components of d--isocon\-nec\-ti\-on see
formulas (12.8) and (12.21)).

The locally adapted components ${\cal R}_{\alpha \beta }={\cal R}ic(D)(%
\widehat{{\delta }}_\alpha ,\widehat{{\delta }}_\beta )$ (we
point that in general on t--isobundles ${\cal R}_{\alpha \beta
}\ne {\cal R}_{\beta \alpha })$ of the {\bf isoricci tensor} are
as follows:
$$
{\cal R}_{ij}=\widehat{{R}}{{_i}^k}_{jk},{\cal R}_{ia}=-{}^{(2)}\widehat{{P}}%
{_{ia}}=-\widehat{{P}}_{i\cdot ka}^{\cdot k}\eqno(14.4)
$$
$$
{\cal R}_{ai}={}\widehat{{P}}{_{ai}}=\widehat{{P}}_{a\cdot ib}^{\cdot b},%
{\cal R}_{ab}=\widehat{{S}}{{_a}^c}_{bc}=\widehat{S}_{ab}.
$$
For scalar curvature, $\overleftarrow{{\ R}}=Sc(\widehat{D})=\widehat{G}%
^{\alpha \beta }\widehat{R}_{\alpha \beta },$ we have
$$
Sc(\widehat{D})=\widehat{R}+\widehat{S},\eqno(14.5)
$$
where $\widehat{R}=\widehat{g}^{ij}{\widehat{R}}_{ij}$ and $\widehat{S}=%
\widehat{h}^{ab}\widehat{S}_{ab}.$

The {\bf isoeinstein--isocartan equations with prescribed
N--iso\-con\-nec\-ti\-on} and h(hh)-- and v(vv)--isotorsions on
v--isobundles (compare with isoeinstein iso\-equa\-ti\-ons
(11.11)) are written as
$$
\widehat{R}^{\alpha \beta }-\frac 12\widehat{g}^{\alpha \beta }(%
\overleftarrow{R}+\widehat{\Theta }-\lambda )={\kappa }_1(\widehat{t}%
^{\alpha \beta }-\widehat{\tau }^{\alpha \beta }),\eqno(14.6)
$$
and
$$
\widehat{T}_{\cdot \beta \gamma }^\alpha +{{G_\beta }^\alpha }\widehat{{T}}{{%
^\tau }_{\gamma \tau }}-{{G_\gamma }^\alpha }\widehat{{T}}{{^\tau
}_{\beta \tau }}={\kappa }_2\widehat{{Q}}{^\alpha }_{\beta \gamma
},\eqno(14.7)
$$
where $\widehat{{Q}}{_{\beta \gamma }^\alpha }$ spin--density of
matter d--isotensors on locally anisotropic and homogeneous
 isospace, ${\kappa }_1$ and ${\kappa }_2$ are the
corresponding interaction constants and ${\lambda }$ is the
cosmological
constant, $\widehat{t}^{\alpha \beta }$ is a source isotensor and $\widehat{%
\tau }^{\alpha \beta }$ is the stress--energy isotensor and there
is satisfied the generalized Freud isoidentity
$$
\widehat{G}_{~\beta }^\alpha -\frac 12\delta _{~\beta }^\alpha (\widehat{%
\Theta }-\lambda )=\widehat{U}_{~\beta }^\alpha +\widehat{\delta
}_\rho \widehat{V}_{\quad \beta }^{\alpha \rho },\eqno(14.8)
$$
where
$$
\widehat{G}_{~\beta }^\alpha =\widehat{R}_{\quad \beta }^\alpha
-\frac 12\delta _{~\beta }^\alpha \overleftarrow{R},
$$
$$
\widehat{U}_{~\beta }^\alpha =-\frac 12\frac{\widehat{\delta }\widehat{%
\Theta }}{\widehat{\partial }(\widehat{D}_\alpha
\widehat{g}^{\gamma \delta })}\widehat{D}_\beta
\widehat{g}^{\gamma \delta }
$$
and
$$
\widehat{V}_{\quad \beta }^{\alpha \rho }=\frac
12[\widehat{g}^{\gamma \delta }\left( \delta _{~\beta }^\alpha
\widehat{\Gamma }_{\alpha \delta }^\rho -\delta _{~\delta
}^\alpha \widehat{\Gamma }_{\alpha \beta }^\rho
\right) +%
$$
$$
\widehat{g}^{\rho \gamma }\widehat{\Gamma }_{\beta \gamma }^\alpha -\widehat{%
g}^{\alpha \gamma }\widehat{\Gamma }_{\beta \gamma }^\rho +\left(
\delta _{~\beta }^\rho \widehat{g}^{\alpha \gamma }-\delta
_{~\beta }^\alpha \widehat{g}^{\rho \gamma }\right)
\widehat{\Gamma }_{\gamma \rho }^\rho ].
$$

By using decompositions (14.1)--(15.5) it is possible an explicit
projection of equations (14.6)--(14.8) into vertical and
horizontal isocomponents (for simplicity we omit such formulas in
this work).

Equations (14.6) constitute the fundamental field equations of
Santilli
 isogravitation  
 [221,222,224,223]
 written in this case for vector isobundles provided with
 compatible N- and d--isoconnection and isometric structures.
The algebraic equations (14.7) have been here added, apparently
 for the first    time for isogravity with isotorsion (see also
 [272,258,259,249,250,251,252,\\ 279,263,275,277,260,265,266,267]
for locally anisotropic gravity and supergravity)
 in order to close the system of
gravitational isofield equations (really we have also to take
into account the system of constraints (14.5) if locally
anisotropic inhomogeneous
 gravitational
isofield is associated to a d--isometric (14.6), or to a
d--isometric (13.5) if the isogravity is modelled on a tangent
isobundle). It should be noted here that the system of
isogravitational field equations (14.8)
 presents a synthesis for
vector isobundles of equations introduced by Anastasiei,
 [9,10]
and
 [160,161], and of equations (11.11) and (11.12) considered in
the Santilli isotheory.

We note that on la--isospaces the divergence
$$
D_\alpha [\widehat{G}_{~\beta }^\alpha -\frac 12\delta _{~\beta }^\alpha (%
\widehat{\Theta }-\lambda )]=\widehat{U}_\beta \eqno(14.9)
$$
does not vanish (this is a consequence of generalized isobianchi
(12.22), or (12.23), and isoricci isoidentities (12.25), or
(12.26)). The problem of nonvanishing of such divergences for
gravitational models on vector bundles provided with nonlinear
connection structures was analyzed in
 [9,10]
and
 [160,161].

The problem of total conservation laws on isospaces has been
studied
in detail in ref. 
 [221] by reformulating  all isospaces
considered in that paper in terms  of the isominkowskian space,
 with consequential elimination of curvature  which permits the
construction of a  universal symmetry and related total
conservation
 laws for all possible isometric.

The latter studies concerning vector isobundles with
 N--isoconnections will be considered in some future works.

We end this subsection by emphasizing that isofield equations of
ty\-pe (14.6)--(14.8) can be similarly introduced for the
particular cases of lo\-cal\-ly anisotropic isospaces with metric
(13.5) on $\tilde {TM}$ with coefficients pa\-ra\-met\-ri\-zed as
for the isolagrange, (13.3), or
 isofinsler, (13.4), isospaces.

\section{Concluding Remarks and Further Possibilities}

One of the most important aspects we attempted to convey in this
Part of
 the monograph is
the possiblity to formulate isotopic variants of extended Finsler
geometry and the application of this isogeometric background in
contemporary theoretical
 and mathematical physics. The approach adopted here provides us an
essentially self--contained, concise and significantly simple
treatment of the material on bundle isospaces enabled with
compatible isotopic nonlinear
 and distinguished isoconnections and isometric structures.

A remarkable features worth recalling is that the considerable
broadening of
 the capabilities of the isotheory via the additional of nonlinear,
 nonlocal and noncanonical effects, is done via the same abstract
 axioms of the conventional formulations.

In this Part we have discussed the basic geometric constructions
for
 isotopic spaces with inhomogeneity and local anisotropy.
 We have computed the distinguished
 isotorsions and isocurvatures. It was shown how to write a manifestly
isotopic model of gravity with locally anisotropic
 and inhomogeneous interactions of isofields.
The assumptions made in deriving the results are similar to those
for
 the geometry of isomanifolds and to the isofield theory.

There are various possible developments of the ideas presented
here. One of the necessary steps is the definition of locally
anisotropic and
 inhomogeneous isotopic spinors
 and explicit constructions of physical models with  isospinor, isogauge and
 isogravitational interactions on locally anisotropic isospaces. The
 problem of formulation of conservation laws on
 locally anisotropic and inhomogeneous isospaces and for
  locally anisotropic and inhomogeneous
isofield interactions presents a substantial intrested for
investigations.
  Here we add
 the theory of isostochastic processes, the supersymmetric extension of
 the concept of isotheory as well possible generalizations of the mentioned
 constructions for higher order anisotropies in string theories. These
tasks remain for future research.

\end{document}